\documentclass[11pt,oneside,english,twoside,reqno]{amsbook}

\usepackage{libertine-type1}

\usepackage[T1]{fontenc}
\usepackage[utf8]{inputenc}
\usepackage{color}
\usepackage{array}
\usepackage{fancybox}
\usepackage{calc}
\usepackage{textcomp}
\usepackage{bbding}
\usepackage{mathrsfs}
\usepackage{amsthm}
\usepackage{amsbsy}
\usepackage{amssymb}
\usepackage{bm}
\usepackage{epsfig}
\usepackage{upgreek}
\usepackage{soul}
\usepackage{graphicx}
\usepackage{setspace}
\usepackage{esint}
\usepackage{standalone}
\usepackage{etoolbox}
\usepackage{epigraph}
\usepackage{geometry}
\usepackage{fancyhdr}
\usepackage{lipsum}
\makeatletter
\patchcmd{\@footnotetext}{\footnotesize}{\small}{}{}
\makeatother
\usepackage{blindtext}
\usepackage{caption}
\DeclareCaptionFormat{myformat}{#1#2#3\hrulefill}
\usepackage[unicode=true,pdfusetitle,
 bookmarks=true,bookmarksnumbered=false,bookmarksopen=false,
 breaklinks=false,pdfborder={0 0 0},backref=false,colorlinks=true]
 {hyperref}
\numberwithin{section}{chapter}
\numberwithin{equation}{section}
\numberwithin{figure}{chapter}
\hypersetup{linkcolor=[rgb]{0,0,0.5}, linktoc=page}
\hypersetup{citecolor=[rgb]{0.5,0,0}}
\hypersetup{urlcolor=[rgb]{0,0.45,0}}

\usepackage[backend=biber,refsegment=chapter, defernumbers=true,style=phys,articletitle=true,citestyle=authoryear-comp,doi=false,url=false]{biblatex}
\AtEveryBibitem{\clearfield{issn}}
\AtEveryBibitem{\clearfield{isbn}}
\AtEveryBibitem{\clearlist{language}}
\AtEveryBibitem{\clearfield{month}}
\AtEveryBibitem{\clearfield{day}}
\renewcommand\headrulewidth{0.5pt}
\def \nn   {\!\!\!\!\!}

\usepackage{enumitem}
\newcommand{\icol}[1]{
  \left[\begin{smallmatrix}#1\end{smallmatrix}\right]%
}

\graphicspath{ {figures/} }
\begin{document}

\pagenumbering{roman}
\thispagestyle{empty}

\begin{spacing}{1.05}

~

\begin{center}
\vspace{-2.35cm}
\includegraphics[scale=0.56]{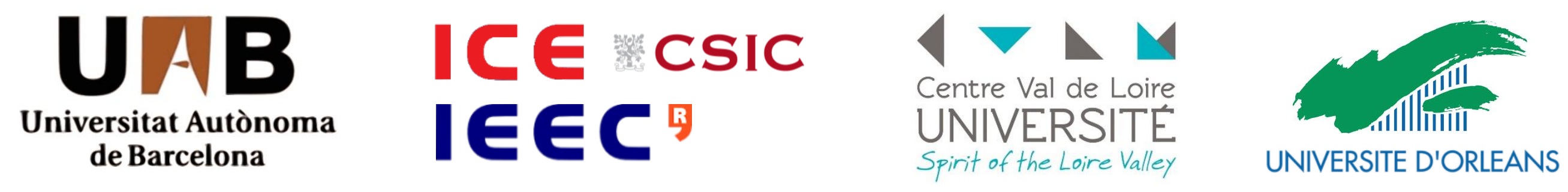}
\par

\vspace{0.3cm}

\textbf{\textit{UNIVERSITAT AUTÒNOMA DE BARCELONA\\
DEPARTAMENT DE FÍSICA, PROGRAMA DE DOCTORAT RD 99/2011}}

\textit{Institut de Ciències de l'Espai (ICE, CSIC)\\
Institut d'Estudis Espacials de Catalunya (IEEC)}

\vspace{0.25cm}

\textbf{\textit{UNIVERSITÉ D'ORLÉANS\\
\scalebox{.95}[1.0]{ÉCOLE DOCTORALE ÉNERGIE, MATÉRIAUX, SCIENCES DE LA TERRE ET DE L’UNIVERS}}}

\textit{Observatoire des Sciences de l'Univers en région Centre (OSUC)\\
Centre National de la Recherche Scientifique (CNRS)\\
Laboratoire de Physique et Chimie de l'Environnement et de l'Espace (LPC2E)}

\vspace{0.45cm}

\textbf{Tesi en cotutela internacional} /\\
\textbf{Thèse en cotutelle internationale}\\
presentada per / présentée par\\
\vspace{0.45cm}
\scalebox{1.35}[1.35]{\textbf{Marius \textsc{Oltean}}}

\vspace{0.45cm}

Defensada el / soutenue le :\\ 
\textbf{24 octubre / octobre 2019}

Per obtenir el grau de / pour obtenir le grade de :\\
\scalebox{0.95}[1.0]{\textbf{Doctor de la Universitat Autònoma de Barcelona \& Docteur de l'Université d'Orléans}}

Disciplina i especialitat / discipline et spécialité :\\ 
\textbf{Física, relativitat general / physique, relativité générale}

\vspace{0.25cm}

\noindent{\fboxrule 0.5pt\fboxsep 4pt\fbox{\parbox[t]{1\textwidth - 2\fboxsep - 2\fboxrule}{%
\begin{spacing}{1.5}
\begin{center}
\scalebox{1.5}[1.5]{\textbf{Study of the Relativistic Dynamics of}}
\par\end{center}
\begin{center}
\scalebox{1.5}[1.5]{\textbf{Extreme-Mass-Ratio Inspirals}}\vspace{-0.2cm}
\par\end{center}
\end{spacing}%
}}}

\end{center}

\vspace{0.5cm}
\noindent \textbf{Tesi dirigida per / thèse dirigée par : }\\ 
\vspace{-0.35cm}\begin{spacing}{0.75}
\noindent\begin{tabular}{ll}
\textbf{\,\,Carlos \textsc{Fernández Sopuerta\,\,\,\,\,\,\,\,\,\,\,\,\,\,\,}} & Científic Titular, \textit{ICE, CSIC \& IEEC}\\
\textbf{\,\,Alessandro D. A. M. \textsc{Spallicci}} & Professeur, \textit{Université d'Orléans}\\
\end{tabular}
\end{spacing}
\vspace{0.5cm}

\noindent \textbf{\scalebox{1}[1.0]{Tutor / tuteur : Diego \textsc{Pavón Coloma}}}\hspace{0.31cm}Professor, \textit{Universitat Autònoma de Barcelona}

\vspace{-0.1cm}




\noindent\hrulefill

\vspace{0.1cm}

\noindent \textbf{Tribunal / jury : }\\ 
\vspace{-0.35cm}\begin{spacing}{0.8}
\noindent\begin{tabular}{ll}
\textbf{\,\,Leor \scalebox{1}[1.0]{\textsc{Barack}}} & Professor, \textit{University of Southampton} - Vocal\\
\textbf{\,\,Roberto  \scalebox{1}[1.0]{\textsc{Emparan}}} & Professor, \textit{Universitat de Barcelona} - President\\
\textbf{\,\,Sascha   \scalebox{1}[1.0]{\textsc{Husa}}} & Professor \scalebox{0.95}[1.0]{Contractat Doctor}, \scalebox{0.95}[1.0]{\textit{Universitat de les Illes Balears}} - Vocal\\
\textbf{\,\,Oriol    \scalebox{1}[1.0]{\textsc{Pujolàs}}} & Investigador Titular, \scalebox{0.99}[1.0]{\textit{Institut de Física d’Altes Energies}} - Secretari\\
\textbf{\,\,\scalebox{0.95}[1.0]{Narciso  \textsc{Román Roy}}} & Catedràtic, \textit{Universitat Politècnica de Catalunya} - Vocal\\
\end{tabular}
\end{spacing}
\vspace{0.25cm}

\end{spacing}

\pagestyle{fancy}
\fancyhf{}
\renewcommand\headrulewidth{0pt}
\cfoot{\thepage}

\tableofcontents{}

\chapter*{General Thesis Summary}
\label{0-summary}
\newrefsegment

The principal subject of this thesis is the gravitational two-body problem in the extreme-mass-ratio regime---that is, where one mass is significantly smaller than the other---in the full context of our contemporary theory of gravity, general relativity. We divide this work into two broad parts: the first provides an overview of the theory of general relativity along with the basic mathematical methods underlying it, focusing on its canonical formulation and perturbation techniques; the second is dedicated to a presentation of our novel work in these areas, focusing on the problems of entropy, motion and the self-force in general relativity.
 
We begin in Part \ref{part1}, accordingly, by offering a historical introduction to general relativity as well as a discussion on current motivation from gravitational wave astronomy in Chapter \ref{1-intro}. Then, in Chapter \ref{2-canonical}, we turn to a detailed technical exposition of this theory, focusing on its canonical (Hamiltonian) formulation. We end this part of the thesis with a rigorous development of perturbation methods in Chapter \ref{3-perturbations}. For the convenience of the reader, we summarize some basic concepts in differential geometry needed for treating these topics in Appendix \ref{a-geometry}.
 
In Part \ref{part2}, we begin with a study of entropy theorems in classical Hamiltonian systems in Chapter \ref{4-entropy}, and in particular, the issue of the second law of thermodynamics in classical mechanics and general relativity, with a focus on the gravitational two-body problem. Then in Chapter \ref{5-motion}, we develop a general approach based on conservation laws for calculating the correction to the motion of a sufficiently small object due to gravitational perturbations in general relativity. When the perturbations are attributed to the small object itself, this effect is known as the gravitational self-force. It is what drives the orbital evolution of extreme-mass-ratio inspirals: compact binary systems where one mass is much smaller than---thus effectively orbiting and eventually spiralling into---the other, expected to be among the main sources for the future space-based gravitational wave detector LISA. In Chapter \ref{6-fd}, we present some work on the numerical computation of the scalar self-force---a helpful testbed for the gravitational case---for circular orbits in the frequency domain, using a method for tackling distributional sources in the field equations called the Particle-without-Particle method. We include also, in Appendix \ref{b-pwp}, some work on the generalization of this method to general partial differential equations with distributional sources, including also applications to other areas of applied mathematics. We summarize our findings in this thesis and offer some concluding reflections in Chapter \ref{7-concl}.
 
\section*{Resum General de la Tesi\\ \normalfont{(translation in Catalan)}}
 
El tema principal d'aquesta tesi és el problema gravitacional de dos cossos en el règim de raons de masses extremes - és a dir, on una massa és significativament més petita que l'altra - en el context complet de la nostra teoria contemporània de la gravetat, la relativitat general. Dividim aquest treball en dues grans parts: la primera proporciona una visió general de la teoria de la relativitat general juntament amb els mètodes bàsics matemàtics en què s’hi basa, centrant-se en la seva formulació canònica i les tècniques de pertorbació; la segona està dedicada a presentar la nostra contribució en aquests àmbits, centrada en els problemes de l’entropia, el moviment i la força pròpia en la relativitat general.
 
Comencem a la part \ref{part1}, en conseqüència, oferint una introducció històrica a la relativitat general, així com una discussió sobre la motivació actual a partir de l'astronomia d'ones gravitacionals al capítol \ref{1-intro}. A continuació, al capítol \ref{2-canonical}, passem a una exposició tècnica detallada d'aquesta teoria, centrada sobre la seva formulació canònica (hamiltoniana). Acabem aquesta part de la tesi amb un desenvolupament rigorós de mètodes de pertorbació al capítol \ref{3-perturbations}. Per a la comoditat del lector, resumim alguns conceptes bàsics en geometria diferencial necessaris per a tractar aquests temes a l'apèndix \ref{a-geometry}.
 
A la part \ref{part2}, comencem amb un estudi dels teoremes d’entropia en sistemes clàssics hamiltonians al capítol \ref{4-entropy}, i en particular, la qüestió de la segona llei de la termodinàmica en la mecànica clàssica i la relativitat general, amb el focus en el problema gravitatori de dos cossos. Al capítol \ref{5-motion}, desenvolupem una anàlisi general basada en lleis de conservació per a calcular la correcció en el moviment d’un objecte prou petit a causa de les pertorbacions gravitacionals de la relativitat general. Quan les pertorbacions s’atribueixen al propi objecte petit, aquest efecte es coneix com a força pròpia gravitacional. És el que impulsa l'evolució orbital de les caigudes en espiral amb raó de masses extrema: sistemes binaris compactes on una massa és molt menor que - i per tant, efectivament orbita i, finalment, fa espirals cap a - l'altre. Es preveu que siguin una de les principals fonts del futur detector d’ones gravitacionals LISA, situat en l’espai. Al capítol \ref{6-fd}, es presenta un treball sobre el càlcul numèric de la força pròpia escalar - una prova útil per al cas gravitatori - per òrbites circulars en el domini de freqüència, utilitzant un mètode per abordar fonts de distribució en les equacions de camp anomenat el mètode Partícula-sense-Partícula. Incloem també, en l'apèndix \ref{b-pwp}, alguns treballs sobre la generalització d’aquest mètode a equacions diferencials parcials generals amb fonts distribucionals, incloent també aplicacions a altres àrees de matemàtiques aplicades. Resumim els nostres resultats en aquesta tesi i oferim algunes reflexions finals al capítol \ref{7-concl}.
 
\section*{Résumé Général de la Thèse\\ \normalfont{(translation in French)}}

Le sujet principal de cette thèse est le problème gravitationnel à deux corps dans le régime des quotients extrêmes des masses - c’est-à-dire où une masse est nettement plus petite que l’autre - dans le contexte complet de notre théorie contemporaine de la gravité, la relativité générale. Nous divisons ce travail en deux grandes parties : la première fournit un aperçu de la théorie de la relativité générale ainsi que des méthodes mathématiques de base qui la sous-tendent, en mettant l'accent sur sa formulation canonique et les techniques de perturbation; la seconde est consacrée à une présentation de notre travail novateur dans ces domaines, en se concentrant sur les problèmes de l'entropie, du mouvement et de la force propre dans la relativité générale.

Nous commençons par la partie \ref{part1} en proposant une introduction historique à la relativité générale ainsi qu’une discussion sur la motivation actuelle à partir de l’astronomie des ondes gravitationnelles au chapitre \ref{1-intro}. Ensuite, au chapitre \ref{2-canonical}, nous abordons un exposé technique détaillé de cette théorie, en nous concentrant sur sa formulation canonique (hamiltonienne). Nous terminons cette partie de la thèse par un développement rigoureux des méthodes de perturbation au chapitre \ref{3-perturbations}. Pour la commodité du lecteur, nous résumons quelques concepts de base de la géométrie différentielle nécessaires pour traiter ces sujets dans l’annexe \ref{a-geometry}.

Dans la partie \ref{part2}, nous commencerons par une étude des théorèmes de l'entropie dans les systèmes hamiltoniens classiques au chapitre \ref{4-entropy}, et en particulier par la question de la deuxième loi de la thermodynamique dans la mécanique classique et la relativité générale, en mettant l'accent sur le problème gravitationnel à deux corps. Ensuite, au chapitre \ref{5-motion}, nous développons une analyse générale basée sur les lois de conservation pour calculer la correction au mouvement d'un objet suffisamment petit dues aux perturbations gravitationnelles dans la relativité générale. Lorsque les perturbations sont attribuées au petit objet lui-même, cet effet s’appelle la force propre gravitationnelle. C’est ce que détermine l’évolution orbitale des inspirals avec quotients extrêmes des masses : des systèmes binaires compacts dans lesquels une masse est beaucoup plus petite que - effectivement orbitant et finissant en faire des spirales dans - l’autre. On s’attend à ce qu’elles soient l’une des principales sources et parmi les plus intéressantes pour le futur détecteur spatial d'ondes gravitationnelles LISA. Au chapitre \ref{6-fd}, nous présentons quelques travaux sur le calcul numérique de la force propre scalaire - un test utile pour le cas gravitationnel - pour les orbites circulaires dans le domaine fréquentiel, en utilisant une méthode pour traiter les sources distributionnelles dans les équations de champ appelée la méthode Particule-sans-Particule. Nous incluons également, dans l’annexe \ref{b-pwp}, des travaux sur la généralisation de cette méthode aux équations aux dérivées partielles générales avec sources distributionnelles, ainsi que des applications à d’autres domaines des mathématiques appliquées. Nous résumons nos résultats de cette thèse et proposons quelques réflexions finales au chapitre \ref{7-concl}.


\chapter*{List of Original Contributions}
\label{0-contributions}
\newrefsegment

Here we list the chapters of this thesis containing original contributions, along with their corresponding publication/preprint information as they appear also in the bibliography.

~

\begin{itemize}

\item \textbf{Chapter \ref{4-entropy}}: [\cite{oltean_entropy_2016}] M. Oltean, L. Bonetti, A. D. A. M. Spallicci, and C. F. Sopuerta, ``Entropy theorems in
classical mechanics, general relativity, and the gravitational two-body problem'', \href{https://journals.aps.org/prd/abstract/10.1103/PhysRevD.94.064049}{Physical
Review D \textbf{94}, 064049 (2016)}.
\vspace{0.3cm}

\item \textbf{Chapter \ref{5-motion}}: [\cite{oltean_motion_2019}] M. Oltean, R. J. Epp, C. F. Sopuerta, A. D. A. M. Spallicci, and R. B. Mann, ``The motion
of localized sources in general relativity: gravitational self-force from quasilocal
conservation laws'', \href{https://arxiv.org/abs/1907.03012}{arXiv:1907.03012 [astro-ph, physics:gr-qc, physics:hep-th] (2019)} (v2). Submitted to Physical Review D.
\vspace{0.3cm}

\item \textbf{Chapter \ref{6-fd}}: [\cite{oltean_frequency-domain_2017}] M. Oltean, C. F. Sopuerta, and A. D. A. M. Spallicci, ``A frequency-domain implementation
of the particle-without-particle approach to EMRIs'', \href{https://iopscience.iop.org/article/10.1088/1742-6596/840/1/012056}{Journal of Physics: Conference
Series \textbf{840}, 012056 (2017)}.
\vspace{0.3cm}

\item \textbf{Appendix \ref{b-pwp}}: [\cite{oltean_particle-without-particle:_2019}] M. Oltean, C. F. Sopuerta, and A. D. A. M. Spallicci, ``Particle-without-Particle: A Practical
Pseudospectral Collocation Method for Linear Partial Differential Equations with
Distributional Sources'', \href{https://link.springer.com/article/10.1007\%2Fs10915-018-0873-9}{Journal of Scientific Computing \textbf{79}, 827 (2019)}.

\end{itemize}


\chapter*{Notation and Conventions}
\label{0-notation}
\newrefsegment

Here we summarize the basic notation and mathematical conventions used throughout this thesis. For more details, definitions and useful results, see Appendix \ref{a-geometry}.

We reserve script upper-case
letters ($\mathscr{A}$, $\mathscr{B}$, $\mathscr{C}$, ...) for denoting mathematical spaces (manifolds, curves \textit{etc.}).
The $n$-dimensional Euclidean space is denoted as usual by $\mathbb{R}^{n}$,
the $n$-sphere of radius $r$ by $\mathbb{S}_{r}^{n}$, and the unit
$n$-sphere by $\mathbb{S}^{n}=\mathbb{S}_{1}^{n}$. For any two spaces
$\mathscr{A}$ and $\mathscr{B}$ that are topologically equivalent
(\textit{i.e.} homeomorphic), we indicate this by writing $\mathscr{A}\simeq\mathscr{B}$.

The set of $(k,l)$-tensors (tensors with $k$ covariant indices and $l$ contravariant indices) on any manifold $\mathscr{U}$ is denoted
by $\mathscr{T}^{k}\,_{l}(\mathscr{U})$. In particular, $T\mathscr{U}=\mathscr{T}^{1}\,_{0}(\mathscr{U})$
is the tangent bundle and $T^{*}\mathscr{U}=\mathscr{T}^{0}\,_{1}(\mathscr{U})$
the dual thereto, \textit{i.e.} the cotangent bundle. 

A spacetime is a $(3+1)$-dimensional Lorentzian manifold, typically denoted by $\mathscr{M}$. 
We work in the $(-,+,+,+)$ signature. Any $(k,l)$-tensor in $\mathscr{M}$ is equivalently denoted either
using the (boldface) index-free notation $\bm{A}\in\mathscr{T}^{k}\,_{l}(\mathscr{M})$
following the practice of, \textit{e.g.}, [\cite{misner_gravitation_1973}; \cite{hawking_large_1975}],
or the abstract index notation $A^{a_{1}\cdots a_{k}}\,_{b_{1}\cdots b_{l}}\in\mathscr{T}^{k}\,_{l}(\mathscr{M})$
following that of, \textit{e.g.}, [\cite{wald_general_1984}]; that is,
depending upon convenience, we equivalently write
\begin{equation}
\bm{A}=A^{a_{1}\cdots a_{k}}\,_{b_{1}\cdots b_{l}}\in\mathscr{T}^{k}\,_{l}(\mathscr{M})\,,\label{eq:tensor}
\end{equation}
with Latin letters from the beginning of the alphabet ($a$, $b$,
$c$, ...) being used for abstract spacetime indices ($0,1,2,3$). The components of $\bm{A}$
in a particular choice of coordinates $\{x^{\alpha}\}_{\alpha=0}^{3}$
are denoted by $A^{\alpha_{1}\cdots\alpha_{k}}\,_{\beta_{1}\cdots\beta_{l}}$,
using Greek (rather than Latin) letters from the beginning of the
alphabet ($\alpha$, $\beta$, $\gamma$, ...). Spatial indices on
an appropriately defined (three-dimensional Riemannian spacelike)
constant time slice of $\mathscr{M}$ are denoted using Latin letters
from the middle third of the alphabet in Roman font: in lower-case
($i$, $j$, $k$, ...) if they are abstract, and in upper-case ($I$,
$J$, $K$, ...) if a particular choice of coordinates $\{x^{I}\}_{I=1}^{3}$
has been made.

More generally, when discussing any $n$-dimensional manifold of interest, we may write this as a collection of objects $(\mathscr{U},\bm{g}_{\mathscr{U}}^{\,},\bm{\nabla}_{\mathscr{U}}^{\,})$, where $\mathscr{U}$ is the manifold itself, $\bm{g}_{\mathscr{U}}^{\,}$ is a metric defined on it, and $\bm{\nabla}_{\mathscr{U}}^{\,}$ the derivative operator compatible with this metric. Its natural volume form is given by
\begin{equation}
\bm{\epsilon}_{\mathscr{U}}^{\,}=\sqrt{\left|{\rm det}\left(\bm{g}_{\mathscr{U}}^{\,}\right)\right|}\;{\rm d}x^{1}\wedge\cdots\wedge{\rm d}x^{n}\,,\label{eq:vol_form}
\end{equation}
where $\wedge$ is the wedge product.

Let $\mathscr{S}\simeq\mathbb{S}^{2}$ be any
(Riemannian) closed two-surface that is topologically a two-sphere. Latin letters from the middle
third of the alphabet in Fraktur font ($\mathfrak{i}$, $\mathfrak{j}$,
$\mathfrak{k}$, ...) are reserved for indices of tensors in $\mathscr{T}^{k}\,_{l}(\mathscr{S})$.
In particular, for $\mathbb{S}^{2}$ itself, $\mathfrak{S}_{\mathfrak{ij}}$
is the metric, $\mathfrak{D}_{\mathfrak{i}}$ the associated derivative
operator, and $\epsilon_{\mathfrak{ij}}^{\mathbb{S}^{2}}$ the volume
form; in standard spherical coordinates $\{\theta,\phi\},$ the latter
is simply given by 
\begin{equation}
\bm{\epsilon}_{\mathbb{S}^{2}}^ {}=\sin\theta\,{\rm d}\theta\wedge{\rm d}\phi\,.\label{eq:S2_volume_form}
\end{equation}

Contractions are indicated in the usual way in the abstract index
notation: \textit{e.g.}, $U^{a}V_{a}$ is the contraction of $\bm{U}$ and
$\bm{V}$. Equivalently, when applicable, we may simply use the ``dot
product'' in the index-free notation, \textit{e.g.} $U^{a}V_{a}=\bm{U}\cdot\bm{V}$,
$A_{ab}B^{ab}=\bm{A}:\bm{B}$ \textit{etc.} We must keep in mind that such
contractions are to be performed using the metric of the space on
which the relevant tensors are defined. Additionally, often we find it convenient to denote the component (projection) of a tensor in a certain direction determined by a vector by simply replacing its pertinent abstract index therewith: \textit{e.g.}, we equivalently write $U^{a}V_{b}=\bm{U}\cdot\bm{V}=U_{\bm{V}}=V_{\bm{U}}$,
$A_{ab}U^{a}=A_{\bm{U}b}$, $A_{ab}U^{a}V^{b}=A_{\bm{U}\bm{V}}$ \textit{etc.}
For any $(0,2)$-tensor $A_{ab}$, we usually write its trace (in
non-boldface) as $A=A_{a}\,^{a}={\rm tr}(\bm{A})$, except if $\bm{A}$ is a metric, in which case $A$ is typically reserved for denoting the determinant.

Finally, let $\mathscr{U}$ and $\mathscr{V}$ be any two diffeomorphic
manifolds and let $f:\mathscr{U}\rightarrow\mathscr{V}$ be a map
between them. This naturally defines a map between tensors on the
two manifolds, which we denote by $f_{*}:\mathscr{T}^{k}\,_{l}(\mathscr{U})\rightarrow\mathscr{T}^{k}\,_{l}(\mathscr{V})$
and its inverse $(f^{-1})_{*}=f^{*}:\mathscr{T}^{k}\,_{l}(\mathscr{V})\rightarrow\mathscr{T}^{k}\,_{l}(\mathscr{U})$.
We generically refer to any map of this sort as a tensor transport
[\cite{felsager_geometry_2012}]. It is simply the generalization to
arbitrary tensors of the pushforward $f_{*}:T\mathscr{U}\rightarrow T\mathscr{V}$
and pullback $f^{*}:T^{*}\mathscr{V}\rightarrow T^{*}\mathscr{U}$,
the action of which is defined in the standard way—see, \textit{e.g.}, Appendix
C of [\cite{wald_general_1984}]. (Note that our convention
of sub-/super-scripting the star is the generally more common one
used in geometry [\cite{felsager_geometry_2012,lee_introduction_2002}];
it is sometimes opposite to and sometimes congruous with that used
in the physics literature, \textit{e.g.} [\cite{wald_general_1984}] and [\cite{carroll_spacetime_2003}] respectively).


\listoffigures

\chapter*{Acknowledgements}
\label{0-acknowledgements}
\newrefsegment

I am indebted most deeply to the guidance of my two advisors, Carlos
F. Sopuerta and Alessandro Spallicci. Their patience and openness
extend as far as any student could ask, and their knowledge of physics
has always been a great pleasure from which to learn. I am especially
grateful for the freedom they have allowed me in the topics pursued
for this work. I thank them also for their invaluable help in its
redaction at all stages, including Alessandro with the translations
to French, and more generally in dealing with all the adventures of
organizing a joint doctoral degree between Barcelona and Orléans.

A special thanks goes to my collaborator and former advisor, Robert
Mann in Waterloo, Canada. Not that many years ago I took my first steps into gravitational
physics research thanks to him, and I am deeply grateful for his continued
involvement and guidance ever since.

I also greatly thank my collaborator Richard Epp, also in Waterloo, who was the first
person to truly show me how behind the
arcane mathematics of relativity there is always an intuitive physical
picture to tell the story. An extra thanks is due to my former colleague
Paul McGrath, whose initial work on quasilocal frames during his own
doctorate laid the foundation for a good part of ours here.

The places and people which have hosted me during the carrying out
of this thesis have truly given me a sense of home. I thank immensely
the colleagues and friends I have made over the past years here in
Barcelona. I cannot imagine this time without Eric Brown, Angelo Gambino,
Marcela Gus, Rafael Murrieta and Aurélien Severino. Graciès especialment
al meu estimat amic Xavi Viader, inclòs amb les traduccions aquí al
català. It has been a pleasure to work in the ICE-IEEC gravitational
waves group with Pau Amaro, Lluís Gesa, Ferran Gibert, Jordina Ho
Zhang, Ivan Lloro, Juan Pedro López, Víctor Martín, Miquel Nofrarias,
Francisco Rivas, and Daniel Santos, to whom I owe a special thanks
for initially helping me to settle. From the cosmology group here,
I thank Enrique Gaztañaga as well as my good friend Esteban González.
I also convey my very warm gratitude to my friends and colleagues
at ESADE Business School, where I have been happy to alternate a bit
from the completion of this work to teaching; I am especially
grateful to Núria Agell, Gerard Gràcia, Jordi Montserrat, Xari Rovira
and Marc Torrens. 

From my first year of this thesis spent in Orléans, I thank my former
colleague and collaborator Luca Bonetti, and my friends Pratik Hardikar
and Saketh Sistla. I would be remiss not to also add my friends and
former colleagues from my earlier time in Canada prior to this thesis,
especially Adam Bognat, Farley Charles, Heiko Santosh, Billy Tran
and Anson Wong.

It has been a great privilege during this doctorate to have the occasion
to travel and interact with many scientists outside of my home institutions.
For interesting discussions and kind encouragements, I thank in particular
Leor Barack, Abraham Harte, Ulrich Sperhake, Helvi Witek and Miguel Zilhão. For
many pleasant and instructive visits for workshops and summer schools over the past years,
I am particularly grateful to the hospitality of the Pedro Pascual Science Center
in Benasque, Spain.

It is also a pleasure to thank my friend and former colleague Hossein
Bazrafshan Moghaddam. Our conversations about physics always teach
me something new. I thank him also for helping me organize a wonderful
visit to present some of this work at IPM Tehran---additional thanks
also to the hospitality there of Hassan Firouzjahi---and the University
of Mashhad, Iran.

A dedication goes to the memory of Milton B. Zysman (1936-2019), polymath
and friend, who taught me much about the history of science and whose
influence towards inquisitive skepticism is, I hope, well alive in
the pages that follow. 

I am deeply thankful for the support of my family. V\u{a} mul\c{t}umesc la
to\c{t}i, \^{i}n special bunicii, p\u{a}rin\c{t}ii, \c{s}i sora mea Andreea. Nagyon köszönöm
jó barátom, Radu St\u{a}nil\u{a}.

Financial support for the realization of this thesis was provided
by the Natural Sciences and Engineering Research Council of Canada (NSERC)
through a Postgraduate Scholarship - Doctoral, Application No. PGSD3
- 475015 - 2015; by Campus France through an Eiffel Bourse d'Excellence,
Grant No. 840856D, awarded for carrying out an international joint doctorate; by LISA CNES funding; and by the Ministry of Economy and Business of
Spain (MINECO) through contracts ESP2013-47637-P, ESP2015-67234-P
and ESP2017-90084-P.

$\,$

$\,$

$\,$

$\,$

$\,$

$\,$

$\,$

$\,$

$\,$

$\,$

$\,$

$\,$

$\,$

$\,$


\fancyfoot{}

\newrefsegment

\thispagestyle{empty}
\begin{quote}
\begin{spacing}{1.05}
{\small{}To clear the way leading from theory to experiment of unnecessary and artificial
assumptions, to embrace an ever-wider region of facts, we must make the chain
longer and longer. The simpler and more fundamental our assumptions become,
the more intricate is our mathematical tool of reasoning; the way from theory
to observation becomes longer, more subtle, and more complicated. Although it
sounds paradoxical, we could say: Modern physics is simpler than the old physics
and seems, therefore, more difficult and intricate.
\begin{flushright}
[\cite{einstein_evolution_1938}]
\par\end{flushright}}{\small \par}
\end{spacing}
\end{quote}

$\,$

$\,$

$\,$

$\,$

$\,$

$\,$

$\,$

$\,$

$\,$

$\,$

$\,$

$\,$

$\,$

$\,$

$\,$

$\,$

$\,$

$\,$

$\,$

$\,$

$\,$

$\,$

$\,$

$\,$

$\,$

$\,$

$\,$

$\,$

$\,$

$\,$

$\,$

$\,$

$\,$

$\,$

$\,$

$\,$

$\,$

$\,$

$\,$

$\,$

$\,$

$\,$

$\,$

$\,$

$\,$

$\,$

$\,$

$\,$


\newpage{}

\pagenumbering{arabic}
\pagestyle{fancy}
\fancyhf{}
\renewcommand\headrulewidth{0.5pt}
\fancyhead[RO,LE]{\rule[-1ex]{0pt}{1ex} \fontsize{11}{12}\selectfont \thepage}
\fancyhead[RE]{\fontsize{11}{12}\sl\selectfont\nouppercase \leftmark}
\fancyhead[LO]{\fontsize{11}{12}\sl\selectfont\nouppercase \rightmark} 
\renewcommand\thepart{\Roman{part}}

\part{Fundamentals of General Relativity:\\ \normalfont{Introduction, Canonical Formulation and Perturbation Theory}\label{part1}}

\chapter{Introduction\label{1-intro}}
\newrefsegment

\subsection*{Chapter summary}

In this introduction, we present a brief history of the gravitational two-body problem and of the conception of gravitation in physics more generally, as well as a discussion of the current relevance of this problem---focusing on the extreme-mass-ratio-regime---in the era of gravitational wave astronomy.
 
We begin in Section \ref{sec:1-1} with a historical discussion of the gravitational two-body problem in pre-relativistic physics. Newton's work, especially the \textit{Principia}, is undeniably regarded as constituting the first true solution to this problem. We discuss its relevance, including Newton's own views on gravity, as well as the path immediately leading to it, especially the work of Kepler.
 
In Section \ref{sec:1-2}, we provide an account of the development of general relativity, our contemporary theory of gravity, including extracts from Einstein’s own papers summarizing the essential content of the theory. With this occasion, we define and establish the notation we use in this thesis for the most basic mathematical objects.
 
In Section \ref{sec:1-3}, we then discuss the interpretation of general relativity, and especially Einstein's views. Instead of the general idea of ``gravity as geometry'', an interpretation he seems to have found rather uninteresting due to its generality, he was much more fascinated with the connection between gravity and inertia, in particular, as established through the equation of motion for idealized particles, the geodesic equation.
 
This leads us, in Section \ref{sec:1-4}, to a discussion of the current relevance of the problem of motion in general relativity thanks to the opportunities presented by the advent of gravitational-wave astronomy. In particular, we focus on systems called extreme-mass-ratio inspirals (EMRIs): these are compact binary systems where one object is much less massive than---thus effectively orbiting and eventually spiraling into---the other. Usually, the latter is a (super-) massive black hole at a galactic center, and the former is a stellar-mass black hole or a neutron star. It is anticipated that these will be one of the main sources for space-based gravitational wave detectors, specifically for the LISA mission expected to launch in the 2030s.

Finally, in Section \ref{sec:1-5}, we enter into a bit of detail on the technical problem of modeling EMRIs. This involves calculating the correction to the motion, away from geodesic, caused by the backreaction of (the mass of) the orbiting object upon the gravitational field. This phenomenon is known as the gravitational self-force, and will be one of the major themes of this thesis.

\subsection*{Introducció \normalfont{(chapter summary translation in Catalan)}}

En aquesta introducció, presentem una breu història del problema gravitatori de dos cossos i de la concepció de la gravitació en física més generalment, així com una discussió de la rellevància actual d’aquest problema - centrat en el règim de raons de masses extremes - en l'era de l'astronomia de les ones gravitacionals.
 
Comencem a la secció \ref{sec:1-1} amb una discussió històrica del problema gravitatori de dos cossos en física pre-relativista. L'obra de Newton, especialment els \textit{Principia}, és considerada la primera veritable solució a aquest problema. Es discuteix la seva rellevància, incloent les opinions pròpies de Newton sobre la gravetat, així com el camí que hi dirigeix directament, especialment el treball de Kepler.
 
A la secció \ref{sec:1-2}, exposem el desenvolupament de la relativitat general, la nostra teoria contemporània de la gravetat, inclosos extractes dels propis treballs d'Einstein que resumeixen el contingut essencial de la teoria. Amb aquesta ocasió, definim i establim la notació que fem servir en aquesta tesi per als objectes matemàtics més bàsics.
 
A la secció \ref{sec:1-3}, es discuteix la interpretació de la relativitat general, i especialment les opinions d’Einstein. Enlloc de la idea general de la ``gravetat com a geometria'', una interpretació que sembla haver trobat poc interessant per la seva generalitat, estava molt més fascinat per la connexió entre la gravetat i la inèrcia, en particular, com es va establir mitjançant l'equació del moviment per partícules idealitzades, l'equació geodèsica.
 
Això ens porta, a la secció \ref{sec:1-4}, a una discussió sobre la rellevància actual del problema del moviment en la relativitat general gràcies a les oportunitats que presenta l'arribada de l'astronomia d'ones gravitacionals. En particular, ens centrem en caigudes en espiral amb raó de masses extrema (\textit{extreme-mass-ratio inspirals}, EMRIs): es tracta de sistemes binaris compactes on un objecte és molt menys massiu que - de manera que orbita i, finalment, fa espirals cap a - l'altre. Normalment, aquest últim és un forat negre (super) massiu en un centre galàctic, i el primer és un forat negre de massa estel·lar o una estrella de neutrons. Es preveu que aquesta sigui una de les principals fonts per als detectors d'ones gravitacionals basades en l'espai, en particular per a la missió LISA que es preveu llançar a la dècada dels 2030.
 
Finalment, a la secció \ref{sec:1-5}, introduïm una mica de detall sobre el problema tècnic de modelar EMRIs. En particular, es tracta de calcular la correcció al moviment, allunyada del geodèsic, causada per la retroacció de (la massa de) l'objecte orbitant sobre el camp gravitatori. Aquest fenomen es coneix com la força pròpia gravitacional i serà un dels principals temes d'aquesta tesi.
 
\subsection*{Introduction \normalfont{(chapter summary translation in French)}}
 
Dans cette introduction, nous présentons un bref historique du problème gravitationnel à deux corps et de la conception plus générale de la gravitation dans la physique, ainsi qu'une discussion sur la pertinence actuelle de ce problème - en se concentrant sur le régime des quotients extrêmes des masses - dans l'ère de l'astronomie des ondes gravitationnelles.

Nous commençons à la section \ref{sec:1-1} avec une discussion historique sur le problème gravitationnel à deux corps dans la physique pré-relativiste. Les travaux de Newton, en particulier le \textit{Principia}, sont indéniablement considérés comme constituant la première véritable solution à ce problème. Nous discutons de sa pertinence, y compris de la propre vision de Newton sur la gravité, ainsi que du chemin qui y conduit immédiatement, en particulier les travaux de Kepler.

Dans la section \ref{sec:1-2}, nous décrivons l'évolution de la relativité générale, notre théorie contemporaine de la gravité, avec des extraits d’articles d’Einstein résumant le contenu essentiel de la théorie. À cette occasion, nous définissons et établissons la notation que nous utilisons dans cette thèse pour les objets mathématiques les plus fondamentaux.

Dans la section \ref{sec:1-3}, nous discutons ensuite de l'interprétation de la relativité générale et en particulier des points de vue d'Einstein. Au lieu de l'idée générale de la ``gravité en tant que géométrie'', une interprétation qu'il semble avoir trouvée pas assez inintéressante en raison de sa généralité, il était beaucoup plus fasciné par le lien entre la gravité et l'inertie, en particulier, établi par l'équation du mouvement de particules idéalisées, l'équation géodésique.

Ceci nous amène, dans la section \ref{sec:1-4}, à une discussion sur la pertinence actuelle du problème du mouvement en relativité générale, grâce aux possibilités offertes par l'avènement de l'astronomie des ondes gravitationnelles. En particulier, nous nous concentrons sur les systèmes appelés inspirals avec quotients extrêmes des masses (\textit{extreme-mass-ratio inspirals}, \textit{EMRIs} en italique car mot non français dans le contexte ; encore une fois, voire si non déjà dit) : il s'agit de systèmes binaires compacts où un objet est beaucoup moins massif que - ce qui permet effectivement une orbite et au final en spirallant dans l'autre. Habituellement, le dernier est un trou noir (super) massif à un centre galactique et le premier est un trou noir à masse stellaire ou une étoile à neutrons. On s'attend à ce qu'ils soient l'une des principales sources de détecteurs d'ondes gravitationnelles situés dans l'espace, en particulier pour la mission LISA qui devrait être lancée dans les années 2030.

Enfin, dans la section \ref{sec:1-5}, nous entrons dans les détails sur le problème technique de la modélisation des \textit{EMRIs}. En particulier, il s'agit de calculer la correction du mouvement, loin de la géodésique, provoquée par la réaction en arrière (de la masse) de l'objet en orbite dans le champ gravitationnel. Ce phénomène est connu sous le nom de la force propre gravitationnelle et il constituera l'un des thèmes majeurs de cette thèse.

\section{\label{sec:1-1}\textit{Gravitatio mundi}, a brief historical prelude}

\newrefsegment

The gravitational two-body problem, in its broadest form, has always
occupied a role apart in the historical development of physics, astronomy,
mathematics, and even philosophy: \emph{How does one massive object
move around another, and why that particular motion?} From labyrinthine
epicycles, to Keplerian orbits, to the notion of a universal gravitational
``force'' and beyond, the centuries-old struggle to tackle this
question directly precipitated---more so, arguably, than any other
single physical problem---the emergence of modern scientific thought
around the turn of the 18th century. Up to the present day, with vast
opportunities currently presented by the revolutionary expansion of
observational astronomy into the domain of gravitational waves, understanding
and solving this problem has remained as galvanizing an incentive
as ever for both technical as well as conceptual advances.

From our contemporary point of view, the two parts of the problem
as formulated above---on the one hand, the empirical question of\emph{
how} motion occurs in a gravitational two-body system, and on the
other hand, the theoretical question of \emph{why} it is that (rather
than any other conceivable) motion---are indisputably regarded as
having reached their first true synthesis in the work of Newton, above
all in the \emph{Principia} [\cite{newton_philosophiae_1687}]\footnote{~For an English translation with excellent accompanying commentary
by Chandrasekhar for today's ``common reader'', see [\cite{chandrasekhar_newtons_2003}].}. Certainly, hardly any of Newton's preeminent predecessors, from
the ancient Greeks to the astronomers of the Renaissance, fell short
of taking an avid interest in not only \emph{how} the Moon and the
planets moved, but \emph{why} they moved so---or, perhaps offering
a better sense of the epochal mindset, ``\emph{what}'' moved them
so. Still, pre-Newtonian ``explanations'' of heavenly mechanics generally appear to us today to rest rather closer to the realm of myth than to that of scientific theory.

\begin{figure}
\begin{centering}
\includegraphics[scale=0.6]{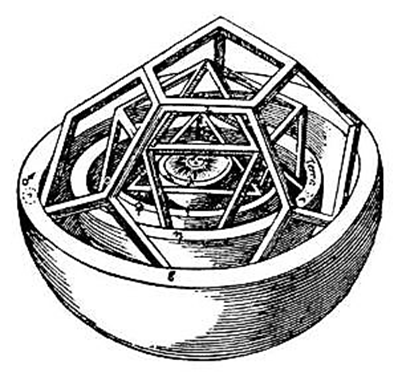}
\par\end{centering}
\caption{Detail of Kepler's model of Solar System motion based on Pythagorean solids, taken from [\cite{koestler_sleepwalkers_1959}] (adapted from \textit{Mysterium
Cosmographicum} [\cite{kepler_prodromus_1596}]). 
A property of all Pythegorean solids, of which five exist, is that they can be exactly
inscribed into---as well as circumscribed around---spheres. As only
six planets were then known (from Mercury to Jupiter), this seemed
to leave room for placing exactly these five perfect solids between
their orbits (determined as an appropriate cross-section through the inscribing/circumscribing spheres). This figure shows the orbits of the planets up to Mars inclusive.}\label{1-kepler-perfect}
\end{figure}

\begin{figure}
\begin{centering}
\includegraphics[scale=0.3]{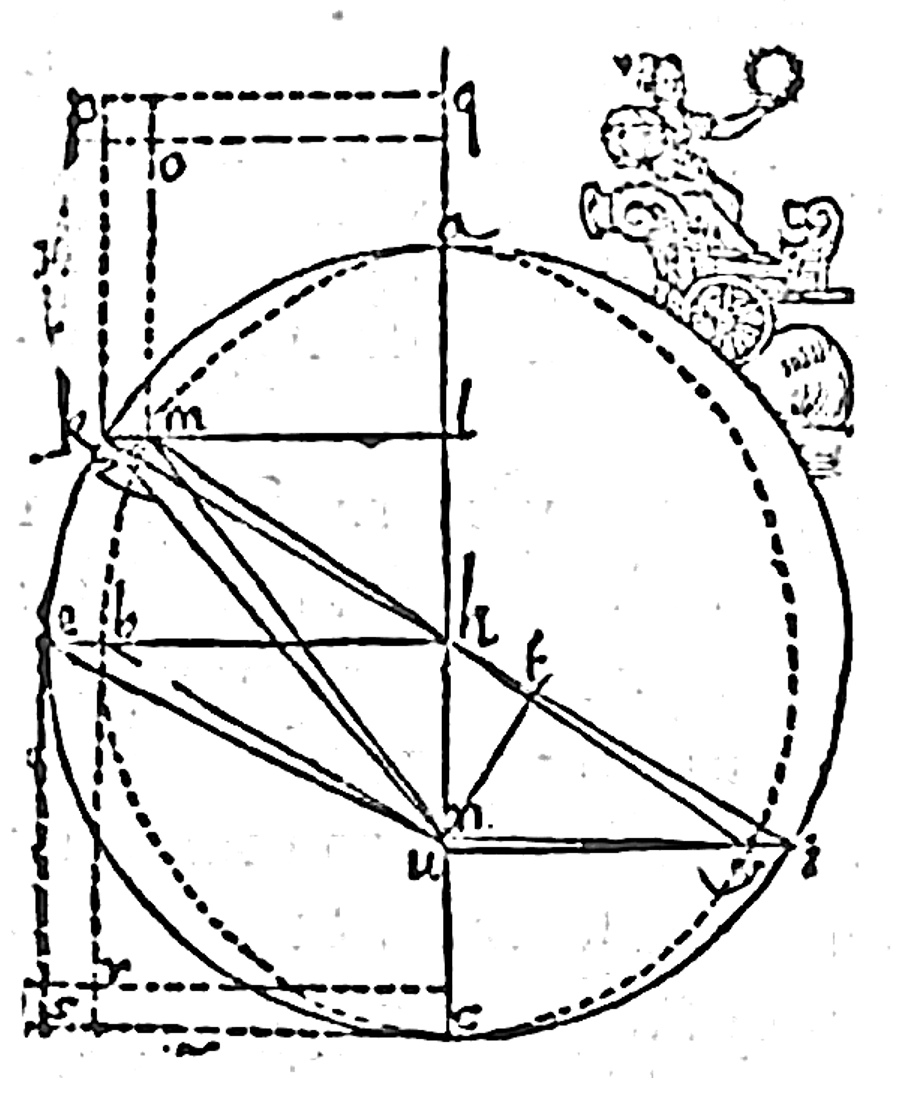}
\par\end{centering}
\caption{A figure of an ellipse (dotted oval) circumscribed by a circle from \textit{Astronomia Nova} [\cite{kepler_astronomia_1609}].}\label{1-kepler-ellipse}
\end{figure}

The figure which stood at the point of highest inflection in the evolution
of the intellectual mentality towards answering
this latter, theoretical type of question was at the same time one of the greatest empiricists and mystics---Johannes Kepler (1571-1630).
A restlessly contradictory character throughout his life, we can glean
a brief sense of the dramatic psychological fluxes that marked it---and
therethrough, ultimately, his entire era---by simply recalling Kepler's
two most famous theoretical models for Solar System motion [\cite{koestler_sleepwalkers_1959}]. When he
was in his mid-20s, he developed in a book called \textit{Mysterium
Cosmographicum} [\cite{kepler_prodromus_1596}] a model in which the
orbits of the planets around the Sun are determined by a particular embedding of
Pythagorean solids\footnote{~Also known as Platonic solids, or perfect solids, these are the set
of three-dimensional solids with identical faces (regular, convex polyhedra). It was shown by Euclid that only five such solids
exist. They are [\cite{koestler_sleepwalkers_1959}]:
\begin{center}
\includegraphics[scale=0.55]{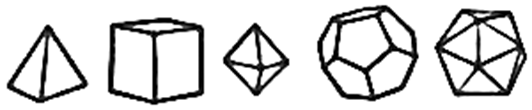}
\par\end{center}
\begin{quote}
(1) the tetrahedron (pyramid) bounded by four equilateral triangles;
(2) the cube; (3) the octahedron (eight equilateral triangles); (4)
the dodecahedron (twelve pentagons) and (5) the icosahedron (twenty
equilateral triangles).
\end{quote}
The Pythagoreans were fascinated with these, and associated four of
them (1,2,3, and 5, in the above numbering) with the ``elements''
(fire, earth, air, and water, respectively) and the remaining one
(4, the dodecahedron) with quintessence, the substance of heavenly
bodies. The latter was considered dangerous, and so ``{[}o{]}rdinary
people were to be kept ignorant of the dodecahedron'' [\cite{sagan_cosmos_1980}]. }  centered thereon, see Fig. \ref{1-kepler-perfect}. Then, a little over a decade later, in \textit{Astronomia
Nova} [\cite{kepler_astronomia_1609}], he put forth an empirical
model of elliptical orbits, based on the observations of Tycho Brahe,
establishing what we nowadays refer to as Kepler's laws of planetary
motion\footnote{~In fact, only the first two of what we today refer to as the three Keplerian
laws of planetary motion were proposed in this work (the third he found a bit later):
(1) the orbits of planets are ellipses with the Sun at a focus; (2)
the planets move such that equal areas in the orbital plane are ``swept
out'', by a straight line with the Sun, in equal time. It is interesting to remark
that these were actually discovered in reverse order. For a detailed
historical account, see Part Four, Chapter 6 of [\cite{koestler_sleepwalkers_1959}].}. See Fig \ref{1-kepler-ellipse}. What may be called the (neo-) Platonic basis of ``explanation''
underlying the former stands, to the modern reader, in radically sharp
contrast with the manifestly quasi-mechanistic one
at the basis of the latter. This reasoning is brought by Kepler to its logical end
in a letter to Herwart, which he wrote as \textit{Astronomia Nova}
was nearing completion (taken from [\cite{koestler_sleepwalkers_1959}]): 
\begin{quote}
\begin{spacing}{1.05}
{\small{}My aim is to show that the heavenly machine is not a kind
of divine, live being, but a kind of clockwork {[}...{]} insofar as
nearly all the manifold motions are caused by a most simple {[}...{]}
and material force, just as all motions of the clock are caused by
a simple weight. And I also show how these physical causes are to
be given numerical and geometrical expression. }{\small \par}
\end{spacing}
\end{quote}
One discerns in these lines an approach towards the sort of thinking
that ultimately led to the paradigmatic Newtonian explanation
of the elliptical shapes of the planetary orbits.

Arthur Koestler, in his authoritative history of pre-Newtonian cosmology
\textit{The Sleepwalkers} [\cite{koestler_sleepwalkers_1959}], to
which we have referred so far a few times, traces out in great detail
the work of Kepler and especially his ``giving of the laws'' of
planetary motion. He summarizes their significance:
\begin{quote}
\begin{spacing}{1.05}
{\small{}
Some of the greatest discoveries [...] consist mainly in the
clearing away of psychological road-blocks which obstruct the approach
to reality; which is why, }\textit{\small{}post factum}{\small{},
they appear so obvious. In a letter to Longomontanus$^{\textrm{33}}$ Kepler
qualified his own achievement as the ``cleansing of the Augean stables''. }{\small \par}

{\small{}But Kepler not only destroyed the antique edifice; he erected
a new one in its place. His Laws are not of the type which appear
self-evident, even in retrospect (as, say, the Law of Inertia appears
to us); the elliptic orbits and the equations governing planetary
velocities strike us as ``constructions'' rather than ``discoveries''.
In fact, they make sense only in the light of Newtonian Mechanics.
From Kepler's point of view, they did not make much sense; he saw
no logical reason why the orbit should be an ellipse instead of an
egg. Accordingly, he was more proud of his five perfect solids than
of his Laws; and his contemporaries, including Galileo, were equally
incapable of recognizing their significance. The Keplerian discoveries
were not of the kind which are ``in the air'' of a period, and which
are usually made by several people independently; they were quite
exceptional one-man achievements. That is why the way he arrived at
them is particularly interesting.}{\small \par}
\end{spacing}
\end{quote}
Nonetheless, the basic new concepts involved in articulating this new, clockwork-type
worldview presented great conceptual difficulties. In the \textit{Astronomia
Nova}, for example, Kepler wrestled profusely with the concept of
the ``force'' causing the motions in his imagined clockwork universe
[\cite{kepler_astronomia_1609}] (taken from [\cite{koestler_sleepwalkers_1959}]):
\begin{quote}
\begin{spacing}{1.05}
{\small{}This kind of force {[}...{]} cannot be regarded as something
which expands into the space between its source and the movable body,
but as something which the movable body receives out of the space
which it occupies... It is propagated through the universe ... but
it is nowhere received except where there is a movable body, such
as a planet. The answer to this is: although the moving force has
no substance, it is aimed at substance, i.e., at the planet-body to
be moved...}{\small \par}
\end{spacing}
\end{quote}
Koestler remarks, interestingly, that Kepler's description above actually seems to be ``closer to the modern notion of the gravitational or electro-magnetic
\textit{field} than to the classic Newtonian concept of \textit{force}'' [\cite{koestler_sleepwalkers_1959}].

\begin{figure}
\begin{centering}
\includegraphics[scale=0.27]{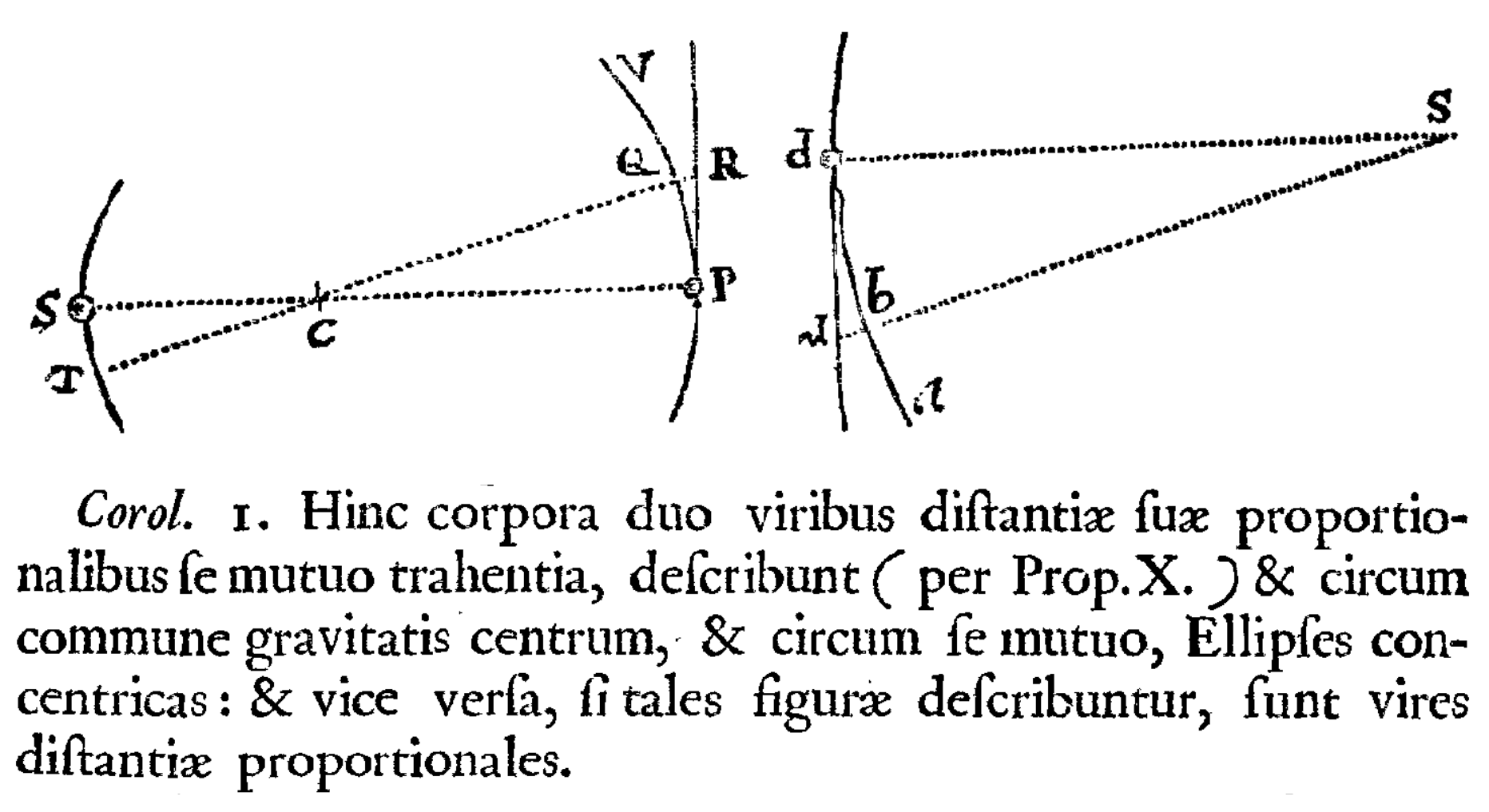}
\par\end{centering}
\caption{Newton's solution of the two-body problem in the \textit{Principia}.
Extracted here are the figure used for his Proposition LVIII , as
well as his Corollary 1 to this proposition [\cite{newton_philosophiae_1687}]:
``Hence two bodies attracting each other with forces proportional
to their distance, describe (by Prop. X), both round their common
centre of gravity, and round each other, concentric ellipses; and,
conversely, if such figures are described, the forces are proportional
to the distances.'' [\cite{chandrasekhar_newtons_2003}]}\label{1-principia}
\end{figure}

With Newton's arrival on the scene, the vision of a mechanistic
clockwork universe took definitive shape in the form of three laws
of motion and the inverse-square law of universal gravitation---with
Kepler's three laws recovered from these as particular consequences [\cite{newton_philosophiae_1687}].
What was particularly crucial here was the veritable introduction---or, at the very least, the unprecedented clarification---of a new sort of reasoning, one rooted in the idea that any useful description of fundamental physical phenomena must assume a universal and mathematical\footnote{~The specific mathematical form of such descriptions---invented by Newton himself and, since then, amply developed but still lying at the basis of all physical laws formulated to this day---is that of the differential equation.} character---a sort of reasoning then called \emph{natural philosphy}, and which later came to be referred to more commonly as \emph{science}.
Koestler once again does better than we can to contextualize the relevance of this moment [\cite{koestler_sleepwalkers_1959}]: 
\begin{quote}
\begin{spacing}{1.05}
{\small{}It is only by bringing into the open the inherent contradictions,
and the metaphysical implications of Newtonian gravity, that one is
able to realize the enormous courage \textendash{} or sleepwalker's
assurance \textendash{} that was needed to use it as the basic concept
of cosmology. In one of the most reckless and sweeping generalizations
in the history of thought, Newton filled the entire space of the universe
with interlocking forces of attraction, issuing from all particles
of matter and acting on all particles of matter, across the boundless
abysses of darkness. }{\small \par}

{\small{}But in itself this replacement of the }\textit{\small{}anima
mundi}{\small{} by a }\textit{\small{}gravitatio mundi}{\small{} would
have remained a crank idea or a poet's cosmic dream; the crucial achievement
was to express it in precise mathematical terms, and to demonstrate
that the theory fitted the observed behaviour of the cosmic machinery
\textendash{} the moon's motion round the earth and the planets' motions
round the sun.}{\small \par}
\end{spacing}
\end{quote}
Newton, of course, was famously aware of the ``inherent contradictions'' to which Koestler is referring.
While comments to this effect appear in the \textit{Principia} itself
[\cite{newton_philosophiae_1687}], in a letter to Bentley just
a few years later, he could not have been clearer \textit{vis-à-vis} what he
thought about his proposed theory---and in particular, the physical
conception of gravitation offered by it (taken from [\cite{koestler_sleepwalkers_1959}]):
\begin{quote}
\begin{spacing}{1.05}
{\small{}It is inconceivable, that inanimate brute matter should,
without the mediation of something else which is not material, operate
upon and affect other matter without mutual contact, as it must be,
if gravitation in the sense of Epicurus, be essential and inherent
in it. And this is one reason why I desired you would not ascribe
innate gravity to me. That gravity should be innate, inherent, and
essential to matter, so that one body may act upon another at a distance
through a vacuum, without the mediation of anything else, by and through
which their action and force may be conveyed from one to another,
is to me so great an absurdity that I believe no man who has in philosophical
matters a competent faculty of thinking can ever fall into it. Gravity
must be caused by an agent acting constantly according to certain
laws; but whether this agent be material or immaterial, I have left
open to the consideration of my readers.}{\small \par}
\end{spacing}
\end{quote}
No less difficult for the consideration of Newton's readers at that
time was the new mathematics describing this metaphysically mysterious
``agent''. In fact, Newton notoriously avoided publishing his work
on calculus---which he referred to as the ``method of fluxions''---for
decades, leading to the
infamous controversy with Leibnitz over its discovery [\cite{gleick_isaac_2004}].
Meanwhile, the \textit{Principia} [\cite{newton_philosophiae_1687}], though clearly bearing the basic
elements of the infinitesimal analysis at the basis of calculus, was
written essentially, one might say in  ``brute-force'' style, in the technical language then
commonly understood: Euclindean geometry. Newton presented his solution
of the two-body problem---the proof of elliptical planetary motion
as a consequence of his laws---in the \textit{Principia}\textit{\emph{,}}
Book I, Section XI, Propositions LVII-LXIII [\cite{newton_philosophiae_1687}].
See Fig. \ref{1-principia}.

Soon afterward, the issue of perturbations to a two-body orbit from
a third body, and more generally the question of the stability of
the entire Solar System, quickly gained interest. Newton also
raised this problem, and seems to have been doubtful about the possibility
of long-term Solar System stability. Subsequent investigation into
this issue went hand in hand with the development of perturbation
theory, especially thanks to the work of Lagrange and Laplace. See
[\cite{laskar_is_2013}] for a detailed review of the history and
the current status of this problem, including the discovery in the
last few decades of chaos in Solar System dynamics.

\section{\label{sec:1-2}The advent of relativity}

While there certainly existed some known empirical discrepancies with
Newton's theory by the end of the 19th century---among the most notable
being, especially in view of the two-body problem, the perihelion
precession of Mercury known since 1859---what primarily led to its
overthrow had, at least in the vision of its chief perpetrator, much
more to do with its eminently long-standing ``inherent contradictions''.
Einstein, indeed, often regarded his development of relativity\footnote{~In fact, Einstein wished to call it the ``theory of invariance'' (to highlight the invariance of the speed of light and that of physical laws in different reference frames), but the term ``theory of relativity'' coined by Max Planck and Max Abraham in 1906 quickly became, to Einstein's dissatisfaction, the more popular nomenclature, and the one which has persisted to this day [\cite{galison_roots_2001}].}
as merely the proverbial cleansing of the Newtonian stables [\cite{einstein_what_1954}]:
\begin{quote}
\begin{spacing}{1.05}
{\small{}Let no one suppose {[}...{]} that the mighty work of Newton
can really be superseded by {[}general relativity{]} or any other
theory. His great and lucid ideas will retain their unique significance
for all time as the foundation of our whole modern conceptual structure
in the sphere of natural philosophy.}{\small \par}
\end{spacing}
\end{quote}

There is a good deal of difference between the circumstances surrounding
the emergence of general relativity compared with that of the Newtonian theory.
While the latter went hand in hand with strong empirical contingencies---primary
among these being, as we have seen, solving the two-body problem---the
former was driven much more by basic conceptual and logical questions.
Cornelius Lanczos, a mathematician contemporary with Einstein, comments [\cite{lanczos_variational_1949}]:
\begin{quote}
\begin{spacing}{1.05}
{\small{}Einstein\textquoteright s Theory of General Relativity {[}...{]}
was obtained by mathematical and philosophical speculation of the
highest order. Here was a discovery made by a kind of reasoning that
a positivist cannot fail to call \textquotedblleft metaphysical,\textquotedblright{}
and yet it provided an insight into the heart of things that mere
experimentation and sober registration of facts could never have revealed.}{\small \par}
\end{spacing}
\end{quote}
Viewed from such a standpoint, the local effects of special relativity---time
dilation, length contraction and all the rest---as well as the globally
curved (non-flat) geometry of the spacetime we inhabit can be regarded
as following, essentially, as logical consequences from: \textit{(i)}
on the one hand, demanding consistency between the physical laws then
known (in particular, as concerns the Maxwellian theory of electromagnetism),
and \textit{(ii)} on the other hand, dispensing with what appeared to be the most unnecessary
assumptions causing the ``inherent contradictions'' of the Newtonian theory: in particular, the notion of absolute
space, and connected with this, the formulation of physical laws in a privileged---the
so-called inertial---class of coordinate reference frames. It is
quite remarkable how what may look from this point of view as a sort
of exercise in logic has ultimately produced such wonderfully diverse physical
insights into the nature of gravity, and even---though this generally
took longer to understand---the sorts of basic objects that can exist in
our Universe, such as black holes and gravitational waves.

An issue that attracted much of Einstein's attention throughout his
development of general relaitivity was that of the motion of an idealized
``test'' mass, that is, one provoking no backreaction in the field equations
of the theory [\cite{renn_genesis_2007}; \cite{lehmkuhl_why_2014}].
Already in 1912, in a note added in proof to [\cite{einstein_zur_1912}],
he stated for the first time that the \emph{geodesic equation}, that is, the extremization
of curve length,
\begin{equation}
\delta\int{\rm d}s=0\,,\label{eq:1-geo}
\end{equation}
is the equation of motion of point particles \textquotedblleft not
subject to external forces\textquotedblright . In this case, ${\rm d}s$
is an infinitesimal distance element in any curved four-dimensional
spacetime. By this point, Einstein understood that the basic mathematical
methods for studying spacetime curvature, logically identified as
gravity, were those of differential geometry pioneered during the
previous century by Gauss, the Bolyais (Farkas and his son János), Lobachevsky, Riemann,
Ricci and Levi-Civita, to name a few of the main players\footnote{~Much of the development of differential geometry had to do with attempts
to prove Euclid's famous fifth postulate. Ever since the appearance
of the \textit{Elements}, which is based on five postulates, there
had been skepticism regarding the necessity of the last of these.
In its original form it is much more complicated to state than the
first four, but it is equivalent to the statement that the sum of
the three angles of a triangle is always equal to two right angles.
The advent of differential (``non-Euclidean'') geometry is essentially
related to the relaxation of this condition, permitting the description
of globally curved surfaces. See [\cite{aczel_gods_2000}] for a brief
history.}. Thus the
basic object, in a theory fundamentally concerned with length measurements
(in the broadest sense), is the metric tensor, denoted $g_{\mu\nu}$
in Einstein's original notation. This object defines the notion of
infinitesimal distance ${\rm d}s$, and hence also that of motion
[\cite{einstein_zum_1913}] (taken from [\cite{lehmkuhl_why_2014}]):
\begin{quote}
\begin{spacing}{1.05}
{\small{}A free mass point moves in a straight and uniform line according
to [our Eq. (\ref{eq:1-geo})], where
\begin{equation}
{\rm d}s^{2}=\sum_{\mu\nu}g_{\mu\nu}{\rm d}x_{\mu}{\rm d}x_{\nu}\,.\label{eq:line-element}
\end{equation}
{[}\dots {]} In general, every gravitational field is going to be
defined by ten components $g_{\mu\nu}$, which are functions of {[}local
coordinates{]} $x_{1}$, $x_{2}$, $x_{3}$, $x_{4}$. }{\small \par}
\end{spacing}
\end{quote}
In 1914, Einstein actually used the word ``geodesic'' for the first time to refer
to length-extremizing curves [\cite{lehmkuhl_why_2014}].
Then, in 1915, the theory was completed with the promulgation of the
final form of the gravitational field equations governing his $g_{\mu\nu}$ [\cite{einstein_feldgleichungen_1915}]
(English translation in [\cite{einstein_field_1996}]). In a paper
consolidating the theory the following year, Einstein summarizes the main ideas [\cite{einstein_grundlage_1916}]
(taken from [\cite{einstein_foundation_1996}]):
\begin{quote}
\begin{spacing}{1.05}
{\small{}We make a distinction hereafter between \textquotedblleft gravitational
field\textquotedblright{} and \textquotedblleft matter\textquotedblright{}
in this way, that we denote everything but the gravitational field
as \textquotedblleft matter.\textquotedblright{} Our use of the word
therefore includes not only matter in the ordinary sense, but the
electromagnetic field as well.}{\small \par}

{\small{}{[}...{]} {[}We{]} require for the matter-free gravitational
field that the symmetrical {[}Ricci tensor{]}, derived from the {[}Riemann
tensor{]}, shall vanish. Thus we obtain ten equations for the ten
quantities $g_{\mu\nu}$ {[}...{]}. }{\small \par}

{\small{}{[}...{]} It must be pointed out that there is only a minimum
of arbitrariness in the choice of these equations. For besides {[}the
Ricci tensor{]} there is no tensor of second rank which is formed
from the $g_{\mu\nu}$ and its derivatives, contains no derivations
higher than second, and is linear in these derivatives.{*} }{\small \par}

{\small{}These equations, which proceed, by the method of pure mathematics,
from the requirement of the general theory of relativity, give us,
in combination with the equations of motion {[}our Eq. (\ref{eq:1-geo}){]}, to
a first approximation Newton\textquoteright s law of attraction, and
to a second approximation the explanation of the motion of the perihelion
of the planet Mercury discovered by Leverrier (as it remains after
corrections for perturbation have been made). These facts must, in
my opinion, be taken as a convincing proof of the correctness of the
theory. }{\small \par}

\noindent{\small{}\_\_\_\_\_\_\_\_\_\_\_\_}{\small \par}
\tiny{}$\,$
\end{spacing}
\begin{spacing}{0.92}

\noindent{\small{}{*} Properly speaking, this can be affirmed only of {[}a
linear combination of the Ricci tensor and the metric times the Ricci
scalar{]}. {[}...{]}}
{\small \par}
\end{spacing}
\end{quote}
We could have done no better ourselves to summarize the essential
content of the theory of general relativity, modulo the inclusion
of matter, which we address momentarily. 

While so far we have been quoting Einstein directly along with his
still widely used notation for spacetime indices---that is, with
Greek letters---we shall, in our notation throughout this thesis,
use Latin letters for spacetime indices instead, broadly following
the conventions of [\cite{wald_general_1984}]. (In principle, these
are to be understood as \emph{abstract} indices. While this may be
slightly abused sometimes if convenient and understood, we typically
try to indicate when a particular choice of coordinates is employed
by explicitly changing the indexing style, as further elaborated in
the Notation and Conventions.) Thus, we denote by $g_{ab}$ ($a,b=0,1,2,3)$ the
spacetime metric, and we work in the $(-+++)$ signature. Using
the summation convention that Einstein introduced not too long after
Eq. (\ref{eq:line-element}), the square of the infinitesimal line element ${\rm d}s$
in terms of the metric components is given by: 
\begin{equation}
{\rm d}s^{2}=g_{ab}{\rm d}x^{a}{\rm d}x^{b}\,,
\end{equation}
where $x^{a}=\{x^{0},x^{1},x^{2},x^{3}\}$ are local coordinates.

We will often find it convenient to talk about tensors without always
having to explicitly write their indices. For example, for referring
to the mathematical object (metric tensor) $g_{ab}$ we sometimes
use, when more convenient, the completely equivalent (slanted boldface)
index-free notation\footnote{~We make this choice as often, at least in physics, the non-boldface
$g$ is used to refer to something else, in this case the determinant
of the metric tensor; when not involving a metric tensor, but nonetheless
still a rank-2 contravariant tensor, this notation is then typically
reserved for referring instead to the trace.} $\bm{g}$, following the classic conventions of [\cite{misner_gravitation_1973}; \cite{hawking_large_1975}].
This is the same idea as using either $v^{i}$ ($i=1,2,3$) or, equivalently, $\vec{v}$
to represent abstractly the same object (in this case, a vector, \textit{e.g.}
a velocity) in classical physics. Interchanging between these two
notations will make our expressions more compact and readable, and our language
more fluid. For example, with a vector $v^{a}$ and another $w^{a}$,
index-free denoted $\bm{v}$ and $\bm{w}$ respectively, we write
their inner product (with respect to the metric $\bm{g}$) as $\bm{v}\cdot\bm{w}=g_{ab}v^{a}w^{b}=v^{a}w_{a}$.
We sometimes use similar notation for the ``double'' inner product,
\textit{i.e.} double contractions, \textit{e.g.} for two rank-2 contravariant tensors
$A_{ab}$ and $B_{ab}$, we write $\bm{A}:\bm{B}=g^{ac}g^{bd}A_{ab}B_{cd}=A^{ab}B_{ab}$.

Another helpful piece of notation we shall frequently employ is the
use of index-free vectors (written in slanted bold) ``as indices''. This
is meant to indicate projection in that index into the direction of
the corresponding vector. For example, for any rank-2 contravariant
tensor $\bm{A}$, we can write some of its projections in the directions
of the vectors $\bm{v}$ and $\bm{w}$ as: $A_{\bm{v}b}=A_{ab}v^{a}$
(which is now a rank-1 contravariant tensor), $A_{\bm{v}\bm{w}}=A_{ab}v^{a}w^{a}$
(which is a scalar) \textit{etc.} 

These conventions naturally generalize to tensors of higher rank.
For more details, see the Notation and Conventions as well as Appendix \ref{a-geometry}.

The next basic object we need to define is the metric-compatible \emph{derivative
operator} or \emph{connection} $\nabla_{a}$ (or $\bm{\nabla}$). It is the
unique derivative operator the action of which on the metric makes
the latter vanish, \textit{i.e.} $\nabla_{a}g_{bc}=0$ (or $\bm{\nabla}\bm{g}=0)$.
Equivalently, its action on an arbitrary vector field $\bm{v}$ is
$\nabla_{a}v^{b}=\partial_{a}v^{b}+\Gamma^{b}\,_{ac}v^{c}$, with
$\partial_{a}$ denoting the usual partial derivative and $\Gamma^{c}\,_{ab}=\frac{1}{2}g^{cd}(\partial_{a}g_{bd}+\partial_{b}g_{ad}-\partial_{d}g_{ab})$
the Christoffel symbols. (If we had any other derivative operator
on the RHS instead of the partial derivative, the latter are generally
referred to as connection coefficients.) The action of $\bm{\nabla}$
on arbitrary tensors can be generalized from this.

We typically denote a spacetime, that is, a four-dimensional Lorentzian
manifold, by $\mathscr{M}$. If a metric $\bm{g}$ and a compatible
derivative $\bm{\nabla}$ are also defined on the manifold $\mathscr{M}$,
then a spacetime more formally refers to the collection of objects
\begin{equation}
\left(\mathscr{M},\bm{g},\bm{\nabla}\right)\,,
\end{equation}
such that $\bm{\nabla}\bm{g}=0$ in $\mathscr{M}$. We denote the set of all $(k,l)$-tensors in $\mathscr{M}$ (tensors with $k$ covariant and $l$ contravariant indices) by $\mathscr{T}^{k}\,_{l}(\mathscr{M})$.

In our notation, the geodesic equation {[}Eq. (\ref{eq:1-geo}){]} is the same,
and equivalent to the condition that the four-velocity $u^{a}$ of
the curve defined as a geodesic is parallel-transported therealong, \textit{i.e.} it satisfies $u^{a}\nabla_{a}u^{b}=0$, or in index-free
notation,
\begin{equation}
\nabla_{\bm{u}}\bm{u}=0\,.
\end{equation}
In local coordinates $x^{a}$, this in turn is equivalent to
\begin{equation}
\ddot{x}^{a}+\Gamma^{a}\,_{bc}\dot{x}^{b}\dot{x}^{c}=0\,,
\end{equation}
where an overdot indicates a total derivative with respect to the
(affine) parameter of the curve.

The notion of curvature is encoded in the \emph{Riemann tensor} $R_{abc}\,^{d}$,
defined from the derivative operator $\boldsymbol{\nabla}$ in the
usual way: for any dual vector $\omega_{a}$,
\begin{equation}
\left(\nabla_{a}\nabla_{b}-\nabla_{b}\nabla_{a}\right)\omega_{c}=R_{abc}\,^{d}\omega_{d}\,.
\end{equation}
Moreover, $R_{ac}=R_{abc}\,^{b}$ is the \emph{Ricci tensor}\footnote{~Note the very usual but notationally unfortunate use of the same symbol
for denoting the these two tensors. In the index-free notation, we
will reserve $\bm{R}$ usually to refer to the Ricci tensor ($R_{ac})$,
and when we are talking about the Riemann tensor ($R_{abc}\,^{d}$)
we shall make it clear.} and, as usual, $R=\textrm{tr}(\bm{R})=g^{ab}R_{ab}=R_{a}\,^{a}$
is its trace, the \emph{Ricci scalar}. 

Defining the \emph{Einstein tensor} as 
\begin{equation}
\boldsymbol{G}=\boldsymbol{R}-\frac{1}{2}R\boldsymbol{g}\,,
\end{equation}
the field equation of general relativity for the matter-free gravitational
field, as Einstein introduced it above, is 
\begin{equation}
\boldsymbol{G}=0\,.
\end{equation}
(This is equivalent to $\boldsymbol{R}=0$.) 

Matter, in the precise sense defined above by Einstein and the one
to which we shall also adhere, is described by a \emph{stress-energy-momentum}
(symmetric rank-2 contravariant) tensor $T_{ab}$. If a matter action
$S_{\textrm{M}}$ is known (constructed from a Lagrangian yielding
the correct matter field equations), $\bm{T}$ is simply defined,
up to a factor, as the functional derivative of this action with respect
to the metric,
\begin{equation}
\bm{T}=-\frac{2}{\sqrt{-g}}\frac{\delta S_{\textrm{M}}}{\delta\bm{g}}\,.\label{eq:intro-Tab}
\end{equation}

The \emph{Einstein equation} in general states that this sources the Einstein
tensor,
\begin{equation}
\bm{G}=\kappa\bm{T}\quad\textrm{in}\,\,\,\mathscr{M}\,,\label{eq:1-ee}
\end{equation}
where $\kappa=8\pi G_{\textrm{N}}/c^{4}$ is the Einstein constant,
with $G_{{\rm N}}$ the Newton constant and $c$ the speed of light.
Note that we sometimes use the  interchangeable
nomenclature ``Einstein equations'' (in plural) to refer to the (ten)
components of Eq. (\ref{eq:1-ee}).

\section{\label{sec:1-3}Geometry, gravity and motion}

While Newton brazenly left ``open to the consideration of {[}his{]}
readers'' the task of contemplating the nature his omnipresent gravitational
``agent'', Einstein had significantly more to say on this topic
in the light and context of his own theory. A simplification of the
main message of general relativity---reflected, at the most basic
level, in the interpretation of the spacetime metric $g_{ab}$ as
the ``gravitational field''---is that gravity ought to be conceived of
as nothing more than the manifestation of curvature in the geometry
of spacetime. This is quite a generally accepted point of view today,
and Einstein himself seems to have endorsed it at least at some ``operational''
level. 

However, it seems that, to Einstein, the essence of the theory was
more subtle than simply ``reducing physics to geometry'', a phrase
to which he oftentimes attributed no, or otherwise completely tautological, meaning---insofar as the basic mathematical language of any theory of physics,
at least in the post-Newtonian paradigm, lends itself to \emph{some}
level of geometric representation by virtue of its ultimate association
to our spatial experiences: from the field lines of Maxwell's theory
that we have all seen plainly materialized, for example, in the orientation
of iron shavings on a sheet of paper underneath a magnet, to more
abstract constructs like the Hamiltonian phase space, described (as
we shall see in ample detail in the next chapter of this thesis) by
beautiful geometrical ideas of their own, in this case those of symplectic
geometry. Indeed, Einstein explicitly complains about this in a letter
to Lincoln Barnett towards the end of this life [\cite{lehmkuhl_why_2014}]:
\begin{quote}
\begin{spacing}{1.05}
{\small{}I do not agree with the idea that the general theory of relativity
is geometrizing Physics or the gravitational field. The concepts of
Physics have always been geometrical concepts and I cannot see why
the $g_{ik}$ field should be called more geometrical than f.{[}or{]}
i.{[}nstance{]} the electromagnetic field or the distance of bodies
in Newtonian Mechanics}\footnote{~Note that this is reminiscent of some current ideas such as shape
dynamics, pioneered by Barbour. See \textit{e.g.} [\cite{barbour_shape_2012}]
for a review.}{\small{}. The notion comes probably from the fact that the mathematical
origin of the $g_{ik}$ field is the Gauss-Riemann theory of the metrical
continuum which we wont look at as a part of geometry. I am convinced,
however, that the distinction between geometrical and other kinds
of fields is not logically found.}{\small \par}
\end{spacing}
\end{quote}
Instead, both during and after the development of general relativity,
Einstein was much more fascinated with the connection implied by his
theory between gravity and inertia---in particular, through the geodesic
equation. The seeds of this lie in the equivalence principle, and
specifically in his famous ``fortunate thought'' of 1907, which
he recollects in 1920 (from [\cite{lehmkuhl_why_2014}]):
\begin{quote}
\begin{spacing}{1.05}
{\small{}Then I had the most fortunate thought of my life in the following
form: The gravitational field only has a relative existence in a manner
similar to the electric field generated by electro-magnetic induction.
}\textit{\small{}Because for an observer in free-fall from the roof
of a house, there is during the fall}{\small{}\textemdash at least
in his immediate vicinity\textemdash }\textit{\small{}no gravitational
field}{\small{}. Namely, if the observer lets go of any bodies, they
remain, relative to him, in a state of rest or uniform motion, independent
of their special chemical or physical nature.}{\small \par}
\end{spacing}
\end{quote}
In a series of lectures a year later, he elaborates upon his thoughts
connecting these ideas [\cite{einstein_four_1922}] (from [\cite{einstein_four_2002}]):
\begin{quote}
\begin{spacing}{1.05}
{\small{}A material particle upon which no force acts moves, according
to the principle of inertia, uniformly in a straight line. {[}...{]}
The natural, that is, the simplest, generalization of the straight
line which is meaningful in the system of concepts of the general
(Riemannian) theory of invariants is that of the straightest, or geodesic,
line. We shall accordingly have to assume, in the sense of the principle
of equivalence, that the motion of a material particle, under the
action only of inertia and gravitation, is described by the equation,
\begin{equation}
\frac{{\rm d}^{2}x_{\mu}}{{\rm d}s^{2}}+\Gamma^{\mu}\,_{\alpha\beta}\frac{{\rm d}x_{\alpha}}{{\rm d}s}\frac{{\rm d}x_{\beta}}{{\rm d}s}=0\,.\label{eq:1-geod}
\end{equation}
In fact, this equation reduces to that of a straight line if all the
components, $\Gamma^{\mu}\,_{\alpha\beta}$, of the gravitational
field vanish. }{\small \par}

{\small{}{[}...{]} {[}The above equations, our Eq. (\ref{eq:1-geod}){]}
express the influence of inertia and gravitation upon the material
particle. The unity of inertia and gravitation is formally expressed
by the fact that the whole left-hand side of {[}our Eq. (\ref{eq:1-geod}){]}
has the character of a tensor (with respect to any transformation
of co-ordinates), but the two terms taken separately do not have tensor
character. In analogy with Newton\textquoteright s equations, the
first term would be regarded as the expression for inertia, and the
second as the expression for the gravitational force. }{\small \par}
\end{spacing}
\end{quote}
It is worth underlining that the latter is only an analogy---and
one that comes about only if one happened to be concerned with comparing
general relativity to another specific theory, in this case the Newtonian
one. To some extent, Einstein seems to have appreciated the unification
of gravity and inertia in his theory, through the geodesic equation,
similarly to that achieved between electricity and magnetism through
Maxwell's equations. The interesting discrepancy, nevertheless, is
that Einstein perceived his unification to lie not in the field equations
of the theory itself (\textit{i.e.} the Einstein equation), as had been the
case with the electromagnetic unification, but rather in the equation
of motion of idealized ``test'' particles.

Not surprisingly, perhaps, Einstein as well as others eventually became
interested in the question of whether the geodesic equation could
actually be obtained, under suitable conditions, as a consequence
of the gravitational field equations---in \textit{lieu} of postulating it
as an independent, additional assumption. The first results in this
direction began to arrive in the 1930s. In one of Einstein's first
seminal papers specifically focused on this issue, co-written with
Rosen, we see articulated the view that, indeed, the ``field'' concept
ought to lie at the basis of motion [\cite{einstein_particle_1935}]:
\begin{quote}
\begin{spacing}{1.05}
{\small{}The main value of the considerations we are presenting consists
in that they point the way to a satisfactory treatment of gravitational
mechanics. One of the imperfections of the original relativistic theory
of gravitation was that as a field theory it was not complete; it
introduced the independent postulate that the law of motion of a particle
is given by the equation of the geodesic. A complete field theory
knows only fields and not the concepts of particle and motion. For
these must not exist independently from the field but are to be treated
as part of it. On the basis of the description of a particle without
singularity, one has the possibility of a logically more satisfactory
treatment of the combined problem: The problem of the field and that
of the motion coincide.}{\small \par}
\end{spacing}
\end{quote}
Einstein continued working on this problem [\cite{einstein_gravitational_1938}],
and by 1946 when he wrote the appendix to the third edition of his
\textit{Meaning of Relativity} [\cite{einstein_meaning_2003}], he
was satisfied that it ``has been shown that this law of motion\textemdash generalized
to the case of arbitrarily large gravitating masses\textemdash can
be derived from the field equations of empty space alone''.

Since then, work in this direction has continued, and a variety of
proofs have been put forward over the decades for the geodesic equation
as the equation of motion of idealized ``test'' particles following
from the field equations of general relativity. See [\cite{geroch_motion_1975}; \cite{ehlers_equation_2004}]
for some of the most famous such proofs. See also [\cite{weatherall_geometry_2018}]
for a recent general review of the most widely used approaches, as
well as an interesting novel proposal.

These proofs often---though certainly not always---involve, in some way,
the modeling of the ``test'' particle as matter concentrated at
a spatial point. In other words, one has a stress-energy-momentum
tensor $T_{ab}^{\textrm{PP}}$ for a point-particle, which is given
by [\cite{gralla_rigorous_2008}]:
\begin{equation}
T_{ab}^{\textrm{PP}}=m\frac{u_{a}u_{b}}{\sqrt{-g}}\delta\left(x^{i}-z^{i}\right)\,,\label{eq:T_abPP_intro}
\end{equation}
where $m$ is the mass, $\delta$ is in this case the three-dimensional
Dirac delta function with $x^{i}$ denoting spatial coordinates, $z^{i}$
is the parametrization of the worldline in terms of proper time and
$u^{a}$ is the four-velocity (understood here as a function of proper
time).

The next logical step from here is to pose the following problem: If a
matter stress-energy-momentum tensor such as (\ref{eq:T_abPP_intro})
is used to model the moving ``small'' mass $m$, this will source
a similarly ``small'' correction to the spacetime metric $g_{ab}$ through
the Einstein equation. This, in turn, will induce a ``small'' correction
to the (geodesic) motion. For historical reasons, mostly having to
do with an analogous phenomenon in classical electromagnetism (see, \textit{e.g.}, Chapter 16 of [\cite{jackson_classical_1999}]), this
effect is referred to as the gravitational \emph{self-force}. The analogy
at the root of this nomenclature, of course, should be as clear as
that of Einstein when talking, \textit{vis-à-vis} the Newtonian theory,
of the Christoffel symbols expressing ``the gravitational force''
[\cite{einstein_four_1922}].

Until the last quarter century or so, the issue of the gravitational
self-force had not been extensively studied. Concordantly, there had
not been any truly compelling empirical opportunities available in
astrophysics where self-force effects might be seen to play an important
role. Now however, with the recent discovery of gravitational waves,
a new window has been opened upon a wide variety of astrophysical
phenomena, and especially binary systems in very strong gravitational
regimes---including, prospectively, ones where the self-force plays a protagonistic role.

\section{\label{sec:1-4}Gravitational waves and extreme-mass-ratio inspirals}

The recent advent of gravitational wave astronomy---propelled
by the ground-based direct detections achieved by the LIGO/Virgo collaboration (see~[\cite{abbott_et_al._gwtc-1:_2018}] for the detections during the O1 and O2 observing runs),
the success of the LISA Pathfinder mission as a proof of principle
for future space-based interferometric detectors [\cite{armano_et_al._sub-femto-g_2016,armano_et_al._beyond_2018}],
and the subsequent approval of the LISA mission for launch in the
2030s [\cite{amaro-seoane_et_al._gravitational_2013,amaro-seoane_et_al._laser_2017}]---has generally brought a multitude 
of both practical and foundational problems to the foreground of gravitational
physics today. While a plethora of possibilities for gravitational wave sources are
actively being investigated theoretically and anticipated to become accessible
observationally, both on the Earth as well as in space, the most ubiquitous
class of such sources has manifestly been---and foreseeably will
remain---the coalescence of compact object binaries [\cite{colpi_gravitational_2016,celoria_lecture_2018}].
These are two-body systems consisting of a pair of compact objects, say of masses $M_{1}$
and $M_{2}$, orbiting and eventually spiraling into each other.
Each of these is, usually, either a \emph{black hole} (BH) or a neutron star. There are also more general possibilities being investigated, including that of having a brown dwarf as one of the objects [\cite{amaro-seoane_x-mris:_2019}].

The LIGO/Virgo detections during the first scientific runs~[\cite{abbott_et_al._gwtc-1:_2018}], O1 and O2, have all involved binaries of \emph{stellar-mass
compact objects} (SCOs) located in our local neighbourhood. These have
comparable masses, of the order of a few tens of solar masses each
($M_{1}\sim M_{2}\sim10^{0-2}M_{\odot}$). 
In addition second- and third- generation terrestrial detectors can also eventually see \emph{intermediate-mass-ratio
inspirals}, binaries consisting of an intermediate-mass BH, of $10^{2-4}M_{\odot}$, and an SCO. While there is as yet no direct evidence for the existence of the former sorts of objects, there are good reasons to anticipate their detection (through gravitational waves) most likely at the centers of globular clusters, and their study provides an essential link to the strongly perturbative regime of compact object binary dynamics.

\begin{figure}
\begin{centering}
\includegraphics[scale=0.65]{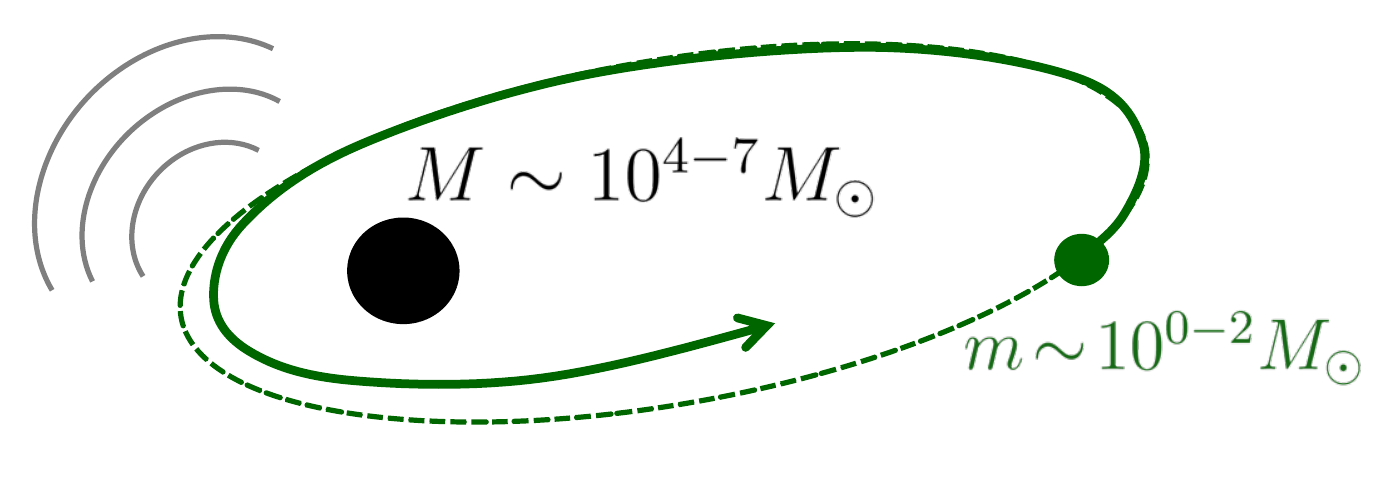}
\par\end{centering}
\caption{Sketch of an extreme-mass-ratio inspiral (EMRI), a two-body system consisting of a stellar-mass compact object (SCO), usually a stellar-mass black hole, of mass $m\sim10^{0-2}M_{\odot}$, orbiting and eventually spiralling into a (super-) massive black hole (MBH), of mass $M\sim10^{4-7}M_{\odot}$, and emitting gravitational waves in the process.}\label{fig-emri}
\end{figure}

It is even further in this direction that future space-based
detectors such as LISA are anticipated to take us. In particular, LISA is expected to see \emph{extreme-mass-ratio
inspirals} (EMRIs) [\cite{amaro-seoane_relativistic_2018}], compact binaries where $M_{1}\gg M_{2}$. An elementary sketch is depicted in Figure \ref{fig-emri}. The
more massive object could be a (\emph{super}-) \emph{massive}\textit{
}black hole (MBH) of mass $M_{1}=M\sim10^{4-7}\,M_{\odot}$ located
at a galactic center, with the significantly less massive object---effectively
orbiting and eventually spiraling into the MBH---being an SCO: either
a stellar-mass black hole or a neutron star, with $M_{2}=m\sim10^{0-2}M_{\odot}$.

Average estimates indicate that LISA will be able to see on the order
of hundreds of EMRI events per year~[\cite{babak_science_2017}], with an expectation of observing, for each,
thousands of orbital cycles over a period on the order of one year before the final plunge [\cite{barack_self-force_2018}].
The trajectories defining these cycles and the gravitational wave signals produced
by them will generally look much more complex than the relatively
generic signals from mergers of stellar-mass black holes of comparable masses as observed, for
example, by LIGO/Virgo.

EMRIs will therefore offer an ideal experimental milieu for strong
gravity: the complicated motion of the SCO around the MBH will effectively
``map out'' the geometry---that is, the gravitational field---around
the MBH, thus presenting us with an unprecedented opportunity for
studying gravity in the very strong regime~[\cite{babak_science_2017,berry_unique_2019}]. In particular, among the
possibilities offered by EMRIs are the measurement of the mass and
spin of the MBH to very high accuracy, testing the validity of the
Kerr metric as the correct description of BHs within general relativity
(GR), and testing GR itself as the correct theory of gravity.

\begin{figure}
\begin{centering}
\includegraphics[scale=0.55]{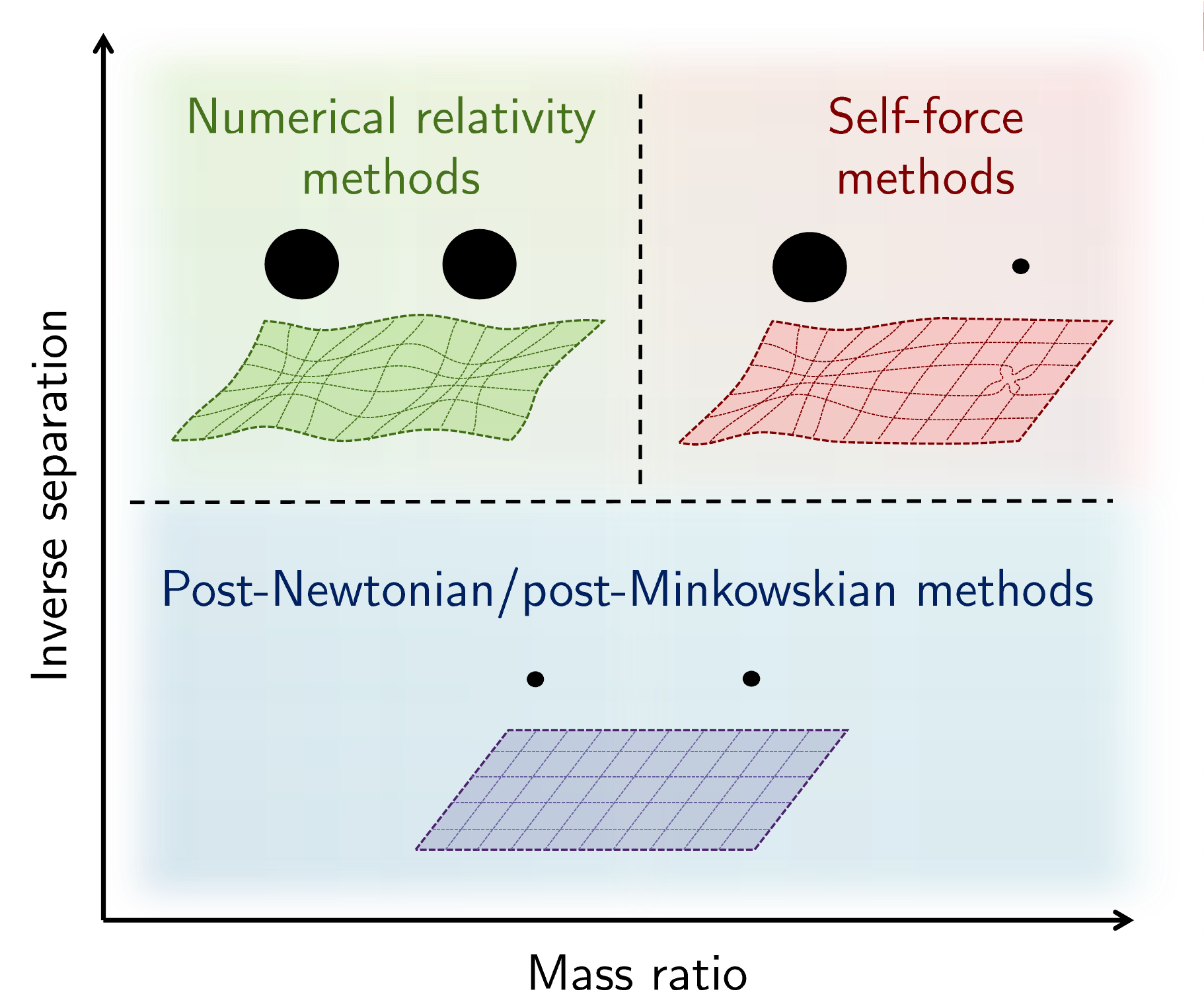}
\par\end{centering}
\caption{The main approaches used in practice for the modeling of compact object binaries as a function of the mass ratio (increasing from $1$) and the inverse separation involved. For high separations between the bodies, post-Newtonian and post-Minkowskian methods are used. For low separations and low mass ratios, numerical relativity is used. For low separations and extreme mass ratios, as the scale of a numerical grid would have to span orders of magnitude thus rendering it impracticable,  perturbation theory must be used---in particular, self-force methods.}\label{fig-pn-nr-sf}
\end{figure}

Yet, the richness of the observational opportunities presented by EMRIs
comes with an inexorable cost: that is, a significant and as yet ongoing
technical challenge in their theoretical modeling. This is all the
more pressing as the EMRI signals expected from LISA are anticipated
to be much weaker than the instrumental noise of the detector. Effectively,
what this means is that extremely accurate models are necessary in
order to produce the waveform templates that can be used to extract
the relevant signals from the detector data stream. At the theoretical
level, the problem of EMRI modeling cannot be tackled directly with
numerical relativity (used for the LIGO/Virgo detections), simply
due to the great discrepancy in (mass/length) scales; however, for
the same reason, the approach that readily suggests itself is perturbation
theory. See Figure \ref{fig-pn-nr-sf} for a graphic depicting the main methods used for compact object binary modeling in the different regimes. In particular, modeling the strong gravity, extreme mass ratio regime turns out to be equivalent to a general
and quite old problem which can be posed in any (not just gravitational)
classical field theory: the so-called \emph{self-force} problem.

\section{\label{sec:1-5}The self-force problem}

Suppose we are dealing with a theory for a field $\psi(x)$ in some
spacetime. If the theory admits a Lagrangian formulation, we can usually
assume that the field equations have the general form 
\begin{equation}
L[\psi\left(x\right)]=S\left(x\right)\,,\label{eq:general_field_eqn}
\end{equation}
where $L$ is a (partial, possibly nonlinear and typically second-order)
differential operator, and we refer to $S$ as the \emph{source} of
the field $\psi$. Broadly speaking, the problem of the self-force
is to find solutions $\psi(x)$ satisfying (\ref{eq:general_field_eqn})
when $S$ is ``localized'' in spacetime. Intuitively, it is the
question of how the existence of a dynamical (field-generating) ``small
object'' (a mass, a charge \textit{etc.}) backreacts upon the total field
$\psi$, and hence in turn upon its own future evolution subject to
that field. Thus, an essential part of any detailed self-force analysis
is a precise specification of what exactly it means for $S$ to be
localized. In standard approaches, one typically devises a perturbative
procedure whereby $S$ ends up being approximated as a distribution,
usually a Dirac delta, compactly supported on a worldline---that
is, the \emph{background} (zeroth perturbative order) worldline of
the small object. However, this already introduces a nontrivial mathematical
issue: if $L$ is non-linear (in the standard PDE sense), then the
problem (\ref{eq:general_field_eqn}) with a distributional source $S$ is
mathematically ill-defined, at least within the classical theory
of distributions [\cite{schwartz_theorie_1957}] where products of distributions
do not make sense [\cite{schwartz_sur_1954}]\footnote{~Nonlinear theories of distributions are being actively investigated
by mathematicians [\cite{li_review_2007,colombeau_nonlinear_2013,bottazzi_grid_2019}]. Some work has been done to apply these to the electromagnetic self-force problem [\cite{gsponer_classical_2008}] and to study their general applicability in GR [\cite{steinbauer_use_2006}], however at this point, to our knowledge, their potential usefulness
for the gravitational self-force problem has not been contemplated to any significant extent.}. 

One might therefore worry that nonlinear physical theories, such as
GR, would a priori not admit solutions sourced by distributions, and
we refer the interested reader to [\cite{geroch_strings_1987}]
for a classic detailed discussion of this topic. The saving point
is that, while the full field equation (in this case, the Einstein
equation) may indeed be generally non-linear, if we devise a perturbative
procedure (where the meaning of the perturbation is prescribed
in such a way as to account for the presence of the small object itself),
then the first-order field equation is, by construction, linear in
the (first-order) perturbation $\delta\psi$ of $\psi$. Thus, assuming
the background field is a known exact solution of the theory,
it always makes sense to seek solutions $\delta\psi$ to
\begin{equation}
\delta L[\delta\psi\left(x\right)]=S\left(x\right)\,,\label{eq:general_perturbed_field_eqn}
\end{equation}
for a distributional source $S$, where $\delta L$ indicates the
first-order part of the operator $L$ in the full field equation (\ref{eq:general_field_eqn}).
As this only makes sense for the (linear) first-order problem, such
an approach becomes again ill defined if we begin to ask about the
(nonlinear) second- or any higher-order problem. Additional technical
constructions are needed to deal with these, the most common of which
for the gravitational self-force has been the so-called ``puncture'' (or ``effective source'')
method [\cite{barack_scalar-field_2007,barack_m-mode_2007,vega_regularization_2008,barack_self-force_2018}]; similar ideas have proven to be very useful also in numerical relativity [\cite{campanelli_accurate_2006,baker_gravitational-wave_2006}]. For work on the second-order equation of motion for the gravitational self-force problem, see \textit{e.g.}  [\cite{pound_second-order_2012,gralla_second-order_2012,pound_nonlinear_2017}]. For now, we assume that
we are interested here in the first-order self-force problem (\ref{eq:general_perturbed_field_eqn}) only.

\begin{figure}
\begin{centering}
\includegraphics[scale=0.6]{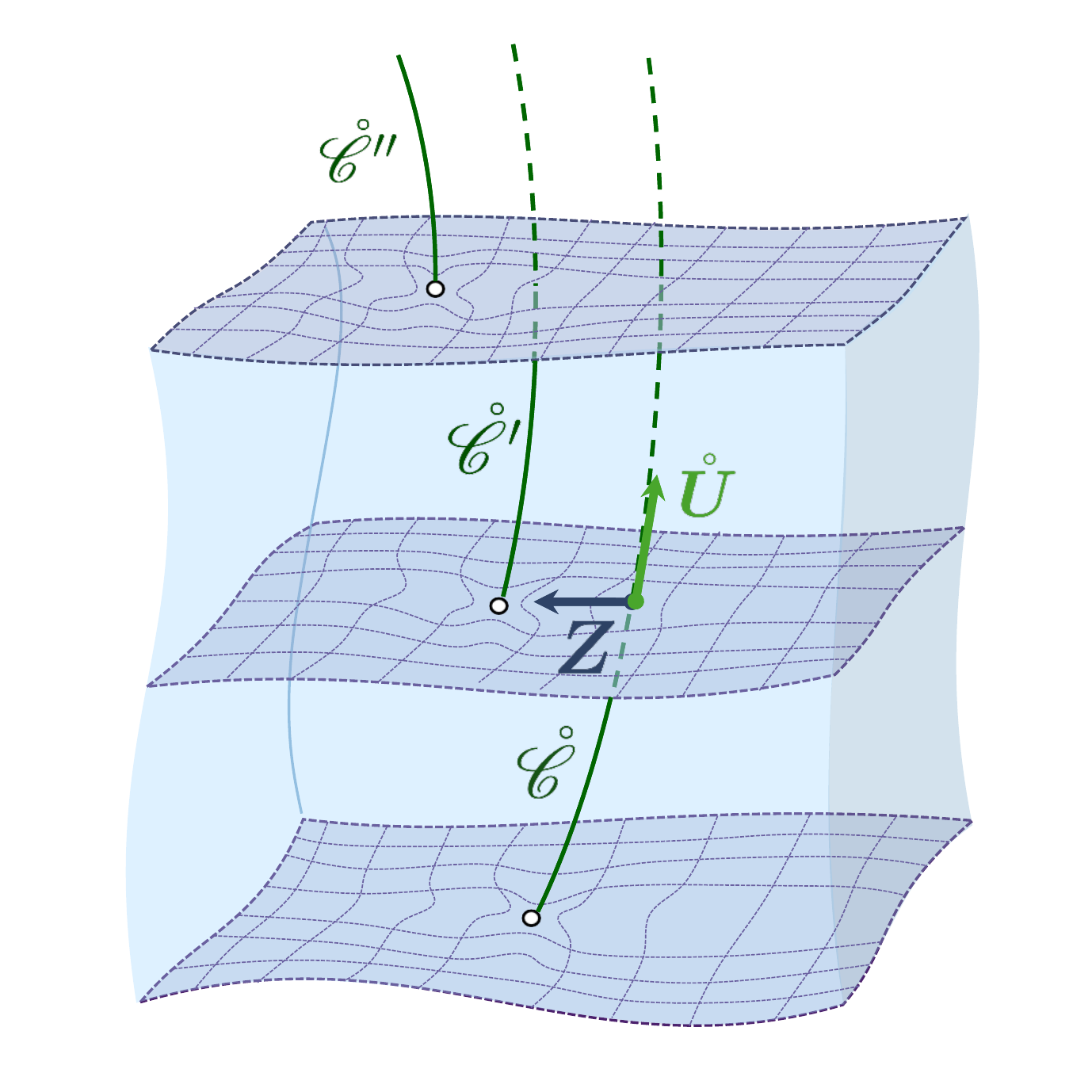}
\par\end{centering}
\caption{A depiction of the perturbative problem for the gravitational self-force (GSF). In particular, this represents one of the most popular conceptions of a so-called ``self-consistent'' approach [\cite{gralla_rigorous_2008}]: at a given step (on a given Cauchy surface) in the time evolution problem, one computes the ``correction to the motion'' away from geodesic ($\mathring{\mathscr{C}}$) in the form of a deviation vector $Z^{a}$, determined by the GSF. Then, at the next time step, one begins on a new (``corrected'') geodesic ($\mathring{\mathscr{C}}'$), computes the new deviation vector, and so on.}\label{fig-sf}
\end{figure}

Now concretely, in GR, our physical field $\psi$ is simply the spacetime metric $g_{ab}$ (where Latin letters from the beginning of the alphabet indicate spacetime indices), and following standard convention we
denote a first-order perturbation thereof by $\delta g_{ab}=h_{ab}$.
The problem (\ref{eq:general_perturbed_field_eqn}) is then just the
first-order Einstein equation,
\begin{equation}
\delta G_{ab}[h_{cd}]=\kappa T_{ab}^{\textrm{PP}}\,,\label{eq:first_order_EFE}
\end{equation}
where $G_{ab}$ is the Einstein tensor, $\kappa=8\pi$ (in geometrized
units $c=G=1$) is the Einstein constant, and $T_{ab}^{\textrm{PP}}$ the energy-momentum tensor 
of a ``point particle'' (PP) compactly supported on a given worldline $\mathring{\mathscr{C}}$. We will return later
to discussing this more precisely, but in typical approaches, $\mathring{\mathscr{C}}$
turns out to be a geodesic---that is, the ``background motion''
of the small object, which is in this case a small mass\footnote{~We consider later in this thesis (Chapter \ref{5-motion}) in detail one approach
to the gravitational self-force which also proves geodesic motion as the ``background'' motion of point particles in GR.
}. Thus, simply solving (\ref{eq:first_order_EFE}) for $h_{ab}$ assuming
a fixed $\mathring{\mathscr{C}}$ for all time, though mathematically
well-defined, is by itself physically meaningless: it would simply
give us the metric perturbations caused by a small object eternally
moving on the same geodesic. Instead what we would ultimately like
is a way to take into account how $h_{ab}$ modifies the motion of
the small object itself. Thus in addition to the field equation (\ref{eq:first_order_EFE}),
any self-force analysis must be supplemented by an \emph{equation
of motion} (EoM) telling us, essentially, how to move from a given
background geodesic $\mathring{\mathscr{C}}$ at one step in the (ultimately
numerical) time evolution problem to a new background geodesic $\mathring{\mathscr{C}}'$
at the next time step---with respect to which the field equation
(\ref{eq:first_order_EFE}) is solved anew, and so on. This is sometimes referred to as a ``self-consistent'' approach. See Fig. \ref{fig-sf} for a visual depiction. 

The first proposal for an EoM for the \emph{gravitational self-force} (GSF)
problem was put forward in the late 1990s, since known as the MiSaTaQuWa
equation after its authors [\cite{mino_gravitational_1997,quinn_axiomatic_1997}].
On any $\mathring{\mathscr{C}}$, it reads:
\begin{equation}
\ddot{Z}^{a}=-\mathring{E}_{b}\,^{a}Z^{b}+F^{a}[h_{cd}^{\textrm{tail}};\mathring{U}^{e}]\,.\label{eq:intro_MiSaTaQuWa}
\end{equation}
The LHS is a second (proper) time derivative of a \emph{deviation
vector} $Z^{a}$ on $\mathring{\mathscr{C}}$ pointing in the direction
of the ``true motion'' (away from $\mathring{\mathscr{C}}$), to
be defined more precisely later. On the RHS, $\mathring{E}_{ab}$
is the electric part of the Weyl tensor on $\mathring{\mathscr{C}}$,
such that the first term is a usual ``geodesic deviation'' term.
The second term on the RHS is the one usually understood as being responsible for self-force effects: $F^{a}[\cdot;\cdot]$ is a four-vector functional of a symmetric
rank-$2$ contravariant tensor and a vector, to which we refer in general (for any
arguments) as the \emph{GSF functional}. In any spacetime with a given
metric $\mathring{g}_{ab}$ and compatible derivative operator $\mathring{\nabla}_{a}$,
it is explicitly given by the following simple action of a first-order
differential operator:
\begin{equation}
F^{a}[H_{bc};V^{d}]=-\left(\mathring{g}^{ab}+V^{a}V^{b}\right)\left(\mathring{\nabla}_{c}H_{bd}-\frac{1}{2}\mathring{\nabla}_{b}H_{cd}\right)V^{c}V^{d}\,.\label{eq:intro_GSF_functional}
\end{equation}
While this is easy enough to calculate once one knows the arguments,
the main technical challenge in using the MiSaTaQuWa equation (\ref{eq:intro_MiSaTaQuWa})
lies precisely in the determination thereof: in particular, $h_{ab}^{\textrm{tail}}$
is not the \emph{full} metric perturbation $h_{ab}$ which solves the field
equation (\ref{eq:first_order_EFE}), but instead represents what
is called the ``tail'' integral of the Green functions of $h_{ab}$ [\cite{poisson_motion_2011}].
This quantity is well defined, but difficult to calculate in practice
and usually requires the fixing of a perturbative gauge---typically
the \emph{Lorenz} gauge, $\mathring{\nabla}^{b}(h_{ab}-\frac{1}{2}h_{cd}\mathring{g}^{cd}\mathring{g}_{ab})=0$. Physically, $h_{ab}^{\textrm{tail}}$ can be thought of as the part of the full perturbation $h_{ab}$ which is scattered back by the spacetime curvature. (In this way, $h_{ab}$ can be regarded as the sum of $h_{ab}^{\textrm{tail}}$ and the remainder, what is sometimes called the ``instantaneous'' or ``direct'' part $h_{ab}^{\textrm{direct}}$, responsible for waves radiated to infinity [\cite{spallicci_self-force_2014}].)

An alternative, equivalent GSF EoM was proposed by Detweiler and Whiting
in the early 2000s [\cite{detweiler_self-force_2003}]. It relies upon
a regularization procedure for the metric perturbations, \textit{i.e.} a choice
of a decomposition for $h_{ab}$ [the full solution of the field equation
(\ref{eq:first_order_EFE})] into the sum of two parts: one
which diverges---in fact, one which contains \emph{all} divergent
contributions\emph{\emph{---}}on $\mathring{\mathscr{C}}$, denoted
$h_{ab}^{\textrm{S}}$ (the so-called ``singular'' field, related to the ``direct'' part of the metric perturbation), and a
remainder which is finite, $h_{ab}^{\textrm{R}}$ (the so-called ``regular''
field, related to the ``tail'' part), so that one writes $h_{ab}=h_{ab}^{\textrm{S}}+h_{ab}^{\textrm{R}}$.
An analogy with the self-force problem in electromagnetism gives some
physical intuition behind how to interpret the meaning of this decomposition
[\cite{barack_self-force_2018}], with $h_{ab}^{\textrm{S}}\sim m/r$
having the heuristic form of a ``Coulombian self-field''. However,
no procedure is known for obtaining the precise expression of $h_{ab}^{\textrm{S}}$
in an arbitrary perturbative gauge, and moreover, once a gauge is
fixed (again, usually the Lorenz gauge), this splitting is not unique [\cite{barack_self-force_2018}].
Nevertheless, if and when such an $h_{ab}^{\textrm{S}}$ is obtained
(from which we thus also get $h_{ab}^{\textrm{R}}=h_{ab}-h_{ab}^{\textrm{S}}$),
the Detweiler-Whiting EoM for the GSF reads:
\begin{equation}
\ddot{Z}^{a}=-\mathring{E}_{b}\,^{a}Z^{b}+F^{a}[h_{cd}^{\textrm{R}};\mathring{U}^{e}]\,.\label{eq:intro_Detweiler-Whiting}
\end{equation}

The EoMs (\ref{eq:intro_MiSaTaQuWa}) and (\ref{eq:intro_Detweiler-Whiting})
are equivalent in the Lorenz gauge and form the basis of the two most popular
methods used today for the numerical computation of the GSF. Yet a
great deal of additional technical machinery is required for handling
gauge transformations. This is essential because, in the EMRI problem,
the background spacetime metric---that of the MBH---is usually assumed
to be Schwarzschild-Droste or Kerr. Perturbation theory for such spacetimes
has been developed and is most easily carried out in, respectively,
the so-called Regge-Wheeler and radiation gauges; in other words,
in practice, it is often difficult (though not infeasible---see, \textit{e.g.}, [\cite{barack_perturbations_2005}]) to compute $h_{ab}$ directly
in the Lorenz gauge for use in (\ref{eq:intro_MiSaTaQuWa}) or (\ref{eq:intro_Detweiler-Whiting}).

A proposal for an EoM for the GSF problem that is valid in a wider class of
perturbative gauges was presented by Gralla in 2011 [\cite{gralla_gauge_2011}].
It was therein formulated in what are called ``parity-regular''
gauges, \textit{i.e.} gauges satisfying a certain parity condition. This condition
ultimately has its origins in the Hamiltonian analysis of Regge and
Teitleboim in the 1970s [\cite{regge_role_1974}], wherein the authors
introduce it in order to facilitate the vanishing of certain surface
integrals and thus to render certain general-relativistic Hamiltonian
notions, such as multipoles and ``center of mass'', well-defined
mathematically. In parity-regular gauges (satisfying the Regge-Teitleboim
parity condition), the Gralla EoM---mathematically equivalent, in
the Lorenz gauge, to the MiSaTaQuWa and the Detweiler-Whiting EoMs---is:
\begin{equation}
\ddot{Z}^{a}=-\mathring{E}_{b}\,^{a}Z^{b}+\frac{1}{4\pi}\lim_{r\rightarrow0}\intop_{\mathbb{S}_{r}^{2}}\bm{\epsilon}^{}_{\mathbb{S}^{2}}\,F^{a}[h_{cd};\mathring{U}^{e}]\,.\label{eq:intro_Gralla}
\end{equation}
The GSF (last) term on the RHS is obtained in this approach by essentially
relating the deviation vector $Z^{a}$ (the evolution of which is expressed
by the LHS) with a gauge transformation vector and then performing
an ``angle average'' over a ``small'' $r$-radius two-sphere $\mathbb{S}_{r}^{2}$,
with $\bm{\epsilon}^{}_{\mathbb{S}^{2}}$ the volume form of the unit two-sphere, of the so-called ``bare''
GSF, $F^{a}[h_{bc};\mathring{U}^{d}]$. The latter is just the GSF
functional [Eq. (\ref{eq:intro_GSF_functional})] evaluated directly using the \emph{full} metric perturbatiuon $h_{ab}$ (\textit{i.e.} the ``tail'' plus ``direct'' parts, or equivalently, the ``regular'' plus ``singular'' parts),
around (rather than at the location of) the distributional source. 
The point therefore is that this formula never requires the evaluation
of $h_{ab}$ on $\mathring{\mathscr{C}}$ itself, where it is divergent
by construction; instead, away from $\mathring{\mathscr{C}}$ it is
always finite\footnote{~This is true provided one does not transform to a perturbative gauge wherein any of the multipole moments of $\bm{h}$ diverge away from $\mathring{\mathscr{C}}$ as a consequence of the gauge definition. Examples of gauges leading to such divergences are the ``half-string'' and ``full-string'' radiation gauges of [\cite{pound_gravitational_2014}], which exhibit string-like singularities in $\bm{h}$ along a radial direction. Nevertheless, in this work it was also shown that one can define a ``no-string'' radiation gauge which is in the parity-regular class, and where the singularities in $\bm{h}$ remain only on $\mathring{\mathscr{C}}$ thus rendering the integral (\ref{eq:intro_GSF_functional}) well-defined.}, and (\ref{eq:intro_Gralla}) says that it suffices
to compute the GSF functional (\ref{eq:intro_GSF_functional}) with
$h_{ab}$ directly in the argument (requiring no further transformations),
and integrate it over a small sphere.

The manifest advantage of (\ref{eq:intro_Gralla}) relative to (\ref{eq:intro_MiSaTaQuWa})
or (\ref{eq:intro_Detweiler-Whiting}) is that no computations of
tail integrals or regularizations of the metric perturbations are
needed at all. Yet, to our knowledge, there has thus far been no attempted
numerical computation of the GSF using this formula. One of the issues
with this remains that of the perturbative gauge: depending upon the detailed setup of the problem, one may still not easily be able to compute $h_{ab}$ directly in a parity-regular gauge (although manifestly, working in the parity-regular ``no-string'' radiation gauge [\cite{pound_gravitational_2014}] may be useful for a GSF calculation in Kerr), \textit{i.e.} a gauge in which (\ref{eq:intro_Gralla}) is strictly valid, and so further gauge transformations may be needed.  Aside from the practical issues with a possible numerical
implementation of this, there is also a conceptual issue: this formula
originates from an essentially mathematical argument---by a convenient ``averaging'' over the angles---so as to make it well-defined in a Hamiltonian setting via a relation to a canonical
definition of the center of mass. Yet its general form as a closed two-surface
integral suggestively hints at the possibility of interpreting it
not merely as a convenient mathematical relation, but as a real physical
flux of (some notion of) ``gravitational momentum''. We contend
and will demonstrate in this thesis (specifically, in Chapter \ref{5-motion}) that indeed an even more general
version of (\ref{eq:intro_Gralla}), not restricted by any perturbative gauge choice (so long as one does not construct it in such a way that produces divergences in $\bm{h}$ away from $\mathring{\mathscr{C}}$), results from the consideration
of momentum conservation laws in GR.


\chapter{Canonical General Relativity\label{2-canonical}}
\newrefsegment

\subsection*{Chapter summary}

The aim of this chapter is to introduce the basic mathematical language and technical machinery of the theory of general relativity following variational methods. We focus especially on developing the canonical formulation of general relativity, also known as the Hamiltonian or $(3+1)$ formulation. In essence, this provides a way of turning the second-order field equations of the theory for the spacetime metric into a first-order time-evolution problem for the induced spatial three-metric and its conjugate momentum. There are, however, also constraint equations in addition to these (constraining permissible initial conditions as well as their development subject to the dynamical equations), and their proper handling requires a great deal of subtlety. The existence of constraints in general relativity is in fact directly related to---and offers fruitful insight on---the gauge freedom available in the theory.
 
We begin in Section \ref{sec:2.1-intro} with a brief introduction. We generally discuss four broad areas of application for canonical general relativity: mathematical relativity, numerical relativity, quantum gravity, and the issue of gravitational energy-momentum. We return to each of these in the final section of this chapter with specific examples, once we have developed the mathematical tools in detail.
 
In Section \ref{sec:2.2-lagrangian}, we present the Lagrangian formulation of classical field theories in general, and then general relativity in particular, forming the typical starting point of any canonical analysis. We comment on the appearance of constraint equations already at the Lagrangian level, a proper explanation of which requires the canonical picture.
 
In Section \ref{sec:2.3-canonical-general}, we develop the canonical formulation of field theories in general, with a careful accounting of the issue of constraints. To this end, we prescribe here the general recipe for foliating spacetime into constant-time spatial (Cauchy) three-surfaces, such that a notion of time evolution in spacetime can be defined. We also define the canonical phase space and the Hamiltonian equations of motion for general field theories, introducing the basic mathematical methods of symplectic geometry.
 
Then, in section \ref{sec:2.4-canonical-GR}, we proceed to apply this formalism to general relativity in order to formulate it as a canonical theory. In particular, the canonical variables are the induced three-metric on each spatial slice (associated with the choice of foliation), as well as the lapse function and the shift vector (associated with a choice of a time flow vector field), plus their respective conjugate momenta. The lapse and shift are not dynamical variables: their associated equations are first-order in time, and their conjugate momenta vanish. These constitute the constraints of general relativity.
 
Finally, in Section \ref{sec:2.5-applications}, we return in greater detail to the four broad areas of application of canonical general relativity enumerated in the introductory section, and we provide illustrations with explicit examples.
 
\subsection*{Relativitat general canònica \normalfont{(chapter summary translation in Catalan)}}

L’objectiu d’aquest capítol és introduir el llenguatge matemàtic bàsic i la maquinària tècnica de la teoria de la relativitat general seguint mètodes variacionals. Ens centrem especialment en desenvolupar la formulació canònica de la relativitat general, també coneguda com la formulació hamiltoniana o $(3+1)$. En essència, això proporciona una manera de convertir les equacions de camp de segon ordre de la teoria per al tensor mètric de l’espai-temps en un problema d’evolució temporal de primer ordre per al tensor mètrica espacial tridimensional induït i el seu moment conjugat. Tanmateix, també hi ha equacions de restricció a més d'aquestes (restringint les condicions inicials admissibles, així com el seu desenvolupament subjecte a les equacions dinàmiques), i el seu correcte maneig requereix molta subtilesa. L'existència de restriccions en la relativitat general està directament relacionada amb - i ofereix una visió útil sobre - la llibertat de mesura disponible en la teoria.
 
Comencem a la Secció \ref{sec:2.1-intro} amb una breu introducció. Generalment es discuteixen quatre àmplies àrees d’aplicació de la relativitat general canònica: la relativitat matemàtica, la relativitat numèrica, la gravetat quàntica i el tema de l’energia i la quantitat de moviment gravitatòria. Tornem a cadascun d’aquests a la secció final d’aquest capítol amb exemples específics, un cop desenvolupades les eines matemàtiques en detall.
 
A la secció \ref{sec:2.2-lagrangian}, presentem la formulació lagrangiana de les teories de camps clàssics en general, i la relativitat general en particular, formant el punt de partida típic de qualsevol anàlisi canònica. Comentem l’aparició d’equacions de restricció ja a nivell lagrangià, una explicació adequada de la qual es requereix la formulació canònica.
 
A la secció \ref{sec:2.3-canonical-general}, desenvolupem la formulació canònica de les teories de camps en general, amb una acurada explicació del problema de les restriccions. Amb aquesta finalitat, prescrivim aquí la recepta general per foliar l’espai-temps en superfícies espacials tridimensionals de temps constant (superfícies Cauchy), de manera que es pot definir una noció d’evolució en el temps en l’espai-temps. També definim l’espai de fase canònica i les equacions de moviment hamiltonianes per a les teories generals de camp, introduint els mètodes matemàtics bàsics de la geometria simplectica.
 
A continuació, a la secció \ref{sec:2.4-canonical-GR}, procedim a aplicar aquest formalisme a la relativitat general per tal de formular-lo com a teoria canònica. En particular, les variables canòniques són el tensor mètric tridimensional induït en cada llesca espacial (associat a l’elecció de la foliació), així com la funció de lapse i el vector de shift (associats amb una elecció d’un camp vectorial de flux de temps), més els seus moments conjugats respectius. El lapse i el shift no són variables dinàmiques: les seves equacions associades són de primer ordre en el temps i els seus moments conjugats desapareixen. Aquests constitueixen les restriccions de la relativitat general.
 
Finalment a la secció \ref{sec:2.5-applications}, tornem amb més detall a les quatre àmplies àrees d’aplicació de la relativitat general canònica enumerades a la secció introductòria, i proporcionem il·lustracions amb exemples explícits.
 
\subsection*{Relativité générale canonique  \normalfont{(chapter summary translation in French)}}

Le but de ce chapitre est de présenter le langage mathématique de base et la machinerie technique de la théorie de la relativité générale suivant les méthodes variationnelles. Nous nous concentrons particulièrement sur le développement de la formulation canonique de la relativité générale, également connue sous le nom de la formulation hamiltonienne ou $(3+1)$. En substance, cela fournit un moyen de transformer les équations de champ de deuxième ordre de la théorie pour le tenseur métrique de l’espace-temps en un problème d'évolution temporelle de premier ordre pour le tenseur métrique spatial trois-dimensionnel induit et son moment conjugué. Cependant, il existe également des équations de contrainte (restreignant les conditions initiales admissibles ainsi que leur développement en fonction des équations dynamiques), et leur traitement correct nécessite beaucoup de subtilité. L’existence de contraintes dans la relativité générale est en fait directement liée à - et offre un éclairage utile sur - la liberté de jauge disponible dans la théorie.

Nous commençons à la section \ref{sec:2.1-intro} avec une brève introduction. Nous traitons généralement quatre grands domaines d’application de la relativité générale canonique : la relativité mathématique, la relativité numérique, la gravitation quantique et la question de l’énergie et la quantité de mouvement gravitationnelles. Nous reviendrons sur chacun d’eux dans la dernière section de ce chapitre avec des exemples spécifiques, une fois que nous aurons développé les outils mathématiques en détail.

Dans la section \ref{sec:2.2-lagrangian}, nous présentons la formulation lagrangienne des théories de champ classiques en général, puis de la relativité générale en particulier, constituant le point de départ typique de toute analyse canonique. Nous commentons sur l’apparition d’équations de contraintes déjà au niveau lagrangien, dont l’explication correcte nécessite la formulation canonique.

Dans la section \ref{sec:2.3-canonical-general}, nous développons la formulation canonique des théories de champ en général, tenant en compte la question des contraintes. À cette fin, nous prescrivons ici la recette générale de feuilleter l'espace-temps en surfaces spatiales trois-dimensionnelles à temps constant (surfaces de Cauchy), de manière à pouvoir définir une notion d'évolution temporelle dans l'espace-temps. Nous définissons également l’espace des phases canonique et les équations hamiltoniennes du mouvement pour les théories générales du champ, en introduisant les méthodes mathématiques de base de la géométrie symplectique.

Ensuite, dans la section \ref{sec:2.4-canonical-GR}, nous appliquons ce formalisme à la relativité générale afin de la formuler en tant que théorie canonique. En particulier, les variables canoniques sont le tenseur métrique trois-dimensionnel induite sur chaque tranche spatiale (associée au choix de la foliation), et la fonction de déchéance (\textit{lapse}) et le vecteur de décalage (\textit{shift}) (associés au choix d’un champ de vecteurs de flux temporel), ainsi que leurs moments conjugués respectifs. La déchéance et le décalage ne sont pas de variables dynamiques : leurs équations associées sont du premier ordre dans le temps et leurs moments conjugués disparaissent. Celles-ci constituent les contraintes de la relativité générale.

Enfin, dans la section \ref{sec:2.5-applications}, nous reviendrons plus en détail sur les quatre grands domaines d’application de la relativité générale canonique énumérés dans la section introductive et nous fournissons des illustrations avec des exemples explicites.

\section{Introduction\label{sec:2.1-intro}}

There
are a number of diverse motivations for casting GR into a canonical
form, and for our choice to introduce the topic in this way. We begin by enumerating
four broad areas of interest, and comment more on each, focusing on specific examples
of applications, in the final section of this chapter.
\begin{enumerate}
\item Mathematically, canonical methods provide a very useful way to develop
the sort of geometrical tools generally used for studying subsets
of spacetimes, in particular by splitting them up into (usually, families of lower-dimensional)
hypersurfaces via some established procedure. The classical Hamiltonian
approach splits spacetime into spatial slices defined, for example,
by the constancy of a time function, and is thus adapted to studying
``the entire space'' at different instants of time. Similar
technical constructions can be employed for studying the dynamics
of finite (bounded) systems within a spacetime throughout some span of time; in such a case, one could foliate spacetime, for instance, by
the constancy of a radial function in order to study the dynamics of the resulting worldtubes. (We shall see more along these lines in Chapter \ref{5-motion}.) Spacetime splittings
of this sort, and especially $(3+1)$ splittings, have supplied the
basic framework for many important results in the mathematics of GR.\\
~
\item Practically, canonical methods form the basis of numerical relativity---that
is, formulating the Einstein equation as a suitable set of time-dependent
partial differential equations (PDEs) which, given some appropriate initial data, can be evolved on
computers to obtain numerical solutions. Simulations of strongly dynamical
gravitational systems rely critically on methods of this sort. \\
~
\item In going beyond GR, in particular in seeking theories of quantum gravity,
canonical methods are often regarded as a key connection between the
languages of GR and quantum mechanics. Indeed, typical canonical quantization
procedures follow some variant of transforming classical canonical variables
into operators on a Hilbert space of quantum states. Loop quantum gravity, for example,
is a candidate theory of quantum gravity which essentially attempts
to do this for gravitational canonical variables defined in a suitable way.
\\
~
\item Last but not at all least, from a physical point of view, canonical
methods form the traditional starting point for understanding the
notion of gravitational energy-momentum. In particular, they can supply
definitions of gravitational energy-momentum of an entire spacetime
under some specific conditions, \textit{e.g.} the Arnowitt-Deser-Misner (ADM) energy-momentum for asymptotically-flat 
spacetimes. On the other hand, canonical methods are not designed
to be able to say much more than this, and in particular, anything
about the gravitational energy-momentum of a \emph{finite} spatial region
within some spacetime. For the latter, methods such as the worldtube
boundary splittings mentioned in (1) are better designed, yet to this
day no general consensus exists among relativists on the ``best''
way to do this. We will comment more on this in later chapters of
this thesis, but we end here by remarking that, nevertheless, any
proposed definition for gravitational energy-momentum of a finite
system is generally expected to agree with, \textit{e.g.}, the ADM energy-momentum
in the flat asymptotic limit.
\\
~
\end{enumerate}
Canonical GR encompasses a variety of possible formulations of GR
in terms of some chosen canonical variables (configurations and their
conjugate momenta). The first canonical formulation of GR was achieved in 1950,
following a quantum gravity motivation, by [\cite{pirani_quantization_1950}; \cite{pirani_quantization_1951}],
and independently the following year by [\cite{anderson_constraints_1951}]. 

Then, 
[\cite{dirac_generalized_1958,dirac_theory_1958}] formulated the general framework for working with constrained Hamiltonian
systems, a topic we systematically develop in this chapter before
applying it to GR. Beginning in the following year, Arnowitt, Deser and Misner [\cite{arnowitt_dynamical_1959}] devised the first coordinate-independent
canonical formulation of GR, since then known eponymously as the \emph{ADM formulation}. Following a series of further papers in the ensuing years, the authors summarized their results in a 1962 review article [\cite{arnowitt_dynamics_1962}], republished more recently [\cite{arnowitt_republication_2008}]. Today, it remains undoubtedly the most famous basic canonical formulation
of GR. 

Over the decades, other formulations have been developed in response
to the application needs---\textit{e.g.}, the Ashtekar variable formulation
used in quantum gravity, the Baumgarte-Shapiro-Shibata-Nakamura (BSSN) and generalized harmonic formulations used widely in numerical
relativity (upon which we will elaborate further in the final section),
and numerous others.

Our presentation of canonical GR in this chapter is based, in its broadest outlines, on Chapter 10 and Appendix E of  [\cite{wald_general_1984}], in combination with Chapter 3 of [\cite{bojowald_canonical_2011}] (especially for the general formulation canonical theories and constraints). See also Chapter 4 of [\cite{poisson_relativists_2007}] for many of the the step-by-step computations,
largely omitted here in \textit{lieu} of directly stating the main results. 

For mathematical clarifications, we generally refer to [\cite{lee_introduction_2002}] for geometry (see also [\cite{nakahara_geometry_2003}], written with more of a view towards physics), and [\cite{evans_partial_1998}] for PDE theory.


\section{Lagrangian formulation\label{sec:2.2-lagrangian}}

\subsection{Lagrangian formulation of general field theories}
Let $(\mathscr{M},\bm{g},\bm{\nabla})$ be any $(3+1)$-dimensional spacetime. Suppose
that we are interested in a theory describing a collection of fields
$\psi=\{\psi^{A}(x^{a})\}_{A\in\mathscr{I}}$ in $\mathscr{M}$, where
$A\in\mathscr{I}$ is a general (possibly multi-) index for the fields
$\psi^{A}$ (and will be accordingly omitted if understood), \textit{i.e.} it can include tensor indices, field indices \textit{etc.} For example, if we are considering only gravity, then $\psi=\bm{g}$, \textit{i.e.} our collection of fields includes only the spacetime metric $g_{ab}$. If we are considering gravity coupled to a matter field, for example Maxwellian electromagnetism, then $\psi=\{\bm{g},\bm{F}\}$, where $F_{ab}$ is the Faraday tensor.

Let $S[\psi]$ be a functional of $\psi$. Let $\{\psi_{(\lambda)}\}_{\lambda\in\mathbb{R}}$
be a smooth one-parameter family of field values and let $\delta\psi^{A}=(\partial_{\lambda}\psi_{(\lambda)}^{A})|_{\lambda=0}$.
For all such families, suppose moreover that $(\partial_{\lambda}S[\psi_{(\lambda)}])|_{\lambda=0}$
exists and also that there exists a smooth field $\chi_{A}$ dual
to $\psi^{A}$ (meaning that if $\psi^{A}\in\mathscr{T}^{k}\,_{l}(\mathscr{M})$,
then $\chi_{A}\in\mathscr{T}^{l}\,_{k}(\mathscr{M})$), such that
\begin{equation}
\left.\left(\frac{\partial S}{\partial\lambda}\right)\right|_{\lambda=0}=\int_{\mathscr{M}}\bm{e}\,\chi_{A}\delta\psi^{A}\,.\label{eq:dS/dlambda}
\end{equation}
Here, for reasons that will become transparent shortly, we choose
to write the integral with respect to the Minkowski volume form\footnote{~In fact, $\bm{e}$ can be chosen to be \emph{any} Lorentzian volume form the non-vanishing components
of which take the values $\pm1$. See Appendix E of [\cite{wald_general_1984}] for a more detailed discussion
as to why the present construction is actually independent of the
choice of a volume form satisfying this property.}, 
\begin{equation}
\bm{e}={\rm d}x^{0}\wedge{\rm d}x^{1}\wedge{\rm d}x^{2}\wedge{\rm d}x^{3}={\rm d}^{4}x\,.\label{eq:vol_form_Minkowski}
\end{equation}
The factor of $\sqrt{-g}$, with $g={\rm det}(\bm{g})$, multiplying the above to yield the volume
form of $\mathscr{M}$, 
\begin{equation}
\bm{\epsilon}_{\mathscr{M}}^{\,}=\sqrt{-g}\,\bm{e}\,,\label{eq:vol_form_M}
\end{equation}
is absorbed into the definition of $\chi_{A}$. Then,
$S[\psi]$ is said to be \emph{functionally differentiable} at $\psi=\psi_{(0)}$
and its \emph{functional derivative} is defined as
\begin{equation}
\left.\left(\frac{\delta S}{\delta\psi^{A}}\right)\right|_{\psi_{(0)}}=\chi_{A}\,.
\end{equation}

\begin{figure}
\begin{centering}
\includegraphics[scale=0.6]{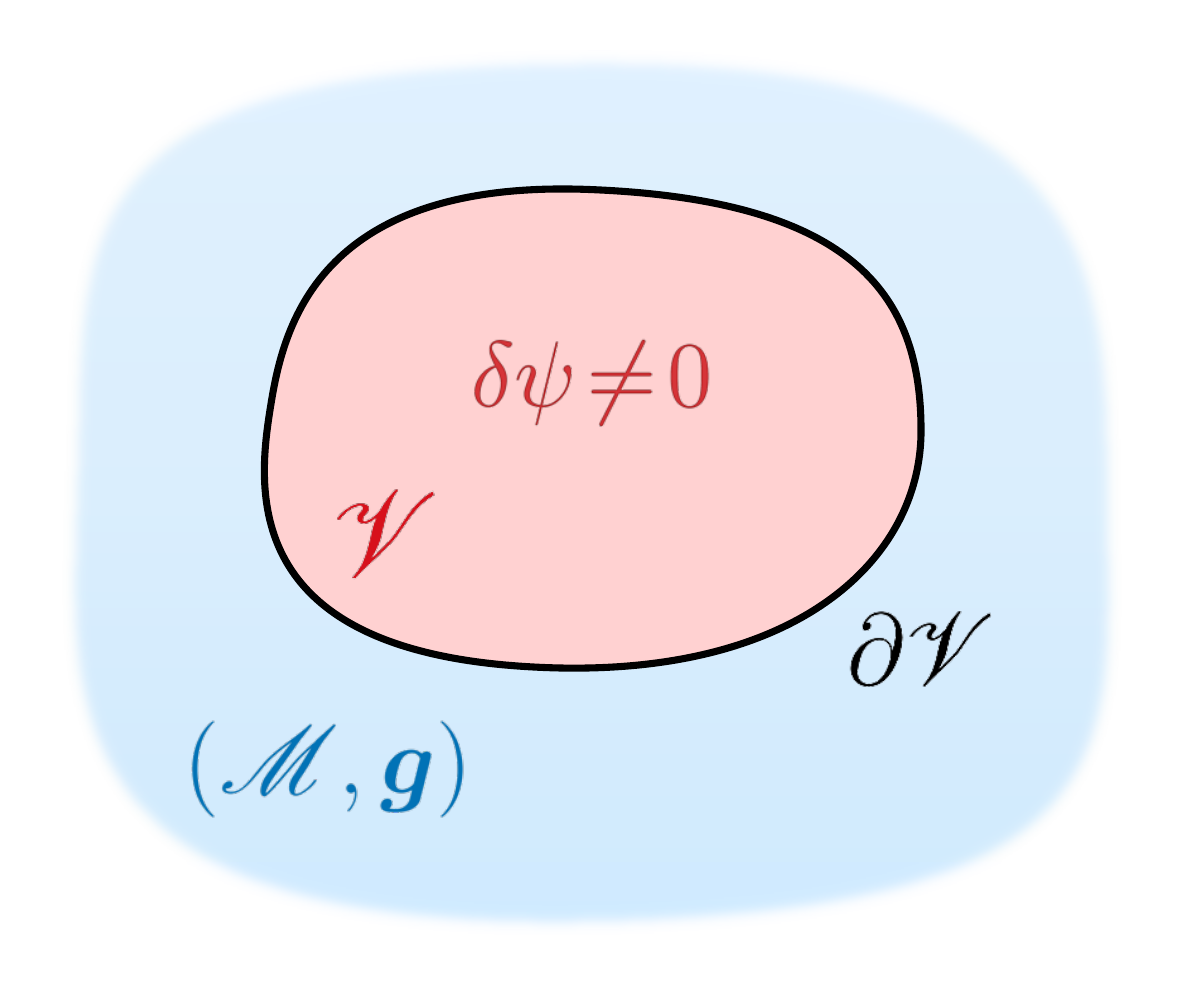}
\par\end{centering}
\caption{A compact region $\mathscr{V}$ in a spacetime $\mathscr{M}$ where the variation of physical fields are non-zero.}\label{2-fig-V}
\end{figure}

We will now focus our attention upon a certain class of such functionals
$S$. Let $\mathscr{V}$ be a compact region in $\mathscr{M}$
such that ${\rm supp}(\delta\psi^{A})\subset\mathscr{V}$ (\textit{i.e.} $\delta\psi^{A}$
takes non-zero values only in the interior of $\mathscr{V}$). See Fig. \ref{2-fig-V}. We assume that $S$ has the form
\begin{equation}
S\left[\psi\right]=\int_{\mathscr{V}}\bm{e}\,\mathcal{L}\left[\psi\right]\,,\label{eq:action_general}
\end{equation}
such that 
\begin{equation}
\mathcal{L}\left[\psi\right]=\sqrt{-g}f\left(\psi^{A},\bm{\nabla}\psi^{A},\ldots,\bm{\nabla}\cdots\bm{\nabla}\psi^{A}\right)
\end{equation}
where $f$ is a local function of $\psi^{A}$ and a finite number
of its derivatives. If $S$ is functionally differentiable
and extremized at the field values $\psi^{A}$ which are solutions
to the field equations of the theory, then $S$ is referred
to as an \emph{action}. We then refer to $\mathcal{L}$ as the Lagrangian
density, and the specification of such an $\mathcal{L}$ is what is
meant by a Lagrangian formulation of the theory. Note that we may
have $\bm{g}\in\psi$ (\textit{i.e.} the gravitational field may be included
in the theory, as in GR), and this is the reason for which we have preferred
to simply absorb the $\sqrt{-g}$ (in this case, $\psi$ dependent)
factor into $\mathcal{L}$, and thus to write $S$ as an integral
with respect to the flat volume form $\bm{e}$ instead of the natural spacetime volume form $\bm{\epsilon}_{\mathscr{M}}$.

All major theories of classical physics, including GR, admit a Lagrangian formulation.
In other words, their field equations are equivalent to the extremization
of an action $S[\psi]$ with respect to their physical fields $\psi$, which in turn
can be shown to be equivalent to a system of PDEs known as the \emph{Euler-Lagrange
equations}. For field theories of typical interest, including GR and Maxwellian
electromagnetism (EM), $\mathcal{L}$ depends on $\psi^{A}$ and its first derivatives
only, \textit{i.e.} $\mathcal{L}=\sqrt{-g}f(\psi^{A},\bm{\nabla}\psi^{A})$. In this case, these equations are
\begin{equation}
0=\frac{\delta S}{\delta\psi^{A}}\Leftrightarrow0=\frac{\partial\mathcal{L}}{\partial\psi^{A}}+\frac{1}{2}\frac{\partial\mathcal{L}}{\partial(\nabla_{a}\psi^{A})}\nabla_{a}\ln\left(-g\right)-\nabla_{a}\left(\frac{\partial\mathcal{L}}{\partial(\nabla_{a}\psi^{A})}\right)\,.
\end{equation}
These are second-order PDEs for $\psi^{A}$.

Now fix a coordinate system $\{x^{\alpha}\}=\{t,x^{i}\}$. Clearly, all
terms which are second order in the derivatives of $\psi$ will emerge
from the final term of the above equation: by implicit differentiation,
this is
\begin{equation}
\nabla_{\alpha}\left(\frac{\partial\mathcal{L}}{\partial(\nabla_{\alpha}\psi^{A})}\right)=\frac{\partial^{2}\mathcal{L}}{\partial(\nabla_{\alpha}\psi^{A})\partial\psi^{B}}\nabla_{\alpha}\psi^{B}+\frac{\partial^{2}\mathcal{L}}{\partial(\nabla_{\alpha}\psi^{A})\partial(\nabla_{\beta}\psi^{B})}\nabla_{\alpha}\nabla_{\beta}\psi^{B}\,.
\end{equation}
The coefficient of the second term on the RHS above is known as the
\emph{principal symbol} of the PDE; the $\alpha=t=\beta$ component of
this term contains all the second time derivatives of the fields.
Thus, the Euler-Lagrange equations have the form
\begin{equation}
0=W_{AB}\partial_{t}^{2}\psi^{B}+l_{A}\,,
\end{equation}
where we have defined
\begin{equation}
W_{AB}=\frac{\partial^{2}\mathcal{L}}{\partial(\nabla_{t}\psi^{A})\partial(\nabla_{t}\psi^{B})}\,,\label{eq:W_AB-Lagrangian}
\end{equation}
and $l_{A}$ indicates lower (\textit{i.e.} first and zeroth) order terms in
time derivatives. It is thus apparent that if and only if $W_{AB}$
is non-degenerate in the indices $A,B\in\mathscr{I}$ will we be able
to obtain a complete set of solutions to the coupled set of equations, \textit{i.e.} a set of $n={\rm card}(\mathscr{I})$ (the cardinality of
$\mathscr{I}$) solutions. If that
is the case, we would be able to invert $W_{AB}$ and write the complete
system explicitly as
\begin{equation}
0=\partial_{t}^{2}\psi^{B}+\left(W^{-1}\right)^{AB}l_{A}\,.
\end{equation}

If $W_{AB}$ is degenerate, however, then $\psi^{A}$ and its derivatives
up to first order in time and second order in space (and in particular,
their initial conditions for the time evolution problem) cannot take
arbitrary values. Specifically, they must yield an $l_{A}$ which
is in the image of $W_{AB}$ seen as a linear mapping between vector
spaces, the dimension of which is thus less than $n$. Equivalently,
$W_{AB}$ can be seen as a matrix (with indices in $\mathscr{I}$) that has ($n-{\rm rank}(W_{AB})$) null eigenvectors
$v_{\mathsf{j}}^{A}$, with $\mathsf{j}$ (in serif font) used to label the set of these eigenvectors (each having components labelled by $A\in\mathscr{I}$). In other words, $v_{\mathsf{j}}^{A}$ are such that  $v_{\mathsf{j}}^{A}W_{AB}=0$.
Multiplying the Euler-Lagrange equation by $v_{\mathsf{j}}^{A}$ on
the left thus yields the independent equations,
\begin{equation}
0=v_{\mathsf{j}}^{A}l_{A}\,.\label{eq:Lagrangian_constr}
\end{equation}
These equations are known as constraints, since they do not involve
second time derivatives of the fields and are thus not regarded as
``dynamical'' \textit{i.e.} they do not prescribe the time evolution. We
will gain a deeper appreciation for what this means once we pass to
the Hamiltonian picture, but before we do, let us apply our ideas so
far to GR.

\subsection{Lagrangian formulation of GR}

In vacuum, our only field is the gravitational field, $\psi=\bm{g}$.
If, in addition to the requirement that ${\rm supp}(\delta\bm{g})\subset\mathscr{V}$
we also assume that ${\rm supp}(\bm{\nabla}\delta\bm{g})\subset\mathscr{V}$,
then the Einstein equation can be recovered fully from an action of
the form (\ref{eq:action_general}). In particular,
\begin{equation}
S_{\textrm{EH}}\left[\bm{g}\right]=\int_{\mathscr{V}}\bm{e}\,\mathcal{L}_{\textrm{EH}}\left[\bm{g}\right]\,,\quad\mathcal{L}_{\textrm{EH}}=\frac{1}{2\kappa}\sqrt{-g}R\,,\label{eq:EH_action}
\end{equation}
are the Einstein-Hilbert action and Lagrangian respectively, with $\kappa$ denoting the Einstein constant, $\kappa=8\pi G/c^{4}=8\pi$ in units of $G=1=c$. This formulation of GR was first proposed by [\cite{hilbert_grundlagen_1915}].

In modern Lagrangian formulations of GR, however, it is typical not to assume anything about the support of
$\bm{\nabla}\delta\bm{g}$, and in particular its values on the bounday
$\partial\mathscr{V}$. Equivalently, only the metric components (and not the derivatives thereof) are to be regarded as being ``held fixed on the boundary'' when one ``varies the action''. In this case, one must add a boundary term
to (\ref{eq:EH_action}) in order to cancel contributions involving $\bm{\nabla}\delta\bm{g}$. In
particular, this is known as the \emph{Gibbons-Hawking-York boundary term}, first proposed by [\cite{york_role_1972}] and later developed by [\cite{gibbons_action_1977}].
It is given by the integral of the trace of the \emph{extrinsic
curvature} (or \emph{second fundamental form}) of $\partial\mathscr{V}$,
\begin{equation}
K_{ab}=\gamma_{ac}\nabla^{c}n_{b}\,,\label{eq:K-ext-curv}
\end{equation}
where $\bm{\gamma}=\bm{g}|_{\partial\mathscr{V}}$ is the induced
metric on $\partial\mathscr{V}$ and $\bm{n}$ is the normal vector thereto. Thus, including a (boundary) integral of $K={\rm tr}(\bm{K})$ in the action with the appropriate
numerical factor, we will henceforth take
\begin{equation}
S_{\textrm{G}}\left[\bm{g}\right]=S_{\textrm{EH}}+\frac{1}{\kappa}\int_{\partial\mathscr{V}}\bm{\epsilon}^{\,}_{\partial\mathscr{V}}\,K\label{eq:G_action}
\end{equation}
to be the total gravitational action, \textit{i.e.} the action of GR. For more on this, see also [\cite{brown_action_2002}] and Chapter 12 of [\cite{padmanabhan_gravitation:_2010}]. 

By direct computation, 
prior to imposing $\delta\bm{g}|_{\partial\mathscr{V}}=0$, one finds that for a family $\bm{g}_{(\lambda)}$ in the
sense of the previous subsection, 
\begin{equation}
\left.\left(\frac{\partial}{\partial\lambda}S_{\textrm{G}}\left[\bm{g}_{(\lambda)}\right]\right)\right|_{\lambda=0}=\frac{1}{2\kappa}\left(\int_{\mathscr{V}}\bm{\epsilon}^{\,}_{\mathscr{M}}\,\bm{G}:\delta\bm{g}+\int_{\partial\mathscr{V}}\bm{\epsilon}^{\,}_{\partial\mathscr{V}}\,\bm{\Pi}:\delta\bm{\gamma}\right)\,.\label{eq:GR_action_variation}
\end{equation}
Here, $\bm{G}$ is the Einstein tensor of $\bm{g}$ and $\bm{\Pi}$ is the so-called
\emph{canonical momentum} of $\partial\mathscr{V}$ (the nomenclature of which will become clearer
when we pass to the Hamiltonian formulation in the next section), given in terms of the extrinsic curvature by $\bm{\Pi}=\bm{K}-K\bm{\gamma}$.
Thus, from (\ref{eq:GR_action_variation}), the stationary action principle yields the vacuum Einstein equation,
\begin{equation}
\bm{G}=0\,,\label{eq:EE}
\end{equation}
provided only that $\delta\bm{\gamma}|_{\partial\mathscr{V}}=0$.

The appearance of constraints can already be seen manifestly at the level of
the full Einstein equation (\ref{eq:EE}). Choose a coordinate system $\{x^{\alpha}\}=\{t,x^{i}\}$.
Then, by direct computation (writing the components of the Einstein
tensor $G_{\alpha\beta}$ purely in terms of those of the metric $\bm{g}$),
one finds that 
\begin{align}
G^{t}\,_{t}=\, & l\,,\label{eq:Gtt}\\
G^{t}\,_{i}=\, & l_{i}\,,\label{eq:Gti}\\
G_{ij}=\, & -\frac{1}{2}g^{tt}\partial_{t}^{2}g_{ij}-\frac{1}{2}g_{ij}(g^{tk}g^{tl}-g^{tt}g^{kl})\partial_{t}^{2}g_{kl}+l_{ij}\,,\label{eq:Gij}
\end{align}
where $l$, $l_{i}$ and $l_{ij}$ (the $l_{A}$ in the previous subsection)
indicate terms lower than second (\textit{i.e.} first or zeroth) order in time
derivatives of $g_{\alpha\beta}$. So we see that the ``time-time''
and ``time-space'' Einstein equations, (\ref{eq:Gtt}) and (\ref{eq:Gti}) respectively, must be constraints, and the
``space-space'' Einstein equations (\ref{eq:Gij}) are the only true dynamical field equations.
This already gives us a crude idea that it is somehow only the ``spatial''
part of the spacetime metric that plays a dynamical role when the
Einstein equation is regarded as a time-dependent problem. To understand
what is meant by this precisely, and to gain a proper appreciation
for the meaning of the constraints, we must pass to the canonical
(Hamiltonian) formulation of the theory.

\section{Canonical formulation of general field theories\label{sec:2.3-canonical-general}}
\subsection{Spacetime foliation\label{sec:2.3.1-foliation}}

A canonical formulation of a field theory in a spacetime $(\mathscr{M},\bm{g},\bm{\nabla})$
presupposes an assumption called \emph{global hyperbolicity}. There
exist a number of equivalent definitions for what this means, but
for our purposes the most suggestive one to state is the following:
$\mathscr{M}$ is said to be globally hyperbolic if and only if it
admits the following topology: 
\begin{equation}
\mathscr{M}\simeq\mathbb{R}\times\Sigma\,,\label{eq:global-hyp}
\end{equation}
where $\Sigma$ is a \emph{Cauchy surface}\textemdash a closed set
which does not intersect its chronological future, and the domain
of dependence of which is the entire spacetime. As we wish to refrain
from entering here into any further technicalities pertaining to general-relativistic
causal structure, a broad and important subject in its own right,
we refer the interested reader to [\cite{hawking_large_1975}] and
Chapter 8 of [\cite{wald_general_1984}] for precise topological
definitions of these terms. Physically, $\Sigma$ can be thought of
as representing the entire (three-dimensional) space at a given instant of time. 

\begin{figure}
\begin{centering}
\includegraphics[scale=0.6]{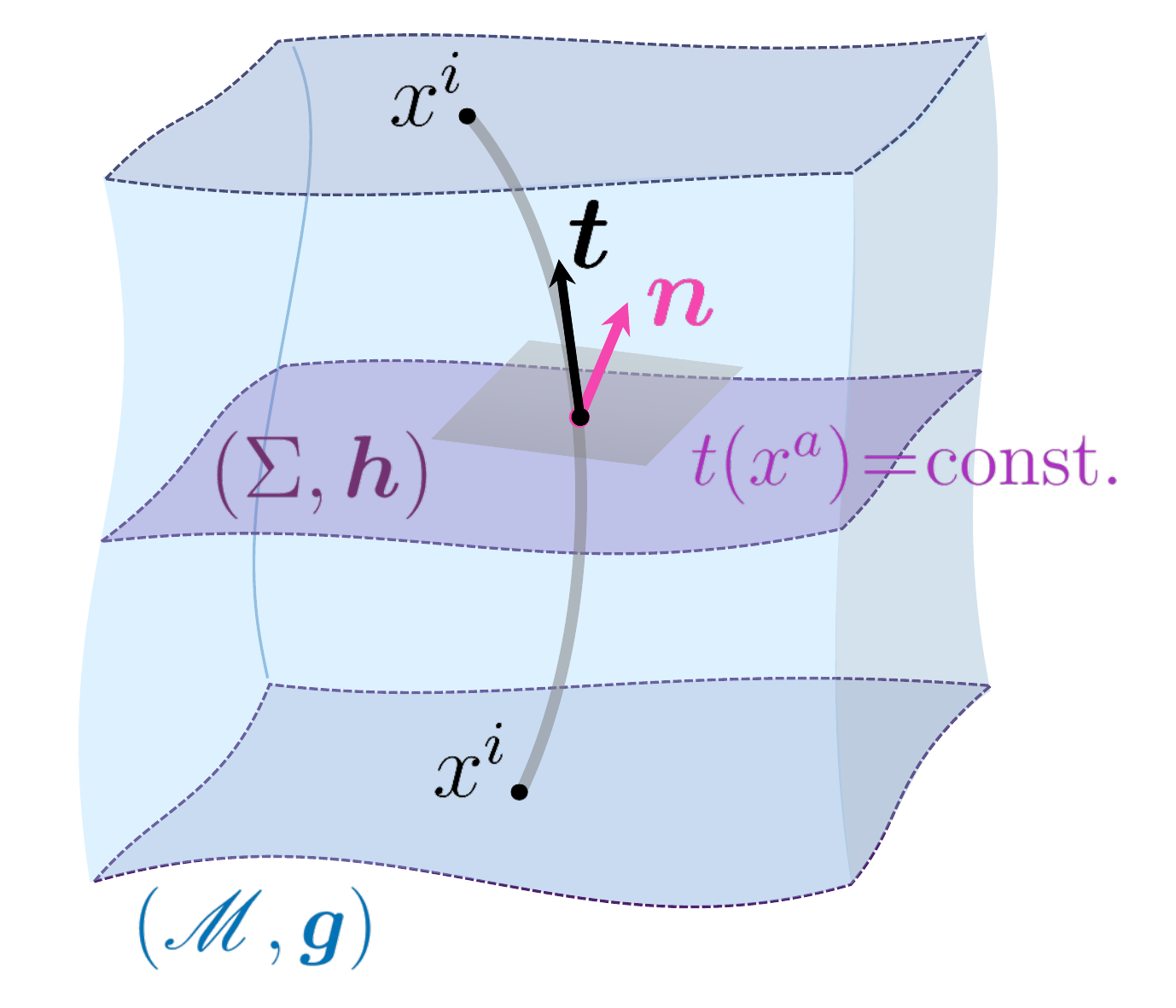}
\par\end{centering}
\caption{A depiction (in $(2+1)$ dimensions) of the foliation of a spacetime $(\mathscr{M},\bm{g})$ into Cauchy surfaces $(\Sigma,\bm{h})$, where $\bm{h}$ is the metric induced on $\Sigma$ by $\bm{g}$. These surfaces are defined by the constancy of a time function, $t(x^{a})=\rm{const.}$, which uniquely determine a normal vector $\bm{n}$. Additionally, one must define a time flow vector field $\bm{t}$ on $\mathscr{M}$ the integral curves of which intersect the``same spatial point'' (with the same coordinates $x^{i}$) on different slices.}\label{2-fig-Sigma}
\end{figure}

Thus, given (\ref{eq:global-hyp}), a canonical formulation must begin
with a specification of the meaning of ``time'' and ``change in
time''. This means that one must specify a choice of a foliation
of spacetime into ``constant time'' Riemannian three-slices (``instants
of time''), as well as a ``time direction'' (indicating how one
identifies spatial points on the slices at ``different times'').
Typically one does this by introducing, respectively, a \emph{time
function} $t(x^{a})$ on $\mathscr{M}$ such that $\nabla^{a}t$ is
everywhere timelike (which is always possible if $\mathscr{M}$ is
globally hyperbolic), along with a \emph{time flow vector field} $t^{a}$
in $T\mathscr{M}$ such that $t^{a}\nabla_{a}t=1$ (intuitively ensuring
that the interpretation of ``time'' implied by these two objects
is consistent). These are shown in Fig. \ref{2-fig-Sigma}. 

In this way, the surfaces of constant $t$ in $\mathscr{M}$ are spacelike
Cauchy surfaces, and the integral curves of $\bm{t}$ define a mapping
between spatial slices as follows: one identifies the intersections
of any particular integral curve of $\bm{t}$ with all constant $t$
slices as being the ``same spatial point'' (\textit{i.e.} as corresponding
to the same ``spatial'' coordinate $x^{i}$). The condition $\nabla_{\bm{t}}t=1$
guarantees that any integral curve of $\bm{t}$ will intersect any
constant $t$ slice exactly once, making the identification well-defined.

\begin{figure}
\begin{centering}
\includegraphics[scale=0.6]{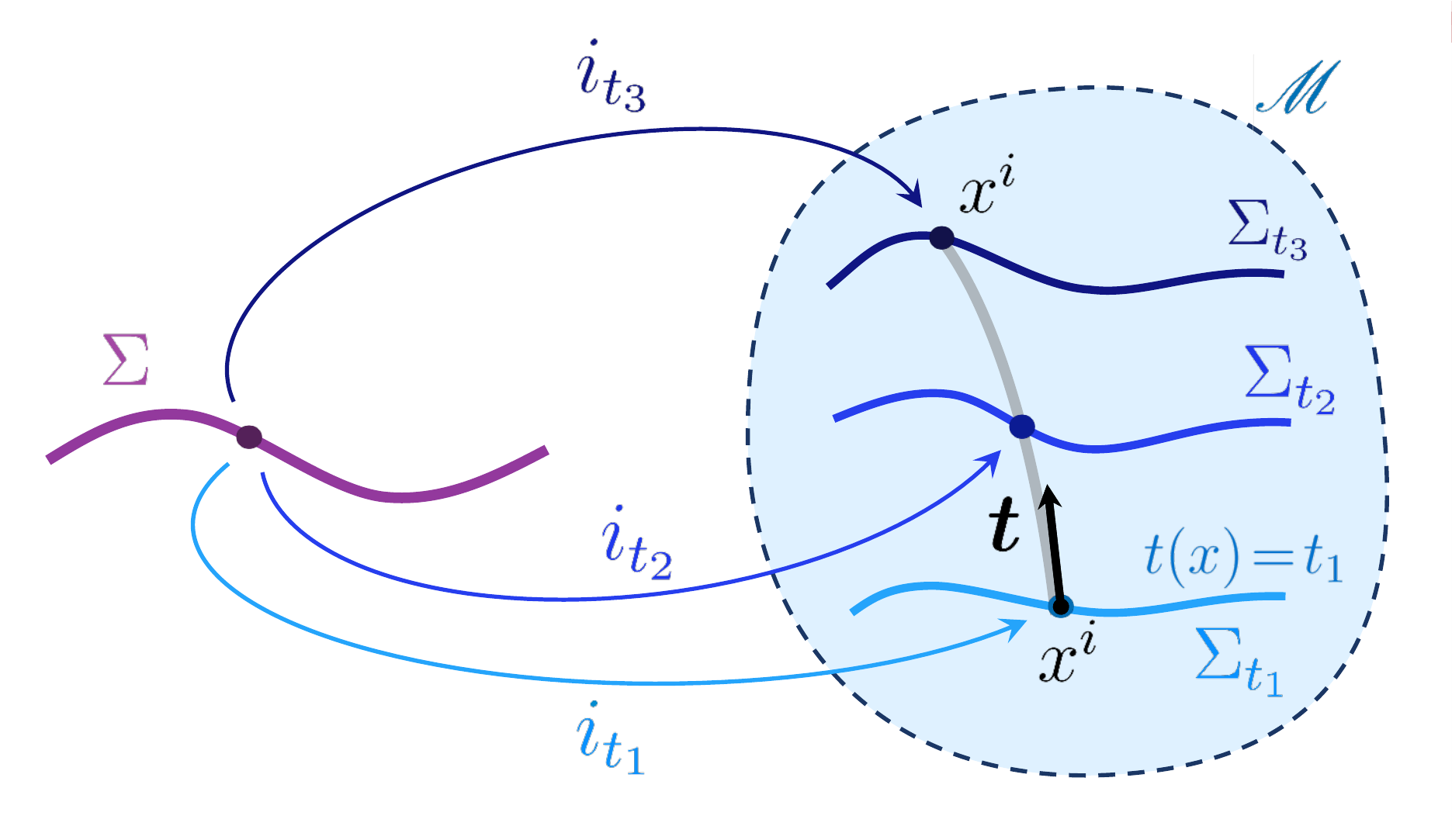}
\par\end{centering}
\caption{A depiction (in $(1+1)$ dimensions) of the spacetime
$\mathscr{M}$ as constituted by a family of embedded submanifolds
$\Sigma_{t}$ obtained from an embedding map $i_{t}:\Sigma\rightarrow\mathscr{M}$.
Three such submanifolds are shown at three different times, with the
time flow vector field identifying the spatial coordinates between
them.}\label{2-fig-i_t}
\end{figure}

A very useful equivalent picture of this construction can be phrased
in the language of the theory of embeddings. (See, \textit{e.g.}, [\cite{giulini_dynamical_2014}]
for more on this.) In particular, the global hyperbolicity condition
(\ref{eq:global-hyp}) implies the existence of a one-parameter family
of embedding maps 
\begin{equation}
i_{t}:\Sigma\rightarrow\mathscr{M}\label{eq:Cauchy_embedding}
\end{equation}
of a (``time-evolving'') Cauchy surface $\Sigma$ into the spacetime
(``at different times''), such that $\Sigma_{t}=i_{t}(\Sigma)\subset\mathscr{M}$
constitute the (spacelike) Riemannian slices of $\mathscr{M}$. In
particular, for any spatial point $p\in\Sigma$, the two spacetime
points $i_{t_{1}}(p)\in\Sigma_{t_{1}}$ and $i_{t_{2}}(p)\in\Sigma_{t_{2}}$
are identified as the ``same spatial point at different times''.
Specifying such a one-parameter family (\ref{eq:Cauchy_embedding})
of embeddings is mathematically equivalent to specifying a time function
and a time flow vector field in $\mathscr{M}$. See Fig. \ref{2-fig-i_t} for a visual
representation. 

Before defining additional structures, we can already use the above
to specify what we mean precisely by the ``time derivative'' of
a given quantity. Let $\bm{A}\in\mathscr{T}^{k}\,_{l}(\mathscr{M})$
be any tensor in our spacetime. Then we define its time rate of change,
or ``time derivative'', $\dot{\bm{A}}$ simply as its Lie derivative
$\mathcal{L}$ along the time flow vector field $\bm{t}$, 
\begin{equation}
\dot{\bm{A}}=\mathcal{L}_{\bm{t}}\bm{A}\,.
\end{equation}

Let $n^{a}$ be the unit (future-oriented) normal to $\Sigma$. Note
that this is uniquely determined once $t(x^{a})$ is specified. (In
particular, $\bm{n}=-(\bm{\nabla}t)/\sqrt{(\bm{\nabla}t)\cdot(\bm{\nabla}t)}$,
with the minus sign ensuring the future orientation.) We thus have
a natural spatial volume (three-) form $\epsilon_{abc}^{\Sigma}=\epsilon_{abcd}^{\mathscr{M}}n^{d}$
induced by the spacetime volume form $\bm{\epsilon}_{\mathscr{M}}^{\,}$
(through projection with $\bm{n}$). However, as in the Lagranigian
formulation, it will be convenient to work instead with flat volume
forms: in particular, $\bm{e}$ on $\mathscr{M}$ (as before), and
${\bf e}$ on $\Sigma$. The latter is notationally distinguished
from the former by writing it in upright rather than italic font (though
the context usually leaves little danger for confusion), and its components
are obtained by projecting $\bm{e}$ with $\bm{t}$, \textit{i.e.} ${\rm e}_{abc}=e_{abcd}t^{d}$.

Although we will endeavor to refrain from entering here into excessive
geometrical technicalities, a few further definitions and citations
of some mathematical results are useful before returning---as we
promise we shall, and thereby in a very illustrative way---to physics. 

First, let $\bm{h}=\bm{g}|_{\Sigma}$ be the metric induced by $\bm{g}$
on $\Sigma$. In abstract index notation, we could equivalently translate
this into the object $h_{ij}$ \textit{i.e.} as a tensor in $\Sigma$ (an embedded
submanifold of $\mathscr{M}$), or as $h_{ab}$, \textit{i.e.} a tensor on
the full spacetime $\mathscr{M}$. In principle and unless otherwise
made explicit, we prefer to retain the meaning of all abstract geometric
expressions as referent to quantities living in $\mathscr{M}$ (so,
\textit{e.g.}, $\bm{h}$ a priori equates to the spacetime field $h_{ab}$).
One can always project these into any submanifold $\mathscr{U}\subset\mathscr{M}$,
in particular by contracting the expressions (over all indices) with
the induced metric $\bm{g}_{\mathscr{U}}^{\,}$ corresponding thereto,
whenever desired. Henceforth we use the notation $(\cdot)|_{\mathscr{U}}$
to refer precisely to such a projection of any quantity $(\cdot)$
onto a submanifold $\mathscr{U}$.

Now, the spatial metric $\bm{h}$ naturally determines a compatible
derivative operator $\bm{\mathcal{D}}$ on $\Sigma$. In turn, $\bm{\mathcal{D}}$
defines in the usual way the (spatial) Riemann tensor $\mathcal{R}_{abcd}$
of $\bm{h}$ on $\Sigma$, written in calligraphic font to distinguish
it from the (Roman font) spacetime Riemann tensor $R_{abcd}$ of $\bm{g}$.
The extrinsic curvature of $\Sigma$ is also defined in the usual way, as the projected spacetime derivative
of the normal vector; equivalently, it can also be shown to equal
half the normal Lie derivative of the spatial metric: 
\begin{equation}
K_{ab}=h_{ac}\nabla_{b}n^{c}=\frac{1}{2}\mathcal{L}_{\bm{n}}h_{ab}\,.\label{eq:extrinsic-curvature}
\end{equation}

Using these definitions, one can prove by direct computation the following
relations for the projections onto $\Sigma$ of the spacetime Riemann
tensor and one normal projection of the spacetime Riemann tensor (see,
\textit{e.g.}, Chapter 3 of [\cite{bojowald_canonical_2011}] for the step-by-step
computations): 
\begin{align}
\left.R_{abcd}\right|_{\Sigma}=\, & \mathcal{R}_{abcd}-K_{ad}K_{bc}+K_{ac}K_{bd}\,,\label{eq:CG1}\\
\left.R_{\boldsymbol{n}abc}\right|_{\Sigma}=\, & \mathcal{D}_{c}K_{ab}-\mathcal{D}_{b}K_{ac}\,.\label{eq:GC2}
\end{align}
The first equation (\ref{eq:CG1}) is usually called the \emph{Gauss
equation}, and the second equation (\ref{eq:CG1}) is called the \emph{Peterson-Mainardi-Codazzi
equation} or (especially common in physics, although historically unfair,
as we will shortly clarify) simply the \emph{Codazzi equation}.
These are classic results in the theory of embeddings, first discovered
in the pioneering days of differential geometry in the early-to-mid
19th century. See [\cite{abbena_modern_2006}] for more historical
and mathematical details.

The Gauss equation (\ref{eq:CG1}) was first obtained by its eponym
[\cite{gauss_disquisitiones_1827}] in two dimensions. It became known as the \textit{Theorema
Egregium} (``remarkable theorem''), and has since then remained
one of the most famous results in geometry. It tells us how the curvature---that is, the Riemann
tensor---of the embedded surface ($\Sigma$) relates to that of the
entire manifold ($\mathscr{M}$) through the extrinsic curvature ($\bm{K}$). 

The Peterson-Mainardi-Codazzi equation (\ref{eq:GC2}) was first obtained
by [\cite{peterson_uber_1853}], and later independently by [\cite{mainardi_sulle_1856}]
and [\cite{codazzi_sulle_1868}]. It expresses the projection onto
the hypersurface ($\Sigma$) of one normal projection of the Riemann
tensor of the entire manifold ($\mathscr{M}$) in terms of derivatives
of the extrinsic curvature ($\bm{K}$). 

These are completely general conditions that are satisfied by any
embedding---in this case, of a three-dimensional spacelike hypersurface
into a spacetime. It will be especially interesting for our purposes
to consider the contracted form of these equations: in particular,
by contracting an appropriate pair of spacetime indices in (\ref{eq:CG1})-(\ref{eq:GC2})
and re-expressing the results in terms of the Einstein tensor $\bm{G}=\bm{R}-\frac{1}{2}R\bm{g}$,
one finds by direct computation that 
\begin{align}
2\left.G_{\bm{n}\bm{n}}\right|_{\Sigma}=\, & \mathcal{R}-\boldsymbol{K}:\boldsymbol{K}+K^{2}\,,\label{eq:contractedGC1}\\
\left.G_{\bm{n}a}\right|_{\Sigma}=\, & \mathcal{D}^{b}K_{ab}-\mathcal{D}_{a}K\,.\label{eq:contractedGC2}
\end{align}

We stress once again that these are purely geometrical requirements
that the embedding must satisfy. Yet, though we have apparently said
nothing so far about physics, the cognizant reader will appreciate
that setting the above equations to zero gives precisely the ``time-time''
and ``time-space'' Einstein equations in typical canonical form,
and we already know from our earlier discussion at the end of the
previous section that these contain no second time derivatives. Hence
it seems that we have already obtained the canonical constraints of
GR without yet essentially doing anything (except anticipating the
vacuum Einstein equation) from the point of view of physics! All that
we have done is to set up the hypersurface embedding, as is needed
for our subsequent canonical formulation (of any field theory).

Finally, it is worth pointing out here one final remark which follows directly
from these geometrical identities, and which also sheds some
insight into physics. In particular, one can contract the Gauss equation
(\ref{eq:CG1}) to express the Ricci scalar $R$ of $\bm{g}$ in terms
of the Ricci scalar $\mathcal{R}$ of $\bm{h}$ and the extrinsuc
curvature $\bm{K}$ of $\Sigma$. The result is [\cite{giulini_dynamical_2014}]:
\begin{equation}
R=\mathcal{R}+\left(\bm{K}:\bm{K}-K^{2}\right)+2\bm{\nabla}\cdot\left(K\boldsymbol{n}-\boldsymbol{\nabla}_{\bm{n}}\bm{n}\right)\,.
\end{equation}
Up to a factor, this is of course simply the Lagrangian of GR in the
bulk (\textit{i.e.} the Einstein-Hilbert Lagrangian). As the final term is
a divergence and hence will only contribute a boundary term, we see
from this that the gravitational action in the bulk is simply:
\begin{equation}
S_{\textrm{G}}|_{{\rm int}(\mathscr{V})}=\frac{1}{2\kappa}\int_{\mathscr{V}}\bm{\epsilon}_{\mathscr{M}}^{\,}\,\left[\mathcal{R}+\left(\bm{K}:\bm{K}-K^{2}\right)\right]\,.
\end{equation}
We have, in this way, a heuristic conceptual link to the meaning of the Lagrangian
in classical particle mechanics as the ``kinetic minus potential
energy'': the spatial curvature scalar $\mathcal{R}$ can be regarded
as minus the ``gravitational potential energy'' (so that, the greater
the curvature, the greater the magnitude of the ``potential
energy'', negatively-signed overall) and the extrinsic curvature
terms $(\bm{K}:\bm{K}-K^{2})$ as the ``gravitational kinetic energy''
(an analogy that becomes clearer later when we see how the extrinsic
curvature is essentially equivalent to the canonical gravitational
momentum, such that these are ``momentum squared'' terms). Of course
this analogy is rather vague and not meant to be taken too formally; we will carefully treat
the basic questions surrounding notions of ``gravitational energy''
at the end of this chapter once we have established the full canonical
formulation.

\subsection{Phase space\label{sec:2.3.2-phase-space}}

Now that we have a splitting of our spacetime, 
\begin{equation}
\mathscr{M}=\bigcup_{t}\Sigma_{t}\simeq\mathbb{R}\times\Sigma\,,\label{eq:splitting}
\end{equation}
into Cauchy surfaces $\Sigma$ with all major geometrical constructions
that will be needed in place, we may confidently return to physics. 

The next step in the canonical formulation of a theory is to introduce
the fields by prescribing what is referred to as a \emph{configuration}
$\varphi=\{\varphi^{A}(x^{i})\}_{A}$ on $\Sigma$. Physically, this
is understood to describe the ``instantaneous'' configuration of
the spacetime fields $\psi(x^{a})$, at a particular ``time'' $t$
(correspondent to a particular $\Sigma_{t}$ in the spacetime foliation).
Thus it is usually (though by no means necessarily, as one has freedom
in how exactly to proceed) defined simply by a direct projection onto
$\Sigma$ of $\psi$, \textit{i.e.} $\varphi=\psi|_{\Sigma}$. (In fact, while
this is a natural starting point, often this definition does not by
itself strictly suffice; in particular, there may be important degrees
of freedom lost in the projection, and one must devise a procedure
for taking these into account too. As we shall see and elaborate upon,
this happens in GR.) 

Once the configuration variables $\varphi$ of the theory have all
been defined, one defines 
\begin{equation}
\mathscr{Q}=\left\{ \varphi\left(x^{i}\right)\right\} \label{eq:configuration-space}
\end{equation}
to be the set of \emph{all possible} (\textit{i.e.}, physically/mathematically
permissible) configurations of the collection of fields $\varphi^{A}$.
This is a functional space referred to as the \emph{configuration
space} of the theory.

Now one must also prescribe what are referred to as the \emph{canonical}
(or \emph{conjugate})\emph{ momenta} $\pi=\{\pi_{A}(x^{i})\}_{A}$
of the fields $\varphi^{A}$, such that $\pi_{A}$ is dual to $\varphi^{A}$
in all indices. We will see momentarily how the Lagrangian $\mathcal{L}[\psi]$
can be used to devise such a prescription (given the definition of
$\varphi$). Once this is in hand, the set 
\begin{equation}
\mathscr{P}=\left\{ \left(\varphi,\pi\right)\right\} =T^{*}\mathscr{Q}\label{eq:phase-space}
\end{equation}
of all possible configurations and momenta taken together will simply
constitute the cotangent bundle of the configuration space, $T^{*}\mathscr{Q}$.
This is called the \emph{phase space}\footnote{~The origin of the nomenclature is from statistical mechanics, where
many of these methods were first developed. See [\cite{davies_physics_1977}; \cite{sklar_physics_1995}; \cite{brown_boltzmanns_2009}]
for good historical accounts.} of the theory.

Phase space (for any field theory) is a particular example of what
is referred to as a \emph{symplectic manifold}. Such objects have
been extensively studied by geometers, and today symplectic geometry
is a broad and fruitful area of mathematics in its own right. (See,
\textit{e.g.}, the reviews/books [\cite{silva_symplectic_2006, silva_lectures_2008}; \cite{hofer_symplectic_2011}].)
Historically, this field originated precisely from the advent of classical
Hamiltonian particle mechanics (much as variational calculus and functional
analysis were heavily precipitated by classical Lagrangian particle
mechanics), and so it is worth our time to briefly offer a description
of phase space in symplectic language before moving on to formulate
the Hamiltonian equations of motion. 

In general, a symplectic manifold $(\mathscr{W};\bm{\omega})$ is
any $2n$-dimensional manifold $\mathscr{W}$ equipped with a two-form
$\bm{\omega}$, called a \emph{symplectic form}, provided that the
latter satisfies two conditions: \emph{(i)} $\bm{\omega}$ is closed
(\textit{i.e.} ${\rm d}_{\mathscr{W}}\bm{\omega}=0$, where ${\rm d}_{\mathscr{W}}$
is the exterior derivative on $\mathscr{W}$); \emph{(ii)} $\bm{\omega}$
is non-degenerate (\textit{i.e.} at any point $p\in\mathscr{W}$ and for any
vectors $\boldsymbol{X}_{p},\boldsymbol{Y}_{p}\in T_{p}\mathscr{W}$, if $\imath_{\bm{X}_{p}}\imath_{\bm{Y}_{p}}\bm{\omega}_{p}=0$
$\forall\bm{Y}_{p}\in T_{p}\mathscr{W}$ where $\imath$ is the interior
product on $\mathscr{W}$, then $\bm{X}_{p}=0$)\footnote{~A natural generalization of this notion which has been developed more
recently is that of a \emph{multisymplectic} form: this is defined
as a $k$- (not necessarily two-) form $\bm{\mu}\in\Omega^{k}(\mathscr{V})$
on a manifold $\mathscr{V}$ of any (not necessarily even) dimension,
satisfying the same first property as in the usual symplectic case
(\textit{i.e.} it is closed, ${\rm d}_{\mathscr{V}}\bm{\mu}=0$), along with
a slightly more general version of the second (nondegeneracy) property, called \emph{1-nondegeneracy}:
that is, one requires, for $\bm{X}_{p}\in T_{p}\mathscr{V}$,
that $\imath_{\bm{X}_{p}}\bm{\eta}_{p}=0$ if and only if $\bm{X}_{p}=0$.
For $k=2$ and an even-dimensional $\mathscr{V}$, this recovers the
usual notion of a symplectic form. See [\cite{roman-roy_properties_2019}; \cite{ryvkin_invitation_2019}]
and references therein for more mathematical details. Multisymplectic
geometry has been usefully applied in recent years to the formulation
of canonical field theories, including GR; see [\cite{gaset_multisymplectic_2018,gaset_multisymplectic_2019}; \cite{nissenbaum_multisymplectic_2017}]
and references therein.}.

Having claimed that the phase space $\mathscr{P}=\{(\varphi^{A}(x^{i}),\pi_{A}(x^{i}))\}$
is an example of a symplectic manifold, we ought to show it by producing
the symplectic form. In order to do this, we first need therefore
to further clarify what we mean by an exterior derivative ${\rm d}_{\mathscr{P}}$
on $\mathscr{P}$. Because we are dealing here with functional spaces,
we must use the \emph{functional exterior derivative} on $\mathscr{P}$
which we denote by $\delta={\rm d}_{\mathscr{P}}$ (not to be confounded
in meaning with the $\delta\psi^{A}$ on $\mathscr{M}$ from the Lagrangian
analysis); see [\cite{crnkovic_symplectic_1987}; \cite{crnkovic_covariant_1989}]
for more technical details on this. For example, $\delta\varphi^{A}(x^{i})$
and $\delta\pi_{A}(x^{i})$ are one-forms on $\mathscr{P}$, and so
for any functional (zero-form) $F[\varphi^{A},\pi_{A}]$ on $\mathscr{P}$,
for example, we have that the action of $\delta$ yields the one-form
\begin{equation}
\delta F\left[\varphi,\pi\right]=\int_{\Sigma}{\bf e}\,\left(\frac{\delta F}{\delta\varphi^{A}(x^{i})}\delta\varphi^{A}(x^{i})+\frac{\delta F}{\delta\pi_{A}(x^{i})}\delta\pi_{A}(x^{i})\right)\,,
\end{equation}
where $\delta F/\delta f(x^{i})$ indicates the functional derivative
of $F$, as defined in the previous section and restricted to functionals
on $\Sigma$. A wedge product can be naturally defined to obtain $p$-forms,
and the action of $\delta$ also easily generalizes thereto.

We are now ready to write down the symplectic form $\bm{\omega}$
for our phase space $(\mathscr{P};\bm{\omega})$. It is possible to
show (a result known generally as the Darboux theorem) that, at least
locally, $\bm{\omega}$ is always given by: 
\begin{equation}
\bm{\omega}=\int_{\Sigma}{\bf e}\,\delta\pi_{A}\wedge\delta\varphi^{A}\,,\label{eq:symplectic-form-general}
\end{equation}
which can be checked to satisfy the symplectic form conditions\footnote{~$\left(\varphi,\pi\right)$ are then referred to as Darboux coordinates,
symplectic coordinates, or canonical coordinates.}. In the case that there is only one (tensorial) field variable in
$\varphi$, it can furthermore be proved that the symplectic form $\bm{\omega}$
is also the volume form $\bm{\epsilon}_{\mathscr{P}}^{\,}$, conventionally
denoted as $\bm{\epsilon}_{\mathscr{P}}^{\,}=\bm{\Omega}$, on $\mathscr{P}$,
that is to say, we have $\bm{\Omega}=\bm{\omega}$. If there are $N$
fields in $\varphi$, then a volume form $\bm{\Omega}$ on $\mathscr{P}$
is given simply by the $N$-th exterior power of the symplectic form,
in particular $\bm{\Omega}=[(-1)^{N(N-1)/2}/N!]\bm{\omega}^{\wedge N}$.

We will need just a few more definitions and results before proceeding
to the equations of motion. Let $\mathscr{F}(\mathscr{P})=\{F:\mathscr{P}\rightarrow\mathbb{R}\}$
be the set of all functionals (zero-forms) on the phase space $(\mathscr{P};\bm{\omega})$.
For any $F[\varphi^{A},\pi_{A}]\in\mathscr{F}(\mathscr{P})$, the
symplectic form $\bm{\omega}$ defines a vector field $\bm{X}_{F}\in T\mathscr{P}$
associated to $F$, referred to as the \emph{Hamiltonian vector field}
(HVF) of $F$, via the relation 
\begin{equation}
\imath_{\bm{X}_{F}}\bm{\omega}=-\delta F\,,\label{eq:HVF-general}
\end{equation}
where $\imath$ is the interior product on $\mathscr{P}$. We furthermore
define an operation $\{\cdot,\cdot\}:\mathscr{F}\times\mathscr{F}\rightarrow\mathbb{R}$
known as the \emph{Poisson bracket}, which for any two functionals
$F,G\in\mathscr{F}$ is given by: 
\begin{equation}
\left\{ F,G\right\} =\int_{\Sigma}{\bf e}\,\left(\frac{\delta F}{\delta\varphi^{A}\left(x^{i}\right)}\frac{\delta G}{\delta\pi_{A}\left(x^{i}\right)}-\frac{\delta F}{\delta\pi_{A}\left(x^{i}\right)}\frac{\delta G}{\delta\varphi^{A}\left(x^{i}\right)}\right)\,.\label{eq:Poisson-general}
\end{equation}

Finally, let $\bm{Y}\in T\mathscr{P}$ be any vector field in $\mathscr{P}$.
We can use it to define the action of a map $\Phi_{t}^{(\bm{Y})}:\mathscr{P}\times\mathbb{R}\rightarrow\mathscr{P}$
by requiring that $\Phi_{t}^{(\bm{Y})}$ moves points around in $\mathscr{P}$
along the integral curves of $\bm{Y}$ (as a function of the parameter
$t\in\mathbb{R}$). Mathematically, this means that such a map is
defined by the ordinary differential equation: 
\begin{equation}
\frac{{\rm d}\Phi_{t}^{(\bm{Y})}}{{\rm d}t}=\bm{Y}\circ\Phi_{t}^{(\bm{Y})}\,.\label{eq:flow-general}
\end{equation}
Any map satisfying (\ref{eq:flow-general}) is called a \emph{flow},
and $\bm{Y}$ is called its (infinitesimal) \emph{generator}. If,
additionally, $\Phi_{t}^{(\bm{Y})}$ preserves the symplectic form
under the pullback, \textit{i.e.} if $(\Phi_{t}^{(\bm{Y})})^{*}\bm{\omega}=\bm{\omega}$,
then $\Phi_{t}^{(\bm{Y})}$ is called a \emph{canonical transformation}
(or, in geometry, a \textit{symplectomorphism}).
It is useful and easy to prove that $\Phi_{t}^{(\bm{Y})}$ is a canonical
transformation if and only if $\bm{Y}$ is an HVF\footnote{~Let $\Phi_{t}^{(\bm{Y})}$ be such that $(\Phi_{t}^{(\bm{Y})})^{*}\bm{\omega}=\bm{\omega}$.
Equivalently, $\mathcal{L}_{\bm{Y}}\bm{\omega}=0$. Recall Cartan's
``magic formula'', which tells us that the action of the Lie derivative when acting on forms
can be expressed as $\mathcal{L}_{\bm{Y}}=\imath_{\bm{Y}}\circ{\rm d}+{\rm d}\circ\imath_{\bm{Y}}$. In our case, as we have seen, ${\rm d}$ is
the functional exterior derivative $\delta$, so we have $0=\mathcal{L}_{\bm{Y}}\bm{\omega}\Leftrightarrow0=\imath_{\bm{Y}}(\delta\bm{\omega})+\delta(\imath_{\bm{Y}}\bm{\omega})$.
But $\bm{\omega}$ is closed (\textit{i.e.} $\delta\bm{\omega}=0$, the first
symplectic form property), so this is equivalent to $0=\delta(\imath_{\bm{Y}}\bm{\omega})\Leftrightarrow\imath_{\bm{Y}}\bm{\omega}=\delta F$
for some $F\in\mathscr{F}(\mathscr{P})$. This is the same as the
definition (\ref{eq:HVF-general}), \textit{i.e.} it is equivalent to saying
that $\bm{Y}$ is an HVF (specifically, $\bm{Y}=\bm{X}_{-F}$).}.

\subsection{The Hamiltonian and the equations of motion\label{sec:2.3.1-hamiltonian-eom}}

We define the \emph{Hamiltonian functional} $H\in\mathscr{F}(\mathscr{P})$,
henceforth simply the \emph{Hamiltonian}, to be a phase space functional
of the form 
\begin{equation}
H\left[\varphi^{A},\pi_{A}\right]=\int_{\Sigma}{\bf e}\,\mathcal{H}\left(\varphi^{A},\pi_{A}\right)\,,
\end{equation}
where $\mathcal{H}$ is a local function of $(\varphi^{A},\pi_{A})$,
referred to as the \emph{Hamiltonian density}\textemdash which, provided
no confusion is created, we will also simply call the Hamiltonian\textemdash such
that the field equations of the original spacetime field theory (for
$\psi$ in $\mathscr{M}$) are equivalent to the \emph{canonical}
(or \emph{Hamiltonian})\emph{ equations of motion} for the phase space
variables, 
\begin{align}
\dot{\varphi}^{A}=\, & \left\{ \varphi^{A},H\right\} =\frac{\delta H}{\delta\pi_{A}}\,,\label{eq:phi-canonical-eqn}\\
\dot{\pi}^{A}=\, & \left\{ \pi^{A},H\right\} =-\frac{\delta H}{\delta\varphi^{A}}\,,\label{eq:pi-canonical-eqn}
\end{align}
with the last equality in each line following from the general definition
of the Poisson bracket (\ref{eq:Poisson-general}).

More generally, the time derivative (Lie derivative along the time
flow vector field) of any functional $F\in\mathscr{F}(\mathscr{P})$
is given by the Poisson bracket with the Hamiltonian $H$, which we
can write as
\begin{equation}
\dot{F}=\left\{ F,H\right\} =\bm{X}_{H}\left(F\right)\,,
\end{equation}
where $\bm{X}_{H}\in T\mathscr{P}$ is the HVF of the Hamiltonian,
\begin{equation}
\bm{X}_{H}=\int_{\Sigma}{\bf e}\,\left(\frac{\delta H}{\delta\pi_{A}}\frac{\delta}{\delta\varphi^{A}}-\frac{\delta H}{\delta\varphi^{A}}\frac{\delta}{\delta\pi_{A}}\right)\,.
\end{equation}

Put differently, the time evolution of any quantity through $\mathscr{P}$
is represented by the integral curves of the HVF of the Hamiltonian,
$\bm{X}_{H}$. We refer to the flow $\Phi_{t}^{(\bm{X}_{H})}:\mathscr{P}\times\mathbb{R}\rightarrow\mathscr{P}$
generated thereby as the \emph{Hamiltonian flow}, and for simplicity
we will henceforth denote it simply as $\Phi_{t}=\Phi_{t}^{(\bm{X}_{H})}$.
Following our discussion at the end of the last subsection, the fact
that $\bm{X}_{H}$ is an HVF guarantees that $\Phi_{t}$ is a canonical
transformation. In particular, it preserves the symplectic (as well
as volume) form: 
\begin{equation}
\mathcal{L}_{\bm{X}_{H}}\bm{\omega}=0=\mathcal{L}_{\bm{X}_{H}}\bm{\Omega}\,.
\end{equation}
This result is commonly known as is \emph{Louville's theorem}.

We have expended much effort so far on developing the technical machinery
for a canonical analysis without yet prescribing the recipe for explicitly
computing the most important pieces: the canonical momenta and the
Hamiltonian itself! Assuming a Lagrangian formulation of the theory
exists, the definition typically ascribed to the former is: 
\begin{equation}
\pi_{A}=\frac{\partial\mathcal{L}}{\partial\dot{\varphi}^{A}}\,,\label{eq:pi_A-general}
\end{equation}
and to the latter: 
\begin{equation}
\mathcal{H}=\dot{\varphi}^{A}\pi_{A}-\mathcal{L}\,,\label{eq:H-general}
\end{equation}
called the \emph{Legendre transform} of the Lagrangian.

Let us consider in turn the subtleties presented by these definitions.
First, the definition of the canonical momenta (\ref{eq:pi_A-general})
may be seen as expressing $\pi_{A}$ as a function of $(\varphi^{A},\dot{\varphi}^{A})$,
given explicitly by the $\dot{\varphi}^{A}$ partial of $\mathcal{L}(\varphi^{A},\dot{\varphi}^{A})$
on the RHS. A one-to-one correspondence between the configuration
time derivatives $\dot{\varphi}^{A}$ and the momenta $\pi_{A}$ exists
if and only if it is possible to invert this mapping, \textit{i.e.} to write
all $\dot{\varphi}^{A}$ as functions of $\pi_{A}$ (and possibly
$\varphi^{A}$). If this is \emph{not} possible (as we will see in
GR), then some of the $\dot{\varphi}^{A}$ will not represent true
``dynamical'' degrees of freedom, but instead will define the constraints.

Let us see how this works in a bit more detail. Let $f:T\mathscr{Q}\rightarrow T^{*}\mathscr{Q}$
denote the mapping taking points from the set $\{(\varphi,\dot{\varphi})\}$,
which is simply the tangent space $T\mathscr{Q}$ of the configuration
space, into the phase space variables $\{(\varphi,\pi)\}$ according
to the rule (\ref{eq:pi_A-general}), \textit{i.e.} 
\begin{align}
f:T\mathscr{Q}\rightarrow & T^{*}\mathscr{Q}\nonumber \\
\left(\varphi^{A},\dot{\varphi}^{A}\right)\mapsto & \left(\varphi^{A},\pi_{A}\left(\varphi^{A},\dot{\varphi}^{A}\right)\right)=\left(\varphi^{A},\partial_{\dot{\varphi}^{A}}\mathcal{L}\left(\varphi^{A},\dot{\varphi}^{A}\right)\right)\,.\label{eq:f-canonical}
\end{align}
The inverse function theorem tells us that for $f^{-1}:T^{*}\mathscr{Q}\rightarrow T\mathscr{Q}$
to exist, we must have ${\rm det}({\rm Jac}(f))\neq0$. In particular,
this requires ${\rm det}(W_{AB})\neq0$, \textit{i.e.} the non-degeneracy of
the matrix $W_{AB}$ given by 
\begin{equation}
W_{AB}=\frac{\partial\pi_{A}}{\partial\dot{\varphi}^{B}}=\frac{\partial^{2}\mathcal{L}}{\partial\dot{\varphi}^{B}\partial\dot{\varphi}^{A}}\,,\label{eq:W_AB-canonical}
\end{equation}
where in the last equality we have used the momentum definition (\ref{eq:pi_A-general}).
It should not surprise the reader that this is essentially the same
as the principal symbol of the general Euler-Lagrange equations (\ref{eq:W_AB-Lagrangian})
that we encountered earlier!

Thus, if the function (\ref{eq:f-canonical}) does not have an inverse
(that is, it has an inverse only on a restriction of its domain),
or equivalently the matrix (\ref{eq:W_AB-canonical}) is degenerate,
then not all $n$ of the $\dot{\varphi}^{A}$ can be solved for in
terms of the $\pi_{A}$, and those which cannot will consequently
avoid picking up an additional time derivative in the equations of
motion. These therefore define constraints on the second-order equations
for the true ``dynamical'' degrees of freedom in $(\varphi^{A},\pi_{A})$.

Suppose $f$ has $\tilde{m}$ degeneracy directions, \textit{i.e.} $\tilde{m}$
of the $\dot{\varphi}^{A}$ map trivially onto the $\pi_{A}$. This is
equivalent to the existence of $\tilde{m}$ phase space functionals
$\tilde{\zeta}_{\tilde{\mathsf{j}}}\in\mathscr{F}(\mathscr{P})$,
$\forall1\leq\tilde{\mathsf{j}}\leq\tilde{m}$, which identically
vanish for solutions satisfying the equations of motion of the theory,
\textit{i.e.} 
\begin{equation}
0=\tilde{\zeta}_{\tilde{\mathsf{j}}}\,.\label{eq:primary-constraints}
\end{equation}
Such constraints are called \emph{primary} constraints. 

As we will see (and as the notation using tildes anticipates), these
may actually not be the only phase space constraints for our theory;
in particular, consistency conditions involving the primary constraints
$\tilde{\zeta}_{\tilde{\mathsf{j}}}$ may imply additional, \emph{independent}
constraints---and we shall concretely see how so in the next subsection.
For the moment, let us simply assume henceforth that \emph{all} the
constraints of the theory\textemdash however they are obtained\textemdash are
collected into the set $\zeta=\{\zeta_{\mathsf{j}}\}_{\mathsf{j}}$,
with the index $\mathsf{j}$ here running over a possibly larger range
than just from $1$ to $\tilde{m}$, and with the first $\tilde{m}$
of them being the primary constraints just discussed (and notationally
identified with tildes), \textit{i.e.} $\zeta_{\mathsf{j}}=\tilde{\zeta}_{\tilde{\mathsf{j}}}$
for $1\leq\mathsf{j}=\tilde{\mathsf{j}}\leq\tilde{m}$.

\begin{figure}
\begin{centering}
\includegraphics[scale=0.7]{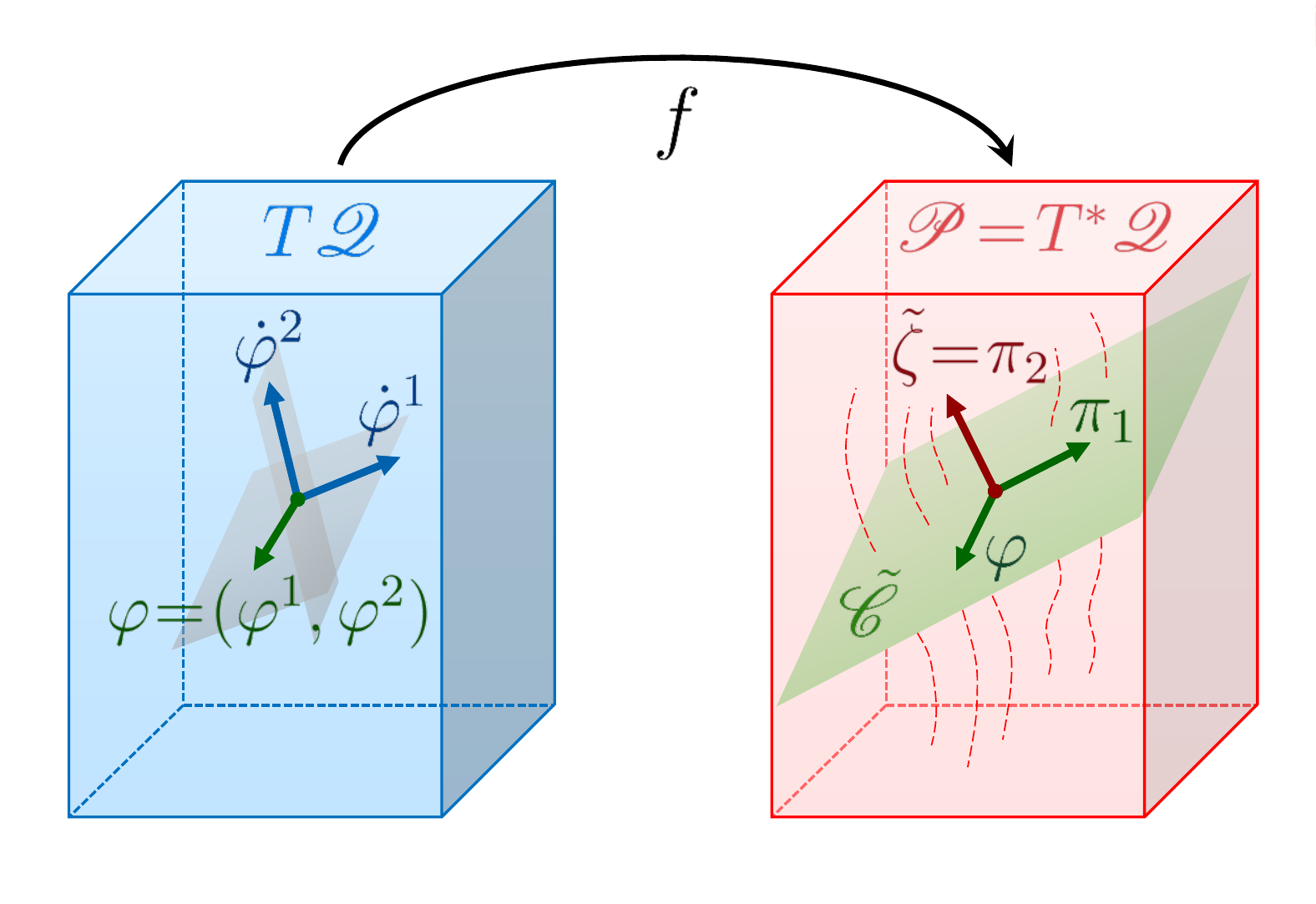}
\par\end{centering}
\caption{An illustration of the transformation $f:T\mathscr{Q}\rightarrow T^{*}\mathscr{Q}$
from the configuration space tangent bundle into the phase space $\mathscr{P}=T^{*}\mathscr{Q}$.
For example, suppose we have a two-dimensional configuration $\varphi=\{\varphi^{1},\varphi^{2}\}$
(visually represented as one dimension), and correspondingly $\dot{\varphi}=\{\dot{\varphi}^{1},\dot{\varphi}^{2}\}$.
Suppose that here, $\dot{\varphi}^{2}$ is in the kernel of this
map, \textit{i.e.} it maps trivially to $\pi_{2}$ such that the only primary
constraint is $0=\pi_{2}=\tilde{\zeta}$. The primary constraint surface
$\tilde{\mathscr{C}}$ thus has coordinates $\{\varphi,\pi_{1}\}$.}\label{2-fig-fTQ}
\end{figure}

The $\tilde{\zeta}_{\tilde{\mathsf{j}}}$ may be regarded as coordinates
locally orthogonal to the image of the function (\ref{eq:f-canonical}),
such that locally $\mathscr{P}$ has a complete set of coordinates
given by $(\varphi^{A},\pi_{A},\tilde{\zeta}_{\tilde{\mathsf{j}}})$.
This is illustrated visually in Fig \ref{2-fig-fTQ}. For convenience, we henceforth
redefine the set $\pi$ in $\mathscr{P}$ to include not only those
momenta $\pi_{A}$ which can be solved for, but also the primary constraints
$\tilde{\zeta}_{\tilde{\mathsf{j}}}$, \textit{i.e.} $\pi=\{\pi_{A},\tilde{\zeta}_{\tilde{\mathsf{j}}}\}$,
such that an arbitrary point in $\mathscr{P}$ is still labeled as
$(\varphi,\pi)$. Henceforth, though we will continue to simply call
it the ``phase space'' if the context is clear enough, we will formally
refer to $\mathscr{P}=\{(\varphi,\pi)\}$ as the \emph{unconstrained},
or \emph{full} \emph{phase space}.

\begin{figure}
\begin{centering}
\includegraphics[scale=0.6]{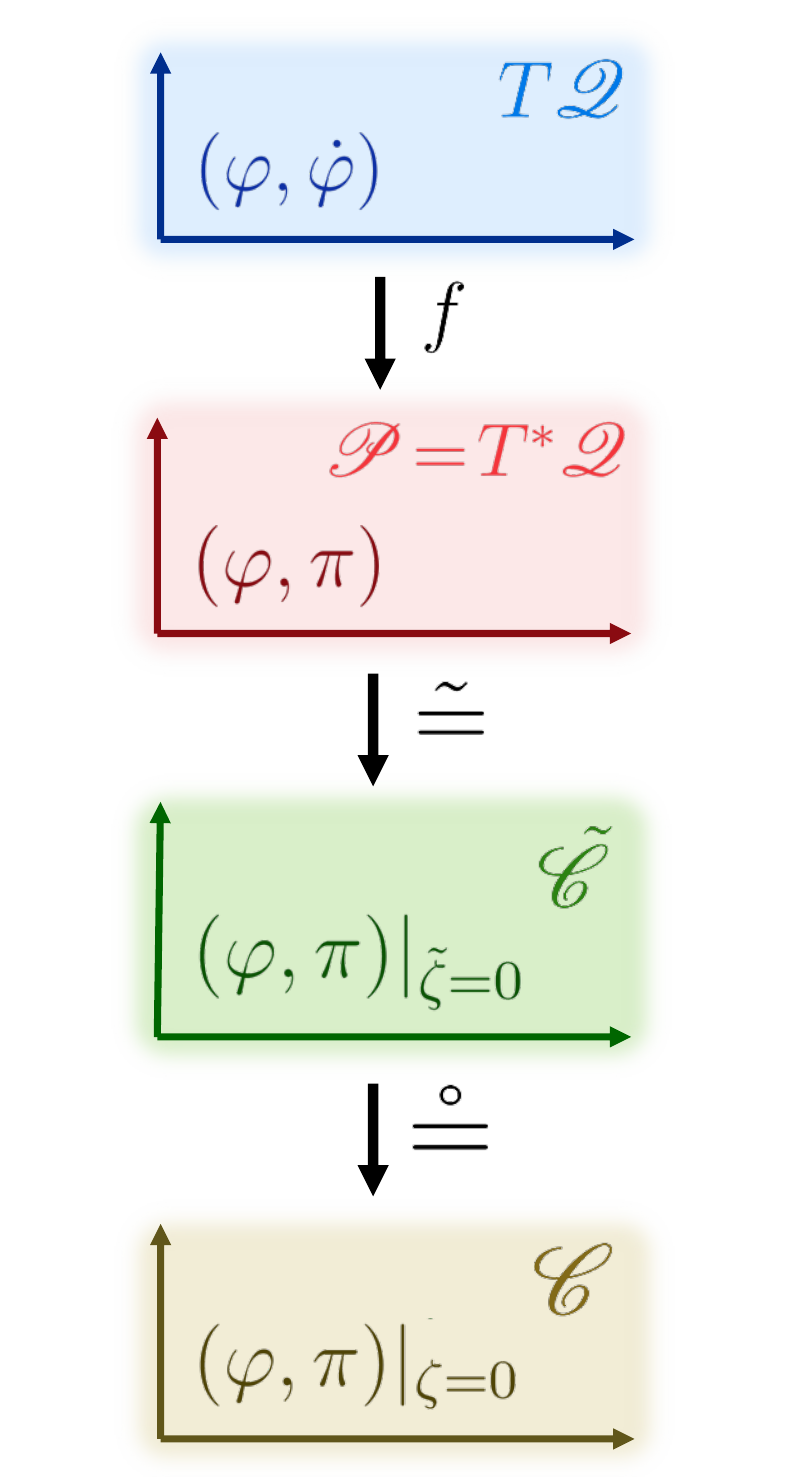}
\par\end{centering}
\caption{An illustration of the configuration space tangent bundle $T\mathscr{Q}$,
the phase space $\mathscr{P}=T^{*}\mathscr{Q}$, the primary constraint
surface $\tilde{\mathscr{C}}$, and the full constraint surface $\mathscr{C}$
with their respective coordinates, and the maps/operations that respectively
transform from one of these spaces to the next.}\label{2-fig-TQPC}
\end{figure}

A few more definitions follow naturally from these considerations
and will be useful for us to establish here before moving on. We offer
in Fig. \ref{2-fig-TQPC} a visual depiction to make the story a little bit easier to
follow. 

We define the \emph{primary constraint surface} $\tilde{\mathscr{C}}\subseteq\mathscr{P}$
as the submanifold of the phase space where only the primary constraints
$\tilde{\zeta}_{\tilde{\mathsf{j}}}$ vanish (with no conditions assumed
on non-primary constraints $\zeta_{\mathsf{j}}$ for $\mathsf{j}\geq\tilde{m}+1$,
if any exist): 
\begin{equation}
\tilde{\mathscr{C}}=\left\{ \left.\left(\varphi,\pi\right)\in\mathscr{P}\right|0=\tilde{\zeta}\right\} \,.\label{eq:primary-constraint-surface}
\end{equation}

Meanwhile, the \emph{constraint surface} or, for emphasis, \emph{full
constraint surface} $\mathscr{C}\subseteq\tilde{\mathscr{C}}$ is
the submanifold of $\mathscr{P}$ where \emph{all} the constraints
(the primary $\tilde{\zeta}$ plus any other constraints, $\zeta$
in total) are satisfied: 
\begin{equation}
\mathscr{C}=\left\{ \left.\left(\varphi,\pi\right)\in\mathscr{P}\right|0=\zeta\right\} \,.\label{eq:full-constraint-surface}
\end{equation}
We will accordingly also find it useful to define operations of equality
under primary and full constraint imposition. In particular, we use
the symbols ``$\tilde{=}$'' and ``$\mathring{=}$'' respectively
to indicate these, such that $a\,\tilde{=}\,b$ means that $a|_{\tilde{\mathscr{C}}}=b|_{\tilde{\mathscr{C}}}$,
and $a\,\mathring{=}\,b$ that $a|_{\mathscr{C}}=b|_{\mathscr{C}}$.

Now, let us turn to a discussion of the definition of the Hamiltonian
(\ref{eq:H-general}) as a Legendre transform of $\mathcal{L}:T\mathscr{Q}\rightarrow\mathbb{R}$.
In particular, we have\textemdash just as in the case of the momentum
definition (\ref{eq:pi_A-general})\textemdash a functional prescription
in terms of $(\varphi,\dot{\varphi})$ (\textit{i.e.}, as a functional on the
tangent bundle $T\mathscr{Q}$). In other words, the definition gives
us $\mathcal{H}[\varphi,\dot{\varphi}]$. How can we know, in general,
that this $\mathcal{H}$ is also---as it ought to be for this procedure
to make sense---a well-defined functional $\mathcal{H}[\varphi,\pi]$
on the phase space (cotangent bundle) $\mathscr{P}=T^{*}\mathscr{Q}$,
\emph{irrespective} of the existence of constraints? (I.e, how can
we know that under the mapping $f:T\mathscr{Q}\rightarrow T^{*}\mathscr{Q}$
given by Eqn. (\ref{eq:f-canonical}), the transformation of $\mathcal{H}$
always ``avoids'' any degeneracy directions that may arise from
its non-invertibility?) An easy way to see this is by computing the
exterior derivative of the Legendre transform $\mathcal{H}=\dot{\varphi}^{A}\pi_{A}-\mathcal{L}$
(\ref{eq:H-general}), immediately yielding 
\begin{align}
\delta\mathcal{H}=\, & \int_{\Sigma}{\bf e}\,\left[\dot{\varphi}^{A}\delta\pi_{A}+\pi_{A}\delta\dot{\varphi}^{A}-\frac{\partial\mathcal{L}}{\partial\varphi^{A}}\delta\varphi^{A}-\frac{\partial\mathcal{L}}{\partial\dot{\varphi}^{A}}\delta\dot{\varphi}^{A}\right]\\
=\, & \int_{\Sigma}{\bf e}\,\left[-\frac{\partial\mathcal{L}}{\partial\varphi^{A}}\delta\varphi^{A}+\dot{\varphi}^{A}\delta\pi_{A}\right]\,,\label{eq:deltaH}
\end{align}
where to obtain the last equality the momentum definition $\pi_{A}=\partial\mathcal{L}/\partial\dot{\varphi}^{A}$
(\ref{eq:pi_A-general}) was once again used, such that no term linear
in $\delta\dot{\varphi}$ remains. Hence the exterior derivative (\ref{eq:deltaH})
of the Hamiltonian functional $\mathcal{H}$ is always a linear combination
\emph{only} of $\delta\varphi^{A}$ and $\delta\pi_{A}$, making it
a well-defined one-form on $\mathscr{P}$, implying that $\mathcal{H}$
itself is a well-defined zero-form (functional) on $\mathscr{P}$.

Now we must verify that this $\mathcal{H}=\dot{\varphi}^{A}\pi_{A}-\mathcal{L}$
indeed gives the correct (canonical) equations of motion (\ref{eq:phi-canonical-eqn})-(\ref{eq:pi-canonical-eqn})
for the theory. For their formulation, it will in fact be more convenient
to first define what is referred to as the \emph{total} Hamiltonian,
which we denote as $\tilde{\mathcal{H}}\in\mathscr{F}(\mathscr{P})$,
given by: 
\begin{equation}
\tilde{\mathcal{H}}=\mathcal{H}-\tilde{\lambda}^{\tilde{\mathsf{j}}}\tilde{\zeta}_{\tilde{\mathsf{j}}}\,,\quad\tilde{H}=\int_{\Sigma}{\bf e}\,\tilde{\mathcal{H}}\,.\label{eq:H-total}
\end{equation}
This is simply the Hamiltonian $\mathcal{H}$ we have been working
with so far (obtained via the Legendre transform), plus an arbitrary
linear combination of the primary constraints $\tilde{\zeta}_{\tilde{\mathsf{j}}}$,
with $\tilde{m}$ coefficient functions $\tilde{\lambda}^{\tilde{\mathsf{j}}}\in\mathscr{F}(\mathscr{P})$
playing the role of Lagrange multipliers. We shall see the usefulness
of this momentarily. We remark for now that by construction, $\tilde{\mathcal{H}}$
will coincide with $\mathcal{H}$ on the (primary and thus also full)
constraint surface, $\tilde{\mathcal{H}}\,\tilde{=}\,\mathcal{H}\,\mathring{=}\,\tilde{\mathcal{H}}$.

Combining $\mathcal{H}=\dot{\varphi}^{A}\pi_{A}-\mathcal{L}$ (\ref{eq:H-general})
and $\tilde{\mathcal{H}}=\mathcal{H}-\tilde{\lambda}\cdot\tilde{\zeta}$
(\ref{eq:H-total}), one can isolate for the Lagrangian as: $\mathcal{L}=\dot{\varphi}^{A}\pi_{A}-\tilde{\mathcal{H}}-\tilde{\lambda}\cdot\tilde{\zeta}$.
Integrating both sides over (a finite range of) $t$, and then taking
their variation (\textit{i.e.} viewing the arguments of both sides as one-parameter
families in $\lambda$, in the sense of the previous section, and
applying $\partial_{\lambda}|_{\lambda=0}$), a straightforward computation
shows that the stationary action principle $0=\delta S[\psi]=\int{\rm d}t{\bf e}\,\mathcal{L}$
is equivalent to the following system of equations in $\mathscr{P}$:
\begin{align}
\dot{\varphi}^{A}=\, & \left\{ \varphi^{A},\tilde{H}\right\} +\tilde{\zeta}_{\tilde{\mathsf{j}}}\frac{\delta\tilde{\lambda}^{\tilde{\mathsf{j}}}}{\delta\pi_{A}}\,,=\frac{\delta\tilde{H}}{\delta\pi_{A}}+\tilde{\zeta}_{\tilde{\mathsf{j}}}\frac{\delta\tilde{\lambda}^{\tilde{\mathsf{j}}}}{\delta\pi_{A}}\,,\label{eq:phi-canonical-eqn-constr}\\
\dot{\pi}^{A}=\, & \left\{ \pi^{A},\tilde{H}\right\} -\tilde{\zeta}_{\tilde{\mathsf{j}}}\frac{\delta\tilde{\lambda}^{\tilde{\mathsf{j}}}}{\delta\varphi^{A}}\,,=-\frac{\delta\tilde{H}}{\delta\varphi^{A}}-\tilde{\zeta}_{\tilde{\mathsf{j}}}\frac{\delta\tilde{\lambda}^{\tilde{\mathsf{j}}}}{\delta\varphi^{A}}\,.\label{eq:pi-canonical-eqn-constr}
\end{align}
So on the constraint surface $\mathscr{C}$, we indeed recover the
standard form of the canonical equations of motion in terms of the
total Hamiltonian: 
\begin{align}
\dot{\varphi}^{A}\,\mathring{=}\, & \left\{ \varphi^{A},\tilde{H}\right\} \,,\\
\dot{\pi}^{A}\,\mathring{=}\, & \left\{ \pi^{A},\tilde{H}\right\} \,.
\end{align}

\subsection{Constraints and the reduced phase space\label{sec:2.3.1-constraints-reduced-phase-space}}

Let us now address the subtleties that the presence of constraints
poses to our Hamiltonian analysis.

First, suppose that all primary constraints are indeed satisfied,
\textit{i.e.} $\tilde{\zeta}_{\tilde{\mathsf{j}}}=0$, so that we are on $\tilde{\mathscr{C}}$
in $\mathscr{P}$. For ease in following the discussion, consulting again
Figs. \ref{2-fig-fTQ} and \ref{2-fig-TQPC} is useful. Observe that the vanishing of $\tilde{\zeta}$
on $\tilde{\mathscr{C}}$ necessarily implies that their time derivatives
$\dot{\tilde{\zeta}}$ vanish thereon too. These are called \emph{consistency
conditions}: 
\begin{equation}
0\,\tilde{=}\,\dot{\tilde{\zeta}}_{\tilde{\mathsf{j}}}\,\mathring{=}\,\left\{ \tilde{\zeta}_{\tilde{\mathsf{j}}},\tilde{H}\right\} \,\mathring{=}\,\left\{ \tilde{\zeta}_{\tilde{\mathsf{j}}},H\right\} -\tilde{\lambda}^{\tilde{\mathsf{k}}}\left\{ \tilde{\zeta}_{\tilde{\mathsf{j}}},\tilde{\zeta}_{\tilde{\mathsf{k}}}\right\} \,,\label{eq:consistency-conditions}
\end{equation}
where in the last equality we have simply used the definition of the
total Hamiltonian $\tilde{\mathcal{H}}=\mathcal{H}-\tilde{\lambda}\cdot\tilde{\zeta}$
(\ref{eq:H-total}).

From this we see that the multiplicative functions (Lagrange multipliers)
$\tilde{\lambda}^{\tilde{\mathsf{j}}}$ can be fully determined (\textit{i.e.}
$\tilde{m}$ independent equations for them can be obtained by inverting
the above equation) if and only if ${\rm det}(\{\tilde{\zeta}_{\tilde{\mathsf{j}}},\tilde{\zeta}_{\tilde{\mathsf{k}}}\})\neq0$.
Otherwise, the system formed by the vanishing of the primary constraints
coupled with the consistency conditions (\ref{eq:consistency-conditions})
is under-determined, and therefore has to be augmented by further
equations, which we can now interpret as constituting the additional
constraints in $\zeta_{\mathsf{j}}$ (those for $\mathsf{j}\geq\tilde{m}+1$). 

To obtain these equations, let $\{\hat{v}_{\hat{\mathsf{j}}}^{\tilde{\mathsf{j}}}\}$
be the set of $\hat{m}\leq\tilde{m}$ null-eigenvectors, indexed in
the set by $1\leq\hat{\mathsf{j}}\leq\hat{m}$ (and with components
indexed by $\tilde{\mathsf{j}}$), of the matrix $\{\tilde{\zeta}_{\tilde{\mathsf{j}}},\tilde{\zeta}_{\tilde{\mathsf{k}}}\}$,
\textit{i.e.} we have $0=\hat{v}_{\hat{\mathsf{j}}}^{\tilde{\mathsf{j}}}\{\tilde{\zeta}_{\tilde{\mathsf{j}}},\tilde{\zeta}_{\tilde{\mathsf{k}}}\}$.
Multiplying the consistency condition (\ref{eq:consistency-conditions})
on the left by $\hat{v}_{\hat{\mathsf{j}}}^{\tilde{\mathsf{j}}}$
yields the \emph{independent} equations 
\begin{equation}
0\,\mathring{=}\,\hat{v}_{\hat{\mathsf{j}}}^{\tilde{\mathsf{j}}}\left\{ \tilde{\zeta}_{\tilde{\mathsf{j}}},H\right\} =\hat{\zeta}_{\hat{\mathsf{j}}}\,,\label{eq:secondary-constraints}
\end{equation}
which we define as $\hat{\zeta}$. These $\hat{m}$ equations are
referred to as \emph{secondary} constraints. 

Assuming that with these, one now obtains a complete system, then
one takes the complete set of ($\tilde{m}+\hat{m}$) constraints $\zeta_{\mathsf{j}}$
to be simply the set of the ($\tilde{m}$) primary and $(\hat{m}$)
secondary constraints $\zeta=\{\tilde{\zeta},\hat{\zeta}\}$. (In
the index notation, we set $\zeta_{\tilde{m}+\mathsf{k}}=\hat{\zeta}_{\hat{\mathsf{k}}}$,
$\forall1\leq\mathsf{k}=\hat{\mathsf{k}}\leq\hat{m}$.) This happens
to be the case in GR (which, as we shall see, has a total of eight
constraints, with $\tilde{m}=4$ primary constraints and $\hat{m}=4$
secondary constraints)\footnote{~If, on the other hand, one still does not have a complete set of equations
by arriving at the secondary constraints, the same process must be
repeated until one does: the secondary constraints imply their own
consistency conditions, those in turn will yield the tertiary constraints,
and so on if necessary.}.

Now we turn to a different and also very useful way of looking at
how to classify the (total set of) constraints $\zeta$. It may be
motivated by reflecting upon the following question: assuming we are
indeed only interested in those field configutations which satisfy
the constraints, \textit{i.e.} only in points $(\varphi,\pi)$ living on $\mathscr{C}$,
could we simply regard $\mathscr{C}$ in some operational sense as
an ``effective'' phase space? The answer in general is \emph{no}:
in particular, $\mathscr{C}$ will \emph{not} in general be a symplectic
manifold. 

To be more precise, let $i:\mathscr{C}\rightarrow\mathscr{P}$ be
the embedding of $\mathscr{C}$ in $\mathscr{P}$. We denote by $\bm{\omega}|_{\mathscr{C}}=i^{*}\bm{\omega}$
the pullback of the symplectic form $\bm{\omega}$ of the phase space
$\mathscr{P}$ to the constraint surface $\mathscr{C}$. In general,
as we will show presently, $\bm{\omega}|_{\mathscr{C}}$ is unfortunately
\emph{not} a symplectic form on $\mathscr{C}$. Thus $\bm{\omega}|_{\mathscr{C}}$
is sometimes instead referred to as the \emph{presymplectic} form.

In order to make progress, the following classification of constraint
functions $\zeta\in\mathscr{F}(\mathscr{P})$ is useful. 
\begin{itemize}
\item A constraint $\zeta\in\mathscr{F}(\mathscr{P})$ is called a \emph{first-class
constraint} if its HVF $\bm{X}_{\zeta}$ is \emph{everywhere} tangent
to $\mathscr{C}$ (\textit{i.e.} $\bm{X}_{\zeta}\in T\mathscr{C}$). 
\item A constraint $\zeta\in\mathscr{F}(\mathscr{P})$ is called a \emph{second-class
constraint} if its HVF $\bm{X}_{\zeta}$ is \emph{nowhere} tangent
to $\mathscr{C}$ (\textit{i.e.} $\nexists p\in\mathscr{P}$ so that $(\bm{X}_{\zeta})_{p}\in T_{p}\mathscr{C}$). 
\end{itemize}
Now suppose that among the constraint functions $\zeta$ in our field
theory, say $\zeta_{1}=f$, is first-class. One finds that the interior
product between its HVF and the presymplectic form vanishes: that
is, $\imath_{\bm{X}_{f}}\bm{\omega}|_{\mathscr{C}}=i^{*}(\imath_{\bm{X}_{f}}\bm{\omega})=i^{*}(\bm{\nabla}f)=0$,
using the definition of the HVF. This means that $\bm{\omega}|_{\mathscr{C}}$
is degenerate (with the degeneracy directions spanned by the HVFs
of the first-class constraints), and hence cannot be a symplectic
form. Conversely, $\bm{\omega}|_{\mathscr{C}}$ will be a symplectic
form (on all of $\mathscr{C}$) if and only if all constraints $\zeta$
are second-class constraints.

Once again it may appear that we are getting too lost in mathematical
abstractions, but this is where we gain a meaningful insight into
physics. It happens that in many classical field theories of interest,
including GR and EM, all constraints $\zeta_{\mathsf{j}}$ turn out
to be first-class. 
The degeneracy directions of the presymplectic form $\bm{\omega}|_{\mathscr{C}}$
in these theories correspond to what at the Lagrangian level are seen
as \emph{gauge transformations}: that is, maps of of the fields $\psi$
which do not change the Lagrangian $\mathcal{L}(\psi,\bm{\nabla}\psi)$.
For example, this corresponds to the $\mathbb{U}(1)$ symmetry of
EM and the diffeomorphism invariance of GR. We will see explicitly
how the latter works when we finally arrive at the canonical formulation
of GR in the next section.

Just before doing so, it is salient to address a slightly more technical
question, one upon which we will not dwell further in this chapter
than the next few paragraphs, but which will reappear in our work
on entropy in Chapter \ref{4-entropy}. Namely, it is the issue of how exactly we
do, in fact, recover a symplectic structure for some ``effective''
subset of the phase space that interests us for the meaningful study
of dynamics---which, so far, has intuitively meant the constraint
surface $\mathscr{C}\subset\mathscr{P}$. In principle, a symplectic
form on $\mathscr{C}$ \textsl{could} be obtained if one directly
factors out the HVFs of the constraints (which span its kernel) living
on $T\mathscr{C}$, by simply identifying all points on the orbits
of their flow in $\mathscr{C}$. These are concordantly called \emph{gauge
orbits}. Thus, one could work with a factor space $\tilde{\mathscr{P}}\subset\mathscr{C}$
defined simply as the space of gauge orbits in $\mathscr{C}$, and
which therefore is, by construction, symplectic (with the factored
presymplectic form).

However, depending on the desired aim of implementing the canonical
construction of a field theory, taking such an approach can turn out
to be problematic. This happens in particular if the theory happens
to be diffeomorphism invariant (such as GR), in which case ``time
evolution'' in the sense defined at the beginning of this subsection
(of the ``instantaneous configuration'' in the Hamiltonian theory)
can equivalently be regarded as effected by spacetime diffeomorphisms
(of the full metric $\bm{g}$ in the Lagrangian theory). Hence moving
to the space of gauge orbits in $\mathscr{C}$ essentially renders
the dynamics nonexistent: they become entirely trivial, because they
are essentially factored out of $\tilde{\mathscr{P}}$, leaving one
with no more sense of ``motion through phase space''. This issue
is developed in clear and lengthy detail in [\cite{schiffrin_measure_2012}].

There exist two possible solutions for ameliorating this difficulty---that
is, for obtaining a symplectic structure out of $\bm{\omega}|_{\mathscr{C}}$
which \textsl{does} still preserve a nontrivial notion of ``time
evolution'':
\begin{enumerate}
\item Instead of passing to the space of gauge orbits, one may instead choose
a representative of each gauge orbit [\cite{schiffrin_measure_2012}].
The idea is that one can find a surface $\mathscr{S}\subset\mathscr{C}$
such that each gauge orbit in $\mathscr{C}$ intersects $\mathscr{S}$
once and only once. (In fact, sometimes a family of such surfaces
that work in localised regions of $\mathscr{C}$ is needed, but we
keep our discussion here simplified.) The choice of $\mathscr{S}$
is not unique, and so taking a different surface $\mathscr{S}'$ effectively
amounts to a change of description\textemdash the freedom of which,
in the context of our spacetime splitting, corresponds to ``time
evolution'' (\textit{i.e.} change of representative Cauchy surface in spacetime)
on one hand and the associated spatial diffeomorphisms on the other.
As we shall discuss further, this is what is effectively encoded in
the constraints of GR. It is beyond our scope here to enter further
into the concrete technicalities of this procedure; for the interested
reader, they are elaborated in [\cite{schiffrin_measure_2012}]. The
key point is essentially that the subspace $\mathscr{S}$ of the constraint
surface $\mathscr{C}$ resulting from such a construction can be shown
to be symplectic,. Therefore, one can work with the symplectic form
$\bm{\omega}|_{\mathscr{S}}$ obtained by pulling back $\bm{\omega}|_{\mathscr{C}}$
to $\mathscr{S}$. \\
~
\item A specific choice of gauge may be explicitly fixed, such that the
combination of the constraints $\zeta_{\mathsf{j}}$ coupled with
the gauge-fixing conditions becomes second-class. This idea is developed
further in Chapter 3 of [\cite{bojowald_canonical_2011}]. One can
by such a procedure obtain a symplectic structure on a subspace of
the constraint surface $\mathscr{S}\subset\mathscr{C}$ where the
(explicitly chosen) gauge-fixing conditions are satisfied, and where
one will thus have a symplectic form $\bm{\omega}|_{\mathscr{S}}$
(for the fixed gauge). \\
~ 
\end{enumerate}
In other to keep our discussion general, unless otherwise stated,
we refer to the symplectic manifold $(\mathscr{S};\bm{\omega}|_{\mathscr{S}})$
as the \emph{reduced phase space} irrespective of whether procedure
(a) or (b) is used to define it. 



\section{Canonical formulation of general relativity\label{sec:2.4-canonical-GR}}

Now that we have established the procedure for producing a canonical
formulation of any field theory given that a Lagrangian formulation
exists, let us apply it to GR. 

\subsection{Canonical variables}

As the only physical field is the metric $\bm{g}$, the first immediately
suggestible choice for a configuration variable $\varphi$ is the
metric $\boldsymbol{h}$ induced by $\bm{g}$ on each Cauchy surface
$\Sigma$, \textit{i.e.} $\bm{h}=\bm{g}|_{\Sigma}$. In index notation, this
is given by 
\begin{equation}
h_{ab}=g_{ab}+n_{a}n_{b}\,.\label{eq:h_ab}
\end{equation}

Now, observe that taking $\boldsymbol{h}$ to be the only gravitational
configuration variable would not suffice: there are ten independent
field variables in $\bm{g}$, and $\boldsymbol{h}$ accounts for only
six of these! We would not obtain a complete set of equations of motion from this
procedure unless all (ten) field variables present in the spacetime
metric are mapped to the same number of field variables in $\varphi$.

Intuitively, the four degrees of freedom ``missing'' from $\bm{h}$ are the ``time-time''
and ``time-space'' components of the spacetime metric $\bm{g}$,
which may be regarded as the projections $g_{\bm{tt}}=g_{ab}t^{a}t^{b}$
and $(\boldsymbol{g\cdot t})|_{\Sigma}=g_{ab}t^{a}h^{bc}$. There
is a one-to-one correspondence between these spacetime metric projections
and the choice of the time flow vector field $\bm{t}$ itself. (In
other words, these four degrees of freedom simply encode the freedom in identifying
spatial points between Cauchy slices.) However, note that we cannot
include $\bm{t}$ as such in $\varphi$ to account for these degrees of freedom,
as $\bm{t}$ of course does not live in $\Sigma$. In particular,
in general it has nonvanishing projections both normally and orthogonally
to a Cauchy slice. As the former is a scalar, $\bm{n}\cdot\bm{t}$,
which may thus be regarded as a function on $\Sigma$, and the latter
is a vector, $\bm{t}|_{\Sigma}=\boldsymbol{t}\cdot\bm{h}$, in the
tangent space of $\Sigma$, we may correspondingly take these quantities
to account for the full set of spacetime metric degrees of freedom in $\varphi$.
They are referred to as the \emph{lapse function} and \emph{shift
vector} respectively: 
\begin{align}
N=\, & -\bm{n}\cdot\bm{t}\in\mathscr{F}(\Sigma)\,,\label{eq:lapse_defn}\\
\bm{N}=\, & \bm{t}|_{\Sigma}=\bm{h}\cdot\bm{t}\in T\Sigma\,.\label{eq:shift_defn}
\end{align}
Note that this implies $\bm{t}=N\bm{n}+\bm{N}$.

Now we have all the pieces for writing down the configuration space
of GR: 
\begin{equation}
\varphi_{\textrm{G}}=(N,\bm{N},\bm{h})\,,
\end{equation}
where all configuration variables are properly defined on $\Sigma$
and ultimately encode the full set of (ten) field variables present
in the spacetime metric $\bm{g}$.

The next step is to write the gravitational action (\ref{eq:G_action})
in terms of $\varphi_{\textrm{G}}=(N,\bm{N},\bm{h})$ and $\dot{\varphi}_{\textrm{G}}=(\dot{N},\dot{\bm{N}},\dot{\bm{h}})$
instead of the spacetime metric $\bm{g}$. The calculation is long
but generally straightforward (using all the definitions we have established),
and we omit writing it explicitly here. The result can be found, \textit{e.g.},
in Chapter 3 of [\cite{bojowald_canonical_2011}].

With $\mathcal{L}_{\textrm{G}}(\varphi_{\textrm{G}},\dot{\varphi}_{\textrm{G}})$
in hand, one can then use it to determine the set of canonical momenta
corresponding to the configuration $\varphi_{\textrm{G}}$. Let $\pi^{(N)}$,
$\pi_{a}^{(\bm{N})}$ and $\pi_{(\bm{h})}^{ab}$ denote, respectively,
the canonical momenta of $N$, $N^{a}$ and $h_{ab}$, such that the
total set of canonical momenta is $\pi_{\textrm{G}}=(\pi^{(N)},\bm{\pi}^{(\bm{N})},\bm{\pi}_{(\bm{h})})$.
We follow standard convention and henceforth drop the ``$(\bm{h})$''
from the canonical momentum of the metric, so we simply write $\pi_{(\bm{h})}^{ab}=\pi^{ab}$
and 
\begin{equation}
\pi_{\textrm{G}}=(\pi^{(N)},\bm{\pi}^{(\bm{N})},\bm{\pi})\,.
\end{equation}
Each of these can be computed as the appropriate partials of $\mathcal{L}_{\textrm{G}}(N,\bm{N},\bm{h},\dot{N},\dot{\bm{N}},\dot{\bm{h}})$
using the general canonical momentum definition (\ref{eq:pi_A-general}):
\begin{align}
\pi^{(N)}=\, & \frac{\partial\mathcal{L}_{\textrm{G}}}{\partial\dot{N}}=0\,,\label{eq:momentum-lapse}\\
\bm{\pi}^{(\bm{N})}=\, & \frac{\partial\mathcal{L}_{\textrm{G}}}{\partial\dot{\bm{N}}}=0\,,\label{eq:momentum-shift}\\
\bm{\pi}=\, & \frac{\partial\mathcal{L}_{\textrm{G}}}{\partial\dot{\bm{h}}}=\sqrt{h}\left(\bm{K}-K\bm{h}\right)\,,\label{eq:momentum-metric}
\end{align}
where $K={\rm tr}(\bm{K})$.

\subsection{Constraints}

Notice that the canonical momenta $\pi^{(N)}$ and $\bm{\pi}^{(\bm{N})}$
corresponding respectively to the lapse and the shift both vanish
identically. These equations, therefore, identify precisely the degeneracy
directions of the map $f:T\mathscr{Q}\rightarrow T^{*}\mathscr{Q}$
(\ref{eq:f-canonical}) discussed earlier, which in this case maps
$(N,\bm{N},\bm{h},\dot{N},\dot{\bm{N}},\dot{\bm{h}})\mapsto(N,\bm{N},\bm{h},\pi_{(N)},\bm{\pi}_{(\bm{N})},\bm{\pi})$.
Consequently, $\pi^{(N)}=0=\bm{\pi}^{(\bm{N})}$ can be taken directly
to be the ($\tilde{m}=4$) primary constraints of GR, $\tilde{\zeta}=\{\frac{1}{\sqrt{h}}\pi^{(N)},\frac{1}{\sqrt{h}}\bm{\pi}^{(\bm{N})}\}$
where we have introduced the (nonvanishing) factors of $\frac{1}{\sqrt{h}}$
for convenience. We write these constraints as 
\begin{align}
\tilde{\zeta}_{\textrm{G}}=\, & \frac{1}{\sqrt{h}}\pi^{(N)}\,,\label{eq:CG-1t}\\
\tilde{\bm{\zeta}}_{\textrm{G}}=\, & \frac{1}{\sqrt{h}}\bm{\pi}^{(\bm{N})}\,,\label{eq:CG-1s}
\end{align}
(with the realization that the indices $\tilde{\mathsf{j}}$ on the
primary constraints that we were using earlier are in this case spacetime
indices). 

Physically, this means that the lapse and shift are not dynamical
variables. Thus, there exists freedom in choosing them. Equivalently,
there is freedom in choosing the time flow vector field in the spacetime
foliation, which in turn translates into the freedom of how to identify
spatial points at different instants of time. In this way, we can
see that these primary constraints (\ref{eq:CG-1t})-(\ref{eq:CG-1s})
are a manifestation of the coordinate freedom of GR.

Next we must ask: are these all the constraints? To investigate, let
us compute their time derivatives and equate them to zero on the primary
constraint surface (amounting to the consistency conditions (\ref{eq:consistency-conditions})):
\begin{align}
0\,\tilde{=}\, & \dot{\tilde{\zeta}}_{\textrm{G}}=\dot{\pi}^{(N)}\,\mathring{=}\,\left\{ \dot{\pi}^{(N)},\tilde{H}_{\textrm{G}}\right\} \,\mathring{=}\,\left\{ \dot{\pi}^{(N)},H_{\textrm{G}}\right\} =-\sqrt{h}C\,,\label{eq:CG-2t}\\
0\,\tilde{=}\, & \dot{\tilde{\boldsymbol{\zeta}}}_{\textrm{G}}=\dot{\boldsymbol{\pi}}^{(\boldsymbol{N})}\,\mathring{=}\,\left\{ \dot{\boldsymbol{\pi}}^{(\boldsymbol{N})},\tilde{H}_{\textrm{G}}\right\} \,\mathring{=}\,\left\{ \dot{\boldsymbol{\pi}}^{(N)},H_{\textrm{G}}\right\} =-\sqrt{h}\bm{C}\,.\label{eq:CG-2s}
\end{align}
The RHS's---whatever they are---have to vanish, and so (up to a
factor of $\sqrt{h}$, extracted for convenience) are identified as
the secondary constraints, typically denoted \textit{vis-à-vis} our earlier
notation as $C=\hat{\zeta}_{\textrm{G}}$ and $\bm{C}=\hat{\boldsymbol{\zeta}}_{\textrm{G}}$
(thus the $\hat{\mathsf{j}}$ indices in $\hat{\zeta}_{\hat{\mathsf{j}}}$
are also spacetime indices here). This is because the equations $\{\dot{\pi}^{(N)},H_{\textrm{G}}\}\,\mathring{=}\,0\,\mathring{=}\,\{\dot{\boldsymbol{\pi}}^{(N)},H_{\textrm{G}}\}$
specify precisely the degeneracy directions of the matrix of Poisson
brackets of the primary constraints, $\{\tilde{\zeta}_{\mathsf{j}}^{\textrm{G}},\tilde{\zeta}_{\mathsf{k}}^{\textrm{G}}\}$.
Direct computation of the brackets $\{\dot{\pi}^{(N)},H_{\textrm{G}}\}$
and $\{\dot{\boldsymbol{\pi}}^{(N)},H_{\textrm{G}}\}$ yields the
expressions:
\begin{align}
C=\, & -\mathcal{R}+\frac{1}{h}\left(\bm{\pi}:\bm{\pi}-\frac{1}{2}\pi^{2}\right)\,,\label{eq:Hamiltonian-constr}\\
\boldsymbol{C}=\, & \bm{\mathcal{D}}\cdot\left(\frac{1}{\sqrt{h}}\bm{\pi}\right)\,,\label{eq:momentum-constr}
\end{align}
where $\mathcal{R}$ is the Ricci scalar of $\bm{h}$ and $\pi={\rm tr}(\bm{\pi})$.

Observe that these are precisely the contracted Gauss-Peterson-Mainardi-Codazzi
equations, (\ref{eq:contractedGC1}) and (\ref{eq:contractedGC2})
respectively, encountered earlier as inevitable geometrical conditions
on the hypersurface embedding! In the present context, they are known
as the \emph{Hamiltonian constraint} (\ref{eq:Hamiltonian-constr})
and \emph{momentum constraint} (\ref{eq:momentum-constr}), and they
complete the set of (eight) constraints of GR. The existence of these
secondary constraints is physically related, just as the primary ones
although via a slightly different mechanism, to coordinate or ``gauge''
freedom in GR. 

In particular, the Hamiltonian constraint implies a time function
redefinition freedom, $t(x)\mapsto t'(x)$ (invariance under a ``change
of time coordinate''). This essentially means that one can perform
the slicing of spacetime into Cauchy surfaces completely as one wishes.
The detailed proof of the equivalence between this freedom and the
Hamiltonian constraint is a bit subtle and involves the introduction
of a few additional technical constructs that we would like to avoid
here. Thus we simply omit it, and we refer the interested reader to
Chapter 3 of [\cite{bojowald_canonical_2011}]. 

The momentum constraint also implies some form of gauge freedom---in
this case, spatial diffeomorphism invariance, \textit{i.e.}, the freedom
to transform the three-metric $\bm{h}$ by the action of a diffeomorphism
$\phi:\Sigma\rightarrow\Sigma$ within the Cauchy surface, \textit{i.e.} $\bm{h}\mapsto\phi_{*}\bm{h}$,
without changing the equations of motion. This may be regarded as
the freedom to choose a different spatial coordinate system (within
$\Sigma$). We leave the proof of this claim to the end of this section,
where we will be able to show it quite succinctly using the symplectic
structure.

\begin{figure}
\begin{centering}
\includegraphics[scale=0.6]{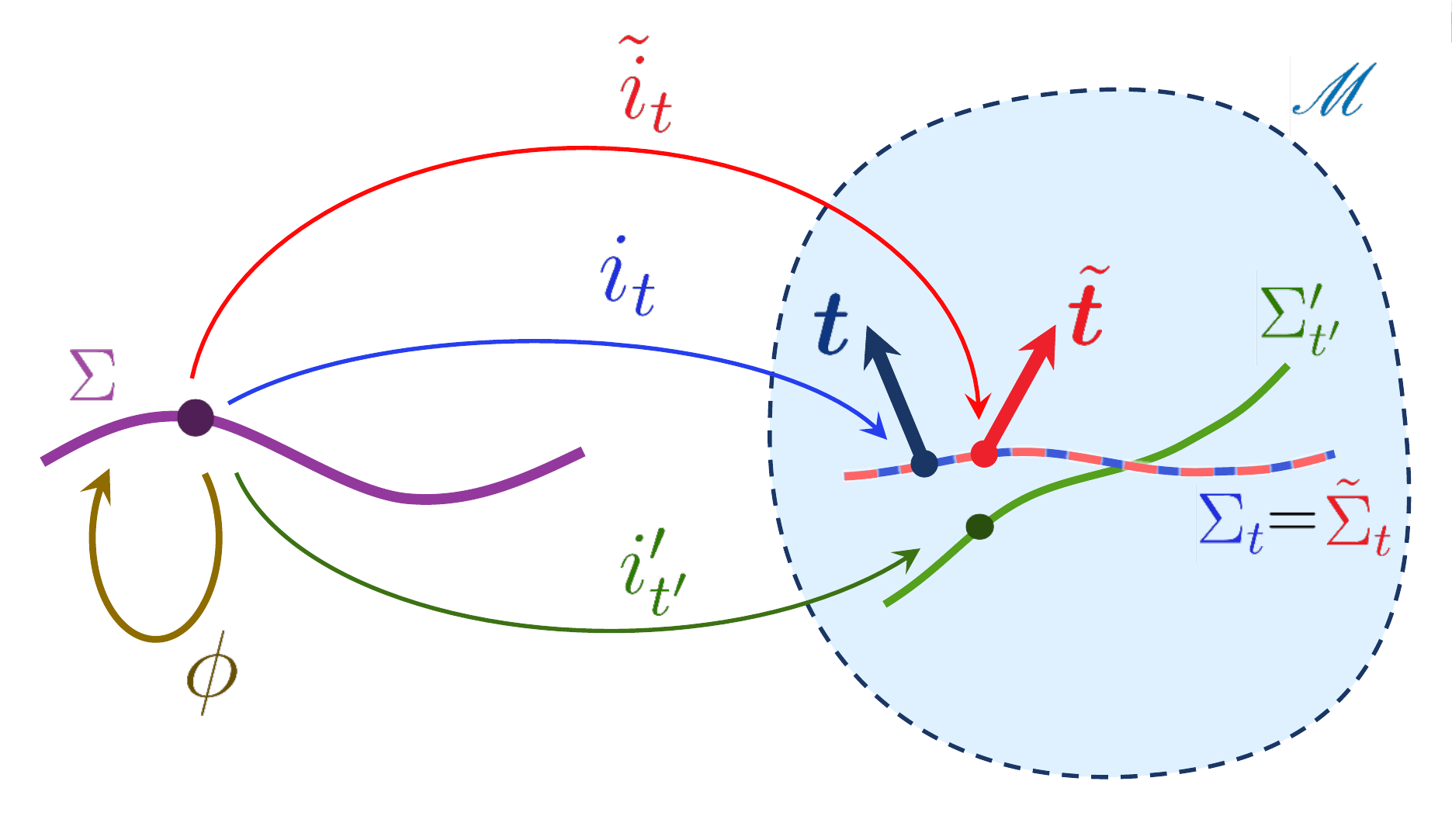}
\par\end{centering}
\caption{A visual representation of the ``gauge freedoms'' of GR. The embedding $i_{t}:\Sigma\rightarrow\mathscr{M}$ is
shown in blue, along with the transformations on this embedding permitted by the constraints, shown in different colours.
In particular, the primary constraints imply that we can change $i_{t}$
to a new embedding $\tilde{i}_{t}$, shown in red, resulting from
a change in the time flow vector field $\bm{t}\mapsto\tilde{\bm{t}}$
(or equivalently, $(N,\bm{N})\mapsto(\tilde{N},\tilde{\bm{N}})$).
The embedded surface itself does not change, \textit{i.e.} $\tilde{\Sigma}_{t}=\tilde{i}_{t}(\Sigma)=i_{t}(\Sigma)=\Sigma_{t}$,
but the identification of spatial coordinates on sequential Cauchy
surfaces in the family of embeddings does. On the other hand, the
Hamiltonian constraint implies the freedom to change from $i_{t}$
to $i'_{t'}$, shown in green, which is a change of foliation, or
time function redefinition $t(x^{a})\mapsto t'(x^{a})$, such that
$\Sigma'_{t'}=i'_{t'}(\Sigma)$ does not coincide with $\Sigma_{t}=i_{t}(\Sigma)$.
Finally, the momentum constraint implies the freedom to map the spatial
metric $\bm{h}$ in $\Sigma$ by a diffeomorphism $\phi$, $\boldsymbol{h}\mapsto\phi_{*}\bm{h}$.}\label{2-fig-i_t_constr}
\end{figure}

We summarize the constraints of GR in Table \ref{2-table_constr}. A visual depiction of
the gauge freedoms related to each of the constraints is shown in
Fig. \ref{2-fig-i_t_constr}.

\begin{table}
\begin{centering}
\begin{tabular}{|c||c|c|c|c|}
\hline 
 & \multicolumn{4}{c|}{The constraints of GR}\tabularnewline
\hline 
Type  & \multicolumn{2}{c|}{\textbf{Primary} } & \multicolumn{2}{c|}{\textbf{Secondary}}\tabularnewline
\hline 
Name  & \emph{Lapse} & \emph{Shift}  & \emph{Hamiltonian}  & \emph{Momentum}\tabularnewline
 & \emph{momentum} & \emph{momentum} & \emph{constraint}  & \emph{constraint}\tabularnewline
 & \emph{constraint}  & \emph{constraint}  &  & \tabularnewline
\hline 
Definition  & $0=\pi^{(N)}$  & $0=\boldsymbol{\pi}^{(\boldsymbol{N})}$  & $0=C$  & $0=\bm{C}$\tabularnewline
\hline 
Eqn.  & (\ref{eq:CG-1t}) & (\ref{eq:CG-1s}) & (\ref{eq:CG-2t}) & (\ref{eq:CG-2s})\tabularnewline
\hline 
DoF  & \multicolumn{2}{c|}{Time flow vector field} & Time function  & Diffeomorphism\tabularnewline
 & \multicolumn{2}{c|}{invariance} & invariance  & invariance on $\Sigma$\tabularnewline
\hline 
Map  & \multicolumn{2}{c|}{$\bm{t}\mapsto\tilde{\bm{t}}$} & $t(x^{a})\mapsto t'(x^{a})$  & $\boldsymbol{h}\mapsto\phi_{*}\bm{h}$\tabularnewline
\hline 
\end{tabular}
\par\end{centering}
\caption{The constraints of GR. These are classified into primary and secondary
constraints, with the name, equation, and DoF (degree of freedom)
associated to each as well as the map permitted by the latter.\label{2-table_constr}}
\end{table}

We end our discussion on constraints here by remarking on the subtle
difference in the meaning of ``coordinate freedom'' in GR implied
by the primary versus the secondary constraints. The secondary constraints---the
momentum and Hamiltonian constraints---are respectively equivalent
to re-labeling spatial coordinates within a Cauchy surface, and re-labeling
the time coordinate (the Cauchy surface foliation within the spacetime).
Instead, the primary constraints---the vanishing of the lapse and
shift momentum---are about the freedom in how one identifies spatial
coordinates on a particular Cauchy surface to coordinates on other
embedded Cauchy surfaces in the future (themselves possessing their
own spatial diffeomorphism invariance), via the time flow vector field.
Nevertheless, these two sorts of freedoms are not completely independent
of each other, a fact which is encoded in the consistency requirement
$\nabla_{\bm{t}}t=1$. 

\subsection{The Hamiltonian and the equations of motion}

The computation of the gravitational Hamiltonian $\mathcal{H}_{\textrm{G}}$,
and hence from this, the total gravitational Hamiltonian $\tilde{\mathcal{H}}_{\textrm{G}}=\mathcal{H}_{\textrm{G}}-\tilde{\lambda}\tilde{\zeta}_{\textrm{G}}-\tilde{\bm{\lambda}}\cdot\tilde{\bm{\zeta}}_{\textrm{G}}$
needed to obtain the equations of motion, now follows by directly applying the recipe
outlined in the previous section for general field theories. We have
laid out all the basic ingredients and from here, conceptually, this
is relatively straightforward, however the computation itself turns
out to be quite lengthy. Detailed step-by-step presentations can be
found, \textit{e.g.}, in Chapter 3 of [\cite{bojowald_canonical_2011}] or
Chapter 4 of [\cite{poisson_relativists_2007}]. The result is 
\begin{equation}
\tilde{H}_{\textrm{G}}\left[\varphi_{\textrm{G}},\pi_{\textrm{G}}\right]=\,\frac{1}{2\kappa}\int_{\Sigma}\mathbf{e}\,\sqrt{h}\left[\lambda\cdot\zeta+2\boldsymbol{\mathcal{D}}\cdot\left(\frac{1}{\sqrt{h}}\boldsymbol{N}\cdot\bm{\pi}\right)\right]\,.\label{eq:HGtotal}
\end{equation}

Notice that, in the bulk (${\rm int}(\Sigma)$), we only have a linear
combination of (all) constraints, $\lambda\cdot\zeta=\lambda^{\mathsf{j}}\zeta_{\mathsf{j}}$.
In particular, these are formed by linear contributions from the primary
constraints $\tilde{\zeta}_{\textrm{G}}$ and $\tilde{\bm{\zeta}}_{\textrm{G}}$
(added, respectively, with multipliers $\tilde{\lambda}$ and $\tilde{\bm{\lambda}}$
to obtain $\tilde{\mathcal{H}}_{\textrm{G}}$ from $\mathcal{H}_{\textrm{G}}$)
and secondary constraints $C$ (\ref{eq:Hamiltonian-constr}) and
$\bm{C}$ (\ref{eq:momentum-constr}) (added with multipliers obtained
from the computation of the Legendre transform): 
\begin{equation}
\lambda\cdot\zeta=-\frac{\tilde{\lambda}}{\sqrt{h}}\pi^{(N)}-\frac{\tilde{\bm{\lambda}}}{\sqrt{h}}\cdot\bm{\pi}^{(\bm{N})}+NC-2\bm{N}\cdot\bm{C}\,.
\end{equation}

The canonical equations of motion, again following a lengthy but straightforward
computation, can now be obtained from what we have established. The
result is:

\begin{align}
\dot{h}_{ab}\,\mathring{=}\,\left\{ h_{ab},\tilde{H}_{\textrm{G}}\right\} \,\mathring{=}\,\, & 2\left[\frac{N}{\sqrt{h}}\left(\pi_{ab}-\frac{1}{2}\pi h_{ab}\right)+D_{(a}N_{b)}\right]\,,\label{eq:hdot}\\
\dot{\pi}^{ab}\,\mathring{=}\,\left\{ \pi^{ab},\tilde{H}_{\textrm{G}}\right\} \,\mathring{=}\,\, & \frac{N}{\sqrt{h}}\left[-\mathcal{R}^{ab}+\frac{1}{2}h^{ab}\left(\mathcal{R}+\bm{\pi}:\bm{\pi}-\frac{1}{2}\pi^{2}\right)-2\pi^{ac}\pi_{c}\,^{b}+\pi\pi^{ab}\right]\nonumber \\
 & +\sqrt{h}\left[D^{a}D^{b}N-h^{ab}D^{2}N+\bm{D}\cdot\left(\frac{\bm{N}\pi^{ab}}{\sqrt{h}}\right)\right]-2\pi^{c(a}D_{c}N^{b)}\,,\label{eq:pidot}
\end{align}
where $D^{2}=\bm{D}\cdot\bm{D}$.

\subsection{Symplectic structure and gauge freedom}

Typically what is referred to as \emph{the symplectic form of GR }is
the symplectic form on the primary constraint surface $\tilde{\mathscr{C}}=\{\left(N,\bm{N},\bm{h},\boldsymbol{\pi}\right)\}$,
denoted (with the correspondent slight abuse of notation) by $\bm{\omega}$.
(A symplectic form on the full GR phase space $\mathscr{P}=\left\{ \left(N,\bm{N},\bm{h},\pi^{(N)},\bm{\pi}^{(\bm{N})},\boldsymbol{\pi}\right)\right\} $
may of course easily be defined by the simple addition of lapse and
shift momentum terms, but for analyzing the symplectic structure one can
assume these automatically to be zero without loss of generality.). In this case, $\bm{\omega}$
is also the volume form $\bm{\Omega}=\bm{\epsilon}_{\tilde{\mathscr{C}}}^{\,}$,
and is given by 
\begin{equation}
\bm{\omega}=\bm{\Omega}=\int_{\Sigma}{\bf e}\,\delta\pi^{ab}\wedge\delta h_{ab}\,.\label{eq:symplectic-form-GR}
\end{equation}
We will see more on this in our work on entropy in Chapter \ref{4-entropy}.

For now, let us use everything we have established to prove that the
momentum constraint $0=\bm{C}=\bm{\mathcal{D}}\cdot(\bm{\pi}/\sqrt{h})$
(\ref{eq:CG-2s}) is related to spatial diffeomorphism degrees of
freedom in the theory. The idea of this proof appears in Appendix
E of [\cite{wald_general_1984}], and we formalize it here by relating
it to the symplectic structure (\ref{eq:symplectic-form-GR}).

Let $\mathcal{G}_{\bm{\xi}}:\tilde{\mathscr{C}}\rightarrow\tilde{\mathscr{C}}$
be an infinitesimal spatial gauge transformation determined by any
vector field $\bm{\xi}\in T\Sigma$. In particular, we know that under
this map, the metric $\bm{h}$ must transform as $\bm{h}\mapsto\bm{h}+\mathcal{L}_{\bm{\xi}}\bm{h}$
where, using the definition of the Lie derivative, $\mathcal{L}_{\bm{\xi}}h_{ab}=\mathcal{D}_{(a}\xi_{b)}$.
Under such a map, the symplectic form transforms as:

\begin{align}
\bm{\omega}=\int_{\Sigma}{\bf e}\,\delta\pi^{ab}\wedge\delta h_{ab}\mapsto\, & \int_{\Sigma}{\bf e}\,\delta\pi^{ab}\wedge\delta\left(h_{ab}+\mathcal{D}_{(a}\xi_{b)}\right)\nonumber \\
=\, & \int_{\Sigma}{\bf e}\,\delta\pi^{ab}\wedge\delta h_{ab}+\int_{\Sigma}{\bf e}\,\delta\pi^{ab}\wedge\delta\mathcal{D}_{(a}\xi_{b)}\nonumber \\
=\, & \bm{\omega}+\int_{\Sigma}{\bf e}\,\delta\pi^{ab}\wedge\mathcal{D}_{(a}\delta\xi_{b)}\,.\label{eq:omega_gauge}
\end{align}
As $\mathcal{G}_{\bm{\xi}}$ is a gauge transformation, it must preserve
the symplectic structure, \textit{i.e.} it must act as the identity on the
symplectic form: $(\mathcal{G}_{\bm{\xi}})^{*}\bm{\omega}=\bm{\omega}$.
Thus, $\bm{\omega}\mapsto\boldsymbol{\omega}$ in (\ref{eq:omega_gauge})
if and only if the integral term in the last equality vanishes. A
simple integration by parts with the appropriate handling of the volume
form and the assumption of vanishing $\bm{\xi}$ on $\partial\Sigma$
turns this into the requirement that 
\begin{equation}
0=\int_{\Sigma}\bm{\epsilon}_{\Sigma}^{\,}\,\left[\mathcal{D}_{(a}\left(\frac{1}{\sqrt{h}}\delta\pi^{ab}\right)\right]\wedge\delta\xi_{b)}\,.
\end{equation}
Since this is true for any vector field $\bm{\xi}$, we find that
this is equivalent to demanding that, everywhere on $\Sigma$,
\begin{equation}
0=\bm{\mathcal{D}}\cdot\left(\frac{1}{\sqrt{h}}\bm{\pi}\right)\,,
\end{equation}
precisely the momentum constraint! This proves that those configurations
of $\bm{h}$ and $\bm{\pi}$ in the primary constraint surface $\tilde{\mathscr{C}}\subset\mathscr{P}$
which enjoy spatial gauge freedom, \textit{i.e.} which are diffeomorphism invariant,
must be those satisfying the momentum constraint.

\section{Applications\label{sec:2.5-applications}}

In this section, we elaborate a bit further on and offer some specific
examples of applications of the canonical formulation of GR. We structure
the discussion into four broad topics: mathematical relativity, numerical
relativity, quantum gravity, and gravitational energy-momentum definitions.

\subsection{Mathematical relativity}

From a mathematical point of view, the vacuum Einstein equation $\bm{G}=0$
on a spacetime $(\mathscr{M},\bm{g},\bm{\nabla})$ constitutes a system
of ten coupled second-order quasi-linear PDEs for the ten components
$g_{\alpha\beta}$ of $\bm{g}$ in some local coordinates $\{x^{\alpha}\}$.
Formulated as a canonical problem, the four secondary constraints
(the Hamiltonian and momentum constraints) form an elliptic system
of PDEs on $\Sigma$, which determines the permissible initial data
for $\bm{h}$ and $\bm{\pi}$. The canonical equations (\textit{i.e.} the time
evolution problem) for $\bm{h}$ and $\bm{\pi}$, together with a
gauge fixing condition (equivalently, a choice of \textbf{$\boldsymbol{t}$},
or $N$ and $\bm{N}$), then form a hyperbolic system of PDEs.

Many important mathematical issues, of intimate concern also for the
physical meaning of GR, can be studied from this perspective. In particular,
while we have devoted much discussion to the formulation of the equations
of motion themselves, we have said little about the general character
of the class of solutions that these equations may admit. We should
of course expect this class to be not only wide enough to include
configurations of the gravitational field actually observed in nature,
but also to avoid ``pathological'' developments of any given (permissible)
initial configurations. Broadly speaking, the aim of \emph{mathematical
relativity} is to rigorously address these sorts of problems via mathematical
analysis and PDE theory. 

A classic problem of in this area, which we outline now briefly, is
that of the well-posedness of GR---or generally, that of a given
field theory. In any theory, if the time evolution of a (complete)
set of canonical fields $\varphi^{A}$ and their conjugate momenta
$\pi_{A}$ is uniquely determined, \textit{i.e.} if solutions to all canonical
equations $\dot{\varphi}=\delta H/\delta\pi$ and $\dot{\pi}=-\delta H/\delta\varphi$
(supplemented with constraints if applicable) exist and are unique,
then the theory is said to possess an \emph{initial value formulation}.

In addition to this, theories in physics are usually also expected
to satisfy the following two properties: 
\begin{enumerate}
\item In a suitable sense, ``small'' changes in initial data $(\varphi|_{\Sigma_{0}},\pi|_{\Sigma_{0}})$
on some $\Sigma_{0}$ should only produce correspondingly ``small''
changes in the solution $(\varphi|_{\Sigma_{t}},\pi|_{\Sigma_{t}})$,
$t>0$, over any fixed compact region of $\mathscr{M}$. 
\item Changes in the initial data $(\varphi|_{\Sigma_{0}},\pi|_{\Sigma_{0}})$
in a given subset of $\Sigma_{0}$ should not produce changes in the
solution outside the causal future of that subset. 
\end{enumerate}
Physically, the first condition is understood to mean that the theory
has basic predictive power, since initial conditions can always be
measured only to finite accuracy. The second condition is essentially
an expression of the relativity principle that information cannot
propagate faster than light.

If a theory possesses an initial value formulation satisfying (1)
and (2), then it is said that the theory is \emph{well-posed}\footnote{~Our definition follows that of [\cite{wald_general_1984}]. For a more technical overview, see [\cite{hilditch_introduction_2013}].}.

Local well-posedness of GR was first proved by Choquet-Bruhat in [\cite{foures-bruhat_theoreme_1952}].
The idea of the proof is to work in the \emph{harmonic} $(\bm{\mathsf{H}})$
gauge, \textit{i.e.} a gauge where locally coordinates $\{x_{\bm{\mathsf{H}}}^{\alpha}\}$,
called harmonic coordinates, satisfy the harmonic equation $\nabla^{2}x_{\bm{\mathsf{H}}}^{\alpha}=0$,
where $\nabla^{2}=\bm{\nabla}\cdot\bm{\nabla}$. In such a gauge,
as shown, \textit{e.g.}, in Chapter 10 of [\cite{wald_general_1984}], the
vacuum Einstein equation is 
\begin{equation}
0=R_{\alpha\beta}^{\bm{\mathsf{H}}}=-\frac{1}{2}g_{\bm{\mathsf{H}}}^{\gamma\delta}\partial_{\gamma}\partial_{\delta}g_{\alpha\beta}^{\bm{\mathsf{H}}}+l_{\alpha\beta}(\bm{g}^{\bm{\mathsf{H}}},\bm{\partial}\bm{g}^{\bm{\mathsf{H}}})\,,\label{eq:EE-harmonic}
\end{equation}
with the final term representing all lower (first and zeroth) order
terms in the PDE, in this case in all (time and space) coordinates. There is then a theorem [\cite{leray_hyperbolic_1952}],
similar in style to the famous Cauchy-Kovalevskaya theorem for linear
hyperbolic PDEs (see, \textit{e.g.}, Chapter 4 of [\cite{evans_partial_1998}])
but applicable to a certain pertinent class of quasilinear hyperbolic
PDEs, including the harmonic gauge Einstein equation (\ref{eq:EE-harmonic}),
from which well-posedness of the latter can be shown.

Since the 1950s, mathematical relativity has developed into a field
in its own right. See, \textit{e.g.}, the extensive textbook [\cite{choquet-bruhat_general_2009}].
Aside from issues of well-posedness, which are of current concern
to modified theories of gravity (where this issue generally complexifies),
another direction of investigation is that of global properties of
solutions\textemdash \textit{e.g.}, global uniqueness of solutions to the Einstein
vacuum equation, first proved in [\cite{choquet-bruhat_global_1969}].
Additionally, there is also the problem of the perturbative stability
of known exact solutions to the Einstein equation. For example, the
(global nonlinear) stability of Minkowski spacetime was famously proved
in [\cite{christodoulou_global_1993}], while the stability of the
Kerr spacetime remains an open problem. See [\cite{coley_open_2017,coley_mathematical_2018}]
for recent comprehensive statements of current open problems in this area.

\subsection{Numerical relativity}

The relevance of obtaining numerical solutions of GR hardly requires
amplification: from investigating the general behaviour and mathematical
character of the Einstein equation and/or (usually quantum/holographic
gravity motivated) modifications thereof (see, \textit{e.g.}, the review [\cite{cardoso_nr/hep:_2012}]),
to direct application in gravitational wave astronomy (see, \textit{e.g.},
the review [\cite{duez_numerical_2018}]); numerical solutions can
give clues not only to the validity of mathematical conjectures, but
also to the sorts of astrophysical observations that may be achievable
and interesting to study. For general recent reviews of the field,
see \textit{e.g.} [\cite{lehner_numerical_2014}; \cite{sarbach_continuum_2012}].

Canonical methods are generally the most preferred approach for numerically
obtaining (especially in the very strong field regime) dynamical solutions
of GR. Another formalism commonly used in numerical relativity aside
from the canonical decomposition is the \emph{null} (or \emph{characteristic})
\emph{decomposition}, involving the choice of one or two of the coordinates
to be null.

Carrying out a numerical evolution in GR in the canonical picture then
essentially begins with a specification of initial data $(\Sigma_{0},\bm{h}_{0},\bm{\pi}_{0})$
for the dynamical fields, along with a lapse $N_{0}$ and shift $\bm{N}_{0}$,
chosen in such a way that the four secondary constraints $C[\varphi_{0}^{\textrm{G}},\pi_{0}^{\textrm{G}}]=0=\boldsymbol{C}[\varphi_{0}^{\textrm{G}},\pi_{0}^{\textrm{G}}]$,
which are four elliptic PDEs on the initial value surface, hold. Then,
one evolves this initial data via the twelve hyperbolic canonical
equations of motion, which can be shown to guarantee the preservation
in time of the secondary constraints. As for the freedom implied by
the primary constraints, this must somehow be taken into account in
the numerical evolution also (recall that this means the freedom to
``re-identify'' spatial points via a transformation of the time
flow vector field from one Cauchy slice to another). In practice this
is typically achieved by some explicit specification either of $(N,\bm{N}$),
or directly of coordinate (``gauge'') conditions. In this way, ensuring
the satisfaction of the primary constraints for numerical solutions
is typically not so problematic.

Instead, what is problematic and what has stifled the progress of
numerical relativity for a long time is dealing with the numerical
propagation of the secondary constraints. While the PDEs ensure these
should remain identically zero, numerically of course one will always
have the accumulation of some non-zero error in time (starting from
initial data exactly satisfying the secondary constraints). A direct
numerical implementation of the standard canonical GR equations in
the exact form developed here, for example, leads to exponentially
growing modes in this error. (These equations are said to be \emph{weakly
hyperbolic}.) To overcome this issue, (so-called \emph{strongly hyperbolic})
reformulations of the initial value problem, friendlier to computational
stability, are required. 

Today, two main such methods are generally in use, which have been
found to keep the secondary constraints under adequate numerical control:
the so-called \emph{generalized harmonic formulation} and the \emph{BSSN}
[\cite{baumgarte_numerical_1998}; \cite{shibata_evolution_1995}]
(or sometimes \emph{BSSNOK}, including the additional authors of [\cite{nakamura_general_1987}])\emph{
formulation}. 

In generalized harmonic formulations, the idea is to work in gauges
similar to the harmonic gauge mentioned in the previous subsection,
generalizing the harmonic equation on the coordinates to include a
source. The first successful simulation of a binary black hole merger
was achieved via a harmonic gauge approach in [\cite{pretorius_evolution_2005}].

The BSSN formulation begins with a conformal rescaling of the spatial
metric $\bm{h}$: in particular, one takes $\tilde{\bm{h}}=\psi^{-4}\bm{h}$
to be the dynamical configuration variable, where $\psi$ is a conformal
factor chosen such that ${\rm det}(\tilde{\bm{h}})=1$. One proceeds
from this to define additional phase space variables via similar rescalings,
and then to obtain the canonical equations of motion for these variables
via the procedures we have outlined in the general canonical (ADM)
case. The equations obtained at this point are still generally numerically
unstable, but what turns out to solve the issue is a clever addition
of the secondary constraints to the dynamical evolution equations.
As the former are identically zero analytically, their addition to
the latter does not affect the solution in theory. Yet it does seem
to affect, in a desirable way, numerical stability in practice. This
has been shown ``empirically'' by its widespread success in strongly
dynamical simulations, and rigorously for perturbations of Minkowski
space [\cite{alcubierre_towards_2000}].

For more on this topic, see the textbooks [\cite{baumgarte_numerical_2010}; \cite{alcubierre_introduction_2012}; \cite{shibata_numerical_2015}].

\subsection{Quantum gravity}

The mathematical formalism of quantum physics has, from its roots
up to modern particle theories, largely taken shape in the basic language
of canonical formulations. Thus modest approaches towards investigating
the question of quantizing gravity have often begun with canonical
formulations of GR. (Indeed, recall that the very first canonical
formulation of GR was developed for this purpose [\cite{pirani_quantization_1950}].)

It turns out that the typical canonical formulation of GR we have
developed in this chapter is not itself easily amenable to standard
(canonical) quantization procedures. That is, an immediately suggestible
idea for quantizing GR would be to turn the set of dynamical classical
phase space variables $(\bm{h},\bm{\pi})|_{\mathscr{C}}$\textemdash \textit{i.e.},
those in the constraint surface (where all the constraints $\zeta=0$
have been solved, and which therefore may be regarded as containing
all observable information without gauge arbitrariness)\textemdash into
quantum Hilbert space operators, with their Poisson bracket relations
becoming commutators (Dirac brackets, with the appropriate factor
of ${\rm i}\hbar$). This turns out to be very difficult to carry
out in practice, as a complete set of observables to characterize
the constraint surface $\mathscr{C}$ of GR seems too difficult to
construct explicitly. Furthermore, even if this is possible, quantizing
only variables in $\mathscr{C}$ by construction leaves out ``off-shell''
information about the solutions (in $\mathscr{P}\backslash\mathscr{C}$),
which may in fact be necessary from a quantum point of view (as, \textit{e.g.},
possible additional contributions in a path integral formulation).

This being the case, one may next contemplate the possibility of working
instead with quantization on the full classical phase space $\mathscr{P}$
(or perhaps just the primary constraint surface $\tilde{\mathscr{C}}$).
Thus, in addition to the dynamical variables, one also promotes the
constraint functions $\zeta_{\mathsf{j}}$ to Hilbert space operators
satisfying the condition that they annihilate any quantum state $|\psi\rangle$
that is a solution of the theory, $\zeta_{\mathsf{j}}|\psi\rangle=0$.
This type of procedure, known as Dirac quantization, also suffers
from a number of technical problems (\textit{e.g.}, ambiguities in the choice
of the order of the factors in the constraint operator, anomalies
\textit{etc.}).

Much greater progress towards the quantization of GR has instead been
made by pursuing \emph{first-order formulations} of the theory, \textit{i.e.}
formulations that produce first-order equations of motion directly
at the Lagrangian level. While of course the Einstein-Hilbert action
$S_{\textrm{EH}}[\bm{g}]=\frac{1}{2\kappa}\int\bm{e}\,\sqrt{-g}R[\bm{g}]$
yields as we have seen second-order equations of motion for the field
$\bm{g}$, a simple example of a first-order Lagrangian formulation
of GR is the \emph{Palatini action}: $S_{\textrm{P}}[\bm{g},\bm{\Gamma}]=\frac{1}{2\kappa}\int\bm{e}\,\sqrt{-g}R[\bm{\Gamma}]$,
where both the metric $\bm{g}$ and the connection $\bm{\Gamma}$
are regarded as physical fields and which, by the vanishing of the
variation with respect to $\bm{g}$, yields first-order equations
of motion for $\bm{\Gamma}$ (which turn out to be Christoffel symbols
by the vanishing of the variation with respect to $\bm{\Gamma}$).

First-order formulations of GR that have proved most useful to quantization
programs have been in the framework of \emph{tetrads}. These can be
very roughly defined as a set of vector fields $\mathfrak{e}_{I}^{a}\in T\mathscr{M}$,
labeled by an ``internal'' index $I=0,1,2,3$, which provides an
orthonormal basis of the tangent space at each point, \textit{i.e.} $\{\mathfrak{e}_{I}^{a}\}_{I=0}^{3}$
are such that 
\begin{equation}
g_{ab}\mathfrak{e}_{I}^{a}\mathfrak{e}_{J}^{b}=\eta_{IJ}={\rm diag}(-1,1,1,1)\label{eq:tetrad}
\end{equation}
is the Minkowski metric in the internal coordinates. The dual to a
tetrad, to which we associate the index-free notation $\bm{\mathfrak{e}}^{I}$,
\begin{equation}
\bm{\mathfrak{e}}^{I}=\mathfrak{e}_{a}^{I}=\eta^{IJ}\mathfrak{e}_{J}^{b}g_{ab}\,,\label{eq:co-tetrad}
\end{equation}
is a one-form on spacetime called a \emph{co-tetrad}; it may be regarded
as encoding the same geometrical/physical information as the metric\textemdash and
thus, one may consider a second order action $S[\bm{\mathfrak{e}}]$
for GR.

Let us see briefly how one can devise a tetrad formulation of GR which
is first order. See Chapter 3 of [\cite{wald_general_1984}] and Chapter
6 of [\cite{bojowald_canonical_2011}] for more details. First, one
can show that the exterior derivative of the co-tetrad (\ref{eq:co-tetrad})
takes the form 
\begin{equation}
{\rm d}\bm{\mathfrak{e}}_{I}=\bm{\mathfrak{e}}^{J}\wedge\bm{\omega}_{IJ}\,,\label{eq:connection-one-forms}
\end{equation}
where the $\bm{\omega}_{IJ}$ are all one-forms (on $\mathscr{M}$)
known as \emph{connection one-forms}\footnote{~We follow standard notational convention for the connection one-forms
in this subsection; these are of course not to be confused with the
symplectic form generally denoted by $\bm{\omega}$.}. Thanks to antisymmetry, all $\bm{\omega}_{IJ}$ can be determined
completely from (\ref{eq:connection-one-forms}) in terms of derivatives
of the co-tetrads $\bm{\mathfrak{e}}^{I}$.

Now consider the spacetime Riemann tensor twice contracted into internal
indices, $R_{abIJ}=R_{abcd}\mathfrak{e}_{I}^{a}\mathfrak{e}_{J}^{b}$.
It is possible to show that this is in fact a collection of spacetime
two-forms $\bm{R}_{IJ}$, labelled by two internal indices. These
can be expressed as a functions of only the connection one-forms $\bm{\omega}$:
\begin{equation}
\bm{R}_{IJ}\left[\bm{\omega}\right]={\rm d}\bm{\omega}_{IJ}+\bm{\omega}_{IK}\wedge\bm{\omega}^{K}\,_{J}\,.\label{eq:Riemann-connection}
\end{equation}
The vacuum Einstein equation can in this case be obtained from the
following action, taken to be a functional of the co-tetrads $\bm{\mathfrak{e}}$
and the connection one-forms $\bm{\omega}$ (and written with respect
to the flat space volume element $\bm{e}$): 
\begin{equation}
S\left[\bm{\mathfrak{e}},\bm{\omega}\right]=\frac{1}{2\kappa}\int e_{IJKL}\bm{\mathfrak{e}}^{I}\wedge\bm{\mathfrak{e}}^{J}\wedge\boldsymbol{R}^{KL}\left[\bm{\omega}\right]\,.\label{eq:tetrad-action}
\end{equation}

In loop quantum gravity, for example, one carries out the quantization
in a canonical formulation derived from an action very similar to
(\ref{eq:tetrad-action}) above, called the \emph{Holst action} [\cite{holst_barberos_1996}].
Essentially, it simply adds an additional topological term which does
not affect the classical equations of motion, but the inclusion of
which turns out to provide crucial space for development of the theory
at the quantum level. The coupling constant of this term, generally
denoted by $\gamma$, is known as the Barbero-Immirzi parameter and,
while freely specifiable at the mathematical level, is thought to
play the role of a physical constant in loop quantum gravity [\cite{barbero_g._real_1995}; \cite{immirzi_real_1997}].

With the addition of this term, a canonical analysis can be carried
out following similar methods as those we have seen in this chapter.
The most useful such formulation for loop quantum gravity was developed
using \emph{Ashtekar variables}, originally introduced in [\cite{ashtekar_new_1987}].
These are closely related to $(\bm{\omega},\bm{\mathfrak{e}})$, and
have proven to be very useful to work with for the purposes of attempting
to quantize the theory.

While much progress has been made in the last few decades following
these lines, a full theory of quantum gravity remains an open problem
in physics today. For more, see the textbook [\cite{rovelli_quantum_2007}]
and the recent review [\cite{ashtekar_general_2015-1}].

\subsection{Gravitational energy-momentum}

The issue of defining gravitational energy-momentum, and conservation
principles in GR more generally, is a notoriously subtle
one for a multitude of reasons. While this will be treated in far
greater detail in Chapter \ref{5-motion}, the key point of this problem has a simple
physical explanation in the equivalence principle [\cite{misner_gravitation_1973}]:
in brief, it is impossible to define a sensible notion of \emph{local}
gravitational energy-momentum (in a similar style as one typically
does for matter), \textit{i.e.} as a volume density, simply because it is always
possible to ``transform away'' any local gravitational field (at
any given spacetime point). Thus a total gravitational energy-momentum
as a volume integral of any local density cannot be meaningfully defined
in GR.

The solution generally accepted to circumvent this problem is instead
to define and work with what are called \emph{quasilocal} definitions
of gravitational energy-momentum: namely, \emph{surface densities}
(rather than volume densities) which, when integrated over the \emph{boundary}
(rather than the interior) of some spatial volume, yield meaningful
definitions of the total energy-momentum of that volume.

Today, there exist a number of proposals for energy-momentum formulas
in this style, often intended to be valid for arbitrary (closed) spatial
regions within a Cauchy surface $\Sigma$ and in agreement with each
other in various limits. See the reviews [\cite{jaramillo_mass_2011}; \cite{szabados_quasi-local_2004}]
for comprehensive summaries. Nevertheless, it is generally expected
that such definitions should agree when applied to the entire Cauchy
surface $\Sigma$, and in particular, that they should recover the
definitions motivated by canonical formulations.

Indeed, as we have seen, canonical methods of the sort developed in
this chapter treat the entire Cauchy surface as the dynamical ``system''
of interest, and are therefore restricted to (possibly) providing
elucidation on the meaning of energy-momentum only for this entire
system, \textit{i.e.} the entire space. For asking questions about ``sub-systems''
of $\Sigma$ (\textit{i.e.} finite spatial regions), further geometrical constructions
are necessary, and often take the form of worldtube boundary splittings
or similar strategies. More on this in Chapter \ref{5-motion}.

For now, let us consider the most classic result for an energy definition
within canonical GR, the \emph{ADM energy}, applicable to a vacuum
spacetime which is \emph{asymptotically flat}. This means that the
spacetime is a development of an initial data set $(\Sigma,\bm{h},\bm{\pi})$
such that, outside some compact subset of $\Sigma$, there exist coordinates
$\{x^{i}\}$ in which the components of $\bm{h}$ and $\bm{\pi}$
satisfy the following ``fall-off'' conditions in terms of $r=(x^{i}x_{i})^{1/2}$:
\begin{align}
h_{ij}-\delta_{ij}=\mathcal{O}\left(r^{-1}\right)\,,\quad & \pi_{ij}=\mathcal{O}\left(r^{-2}\right)\,,\label{eq:AF1}\\
\partial^{n}h_{ij}=\mathcal{O}\left(r^{-(n+1)}\right)\,,\quad & \partial^{n}\pi_{ij}=\mathcal{O}\left(r^{-(n+2)}\right)\,,\quad\forall n\geq1\,.\label{eq:AF2}
\end{align}
This formalizes the idea that, for ``sufficiently large'' $r$,
the spacetime is ``sufficiently close'' to Minkowski. Therefore
such spacetimes can be physically interpreted to describe ``isolated
systems''.

Now consider again the gravitational Hamiltonian $H_{\textrm{G}}=\frac{1}{2\kappa}\int_{\Sigma}\mathbf{e}\,\sqrt{h}[NC-2\bm{N}\cdot\bm{C}+2\boldsymbol{\mathcal{D}}\cdot(\frac{1}{\sqrt{h}}\boldsymbol{N}\cdot\bm{\pi})]$.
On the constraint surface $\mathscr{C}$, only the divergence in the
integrand is in general non-zero. Using Stokes' theorem, one may write
it as a (closed) boundary integral:
\begin{equation}
H_{\textrm{G}}\mathring{=}-\frac{1}{\kappa}\oint_{\mathscr{S}}\boldsymbol{\epsilon}_{\mathscr{S}}^{\,}\left(Nk-\frac{N_{a}r_{b}\pi^{ab}}{N\sqrt{\sigma}}\right)\,,\label{eq:Hsolution}
\end{equation}
where $\mathscr{S}=\partial\Sigma\simeq\mathbb{S}^{2}$ is the Cauchy
surface boundary (topologically a two-sphere), with unit normal $r^{a}$
and induced metric $\bm{\sigma}=\bm{h}|_{\partial\Sigma}$, and where $k={\rm tr}(\boldsymbol{k})$ is the trace of the extrinsic
curvature $\bm{k}$ of $\mathscr{S}$ in $\Sigma$. This $H_{\textrm{G}}|_{\mathscr{C}}$
(\ref{eq:Hsolution}) is sometimes called the ``solution-valued''
Hamiltonian.

For an asymptotically flat $(\Sigma,\bm{h},\bm{\pi})$, \textit{i.e.} satisfying
(\ref{eq:AF1})-(\ref{eq:AF2}), choose a time flow vector field $\boldsymbol{t}$
such that as $r\rightarrow\infty$ we have $\bm{t}\rightarrow\boldsymbol{n}$,
or equivalently $N\rightarrow1$ and $\bm{N}\rightarrow0$ . This
means that asymptotically, spatial points on one time slice of the
spacetime are identified directly along the normal with those on a
future time slice. Such a $\bm{t}$ is said to generate an \emph{asymptotic
time translation}, and so the evaluation of the gravitational Hamiltonian
$H_{\textrm{G}}|_{\mathscr{C}}$ for this type of $\bm{t}$ can be
interpreted physically as the total gravitational energy of $\Sigma$.
It is referred to as the ADM energy [\cite{arnowitt_dynamics_1962}],
and we see by inspection from Eq. (\ref{eq:Hsolution}) that it is
given by:
\begin{equation}
E_{\textrm{ADM}}=-\frac{1}{\kappa}\lim_{r\rightarrow\infty}\oint_{\mathscr{S}}\boldsymbol{\epsilon}_{\mathscr{S}}^{\,}\,k\,,
\end{equation}
the integral of a surface energy density given by $k$. This can be shown to recover,
for example, the mass parameter in exact black hole spacetimes.

A notion of gravitational momentum can also be defined in this setting.
Yet, the asymptotic flatness conditions (\ref{eq:AF1})-(\ref{eq:AF2})
alone do not suffice, as was first analyzed
in detail in [\cite{regge_role_1974}]. In particular, a momentum
definition also requires the \emph{Regge-Teitelboim parity conditions}:
\begin{equation}
\partial^{n}h_{ij}^{\textrm{odd}}=\mathcal{O}\left(r^{-(n+2)}\right)\,,\quad\partial^{n}\pi_{ij}^{\textrm{even}}=\mathcal{O}\left(r^{-(n+3)}\right)\,,\label{eq:RTparity}
\end{equation}
where $f^{\textrm{odd}}(x)=f(x)-f(-x)$ is the odd part of a function
and $f^{\textrm{even}}(x)=f(x)+f(-x)$ the even part. 

If these parity conditions are satisfied, then it is possible to define
an ADM linear momentum, for example, by an application of Noether's
theorem to asymptotic space  translations. (See Chapter 3 of [\cite{bojowald_canonical_2011}].)
The result is:
\begin{equation}
P_{\textrm{ADM}}^{a}=-\frac{1}{\kappa}\lim_{r\rightarrow\infty}\oint_{\mathscr{S}}{\rm d}^{2}x\,r_{b}\pi^{ab}\,.
\end{equation}
A similar formula for an ADM angular momentum may be defined from
asymptotic rotations\footnote{~However, there is arbitrariness in such a formula since it refers,
in asymptotic coordinates, to an origin of rotations which may lie
outside the asymptotic region [\cite{bojowald_canonical_2011}].}. 

It has been proven that data $(\Sigma,\bm{h},\bm{\pi})$ which satisfy
the Regge-Teitleboim parity conditions (\ref{eq:RTparity}) are dense
among asymptotically flat data (satisfying (\ref{eq:AF1})-(\ref{eq:AF2}))
in a suitable weighted Sobolev space [\cite{corvino_asymptotics_2006}].


\chapter{General Relativistic Perturbation Theory\label{3-perturbations}}
\newrefsegment

\subsection*{Chapter summary}

This chapter offers a rigorous presentation of perturbation methods in general relativity. Many problems of interest, in gravitational physics generally, often involve phenomena that are “very close”, in some suitable sense, to a known exact solution of the theory. This permits the expression of quantities of interest in the form of infinite Taylor series about the known, “background” value, and simplifies the problem to that of computing the terms in these series up to the desired order of accuracy. Such tactics form the basis of computing corrections to the motion of a moving object in general relativity, specifically as caused by self-force effects---a topic that we treat in extensive detail in Chapter \ref{5-motion}.
 
We begin in Section \ref{sec:3.1-intro} with a brief introduction, outlining the basic idea behind the general philosophy of perturbation theory in general relativity. Essentially, the view is that one is trying to solve analytically intractable equations defined on an abstract (“perturbed”) spacetime that one cannot construct explicitly, but one that is nonetheless “close enough” to a known exact solution of the theory (the “background”). What must be done, in this case, is to transport these equations to the background manifold under a map---in particular, a diffeomorphism---identifying the different spacetimes, thus turning them into solvable Taylor series on an explicitly known mathematical space. This basic picture, from both a physical and mathematical point of view, lends sensible meaning to the heuristic idea of “adding a perturbation on top of a background”.
 
Section \ref{sec:3.1-general-perturbations} is dedicated to formalizing these ideas mathematically. Special attention is paid to the issue of perturbative gauge freedom. In particular, the choice of the map relating the “perturbed” and “background” spacetimes is not unique, and a change to a different map is shown to correspond to a perturbative gauge transformation.
 
Then in Sections \ref{sec:3.2-SD} and \ref{sec:3.3-Kerr}, we summarize the main ideas and results that have been obtained from the application of perturbation methods to black hole spacetimes. Respectively, these sections consider perturbations to the Schwarzschild-Droste and Kerr spacetimes. The perturbative equations in the former case (the Regge-Wheeler and Zerilli equations) are presented in the context of a canonical analysis, and that in the latter case (the Teukolsky equation) from the point of view of the Newman-Penrose formalism.
 
\subsection*{Teoria general relativista de les pertorbacions \normalfont{(chapter summary translation in Catalan)}}

Aquest capítol ofereix una presentació rigorosa dels mètodes de pertorbació en la relativitat general. Molts problemes d’interès, en la física gravitacional generalment, solen implicar fenòmens “molt propers”, en algun sentit adequat, a una solució exacta coneguda de la teoria. Això permet l’expressió de quantitats d’interès en forma de sèries infinites de Taylor al voltant del valor conegut, de “fons”, i simplifica el problema per computar els termes d’aquestes sèries fins a l’ordre de precisió desitjat. Aquestes tàctiques constitueixen la base del càlcul de correccions al moviment d'un objecte en la relativitat general, concretament causada per efectes d'auto-força, un tema que tractem detalladament al capítol \ref{5-motion}.
 
Comencem a la secció \ref{sec:3.1-intro} amb una breu introducció, que descriu la idea bàsica de la filosofia general de la teoria de les pertorbacions en la relativitat general. Essencialment, es tracta de resoldre equacions analíticament intractables definides en un espai abstracte (``pertorbat'') que no es pot construir explícitament, però que és ``prou proper'' a una solució exacta coneguda de la teoria (el “fons”). El que s’ha de fer, en aquest cas, és transportar aquestes equacions al fons sota un mapa - en concret, un difomorfisme - identificant els diferents espais-temps, convertint-les així en sèries de Taylor solucionables en un espai matemàtic explícitament conegut. Aquesta imatge bàsica, tant des del punt de vista físic com matemàtic, dóna un sentit raonable a la idea heurística “d’afegir una pertorbació en un fons”.
 
La secció \ref{sec:3.1-general-perturbations} es dedica a formalitzar matemàticament aquestes idees. Es presta una atenció especial al problema de la llibertat de mesura pertorbativa. En particular, l’elecció del mapa relacionant els espais-temps “pertorbats” i de “fons” no és única, i es mostra que un canvi a un mapa diferent correspon a una transformació de mesura pertorbativa.

A continuació, a les Seccions \ref{sec:3.2-SD} i \ref{sec:3.3-Kerr}, resumim les idees i resultats principals que s’han obtingut a partir de l’aplicació de mètodes de pertorbació als espais-temps de forats negres. Respectivament, aquestes seccions consideren pertorbacions als espais-temps de Schwarzschild-Droste i Kerr. Les equacions pertorbatives en el primer cas (les equacions Regge-Wheeler i Zerilli) es presenten en el context d’una anàlisi canònica, i en el segon cas (l’equació de Teukolsky) des del punt de vista del formalisme de Newman-Penrose.
 
\subsection*{Théorie générale relativiste des perturbations  \normalfont{(chapter summary translation in French)}}

Ce chapitre propose une présentation rigoureuse des méthodes de perturbation dans la relativité générale. De nombreux problèmes d’intérêt, dans la physique gravitationnelle en général, impliquent souvent des phénomènes « très proches », dans un sens approprié, d’une solution exacte connue de la théorie. Cela permet d’exprimer des quantités d’intérêt sous la forme d’une série infinie de Taylor autour de la valeur « de fond » connue et simplifie le problème en se limitant au calcul des termes de ces séries jusqu’à l’ordre de précision souhaité. De telles tactiques constituent la base du calcul des corrections apportées au mouvement d’un objet en mouvement dans la relativité générale, en particulier à cause des effets de la force propre - un sujet que nous traitons en détail au chapitre \ref{5-motion}.

Nous commençons à la section \ref{sec:3.1-intro} par une brève introduction, décrivant l’idée de base de la philosophie générale de la théorie des perturbations dans la relativité générale. L’essentiel, c’est que l’on essaie de résoudre des équations analytiquement insolubles définies sur un espace-temps abstrait (« perturbé ») qu’on ne peut pas construire explicitement, mais qu’est néanmoins « suffisamment proche » d’une solution exacte connue de la théorie (« le fond »). Ce qui doit être fait, dans ce cas, est de transporter ces équations vers le fond usant une application - en particulier, un difféomorphisme - identifiant les différents espaces-temps, les transformant ainsi en séries de Taylor résolubles sur un espace mathématique explicitement connu. Cette image de base, d’un point de vue physique et mathématique, donne un sens raisonnable à l’idée heuristique « d’ajouter une perturbation au-dessus d’un fond ».

La section \ref{sec:3.1-general-perturbations} est consacrée à la formalisation mathématique de ces idées. Une attention particulière est accordée à la question de la liberté de jauge perturbative. En particulier, le choix de l’application reliant les espaces-temps « perturbé » et « de fond » n’est pas unique et une modification apportée à une carte différente correspond à une transformation perturbative de jauge perturbative.

Ensuite, dans les sections \ref{sec:3.2-SD} et \ref{sec:3.3-Kerr}, nous résumons les idées principales et les résultats qui ont été obtenus à partir de l’application de méthodes de perturbation aux espaces-temps de trous noirs. Respectivement, ces sections traitent des perturbations des espaces-temps de Schwarzschild-Droste et de Kerr. Les équations perturbatives dans le premier cas (les équations de Regge-Wheeler et Zerilli) sont présentées dans le contexte d’une analyse canonique et cela dans le second cas (l’équation de Teukolsky) du point de vue du formalisme de Newman-Penrose.

\section{Introduction\label{sec:3.1-intro}}

Many problems in GR, when exact or fully numerical
solutions cannot be obtained or are impracticable, may be
amenable instead to treatment via perturbation theory. That is, one often
encounters situations where the desired solution to the Einstein equation,
though infeasible to obtain explicitly, is nonetheless ``sufficiently
close'', in some suitable sense, to a known exact solution of the
theory. This allows one then to obtain approximate solutions in the
form of Taylor series about the known exact solution. Undoubtedly
the most famous example of such solutions is that of plane gravitational
waves (in the simplest case, on a flat spacetime background). 

The heuristic notion that one often starts with in thinking about
perturbations is the following. Suppose a background quantity $\mathring{Q}$
(such as the metric), in a background manifold $\mathring{\mathscr{M}}$,
is known explicitly as the solution to an equation of interest $E[\mathring{Q}]=0$
(in $\mathring{\mathscr{M}}$). One then imagines adding a ``small''
perturbation $\delta Q$ to this known background quantity, and then
assuming that the approximate solution which one seeks is $Q\approx\mathring{Q}+\delta Q$,
or more generally $Q=\mathring{Q}+\lambda\delta Q+\mathcal{O}(\lambda^{2})$
where $\lambda$ is a ``small'' expansion parameter. Then, one inserts
this form of $Q$ into the equation of interest $E[Q]=0$ which is
thereby expanded and solved, order by order (up to the desired order),
in $\lambda$.

This point of view of perturbations is in many cases sufficient for
simple calculations, \textit{e.g.} plane gravitational waves can usefully and
simply be thought of as wave-like perturbations ``on top of'' Minkowski for practical purposes.
However, often this perspective is too limiting, and in particular
a careful treatment of the self-force problem---the main topic of
Chapter \ref{5-motion}---requires us to be a bit more precise about exactly what
we mean by ``a perturbation on top of a background''.

Let us begin with a simple question: formally speaking, where (\textit{i.e.}
in what space) does $Q$ live as a mathematical quantity? Clearly
it must live on $\mathring{\mathscr{M}}$, as this is the (exact)
manifold that we know, and in which we know how to carry out calculations.
Nonetheless, $Q$ is the solution to a (``perturbed'') equation
$E[Q]=0$ which is, obviously, \emph{not} the exact equation $E[\mathring{Q}]=0$
for the background solution $\mathring{Q}$ on $\mathring{\mathscr{M}}$.
So where does the equation $E[Q]=0$ really come from?

The answer, of course, is that it is an equation we do not know how
to solve exactly, in a manifold which we also do not know exactly
(and precisely due to which one designs a perturbation procedure to
deal with the problem in the first place). Let $\mathscr{M}_{(\lambda)}$
denote this ``true'', (analytically) unsolvable manifold. The ``true''
quantity $Q_{(\lambda)}$ lives here, and satisfies the equation $E_{(\lambda)}[Q_{(\lambda)}]=0$
on $\mathscr{M}_{(\lambda)}$ exactly. In fact, what we shall ultimately need to work with is a \emph{one-parameter family} of such objects (manifolds and related equations) in $\lambda$; we develop this in detail in the next section, but for the moment, to continue setting out the general idea, it is enough to think of $\mathring{\mathscr{M}}$ and $\mathscr{M}_{(\lambda)}$ as just two manifolds.

Now, the ``true'' manifold $\mathscr{M}_{(\lambda)}$
is assumed to be diffeomorphic to the background $\mathring{\mathscr{M}}$,
so that there exists a diffeomorphism $\varphi:\mathring{\mathscr{M}}\rightarrow\mathscr{M}_{(\lambda)}$
which identifies spacetime points in the background with points in
the ``perturbed'' spacetime. The ``perturbed'' equation $E[Q]=0$
to be solved on $\mathring{\mathscr{M}}$ is then nothing more than
the transport (under $\varphi$) of the ``true'' equation $E_{(\lambda)}[Q_{(\lambda)}]=0$
from $\mathscr{M}_{(\lambda)}$ to $\mathring{\mathscr{M}}$, and
so $Q$ is understood as the transport of $Q_{(\lambda)}$ to the
background, \textit{i.e.} $Q=\varphi^{*}Q_{(\lambda)}$. In other words, what
one is really solving is $\varphi^{*}(E_{(\lambda)}[Q_{(\lambda)}])=E[Q]=0$
on $\mathring{\mathscr{M}}$. If $\lambda$ is ``small'' in some
suitable sense (to be defined more precisely in the next section),
then this produces Taylor series on $\mathring{\mathscr{M}}$ in $\lambda$,
which one then solves order by order.

While this may sound a bit abstract, there is very a sensible physical
meaning to this perspective. The ``background'', strictly speaking,
does not exist as an object of study in the ``real'' world, one
``on top of'' which one ``adds'' perturbations. Rather, the background
is a mathematical idealization---a crucial one, as it provides the
stage upon which we know how to do calculations---which is ``close
enough'' to the ``true'' world as to permit the representation
of quantities of interest in the form of infinite Taylor series thereabout. The latter are just transports to the background
of equations that we do not know how to deal with directly in the
``real'' spacetime, and permit one to arrive at approximations by
truncating the Taylor series at the desired order in the perturbation
parameter.

This geometrical view of perturbation theory not only renders the technical construction conceptually well-motivated, as we shall see, it thereby also avoids running into any
dangerous ambiguities in the interpretation of the perturbation quantities,
especially \textit{vis-à-vis} the delicate issue of perturbative gauge transformations.
Indeed, it is worth remembering that in the history of GR much confusion has
been created by insufficiently careful treatments of general relativistic
perturbations and their related gauge issues, which have often taken a long time to clarify\footnote{~This is particularly
true in the history of cosmological perturbation theory before the
work of [\cite{bardeen_gauge-invariant_1980}], who was the first
to formulate it in terms of gauge-invariant quantities. We will not
comment more on this particular topic in this thesis; see \textit{e.g.} [\cite{mukhanov_theory_1992}; \cite{brandenberger_lectures_2004}].}.

Given the complexity of the self-force problem, and the fact that
gauge issues have proven notoriously difficult therein also, we choose
in this chapter to develop perturbation theory from the geometrical perspective
just outlined. It will prove indispensable for our work on the self-force
in Chapter \ref{5-motion}. 


\section{General formulation of perturbation theory\label{sec:3.1-general-perturbations}}

\subsection{Setup}
Our exposition of perturbation theory in this subsection follows closely
the treatment of {[}\cite{bruni_perturbations_1997}{]}. See also
Chapter 7 of {[}\cite{wald_general_1984}{]} for a simpler treatment
of this topic but following the same philosophy.

\begin{figure}
\begin{centering}
\includegraphics[scale=0.8]{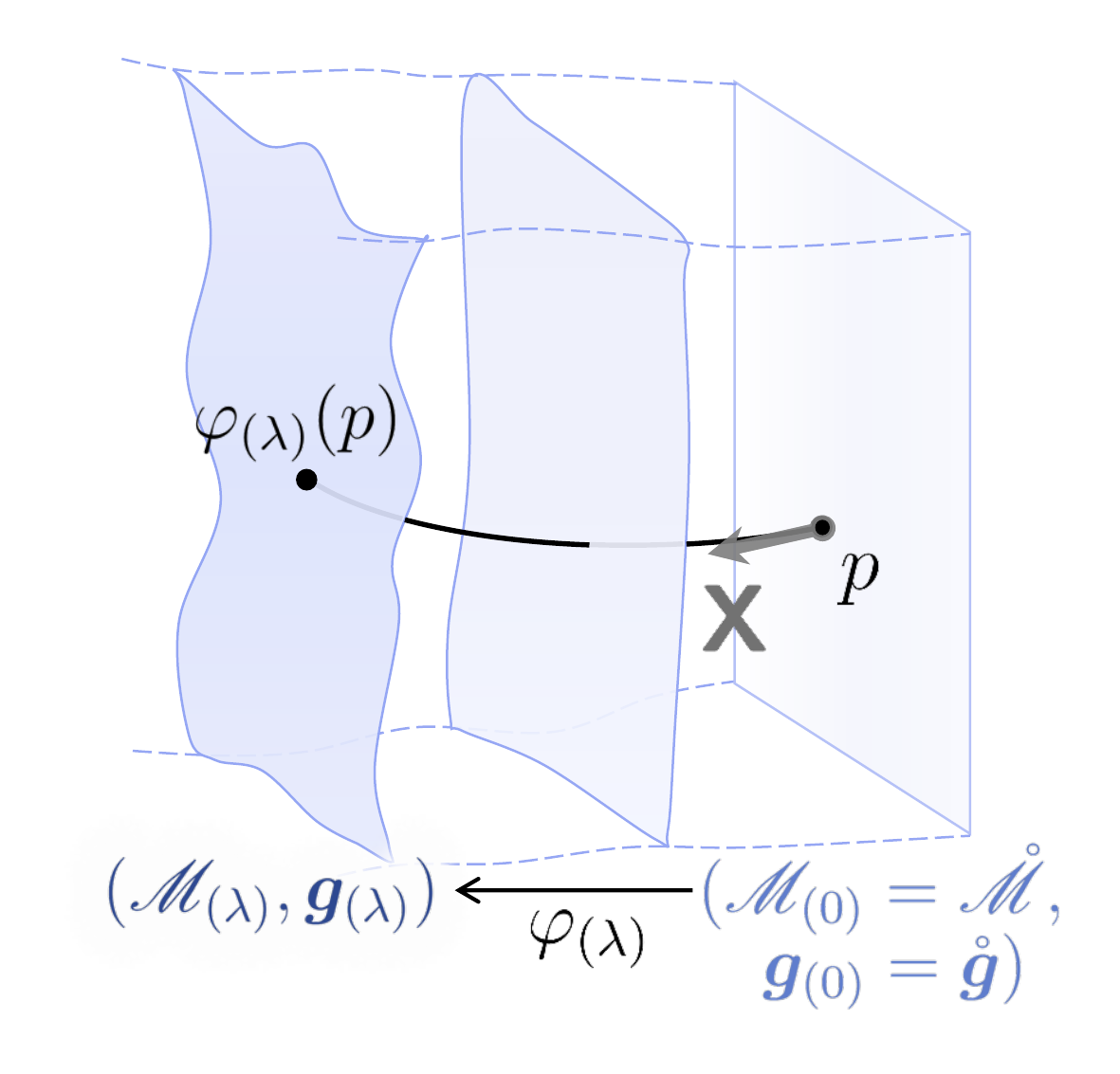} 
\par\end{centering}
\caption{Representation of a one-parameter family of spacetimes $\{\mathscr{M}_{(\lambda)}\}_{\lambda\geq0}$
used for perturbation theory. Each of the $\mathscr{M}_{(\lambda)}$
are depicted visually in $(1+1)$ dimensions, as leaves of a (five-dimensional)
product manifold $\mathscr{N}=\mathscr{M}_{(\lambda)}\times\mathbb{R}$,
with the coordinate $\lambda\geq0$ representing the perturbative
expansion parameter. A choice of a map (or gauge) $\varphi_{(\lambda)}:\mathring{\mathscr{M}}\rightarrow\mathscr{M}_{(\lambda)}$, the flow of which is defined by the integral curves of a vector field $\bm{\mathsf{X}}\in T\mathscr{N}$,
gives us a way of identifying any point $p\in\mathring{\mathscr{M}}=\mathscr{M}_{(0)}$
on the background to one on some perturbed ($\lambda>0$)
spacetime, \textit{i.e.} $p\protect\mapsto\varphi_{(\lambda)}(p)$.}
\label{3-fig-perturbations} 
\end{figure}

Let $\lambda\geq0$ represent our perturbation parameter. It is a
purely formal parameter, in the sense that it should be set equal
to unity at the end of any computation and serves only to indicate
the order of the perturbation. To formalize the ideas outlined in
this chapter introduction, we begin by defining a \emph{one-parameter
family} of spacetimes $\{(\mathscr{M}_{(\lambda)},\bm{g}_{(\lambda)},\bm{\nabla}_{(\lambda)})\}_{\lambda\geq0}$,
where $\bm{\nabla}_{(\lambda)}$ is the connection compatible with
the metric $\bm{g}_{(\lambda)}$ in $\mathscr{M}_{(\lambda)}$, $\forall\lambda\geq0$,
such that $(\mathscr{M}_{(0)},\bm{g}_{(0)},\bm{\nabla}_{(0)})=(\mathring{\mathscr{M}},\mathring{\bm{g}},\mathring{\bm{\nabla}})$
is a known, exact spacetime—the \emph{background}. See Fig. \ref{3-fig-perturbations}
for a visual depiction. For notational convenience, any object with
a sub-scripted ``$(0)$'' (from a one-parameter perturbative family)
is equivalently written with an overset ``$\circ$'' instead. For
the GSF problem, $\mathring{\bm{g}}$ is usually the Schwarzschild-Droste\footnote{~Commonly, this is referred to simply as the ``Schwarzschild metric''. Yet, it has long gone unrecognized that Johannes Droste, then a doctoral student of Lorentz, discovered this metric independently and announced it only four months after Schwarzschild [\cite{droste_het_1916}; \cite{droste_het_1916-1}; \cite{schwarzschild_uber_1916}; \cite{rothman_editors_2002}], so for the sake of historical fairness, throughout this work, we use the nomenclature ``Schwarzschild-Droste metric'' instead.}
or Kerr metric. Then, one should establish a way of smoothly relating
the elements of this one-parameter family (between each other) such
that calculations on any $\mathscr{M}_{(\lambda)}$ for $\lambda>0$—which
may be, in principle, intractable analytically—can be mapped to calculations
on $\mathring{\mathscr{M}}$ in the form of infinite (Taylor) series
in $\lambda$—which, provided $\mathring{\mathscr{M}}$ is chosen
to be a known, exact spacetime, become tractable, order-by-order,
in $\lambda$.

Thus, it is convenient to define a (five-dimensional, Lorentzian)
product manifold 
\begin{equation}
\mathscr{N}=\mathscr{M}_{(\lambda)}\times\mathbb{R}^{\geq}\,\label{eq:stacked},
\end{equation}
the natural differentiable structure of which is given simply by the
direct product of those on $\mathscr{M}_{(\lambda)}$ and the non-negative
real numbers (labeling the perturbation parameter), $\mathbb{R}^{\geq}=\{\lambda\in\mathbb{R}|\lambda\geq0\}$.
For any one-parameter family of $(k,l)$-tensors $\{\bm{A}_{(\lambda)}\}_{\lambda\geq0}$
such that $\bm{A}_{(\lambda)}\in\mathscr{T}^{k}\,_{l}(\mathscr{M}_{(\lambda)})$,
$\forall\lambda\geq0$, we define $\bm{\mathsf{A}}\in\mathscr{T}^{k}\,_{l}(\mathscr{N})$
by the relation 
\begin{equation}
\mathsf{A}^{\alpha_{1}\cdots\alpha_{k}}\,_{\beta_{1}\cdots\beta_{l}}(p,\lambda)=A_{(\lambda)}^{\alpha_{1}\cdots\alpha_{k}}\,_{\beta_{1}\cdots\beta_{l}}(p)\,,\quad \forall p\in\mathscr{M}_{(\lambda)}\, \textrm{ and }\, \forall\lambda\geq0\,.
\end{equation}
Henceforth any such tensor living on the product manifold will be
denoted in serif font—instead of Roman font, which remains reserved
for tensors living on $(3+1)$-dimensional spacetimes. Furthermore,
any spacetime tensor (except for volume forms) or operator written
without a sub- or super-scripted $(\lambda)$ lives on $\mathring{\mathscr{M}}$.
Conversely, any tensor (except for volume forms) or operator living
on $\mathscr{M}_{(\lambda)}$, $\forall\lambda>0$, is indicated via
a sub- or (equivalently, if notationally more convenient) super-scripted
$(\lambda)$, \textit{e.g.} $\bm{A}_{(\lambda)}=\bm{A}^{(\lambda)}\in\mathscr{T}^{k}\,_{l}(\mathscr{M}_{(\lambda)})$
is always tensor in $\mathscr{M}_{(\lambda)}$. The volume form of
any (sub-)manifold $\mathscr{U}$ is always simply denoted by the
standard notation $\bm{\epsilon}_{\mathscr{U}}$ (and is always understood
to live on $\mathscr{U}$).

Let $\Phi_{(\lambda)}^{\bm{\mathsf{X}}}:\mathscr{N}\rightarrow\mathscr{N}$
be a one-parameter group of diffeomorphisms generated by a vector
field $\bm{\mathsf{X}}\in T\mathscr{N}$. (That is to say, the integral
curves of $\bm{\mathsf{X}}$ define a flow on $\mathscr{N}$ which
connects any two leaves of the product manifold.) For notational convenience,
we denote its restriction to maps from the background to a particular
perturbed spacetime (identified by a particular value of $\lambda>0$)
as 
\begin{align}
\varphi_{(\lambda)}^{\bm{\mathsf{X}}}=\Phi_{(\lambda)}^{\bm{\mathsf{X}}}|_{\mathring{\mathscr{M}}}:\mathring{\mathscr{M}}\, & \rightarrow\mathscr{M}_{(\lambda)}\label{eq:phi_map}\\
p\, & \mapsto\varphi_{(\lambda)}^{\bm{\mathsf{X}}}\left(p\right)\,.
\end{align}

The choice of $\bm{\mathsf{X}}$—equivalently, the choice of $\varphi_{(\lambda)}^{\bm{\mathsf{X}}}$—is
not unique; there exists freedom in choosing it, and for this reason,
$\bm{\mathsf{X}}$—equivalently, $\varphi_{(\lambda)}^{\bm{\mathsf{X}}}$—is
referred to as the \emph{perturbative gauge}. We may work with any
different gauge choice $\bm{\mathsf{Y}}$ generating a different map
$\varphi_{(\lambda)}^{\bm{\mathsf{Y}}}:\mathring{\mathscr{M}}\rightarrow\mathscr{M}_{(\lambda)}$.
If we do not need to render the issue of gauge specification explicit,
we may drop the superscript and, instead of $\varphi_{(\lambda)}^{\bm{\mathsf{X}}}$,
we simply write $\varphi_{(\lambda)}$.

Consider now the transport under $\varphi_{(\lambda)}^{\bm{\mathsf{X}}}$
of any tensor $\bm{A}_{(\lambda)}\in\mathscr{T}^{k}\,_{l}(\mathscr{M}_{(\lambda)})$
from a perturbed spacetime to the background manifold. We always denote
the transport of any such tensor by simply dropping the $(\lambda)$
sub- or super-script and optionally including a superscript to indicate
the gauge—that is, $\forall\bm{A}_{(\lambda)}\in\mathscr{T}^{k}\,_{l}(\mathscr{M}_{(\lambda)})$,
\begin{equation}
(\varphi_{(\lambda)}^{\bm{\mathsf{X}}})^{*}\bm{A}_{(\lambda)}=\bm{A}^{\bm{\mathsf{X}}}=\bm{A}\in\mathscr{T}^{k}\,_{l}(\mathring{\mathscr{M}})\,,\label{eq:transport}
\end{equation}
and similarly the transport of $\bm{\nabla}_{(\lambda)}$ to $\mathring{\mathscr{M}}$
is $\bm{\nabla}$. We know, moreover, that we can express any such
$\bm{A}$ as a Taylor series around its background value, $\bm{A}_{(0)}=\mathring{\bm{A}}$
in $\mathring{\mathscr{M}}$. This follows from the Taylor expansion
of $\Phi_{(\lambda)}^{*}\boldsymbol{\mathsf{A}}$ in $\mathscr{N}$
along with the definition of the Lie derivative $\mathcal{L}$ and
the group properties of $\Phi_{(\lambda)}$ {[}\cite{bruni_perturbations_1997}{]}:
\begin{align}
\bm{A}=\, & \mathring{\bm{A}}+\sum_{n=1}^{\infty}\frac{\lambda^{n}}{n!}\mathcal{L}_{\bm{\mathsf{X}}}^{n}\bm{\mathsf{A}}|_{\mathring{\mathscr{M}}}\\
=\, & \mathring{\bm{A}}+\sum_{n=1}^{\infty}\lambda^{n}\delta^{n}\bm{A}\,,\label{eq:A_perturbation_series}
\end{align}
where, in the last equality, we have defined $\delta^{n}\bm{A}=(1/n!)(\partial_{\lambda}^{n}\bm{A})|_{\lambda=0}$
and so the (gauge-dependent) first-order perturbation is $\delta^{1}\bm{A}=\delta\bm{A}=\delta\bm{A}^{\bm{\mathsf{X}}}$.
Note that the symbol $\delta^{n}$, $\forall n$, can be thought of
as an operator $\delta^{n}=(1/n!)\partial_{\lambda}^{n}|_{\lambda=0}$
that acts upon and extracts the $\mathcal{O}(\lambda^{n})$ part of
any tensor in $\mathring{\mathscr{M}}$. We refer to
\begin{equation}
\Delta\bm{A}=\boldsymbol{A}-\mathring{\boldsymbol{A}}=\sum_{n=1}^{\infty}\lambda^{n}\delta^{n}\bm{A}\,\label{eq:tensor perturbation}
\end{equation}
as the (full) perturbation (in the background) of $\bm{A}$.

In particular, we have that the background value of the perturbed
metric $\bm{g}=(\varphi_{(\lambda)}^{\bm{\mathsf{X}}})^{*}\bm{g}_{(\lambda)}$
is $\mathring{\bm{g}}$ and we denote its first-order perturbation
for convenience and according to convention as $\bm{h}=\delta\bm{g}$.
(It is unfortunate that the convention for denoting the spatial three-metric
on a Cauchy slice, as in the previous chapter, is usually the same; we henceforth
clarify which of these two we are talking about if the context does
not make it sufficiently apparent.) Thus we have 
\begin{equation}
\bm{g}=\mathring{\bm{g}}+\lambda\bm{h}+\mathcal{O}(\lambda^{2})\,,\label{eq:metric-expansion}
\end{equation}
where we have omitted explicitly specifying the gauge ($\bm{\mathsf{X}}$)
dependence for now.

Let us define one further piece of notation that we shall later need
to use: let $\mathring{\bm{\Gamma}}$ and $\bm{\Gamma}=(\varphi_{(\lambda)}^{\bm{\mathsf{X}}})^{*}\bm{\Gamma}_{(\lambda)}$
denote the Christoffel symbols (living on $\mathring{\mathscr{M}}$)
associated respectively with $\mathring{\bm{g}}$ and $\bm{g}$, defined
in the usual way (as the connection coefficients between their respective
compatible covariant derivatives and the partial derivative). Then their difference,
\begin{equation}
\bm{C}=(\varphi_{(\lambda)}^{\bm{\mathsf{X}}})^{*}\bm{\Gamma}_{(\lambda)}-\mathring{\bm{\Gamma}}=\bm{\Gamma}-\mathring{\bm{\Gamma}}\,,   
\end{equation}
is the connection coefficient
relating $\bm{\nabla}$ and $\mathring{\bm{\nabla}}$ on $\mathring{\mathscr{M}}$,
which is in fact a tensor. Note that $\mathring{\bm{C}}=0$, i.e.
$\bm{C}=\lambda\delta\bm{C}+\mathcal{O}(\lambda^{2})$. In particular,
it is given by 
\begin{equation}
C^{a}\,_{bc}=\frac{\lambda}{2}\mathring{g}^{ad}\left(\mathring{\nabla}_{b}h_{cd}+\mathring{\nabla}_{c}h_{bd}-\mathring{\nabla}_{d}h_{bc}\right)+\mathcal{O}\left(\lambda^{2}\right)\,.\label{eq:connection_coeff}
\end{equation}

\subsection{Perturbed Einstein equations}
In this setting, then, what one is often interested in is computing
the metric perturbation $\Delta\bm{g}$ for a known background metric
$\mathring{\bm{g}}$. This means transporting the vacuum Einstein
equation $\bm{R}_{(\lambda)}[\bm{g}_{(\lambda)}]=0$ on $\mathscr{M}_{(\lambda)}$
to $\mathscr{M}_{(0)}$. In this way, an \textit{\emph{approximate}}
solution can be expediently obtained for the metric of the spacetime
of interest up to the desired order in $\lambda$. In principle, a
number of technical subtleties must also be kept in mind
whenever a procedure of this sort is implemented {[}\cite{wald_general_1984}{]}:
\textit{(i)} For an $n$-th order approximation (in $\lambda$), it
is in general difficult to estimate the $(n+1)$-th order \textit{error}.
Thus, it may be problematic to determine just how ``small'' $\lambda$
needs to be in order for this perturbative scheme to be valid to sufficient
accuracy. \textit{(ii)} The existence of a one-parameter family $\{\bm{g}_{(\lambda)}\}$
implies the existence of a solution $\bm{h}$ to the linearized field
equation. However, the converse is not true. Thus, merely solving
for $\bm{h}$ does not guarantee that one will have \textit{linearization
stability}, \textit{i.e.} a corresponding exact solution $\bm{g}_{(\lambda)}$
in $\mathscr{M}_{(\lambda)}$.

Now, to transform the vacuum Einstein equation $\bm{R}_{(\lambda)}[\bm{g}_{(\lambda)}]=0$
on $\mathscr{M}_{(\lambda)}$ into an equation on $\mathscr{M}_{(0)}$,
let us begin by considering the definition of the Riemann tensor:
for any $(0,1)$-tensor $\bm{\omega}_{(\lambda)}$ on $\mathscr{M}_{(\lambda)}$,
we have $\nabla_{a}^{(\lambda)}\nabla_{b}^{(\lambda)}\omega_{c}^{(\lambda)}-\nabla_{b}^{(\lambda)}\nabla_{a}^{(\lambda)}\omega_{c}=R_{abc}^{(\lambda)}\,^{d}\omega_{d}^{(\lambda)}$.
The transport of this equation to $\mathscr{M}_{(0)}$, using the fact that the tensor transport commutes with contractions and denoting
$\bm{\omega}=\varphi_{(\lambda)}^{*}\bm{\omega}_{(\lambda)}$, is
simply: 
\begin{equation}
\nabla_{a}\nabla_{b}\omega_{c}-\nabla_{b}\nabla_{a}\omega_{c}=R_{abc}\,^{d}\omega_{d},\label{eq:Riemann formula pulled back}
\end{equation}
where $\bm{\nabla}$ is the transport to to $\mathscr{M}_{(0)}$ of
the derivative operator $\bm{\nabla}_{(\lambda)}$ on $\mathscr{M}_{(\lambda)}$
(compatible with $\bm{g}_{(\lambda)}$). Inserting $\nabla_{a}\omega_{b}=\mathring{\nabla}_{a}\omega_{b}-C^{c}\,_{ab}\omega_{c}$
where $\bm{C}$ is the connection coefficient [Eq. (\ref{eq:connection_coeff})]
into the LHS of the transported Riemann formula {[}Eq. (\ref{eq:Riemann formula pulled back}){]},
a straightforward computation turns this into a relation (in $\mathring{\mathscr{M}}$)
between the perturbed and background values of the Riemann tensor:
\begin{equation}
R_{abc}\,^{d}=\mathring{R}_{abc}\,^{d}-2\mathring{\nabla}_{[a}C^{d}\,_{b]c}+2C^{e}\,_{c[a}C^{d}\,_{b]e}.\label{eq:Riemann w.r.t. Riemann(0)}
\end{equation}
Using the fact that the background metric $\mathring{\bm{g}}$ satisfies
the vacuum Einstein equation, $\mathring{R}_{ac}=0$, we contract
the above to get: 
\begin{equation}
R_{ac}=-2\mathring{\nabla}_{[a}C^{b}\,_{b]c}+2C^{e}\,_{c[a}C^{b}\,_{b]e}.\label{eq:Ricci tensor}
\end{equation}
Inserting the metric expansion {[}Eq. (\ref{eq:metric-expansion}){]}
into the definition of $\bm{C}$ {[}Eq. (\ref{eq:connection_coeff}){]},
and then this into Eq. (\ref{eq:Ricci tensor}), the perturbed vacuum equation
$\bm{R}=0$ on $\mathscr{M}_{(0)}$ becomes an infinite set of equations
at each order in $\lambda$. Carrying out the calculation to second order yields: 
\begin{align}
\mathcal{O}(1):\quad & 0=\mathring{R}_{ac}[\mathring{\bm{g}}]\label{eq:EFE order 1}\\
\mathcal{O}(\lambda):\quad & 0=\delta R_{ac}[\bm{h}]=-\tfrac{1}{2}\mathring{\nabla}_{a}\mathring{\nabla}_{c}h-\tfrac{1}{2}\mathring{\square}h_{ac}+\nabla^{b}\nabla_{(c}h_{a)b}\label{eq:EFE order lambda}\\
\mathcal{O}(\lambda^{2}):\quad & 0=\delta R_{ac}[\delta^{2}\bm{g}]+\delta^{2}R_{ac}[\bm{h}],\label{eq:EFE order lambda^2}
\end{align}
where $\delta^{n}R_{ac}$ is the $n$-th order part of the expansion
of $R_{ac}$ (in powers of $\lambda$), and $\mathring{\square}=\mathring{\nabla}^{b}\mathring{\nabla}_{b}$
is the background wave operator.

\subsection{Gauge transformations}
We now turn to the subtle problem of gauge transformations in perturbation
theory. Thus far, we have been working with a one-parameter group
of diffeomorphisms $\Phi_{(\lambda)}^{\boldsymbol{\mathsf{X}}}:\mathscr{N}\rightarrow\mathscr{N}$
generated by the vector field $\bm{\mathsf{X}}\in T\mathscr{N}$.
What this does, in essence, is to prescribe an identification between
points on the different leaves $\mathscr{M}_{(\lambda)}$ of $\mathscr{N}$
along the integral curves of $\bm{\mathsf{X}}$ (and in particular,
between points on the background and any given perturbed spacetime
via $\varphi_{(\lambda)}^{\bm{\mathsf{X}}}=\Phi_{(\lambda)}^{\bm{\mathsf{X}}}|_{\mathring{\mathscr{M}}}$).
However, this identification is not unique; there is freedom in choosing
the vector field $\boldsymbol{\mathsf{X}}$ (equivalently, the map
$\Phi_{(\lambda)}^{\boldsymbol{\mathsf{X}}})$, referred to as the
\textit{gauge choice}, and a change of this vector field (equivalently,
the associated map) is called a \textit{gauge transformation}.

\begin{figure}
\begin{centering}
\includegraphics[scale=0.8]{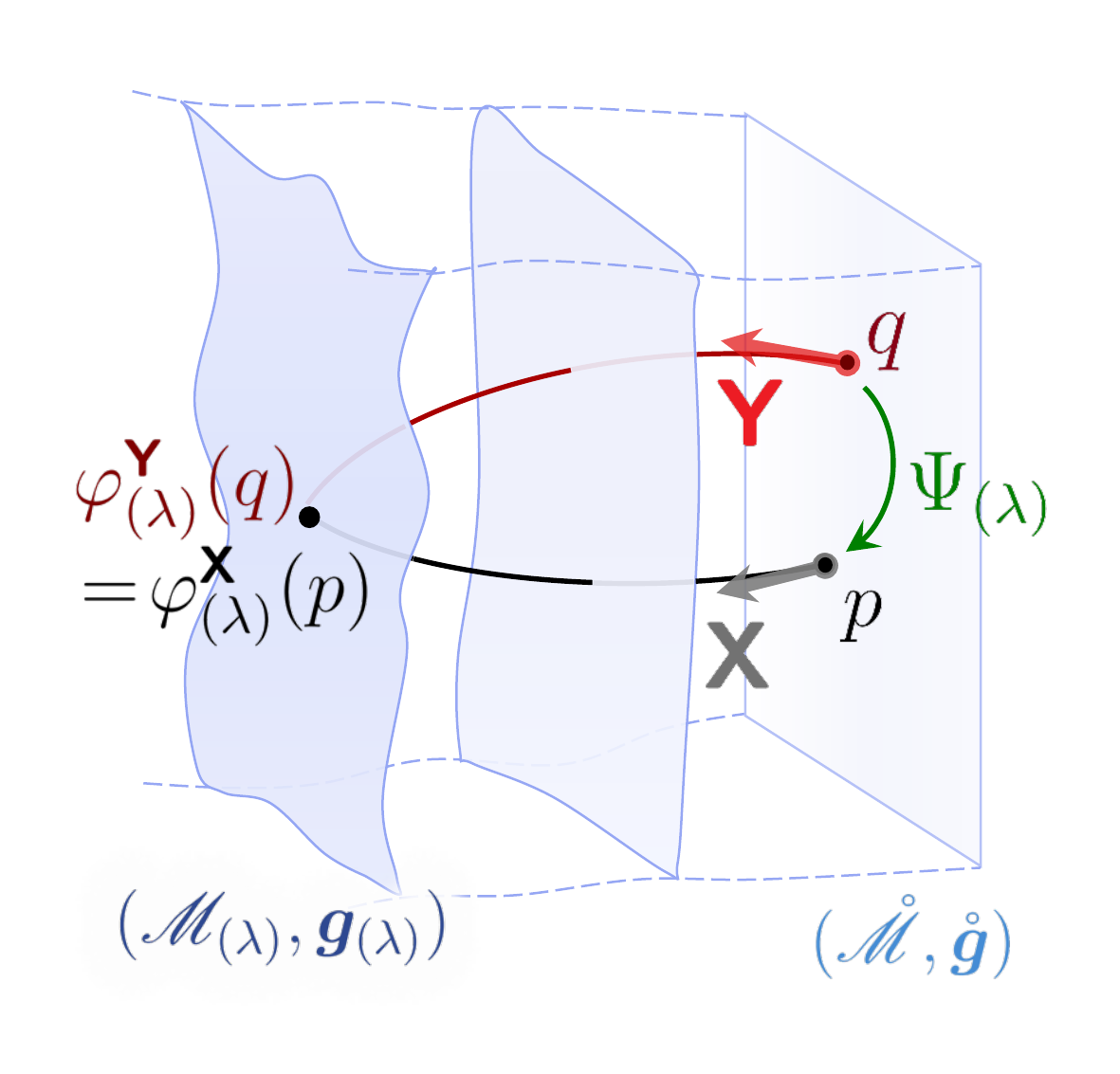} 
\par\end{centering}
\caption{A gauge transformation consists in choosing a different vector field in $T \mathscr{N}$, or equivalently a different associated diffeomorphism, for identifying points between the background and the perturbed spacetimes. In this illustration, the point $p\in \mathring{\mathscr{M}}$ is mapped under the flow of $\bm{\mathsf{X}}$ to the same point in $\mathscr{M}_{(\lambda)}$ as is $q\in \mathring{\mathscr{M}}$ under the flow of $\bm{\mathsf{Y}}$ (for $p\neq q$ and $\bm{\mathsf{X}}\neq \bm{\mathsf{Y}}$). One thus has a gauge transformation on the background $q\mapsto \Psi_{(\lambda)}(q)=p$.}
\label{3-fig-gauge} 
\end{figure}

To understand the effect of performing a gauge transformation, let
$\Phi_{(\lambda)}^{\boldsymbol{\mathsf{X}}}:\mathscr{N}\rightarrow\mathscr{N}$
and $\Phi_{(\lambda)}^{\boldsymbol{\mathsf{Y}}}:\mathscr{N}\rightarrow\mathscr{N}$
be two different (one-parameter groups of) diffeomorphisms, defined
by the integral curves of two different vector fields, $\bm{\mathsf{X}}$
and $\bm{\mathsf{Y}}$ respectively, on $\mathscr{N}$ (such that
$X^{4}=\lambda=Y^{4}$). See Fig. \ref{3-fig-gauge}. 

According to the discussion above, we will obtain two different values of the
perturbation in any $(k,l)$-tensor, $\Delta\bm{A}^{\boldsymbol{\mathsf{X}}}=\boldsymbol{A}^{\boldsymbol{\mathsf{X}}}-\mathring{\bm{A}}$
and $\Delta\bm{A}^{\boldsymbol{\mathsf{Y}}}=\boldsymbol{A}^{\boldsymbol{\mathsf{Y}}}-\mathring{\bm{A}}$
respectively, depending on which map (or vector field) we use. This
is sometimes referred to as \textit{\emph{``gauge ambiguity''}}
in the calculation of the perturbation, and it is said that
$\boldsymbol{\mathsf{A}}$ is (totally) \textit{gauge invariant} if
$\Delta\bm{A}^{\boldsymbol{\mathsf{X}}}=\Delta\bm{A}^{\boldsymbol{\mathsf{Y}}}$
for any $\bm{\mathsf{X}}\neq\boldsymbol{\mathsf{Y}}$. The Stewart-Walker lemma [\cite{stewart_perturbations_1974}] (see also Chapter 1 of [\cite{stewart_advanced_1993}]) tells us that this happens if and only if $\mathring{\bm{A}}$ vanishes, is a constant scalar field, or is a linear combination of products of Kronecker deltas with constant coefficients. In general,
however, this is not necessarily the case, and so it is important to
understand how perturbations change under a gauge transformation.

Let us now define a one-parameter family of diffeomorphisms $\Psi_{(\lambda)}:\mathring{\mathscr{M}}\rightarrow\mathring{\mathscr{M}}$
on the background by: 
\begin{equation}
\Psi_{(\lambda)}=\varphi_{(-\lambda)}^{\bm{\mathsf{X}}}\circ\varphi_{(\lambda)}^{\boldsymbol{\mathsf{Y}}}\,.\label{eq:Psi-gauge}
\end{equation}
What this does is to move points in the background along the integral curves of $\bm{\mathsf{Y}}$ into the perturbed spacetimes, and then along the integral curves of $\bm{\mathsf{X}}$ ``in reverse'', back onto the background.
(Note that this does \textit{not}, in general, form a group.) Then observe
that $\boldsymbol{A}^{\boldsymbol{\mathsf{X}}}$ and $\boldsymbol{A}^{\boldsymbol{\mathsf{Y}}}$
are related by 
\begin{align}
\boldsymbol{A}^{\boldsymbol{\mathsf{Y}}}= & [(\Phi_{(\lambda)}^{\boldsymbol{\mathsf{Y}}})^{*}\boldsymbol{\mathsf{A}}]_{\mathring{\mathscr{M}}} \\
= & [(\Phi_{(\lambda)}^{\boldsymbol{\mathsf{Y}}})^{*}\circ[(\Phi_{(\lambda)}^{\bm{\mathsf{X}}})^{*}]^{-1}\circ(\Phi_{(\lambda)}^{\bm{\mathsf{X}}})^{*}\boldsymbol{\mathsf{A}}]_{\mathring{\mathscr{M}}} \\
= & [(\Phi_{(\lambda)}^{\boldsymbol{\mathsf{Y}}})^{*}\circ(\Phi_{(-\lambda)}^{\bm{\mathsf{X}}})^{*}\circ(\Phi_{(\lambda)}^{\bm{\mathsf{X}}})^{*}\boldsymbol{\mathsf{A}}]_{\mathring{\mathscr{M}}} \\
= & [\Psi_{(\lambda)}^{*}\circ(\Phi_{(\lambda)}^{\bm{\mathsf{X}}})^{*}\boldsymbol{\mathsf{A}}]_{\mathring{\mathscr{M}}} \\
= & \Psi_{(\lambda)}^{*}\boldsymbol{A}^{\boldsymbol{\mathsf{X}}}\,,\label{eq:gauge diffeomorphism}
\end{align}
where in the second line we have introduced the identity, and in the following lines we have applied the definitions established so far.
Now, according to theorems the proofs of which can be found in {[}\cite{bruni_perturbations_1997}{]},
for any one-parameter family of diffeomorphisms $\Psi_{(\lambda)}:\mathring{\mathscr{M}}\rightarrow\mathring{\mathscr{M}}$
{[}Eq. (\ref{eq:Psi-gauge}){]}, there exists an infinite sequence of one-parameter
groups of diffeomorphisms $\{\psi_{(\lambda)}^{(n)}:\mathring{\mathscr{M}}\rightarrow\mathring{\mathscr{M}}\}_{n=1}^{\infty}$
such that $\Psi_{(\lambda)}=\cdots\circ\psi_{(\lambda^{n}/n!)}^{(n)}\circ\cdots\circ\psi_{(\lambda^{2}/2!)}^{(2)}\circ\psi_{(\lambda)}^{(1)}$.
Moreover, the transport under $\Psi_{(\lambda)}$ of any tensor field
$\bm{A}$ on $\mathring{\mathscr{M}}$ has the following series expansion in $\lambda$:
\begin{equation}
\Psi_{(\lambda)}^{*}\bm{A}=\sum_{l_{1},l_{2},l_{3},\ldots=0}^{\infty}\frac{\lambda^{(\sum_{j=1}^{\infty}jl_{j})}}{\prod_{k=1}^{\infty}(k!)^{l_{k}}l_{k}!}\mathcal{L}_{\bm{\xi}_{(1)}}^{l_{1}}\mathcal{L}_{\bm{\xi}_{(2)}}^{l_{2}}\mathcal{L}_{\bm{\xi}_{(3)}}^{l_{3}}\cdots\bm{A}\,,\label{eq:knight diffeomorphism theorem}
\end{equation}
where $\bm{\xi}_{(n)}\in T\mathring{\mathscr{M}}$ is the vector field
in the background generating the flow of each $\psi_{(\lambda)}^{(n)}$.

Applying the above theorem {[}Eq. (\ref{eq:knight diffeomorphism theorem}){]}
to the relation between $\boldsymbol{A}^{\boldsymbol{\mathsf{X}}}$
and $\boldsymbol{A}^{\boldsymbol{\mathsf{Y}}}$ {[}Eq. (\ref{eq:gauge diffeomorphism}){]},
one obtains: 
\begin{equation}
\boldsymbol{A}^{\boldsymbol{\mathsf{Y}}}=\boldsymbol{A}^{\boldsymbol{\mathsf{X}}}+\lambda\mathcal{L}_{\bm{\xi}_{(1)}}\boldsymbol{A}^{\boldsymbol{\mathsf{X}}}+\frac{\lambda^{2}}{2}\left(\mathcal{L}_{\bm{\xi}_{(1)}}^{2}+\mathcal{L}_{\bm{\xi}_{(2)}}\right)\boldsymbol{A}^{\boldsymbol{\mathsf{X}}}+\mathcal{O}(\lambda^{3})\,.\label{eq:gauge diffeomorphism series}
\end{equation}
Substituting series expansions {[}Eq. (\ref{eq:tensor perturbation}){]}
for $\boldsymbol{A}^{\boldsymbol{\mathsf{X}}}$ and $\boldsymbol{A}^{\boldsymbol{\mathsf{Y}}}$
into the above and demanding that the resulting expression holds order
by order yields: 
\begin{align}
\Delta\boldsymbol{A}^{\boldsymbol{\mathsf{Y}}}= & \Delta\boldsymbol{A}^{\boldsymbol{\mathsf{X}}}+\lambda\left\{ \mathcal{L}_{\bm{\xi}_{(1)}}\mathring{\bm{A}}\right\} \nonumber \\
 & +\lambda^{2}\left\{ \frac{1}{2}\left(\mathcal{L}_{\bm{\xi}_{(1)}}^{2}+\mathcal{L}_{\bm{\xi}_{(2)}}\right)\mathring{\bm{A}}+\mathcal{L}_{\bm{\xi}_{(1)}}\left[\left(\mathcal{L}_{\bm{\mathsf{X}}}\boldsymbol{\mathsf{A}}\right)_{\mathring{\mathscr{M}}}\right]\right\} +\mathcal{O}(\lambda^{3}),\label{eq:gauge ambiguity}
\end{align}
where $\bm{\xi}_{(1)}=\bm{Y}-\bm{X}$, and $\bm{\xi}_{(2)}=[\bm{X},\bm{Y}]$. 

To see the relation between a perturbative gauge transformation and
a ``change of coordinates'', let us apply the above to the situation
where $\boldsymbol{\mathsf{A}}$ is simply a coordinate function $x^{\alpha}$.
Then, one can easily check that 
\begin{equation}
\Psi_{(\lambda)}^{*}x^{\alpha}=x^{\alpha}+\lambda\xi_{(1)}^{\alpha}+\frac{\lambda^{2}}{2}\left(\xi_{(1)}^{\beta}\partial_{\beta}\xi_{(1)}^{\alpha}+\xi_{(2)}^{\alpha}\right)+\mathcal{O}(\lambda^{3})\,.\label{eq:coordinate change}
\end{equation}

\subsection{The Lorenz and transverse-traceless gauges}
It is convenient to introduce the \textit{trace-reversed} metric perturbation,
\begin{equation}
\tilde{\bm{h}}=\bm{h}-\frac{1}{2}h\mathring{\bm{g}}\,.    
\end{equation}
Then, under a gauge
transformation generated by some vector field $\bm{\xi}_{(1)}=\bm{\xi}$ (to first
order), we have $\bm{h}^{\boldsymbol{\mathsf{X}}}\mapsto\bm{h}^{\boldsymbol{\mathsf{Y}}}=\bm{h}^{\boldsymbol{\mathsf{X}}}+\mathcal{L}_{\bm{\xi}}\mathring{\bm{g}}$
and so, 
\begin{align}
\tilde{h}_{ab}^{\boldsymbol{\mathsf{Y}}} & =\tilde{h}_{ab}^{\boldsymbol{\mathsf{X}}}+\nabla_{a}\xi_{b}+\nabla_{b}\xi_{a}-(\mathring{\bm{\nabla}}\cdot\bm{\xi})\mathring{g}_{ab}\,,\label{eq:gauge transformation h^tilde}\\
\Rightarrow\mathring{\nabla}^{b}\tilde{h}_{ab}^{\boldsymbol{\mathsf{Y}}} & =\mathring{\nabla}^{b}\tilde{h}_{ab}^{\boldsymbol{\mathsf{X}}}+\mathring{\square}\xi_{a}+\mathring{R}_{ab}\xi^{b}=\mathring{\nabla}^{b}\tilde{h}_{ab}^{\boldsymbol{\mathsf{X}}}+\mathring{\square}\xi_{a}\,,\label{eq: gauge transformation h^tilde derivative}
\end{align}
where in the last line we have used the contraction of the Riemann
formula on $\mathring{\mathscr{M}}$ and finally the fact that $\mathring{\bm{R}}=0$.
If we choose $\bm{\xi}$ to be a solution of the equation $\mathring{\square}\xi_{a}=-\mathring{\nabla}^{b}\tilde{h}_{ab}^{\bm{\mathsf{X}}}$,
then the gauge defined by $\boldsymbol{\mathsf{Y}}$ is known as the
\textit{Lorenz gauge}. In this case we denote $\boldsymbol{\mathsf{Y}}=\boldsymbol{\mathsf{L}}$,
whereupon we have 
\begin{equation}
\mathring{\nabla}^{b}\tilde{h}_{ab}^{\boldsymbol{\mathsf{L}}}=0\,.\label{eq:Lorenz gauge condition}
\end{equation}
This does not, however, completely fix all of the available gauge
freedom, for we could still add to $\bm{\xi}$ any other vector $\bm{\zeta}$
satisfying $\mathring{\square}\bm{\zeta}=0$, and the Lorenz gauge
condition {[}Eq. (\ref{eq:Lorenz gauge condition}){]} would still
hold. Performing a further gauge transformation generated by such
a $\bm{\zeta}$, the trace of $\bm{h}^{\boldsymbol{\mathsf{L}}}$
transforms as $h^{\boldsymbol{\mathsf{L}}}\mapsto h^{\boldsymbol{\mathsf{L}}}+\mathring{\bm{\nabla}}\cdot\bm{\zeta}$.
We can now choose $\bm{\zeta}$ to be a solution of $\bm{\nabla}\cdot\bm{\zeta}=-h^{\boldsymbol{\mathsf{L}}}$.
(Note that this is consistent, since the trace of the linearized Einstein
equation {[}Eq. (\ref{eq:EFE order lambda}){]} assuming the Lorenz
gauge condition {[}Eq. (\ref{eq:Lorenz gauge condition}){]} is $\mathring{\square}h^{\boldsymbol{\mathsf{L}}}=0$,
and so taking the Laplacian of the equation $\mathring{\bm{\nabla}}\cdot\bm{\zeta}=-h^{\boldsymbol{\mathsf{L}}}$
leads to both sides vanishing if $\mathring{\square}\bm{\zeta}=0$.)
In other words, the first-order metric perturbation can also be made traceless;
combined with the Lorenz gauge condition {[}Eq. (\ref{eq:Lorenz gauge condition}){]}
(which now is satisfied also by the metric perturbation itself, since it
equals its trace-reversed part), this leads to the \textit{transverse-traceless}
gauge, defined by the gauge vector $\boldsymbol{\mathsf{T}}$,
\begin{equation}
\mathring{\nabla}^{b}h_{ab}^{\boldsymbol{\mathsf{T}}}=0,\quad h^{\boldsymbol{\mathsf{T}}}=0\,.\label{eq:TT gauge condition}
\end{equation}
In this gauge, the Einstein equation {[}Eq. (\ref{eq:EFE order lambda}){]}
reduces to a very simple wave-type equation: 
\begin{equation}
\mathring{\square}h_{ac}^{\boldsymbol{\mathsf{T}}}+2\mathring{R}_{a}\,^{b}\,_{c}\,^{d}h_{bd}^{\boldsymbol{\mathsf{T}}}=0\,.\label{eq:EFE in TT gauge}
\end{equation}

If the background is Minkowski space, the above simplifies to the
elementary wave equation $\mathring{\square}\boldsymbol{h}^{\bm{\mathsf{T}}}=0$.
This readily admits plane wave solutions; for example, choosing a
coordinate system such that the direction of propagation of the plane
wave is aligned with the Cartesian $z$-direction, one finds the solution:
\begin{equation}
h_{\alpha\beta}^{\bm{\mathsf{T}}}=\left[\begin{array}{cccc}
0 & 0 & 0 & 0\\
0 & h_{+} & h_{\times} & 0\\
0 & h_{\times} & -h_{+} & 0\\
0 & 0 & 0 & 0
\end{array}\right]{\rm exp}\left({\rm i}\bm{k}\cdot\bm{x}\right)\,.\label{eq:GW-solution}
\end{equation}
See Chapter 7 of [\cite{carroll_spacetime_2003}] for more discussion
on this solution and related elementary aspects of gravitational waves.
See also the textbook [\cite{maggiore_gravitational_2007}] for a
more involved development of this topic.

\section{Perturbations of the Schwarzschild-Droste spacetime\label{sec:3.2-SD}}

\subsection{The Schwarzschild-Droste spacetime}

In this section we will consider the problem of perturbations to the
Schwarzschild-Droste spacetime $(\mathring{\mathscr{M}},\mathring{\bm{g}},\mathring{\bm{\nabla}})$
the metric of which is given, in Schwarzschild coordinatres, by:
\begin{equation}
\mathring{\bm{g}}=-f\left(r\right){\rm d}t^{2}+\frac{1}{f\left(r\right)}{\rm d}r^{2}+r^{2}\bm{g}^{\,}_{\mathbb{S}^{2}}\,,\label{eq:g_0_SD}
\end{equation}
where $f(r)=1-2M/r$ is the Schwarzschild function and $\bm{g}^{\,}_{\mathbb{S}^{2}}$
is the metric on the unit two-sphere. We furthermore denote the derivative
operator compatible with $\bm{g}^{\,}_{\mathbb{S}^{2}}$, for ease of readability,
simply as $\bm{\nabla}_{\mathbb{S}^{2}}$ for the remainder of this
section.

It will be instructive to cast this into canonical language, using
the notation and tools established in the previous chapter. Thus,
we reserve here the notation $\bm{h}$ to refer to the spatial three-metric,
and all metric perturbations will be denoted strictly using the $\delta$
notation in order to avoid confusion.

Cauchy surfaces $\mathring{\Sigma}$ for this spacetime can most easily be defined by the constancy of
the Schwarzschild time coordinate, \textit{i.e.} $t={\rm const.}$ in (\ref{eq:g_0_SD}).
In this case the (background) canonical variables can be read off
the metric directly by inspection:
\begin{align}
\mathring{N}=\, & \sqrt{f\left(r\right)}\,,\label{eq:N0_SD}\\
\mathring{\bm{N}}=\, & 0\,,\label{eq:shift0_SD}\\
\mathring{\bm{h}}=\, & \frac{1}{f\left(r\right)}{\rm d}r^{2}+r^{2}\bm{g}^{\,}_{\mathbb{S}^{2}}\,,\label{eq:h0_SD}\\
\mathring{\bm{\pi}}=\, & 0\,.\label{eq:pi0_SD}
\end{align}
Of course, the (background) canonical equations of motion are in this
case trivial.

Now, observe that it is possible to perform a further foliation of
the Schwarzschild Cauchy surfaces $\mathring{\Sigma}$ themselves
each into a set of two-spheres, $\mathring{\Sigma}\simeq\mathbb{R}\times\mathbb{S}^{2}$.
This is in fact a general consequence of the spherical symmetry of
the Schwarzschild-Droste spacetime, and is true irrespective of the
(background) coordinate choice. In particular,
let us define
\begin{equation}
\mathscr{S}_{r}=\left\{ p\in\mathring{\Sigma}:\left.\left(x^{i}x_{i}\right)^{1/2}\right|_{p}=r\right\} \simeq\mathbb{S}_{r}^{2}\,
\end{equation}
to be $r$-radius two-spheres embedded in $\mathring{\Sigma}$, topologically
equivalent to $r$-radius two spheres $\mathbb{S}_{r}^{2}$ embedded
in $\mathbb{R}^{3}$. Then we may write
\begin{equation}
\mathring{\Sigma}=\bigcup_{r>2M}\mathscr{S}_{r}\simeq\bigcup_{r>2M}\mathbb{S}_{r}^{2}\,.
\end{equation}
Furthermore let $r^{a}$ denote the outward-pointing unit vector normal
to $\mathscr{S}_{r}$.

\subsection{Historical development of perturbation approaches}

In principle, the perturbation problem for $\delta\bm{g}$ in a given
background spacetime involves solving for ten degrees of freedom (determined
by the ten linearized Einstein equations [Eq. (\ref{eq:EFE order lambda})]). Yet, just as the symmetries
of the exact Schwarzschild-Droste spacetime---\textit{i.e.}, the fact that
it is static and spherically-symmetric---tremendously reduce the
number of degrees of freedom and simplify the obtention of the explicit
background solution, they do so too for its perturbations.

Historically, the starting point for this problem has been to profit
from these symmetries from the outset by expanding all expressions
of interest in spherical harmonics, thus reducing the problem to studying
the evolution of spherical harmonic modes. We proceed to define these
following [\cite{martel_gravitational_2005}].

Concretely, any function on $\mathring{\mathscr{M}}$ can be expanded
in the form of a series in the standard scalar spherical harmonic
functions $Y^{lm}:\mathbb{S}^{2}\rightarrow\mathbb{R}$, which are
defined by the eigenvalue PDE
\begin{equation}
0=\left[\triangle_{\mathbb{S}^{2}}+l\left(l+1\right)\right]Y^{lm}\,.\label{eq:Ylm_defn}
\end{equation}
This can be used to account for all coordinate components of the metric
perturbation $\delta\bm{g}$ in the time-time, radial-radial, and
time-radial directions, which thanks to our $(2+1+1)$ splitting may
be viewed as well-defined functions on $\mathring{\mathscr{M}}$.
Hence the set of functions $\{Y^{lm}\}$ forms a complete basis for
these. However, we will also have vector and tensor degrees of freedom
induced by $\delta\bm{g}$ on each $\mathscr{S}_{r}$, for which tensorial
generalizations of the scalar spherical harmonics $Y^{lm}$ are needed.

One defines \emph{even-parity} vector harmonics as the derivatives
of the scalars, $\bm{\nabla}_{\mathbb{S}^{2}}Y^{lm}$, and \emph{odd-parity}
vector harmonics by contracting the Levi-Civita symbol $\epsilon_{\mathfrak{ij}}$
with the derivatives, $\epsilon_{\mathfrak{j}}\,^{\mathfrak{i}}\nabla_{\mathbb{S}^{2}}^{\mathfrak{j}}Y^{lm}$.
(We remind the reader that Fraktur indices $\mathfrak{i}$, $\mathfrak{j}$,
$\mathfrak{k}$,... are used throughout for indicating components
of tensors living on topological two-spheres.) The set 
\begin{equation}
\mathscr{B}_{\textrm{VH}}=\left\{ \nabla_{\mathbb{S}^{2}}^{\mathfrak{i}}Y^{lm},\epsilon_{\mathfrak{j}}\,^{\mathfrak{i}}\nabla_{\mathbb{S}^{2}}^{\mathfrak{j}}Y^{lm}\right\} \label{eq:Basis_VH}
\end{equation}
then forms a complete basis for expanding vectors on two-spheres.
Similarly, one defines \emph{even-parity} ($(0,2)$-) tensor harmonics
as the quantities $g_{\mathfrak{ij}}^{\mathbb{S}^{2}}Y^{lm}$ and
$\nabla_{\mathfrak{i}}^{\mathbb{S}^{2}}\nabla_{\mathfrak{j}}^{\mathbb{S}^{2}}Y^{lm}$
(or any two independent linear combinations of these), and \emph{odd-parity}
($(0,2)$-) tensor harmonics as $\epsilon^{\mathfrak{k}}\,_{(\mathfrak{i}}\nabla_{\mathfrak{j})}^{\mathbb{S}^{2}}\nabla_{\mathfrak{k}}^{\mathbb{S}^{2}}Y^{lm}$.
The set 
\begin{equation}
\mathscr{B}_{\textrm{TH}}=\left\{ g_{\mathfrak{ij}}^{\mathbb{S}^{2}}Y^{lm},\nabla_{\mathfrak{i}}^{\mathbb{S}^{2}}\nabla_{\mathfrak{j}}^{\mathbb{S}^{2}}Y^{lm},\epsilon^{\mathfrak{k}}\,_{(\mathfrak{i}}\nabla_{\mathfrak{j})}^{\mathbb{S}^{2}}\nabla_{\mathfrak{k}}^{\mathbb{S}^{2}}Y^{lm}\right\} \label{eq:Basis_TH}
\end{equation}
then forms a complete basis for expanding ($(0,2)$-) tensors on two-spheres.
The basis $\{Y^{lm}\}$ of scalar harmonics is itself also referred
to as \emph{even-parity}. The procedure for generating higher tensorial
harmonics can be continued onward in this fashion, and orthogonality
properties generalize from the classic ones known for the scalar harmonics.
See [\cite{martel_gravitational_2005}; \cite{nagar_gauge-invariant_2005}; \cite{price_developments_2007}]
for more technical details on this.

Then, the basic idea of the spherical harmonic approach to Schwarzschild-Droste
perturbations is to expand the components of the time-radial sector
of $\delta\bm{g}$ in $\{Y^{lm}\}$ (as we have argued that we can),
the two-sphere projection of $\delta\bm{g}$---in this case, as a
full $(0,2)$-tensor---on the two-spheres in the tensor harmonic
basis $\mathscr{B}_{\textrm{TH}}$ (\ref{eq:Basis_TH}), and finally
the time-angular and radial-angular components of $\delta\bm{g}$
(``lost'' in the full two-sphere projection, and reminiscent of
the shift vector in the canonical spacetime splitting), as vectors
on the two spheres in the vector harmonic basis $\mathscr{B}_{\textrm{VH}}$
(\ref{eq:Basis_VH}).

Initial investigations in this direction were pioneered by [\cite{regge_stability_1957}],
who began by considering only the odd-parity modes. Supposing we write
$\delta\bm{g}=\delta\bm{g}_{\textrm{even}}+\delta\bm{g}_{\textrm{odd}}$
to indicate the total splitting of the metric perturbation into even
and odd parity spherical harmonic series expansions, the idea was
thus to consider first $\delta\bm{g}_{\textrm{odd}}$ since clearly
its general form is simpler (completely lacking any contributions
from the time-radial sector) compared with that of $\delta\bm{g}_{\textrm{even}}$. 

Regge and Wheeler actually found a perturbative gauge transformation---now
eponymously named---that reduces the number of odd-parity degrees
of freedom even further, showing that all of the information about
$\delta g_{ab}^{\textrm{odd}}$ can be reconstructed from a single,
gauge-invariant \emph{master function}, also known eponymously as
the \emph{Regge-Wheeler function}, which we denote as $\Phi_{(-)}(t,r,\theta,\phi)$.
To state the exact relationship in this language requires a few further
definitions which we prefer to omit here. Nevertheless, we shall see
explicitly in the following subsection an approach to this problem
from a canonical point of view, where the definition of this master
function is possible to state quite simply in canonical language.

In fact, the exact choice of definition of this master function is
not completely unique. There is freedom in its normalization, and
often it is more convenient to work with what effectively amounts to the time
integral of the Regge-Wheeler function $\Phi_{(-)}$, called the Cunningham-Price-Moncrief
function [\cite{cunningham_radiation_1978}]. The variety of master
functions and normalization conventions which are today often worked with
is detailed, \textit{e.g.}, in [\cite{nagar_gauge-invariant_2005}].

In any case, the harmonic modes $\Psi_{(-)}^{lm}(t,r)$ of the Regge-Wheeler
master function $\Phi_{(-)}(t,r,\theta,\phi)$ (or of the Cunningham-Price-Moncrief
function) completely decouple and satisfy a $(1+1)$-dimensional wave
equation with a radially-dependent potential. It has since been known
as the \emph{Regge-Wheeler equation}. (We shall see it explicitly
in the next subsection.) Remarkably, some years later when the analysis
of the even-parity part of the metric perturbation $\delta\bm{g}_{\textrm{even}}$
was also fully carried out in this style by [\cite{vishveshwara_stability_1970}]
and [\cite{zerilli_gravitational_1970}], it was similarly found that
all the components thereof can also be reconstructed from a gauge-invariant
master function, in this case called the \emph{Zerilli-Moncrief function},
and denoted here as $\Phi_{(+)}(t,r,\theta,\phi)$. Moreover, the
equation satisfied by the modes $\Psi_{(+)}^{lm}(t,r)$ of this function,
known as the \emph{Zerilli equation}, is the same as the one for the
odd-parity modes $\Psi_{(-)}^{lm}(t,r)$, just with a different (slightly
more complicated) radial potential. We now develop this in detail,
following a canonical approach, in the next subsection.

\subsection{The Regge-Wheeler and Zerilli equations via canonical methods}

Here we summarize a derivation of the Schwarzschild-Droste perturbation
equations---the Regge-Wheeler and Zerilli equations---put forward
by [\cite{jezierski_energy_1999}]. In particular, this approach is
based on canonical methods, and will thus reveal a useful synthesis
of many of the main ideas we have developed in this thesis so far.
Furthermore, it will also prove very helpful in our work on entropy
in Chapter \ref{4-entropy}. 

First, in order to avoid notational confusion, let $\delta_{\mathscr{P}}$
denote for this subsection the functional exterior derivative on
the phase space $\mathscr{P}$ (so as not to confuse it with the use
of ``$\delta$'' in our perturbative notation). In [\cite{jezierski_energy_1999}],
the reduced symplectic form of GR for the perturbed Schwarzschild-Droste
spacetime is computed---that is, the pullback of the symplectic form
$\bm{\omega}=\int_{\mathring{\Sigma}}\bm{{\rm e}}\,\delta_{\mathscr{P}}\pi^{ab}\wedge\delta_{\mathscr{P}}h_{ab}$
to the reduced phase space $\mathscr{S}$. (See Chapter \ref{2-canonical}.) 

In this case, both the three-metric and its canonical momentum are
perturbation series in $\lambda$ (themselves transports onto $\mathring{\Sigma}$
in $\mathring{\mathscr{M}}$ from a perturbed Cauchy surface of $\mathscr{M}_{(\lambda)}$),
\textit{i.e.} $\bm{h}=\mathring{\bm{h}}+\lambda\delta\bm{h}+\mathcal{O}(\lambda^{2})$
(with $\mathring{\bm{h}}$ given by (\ref{eq:h0_SD})) and $\bm{\pi}=\lambda\delta\bm{\pi}+\mathcal{O}(\lambda^{2})$
(since here $\mathring{\bm{\pi}}=0$). Then, one can decompose all
of these quantities into ``radial'' and ``angular'' parts according
to the $(2+1)$ spatial decomposition. Doing this turns out to naturally
isolate the components which are pure gauge degrees of freedom. One
then factors these out, and the final result [\cite{jezierski_energy_1999}]
can be expressed solely in terms of two canonical pairs $(\Phi_{(\pm)}(t,r,\theta,\pi),\Pi_{(\pm)}(t,r,\theta,\pi))$,
all simple functions on $\mathring{\mathscr{M}}$, as:
\begin{equation}
\bm{\omega}|_{\mathscr{S}}=\sum_{\varsigma=\pm}\int_{\mathring{\Sigma}}\bm{{\rm e}}\,\delta_{\mathscr{P}}\Pi_{\left(\pm\right)}\wedge\mathbb{D}\delta_{\mathscr{P}}\Phi_{\left(\pm\right)}\,,
\end{equation}
where the operator $\mathbb{D}=\triangle_{\mathbb{S}^{2}}^{-1}(\triangle_{\mathbb{S}^{2}}+2)^{-1}$
is formed from the unit two-sphere Laplacian $\triangle_{\mathbb{S}^{2}}=\bm{\nabla}_{\mathbb{S}^{2}}\cdot\bm{\nabla}_{\mathbb{S}^{2}}$. 

The two configuration variables $\Phi_{(\pm)}$ are in fact, as notationally
anticipated, precisely the Zerilli and Regge-Wheeler master functions
respectively, discussed in the previous subsection. In terms of the
(dynamical) perturbations of the canonical variables $(\delta\bm{h},\delta\bm{\pi})$,
these are given by:
\begin{align}
\Phi_{\left(-\right)}=\, & \frac{2r^{2}}{\mathring{N}\sqrt{\mathring{h}}}\bm{\epsilon}:\left(\bm{\nabla}_{\mathbb{S}^{2}}\left(\bm{r}\cdot\delta\bm{\pi}\right)\right)\,,\label{eq:Phi_minus}\\
\Phi_{\left(+\right)}=\, & r^{2}\bm{\nabla}_{\mathbb{S}^{2}}\cdot\left(\bm{\nabla}_{\mathbb{S}^{2}}\cdot\delta\bm{h}\right)-\left(\triangle_{\mathbb{S}^{2}}+1\right){\rm tr}_{\mathbb{S}_{r}^{2}}\left(\delta\bm{h}\right)\nonumber \\
 & +\triangle_{\left(+\right)}\left(2\left(\bm{r}+r\bm{\nabla}_{\mathbb{S}^{2}}\right)\cdot\left(\bm{r}\cdot\delta\bm{h}\right)-rf\left(r\right)\partial_{r}{\rm tr}_{\mathbb{S}_{r}^{2}}\left(\delta\bm{h}\right)\right)\,,\label{eq:Phi_plus}
\end{align}
where $\epsilon_{\mathfrak{ij}}$ is the Levi-Civita symbol, and the
last line is written using the operator $\triangle_{\left(+\right)}=(\triangle_{\mathbb{S}^{2}}+2)(\triangle_{\mathbb{S}^{2}}+2-\frac{6M}{r})^{-1}$.
Their conjugate momenta are given by
\begin{equation}
\Pi_{\left(\pm\right)}=\frac{\mathring{N}}{\sqrt{\mathring{h}}}\dot{\Phi}_{\left(\pm\right)}\,,\label{eq:Pi_plus-minus}
\end{equation}
with the factor $\mathring{N}/\sqrt{\mathring{h}}=r^{2}\sin\theta/f\left(r\right)$
in Schwarzschild coordinates.

The full set of $\mathcal{O}(\lambda)$ canonical equations of motion
for $(\delta\bm{h},\delta\bm{\pi})$ can then be shown to reduce to
a set of two simple systems of canonical equations for $(\Phi_{(\pm)},\Pi_{(\pm)})$.
Combining these into second-order equations, they can be written together
compactly as:
\begin{equation}
\left(\mathring{\square}+\frac{8M}{r^{3}}\mathfrak{Q}_{\left(\pm\right)}\right)\Phi_{\left(\pm\right)}=0\,,\label{eq:Phi_pm_fullPDE}
\end{equation}
where $\mathring{\square}=\mathring{\bm{\nabla}}\cdot\mathring{\bm{\nabla}}$
is the wave operator on $\mathring{\mathscr{M}}$ and 
\begin{align}
\mathfrak{Q}_{(-)}=\, & 1\,,\\
\mathfrak{Q}_{(+)}=\, & \left(\triangle_{\mathbb{S}^{2}}-1\right)\left(\triangle_{\mathbb{S}^{2}}+2-\frac{3M}{r}\right)\left(\triangle_{\mathbb{S}^{2}}+2-\frac{6M}{r}\right)^{-2}\,.
\end{align}

Now to finally simplify (\ref{eq:Phi_pm_fullPDE}) into the classic
Regge-Wheeler and Zerilli equations, we can introduce the spherical
harmonic decomposition of the master functions (with a conventional
$1/r$ factor), \textit{i.e.} expand them as series in $\{Y^{lm}\}$:
\begin{equation}
\Phi_{\left(\pm\right)}\left(t,r,\theta,\phi\right)=\frac{1}{r}\sum_{l=0}^{\infty}\sum_{m=-l}^{l}Y^{lm}\left(\theta,\phi\right)\Psi_{\left(\pm\right)}^{lm}(t,r)\,.
\end{equation}
Inserting this into (\ref{eq:Phi_pm_fullPDE}) yields the $(1+1)$-dimensional
wave equation with a radially-dependent potential:
\begin{equation}
\left(\partial_{t}^{2}-\partial_{r_{*}}^{2}+V_{\left(\pm\right)}^{l}\left(r\right)\right)\Psi_{\left(\pm\right)}^{lm}\left(t,r\right)=0\,.
\end{equation}
The wave operator is here written in terms of the tortoise coordinate
$r_{*}=r+2M\ln(\frac{r}{2M}-1)$. The potentials are explicitly
\begin{align}
V_{\left(-\right)}^{l}\left(r\right)=\, & f\left(r\right)\frac{\Lambda r-6M}{r^{3}}\,,\\
V_{\left(+\right)}^{l}\left(r\right)=\, & f\left(r\right)\frac{\left(\Lambda-2\right)^{2}\left(\Lambda r^{3}+6Mr^{2}\right)+36\left(\Lambda-2\right)M^{2}r+72M^{3}}{r^{3}\left(\left(\Lambda-2\right)r+6M\right)^{2}}\,,
\end{align}
where $\Lambda=l\left(l+1\right)$ is minus the eigenvalue of $Y^{lm}$.

\section{Perturbations of the Kerr spacetime\label{sec:3.3-Kerr}}

\subsection{The Kerr spacetime and the Newman-Penrose formalism}

The metric of the Kerr spacetime $(\mathring{\mathscr{M}},\mathring{\bm{g}},\mathring{\bm{\nabla}})$,
in Boyer-Lindquist coordinates $\{t,r,\theta,\phi\}$, is given by:
\begin{align}
\mathring{\bm{g}}=\, & -\frac{\Delta\left(r\right)}{\Sigma\left(r,\theta\right)}\left({\rm d}t-a\sin^{2}\theta{\rm d}\phi\right)^{2}+\frac{\sin^{2}\theta}{\Sigma\left(r,\theta\right)}\left(\left(r^{2}+a^{2}\right){\rm d}\varphi-a{\rm d}t\right)^{2}\nonumber \\
 & +\frac{\Sigma\left(r,\theta\right)}{\Delta\left(r\right)}{\rm d}r^{2}+\Sigma\left(r,\theta\right){\rm d}\theta^{2}\,,
\end{align}
where
\begin{align}
\Delta\left(r\right)=\, & r^{2}-2Mr+a^{2}\,,\\
\Sigma\left(r,\theta\right)=\, & r^{2}+a^{2}\cos^{2}\theta\,.
\end{align}
This metric was discovered by [\cite{kerr_gravitational_1963}]. See
[\cite{teukolsky_kerr_2015}] for more details on the history. 

Unlike the Schwarzschild-Droste spacetime, there is no spherical symmetry
here to permit tackling the perturbation problem via methods
such as the $(2+1+1)$ decomposition and spherical harmonic expansions,
as described in the previous section. 

Instead, an approach which has proven very useful in this case is the
\emph{Newman-Penrose formalism} [\cite{newman_approach_1962}]. This
is an approach that can be formulated more generally and powerfully
from the point of view of spinor methods (see Chapter 13 of [\cite{wald_general_1984}], Chapter 2 of [\cite{stewart_advanced_1993}], or the lecture notes [\cite{andersson_geometry_2016}]),
but we describe it here more accessibly as a variant of the tetrad
method introduced during our discussion on quantum gravity in Chapter
\ref{2-canonical} (Section \ref{sec:2.5-applications}). See Chapter 1 of [\cite{chandrasekhar_mathematical_1998}] for
a detailed exposition from this point of view.

The idea is to introduce what is referred to as a \emph{null tetrad},
typically denoted as $\{\mathfrak{e}_{I}^{a}\}_{I=0}^{3}=\{l^{a},n^{a},m^{a},\bar{m}^{a}\}$,
where the vectors $\bm{l}$ and $\bm{n}$ are real, while $\bm{m}$
and $\bar{\bm{m}}$ are complex conjugates. Such a tetrad is defined
by replacing the flat metric in internal coordinates (the condition
$g_{ab}\mathfrak{e}_{I}^{a}\mathfrak{e}_{J}^{b}=\eta_{IJ}$ used earlier
to define tetrads in Section \ref{sec:2.5-applications}) instead with the following:
\begin{equation}
g_{ab}\mathfrak{e}_{I}^{a}\mathfrak{e}_{J}^{b}=\left[\begin{array}{cccc}
0 & -1 & 0 & 0\\
-1 & 0 & 0 & 0\\
0 & 0 & 0 & 1\\
0 & 0 & 1 & 0
\end{array}\right]\,.
\end{equation}
In other words, $\bm{l}\cdot\bm{n}=-1=-\bm{m}\cdot\bar{\bm{m}}$,
with all other inner products vanishing. The latter means that all
vectors are null, and moreover that $\bm{m}$ and $\bar{\bm{m}}$
are orthogonal to $\bm{l}$ and $\bm{n}$. Locally, the complex vectors
$\bm{m}$ and $\bar{\bm{m}}$ can be regarded as complex combinations
of two orthonormal spacelike (real) vectors $X^{a}$ and $Y^{a}$
which are both orthogonal to the two real null vectors $\bm{l}$ and
$\bm{n}$; in particular, $\bm{m}=\frac{1}{\sqrt{2}}(\bm{X}+{\rm i}\bm{Y})$.

Now consider the Weyl tensor $C_{abcd}$, defined to be the trace-free
part of the Riemann tensor $R_{abcd}$. (See, \textit{e.g.}, Chapter 13 of [\cite{misner_gravitation_1973}]).
In dimensions lower than four, $C_{abcd}$ is actually zero (so the
Ricci tensor completely determines the Riemann tensor). In four dimensions,
it is given by:
\begin{equation}
C^{ab}\,_{cd}=R^{ab}\,_{cd}-2\delta^{[a}\,_{[c}R^{b]}\,_{d]}+\frac{1}{3}\delta^{[a}\,_{[c}\delta^{b]}\,_{d]}R\,.\label{eq:Weyl}
\end{equation}
The Riemann tensor has twenty independent components, ten of which
are accounted for by the Ricci scalar and the other ten by the Weyl
tensor. In vacuum spacetimes, the Weyl tensor is thus the same as
the Riemann tensor, and is for this reason that it is often said to
represent the ``purely gravitational degrees of freedom'' of GR. 

The ten independent components of $C_{abcd}$ can be shown to have
a one-to-one correspondence with the following five complex scalars
[\cite{chandrasekhar_mathematical_1998}]:
\begin{align}
\psi_{0}=\, & C_{abcd}l^{a}m^{b}l^{c}m^{d}\,,\label{eq:NPpsi0}\\
\psi_{1}=\, & C_{abcd}l^{a}n^{b}l^{c}m^{d}\,,\label{eq:NPpsi1}\\
\psi_{2}=\, & C_{abcd}l^{a}m^{b}\bar{m}^{c}n^{d}\,,\label{eq:NPpsi2}\\
\psi_{3}=\, & C_{abcd}l^{a}n^{b}\bar{m}^{c}n^{d}\,,\label{eq:NPpsi3}\\
\psi_{4}=\, & C_{abcd}n^{a}\bar{m}^{b}n^{c}\bar{m}^{d}\,.\label{eq:NPpsi4}
\end{align}

As a concrete example before moving on to discussing perturbations
in this context, a commonly used null tetrad which is especially useful
for calculations involving the Kerr spacetime is the Kinnersley tetrad
[\cite{kinnersley_type_1969}]. In Boyer-Lindquist coordinates,
\begin{align}
\boldsymbol{l}=\, & \frac{1}{\Delta\left(r\right)}\left[\left(r^{2}+a^{2}\right)\partial_{t}+\Delta\left(r\right)\partial_{r}+a\partial_{\phi}\right]\,,\\
\bm{n}=\, & \frac{1}{2\Sigma^{2}\left(r,\theta\right)}\left[\left(r^{2}+a^{2}\right)\partial_{t}-\Delta\left(r\right)\partial_{r}+a\partial_{\phi}\right]\,,\\
\bm{m}=\, & \frac{1}{\sqrt{2}\left(r+{\rm i}a\cos\theta\right)}\left[{\rm i}a\sin\theta\partial_{t}+\partial_{\theta}+\frac{{\rm i}}{\sin\theta}\partial_{\phi}\right]\,.
\end{align}
Using this null tetrad in (\ref{eq:NPpsi0})-(\ref{eq:NPpsi4}), one
finds that all complex scalars vanish except for $\psi_{2}$, which
is given by:
\begin{equation}
\psi_{2}=\frac{M}{\left(r-{\rm i}a\cos\theta\right)^{3}}\,.
\end{equation}

\subsection{The Teukolsky equation}

Consider an asymptotically flat background spacetime (not necessarily
Kerr) perturbed by a plane gravitational wave that is outgoing near
future null infinity. Then one can show that the perturbation to $\psi_{4}$,
also typically denoted $\psi_{4}$ (and we concordantly abuse our
notation, where it should be called $\delta\psi_{4}$), is related
to the $+$ and $\times$ (independent) wave polarization modes $h_{+}$
and $h_{\times}$ respectively, according to:
\begin{equation}
\lim_{r\rightarrow\infty}\psi_{4}=\frac{1}{2}\left(\ddot{h}_{+}-{\rm i}\ddot{h}_{\times}\right)\,.
\end{equation}
See, \textit{e.g.}, the review [\cite{sasaki_analytic_2003}]. In this way, the
complex scalar $\psi_{4}$ is regarded as describing all pertinent
information about outgoing radiation. Hence a general equation for
(perturbations of) $\psi_{4}$ is of interest for the study of gravitational
waveforms. 

The idea of the Newman-Penrose approach to Kerr perturbations, then,
is to develop evolution equations for the complex scalars $\psi_{0},\ldots,\psi_{4}$
rather than using the (linearized) Einstein equation [Eq. (\ref{eq:EFE order lambda})] for the metric
perturbation $\delta\bm{g}$ directly. This issue presents a number
of mathematical subtleties \textit{vis-à-vis} accounting for the correspondence
of the various degrees of freedom; we do not wish to enter deeply
into this here, but the basic idea is to use the Einstein equation
in the definition of the Weyl tensor (\ref{eq:Weyl}), written in
terms of the complex scalars, and then to treat the equations for
the connection one-forms (see Section \ref{sec:2.5-applications}) essentially as the equations
of motion.

It turns out that the equation for $\psi_{4}$ is non-separable. However,
Teukolsky discovered that the equation for a re-scaled perturbation
variable,
\begin{equation}
\psi(t,r,\theta,\phi)=\frac{\psi_{4}(t,r,\theta,\phi)}{\rho^{4}(\theta)}\,\quad \textrm{where}\, \quad \rho(\theta)=-(r-{\rm i}a\cos\theta)^{-1}
\end{equation}
led
to a separable equation now eponymously named [\cite{teukolsky_rotating_1972,teukolsky_perturbations_1973}]. In
vacuum, it reads:
\begin{align}
0=\, & \left(\frac{\left(r^{2}+a^{2}\right)^{2}}{\Delta\left(r\right)}-a^{2}\sin^{2}\theta\right)\partial_{t}^{2}\psi+4\frac{Mar}{\Delta\left(r\right)}\partial_{t}\partial_{\phi}\psi+\left(\frac{a^{2}}{\Delta\left(r\right)}-\frac{1}{\sin^{2}\theta}\right)\partial_{\phi}^{2}\psi\nonumber \\
 & -\frac{1}{\Delta^{2}\left(r\right)}\partial_{r}\left(\Delta^{3}\left(r\right)\partial_{r}\psi\right)-\frac{1}{\sin\theta}\partial_{\theta}\left(\sin\theta\partial_{\theta}\psi\right)\nonumber \\
 & -4\left(\frac{a\left(r-M\right)}{\Delta\left(r\right)}+{\rm i}\frac{\cos\theta}{\sin^{2}\theta}\right)\partial_{\phi}\psi-4\left(\frac{M\left(r^{2}-a^{2}\right)}{\Delta\left(r\right)}-r-{\rm i}a\cos\theta\right)\partial_{t}\psi\nonumber \\
 & +\left(4\cot^{2}\theta-2\right)\psi\,.\label{eq:Teukolsky}
\end{align}
(There is a very similar generalization of this equation for scalar, neutrino,
and electromagnetic perturbations). See also [\cite{teukolsky_kerr_2015}]
for more technical as well as historical details.


\part{Novel Contributions:\\ \normalfont{Entropy, Motion and Self-Force in General Relativity}\label{part2}}

\chapter{Entropy Theorems and the Two-Body Problem\label{4-entropy}}
\newrefsegment

\subsection*{Chapter summary}

This chapter is based on the publication [\cite{oltean_entropy_2016}].
 
In general Hamiltonian theories, entropy may be understood either as a statistical property of canonical systems (attributable to epistemic ignorance), or as a mechanical property (that is, as a monotonic function on the phase space along trajectories). In classical mechanics, various theorems have been proposed for proving the nonexistence of entropy in the latter sense. Here we explicate, clarify, and extend the proofs of these theorems to some standard matter (scalar and electromagnetic) field theories in curved spacetime, and then we show why these proofs fail in general relativity. As a concrete application, we focus on the consequences of these results for the gravitational two-body problem.
 
In Section \ref{sec:Intro}, we provide a historical overview of these issues following the development of statistical mechanics at the end of the 19th century. We formulate more exactly the problem of explaining the second law of thermodynamics for entropy in the two---statistical and mechanical---senses mentioned above. For the remainder of this chapter, we treat the notion of entropy in the latter (mechanical) sense.
 
In Section \ref{sec:CM}, we carry out a proof for the nonexistence of entropy in classical mechanics following an idea briefly sketched by Poincaré (and previously never developed into a full proof), following a perturbative approach based on Taylor expansions of Poisson brackets about a hypothetical “thermodynamic equilibrium” point. We also discuss an alternative, topological approach developed by Olsen.
 
The aim of section \ref{sec:GR} is then to examine to what extent these proofs carry over to general relativity. We show that the perturbative approach can be used to prove the nonexistence of entropy in standard non-gravitational (scalar and electromagnetic) field theories on curved spacetime, but fails to apply to general relativity itself. We also discuss the topological approach and its failure to prove nonexistence of entropy in general relativity due to the non-compactness of the phase space of the theory.
 
In Section \ref{sec:2BP}, we focus on the gravitational two-body problem in light of these ideas, and in particular, we prove the non-compactness of the phase space of perturbed Schwarzschild-Droste spacetimes. We thus identify the lack of recurring orbits in phase space as a distinct sign of dissipation and hence entropy production.
 
Section \ref{sec:Conclusions} offers some concluding remarks.
 
\subsection*{Teoremes d’entropia i el problema de dos cossos \normalfont{(chapter summary translation in Catalan)}}
 
Aquest capítol es basa en la publicació [\cite{oltean_entropy_2016}].
 
En teories hamiltonianes generals, l’entropia es pot entendre o bé com una propietat estadística dels sistemes canònics (atribuïble a la ignorància epistèmica), o com a propietat mecànica (és a dir, com a funció monotònica en l’espai de fase al llarg de les trajectòries). En la mecànica clàssica, s’han proposat diversos teoremes per demostrar la inexistència d’entropia en aquest darrer sentit. Aquí expliquem, aclarim i estenem les proves d’aquests teoremes a algunes teories de camps de matèria estàndard (escalar i electromagnètica) en l’espai-temps corbat, i després mostrem per què aquestes proves fracassen en la relativitat general. Com a aplicació concreta, ens centrem en les conseqüències d’aquests resultats sobre el problema gravitatori de dos cossos.
 
A la secció \ref{sec:Intro}, proporcionem una panoràmica històrica d’aquestes qüestions després del desenvolupament de la mecànica estadística a finals del segle XIX. Formulem més exactament el problema d’explicar la segona llei de la termodinàmica per a l’entropia en els dos sentits - estadístic i mecànic - mencionats anteriorment. Per a la resta d’aquest capítol, tractem la noció d’entropia en el darrer sentit (mecànic).
 
A la secció \ref{sec:CM}, realitzem una prova de la inexistència d’entropia en la mecànica clàssica seguint una idea breument esbossada per Poincaré (i mai abans desenvolupada en una prova completa), seguint un mètode pertorbatiu basat en les expansions de Taylor dels claudàtors de Poisson al voltant d’un hipotètic “equilibri termodinàmic”. També discutirem un mètode topològic alternatiu desenvolupat per Olsen.
 
L'objectiu de la secció \ref{sec:GR} és examinar fins a quin punt aquestes proves es traslladen a la relativitat general. Mostrem que el mètode pertorbatiu es pot utilitzar per demostrar la inexistència d’entropia en teories estàndard de camp no gravitacionals (escalars i electromagnètiques) en l’espai-temps corbat, però no s’aplica a la mateixa relativitat general. També discutirem el mètode topològic i el seu impossibilitat de demostrar la inexistència d’entropia en la relativitat general a causa de la no compactitat de l’espai de fase de la teoria.
 
A la secció \ref{sec:2BP}, ens centrem en el problema gravitatori de dos cossos a la vista d’aquestes idees i, en particular, demostrem la no compactitat de l’espai de fase dels espais-temps de Schwarzschild-Droste pertorbats. Així identifiquem la manca d’òrbites recurrents en l’espai de fase com a signe distint de dissipació i per tant de producció d’entropia.

La secció \ref{sec:Conclusions} ofereix algunes observacions finals.
 
\subsection*{Théorèmes d'entropie et le problème à deux corps  \normalfont{(chapter summary translation in French)}}
 
Ce chapitre est basé sur la publication [\cite{oltean_entropy_2016}].
 
Dans les théories hamiltoniennes générales, l'entropie peut être comprise soit comme une propriété statistique des systèmes canoniques (imputable à l'ignorance épistémique), soit comme une propriété mécanique (c'est-à-dire comme une fonction monotone sur l'espace des phases le long des trajectoires). Dans la mécanique classique, différents théorèmes ont été proposés pour démontrer l’absence d’entropie dans ce dernier sens. Ici, nous expliquons, clarifions et étendons les démonstrations de ces théorèmes à certaines théories de champ standard (scalaire et électromagnétique) dans l’espace-temps courbé, puis nous montrons pourquoi ces démonstrations échouent en relativité générale. Comme application concrète, nous nous concentrons sur les conséquences de ces résultats sur le problème gravitationnel à deux corps.

Dans la section \ref{sec:Intro}, nous fournissons un aperçu historique de ces questions suite au développement de la mécanique statistique à la fin du XIXe siècle. Nous formulons plus précisément le problème de l'explication de la deuxième loi de la thermodynamique pour l'entropie dans les deux sens - statistique et mécanique - mentionnés ci-dessus. Pour la suite de ce chapitre, nous traitons la notion d'entropie dans ce dernier sens (mécanique).

Dans la section \ref{sec:CM}, nous effectuons une démonstration de l’absence d’entropie dans la mécanique classique en suivant une idée brièvement exposée par Poincaré (que n’a jamais été transformée en une démonstration complète), en suivant une méthode perturbative basée sur expansions de Taylor, des crochets de Poisson autour d’une hypothétique « équilibre thermodynamique ». Nous discutons également d'une approche topologique alternative développée par Olsen.

Le but de la section \ref{sec:GR} est alors d’examiner dans quelle mesure ces démonstrations se reportent à la relativité générale. Nous montrons que l'approche perturbative peut être utilisée pour prouver l’absence d'entropie dans les théories standard des champs non gravitationnels (scalaires et électromagnétiques) sur l'espace-temps courbé, mais elle ne s'applique pas à la relativité générale elle-même. Nous discutons également de l'approche topologique et de son incapacité à démontrer l’absence d'entropie dans la relativité générale à cause de la non compacité de l'espace des phases de la théorie.

Dans la section \ref{sec:2BP}, nous nous concentrons sur le problème gravitationnel à deux corps a vu de ces idées et en particulier, nous prouvons la non compacité de l'espace des phases d'espaces-temps de Schwarzschild-Droste perturbés. Nous identifions ainsi le manque d'orbites récurrentes dans l'espace des phases comme un signe distinct de dissipation et donc de production d'entropie.

La section \ref{sec:Conclusions} offre quelques remarques de conclusion.

\section{\label{sec:Intro}Introduction}

The problem of reconciling the second law of thermodynamics\footnote{~``It is the only physical theory of universal content concerning which I am convinced that, within the framework of applicability of its basic concepts, it will never be overthrown.'' [\cite{einstein_autobiographical_1949}]} with classical (deterministic) Hamiltonian evolution is among the oldest in fundamental physics [\cite{sklar_physics_1995,davies_physics_1977,brown_boltzmanns_2009}]. In the context of classical mechanics (CM), this question motivated much of the development of statistical thermodynamics in the second half of the 19th century. In the context of general relativity (GR), thermodynamic ideas have occupied---and, very likely, will continue to occupy---a central role in our understanding of black holes and efforts to develop a theory of quantum gravity. Indeed, much work in recent years has been expended relating GR and thermodynamics [\cite{rovelli_general_2012}], be it in the form of ``entropic gravity'' proposals [\cite{verlinde_origin_2010,van_putten_entropic_2012,carroll_what_2016}] (which derive the Einstein equation from entropy formulas), or gravity-thermodynamics correspondences [\cite{freidel_gravitational_2015,freidel_non-equilibrium_2015-1}] (wherein entropy production in GR is derived from conservation equations, in analogy with classical fluid dynamics). And yet, there is presently little consensus on the general meaning of ``the entropy of a gravitational system'', and still less on the question of why---purely as a consequence of the dynamical (Hamiltonian) equations of motion---such an entropy should (strictly) monotonically increase in time, \textit{i.e.} obey the second law of thermodynamics.

However one wishes to approach the issue of defining it, gravitational entropy should in some sense emerge from suitably defined (micro-)states associated with the degrees of freedom \textsl{not} of any matter content in spacetime, but of the gravitational field itself---which, in GR, means the spacetime geometry---or statistical properties thereof. Of course, we know of restricted situations in GR where we not only have entropy definitions which make sense, but which also manifestly obey the second law---that is, in black hole thermodynamics\footnote{~This field was initially pioneered by [\cite{bekenstein_black_1973}] and [\cite{hawking_particle_1975}]. See [\cite{wald_thermodynamics_2001}] for a review.}. In particular, the black hole entropy is identified (up to proportionality) with its area, and hence, we have that the total entropy increases when, say, two initially separated black holes merge---a process resulting, indeed, as a direct consequence of standard evolution of the equations of motion. What is noteworthy about this is that black hole entropy is thus understood not as a statistical idea, but directly as a functional on the phase space of GR (comprising degrees of freedom which are subject to deterministic canonical evolution).

In CM, the question of the statistical nature of entropy dominated many of the early debates on the origin of the second law of thermodynamics during the development of the kinetic theory of gases [\cite{sklar_physics_1995}]. Initial hopes, especially by [\cite{boltzmann_weitere_1872}], were that entropy could in fact be understood as a (strictly monotonic) function on classical phase space. However, many objections soon appeared which rendered this view problematic---the two most famous being the reversibility argument of [\cite{loschmidt_uber_1876}] and the recurrence theorem of [\cite{poincare_sur_1890}].

The Loschmidt reversibility argument, in essence, hinges upon the time-reversal symmetry of the canonical equations of motion, and hence, the ostensibly equal expectation of evolution towards or away from equilibrium. Yet, arguably, this is something which may be circumvented via a sufficiently convincing proposition for identifying the directionality of (some sort of) arrow of time---and in fact, recent work [\cite{barbour_gravitational_2013,barbour_identification_2014}] shows how this can actually be done in the Newtonian $N$-body problem, leading in this context to a clearly defined ``gravitational'' arrow of time. For related work in a cosmological context, see [\cite{sahni_arrow_2015,sahni_cosmological_2012}].

The Poincar\'{e} recurrence argument, on the other hand, relies on a proof that any canonical system in a bounded phase space will always return arbitrarily close to its initial state (and moreover it will do so an unbounded number of times) [\cite{arnold_mathematical_1997,luis_barreira_poincare_2006}]. As the only other assumption needed for this proof is Liouville's theorem (which asserts that, in any Hamiltonian theory, the probability measure for a system to be found in an infinitesimal phase space volume is time independent), the only way for it to be potentially countered is by positing an unbounded phase space for all systems---which clearly is not the case for situations such as an ideal gas in a box.

Such objections impelled the creators of kinetic theory, Maxwell and Boltzmann in particular, to abandon the attempt to understand entropy---in what we may accordingly call a \emph{mechanical} sense---as a phase space function, and instead to conceive of it as a statistical notion the origin of which is \emph{epistemic ignorance}, \textit{i.e.} observational uncertainty
of the underlying (deterministic) dynamics. The famous \emph{H-Theorem} of [\cite{boltzmann_weitere_1872}], which was in fact initially put forth for the purposes of expounding the former, became reinterpreted and propounded in the light of the latter.

Of course, later such a statistical conception of entropy came to be understood in the context of quantum mechanics via the von Neumann entropy (defined in terms of the density matrix of a quantum system) and also in the context of information theory via the Shannon entropy (defined in terms of probabilities of a generic random variable) [\cite{nielsen_quantum_2010}]. Indeed, the meaning of the word ``entropy'' is nowadays often taken to reflect an observer's knowledge (or ignorance) about the microstates of a system.

Thus, the question of why the second law of thermodynamics should hold in a Hamiltonian system may be construed within two possible formulations---on the one hand, a \textsl{mechanical} (or \emph{ontological}), and on the other, a \textsl{statistical} (or \emph{epistemic}) point of view. Respectively, we can state these as follows.

~

\noindent {\bf Problem I} (\emph{mechanical/ontological}):
Does there exist a function (or functional, if we are dealing with a field theory) on phase space which monotonically increases along the orbits of the Hamiltonian flow?

~

\noindent {\bf Problem II} (\emph{statistical/epistemic}):
Does there exist a function of time, defined in a suitable way in terms of a probability density on phase space, which always has a non-negative time derivative in a Hamiltonian system?

~

In the language and notation that we have established in Chapter \ref{2-canonical} for describing general canonical systems, we can state these questions more precisely:

~

\noindent {\bf Problem I}  (\emph{mechanical/ontological}):
Does there exist any function on the (reduced, if there are constraints) phase space $S:\mathscr{S}\rightarrow\mathbb{R}$ which monotonically increases along the orbits of (the Hamiltonian flow) $\Phi_{t}$?

~

\noindent {\bf Problem II} (\emph{statistical/epistemic}):
Does there exist any function of time $S:\mathscr{T}\rightarrow\mathbb{R}$, on some time interval $\mathscr{T}\subseteq\mathbb{R}$, defined in a suitable way in terms of a probability density $\rho:\mathscr{S}\times\mathscr{T}\rightarrow[0,1]$ on the (reduced) phase space, satisfying ${\rm d}S/{\rm d}t\geq0$ in a Hamiltonian system? [Traditionally, the definition taken here for entropy is (a coarse-grained version of) $S\left(t\right)=-\int_{\mathscr{P}}\bm{\Omega}\,\rho\ln\rho$, or its appropriate reduction to $\mathscr{S}$ if there are constraints.]

~

In CM, it is Problem II that has received the most attention since the end of the 19th century. In fact, there has been significant work in recent years by mathematicians [\cite{villani_h-theorem_2008,yau_work_2011}] aimed at placing the statistical formulation of the H-Theorem on more rigorous footing, and thus at proving more persuasively that, using appropriate assumptions, the answer to Problem II is in fact \textsl{yes}. In contrast, after the early Loschmidt reversibility and Poincar\'{e} recurrence arguments, Problem I has received some less well-known responses to the effect of demonstrating (even more convincingly) that the answer to it \textsl{under certain conditions} (to be carefully elaborated) is actually \textsl{no}. In this chapter, we will concern ourselves with two such types of responses to Problem I: first, what we call the \textsl{perturbative} approach, also proposed by [\cite{poincare_sur_1889}]; and second, what we call the \textsl{topological} approach, due to [\cite{olsen_classical_1993}] and related to the recurrence theorem. In the former, one tries to Taylor expand the time derivative of a phase space function, computed via the Poisson bracket, about a hypothetical equilibrium point in phase space, and one obtains contradictions with its strict positivity away from equilibrium. We revisit the original paper of Poincar\'{e}, clarify the assumptions of the argument, and carefully carry out the proof which is---excepting a sketch which makes it seem more trivial than it actually turns out to be---omitted therein. We furthermore extend this theorem to matter fields---in particular, a scalar and electromagnetic field---in curved spacetime. In the topological approach, on the other hand, one uses topological properties of the phase space itself to prove non-existence of monotonic functions. We review the proof of Olsen, and discuss its connections with the recurrence theorem and more recent periodicity theorems in Hamiltonian systems from symplectic geometry.

In GR, one may consider similar lines of reasoning as in CM to attempt to answer Problems I and II. Naively, one might expect the same answer to Problem II, namely \textsl{yes}---however, as we will argue later in greater detail, there are nontrivial mathematical issues that need to be circumvented here even in formulating it. For Problem I, as discussed, one might confidently expect the answer to also be \textsl{yes}. Therefore, although we do not yet know how to define entropy in GR with complete generality, we \textsl{can} at least ask why the proofs that furnish a negative answer to Problem I in CM fail here, and perhaps thereby gain fruitful insight into the essential features we should expect of such a definition. 

Following the perturbative approach, we will show that a Taylor-expanded Poisson bracket does not contain terms which satisfy definite inequalities (as they do in CM). The reason, as we will see, is that the second functional derivatives of the gravitational Hamiltonian can (unlike in CM) be both positive and negative, and so its curvature in phase space cannot be used to constrain (functionals of) the orbits; no contradiction arises here with the second law of thermodynamics. 

Following the topological approach, there are two points of view which may explicate why the proofs in CM do not carry over to GR. Firstly, it is believed that, in general, the phase space of GR is non-compact [\cite{schiffrin_measure_2012}]. Of course, this assertion depends on the nature of the degrees of freedom thought to be available in the spacetime under consideration, but even in very simple situations (such as cosmological spacetimes), it has been shown explicitly that the total phase space measure diverges. Physically, what this non-compactness implies is the freedom of a gravitational system to explore phase space unboundedly, without having to return (again and again) to its initial state. This leads us to the second (related) point of view as to why the topological proofs in CM fail in GR: namely, the non-recurrence of phase space orbits. Aside from trivial situations, solutions to the canonical equations of GR are typically non-cyclic (\textit{i.e.} they do not close in phase space) permitting the existence of functionals which may thus increase along the Hamiltonian flow. In fact, to counter the Poincar\'{e} recurrence theorem in CM, there even exists a ``no-return'' theorem in GR [\cite{tipler_general_1979,tipler_general_1980,newman_compact_1986}] for spacetimes which admit compact Cauchy surfaces and satisfy suitable energy and genericity conditions; it broadly states that the spacetime cannot return, even arbitrarily close, to a previously occupied state. One might nonetheless expect non-recurrence to be a completely general feature of all (nontrivial) gravitational systems, including spacetimes with non-compact Cauchy surfaces.

A setting of particular interest for this discussion is the gravitational two-body problem. With the recent detections and ongoing efforts towards further observations of gravitational waves from two-body systems, the emission of which ought to be closely related to entropy production, a precise understanding and quantification of the latter is becoming more and more salient. In the CM two-body (\textit{i.e.} Kepler) problem, the consideration of Problem I clearly explains the lack of entropy production due to phase space compactness (for a given finite range of initial conditions). In the Newtonian $N$-body problem, where (as we will elaborate) neither the perturbative nor the topological proofs are applicable, the answer to Problem I was actually shown to be \textsl{yes} in [\cite{barbour_gravitational_2013,barbour_identification_2014}]. In GR, the two-body problem may be considered in the context of perturbed Schwarzschild-Droste (SD) spacetimes
(as is relevant, for instance, in the context of extreme-mass-ratio inspirals). Here, the phase space volume (symplectic) form has been explicitly computed in [\cite{jezierski_energy_1999}]. We will use this in this chapter to show that in such spacetimes, the phase space is non-compact; hence there are no contradictions with non-recurrence or entropy production.

\section{\label{sec:CM}Entropy theorems in classical mechanics}

\subsection{\label{sec:3.1}Setup}

Classical particle mechanics with $N$ degrees of freedom~[\cite{arnold_mathematical_1997}] can be formulated
as a Lagrangian theory with an $N$-dimensional configuration space
$\mathscr{Q}$. This means that we will have a canonical theory\footnote{~Note that the formulation here proceeds essentially exactly as in the case of field theories, elaborated at length in Chapter \ref{2-canonical}. In fact, the treatment of particle mechanics can be regarded mathematically as just a special case of the general treatment of field theories, in particular by using a collection of ($N$) fields which are all distributional (Dirac delta functions in three-dimensional space) and supported at the respective particle locations.} on
a $2N$-dimensional phase space $\mathscr{P}$. We can choose canonical
coordinates $\left(q_{1},...,q_{N}\right)$ with conjugate momenta
$\left(p_{1},...,p_{N}\right)$ such that the symplectic form on $\mathscr{P}$
is given by 
\begin{equation}
\bm{\omega}=\sum_{j=1}^{N}{\rm d}p_{j}\wedge{\rm d}q_{j}.\label{eq:CM_symplectic_form}
\end{equation}
Then, the volume form on $\mathscr{P}$ is simply the $N$-th
exterior power of the symplectic form, in particular $\bm{\Omega}=[(-1)^{N(N-1)/2}/N!]\bm{\omega}^{\wedge N}$, and $\bm{X}_{H}$ is here given in coordinates by 
\begin{equation}
\bm{X}_{H}=\sum_{j=1}^{N}\left(\frac{\partial H}{\partial p_{j}}\frac{\partial}{\partial q_{j}}-\frac{\partial H}{\partial q_{j}}\frac{\partial}{\partial p_{j}}\right).\label{eq:CM_time-evolution_vector}
\end{equation}
The action of $\bm{X}_{H}$ on any phase space function
$F:\mathscr{P}\rightarrow\mathbb{R}$, called the Poisson bracket,
gives its time derivative: 
\begin{equation}
\dot{F}=\frac{{\rm d}F}{{\rm d}t}=\bm{X}_{H}\left(F\right)=\left\{ F,H\right\} .\label{eq:CM_Poisson_bracket_definition}
\end{equation}
We obtain from this $\dot{q}_{j}=\{q_{j},H\}=\partial H/\partial p_{j}$
and $\dot{p}_{j}=\{p_{j},H\}=-\partial H/\partial q_{j}$, which are the canonical equations of motion. Moreover, we have that the symplectic form of $\mathscr{P}$, and hence its volume form, are preserved along $\Phi_{t}$; in other words, we have $\mathcal{L}_{\bm{X}_{H}}\bm{\omega}=0=\mathcal{L}_{\bm{X}_{H}}\bm{\Omega}$, which is known as Liouville's theorem.

\begin{figure}
\begin{centering}
\includegraphics[scale=0.3]{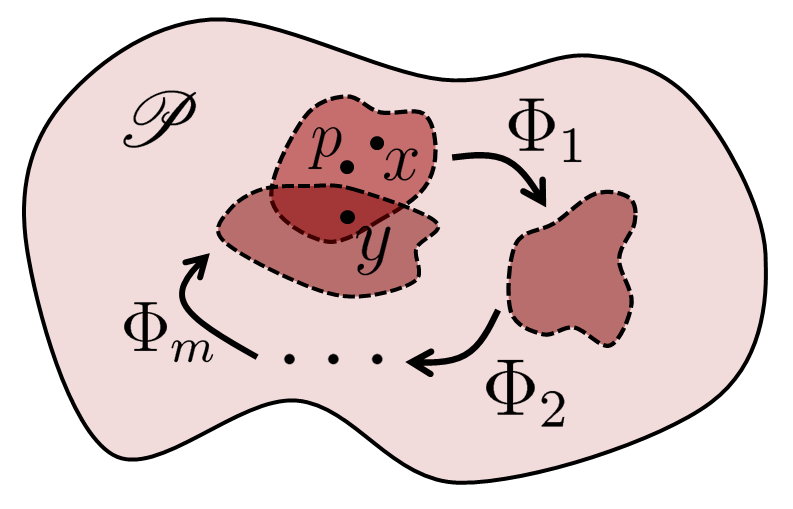}
\par\end{centering}

\protect\caption{\label{fig:recurrence}The idea of the proof for the Poincar\'{e} recurrence theorem.}

\end{figure}

We now turn to addressing Problem I in CM---that is, the question of whether there exists a function
$S:\mathscr{P}\rightarrow\mathbb{R}$ that behaves like entropy in
a classical Hamiltonian system.
Possibly the most well-known answer given to this is the Poincar\'{e} recurrence theorem. We can easily offer a proof of this, shown pictorially in Figure \ref{fig:recurrence} (see section 16 of [\cite{arnold_mathematical_1997}]): Assume $\mathscr{P}$ is compact and $\Phi_{t}(\mathscr{P})=\mathscr{P}$.
Let $\mathscr{U}\subset\mathscr{P}$ be the neighborhood of any point
$p\in\mathscr{P}$, and consider the sequence of images $\{\Phi_{n}(\mathscr{U})\}_{n=0}^{\infty}$.
Each $\Phi_{n}(\mathscr{U})$ has the same measure $\int_{\Phi_{n}(\mathscr{U})}\bm{\Omega}$ (because of Liouville's theorem),
so if they never intersected, $\mathscr{P}$ would have infinite measure.
Therefore there exist $k$, $l$ with $k>l$ such that $\Phi_{k}(\mathscr{U})\cap\Phi_{l}(\mathscr{U})\neq\varnothing$,
implying $\Phi_{m}(\mathscr{U})\cap\mathscr{U}\neq\varnothing$ where
$m=k-l$. For any $y\in\Phi_{m}(\mathscr{U})\cap\mathscr{U}$, there
exists an $x\in\mathscr{U}$ such that $y=\Phi_{m}(x)$. Thus, any
point returns arbitrarily close to the initial conditions in a compact and invariant phase space. 

Let us now discuss, in turn, the perturbative and topological approaches to this problem.

\subsection{\label{sec:3.2}Perturbative approach}

We revisit and carefully explicate, in this subsection, the argument given by [\cite{poincare_sur_1889}] to the effect that an entropy function $S:\mathscr{P}\rightarrow\mathbb{R}$ does not exist.
First, we will clarify the assumptions that need to go into it, \textit{i.e.} the conditions we must impose both on the entropy $S$ as well as on the Hamiltonian $H$, and then we will supply a rigorous proof.

\subsubsection{\label{sec:3.2.1}Review of Poincar\'{e}'s idea for a proof}

In his original paper [\cite{poincare_sur_1889}] (translated into English in~[\cite{olsen_classical_1993}]), the argument given by Poincar\'{e} (expressed using the contemporary notation that we employ here) for the non-existence
of such a function $S:\mathscr{P}\rightarrow\mathbb{R}$ is the following:
if $S$ behaves indeed like entropy, it should satisfy 
\begin{equation}
\dot{S}=\left\{ S,H\right\} =\sum_{k=1}^{N}\left(\frac{\partial H}{\partial p_{k}}\frac{\partial S}{\partial q_{k}}-\frac{\partial H}{\partial q_{k}}\frac{\partial S}{\partial p_{k}}\right)>0\label{eq:CM_Sdot_Poisson_Poincare}
\end{equation}
around a hypothetical equilibrium point in $\mathscr{P}$. Taylor
expanding each term and assuming all first partials of $S$ and
$H$ vanish at this equilibrium, we obtain a quadratic form (in the
distances away from equilibrium) plus higher-order terms. If we are
``sufficiently close'' to equilibrium, we may ignore the higher-order
terms and simply consider the quadratic form, which thus needs to
be positive definite for the above inequality [Eq.~\eqref{eq:CM_Sdot_Poisson_Poincare}]
to hold. But here Poincar\'{e}, without presenting any further explicit
computations, simply asserts that ``it is easy to satisfy oneself
that this is impossible if one or the other of the two \textit{quadratic forms $S$
and $H$} is definite, which is the case here''. (Our modern language modification is in italic.)

Neither the casual dismissal of the higher-order terms, nor, even
more crucially, the fact that ``it is easy to satisfy oneself''
of the impossibility of this quadratic form to be positive definite
is immediately apparent from this discussion. In fact, all of the
points in this line of reasoning require a careful statement of the
necessary assumptions as well as some rather non-trivial details of
the argumentation required to obtain the conclusion (that $\dot{S}=0$).

In what follows, we undertake precisely that. First we look at the
assumptions needed for this method to yield a useful proof, and then
we carry out the proof in full detail and rigor.

\subsubsection{\label{sec:3.2.2}Entropy conditions}

A function $S:\mathscr{P}\rightarrow\mathbb{R}$ can be said to behave
like entropy insofar as it satisfies the laws of thermodynamics. In
particular, it should conform to two assumptions: first, that it should
have an equilibrium point, and second, that it should obey the second
law of thermodynamics---which heuristically states that it should be
increasing in time everywhere except at the equilibrium point, where
it should cease to change. We state these explicitly as follows: 

~

\noindent{\bf S1} \textsl{(Existence of equilibrium)}{\bf:} We assume there exists a point
in phase space, $x_{0}\in\mathscr{P}$, henceforth referred to as
the ``equilibrium'' configuration of the system, which is a stationary
point of the entropy $S$, \textit{i.e.} all first partials thereof should
vanish when evaluated there: 
\begin{equation}
\left(\frac{\partial S}{\partial q_{j}}\right)_{0}=0=\left(\frac{\partial S}{\partial p_{j}}\right)_{0},\label{eq:CM_Equilibrium}
\end{equation}
where, for convenience, we use the notation $\left(\cdot\right)_{0}=\left.\left(\cdot\right)\right|_{x_{0}}$
to indicate quantities evaluated at equilibrium. Note that by the
definition of the Poisson bracket [Eq.~\eqref{eq:CM_Poisson_bracket_definition}],
this implies $(\dot{S})_{0}=0$. 

~

\noindent{\bf S2} \textsl{(Second law of thermodynamics)}{\bf:} A common formulation of the
second law asserts that the entropy $S$ is always increasing in time
when the system is away from equilibrium (\textit{i.e.} $\dot{S}>0$ everywhere
in $\mathscr{P}\backslash x_{0}$), and attains its maximum value
at equilibrium, where it ceases to change in time (\textit{i.e.} $\dot{S}=0$
at $x_{0}$, as implied by the first condition). We need to work,
however, with a slightly stronger version of the second law: namely,
the requirement that the Hessian matrix of $\dot{S}$, 
\begin{equation}
\mathbf{Hess}(\dot{S})=
\left[
\begin{array}{c|c}
\underset{}{{\displaystyle \frac{\partial^{2}\dot{S}}{\partial q_{i}\partial q_{j}}}} & \underset{}{{\displaystyle \frac{\partial^{2}\dot{S}}{\partial q_{i}\partial p_{j}}}} \\
\hline
\overset{}{{\displaystyle \frac{\partial^{2}\dot{S}}{\partial p_{i}\partial q_{j}}}} & \overset{}{{\displaystyle \frac{\partial^{2}\dot{S}}{\partial p_{i}\partial p_{j}}}}
\end{array}
\right],\label{eq:CM_Hessian_S_dot}
\end{equation}
is positive definite when evaluated at equilibrium, \textit{i.e.} $(\mathbf{Hess}(\dot{S}))_{0}\succ0$. 

~

We make now a few remarks about these assumptions. Firstly, S2 is a sufficient---though not strictly necessary---condition to
guarantee $\dot{S}>0$ in $\mathscr{P}\backslash x_{0}$ and $\dot{S}=0$
at $x_{0}$. However, the assumption of positive definiteness of the Hessian
of the entropy $S$ itself at equilibrium is often used in statistical
mechanics [\cite{abad_principles_2012}], and so it may not be objectionable
to extend this supposition to $\dot{S}$ as well. (In any case, this
leaves out only special situations where higher-order derivative tests
are needed to certify the global minimization of $\dot{S}$ at equilibrium,
which arguably are more of mathematical rather than physical interest;
we may reasonably expect the entropy as well as its time derivative
to be quadratic in the phase space variables as a consequence of its
ordinary statistical mechanics definitions in terms of energy.)

Secondly, the above two conditions omit the consideration of functions on $\mathscr{P}$
which are \textsl{everywhere} strictly monotonically increasing in
time, \textit{i.e.} the time derivative of which is always positive with no equilibrium
point. The topological approaches to Problem I, which we will turn to in the next subsection, do accommodate the possibility such
functions.

Thirdly, the equilibrium point $x_{0}\in\mathscr{P}$, though usually (physically)
expected to be unique, need not be for the purposes of what follows,
so long as it obeys the two conditions S1 and S2. In other words,
it suffices that there exists at least one such point in $\mathscr{P}$.

Fourthly, there is no topological requirement being imposed on
the phase space $\mathscr{P}$. It is possible, in other words, for
its total measure $\mu\left(\mathscr{P}\right)=\int_{\mathscr{P}}\bm{\Omega}$
to diverge. This means that the theorem applies to systems which can, in principle, explore phase space unboundedly, without any limits being imposed (either physically or mathematically) thereon.

\subsubsection{\label{sec:3.2.3}Hamiltonian conditions}

Next, we make a few assumptions about the Hamiltonian $H:\mathscr{P}\rightarrow\mathbb{R}$
which we need to impose in order to carry out our proof. The first
two assumptions are reasonable for any typical Hamiltonian in classical
mechanics, as we will discuss. The third, however, is stronger than
necessary to account for all Hamiltonians in general---and indeed,
as we will see, unfortunately leaves out certain classes of Hamiltonians
of interest. However, we regard it as a necessary assumption which
we cannot relax in order to formulate the proof according to this
approach. Our assumptions on $H$ are thus as follows: 

~

\noindent {\bf H1} \textsl{(Kinetic terms)}{\bf:} With regards to the second partials of $H$
with respect to the momentum variables, we assume the following:


(a) We can make a choice of coordinates so as to diagonalize (\textit{i.e.} decouple) the kinetic terms. In other words, we can choose to write $H$ in such a form that we have: 
\begin{equation}
\frac{\partial^{2}H}{\partial p_{i}\partial p_{j}}=\delta_{ij}\frac{\partial^{2}H}{\partial p_{j}^{2}}\,. \label{eq:CM_H_pipj}
\end{equation}


(b) Additionally, the second partials of $H$ with respect to each momentum
variable, representing the coefficients of the kinetic terms, should
be non-negative: 
\begin{equation}
\frac{\partial^{2}H}{\partial p_{j}^{2}}\geq0\,. \label{eq:CM_H_pp}
\end{equation}


\noindent {\bf H2} \textsl{(Mixed terms)}{\bf:} We assume that we can decouple the terms that
mix kinetic and coordinate degrees of freedom (via performing integrations
by parts, if necessary, in the action out of which the Hamiltonian
is constructed), such that $H$ can be written in a form where: 
\begin{equation}
\frac{\partial^{2}H}{\partial p_{i}\partial q_{j}}=0\,. \label{eq:CM_H_pq}
\end{equation}


\noindent {\bf H3} \textsl{(Potential terms)}{\bf:} We need to restrict our consideration to
Hamiltonians the partial Hessian of which, with respect to the coordinate
variables, is positive semidefinite at the point of equilibrium (assuming
it exists), \textit{i.e.} $[\partial^{2}H/\partial q_{i}\partial q_{j}]_{0}\succeq0$.
In fact we need to impose a slightly stronger (sufficient, though not strictly necessary) condition: that any of the row sums of $[\partial^{2}H/\partial q_{i}\partial q_{j}]_{0}$
are non-negative. That is to say, we assume: 
\begin{equation}
\sum_{i=1}^{N}\left(\frac{\partial^{2}H}{\partial q_{i}\partial q_{j}}\right)_{0}\geq0.\label{eq:CM_rowsum_V}
\end{equation}

We can make a few remarks about these assumptions. Firstly, H1 and H2 are manifestly satisfied for the most typically-encountered
form of the Hamiltonian in CM, 
\begin{equation}
H=\sum_{j=1}^{N}\frac{p_{j}^{2}}{2m_{j}}+V\left(q_{1},...,q_{N}\right),\label{eq:CM_typical_Hamiltonian}
\end{equation}
where $m_{j}$ are the masses associated with each degree of freedom
and $V$ is the potential (a function of only the configuration variables,
and not the momenta). Indeed, H1(a) is satisfied since we have $\partial^{2}H/\partial p_{i}\partial p_{j}=0$
unless $i=j$, regardless of $V$. For H1(b), we clearly have $\partial^{2}H/\partial p_{j}^{2}=1/m_{j}>0$
assuming masses are positive. 
Finally, H2 holds as $\partial^{2}H/\partial p_{i}\partial q_{j}=0$ is satisfied by construction.

Secondly, For typical Hamiltonians [Eq.~\eqref{eq:CM_typical_Hamiltonian}],
H3 translates into a condition on the potential $V$, \textit{i.e.} the requirement that $\sum_{i=1}^{N}(\partial^{2}V/\partial q_{i}\partial q_{j})_{0}\geq0$.
This is not necessarily satisfied in general in CM, though it is for
many systems. For example, when we have just one degree of freedom,
$N=1$, this simply means that the potential $V(q)$ is concave upward
at the point of equilibrium (thus regarded as a \emph{stable} equilibrium), \textit{i.e.} $({\rm d}^{2}V(q)/{\rm d}q^{2})_{0}\geq0$, which is reasonable
to assume. As another example, for a system of harmonic oscillators
with no interactions, $V=\frac{1}{2}\sum_{j=1}^{N}m_{j}\omega_{j}^{2}q_{j}^{2}$,
we clearly have $\sum_{i=1}^{N}\partial^{2}V/\partial q_{i}\partial q_{j}=m_{j}\omega_{j}^{2}>0$
for positive masses. Indeed, even introducing interactions does not
change this so long as the couplings are mostly non-negative. (In
other words, if the negative couplings do not dominate in strength
over the positive ones.) Higher (positive) powers of the $q_{j}$
variables in $V$ are also admissible under a similar argument. However,
we can see that condition H3 [Eq.~\eqref{eq:CM_rowsum_V}] excludes
certain classes of inverse-power potentials. Most notably, it excludes
the Kepler (gravitational two-body) Hamiltonian, $H=\left(1/2m\right)(p_{1}^{2}+p_{2}^{2})-GMm/(q_{1}^{2}+q_{2}^{2})^{1/2}$,
where $q_{j}$ are the Cartesian coordinates in the orbital plane,
and $p_{j}$ the associated momenta. In this case, we have $\det([\partial^{2}H/\partial q_{i}\partial q_{j}])=-2\left(GMm\right)^{2}/(q_{1}^{2}+q_{2}^{2})^{3}<0$,
hence $[\partial^{2}H/\partial q_{i}\partial q_{j}]$ is negative
definite everywhere and therefore cannot satisfy H3 [Eq.~\eqref{eq:CM_rowsum_V}].

\subsubsection{\label{sec:3.2.4}Our proof}

\begin{figure}
\begin{centering}
\includegraphics[scale=0.3]{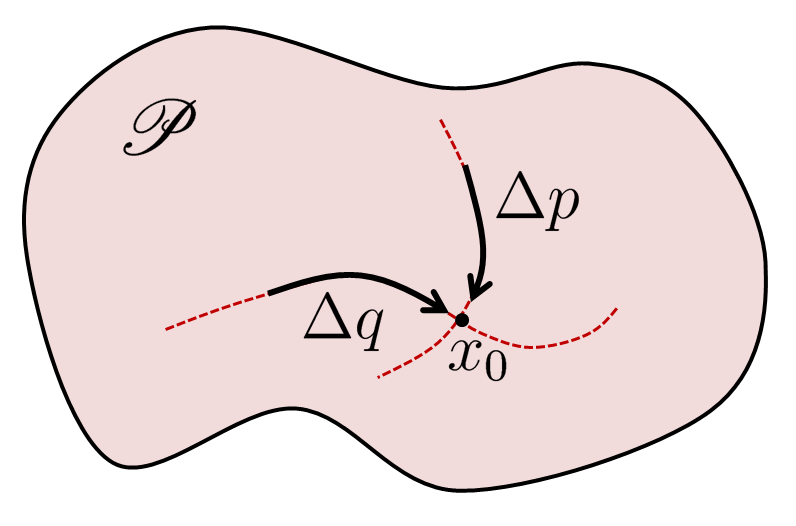}
\par\end{centering}

\protect\caption{\label{fig:perturbative}The idea of the perturbative approach is to evaluate $\dot{S}$ along different directions in phase space away from equilibrium, and arrive at a contradiction with its strict positivity.}

\end{figure}

We will now show that there cannot exist a function $S:\mathscr{P}\rightarrow\mathbb{R}$
satisfying the assumptions S1-S2 of subsubsection \ref{sec:3.2.2}
in a Hamiltonian system that obeys the assumptions H1-H3 of subsubsection
\ref{sec:3.2.3} on $H:\mathscr{P}\rightarrow\mathbb{R}$. We do this
by simply assuming that such a function exists, and we will show that
this implies a contradiction. For a pictorial representation, see Figure~\ref{fig:perturbative}.

~

\noindent$\bm{N=1}${\bf:} Let us first carry out the proof for $N=1$ degree of freedom so as to
make the argument for general $N$ easier to follow. Let $S:\mathscr{P}\rightarrow\mathbb{R}$
be any function on the configuration space $\mathscr{P}$ satisfying
assumptions S1-S2 of subsubsection \ref{sec:3.2.2}, \textit{i.e.} it has an
equilibrium point and the Hessian of its time derivative is positive
definite there. We know that its time derivative at any point $x=(q,p)\in\mathscr{P}$
can be evaluated, as discussed in subsection \ref{sec:3.1}, via the
Poisson bracket: 
\begin{equation}
\dot{S}=\frac{\partial H}{\partial p}\frac{\partial S}{\partial q}-\frac{\partial H}{\partial q}\frac{\partial S}{\partial p}.\label{eq:CM_Sdot_Poisson-1}
\end{equation}
Let us now insert into this the Taylor series for each term expanded
about the equilibrium point $x_{0}=(q_{0},p_{0})$. Denoting $\Delta q=q-q_{0}$
and $\Delta p=p-p_{0}$, and using $\mathcal{O}(\Delta^{n})$ to represent $n$-th order terms in products of $\Delta q$ and $\Delta p$, we have: 
\begin{equation}
\frac{\partial H}{\partial q}=\left(\frac{\partial H}{\partial q}\right)_{0}+\left(\frac{\partial^{2}H}{\partial q^{2}}\right)_{0}\Delta q+\left(\frac{\partial^{2}H}{\partial p\partial q}\right)_{0}\Delta p+\mathcal{O}\left(\Delta^{2}\right),\label{eq:CM_dHdq-1}
\end{equation}
and similarly for the $p$ partial of $H$, while 
\begin{equation}
\frac{\partial S}{\partial q}=\left(\frac{\partial^{2}S}{\partial q^{2}}\right)_{0}\Delta q+\left(\frac{\partial^{2}S}{\partial p\partial q}\right)_{0}\Delta p+\mathcal{O}\left(\Delta^{2}\right),\label{eq:CM_dSdq-1}
\end{equation}
and similarly for the $p$ partial of $S$, where we have used the
condition S1 [Eq.~\eqref{eq:CM_Equilibrium}] which entails that
the zero-order term vanishes. Inserting all Taylor series into the
Poisson bracket [Eq.~\eqref{eq:CM_Sdot_Poisson-1}] and collecting
terms, we obtain the following result: 
\begin{equation}
\dot{S}=\!\left[\begin{array}{cc}
a & b\end{array}\right]\left[\begin{array}{c}
\Delta q\\
\Delta p
\end{array}\right]+\left[\begin{array}{cc}
\Delta q & \Delta p\end{array}\right]\left[\begin{array}{cc}
A & B\\
B & C
\end{array}\right]\left[\begin{array}{c}
\Delta q\\
\Delta p
\end{array}\right]+\mathcal{O}\left(\Delta^{3}\right)\!,\label{eq:CM_Sdot_QuadForm-1}
\end{equation}
where: 
\begin{align}
a= & \left(\frac{\partial H}{\partial p}\right)_{0}\left(\frac{\partial^{2}S}{\partial q^{2}}\right)_{0}-\left(\frac{\partial H}{\partial q}\right)_{0}\left(\frac{\partial^{2}S}{\partial q\partial p}\right)_{0},\label{eq:CM_ai-1}\\
b= & \left(\frac{\partial H}{\partial p}\right)_{0}\left(\frac{\partial^{2}S}{\partial p\partial q}\right)_{0}-\left(\frac{\partial H}{\partial q}\right)_{0}\left(\frac{\partial^{2}S}{\partial p^{2}}\right)_{0},\label{eq:CM_bi-1}
\end{align}
and: 
\begin{align}
A= & \left(\frac{\partial^{2}H}{\partial q\partial p}\right)_{0}\left(\frac{\partial^{2}S}{\partial q^{2}}\right)_{0}-\left(\frac{\partial^{2}H}{\partial q^{2}}\right)_{0}\left(\frac{\partial^{2}S}{\partial q\partial p}\right)_{0},\label{eq:CM_Aij-1}\\
B= & \frac{1}{2}\Bigg[\left(\frac{\partial^{2}H}{\partial p^{2}}\right)_{0}\left(\frac{\partial^{2}S}{\partial q^{2}}\right)_{0}-\left(\frac{\partial^{2}H}{\partial q^{2}}\right)_{0}\left(\frac{\partial^{2}S}{\partial p^{2}}\right)_{0}\Bigg],\label{eq:CM_Bij-1}\\
C= & \left(\frac{\partial^{2}H}{\partial p^{2}}\right)_{0}\left(\frac{\partial^{2}S}{\partial p\partial q}\right)_{0}-\left(\frac{\partial^{2}H}{\partial p\partial q}\right)_{0}\left(\frac{\partial^{2}S}{\partial p^{2}}\right)_{0}.\label{eq:CM_Cij-1}
\end{align}
By assumption S2, we have that $\dot{S}$ as given above [Eq.~\eqref{eq:CM_Sdot_QuadForm-1}]
is strictly positive for any $x\neq x_{0}$ in $\mathscr{P}$. In
particular, let $\delta>0$ and let us consider $\dot{S}$ [Eq.~\eqref{eq:CM_Sdot_QuadForm-1}] evaluated at the sequence of points
$\left\{ x_{n}^{\pm}\right\} _{n=1}^{\infty}$, where $x_{n}^{\pm}=(q_{0}\pm\delta/n,p_{0})$,
such that the only deviation away from equilibrium is along the direction
$\Delta q=\pm\delta/n$, with all other $\Delta p$ vanishing. Then,
for any $n$, we must have according to our expression for $\dot{S}$
[Eq.~\eqref{eq:CM_Sdot_QuadForm-1}]: 
\begin{align}
\dot{S}\left(x_{n}^{+}\right) & =a\frac{\delta}{n}+A\frac{\delta^{2}}{n^{2}}+\mathcal{O}\left(\frac{\delta^{3}}{n^{3}}\right)>0,\label{eq:CM_Sdotplus-1}\\
\dot{S}\left(x_{n}^{-}\right) & =-a\frac{\delta}{n}+A\frac{\delta^{2}}{n^{2}}+\mathcal{O}\left(\frac{\delta^{3}}{n^{3}}\right)>0.\label{eq:CM_Sdotminus-1}
\end{align}
Taking the $n\rightarrow\infty$ limit of the first inequality implies
$a\geq0$, while doing the same for the second inequality implies
$a\leq0$. Hence $a=0$. A similar argument (using $\Delta p=\pm\delta/n$)
implies $b=0$. Thus, $S$ needs to satisfy the constraints 
\begin{align}
0= & \left(\frac{\partial H}{\partial p}\right)_{0}\left(\frac{\partial^{2}S}{\partial q^{2}}\right)_{0}-\left(\frac{\partial H}{\partial q}\right)_{0}\left(\frac{\partial^{2}S}{\partial q\partial p}\right)_{0},\label{eq:CM_ai0-1}\\
0= & \left(\frac{\partial H}{\partial p}\right)_{0}\left(\frac{\partial^{2}S}{\partial p\partial q}\right)_{0}-\left(\frac{\partial H}{\partial q}\right)_{0}\left(\frac{\partial^{2}S}{\partial p^{2}}\right)_{0},\label{eq:CM_bi0-1}
\end{align}
and this leaves us with 
\begin{equation}
\dot{S}=\left[\begin{array}{cc}
\Delta q & \Delta p\end{array}\right]\left[\begin{array}{cc}
A & B\\
B & C
\end{array}\right]\left[\begin{array}{c}
\Delta q\\
\Delta p
\end{array}\right]+\mathcal{O}\left(\Delta^{3}\right).\label{eq:CM_Sdot_QuadForm2-1}
\end{equation}

Now, imposing the Hamiltonian assumptions H1(a) and H2 {[}eqs. (\ref{eq:CM_H_pipj})
and (\ref{eq:CM_H_pq}) respectively{]} simplifies $A$ and $C$,
from the above {[}eqs. (\ref{eq:CM_Aij-1}) and (\ref{eq:CM_Bij-1})
respectively{]} to: 
\begin{align}
A=\, & -\left(\frac{\partial^{2}H}{\partial q^{2}}\right)_{0}\left(\frac{\partial^{2}S}{\partial q\partial p}\right)_{0},\label{eq:CM_Aij_simplified-1}\\
C=\, & \left(\frac{\partial^{2}H}{\partial p^{2}}\right)_{0}\left(\frac{\partial^{2}S}{\partial p\partial q}\right)_{0}.\label{eq:CM_Cij_simplified-1}
\end{align}
Positive-definiteness of $(\mathbf{Hess}(\dot{S}))_{0}$ (assumption
S2) implies that the quadratic form above [Eq.~\eqref{eq:CM_Sdot_QuadForm2-1}]
should be positive definite. This means that we cannot have $(\partial^{2}H/\partial p^{2})_{0}=0$,
since then $C$ would not be strictly positive and we would get a
contradiction. This, combined with assumption H1(b) [Eq.~\eqref{eq:CM_H_pp}],
implies that $(\partial^{2}H/\partial p^{2})_{0}>0$. This in combination
with $C>0$ means that $(\partial^{2}S/\partial p\partial q)_{0}>0$.
But $A>0$ also, in order to have positive-definiteness of the quadratic
form [Eq.~\eqref{eq:CM_Sdot_QuadForm2-1}], and this combined
with assumption H3 [Eq.~\eqref{eq:CM_rowsum_V}], \textit{i.e.} $(\partial^{2}H/\partial q^{2})_{0}\geq0$,
implies $(\partial^{2}S/\partial p\partial q)_{0}<0$. Thus we get
a contradiction, and so no such function $S$ exists.

~

\noindent{\bf General} $\bm{N}${\bf:} The extension of the proof to general $N$ follows similar lines,
though with a few added subtleties. Let us now proceed with it. As
before, suppose $S:\mathscr{P}\rightarrow\mathbb{R}$ is any function
on $\mathscr{P}$ satisfying S1-S2. Its time derivative at any point
$x=(q_{1},...,q_{N},p_{1},...,p_{N})\in\mathscr{P}$ can be evaluated
via the Poisson bracket: 
\begin{equation}
\dot{S}=\sum_{k=1}^{N}\left(\frac{\partial H}{\partial p_{k}}\frac{\partial S}{\partial q_{k}}-\frac{\partial H}{\partial q_{k}}\frac{\partial S}{\partial p_{k}}\right).\label{eq:CM_Sdot_Poisson}
\end{equation}
Let us now insert into this the Taylor series for each term expanded
about the equilibrium point $x_{0}=((q_{0})_{1},...,(q_{0})_{N},(p_{0})_{1},...,(p_{0})_{1})$.
Denoting $\Delta q_{i}=q_{i}-(q_{0})_{i}$ and $\Delta p_{i}=p_{i}-(p_{0})_{i}$, and using $\mathcal{O}(\Delta^{n})$ to represent $n$-th order terms in products of $\Delta q_{i}$ and $\Delta p_{i}$,
we have: 
\begin{equation}
\frac{\partial H}{\partial q_{k}}=\left(\frac{\partial H}{\partial q_{k}}\right)_{0}+\sum_{i=1}^{N}\Bigg[\left(\frac{\partial^{2}H}{\partial q_{i}\partial q_{k}}\right)_{0}\Delta q_{i}+\left(\frac{\partial^{2}H}{\partial p_{i}\partial q_{k}}\right)_{0}\Delta p_{i}\Bigg]+\mathcal{O}\left(\Delta^{2}\right),\label{eq:CM_dHdq}
\end{equation}
and similarly for the $p_{k}$ partial of $H$, while 
\begin{equation}
\frac{\partial S}{\partial q_{k}}\!=\sum_{i=1}^{N}\left[\left(\frac{\partial^{2}S}{\partial q_{i}\partial q_{k}}\right)_{0}\Delta q_{i}+\left(\frac{\partial^{2}S}{\partial p_{i}\partial q_{k}}\right)_{0}\Delta p_{i}\right]+\mathcal{O}\left(\Delta^{2}\right)\!,\label{eq:CM_dSdq}
\end{equation}
and similarly for the $p_{k}$ partial of $S$, where we have used
the condition S1 [Eq.~\eqref{eq:CM_Equilibrium}] which entails
that the zero-order term vanishes. Inserting all Taylor series into
the Poisson bracket [Eq.~\eqref{eq:CM_Sdot_Poisson}] and collecting
terms, we obtain the following result: 
\begin{equation}
\dot{S}=\left[\begin{array}{cc}
\!\mathbf{a}^{{\rm T}} & \mathbf{b}^{{\rm T}}\end{array}\!\right]\left[\!\begin{array}{c}
\Delta q_{1}\\
\vdots\\
\Delta p_{N}
\end{array}\!\right]+\left[\!\begin{array}{ccc}
\Delta q_{1} & \cdots & \Delta p_{N}\end{array}\!\right]\left[\!\begin{array}{cc}
\mathbf{A} & \mathbf{B}\\
\mathbf{B}^{{\rm T}} & \mathbf{C}
\end{array}\!\right]\left[\!\begin{array}{c}
\Delta q_{1}\\
\vdots\\
\Delta p_{N}
\end{array}\!\right]+\mathcal{O}\left(\Delta^{3}\right),\label{eq:CM_Sdot_QuadForm}
\end{equation}
where we have the following components for the $N$-dimensional vectors:
\begin{equation}
a_{i}=\sum_{k=1}^{N}\Bigg[\left(\frac{\partial H}{\partial p_{k}}\right)_{0}\!\!\left(\frac{\partial^{2}S}{\partial q_{i}\partial q_{k}}\right)_{0}\!\!-\left(\frac{\partial H}{\partial q_{k}}\right)_{0}\!\!\left(\frac{\partial^{2}S}{\partial q_{i}\partial p_{k}}\right)_{0}\!\!\Bigg],\label{eq:CM_ai}
\end{equation}
and
\begin{equation}
b_{i}=\sum_{k=1}^{N}\Bigg[\left(\frac{\partial H}{\partial p_{k}}\right)_{0}\!\!\left(\frac{\partial^{2}S}{\partial p_{i}\partial q_{k}}\right)_{0}\!\!-\left(\frac{\partial H}{\partial q_{k}}\right)_{0}\!\!\left(\frac{\partial^{2}S}{\partial p_{i}\partial p_{k}}\right)_{0}\!\!\Bigg],\label{eq:CM_bi}
\end{equation}
and for the $N\times N$ matrices:
\begin{align}
A_{ij}=&\frac{1}{2}\sum_{k=1}^{N}\Bigg[\left(\frac{\partial^{2}H}{\partial q_{i}\partial p_{k}}\right)_{0}\!\!\left(\frac{\partial^{2}S}{\partial q_{j}\partial q_{k}}\right)_{0}\!\!+\left(\frac{\partial^{2}H}{\partial q_{j}\partial p_{k}}\right)_{0}\!\!\left(\frac{\partial^{2}S}{\partial q_{i}\partial q_{k}}\right)_{0}\nonumber \\
 & -\left(\frac{\partial^{2}H}{\partial q_{i}\partial q_{k}}\right)_{0}\!\!\left(\frac{\partial^{2}S}{\partial q_{j}\partial p_{k}}\right)_{0}\!\!-\left(\frac{\partial^{2}H}{\partial q_{j}\partial q_{k}}\right)_{0}\!\!\left(\frac{\partial^{2}S}{\partial q_{i}\partial p_{k}}\right)_{0}\!\!\Bigg],\label{eq:CM_Aij}\\
B_{ij}=&\frac{1}{2}\sum_{k=1}^{N}\Bigg[ \left(\frac{\partial^{2}H}{\partial q_{i}\partial p_{k}}\right)_{0}\!\!\left(\frac{\partial^{2}S}{\partial p_{j}\partial q_{k}}\right)_{0}\!\!+\left(\frac{\partial^{2}H}{\partial p_{j}\partial p_{k}}\right)_{0}\!\!\left(\frac{\partial^{2}S}{\partial q_{i}\partial q_{k}}\right)_{0}\nonumber \\
 & -\left(\frac{\partial^{2}H}{\partial q_{i}\partial q_{k}}\right)_{0}\!\!\left(\frac{\partial^{2}S}{\partial p_{j}\partial p_{k}}\right)_{0}\!\!-\left(\frac{\partial^{2}H}{\partial p_{j}\partial q_{k}}\right)_{0}\!\!\left(\frac{\partial^{2}S}{\partial q_{i}\partial p_{k}}\right)_{0}\!\!\Bigg],\label{eq:CM_Bij}\\
C_{ij}=&\frac{1}{2}\sum_{k=1}^{N}\Bigg[\left(\frac{\partial^{2}H}{\partial p_{i}\partial p_{k}}\right)_{0}\!\!\left(\frac{\partial^{2}S}{\partial p_{j}\partial q_{k}}\right)_{0}\!\!+\left(\frac{\partial^{2}H}{\partial p_{j}\partial p_{k}}\right)_{0}\!\!\left(\frac{\partial^{2}S}{\partial p_{i}\partial q_{k}}\right)_{0}\nonumber \\
 & -\left(\frac{\partial^{2}H}{\partial p_{i}\partial q_{k}}\right)_{0}\!\!\left(\frac{\partial^{2}S}{\partial p_{j}\partial p_{k}}\right)_{0}\!\!-\left(\frac{\partial^{2}H}{\partial p_{j}\partial q_{k}}\right)_{0}\!\!\left(\frac{\partial^{2}S}{\partial p_{i}\partial p_{k}}\right)_{0}\!\!\Bigg].\label{eq:CM_Cij}
\end{align}
By assumption S2, we have that $\dot{S}$ as given above [Eq.~\eqref{eq:CM_Sdot_QuadForm}]
is strictly positive for any $x\neq x_{0}$ in $\mathscr{P}$. In
particular, let $\delta>0$ and let us consider $\dot{S}$ [Eq.~\eqref{eq:CM_Sdot_QuadForm}] evaluated at the sequence of points
$\left\{ x_{n}^{\pm}\right\} _{n=1}^{\infty}$, where $x_{n}^{\pm}=((q_{0})_{1},...,(q_{0})_{l}\pm\delta/n,...,(q_{0})_{N},(p_{0})_{1},...,(p_{0})_{1})$,
for any $l$, such that the only deviation away from equilibrium is
along the direction $\Delta q_{l}=\pm\delta/n$, with all other $\Delta q_{i}$
and $\Delta p_{i}$ vanishing. Then, for any $n$, we must have according
to the above expression for $\dot{S}$ [Eq.~\eqref{eq:CM_Sdot_QuadForm}]:
\begin{align}
\dot{S}\left(x_{n}^{+}\right) & =a_{l}\frac{\delta}{n}+A_{ll}\frac{\delta^{2}}{n^{2}}+\mathcal{O}\left(\frac{\delta^{3}}{n^{3}}\right)>0,\label{eq:CM_Sdotplus}\\
\dot{S}\left(x_{n}^{-}\right) & =-a_{l}\frac{\delta}{n}+A_{ll}\frac{\delta^{2}}{n^{2}}+\mathcal{O}\left(\frac{\delta^{3}}{n^{3}}\right)>0.\label{eq:CM_Sdotminus}
\end{align}
Taking the $n\rightarrow\infty$ limit of the first inequality implies
$a_{l}\geq0$, while doing the same for the second inequality implies
$a_{l}\leq0$. Hence $a_{l}=0$. Since $l$ is arbitrary, this means
that $a_{i}=0$, $\forall i$. A similar argument (using $\Delta p_{l}=\pm\delta/n$)
implies $b_{i}=0$, $\forall i$. Thus, $S$ needs to satisfy the
constraints 
\begin{align}
0= & \sum_{k=1}^{N}\Bigg[\left(\frac{\partial H}{\partial p_{k}}\right)_{0}\left(\frac{\partial^{2}S}{\partial q_{i}\partial q_{k}}\right)_{0}-\left(\frac{\partial H}{\partial q_{k}}\right)_{0}\left(\frac{\partial^{2}S}{\partial q_{i}\partial p_{k}}\right)_{0}\Bigg],\label{eq:CM_ai0}\\
0= & \sum_{k=1}^{N}\Bigg[\left(\frac{\partial H}{\partial p_{k}}\right)_{0}\left(\frac{\partial^{2}S}{\partial p_{i}\partial q_{k}}\right)_{0}-\left(\frac{\partial H}{\partial q_{k}}\right)_{0}\left(\frac{\partial^{2}S}{\partial p_{i}\partial p_{k}}\right)_{0}\Bigg],\label{eq:CM_bi0}
\end{align}
and this leaves us with 
\begin{equation}
\dot{S}=\left[\begin{array}{ccc}
\Delta q_{1} & \cdots & \Delta p_{N}\end{array}\right]\left[\begin{array}{cc}
\mathbf{A} & \mathbf{B}\\
\mathbf{B}^{{\rm T}} & \mathbf{C}
\end{array}\right]\left[\begin{array}{c}
\Delta q_{1}\\
\vdots\\
\Delta p_{N}
\end{array}\right]+\mathcal{O}\left(\Delta^{3}\right).\label{eq:CM_Sdot_QuadForm2}
\end{equation}

Now, imposing the Hamiltonian assumptions H1(a) and H2 {[}eqs. (\ref{eq:CM_H_pipj})
and (\ref{eq:CM_H_pq}) respectively{]} simplifies $\mathbf{A}$ and
$\mathbf{C}$, from the above {[}eqs. (\ref{eq:CM_Aij}) and (\ref{eq:CM_Cij})
respectively{]} to: 
\begin{align}
A_{ij}=\, & -\frac{1}{2}\sum_{k=1}^{N}\Bigg[\left(\frac{\partial^{2}H}{\partial q_{i}\partial q_{k}}\right)_{0}\left(\frac{\partial^{2}S}{\partial q_{j}\partial p_{k}}\right)_{0}+\left(\frac{\partial^{2}H}{\partial q_{j}\partial q_{k}}\right)_{0}\left(\frac{\partial^{2}S}{\partial q_{i}\partial p_{k}}\right)_{0}\Bigg],\label{eq:CM_Aij_simplified}\\
C_{ij}=\, & \frac{1}{2}\Bigg[\left(\frac{\partial^{2}H}{\partial p_{i}^{2}}\right)_{0}\!\!\left(\frac{\partial^{2}S}{\partial p_{j}\partial q_{i}}\right)_{0}\!\!+\left(\frac{\partial^{2}H}{\partial p_{j}^{2}}\right)_{0}\!\!\left(\frac{\partial^{2}S}{\partial p_{i}\partial q_{j}}\right)_{0}\!\!\Bigg].\label{eq:CM_Cij_simplified}
\end{align}
Positive-definiteness of $(\mathbf{Hess}(\dot{S}))_{0}$ implies that
the quadratic form above [Eq.~\eqref{eq:CM_Sdot_QuadForm2}] should
be positive definite. This means that we cannot have $(\partial^{2}H/\partial p_{j}^{2})_{0}=0$,
$\forall j$, since then $\mathbf{C}$ would not be positive definite
and we would get a contradiction. This, combined with assumption H1(b)
[Eq.~\eqref{eq:CM_H_pp}], implies that $(\partial^{2}H/\partial p_{j}^{2})_{0}>0$,
$\forall j$. Moreover, we also have: 
\begin{equation}
\sum_{i,j=1}^{N}C_{ij}=\sum_{i,j=1}^{N}\left(\frac{\partial^{2}H}{\partial p_{j}^{2}}\right)_{0}\left(\frac{\partial^{2}S}{\partial p_{i}\partial q_{j}}\right)_{0}>0.\label{eq:CM_Cij_posdef}
\end{equation}
The reason for this is easily seen by noting that positive-definiteness of $\mathbf{C}$, by definition, means that its product with any nonzero
vector and its transpose should be positive, \textit{i.e.} $\mathbf{z}^{{\rm T}}\mathbf{C}\mathbf{z}>0$
for any nonzero vector $\mathbf{z}$; in particular, $\mathbf{z}=(1,1,...,1)^{{\rm T}}$
achieves the above inequality [Eq.~\eqref{eq:CM_Cij_posdef}].
But then, let us consider $\sum_{i,j=1}^{N}A_{ij}$. Positive-definiteness
of $(\mathbf{Hess}(\dot{S}))_{0}$ (\textit{i.e.} of the quadratic form [Eq.~\eqref{eq:CM_Sdot_QuadForm}]) implies, just as in the case of $\mathbf{C}$,
that $\sum_{i,j=1}^{N}A_{ij}>0$, or 
\begin{equation}
\sum_{i,j=1}^{N}(-A_{ij})<0\label{CM_Aij_ineq}.
\end{equation}
At the same time, we have: 
\begin{equation}
\sum_{i,j=1}^{N}\left(-A_{ij}\right)=\sum_{i,j,k=1}^{N}\left(\frac{\partial^{2}H}{\partial q_{i}\partial q_{k}}\right)_{0}\left(\frac{\partial^{2}S}{\partial q_{j}\partial p_{k}}\right)_{0}.\label{CM_Aij_eq}
\end{equation}
Taking the minimum over the $k$ index in the term with the $H$ partials,
\begin{equation}
\sum_{i,j=1}^{N}\left(-A_{ij}\right)\geq\sum_{i,j,k=1}^{N}\left[\min_{1\leq l\leq N}\left(\frac{\partial^{2}H}{\partial q_{i}\partial q_{l}}\right)_{0}\right]\left(\frac{\partial^{2}S}{\partial q_{j}\partial p_{k}}\right)_{0},\label{eq:CM_-Aij_1}
\end{equation}
This means that the sums can be separated, and after relabelling,
the above [Eq.~\eqref{eq:CM_-Aij_1}] becomes: 
\begin{equation}
\sum_{i,j=1}^{N}\left(-A_{ij}\right)\geq\left[\min_{1\leq l\leq N}\sum_{k=1}^{N}\left(\frac{\partial^{2}H}{\partial q_{k}\partial q_{l}}\right)_{0}\right]\sum_{i,j=1}^{N}\left(\frac{\partial^{2}S}{\partial p_{i}\partial q_{j}}\right)_{0}.\label{eq:CM_-Aij_2}
\end{equation}
Now, insert the identity $1=(\partial^{2}H/\partial p_{j}^{2})_{0}/(\partial^{2}H/\partial p_{j}^{2})_{0}$
into the $i,j$ sum, and maximize over the denominator to get:
\begin{align}
\sum_{i,j=1}^{N}\left(-A_{ij}\right)\geq & \left[\min_{1\leq l\leq N}\sum_{k=1}^{N}\left(\frac{\partial^{2}H}{\partial q_{k}\partial q_{l}}\right)_{0}\right]\sum_{i,j=1}^{N}\frac{\left(\partial^{2}H/\partial p_{j}^{2}\right)_{0}}{\left(\partial^{2}H/\partial p_{j}^{2}\right)_{0}}\left(\frac{\partial^{2}S}{\partial p_{i}\partial q_{j}}\right)_{0}\label{eq:CM_-Aij_3}\\
\geq & \left[\min_{1\leq l\leq N}\sum_{k=1}^{N}\!\left(\frac{\partial^{2}H}{\partial q_{k}\partial q_{l}}\right)_{0}\right]\!\sum_{i,j=1}^{N}\!\left[\max_{1\leq m\leq N}\left(\frac{\partial^{2}H}{\partial p_{m}^{2}}\right)_{0}\right]^{-1}\!\!\left(\frac{\partial^{2}H}{\partial p_{j}^{2}}\right)_{0}\!\!\left(\frac{\partial^{2}S}{\partial p_{i}\partial q_{j}}\right)_{0}\label{eq:CM_-Aij_4}\\
= & \left\{ \left[\min_{1\leq l\leq N}\sum_{k=1}^{N}\left(\frac{\partial^{2}H}{\partial q_{k}\partial q_{l}}\right)_{0}\right]\left[\max_{1\leq m\leq N}\left(\frac{\partial^{2}H}{\partial p_{m}^{2}}\right)_{0}\right]^{-1}\right\} \sum_{i,j=1}^{N}C_{ij}\label{eq:CM_-Aij_5}\\
\geq & \,0,\label{eq:CM_-Aij_6}
\end{align}
since the term in curly brackets is non-negative (because of assumption
H3 on the Hamiltonian), and we had earlier $\sum_{i,j=1}^{N}C_{ij}>0$.
But we also had $\sum_{i,j=1}^{N}(-A_{ij})<0$. Hence we get a contradiction.
Therefore, no such function $S$ exists. This concludes our proof.

\subsection{\label{sec:3.3}Topological approach}

We now turn to the topological approach to answering Problem I in CM.
First we review
the basic ideas of Olsen's line of argumentation [\cite{olsen_classical_1993}], then we discuss their connections with the periodicity of phase space orbits.

\subsubsection{\label{sec:3.3.1}Review of Olsen's proof}

\begin{figure}
\begin{centering}
\includegraphics[scale=0.3]{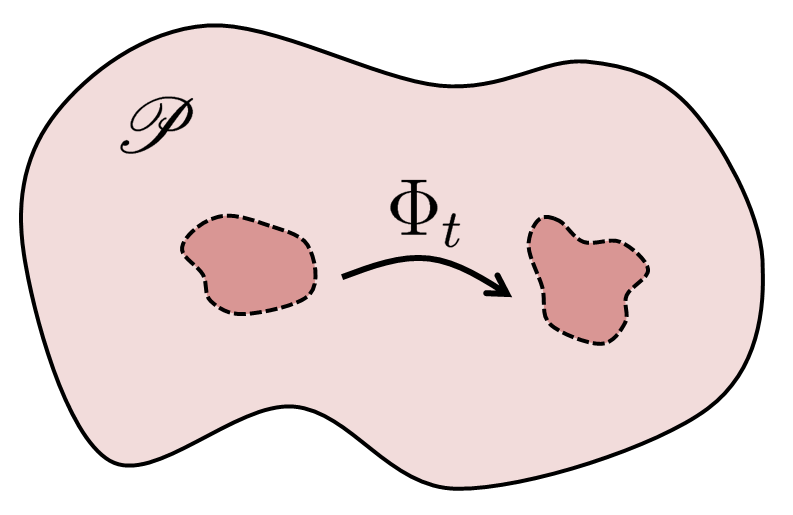}
\par\end{centering}

\protect\caption{\label{fig:topological}The topological approach relies on phase space compactness and Liouville's theorem, \textit{i.e.} the fact that the Hamiltonian flow is volume-preserving.}

\end{figure}

The assumptions made on $S:\mathscr{P}\rightarrow\mathbb{R}$ are
in this case not as strict as in the perturbative approach. See Figure \ref{fig:topological} for a pictorial representation.

In effect, we simply
need to assume that $S$ is nondecreasing along trajectories, which
are confined to an invariant closed space $\mathscr{P}$. Under these conditions, Olsen furnishes two proofs [\cite{olsen_classical_1993}] for why $S$ is necessarily a constant. In the first one, the
essential idea is that the volume integral of $S$ in $\mathscr{P}$ can be written after a change of variables
as 
\begin{equation}
\int_{\mathscr{P}}\bm{\Omega}\, S=\int_{\mathscr{P}}\bm{\Omega}\, \left(S\circ\Phi_{t}\right),\label{eq:CM_Olsen_pf}
\end{equation}
owing to the fact that $\mathscr{P}$ is left invariant by the Hamiltonian
flow $\Phi_{t}$ generated by the Hamiltonian vector field $\bm{X}_{H}$,
and that $\mathcal{L}_{\bm{X}_{H}}\bm{\Omega}=0$. Because the above
expression [Eq.~\eqref{eq:CM_Olsen_pf}] is time-independent,
$S$ must be time-independent, hence constant along trajectories. The second proof (based on the same assumptions) is rather more technical, but relies also basically on topological ideas; in fact, it is more related to the Poincar\'{e} recurrence property [\cite{luis_barreira_poincare_2006}].

We can make a few remarks. Firstly, there is in this case no requirement on the specific form of the Hamiltonian
function $H:\mathscr{P\rightarrow\mathbb{R}}$. In fact, $H$ can
even contain explicit dependence on time and the proof still holds.

Secondly, the essential ingredient here is the compactness of the phase space
$\mathscr{P}$. Indeed, even in Poincar\'{e}'s original recurrence theorem [\cite{poincare_sur_1890}], as we saw in Section \ref{sec:CM}, the only necessary assumptions were also phase space compactness and invariance along with Liouville's theorem. 

\subsubsection{\label{sec:3.3.2}Periodicity in phase space}

Even more can be said about the connection between phase space compactness and the recurrence of orbits than the Poincar\'{e} recurrence theorem. There are recent theorems in symplectic geometry which show that exact periodicity of orbits can exist in compact phase spaces.

For example, let us assume the Hamiltonian is of typical form [Eq.~\eqref{eq:CM_typical_Hamiltonian}]. Then, there is a theorem [\cite{hofer_symplectic_2011}] which
states that for a compact configuration space $\mathscr{Q}$, we have
periodic solutions of $\bm{X}_{H}$. In fact, it was even shown [\cite{suhr_linking_2016}] that
we have periodic solutions provided certain conditions on the potential
$V$ are satisfied and $\mathscr{Q}$ just needs to have bounded geometry
(\textit{i.e.} to be geodesically complete and to have the scalar curvature
and derivative thereof bounded).

Thus, under the assumption of compactness or any other condition which
entails closed orbits, we cannot have a function which behaves like
entropy in this sense for a very simple reason. Assume $S:\mathscr{P}\rightarrow\mathbb{R}$
is nondecreasing along trajectories and let us consider an orbit $\gamma:\mathbb{R}\rightarrow\mathscr{P}$
in phase space (satisfying ${\rm d}\gamma\left(t\right)/{\rm d}t=\bm{X}_{H}(\gamma\left(t\right))$)
which is closed. This means that for any $x\in\mathscr{P}$ on the
orbit, there exist $t_{0},T\in\mathbb{R}$ such that $x=\gamma(t_{0})=\gamma(t_{0}+T)$.
Hence, we have $S(x)=S(\gamma(t_{0}))=S(\gamma(t_{0}+T))=S(x)$, so $S$ is
constant along the orbit and therefore cannot behave like entropy.

\section{\label{sec:GR}Entropy theorems in general relativity}

We now turn to addressing the question of why these theorems do not carry over from CM to GR. We follow the notation and general setup presented in Chapter \ref{2-canonical}. The only notable exception is that we write the general (possibly multi-) index $A$ on the configuration variables $\varphi$ as a subscript ($\varphi_{A}$) instead of a superscript ($\varphi^{A}$, as before), and we indicate summation over such indices explicitly. Moreover, we write integrals over Cauchy surfaces with respect to the flat volume form, $\mathbf{e}={\rm d}^{3}x={\rm d}x^{1}\wedge{\rm d}x^{2}\wedge{\rm d}x^{3}$.

\subsection{\label{sec:4.2}Perturbative approach}

We wish to investigate under what conditions the CM no-entropy proof
of subsection \ref{sec:3.2} transfers over to field theories in curved
spacetime. To this effect, we consider the equivalent setup: broadly
speaking, we ask whether there exists a phase space functional $S:\mathscr{P}\rightarrow\mathbb{R}$
which is increasing in time everywhere except at an ``equilibrium''
configuration. In particular, we use the following two entropy conditions
in analogy with those of subsubsection \ref{sec:3.2.3} in CM: 

~

\noindent{\bf S1} \textsl{(Existence of equilibrium)}{\bf:} We assume there exists a point
$x_{0}=\left(\mathring{\varphi},\mathring{\pi}\right)\in\mathscr{P}$,
where $S$ is stationary, and (to simplify the analysis) $H$ is stationary
as well: 
\begin{equation}
\frac{\delta S\left[\mathring{\varphi},\mathring{\pi}\right]}{\delta\mathring{\varphi}_{A}\left(x\right)}=\frac{\delta S\left[\mathring{\varphi},\mathring{\pi}\right]}{\delta\mathring{\pi}_{A}\left(x\right)}=0=\frac{\delta H\left[\mathring{\varphi},\mathring{\pi}\right]}{\delta\mathring{\varphi}_{A}\left(x\right)}=\frac{\delta H\left[\mathring{\varphi},\mathring{\pi}\right]}{\delta\mathring{\pi}_{A}\left(x\right)}.\label{eq:GR_stationarity_assumption_general}
\end{equation}
This implies $\dot{S}[\mathring{\varphi},\mathring{\pi}]=0=\dot{H}[\mathring{\varphi},\mathring{\pi}]$. 

~

\noindent{\bf S2} \textsl{(Second law of thermodynamics)}{\bf:} We assume that the Hessian
of $\dot{S}$ is positive definite at equilibrium, \textit{i.e.} $\mathbf{Hess}(\dot{S}[\mathring{\varphi},\mathring{\pi}])\succ0$.
This is a sufficient condition to ensure that $\dot{S}>0$ in $\mathscr{P}\backslash x_{0}$,
and $\dot{S}=0$ at $x_{0}$.

~

We then follow the same procedure as in subsubsection \ref{sec:3.2.4}:
we insert into the Poisson bracket 
\begin{equation}
\dot{S}\!=\!\int_{\Sigma}\!{\rm d}^{3}x\sum_{A}\left(\frac{\delta H\left[\varphi,\pi\right]}{\delta\pi_{A}\left(x\right)}\frac{\delta S\left[\varphi,\pi\right]}{\delta\varphi_{A}\left(x\right)}-\frac{\delta H\left[\varphi,\pi\right]}{\delta\varphi_{A}\left(x\right)}\frac{\delta S\left[\varphi,\pi\right]}{\delta\pi_{A}\left(x\right)}\right)\label{eq:GR_poisson_bracket_S_general}
\end{equation}
the functional Taylor series [\cite{dreizler_density_2011}] for each term about $\left(\mathring{\varphi},\mathring{\pi}\right)$,
denoting $\Delta\varphi_{A}\left(x\right)=\varphi_{A}\left(x\right)-\mathring{\varphi}\left(x\right)$ and
$\Delta\pi_{A}\left(x\right)=\pi_{A}\left(x\right)-\mathring{\pi}\left(x\right)$:
\begin{align}
\frac{\delta H\left[\varphi,\pi\right]}{\delta\pi_{A}\left(x\right)}=\frac{\delta H\left[\mathring{\varphi},\mathring{\pi}\right]}{\delta\mathring{\pi}_{A}\left(x\right)}+\int_{\Sigma}{\rm d}^{3}y\sum_{B}\Bigg\{ &  \frac{\delta^{2}H\left[\mathring{\varphi},\mathring{\pi}\right]}{\delta\mathring{\varphi}_{B}\left(y\right)\delta\mathring{\pi}_{A}\left(x\right)}\Delta\mathring{\varphi}_{B}\left(y\right)\nonumber\\
 & +\frac{\delta^{2}H\left[\mathring{\varphi},\mathring{\pi}\right]}{\delta\mathring{\pi}_{B}\left(y\right)\delta\mathring{\pi}_{A}\left(x\right)}\Delta\mathring{\pi}_{B}\left(y\right)\Bigg\} +\mathcal{O}\left(\Delta^{2}\right),\label{eq:GR_taylor_series_general}
\end{align}
and similarly for the other terms. Then we apply S1 in this case [Eq.~\eqref{eq:GR_stationarity_assumption_general}], which makes all
zero-order terms vanish. Finally, the Poisson bracket in this case
[Eq.~\eqref{eq:GR_poisson_bracket_S_general}] becomes: 
\begin{align}
\dot{S}=\, & \int_{\Sigma}{\rm d}^{3}x\int_{\Sigma}{\rm d}^{3}y\int_{\Sigma}{\rm d}^{3}z\sum_{A,B,C}\nonumber\\
 & \big\{\Bigg[\frac{\delta^{2}H\left[\mathring{\varphi},\mathring{\pi}\right]}{\delta\mathring{\varphi}_{B}\left(y\right)\delta\mathring{\pi}_{A}\left(x\right)}\frac{\delta^{2}S\left[\mathring{\varphi},\mathring{\pi}\right]}{\delta\mathring{\varphi}_{C}\left(z\right)\delta\mathring{\varphi}_{A}\left(x\right)}-\frac{\delta^{2}H\left[\mathring{\varphi},\mathring{\pi}\right]}{\delta\mathring{\varphi}_{B}\left(y\right)\delta\mathring{\varphi}_{A}\left(x\right)}\frac{\delta^{2}S\left[\mathring{\varphi},\mathring{\pi}\right]}{\delta\mathring{\varphi}_{C}\left(z\right)\delta\mathring{\pi}_{A}\left(x\right)}\Bigg]\nonumber\\
 & \hspace{9.5cm}\times\Delta\varphi_{B}\left(y\right)\Delta\varphi_{C}\left(z\right)\nonumber\\
+ & \Bigg[\frac{\delta^{2}H\left[\mathring{\varphi},\mathring{\pi}\right]}{\delta\mathring{\varphi}_{B}\left(y\right)\delta\mathring{\pi}_{A}\left(x\right)}\frac{\delta^{2}S\left[\mathring{\varphi},\mathring{\pi}\right]}{\delta\mathring{\pi}_{C}\left(z\right)\delta\mathring{\varphi}_{A}\left(x\right)}+\frac{\delta^{2}H\left[\mathring{\varphi},\mathring{\pi}\right]}{\delta\mathring{\pi}_{C}\left(z\right)\delta\mathring{\pi}_{A}\left(x\right)}\frac{\delta^{2}S\left[\mathring{\varphi},\mathring{\pi}\right]}{\delta\mathring{\varphi}_{B}\left(y\right)\delta\mathring{\varphi}_{A}\left(x\right)}\nonumber\\
 & -\frac{\delta^{2}H\left[\mathring{\varphi},\mathring{\pi}\right]}{\delta\mathring{\varphi}_{B}\left(y\right)\delta\mathring{\varphi}_{A}\left(x\right)}\frac{\delta^{2}S\left[\mathring{\varphi},\mathring{\pi}\right]}{\delta\mathring{\pi}_{C}\left(z\right)\delta\mathring{\pi}_{A}\left(x\right)}-\frac{\delta^{2}H\left[\mathring{\varphi},\mathring{\pi}\right]}{\delta\mathring{\pi}_{C}\left(z\right)\delta\mathring{\varphi}_{A}\left(x\right)}\frac{\delta^{2}S\left[\mathring{\varphi},\mathring{\pi}\right]}{\delta\mathring{\varphi}_{B}\left(y\right)\delta\mathring{\pi}_{A}\left(x\right)}\Bigg]\nonumber\\
 & \hspace{9.5cm}\times\Delta\varphi_{B}\left(y\right)\Delta\pi_{C}\left(z\right)\nonumber\\
+ & \Bigg[\frac{\delta^{2}H\left[\mathring{\varphi},\mathring{\pi}\right]}{\delta\mathring{\pi}_{B}\left(y\right)\delta\mathring{\pi}_{A}\left(x\right)}\frac{\delta^{2}S\left[\mathring{\varphi},\mathring{\pi}\right]}{\delta\mathring{\pi}_{C}\left(z\right)\delta\mathring{\varphi}_{A}\left(x\right)}-\frac{\delta^{2}H\left[\mathring{\varphi},\mathring{\pi}\right]}{\delta\mathring{\pi}_{B}\left(y\right)\delta\mathring{\varphi}_{A}\left(x\right)}\frac{\delta^{2}S\left[\mathring{\varphi},\mathring{\pi}\right]}{\delta\mathring{\pi}_{C}\left(z\right)\delta\mathring{\pi}_{A}\left(x\right)}\Bigg]\nonumber \\
 & \hspace{7.25cm}\times\Delta\pi_{B}\left(y\right)\Delta\pi_{C}\left(z\right)\big\}+\mathcal{O}\left(\Delta^{3}\right).\label{eq:GR_Sdot_quadratic_form_general}
\end{align}
We compute this, in turn, for a scalar field in curved spacetime,
for EM in curved spacetime, and for GR. We will show
that no function $S$ obeying the conditions S1-S2 given here exists
in the case of the first two, but that the same cannot be said of
the latter.


\subsubsection{\label{sec:4.2.1}Scalar field}

Let us consider a theory for a scalar field $\phi\left(x\right)$
in a potential $V\left[\phi\left(x\right)\right]$, defined by the
Lagrangian 
\begin{equation}
\mathcal{L}=\sqrt{-g}\left(-\frac{1}{2}g^{ab}\nabla_{a}\phi\nabla_{b}\phi-V\left[\phi\right]\right).\label{eq:GR_scalar_Lagrangian}
\end{equation}
There are no constraints in this case. For turning the above [Eq.~\eqref{eq:GR_scalar_Lagrangian}] into a canonical theory, let us
choose a foliation of $\mathscr{M}$ such that $\bm{N}=0$. The canonical
measure [\cite{crnkovic_covariant_1989}] is then simply given by $\bm{\Omega}=\int_{\Sigma}{\rm d}^{3}x\,\delta\dot{\phi}\wedge\delta\phi$,
and the Hamiltonian [\cite{poisson_relativists_2007}] is 
\begin{equation}
H\left[\phi,\pi\right]\!=\!\int_{\Sigma}\!{\rm d}^{3}x\, N\!\left(\frac{\pi^{2}}{2\sqrt{h}}+\frac{\sqrt{h}}{2}h^{ab}\nabla_{a}\phi\nabla_{b}\phi+\sqrt{h}V\left[\phi\right]\right)\!,\label{eq:GR_scalar_Hamiltonian}
\end{equation}
where $\pi=(\sqrt{h}/N)\dot{\phi}$ is the canonical momentum.

Let us compute the second functional derivatives of $H$. We have:
\begin{align}
\frac{\delta^{2}H\left[\phi,\pi\right]}{\delta\phi\left(y\right)\delta\phi\left(x\right)}= & N\left(x\right)\sqrt{h\left(x\right)}V''\left[\phi\left(x\right)\right]\delta\left(x-y\right)\nonumber\\
 & -\partial_{a}\left(N\left(x\right)\sqrt{h\left(x\right)}h^{ab}\left(x\right)\partial_{b}\delta\left(x-y\right)\right),\label{eq:GR_scalar_d2H_phi}\\
\frac{\delta^{2}H\left[\phi,\pi\right]}{\delta\pi\left(y\right)\delta\pi\left(x\right)}= & \frac{N\left(x\right)}{\sqrt{h\left(x\right)}}\delta\left(x-y\right),\label{eq:GR_scalar_d2H_pi}
\end{align}
and the mixed derivatives $\delta^{2}H\left[\phi,\pi\right]/\delta\pi\left(y\right)\delta\phi\left(x\right)$
vanish.

We now proceed as outlined above: We assume there exists an entropy
function $S:\mathscr{P}\rightarrow\mathbb{R}$ obeying S1-S2 with
an equilibrium field configuration $(\mathring{\phi},\mathring{\pi})$,
and we will show that there is a contradiction with $\dot{S}>0$.
Additionally, we assume that $V''[\mathring{\phi}]\geq0$; in other
words, the equilibrium field configuration is one where the potential
is concave upwards, \textit{i.e.} it is a stable equilibrium.

According to the above expression for $\dot{S}$ [Eq.~\eqref{eq:GR_Sdot_quadratic_form_general}],
we have that entropy production in this case is given by 
\begin{align}
\dot{S}=\, & \int_{\Sigma}{\rm d}^{3}x\,{\rm d}^{3}y\,{\rm d}^{3}z\,\Bigg\{\Bigg[-\frac{\delta^{2}H[\mathring{\phi},\mathring{\pi}]}{\delta\mathring{\phi}\left(y\right)\delta\mathring{\phi}\left(x\right)}\frac{\delta^{2}S[\mathring{\phi},\mathring{\pi}]}{\delta\mathring{\phi}\left(z\right)\delta\mathring{\pi}\left(x\right)}\Bigg]\Delta\phi\left(y\right)\Delta\phi\left(z\right)\nonumber \\
+ & \Bigg[\frac{\delta^{2}H[\mathring{\phi},\mathring{\pi}]}{\delta\mathring{\pi}\left(z\right)\delta\mathring{\pi}\left(x\right)}\frac{\delta^{2}S[\mathring{\phi},\mathring{\pi}]}{\delta\mathring{\phi}\left(y\right)\delta\mathring{\phi}\left(x\right)}-\frac{\delta^{2}H[\mathring{\phi},\mathring{\pi}]}{\delta\mathring{\phi}\left(y\right)\delta\mathring{\phi}\left(x\right)}\frac{\delta^{2}S[\mathring{\phi},\mathring{\pi}]}{\delta\mathring{\pi}\left(z\right)\delta\mathring{\pi}\left(x\right)}\Bigg]\Delta\phi\left(y\right)\Delta\pi\left(z\right)\nonumber \\
+ & \Bigg[\frac{\delta^{2}H[\mathring{\phi},\mathring{\pi}]}{\delta\mathring{\pi}\left(y\right)\delta\mathring{\pi}\left(x\right)}\frac{\delta^{2}S[\mathring{\phi},\mathring{\pi}]}{\delta\mathring{\pi}\left(z\right)\delta\mathring{\phi}\left(x\right)}\Bigg]\Delta\pi\left(y\right)\Delta\pi\left(z\right)\Bigg\}+\mathcal{O}\left(\Delta^{3}\right),\label{eq:GR_scalar_Sdot_quadratic_form}
\end{align}
where we have used the fact that the mixed derivatives vanish. Let
us now evaluate $\dot{S}$ along different directions in $\mathscr{P}$
away from $(\mathring{\phi},\mathring{\pi})$. Suppose $\Delta\pi$
is nonzero everywhere on $\Sigma$, and $\Delta\phi$ vanishes everywhere
on $\Sigma$. Then, using the second momentum derivative of $H$ [Eq.~\eqref{eq:GR_scalar_d2H_pi}], $\dot{S}$ [Eq.~\eqref{eq:GR_scalar_Sdot_quadratic_form}]
becomes: 
\begin{align}
\dot{S}= & \int_{\Sigma}{\rm d}^{3}x\,{\rm d}^{3}y\,{\rm d}^{3}z\,\frac{N\left(x\right)}{\sqrt{h\left(x\right)}}\delta\left(x-y\right)\frac{\delta^{2}S[\mathring{\phi},\mathring{\pi}]}{\delta\mathring{\pi}\left(z\right)\delta\mathring{\phi}\left(x\right)}\Delta\pi\left(y\right)\Delta\pi\left(z\right)+\mathcal{O}\left(\Delta^{3}\right)\label{eq:GR_scalar_Sdot_dpi_1}\\
= & \int_{\Sigma}{\rm d}^{3}y\,{\rm d}^{3}z\,\frac{N\left(y\right)}{\sqrt{h\left(y\right)}}\frac{\delta^{2}S[\mathring{\phi},\mathring{\pi}]}{\delta\mathring{\pi}\left(z\right)\delta\mathring{\phi}\left(y\right)}\Delta\pi\left(y\right)\Delta\pi\left(z\right)+\mathcal{O}\!\left(\Delta^{3}\right)\label{eq:GR_scalar_Sdot_dpi_2}\\
\leq & \left\{ \max_{x\in\Sigma}\frac{N\left(x\right)}{\sqrt{h\left(x\right)}}\left(\Delta\pi\left(x\right)\right)^{2}\right\} \int_{\Sigma}{\rm d}^{3}y\,{\rm d}^{3}z\,\frac{\delta^{2}S[\mathring{\phi},\mathring{\pi}]}{\delta\mathring{\pi}\left(z\right)\delta\mathring{\phi}\left(y\right)}\nonumber+\mathcal{O}\left(\Delta^{3}\right)\,.\label{eq:GR_scalar_Sdot_dpi_3}\\
\end{align}
The requirement that the LHS of the first line above [Eq.~\eqref{eq:GR_scalar_Sdot_dpi_1}]
is strictly positive, combined with the strict positivity of the term
in curly brackets in the third line [Eq.~\eqref{eq:GR_scalar_Sdot_dpi_3}]
and the assumption (S2) of the definiteness of the Hessian of $\dot{S}$
at $(\mathring{\phi},\mathring{\pi})$, altogether mean that the above
[Eqs.~\eqref{eq:GR_scalar_Sdot_dpi_1}-\eqref{eq:GR_scalar_Sdot_dpi_3}]
imply: 
\begin{equation}
\int_{\Sigma}{\rm d}^{3}y\,{\rm d}^{3}z\,\frac{\delta^{2}S[\mathring{\phi},\mathring{\pi}]}{\delta\mathring{\pi}\left(z\right)\delta\mathring{\phi}\left(y\right)}>0.\label{eq:GR_scalar_Spiphi}
\end{equation}
Now let us evaluate $\dot{S}$ in a region of $\mathscr{P}$ where
$\Delta\phi$ is nonzero everywhere on $\Sigma$, while $\Delta\pi$
vanishes everywhere on $\Sigma$. Then, using the second field derivative
of $H$ [Eq.~\eqref{eq:GR_scalar_d2H_phi}], the negative of the
above expression for $\dot{S}$ [Eq.~\eqref{eq:GR_scalar_Sdot_quadratic_form}]
becomes: 
\begin{align}
-\dot{S}= & \int_{\Sigma}{\rm d}^{3}x\,{\rm d}^{3}y\,{\rm d}^{3}z\,\bigg\{ N\left(x\right)\sqrt{h\left(x\right)}V''[\mathring{\phi}\left(x\right)]\delta\left(x-y\right)\nonumber \\
 & -\partial_{a}\left(N\left(x\right)\sqrt{h\left(x\right)}h^{ab}\left(x\right)\partial_{b}\delta\left(x-y\right)\right)\bigg\}\nonumber\\
 & \times\frac{\delta^{2}S[\mathring{\phi},\mathring{\pi}]}{\delta\mathring{\phi}\left(z\right)\delta\mathring{\pi}\left(x\right)}\Delta\phi\left(y\right)\Delta\phi\left(z\right)+\mathcal{O}\left(\Delta^{3}\right).\label{eq:GR_scalar_Sdot_dphi_1}
\end{align}
Now, observe that 
\begin{multline}
\int_{\Sigma}{\rm d}^{3}x\,{\rm d}^{3}y\,{\rm d}^{3}z\,\left\{ \partial_{a}\left(N\left(x\right)\sqrt{h\left(x\right)}h^{ab}\left(x\right)\partial_{b}\delta\left(x-y\right)\right)\right\}\\
\times\frac{\delta^{2}S[\mathring{\phi},\mathring{\pi}]}{\delta\mathring{\phi}\left(z\right)\delta\mathring{\pi}\left(x\right)}\Delta\phi\left(y\right)\Delta\phi\left(z\right)\label{eq:GR_scalar_boundary_term}
\end{multline}
is simply a boundary term. This can be seen by integrating by parts
until the derivative is removed from the delta distribution, the definition
of the latter is applied to remove the $x$ integration, and the result
is a total derivative in the integrand. Assuming asymptotic decay
properties sufficient to make this boundary term vanish, the above
$-\dot{S}$ [Eq.~\eqref{eq:GR_scalar_Sdot_dphi_1}] simply becomes:
\begin{align}
-\dot{S}= & \int_{\Sigma}{\rm d}^{3}y\,{\rm d}^{3}z\, N\left(y\right)\sqrt{h\left(y\right)}V''[\mathring{\phi}\left(y\right)]\frac{\delta^{2}S[\mathring{\phi},\mathring{\pi}]}{\delta\mathring{\phi}\left(z\right)\delta\mathring{\pi}\left(y\right)}\Delta\phi\left(y\right)\Delta\phi\left(z\right)+\mathcal{O}\left(\Delta^{3}\right)\label{eq:GR_scalar_Sdot_dphi_2}\\
\geq & \left\{ \min_{x\in\Sigma}N\left(x\right)\sqrt{h\left(x\right)}V''[\mathring{\phi}\left(x\right)]\left(\Delta\phi\left(x\right)\right)^{2}\right\}\int_{\Sigma}{\rm d}^{3}y\,{\rm d}^{3}z\,\frac{\delta^{2}S[\mathring{\phi},\mathring{\pi}]}{\delta\mathring{\phi}\left(z\right)\delta\mathring{\pi}\left(y\right)}+\mathcal{O}\left(\Delta^{3}\right).\label{eq:GR_scalar_Sdot_dphi_3}
\end{align}
The LHS of the first line [Eq.~\eqref{eq:GR_scalar_Sdot_dphi_2}]
should be strictly negative, and the term in curly brackets in the
second line [Eq.~\eqref{eq:GR_scalar_Sdot_dphi_3}] is strictly
positive. Hence, owing to the definiteness of the Hessian of $\dot{S}$
at $(\mathring{\phi},\mathring{\pi})$, and using the symmetry of
the arguments in the integrand and equality of mixed derivatives,
the above [Eqs.~\eqref{eq:GR_scalar_Sdot_dphi_2}-\eqref{eq:GR_scalar_Sdot_dphi_3}]
imply: 
\begin{equation}
\int_{\Sigma}{\rm d}^{3}y\,{\rm d}^{3}z\,\frac{\delta^{2}S[\mathring{\phi},\mathring{\pi}]}{\delta\mathring{\pi}\left(z\right)\delta\mathring{\phi}\left(y\right)}<0.\label{eq:GR_scalar_Spiphi_contradiction}
\end{equation}
This is a contradiction with the inequality obtained previously [Eq.~\eqref{eq:GR_scalar_Spiphi}]. Therefore, we have no function $S$
for a scalar field theory that behaves like entropy according to assumptions
S1-S2.

We remark that in this case, we get the conclusion $\dot{S}=0$ using
the perturbative approach despite the fact that the topological one would
not work in the case of a non-compact Cauchy surface. The reason is
that:
\begin{align}
\mu\left(\mathscr{P}\right)=\int_{\mathscr{P}}\!\!\bm{\Omega} & =\int_{\mathscr{P}}\int_{\Sigma}{\rm d}^{3}x\,\delta\dot{\phi}\left(x\right)\wedge\delta\phi\left(x\right)\label{eq:GR_scalar_mu1}\\
 & \geq\int_{\mathscr{P}}\int_{\Sigma}{\rm d}^{3}x\,\min_{y\in\Sigma}\left[\delta\dot{\phi}\left(y\right)\wedge\delta\phi\left(y\right)\right]\label{eq:GR_scalar_mu2}\\
 & =\int_{\mathscr{P}}\left\{ \min_{y\in\Sigma}\left[\delta\dot{\phi}\left(y\right)\wedge\delta\phi\left(y\right)\right]\right\} \left[\int_{\Sigma}{\rm d}^{3}x\right]\!,\label{eq:GR_scalar_mu3}
\end{align}
which diverges if $\Sigma$ is non-compact. (N.B. The reason why the
term in curly brackets is finite but non-zero is that the field and
its time derivative cannot be always vanishing at any given point, for if they
were it would lead only to the trivial solution.) Thus, only the perturbative
approach is useful here for deducing lack of entropy production for
spacetimes with non-compact Cauchy surfaces.


\subsubsection{\label{sec:4.2.2}Electromagnetism}

Before we inspect EM in curved spacetime, let us carry out the analysis
in flat spacetime ($N=1$, $\bm{N}=\bm{0}$, and $\bm{h}={}^{(3)}\bm{\delta}={\rm diag(0,1,1,1)}$),
for massive (or de Broglie-Proca) EM [\cite{prescod-weinstein_extension_2014}], defined by the Lagrangian 
\begin{equation}
\mathcal{L}=-\frac{1}{4}\bm{F}:\bm{F}-\frac{1}{2}m^{2}\bm{A}\cdot\bm{A}+\bm{A}\cdot\bm{J},\label{eq:GR_massive_EM_Lagrangian}
\end{equation}
where $F_{ab}=\partial_{a}A_{b}-\partial_{b}A_{a}$, $A_{a}$ is the
electromagnetic potential (Faraday tensor), and $J_{a}$ is an external source.

We have a constrained Hamiltonian system in this case. In particular,
the momentum canonically conjugate to $A_{0}=V$ vanishes identically.
This means that instead of $A_{a}$, we may take (its spatial part)
$\mathcal{A}_{a}={}^{(3)}\delta_{ab}A^{b}$ along with its conjugate
momentum, $\pi^{a}=\dot{\mathcal{A}}^{a}-\partial^{a}V$, to be the
phase space variables---while appending to the canonical equations
of motion resulting from $H\left[\bm{\mathcal{A}},\bm{\pi}\right]$
the constraint $0=\delta H/\delta V$. In particular, we have [\cite{prescod-weinstein_extension_2014}]:
\begin{align}
H\left[\bm{\mathcal{A}},\bm{\pi}\right]=\!\!\int_{\Sigma}\!\!{\rm d}^{3}x & \bigg(\frac{1}{4}\bm{\mathcal{F}}:\bm{\mathcal{F}}+\!\frac{m^{2}}{2}\left(\bm{\mathcal{A}}\cdot\bm{\mathcal{A}}-V^{2}\right)\!-\bm{\mathcal{A}}\cdot\bm{\mathcal{J}}\nonumber \\
 & +\frac{1}{2}\bm{\pi}\cdot\bm{\pi}-\left(\partial_{a}\pi^{a}+\rho\right)V+\partial_{a}\left(V\pi^{a}\right)\!\bigg)\!,\label{eq:GR_massive_EM_Hamiltonian}
\end{align}
where $\mathcal{F}_{ab}={}^{(3)}\delta_{ac}{}^{(3)}\delta_{bd}F^{cd}$,
$\rho=J^{0}$ and $\mathcal{J}^{a}={}^{(3)}\delta^{ab}J_{b}$.

The Poisson bracket [Eq.~\eqref{eq:GR_Sdot_quadratic_form_general}]
is, in this case: 
\begin{align}
\dot{S}=\int_{\Sigma}{\rm d}^{3}x\,{\rm d}^{3}y\,{\rm d}^{3}z\,\Bigg\{ & \Bigg[-\frac{\delta^{2}H[\mathring{\bm{\mathcal{A}}},\mathring{\bm{\pi}}]}{\delta\mathring{\mathcal{A}}_{b}\left(y\right)\delta\mathring{\mathcal{A}}_{a}\left(x\right)}\frac{\delta^{2}S[\mathring{\bm{\mathcal{A}}},\mathring{\bm{\pi}}]}{\delta\mathring{\mathcal{A}}_{c}\left(z\right)\delta\mathring{\pi}^{a}\left(x\right)}\Bigg]\Delta\mathcal{A}_{b}\left(y\right)\Delta\mathcal{A}_{c}\left(z\right)\nonumber \\
+ & \Bigg[\frac{\delta^{2}H[\mathring{\bm{\mathcal{A}}},\mathring{\bm{\pi}}]}{\delta\mathring{\pi}_{c}\left(z\right)\delta\mathring{\pi}_{a}\left(x\right)}\frac{\delta^{2}S[\mathring{\bm{\mathcal{A}}},\mathring{\bm{\pi}}]}{\delta\mathring{\mathcal{A}}_{b}\left(y\right)\delta\mathring{\mathcal{A}}^{a}\left(x\right)}\nonumber \\
 & -\frac{\delta^{2}H[\mathring{\bm{\mathcal{A}}},\mathring{\bm{\pi}}]}{\delta\mathring{\mathcal{A}}_{b}\left(y\right)\delta\mathring{\mathcal{A}}_{a}\left(x\right)}\frac{\delta^{2}S[\mathring{\bm{\mathcal{A}}},\mathring{\bm{\pi}}]}{\delta\mathring{\pi}_{c}\left(z\right)\delta\mathring{\pi}^{a}\left(x\right)}\Bigg]\Delta\mathcal{A}_{b}\left(y\right)\Delta\pi_{c}\left(z\right)\nonumber \\
+ & \Bigg[\frac{\delta^{2}H[\mathring{\bm{\mathcal{A}}},\mathring{\bm{\pi}}]}{\delta\mathring{\pi}_{b}\left(y\right)\delta\mathring{\pi}_{a}\left(x\right)}\frac{\delta^{2}S[\mathring{\bm{\mathcal{A}}},\mathring{\bm{\pi}}]}{\delta\mathring{\pi}_{c}\left(z\right)\delta\mathring{\mathcal{A}}^{a}\left(x\right)}\Bigg]\Delta\pi_{b}\left(y\right)\Delta\pi_{c}\left(z\right)\Bigg\}+\mathcal{O}\left(\Delta^{3}\right),\label{eq:GR_EM_Sdot_quadratic_form}
\end{align}
where we have used the fact that the mixed derivatives of the Hamiltonian
[Eq.~\eqref{eq:GR_massive_EM_Hamiltonian}] vanish by inspection,
and we compute the second field and momentum derivatives thereof to
be, respectively:
\begin{align}
\frac{\delta^{2}H[\mathring{\bm{\mathcal{A}}},\mathring{\bm{\pi}}]}{\delta\mathring{\mathcal{A}}_{b}\left(y\right)\delta\mathring{\mathcal{A}}_{a}\left(x\right)}\!= & \!-\left\{\! ^{(3)}\delta^{ab}\partial^{c}\partial_{c}\delta\left(x-y\right)-\partial^{b}\partial^{a}\delta\left(x-y\right)\!\right\}+m^{2}\left[^{(3)}\delta^{ab}\delta\left(x-y\right)\right],\label{eq:GR_massive_EM_HAA}\\
\frac{\delta^{2}H[\mathring{\bm{\mathcal{A}}},\mathring{\bm{\pi}}]}{\delta\mathring{\pi}_{b}\left(y\right)\delta\mathring{\pi}_{a}\left(x\right)}\!= & ^{(3)}\delta^{ab}\delta\left(x-y\right).\label{eq:GR_massive_EM_Hpipi}
\end{align}
Analogously with our strategy in the scalar field case, let us
evaluate $\dot{S}$ along different directions away from equilibrium.
In particular, let us suppose $\Delta\pi_{1}$ is nonzero everywhere
on $\Sigma$, and that $\Delta\pi_{2}$, $\Delta\pi_{3}$, and $\Delta\mathcal{A}_{a}$
all vanish everywhere on $\Sigma$. Then, using the second momentum
derivative of $H$ [Eq.~\eqref{eq:GR_massive_EM_Hpipi}], $\dot{S}$
[Eq.~\eqref{eq:GR_EM_Sdot_quadratic_form}] becomes:
\begin{align}
\dot{S}= & \int_{\Sigma}{\rm d}^{3}x\,{\rm d}^{3}y\,{\rm d}^{3}z\,\delta\left(x-y\right)\frac{\delta^{2}S[\mathring{\bm{\mathcal{A}}},\mathring{\bm{\pi}}]}{\delta\mathring{\pi}_{1}\left(z\right)\delta\mathring{\mathcal{A}}_{1}\!\left(x\right)}\Delta\pi_{1}\!\left(y\right)\Delta\pi_{1}\left(z\right)+\mathcal{O}\left(\Delta^{3}\right)\label{eq:GR_massive_EM_Sdot_dpi_1}\\
= & \int_{\Sigma}{\rm d}^{3}y\,{\rm d}^{3}z\,\frac{\delta^{2}S[\mathring{\bm{\mathcal{A}}},\mathring{\bm{\pi}}]}{\delta\mathring{\pi}_{1}\left(z\right)\delta\mathring{\mathcal{A}}_{1}\left(y\right)}\Delta\pi_{1}\left(y\right)\Delta\pi_{1}\left(z\right)+\mathcal{O}\left(\Delta^{3}\right)\label{eq:GR_massive_EM_Sdot_dpi_2}\\
\leq & \left\{ \max_{x\in\Sigma}\left(\Delta\pi_{1}\left(x\right)\right)^{2}\right\} \int_{\Sigma}{\rm d}^{3}y\,{\rm d}^{3}z\,\frac{\delta^{2}S[\mathring{\bm{\mathcal{A}}},\mathring{\bm{\pi}}]}{\delta\mathring{\pi}_{1}\left(z\right)\delta\mathring{\mathcal{A}}_{1}\left(y\right)}+\mathcal{O}\left(\Delta^{3}\right).\label{eq:GR_massive_EM_Sdot_dpi_3}
\end{align}
The argument proceeds as before: the strict positivity of the LHS
of the first line above [Eq.~\eqref{eq:GR_massive_EM_Sdot_dpi_1}],
combined with that of the term in curly brackets in the third line
[Eq.~\eqref{eq:GR_massive_EM_Sdot_dpi_3}] and the assumption
(S2) of the definiteness of the Hessian of $\dot{S}$ at $(\mathring{\bm{\mathcal{A}}},\mathring{\bm{\pi}})$,
altogether mean that the above [Eqs.~\eqref{eq:GR_massive_EM_Sdot_dpi_1}-\eqref{eq:GR_massive_EM_Sdot_dpi_3}]
imply: 
\begin{equation}
\int_{\Sigma}{\rm d}^{3}y\,{\rm d}^{3}z\,\frac{\delta^{2}S[\mathring{\bm{\mathcal{A}}},\mathring{\bm{\pi}}]}{\delta\mathring{\pi}_{1}\left(z\right)\delta\mathring{\mathcal{A}}_{1}\left(y\right)}>0.\label{eq:GR_massive_EM_SpiA}
\end{equation}
Now let us evaluate $\dot{S}$ where $\Delta\mathcal{A}_{1}$ is nonzero
everywhere on $\Sigma$, while $\Delta\mathcal{A}_{2}$, $\Delta\mathcal{A}_{3}$
and $\Delta\pi_{a}$ all vanish everywhere on $\Sigma$. Then, using
the second field derivative of $H$ [Eq.~\eqref{eq:GR_massive_EM_HAA}],
the negative of the above expression for $\dot{S}$ [Eq.~\eqref{eq:GR_EM_Sdot_quadratic_form}]
becomes:
\begin{multline}
-\dot{S}=\int_{\Sigma}{\rm d}^{3}x\,{\rm d}^{3}y\,{\rm d}^{3}z\,\Bigg[\Big(-\big\{ ^{(3)}\delta^{ab}\partial^{c}\partial_{c}\delta\left(x-y\right)\\
-\partial^{b}\partial^{a}\delta\left(x-y\right)\big\}+m^{2}\left[^{(3)}\delta^{ab}\delta\left(x-y\right)\right]\Big)\frac{\delta^{2}S[\mathring{\bm{\mathcal{A}}},\mathring{\bm{\pi}}]}{\delta\mathring{\mathcal{A}}_{c}\left(z\right)\delta\mathring{\pi}^{a}\left(x\right)}\Bigg]\\
\times\Delta\mathcal{A}_{b}\left(y\right)\Delta\mathcal{A}_{c}\left(z\right)+\mathcal{O}\left(\Delta^{3}\right).\label{eq:GR_massive_EM_Sdot_dA}
\end{multline}
The term in curly brackets simply furnishes a (vanishing) boundary
term (up to $\mathcal{O}(\Delta^{3})$). Note that for $m=0$ (corresponding
to Maxwellian EM in flat spacetime) we would thus get an indefinite
Hessian of $\dot{S}$ at $(\mathring{\bm{\mathcal{A}}},\mathring{\bm{\pi}})$,
and hence no function $S$ that behaves like entropy as per S1-S2.
So let us assume $m^{2}>0$. Using the symmetry of the arguments in
the integrand and equality of mixed derivatives, we are thus left
with:
\begin{align}
-\dot{S}= & \int_{\Sigma}{\rm d}^{3}y\,{\rm d}^{3}z\, m^{2}\frac{\delta^{2}S[\mathring{\bm{\mathcal{A}}},\mathring{\bm{\pi}}]}{\delta\mathring{\pi}_{1}\left(z\right)\delta\mathring{\mathcal{A}}_{1}\left(y\right)}\Delta\mathcal{A}_{1}\left(y\right)\Delta\mathcal{A}_{1}\left(z\right)+\mathcal{O}\left(\Delta^{3}\right)\label{eq:GR_massive_EM_Sdot_dA_1}\\
\geq & \left\{ m^{2}\min_{x\in\Sigma}\left(\Delta\mathcal{A}_{1}\left(x\right)\right)^{2}\right\} \int_{\Sigma}{\rm d}^{3}y\,{\rm d}^{3}z\,\frac{\delta^{2}S[\mathring{\bm{\mathcal{A}}},\mathring{\bm{\pi}}]}{\delta\mathring{\pi}_{1}\left(z\right)\delta\mathring{\mathcal{A}}_{1}\left(y\right)}+\mathcal{O}\left(\Delta^{3}\right).\label{eq:GR_massive_EM_Sdot_dA_2}
\end{align}
The LHS of the first line [Eq.~\eqref{eq:GR_massive_EM_Sdot_dA_1}]
should be strictly negative, and the term in curly brackets in the
second line [Eq.~\eqref{eq:GR_massive_EM_Sdot_dA_2}] is strictly
positive. Hence, owing to the definiteness of the Hessian of $\dot{S}$
at $(\mathring{\bm{\mathcal{A}}},\mathring{\bm{\pi}})$, the above [Eqs.~\eqref{eq:GR_massive_EM_Sdot_dA_1}-\eqref{eq:GR_massive_EM_Sdot_dA_2}]
imply:
\begin{equation}
\int_{\Sigma}{\rm d}^{3}y\,{\rm d}^{3}z\,\frac{\delta^{2}S[\mathring{\bm{\mathcal{A}}},\mathring{\bm{\pi}}]}{\delta\mathring{\pi}_{1}\left(z\right)\delta\mathring{\mathcal{A}}_{1}\left(y\right)}<0.\label{eq:GR_massive_EM_SpiA_2}
\end{equation}
This is a contradiction with the previous inequality on the same quantity
[Eq.~\eqref{eq:GR_massive_EM_SpiA}]. Hence there is no function
$S$ that behaves like entropy (according to S1-S2) for a massive
EM field in flat spacetime.

Let us now carry out the proof for a simple Maxwellian EM field in
curved spacetime, defined by the Lagrangian 
\begin{equation}
\mathcal{L}=-\frac{1}{4}\sqrt{-g}\bm{F}:\bm{F},\label{eq:GR_EM_Lagrangian}
\end{equation}
where $F_{ab}=\nabla_{a}A_{b}-\nabla_{b}A_{a}$ and $A_{a}$ is the
electromagnetic potential. As in the scalar field case, we work with
a spacetime foliation such that $\bm{N}=0$.

As with EM in flat spacetime, this is a constrained Hamiltonian system:
the momentum canonically conjugate to $A_{0}=V$ vanishes identically,
meaning again that instead of $A_{a}$, we may take (its spatial part)
$\mathcal{A}_{a}=h_{ab}A^{b}$ along with its conjugate momentum,
$\pi^{a}=(\sqrt{h}/N)h^{ab}(\dot{\mathcal{A}}_{b}-\partial_{b}V)$,
to be the physical phase space variables---appending to the canonical
equations of motion resulting from $H\left[\bm{\mathcal{A}},\bm{\pi}\right]$
the constraint $0=\delta H/\delta V=\partial_{a}\pi^{a}$ (which is
simply Gauss' law). In particular, we have [\cite{prescod-weinstein_extension_2014}]:
\begin{equation}
H\left[\bm{\mathcal{A}},\bm{\pi}\right]=\!\!\int_{\Sigma}\!\!{\rm d}^{3}x\left(\frac{1}{4}N\sqrt{h}\,\bm{\mathcal{F}}:\bm{\mathcal{F}}+\frac{N}{2\sqrt{h}}\bm{\pi}\cdot\bm{\pi}+\pi^{a}\partial_{a}V\!\right)\!,\label{eq:GR_EM_Hamiltonian}
\end{equation}
where $\mathcal{F}_{ab}=h_{ac}h_{bd}F^{cd}=\mathcal{D}_{a}\mathcal{A}_{b}-\mathcal{D}_{b}\mathcal{A}_{a}$, where $\bm{\mathcal{D}}$ is the derivative induced on $\Sigma$.

The Poisson bracket [Eq.~\eqref{eq:GR_Sdot_quadratic_form_general}]
is here given by the same expression as in flat spacetime [Eq.~\eqref{eq:GR_scalar_Sdot_quadratic_form}],
owing to the fact that the mixed derivatives of the Hamiltonian [Eq.~\eqref{eq:GR_EM_Hamiltonian}] vanish. Let us focus on regions in
phase space where $\Delta\bm{\pi}$ vanishes everywhere on $\Sigma,$
but $\Delta\bm{\mathcal{A}}$ is everywhere nonzero. There, 
\begin{multline}
\dot{S}=\int_{\Sigma}{\rm d}^{3}x\,{\rm d}^{3}y\,{\rm d}^{3}z\Bigg[-\frac{\delta^{2}H[\mathring{\bm{\mathcal{A}}},\mathring{\bm{\pi}}]}{\delta\mathring{\mathcal{A}}_{b}\left(y\right)\delta\mathring{\mathcal{A}}_{a}\left(x\right)}\frac{\delta^{2}S[\mathring{\bm{\mathcal{A}}},\mathring{\bm{\pi}}]}{\delta\mathring{\mathcal{A}}_{c}\left(z\right)\delta\mathring{\pi}^{a}\left(x\right)}\Bigg]\\
\times\Delta\mathcal{A}_{b}\left(y\right)\Delta\mathcal{A}_{c}\left(z\right)+\mathcal{O}\left(\Delta^{3}\right).\label{eq:GR_EM_Sdot_dA}
\end{multline}
We compute: 
\begin{align}
\frac{\delta^{2}H[\mathring{\bm{\mathcal{A}}},\mathring{\bm{\pi}}]}{\delta\mathring{\mathcal{A}}_{b}\left(y\right)\delta\mathring{\mathcal{A}}_{a}\left(x\right)}\!= & \!-\!\sqrt{h\left(x\right)}\bigg\{ \!\mathcal{D}^{c}\!\left[N\left(x\right)h^{ab}\left(x\right)\mathcal{D}_{c}\delta\left(x-y\right)\right]\!-\!\mathcal{D}^{b}\left[N\left(x\right)\mathcal{D}^{a}\delta\left(x-y\right)\right]\bigg\} .\label{eq:GR_EM_d2H_dA}
\end{align}
Inserting this into the above expression for $\dot{S}$ [Eq.~\eqref{eq:GR_EM_Sdot_dA}],
we simply get a (vanishing) boundary term (up to $\mathcal{O}(\Delta^{3})$).
We conclude that we have an indefinite Hessian of $\dot{S}$ at $(\mathring{\bm{\mathcal{A}}},\mathring{\bm{\pi}})$,
and hence no function $S$ that behaves like entropy as per S1-S2.

\subsubsection{\label{sec:4.2.3}Gravity}

Here we use the basic notation and phase space construction of Section \ref{sec:2.4-canonical-GR}. We briefly remind the reader that the Hamiltonian of general relativity
is given by
\begin{equation}
H=\frac{1}{2\kappa}\int_{\Sigma}\boldsymbol{\epsilon}_{\Sigma}^{}\left[NC-2\boldsymbol{N}\cdot\bm{C}+2\boldsymbol{\mathcal{D}}\cdot\left(\frac{\boldsymbol{N}\cdot\boldsymbol{\pi}}{\sqrt{h}}\right)\right]\,,
\end{equation}
where $N$ and $\boldsymbol{N}$ are the lapse and shift, $C$ and
$\boldsymbol{C}$ are the secondary constraints, and $\boldsymbol{h}$
and $\boldsymbol{\pi}$ are the induced three-metric on $\Sigma$
and its canonical momentum respectively. See Chapter \ref{2-canonical} for full details.

Following the same procedure as before for a hypothetical entropy
functional $S[\bm{h},\bm{\pi}]$ and an equilibrium configuration
$(\mathring{\bm{h}},\mathring{\bm{\pi}})$ in phase space, we see
that the Poisson bracket [Eq.~\eqref{eq:GR_Sdot_quadratic_form_general}]
in this case has the following form: 
\begin{align}
\dot{S}= & \int_{\Sigma}{\rm d}^{3}x\,{\rm d}^{3}y\,{\rm d}^{3}z \nonumber \\
 & \big\{ \Bigg[\frac{\delta^{2}H[\mathring{\bm{h}},\mathring{\bm{\pi}}]}{\delta\mathring{h}^{cd}\left(y\right)\delta\mathring{\pi}^{ab}\left(x\right)}\frac{\delta^{2}S[\mathring{\bm{h}},\mathring{\bm{\pi}}]}{\delta\mathring{h}_{ef}\left(z\right)\delta\mathring{h}_{ab}\left(x\right)}-\frac{\delta^{2}H[\mathring{\bm{h}},\mathring{\bm{\pi}}]}{\delta\mathring{h}^{cd}\left(y\right)\delta\mathring{h}^{ab}\left(x\right)}\frac{\delta^{2}S[\mathring{\bm{h}},\mathring{\bm{\pi}}]}{\delta\mathring{h}_{ef}\left(z\right)\delta\mathring{\pi}_{ab}\left(x\right)}\Bigg]\nonumber \\
  & \hspace{9cm}\times\Delta h^{cd}\left(y\right)\Delta h_{ef}\left(z\right)\nonumber \\
+ & \Bigg[\frac{\delta^{2}H[\mathring{\bm{h}},\mathring{\bm{\pi}}]}{\delta\mathring{h}^{cd}\left(y\right)\delta\mathring{\pi}^{ab}\left(x\right)}\frac{\delta^{2}S[\mathring{\bm{h}},\mathring{\bm{\pi}}]}{\delta\mathring{\pi}_{ef}\left(z\right)\delta\mathring{h}_{ab}\left(x\right)}+\frac{\delta^{2}H[\mathring{\bm{h}},\mathring{\bm{\pi}}]}{\delta\mathring{\pi}_{ef}\left(z\right)\delta\mathring{\pi}^{ab}\left(x\right)}\frac{\delta^{2}S[\mathring{\bm{h}},\mathring{\bm{\pi}}]}{\delta\mathring{h}^{cd}\left(y\right)\delta\mathring{h}_{ab}\left(x\right)}\nonumber \\
 & -\frac{\delta^{2}H[\mathring{\bm{h}},\mathring{\bm{\pi}}]}{\delta\mathring{h}^{cd}\left(y\right)\delta\mathring{h}^{ab}\left(x\right)}\frac{\delta^{2}S[\mathring{\bm{h}},\mathring{\bm{\pi}}]}{\delta\mathring{\pi}_{ef}\left(z\right)\delta\mathring{\pi}_{ab}\left(x\right)}-\frac{\delta^{2}H[\mathring{\bm{h}},\mathring{\bm{\pi}}]}{\delta\mathring{\pi}_{ef}\left(z\right)\delta\mathring{h}^{ab}\left(x\right)}\frac{\delta^{2}S[\mathring{\bm{h}},\mathring{\bm{\pi}}]}{\delta\mathring{h}^{cd}\left(y\right)\delta\mathring{\pi}_{ab}\left(x\right)}\Bigg]\nonumber \\
  & \hspace{9cm}\times\Delta h^{cd}\left(y\right)\Delta\pi_{ef}\left(z\right)\nonumber \\
+ & \Bigg[\frac{\delta^{2}H[\mathring{\bm{h}},\mathring{\bm{\pi}}]}{\delta\mathring{\pi}^{cd}\left(y\right)\delta\mathring{\pi}^{ab}\left(x\right)}\frac{\delta^{2}S[\mathring{\bm{h}},\mathring{\bm{\pi}}]}{\delta\mathring{\pi}_{ef}\left(z\right)\delta\mathring{h}_{ab}\left(x\right)}-\frac{\delta^{2}H[\mathring{\bm{h}},\mathring{\bm{\pi}}]}{\delta\mathring{\pi}^{cd}\left(y\right)\delta\mathring{h}^{ab}\left(x\right)}\frac{\delta^{2}S[\mathring{\bm{h}},\mathring{\bm{\pi}}]}{\delta\mathring{\pi}_{ef}\left(z\right)\delta\mathring{\pi}_{ab}\left(x\right)}\Bigg]\nonumber \\
  & \hspace{9cm}\times\Delta\pi^{cd}\left(y\right)\Delta\pi_{ef}\left(z\right)\big\}\nonumber \\
+ & \mathcal{O}\left(\Delta^{3}\right).\label{eq:GR_Sdot_quadratic_form}
\end{align}
The difference with the previous cases is that here, in general, none
of the second derivatives of the Hamiltonian vanish, and crucially,
they do not have a definite sign. For example, let us compute the
second derivative of $H$ with respect to the canonical momentum:
\begin{equation}
\frac{\delta^{2}H[\mathring{\bm{h}},\mathring{\bm{\pi}}]}{\delta\mathring{\pi}_{cd}\left(y\right)\delta\mathring{\pi}^{ab}\left(x\right)}=\frac{2\mathring{N}\left(x\right)}{\sqrt{\mathring{h}\left(x\right)}}\left(\delta^{c}{}_{(a}\delta^{d}{}_{b)}-\frac{1}{2}\mathring{h}_{ab}\left(x\right)\delta^{cd}\right)\delta\left(x-y\right).\label{eq:GR_Hpipi}
\end{equation}
In CM or the examples of field theories in curved spacetime we have
considered, the second derivative of $H$ with respect to the momentum
had a definite sign (by virtue of its association with the positivity
of kinetic-type terms). In this case, however, this second derivative
[Eq.~\eqref{eq:GR_Hpipi}] is neither always positive nor always
negative. Thus an argument similar to the previous proofs cannot work
here: the gravitational Hamiltonian [Eq.~\eqref{eq:HGtotal}]
is of such a nature that its concavity in phase space components (as
is, for example, its concavity in the canonical momentum components
[Eq.~\eqref{eq:GR_Hpipi}]) is not independent of the phase space
variables themselves, and cannot be ascribed a definite (positive
or negative) sign. And so, a contradiction cannot arise with the Poisson
bracket of a phase space functional (such as the gravitational entropy)
being non-zero (and, in particular, positive).

\subsection{\label{sec:4.3}Topological approach}

As discussed in subsection \ref{sec:3.3}, the topological proofs
of Olsen for the non-existence of entropy production in CM rely crucially
on the assumption that the phase space $\mathscr{P}$ is compact.
In such a situation, a system has a finite measure of phase space
$\mu(\mathscr{P})$ available to explore, and there cannot exist a
function which continually increases along orbits.

By contrast, in GR, it is believed that the (reduced) phase
space $\mathscr{S}$ is generically noncompact [\cite{schiffrin_measure_2012}]. That is to say, the
measure $\mu(\mathscr{S})=\int_{\mathscr{S}}\bm{\Omega}|_{\mathscr{S}}$
in general diverges, where $\bm{\Omega}|_{\mathscr{S}}$ is (using
the notation of Chapter \ref{2-canonical}) the pullback of the symplectic
form of GR,
\begin{equation}
\boldsymbol{\Omega}=\int_{\Sigma}{\rm d}^{3}x\,\delta\pi^{ab}\wedge\delta h_{ab}\,,
\end{equation}
to $\mathscr{S}$. (See Section \ref{sec:2.4-canonical-GR} for more details.) This
means that the same methods of proof as in CM (subsection \ref{sec:3.3})
cannot be applied.

The connection between a (monotonically increasing) entropy function
in GR and the divergence of its (reduced) phase space measure warrants
some discussion. The latter, it may be noted, is arguably not completely
inevitable. In other words, one may well imagine a space of admissible
solutions to the Einstein equations (or equivalently, the canonical
gravitational equations) the effective degrees of freedom of which are such
that they form a finite-measure phase space. Dynamically-trivial examples
of this might be SD black holes. Thus the assertion that
$\mu(\mathscr{S})$ diverges hinges on the nature of the degrees of
freedom believed to be available in the spacetimes under consideration.
However, it has been explicitly shown [\cite{schiffrin_measure_2012}] that even in very basic dynamically-nontrivial
situations, such as simple cosmological spacetimes, $\mu(\mathscr{S})$
does indeed diverge. In fact, the proof found in [\cite{schiffrin_measure_2012}] is carried out for compact Cauchy surfaces, and the conclusion is therefore in concordance with the no-return theorem [\cite{tipler_general_1979,tipler_general_1980}] which also assumes compact Cauchy surfaces. In the following section, we will show that this
happens for perturbed SD spacetimes as well (where the Cauchy surface is non-compact).

The generic divergence of $\mu(\mathscr{S})$ entails that a gravitational
system has an unbounded region of phase space available to explore.
In other words, it is not confined to a finite region where it would
have to eventually return to a configuration from which it started
(which would make a monotonically increasing entropy function impossible). 

It is moreover worth remarking that this situation creates nontrivial problems
for a statistical (\textit{i.e.} probability-based) general-relativistic definition
of entropy, $S(t)$ (as described in
Section \ref{sec:Intro})---which, indeed, one may also ultimately desire
to work with and relate to the mechanical meaning of entropy mainly
discussed in this chapter. Naively, one might think of defining such
a statistical entropy function as something along the lines of $S=-\sum_{X}P\left(X\right)\ln P\left(X\right)$,
where $X$ denotes a physical property of interest and $P\left(X\right)$
its probability. In turn, the latter might be understood as the relative
size of the phase space region $\mathscr{S}_{X}\subset\mathscr{S}$
possessing the property $X$, \textit{i.e.} $P\left(X\right)=\mu(\mathscr{S}_{X})/\mu(\mathscr{S})$.
In this case, we either have [\cite{schiffrin_measure_2012}]: $P\left(X\right)=0$ if $\mu(\mathscr{S}_{X})$
is finite, $P\left(X\right)=1$ if $\mu(\mathscr{S}\backslash\mathscr{S}_{X})$
is finite, or $P\left(X\right)$ is ill-defined otherwise. Ostensibly,
one would need to invoke a regularization procedure in order to obtain
finite probabilities (in general) according to this. However, different
regularization procedures that have been applied (mainly in the context
of cosmology) have proven to yield widely different results depending
on the method of the procedure being used [\cite{schiffrin_measure_2012}]. Alternatively, a statistical
general-relativistic definition of $S$ in terms of a probability
density $\rho:\mathscr{S}\times\mathscr{T}\rightarrow[0,1]$ (similarly
to CM) as $S=-\int_{\mathscr{S}}\bm{\Omega}|_{\mathscr{S}}\rho\ln\rho$
would likewise face divergence issues. Therefore, any future attempt
to define gravitational entropy in such a context will have to either
devise an unambiguous and well-defined regularization procedure (for
obtaining finite probabilities), or implement a well-justified cutoff
of the (reduced) phase space measure.

We now turn to discussing these issues in a context where we expect
an intuitive illustration of gravitational entropy production---the
two-body problem.

\section{\label{sec:2BP}Entropy in the gravitational two-body problem}

One of the most elementary situations in GR in which we expect the
manifestation of a phenomenon such as entropy production is the gravitational
two-body problem.

In CM, the two-body (or Kepler) problem manifestly involves no increase
in the entropy of a system. The perturbative approach, as discussed
in subsubsection \ref{sec:3.2.3}, involves assumptions on the nature
of the Hamiltonian which preclude any conclusions from it in this
regard. However, the topological approach, elaborated in subsection
\ref{sec:3.3}, is applicable: assuming that Keplerian orbits are
bounded, the configuration space $\mathscr{Q}$ can be considered
to be compact, and therefore the phase space $\mathscr{P}$ obtained
from it (involving finite conjugate momenta) is compact as well. Concordant
with the topological proofs, then, we will have no entropy production
in such a situation. The case of the $N$-body problem however is, as alluded to earlier, not the same: neither the assumptions of the perturbative approach, not of the topological approach (specifically, a compact phase space) are applicable, and it has been shown that a monotonically increasing function on phase space does in fact exist [\cite{barbour_gravitational_2013,barbour_identification_2014}], and hence, a gravitational arrow of time (and entropy production) associated with it.

In GR, we know the two-body problem involves energy loss and therefore
should implicate an associated production of entropy. The no-return theorem [\cite{tipler_general_1979,tipler_general_1980}] is inapplicable here because this problem does not involve a compact Cauchy surface. The perturbative
approach here fails to disprove the second law (as discussed in subsubsection
\ref{sec:4.2.3}), and we will now show that so too does the topological
approach.

The two-body problem in GR where one small body orbits a much larger
body of mass $M$ can be modeled in the context of perturbations
to the SD metric, 
\begin{equation}
g_{ab}{\rm d}x^{a}{\rm d}x^{b}=-f\left(r\right){\rm d}t^{2}+\frac{{\rm d}r^{2}}{f\left(r\right)}+r^{2}\sigma_{ab}{\rm d}x^{a}{\rm d}x^{b},\label{eq:2BP_Schwarzschild}
\end{equation}
where $f\left(r\right)=1-2M/r$ and $\bm{\sigma}={\rm diag}(0,0,1,\sin^{2}\theta)$ is the metric
of the two-sphere $\mathbb{S}^{2}$. According to standard black hole
perturbation theory (see, for example, [\cite{chandrasekhar_mathematical_1998,frolov_black_1998,price_developments_2007}]), and as developed at greater length in Section 3.3, it is possible to choose
a gauge so that the polar and axial parts of perturbations to this
metric are encoded in a single gauge-invariant variable each. In particular,
they are given respectively by 
\begin{equation}
\Phi_{\left(\pm\right)}=\frac{1}{r}\sum_{l=0}^{\infty}\sum_{m=-l}^{l}Y^{lm}\left(\theta,\phi\right)\Psi_{\left(\pm\right)}^{lm}\left(t,r\right),\label{eq:2BP_perturbations}
\end{equation}
where $Y^{lm}$ are spherical harmonics and $\Psi_{(\pm)}^{lm}$ are
called, respectively, the Zerilli and Regge-Wheeler master functions,
which satisfy known wave-like equations and from which the perturbations
to $\bm{g}$ can be reconstructed. In [\cite{jezierski_energy_1999}], the symplectic form of the
reduced phase space $\mathscr{S}$ for such spacetimes is computed:
\begin{equation}
\bm{\Omega}|_{\mathscr{S}}=\sum_{\varsigma=\pm}\int_{\Sigma}{\rm d}^{3}x\,\delta\Upsilon_{\left(\varsigma\right)}\wedge\mathbb{D}\delta\Phi_{\left(\varsigma\right)},\label{eq:2BP_symplectic_form}
\end{equation}
where $\Upsilon_{(\pm)}=[r^{2}\sin\theta/f(r)]\dot{\Phi}_{(\pm)}$,
and $\mathbb{D}=\Delta_{\bm{\sigma}}^{-1}(\Delta_{\bm{\sigma}}+2)^{-1}$
where $\Delta_{\bm{\sigma}}$ is the Laplace operator on $\mathbb{S}^{2}$. See Section 3.3. for more details.

The work [\cite{jezierski_energy_1999}] where this symplectic form [Eq.~\eqref{eq:2BP_symplectic_form}]
was derived simply uses it to define and formulate conservation laws
for energy and angular momentum in perturbed SD spacetimes.
It does not, however, address the question of the total measure of
$\mathscr{S}$. We will now show that the (reduced) phase space measure
$\mu\left(\mathscr{S}\right)=\int_{\mathscr{S}}\bm{\Omega}|_{\mathscr{S}}$
for such spacetimes in fact diverges, preventing any argument based
on phase space compactness for the non-existence of entropy production.

Inserting the definitions of the different variables and suppressing
for the moment the coordinate dependence of the spherical harmonics
and master functions, we have 
\begin{align}
\mu\left(\mathscr{S}\right)=\, & \sum_{\varsigma=\pm}\int_{\mathscr{S}}\int_{\Sigma}{\rm d}^{3}x\,\delta\Upsilon_{\left(\varsigma\right)}\wedge\mathbb{D}\delta\Phi_{\left(\varsigma\right)}\label{eq:2BP_measure_1}\\
=\, & \sum_{\varsigma=\pm}\int_{\mathscr{S}}\int_{\Sigma}{\rm d}^{3}x\,\delta\left(\frac{r^{2}\sin\theta}{f\left(r\right)}\dot{\Phi}_{\left(\varsigma\right)}\right)\wedge\mathbb{D}\delta\Phi_{\left(\varsigma\right)}\label{eq:2BP_measure_2}\\
=\, & \sum_{\varsigma=\pm}\int_{\mathscr{S}}\int_{\Sigma}{\rm d}^{3}x\,\delta\left(\frac{r\sin\theta}{f\left(r\right)}\sum_{l,m}Y^{lm}\dot{\Psi}_{\left(\varsigma\right)}^{lm}\right)\wedge\mathbb{D}\delta\left(\frac{1}{r}\sum_{l',m'}Y^{l'm'}\Psi_{\left(\varsigma\right)}^{l'm'}\right).\label{eq:2BP_measure_3}
\end{align}
Now using the fact that the functional exterior derivative acts only
on the master functions and the operator $\mathbb{D}$ only on the
spherical harmonics, we can write this as 
\begin{align}
\mu\left(\mathscr{S}\right)=\, & \sum_{\varsigma=\pm}\int_{\mathscr{S}}\int_{\Sigma}{\rm d}^{3}x\,\left(\frac{r\sin\theta}{f\left(r\right)}\sum_{l,m}Y^{lm}\delta\dot{\Psi}_{\left(\varsigma\right)}^{lm}\right)\wedge\left(\frac{1}{r}\sum_{l',m'}\left(\mathbb{D}Y^{l'm'}\right)\delta\Psi_{\left(\varsigma\right)}^{l'm'}\right)\label{eq:2BP_measure_4}\\
=\, & \sum_{\varsigma=\pm}\sum_{l,l',m,m'}\int_{\mathscr{S}}\int_{\Sigma}{\rm d}^{3}x\,\left(\frac{r\sin\theta}{f\left(r\right)}Y^{lm}\frac{1}{r}\mathbb{D}Y^{l'm'}\right)\delta\dot{\Psi}_{\left(\varsigma\right)}^{lm}\wedge\delta\Psi_{\left(\varsigma\right)}^{l'm'}.\label{eq:2BP_measure_5}
\end{align}
Writing the Cauchy surface integral in terms of coordinates and collecting
terms, 
\begin{align}
\mu\left(\mathscr{S}\right)=\, & \sum_{\varsigma=\pm}\sum_{l,l',m,m'}\int_{\mathscr{S}}\int_{2M}^{\infty}{\rm d}r\int_{\mathbb{S}^{2}}{\rm d}\theta{\rm d}\phi\,\frac{1}{f\left(r\right)}\left[\left(\sin\theta\right)Y^{lm}\mathbb{D}Y^{l'm'}\right]\delta\dot{\Psi}_{\left(\varsigma\right)}^{lm}\wedge\delta\Psi_{\left(\varsigma\right)}^{l'm'}\label{eq:2BP_measure_6}\\
=\, & \sum_{\varsigma=\pm}\sum_{l,l',m,m'}\int_{\mathscr{S}}\left[\int_{\mathbb{S}^{2}}{\rm d}\theta{\rm d}\phi\,\left(\sin\theta\right)Y^{lm}\mathbb{D}Y^{l'm'}\right]\int_{2M}^{\infty}\frac{{\rm d}r}{f\left(r\right)}\delta\dot{\Psi}_{\left(\varsigma\right)}^{lm}\wedge\delta\Psi_{\left(\varsigma\right)}^{l'm'}\label{eq:2BP_measure_7}\\
=\, & \sum_{\varsigma=\pm}\sum_{l,l',m,m'}A^{ll'mm'}\int_{\mathscr{S}}\int_{2M}^{\infty}\frac{{\rm d}r}{f\left(r\right)}\delta\dot{\Psi}_{\left(\varsigma\right)}^{lm}\wedge\delta\Psi_{\left(\varsigma\right)}^{l'm'},\label{eq:2BP_measure_8}
\end{align}
where $A^{ll'mm'}=\int_{\mathbb{S}^{2}}{\rm d}\theta{\rm d}\phi\,(\sin\theta)Y^{lm}\mathbb{D}Y^{l'm'}$
is a finite integral involving only the spherical harmonics. Restoring
the arguments of the master functions, and recalling that the meaning
of $\delta f(t,r)$ (for any function $f$) is simply that of a one-form
on the phase space at $(t,r)$ in spacetime, we can write from the
above [Eq.~\eqref{eq:2BP_measure_8}]: 
\begin{align}
\mu\left(\mathscr{S}\right)=\, & \sum_{\varsigma=\pm}\sum_{l,l',m,m'}A^{ll'mm'}\int_{\mathscr{S}}\int_{2M}^{\infty}\frac{{\rm d}r}{f\left(r\right)}\left[\delta\dot{\Psi}_{\left(\varsigma\right)}^{lm}\left(t,r\right)\wedge\delta\Psi_{\left(\varsigma\right)}^{l'm'}\left(t,r\right)\right]\label{eq:2BP_measure_9}\\
\geq\, & \sum_{\varsigma=\pm}\sum_{l,l',m,m'}A^{ll'mm'}\int_{\mathscr{S}}\int_{2M}^{\infty}\frac{{\rm d}r}{f\left(r\right)}\left[\min_{\bar{r}\in[2M,\infty)}\delta\dot{\Psi}_{\left(\varsigma\right)}^{lm}\left(t,\bar{r}\right)\wedge\delta\Psi_{\left(\varsigma\right)}^{l'm'}\left(t,\bar{r}\right)\right]\label{eq:2BP_measure_10}\\
=\, & \sum_{\varsigma=\pm}\sum_{l,l',m,m'}A^{ll'mm'}\left[\int_{\mathscr{S}}\min_{\bar{r}\in[2M,\infty)}\delta\dot{\Psi}_{\left(\varsigma\right)}^{lm}\left(t,\bar{r}\right)\wedge\delta\Psi_{\left(\varsigma\right)}^{l'm'}\left(t,\bar{r}\right)\right]\int_{2M}^{\infty}\frac{{\rm d}r}{f\left(r\right)}\label{eq:2BP_measure_11}\\
=\, & \left\{ \sum_{\varsigma=\pm}\sum_{l,l',m,m'}\!\!A^{ll'mm'}\!\left[\min_{\bar{r}\in[2M,\infty)}\int_{\mathscr{S}}\delta\dot{\Psi}_{\left(\varsigma\right)}^{lm}\left(t,\bar{r}\right)\wedge\delta\Psi_{\left(\varsigma\right)}^{l'm'}\left(t,\bar{r}\right)\right]\!\right\} \left[\int_{2M}^{\infty}\frac{{\rm d}r}{f\left(r\right)}\right].\label{eq:2BP_measure_12}
\end{align}
The phase space integral $\int_{\mathscr{S}}\delta\dot{\Psi}_{(\varsigma)}^{lm}(t,\bar{r})\wedge\delta\Psi_{(\varsigma)}^{l'm'}(t,\bar{r})$
is finite but nonzero even when minimised over $\bar{r}$, because
for any nontrivial solutions of the master functions, there will be
no point in spacetime where they will always be vanishing (for all
time). Thus (assuming that the $l,l',m,m'$ sums are convergent),
everything in the curly bracket in the last line above [Eq.~\eqref{eq:2BP_measure_12}]
is nonzero but finite. However, it multiplies $\int_{2M}^{\infty}{\rm d}r/f(r)$
which diverges (at both integration limits). Hence, $\mu(\mathscr{S})$
diverges for such spacetimes.

We can make a few remarks. Firstly, one might be concerned in the above argument, specifically in the
last line [Eq.~\eqref{eq:2BP_measure_12}], about what might happen
in the asymptotic limit of the phase space integral: in other words,
it maybe the case (i) that $\min_{\bar{r}\in[2M,\infty)}\int_{\mathscr{S}}\delta\dot{\Psi}_{(\varsigma)}^{lm}(t,\bar{r})\wedge\delta\Psi_{(\varsigma)}^{l'm'}(t,\bar{r})$
could turn out to be $\lim_{r\rightarrow\infty}\int_{\mathscr{S}}\delta\dot{\Psi}_{(\varsigma)}^{lm}(t,r)\wedge\delta\Psi_{(\varsigma)}^{l'm'}(t,r)$;
and, if so, one might naively worry (ii) that the latter vanishes
due to asymptotic decay properties of the master functions. This will
actually not happen. To see why, suppose (i) is true. The master functions
must obey outgoing boundary conditions at spatial infinity, \textit{i.e.} $0=[\partial_{t}+f(r)\partial_{r}]\Psi_{(\varsigma)}^{lm}$
as $r\rightarrow\infty$. Hence we have
\begin{align}
 & \lim_{r\rightarrow\infty}\int_{\mathscr{S}}\delta\dot{\Psi}_{\left(\varsigma\right)}^{lm}\wedge\delta\Psi_{\left(\varsigma\right)}^{l'm'}\nonumber\\
= & \lim_{r\rightarrow\infty}\int_{\mathscr{S}}\delta\left(-v\partial_{r}\Psi_{\left(\varsigma\right)}^{lm}\right)\wedge\delta\Psi_{\left(\varsigma\right)}^{l'm'}\nonumber\\
= & -\int_{\mathscr{S}}\lim_{r\rightarrow\infty}\delta\left(\partial_{r}\Psi_{\left(\varsigma\right)}^{lm}\right)\wedge\delta\Psi_{\left(\varsigma\right)}^{l'm'},\label{eq:2BP_lim}
\end{align}
which is nonzero, because the vanishing of the master functions and
their radial partials at spatial infinity for all time corresponds
only to trivial solutions. Therefore, we have that $\min_{\bar{r}\in[2M,\infty)}\int_{\mathscr{S}}\delta\dot{\Psi}_{(\varsigma)}^{lm}(t,\bar{r})\wedge\delta\Psi_{(\varsigma)}^{l'm'}(t,\bar{r})$
is always nonzero for nontrivial solutions.

Secondly, if the two-body system in this framework is an extreme-mass-ratio
inspiral, \textit{i.e.} the mass of the orbiting body, or ``particle'', is
orders of magnitude smaller than that of the larger one, and the former
is modeled using a stress-energy-momentum tensor with support only
on its worldline, then it is known that $\Psi_{(\pm)}^{lm}(t,r)$
has a discontinuity at the particle location, and thus, $\dot{\Psi}_{(\pm)}^{lm}(t,r)$
has a divergence there. Hence, the integral over $\mathscr{S}$ even
before our inequality above [Eq.~\eqref{eq:2BP_measure_9}] is
already divergent due to the divergence of $\dot{\Psi}_{(\pm)}^{lm}(t,r)$
in the integrand. However, given that such an approach to describing
these systems (\textit{i.e.} having a stress-energy-momentum tensor of the
particle with a delta distribution) is only an idealization, we regard
the conclusion that $\mu(\mathscr{S})$ diverges as more convincing based on
our earlier argument, which is valid in general---that is, even for possible descriptions of the smaller body that may be more realistic than that using delta distributions.

\section{\label{sec:Conclusions}Conclusions}

We have proven that there does not exist a monotonically increasing function of phase space---which may be identified as (what we have referred to as a ``mechanical'' notion of) entropy---in classical mechanics with $N$ degrees of freedom for certain classes of Hamiltonians, as well as in some (classical) matter field theories in curved (nondynamical) spacetime, \textit{viz.} for standard scalar and electromagnetic fields. To do this, we have followed the procedure for the proof sketched by [\cite{poincare_sur_1889}] (what we have dubbed the \textsl{perturbative} approach), and we have here carried it out in full rigour for classical mechanics and extended it via similar techniques to field theories. What is noteworthy about this perturbative proof---counter (to our knowledge) to all other well-known proofs for the non-existence of entropy (in the ``mechanical'' sense) in classical canonical theories---is that it assumes nothing about the topology of the phase space; in other words, the phase space can be \textsl{non-compact}. Essentially, it relies only on curvature properties (in phase space) of the Hamiltonian of the canonical theory being considered. We have explicated these properties in the case of classical mechanics, and have assumed standard ones for the particular (curved spacetime) matter field theories we have investigated. It would be of interest for future work to determine, in the case of the former, whether they can be made less restrictive (than what we have required for our proof, which thus omits some classes of Hamiltonians of interest such as that for the the gravitational two-body problem), and in the case of the latter, whether they can be generalized or extended to broader classes of field theories. Indeed, it would be in general an interesting question to determine not only the necessary but also---\textsl{if possible}---the sufficient conditions that a Hamiltonian of a generic canonical theory needs to satisfy in order for this theorem to be applicable, \textit{i.e.} in order to preclude ``mechanical'' entropy production. We have seen that it is precisely the curvature properties of the vacuum Hamiltonian of general relativity that prevent this method of proof from being extended thereto, where in fact one does expect (some version of) the second law of thermodynamics to hold.

Topological properties of the phase space can also entail the non-existence of ``mechanical'' entropy, as per the more standard and already well-understood proofs in classical mechanics where the phase space is assumed to be compact [\cite{poincare_sur_1890,olsen_classical_1993}]. However, even for non-gravitational canonical theories this assumption might be too restrictive, and for general relativity, it is believed that in general it is not the case. This renders any of these topological proofs inapplicable in the case of the latter, and moreover, it also significantly complicates any attempt to formulate a sensible ``statistical'' notion of (gravitational) entropy due to the concordant problems in working with finite probabilities (of phase space properties). These must ultimately be overcome (via some regularization procedure or cutoff argument) for establishing a connection between a ``statistical'' and ``mechanical'' entropy in general relativity. While we still lack any consensus on how to define the latter, it may be hoped that in the future, the generic validity of a (general relativistic) second law may be demonstrated on the basis of (perhaps curvature related) properties of the gravitational Hamiltonian---which in turn may enter into a statistical mechanics type definition of gravitational entropy in terms of some suitably defined partition function. In this regard, older work based on field-theoretic approaches [\cite{horwitz_steepest_1983}] and more recent developments such as proposals to relate entropy with a Noether charge (specifically, the Noether invariant associated with an infinitesimal time translation) in classical mechanics [\cite{sasa_thermodynamic_2016}] may provide fruitful hints.

A clear situation in which we anticipate entropy production in general relativity, unlike in classical mechanics, is the gravitational two-body problem. For the latter, as we have discussed, the $N$-body problem actually does also exhibit features of entropy production. We have here shown explicitly that the phase space of perturbed Schwarzschild-Droste spacetimes is non-compact (even without the assumption of self-force). This means that the topological proofs are here inapplicable (but also, on the other hand, so is the ``no-return'' theorem for compact Cauchy surfaces, which by itself cannot be used in this case to understand the non-recurrence of phase space orbits). It is hoped that once a generally agreed upon definition of gravitational entropy is established, one would not only be able to use it to compute the entropy of two-body systems, but also to demonstrate that it should obey the second law (\textit{i.e.} that it should be monotonically increasing in time). In the long run, an interesting problem to investigate is whether an entropy change, once defined and associated to motion in a Lagrangian formulation, could determine the trajectory of a massive and radiating body, moving in a gravitational field.





\chapter[The Motion of Localized Sources in General Relativity]{The Motion of Localized Sources in General Relativity:\\
Gravitational Self-Force from Quasilocal Conservation Laws\label{5-motion}}
\newrefsegment

\subsection*{Chapter summary}

This chapter is based on the preprint [\cite{oltean_motion_2019}].
 
An idealized ``test'' object in general relativity moves along a geodesic. However, if the object has a finite mass, this will create additional curvature in the spacetime, causing it to deviate from geodesic motion. If the mass is nonetheless sufficiently small, such an effect is usually treated perturbatively and is known as the gravitational self-force due to the object. This issue is still an open problem in gravitational physics today, motivated not only by basic foundational interest, but also by the need for its direct application in gravitational-wave astronomy. In particular, the observation of extreme-mass-ratio inspirals by the future space-based detector LISA will rely crucially on an accurate modeling of the self-force driving the orbital evolution and gravitational wave emission of such systems.
 
In this chapter, we present a novel derivation, based on conservation laws, of the basic equations of motion for this problem. They are formulated with the use of a quasilocal (rather than matter) stress-energy-momentum tensor---in particular, the Brown-York tensor---so as to capture gravitational effects in the momentum flux of the object, including the self-force. Our formulation and resulting equations of motion are independent of the choice of the perturbative gauge. We show that, in addition to the usual gravitational self-force term, they also lead to an additional ``self-pressure'' force not found in previous analyses, and also that our results correctly recover known formulas under appropriate conditions. Our approach thus offers a fresh geometrical picture from which to understand the self-force fundamentally, and potentially useful new avenues for computing it practically.
 
We begin in Section \ref{sec:5-Intro} with a brief introductory discussion on the idea of using conservation law approaches for the self-force problem generally, that is of understanding and computing the self-force as a momentum change or flux. While this has proven successful in the past for the electromagnetic self-force problem, such an approach has, up to this work, not been attempted in general in the gravitational case. This mainly has to do with the subtleties involved in properly defining notions of gravitational energy-momentum. These are concepts which do not make sense locally in relativistic physics (\textit{i.e.} as volume densities), and so the typical solution---as in canonical general relativity---is to define them quasilocally (\textit{i.e.} as boundary densities).
 
In Section \ref{sec:qf}, we review the general quasilocal energy-momentum conservation laws for general relativity used in this chapter. These laws have been obtained in recent work based on the Brown-York tensor, account for both gravitational as well as matter fluxes, and are valid in any arbitrary spacetime. They are constructed with the use of a concept called a quasilocal frame: a topological two-sphere of observers tracing out the worldtube boundary of the history of a finite spatial volume (that is, the finite system the fluxes of which we are studying).
 
In Section \ref{sec:general-analysis}, we prove that the correction to the momentum flux of any small spatial region due to any metric perturbations in any spacetime in general relativity always contains the known form of the gravitational self-force. Our analysis also reveals a new term, not found in previous analyses and in principle equally dominant in general, namely one arising from a “self-pressure” effect with no analogy in Newtonian gravity. The appearance of these terms as corrections to the motion is independent of what actually sources the metric perturbations upon which they depend; rather than the “mass” of the small moving object itself, what seems to be fundamentally responsible for self-force effects in our analysis is the mass (or energy) and pressure of the spacetime vacuum.
 
In Section \ref{sec:gralla-wald-analysis} we proceed to apply our analysis to a concrete self-force analysis actually used for computations, that is a specific choice of a perturbative family of spacetimes designed to describe the correction to the motion of a small object. We work with the rigorous approach of Gralla and Wald, and we show how under appropriate conditions our analysis recovers equations of motion comparable with theirs.
 
Finally, Section \ref{sec:concl} offers some conclusions and outlook to future work.

\subsection*{La moció de les fonts localitzades en la relativitat general \normalfont{(chapter summary translation in Catalan)}}
 
Aquest capítol es basa en el preprint [\cite{oltean_motion_2019}].
 
Un objecte de ``prova'' idealitzat en la relativitat general es mou al llarg d'una geodèsica. Tanmateix, si l'objecte té una massa finita, això crearà una corbatura addicional en l'espai-temps, fent que es desviï del moviment geodèsic. Si la massa és tot i això prou petita, aquest efecte se sol tractar de manera pertorbadora i es coneix com a força pròpia gravitacional a causa de l'objecte. Aquesta qüestió continua sent un problema obert en la física gravitatòria actual, motivada no només per l’interès fonamental bàsic, sinó també per la necessitat de la seva aplicació directa en l’astronomia d’ones gravitacionals. En particular, l'observació de caigudes en espiral amb raó de masses extrema per part del futur detector LISA basat en l'espai es basarà crucialment en un modelat precís de la força pròpia impulsant l'evolució orbital i l'emissió d'ones gravitacionals (crec que aquesta és la forma més correcta, tot in que són sinònimes, juntament amb gravitatòries) d'aquests sistemes.
 
En aquest capítol, es presenta una nova derivació, basada en lleis de conservació, de les equacions bàsiques de moviment d’aquest problema. Es formulen amb l'ús d'un tensor de tensió-energia quasilocal (en lloc de material), en particular, el tensor de Brown-York, per tal de captar efectes gravitacionals en el flux de moment de l'objecte, inclòs la força pròpia. La nostra formulació i les equacions de moviment resultants són independents de l’elecció de la mesura pertorbativa. Mostrem que, a més del terme de la força pròpia gravitacional habitual, també condueixen a una força de “pressió pròpia” addicional que no es va trobar en anàlisis anteriors, i també que els nostres resultats recuperen correctament les fórmules conegudes en condicions adequades. El nostre treball ofereix així una nova imatge geomètrica a partir de la qual es pot entendre fonamentalment la força pròpia, i possibles noves vies potencialment útils per a computar-la pràcticament.
 
Comencem a la secció \ref{sec:5-Intro} amb una breu discussió introductòria sobre la idea d’utilitzar els mètodes de lleis de conservació per al problema de la força pròpia en general, és a dir, comprendre i calcular la força pròpia com a canvi o flux d’impuls. Si bé en el passat això ha tingut èxit pel problema de la força pròpia electromagnètica, un anàlisi d’aquest tipus no s’ha intentat, fins a aquest treball, en general en el cas gravitatori. Això té a veure principalment amb les subtileses relacionades amb la definició adequada de les nocions d’energia-moment gravitatòria. Es tracta de conceptes que no tenen sentit localment en la física relativista (és a dir, com a densitats de volum), i per tant, la solució típica - com en la relativitat general canònica - és definir-los de forma quasilocal (és a dir, com a densitats de frontera).
 
A la secció \ref{sec:qf}, revisem les lleis quasilocals generals de conservació d’energia-moment per a la relativitat general utilitzades en aquest capítol. Aquestes lleis s'han obtingut en treballs recents basats en el tensor de Brown-York, tant per a fluxos gravitacionals com per a matèries, i són vàlids en qualsevol espai-temps arbitrari. Es construeixen amb l'ús d'un concepte conegut com a sistema de referència quasilocal: una esfera topològica bidimensional d'observadors que traça la frontera de la història d'un volum espacial finit (és a dir, el sistema finit els fluxos del qual estem estudiant).
 
A la secció \ref{sec:general-analysis}, demostrem que la correcció al flux d’impuls de qualsevol petita regió espacial a causa de pertorbacions mètriques en qualsevol espai-temps en la relativitat general sempre conté la forma coneguda de la força pròpia gravitacional. La nostra anàlisi també revela un nou terme, que no es troba en anàlisis anteriors i, en principi, és igualment dominant en general, és a dir, un efecte de de “pressió pròpia” sense analogia en la gravetat newtoniana. L’aparició d’aquests termes com a correccions al moviment és independent del que realment provoca les pertorbacions mètriques de les quals depenen; més que la ``massa'' del petit objecte en moviment propi, el que sembla ser fonamentalment responsable dels efectes de la força pròpia en la nostra anàlisi és la massa (o energia) i la pressió del buit de l’espai-temps.
 
A la secció \ref{sec:gralla-wald-analysis} procedim a aplicar la nostra anàlisi a una anàlisi concreta de força pròpia utilitzada realment per a càlculs, és a dir, una elecció específica d’una família pertorbadora d’espais-temps dissenyada per descriure la correcció al moviment d’un objecte petit. En particular, treballem amb el formalisme rigorós de Gralla i Wald, i mostrem com en condicions adequades la nostra anàlisi recupera equacions de moviment comparables a les seves.

Finalment, la secció \ref{sec:concl} ofereix algunes conclusions i perspectives per a futurs treballs.

\subsection*{Le mouvement des sources localisées dans la relativité générale  \normalfont{(chapter summary translation in French)}}
 
Ce chapitre est basé sur le pré-impression  [\cite{oltean_motion_2019}].
 
Un objet de « test » idéalisé dans la relativité générale se déplace le long d'une géodésique. Cependant, si l'objet a une masse finie, cela créera une courbure supplémentaire dans l'espace-temps, le faisant s'écarter du mouvement géodésique. Si la masse est néanmoins suffisamment petite, un tel effet est généralement traité de manière perturbative et il est connu comme la force propre gravitationnelle à cause de l'objet. Cette question est toujours un problème ouvert dans la physique gravitationnelle aujourd'hui, motivée non seulement par l’intérêt fondamental, mais également par la nécessité de son application directe dans l'astronomie des ondes gravitationnelles. En particulier, l'observation d’inspirals avec quotients extrêmes des masses par le futur détecteur spatial LISA reposera de manière cruciale sur une modélisation précise de la force propre à l'origine de l'évolution orbitale et de l'émission des ondes gravitationnelles de tels systèmes.

Dans ce chapitre, nous présentons une nouvelle dérivation, basée sur des lois de conservation, des équations de base du mouvement pour ce problème. Ils sont formulés avec l’utilisation d’un tenseur énergie-impulsion quasi-local (plutôt que de la matière) - en particulier, le tenseur de Brown-York - afin de capturer les effets gravitationnels dans le flux d’impulsion de l’objet, y compris la force propre. Notre formulation et les équations de mouvement résultantes sont indépendantes du choix de la jauge perturbative. Nous montrons que, en plus du terme habituel de la force propre gravitationnelle, ils conduisent également à une force de « pression propre » supplémentaire, pas trouvée dans les analyses précédentes et que nos résultats récupèrent correctement les formules connues dans des conditions appropriées. Notre analyse offre donc une nouvelle image géométrique à partir de laquelle on peut comprendre fondamentalement la force propre et de nouvelles voies potentiellement utiles pour la calculer de manière pratique.

Nous commençons à la section \ref{sec:5-Intro} par une brève discussion introductive sur l’idée d’appliquer les lois de conservation au problème de la force propre en général, c’est-à-dire de la compréhension et du calcul de la force propre comme un changement ou un flux en la quantité de mouvement. Bien que cela eût du succès au passé pour le problème de la force propre électromagnétique, une telle analyse n'a jusqu'à présent pas été tentée de manière générale dans le cas de la gravitation. Cela concerne principalement les subtilités impliquées dans la définition correcte des notions d'énergie-impulsion gravitationnelle. Ce sont des concepts qui n’ont pas de sens local dans la physique relativiste (c’est-à-dire, comme densités de volume) et la solution typique - comme dans la relativité générale canonique - consiste à les définir de manière quasilocale (c’est-à-dire, comme densités de frontière).

Dans la section \ref{sec:qf}, nous passons en revue les lois générales quasilocales de conservation d'énergie-impulsion pour la relativité générale utilisées dans ce chapitre. Ces lois ont été obtenues dans des travaux récents basés sur le tenseur de Brown-York, tiennent compte à la fois des flux gravitationnels et des flux de matière, et sont valables dans tout espace-temps arbitraire. Ils sont construits à l'aide d'un concept appelé le référentiel quasilocal (\textit{quasilocal frame}): une sphère topologique bidimensionelle d'observateurs traçant la frontière de l'histoire d'un volume spatial fini (c'est-à-dire du système fini dont nous étudions les flux).

Dans la section \ref{sec:general-analysis}, nous montrons que la correction du flux de la quantité de mouvement de toute petite région spatiale à cause des perturbations métriques dans quelque espace-temps de la relativité générale contient toujours la forme connue de la force propre gravitationnelle. Notre analyse révèle également un nouveau terme, pas retrouvé dans les analyses précédentes et en principe tout aussi dominant en général, à savoir un effet de « pression propre » sans analogie dans la gravité newtonienne. L'apparition de ces termes en tant que corrections du mouvement est indépendante de la source des perturbations métriques dont ils dépendent. Plutôt que la « masse » du petit objet en mouvement lui-même, ce qui semble être fondamentalement responsable pour les effets de force propre, dans notre analyse est la masse (ou énergie) et la pression du vide de l’espace-temps.

Dans la section \ref{sec:gralla-wald-analysis}, nous appliquons notre analyse à une formulation concrète de la force propre réellement utilisée pour les calculs, c’est-à-dire un choix spécifique d’une famille des espaces-temps perturbatifs conçue pour décrire la correction du mouvement d’un petit objet. Nous travaillons en particulier avec le formalisme rigoureux de Gralla et Wald et nous montrons comment, dans des conditions appropriées, notre analyse récupère équations de mouvement comparables aux leurs.

Enfin, la section \ref{sec:concl} propose quelques conclusions et perspectives pour les travaux futurs.

\section{\label{sec:5-Intro}Introduction: the self-force problem via conservation laws}

\begin{figure}
\begin{centering}
\includegraphics[scale=0.8]{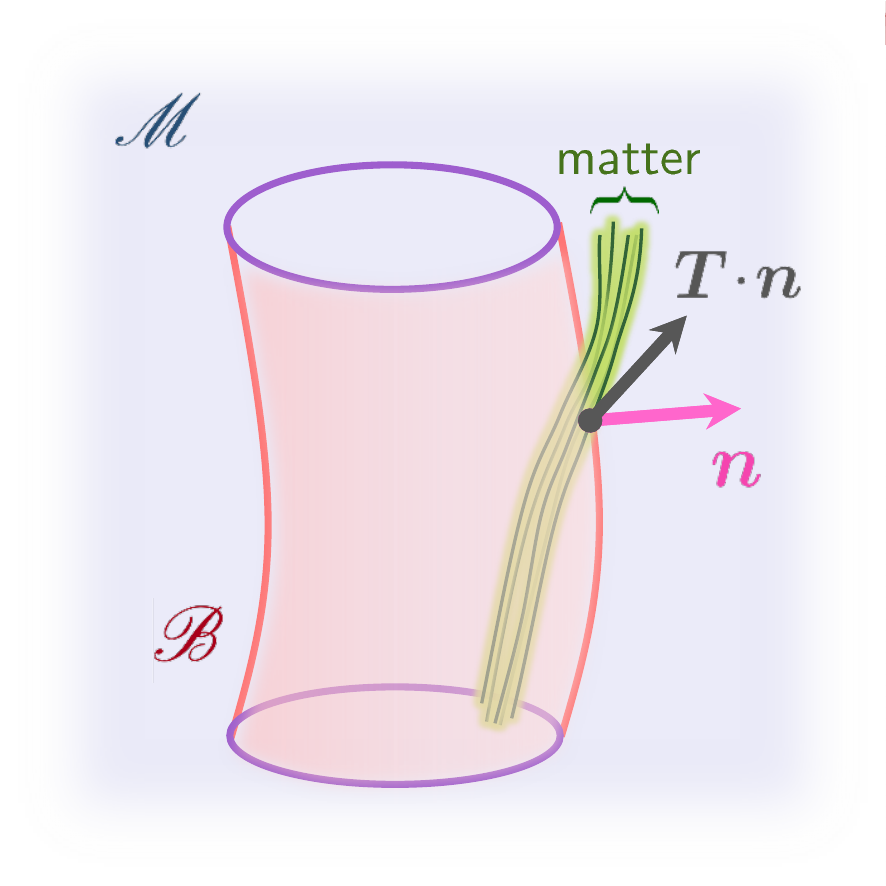}
\par\end{centering}
\caption{A worldtube boundary $\mathscr{B}$
(topologically $\mathbb{R}\times\mathbb{S}^{2}$) in $\mathscr{M}$, with (outward-directed) unit normal
$n^{a}$. The change in matter four-momentum between two constant time slices of this worldtube is given by the flux of the normal projection (in one index) of the matter stress-energy-momentum tensor $T_{ab}$ through the portion of $\mathscr{B}$ bounded thereby.\label{figure-worldtube-boundary}}
\end{figure}

The idea of using conservation laws for tackling the self-force problem
was appreciated and promptly exploited quite early on for the electromagnetic
self-force. In the 1930s, [\cite{dirac_classical_1938}] was the
first to put forward such an analysis in flat spacetime, and later
on in 1960, [\cite{dewitt_radiation_1960}] extended
it to non-dynamically curved spacetimes\footnote{~By this, we mean spacetimes with non-flat but fixed metrics, which do not evolve dynamically (gravitationally) in response to the matter stress-energy-momentum present therein.}. In such approaches, it can
be shown\footnote{~See  [\cite{poisson_introduction_1999}] for a basic and more contemporary
presentation.} that the EoM for the electromagnetic self-force follows from local
conservation expressions of the form
\begin{equation}
\Delta P^{a}=\intop_{\Delta\mathscr{B}}\bm{\epsilon}^{\,}_{\mathscr{B}}T^{ab}n_{b}\,,\label{eq:local_cons}
\end{equation}
where the LHS expresses the flux of \emph{matter} four-momentum $P^{a}$ (associated with $T_{ab}$) between
the ``caps'' of (\textit{i.e.} closed spatial two-surfaces delimiting) a
portion (or ``time interval'') of a thin worldtube boundary $\mathscr{B}$
(topologically $\mathbb{R}\times\mathbb{S}^{2}$), with natural volume
form $\bm{\epsilon}^{\,}_{\mathscr{B}}$ and (outward-directed) unit normal
$n^{a}$ (see Figure~\ref{figure-worldtube-boundary}). In particular, one takes a time derivative of (\ref{eq:local_cons})
to obtain an EoM expressing the time rate of change of momentum in
the form of a closed spatial two-surface integral (by differentiating the worldtube boundary integral). For the electromagnetic
self-force problem, the introduction of an appropriate matter stress-energy-momentum tensor
$T_{ab}$ into Eq.~(\ref{eq:local_cons}) and a bit of subsequent argumentation
reduces the integral expression to the famous Lorentz-Dirac equation;
on a spatial three-slice in a Lorentz frame and in the absence of
external forces, for example, this simply reduces to $\dot{P}^{i}=\frac{2}{3}q^{2}\dot{a}^{i}$ for a charge $q$. Formulations of the scalar and electromagnetic self-forces using generalized Killing fields have more recently been put forward in [\cite{harte_self-forces_2008,harte_electromagnetic_2009}].

The success of conservation law approaches for formulating the electromagnetic
self-force in itself inspires hope that the same may be done
in the case of the gravitational self-force (GSF) problem. In particular,
Gralla's EoM (\ref{eq:intro_Gralla})
strongly hints at the possibility of understanding the RHS
not just as a mathematical (``angle averaging'') device, but as a \emph{true, physical
flux of gravitational momentum} arising from a consideration of conservation
expressions.

Nevertheless, to our knowledge, there has thus far been no proposed
general treatment of the GSF following such an approach. This may, in large
part, be conceivably attributed to the notorious conceptual difficulties
surrounding the very question of the basic formulation of conservation
laws in GR. Local conservation laws, along the lines of Eq.~(\ref{eq:local_cons}) that can
readily be used for electromagnetism, no longer make sense fundamentally once gravity
is treated as dynamical. The reason has a simple explanation in the
equivalence principle (see, \emph{e.g.}, Section 20.4 of [\cite{misner_gravitation_1973}]): one can always find a local frame of reference
with a vanishing local ``gravitational field'' (metric connection
coefficients), and hence a vanishing local ``gravitational energy-momentum'',
irrespective of how one might feel inclined to define the latter\footnote{~It is worth remarking here that, in a perturbative setting, an approach that is sometimes taken is to work with an ``effective'' local gravitational stress-energy-momentum tensor, defined as the RHS of a suitably-rearranged (first-order) Einstein equation. This is a common tactic often used for studying, for example, the energy-momentum of gravitational waves, with some applicable caveats (see, \textit{e.g.}, Chapter 35 of [\cite{misner_gravitation_1973}]). In fact, one of the first formulations of the gravitational self-force---in particular, the derivation of the MiSaTaQuWa EoM (\ref{eq:intro_MiSaTaQuWa}) presented in Section III of [\cite{mino_gravitational_1997}]---made use of the local conservation (vanishing of the spacetime divergence) of a suitably-defined local tensor of such a sort (in analogy with the approach of \cite{dewitt_radiation_1960} to the electromagnetic self-force). We elaborate in the remainder of this subsection and at greater length in Section \ref{sec:qf} on why such a notion of gravitational conservation principles, while demonstrably useful for operational computations in some situations, cannot in general be expected to capture the fundamentally quasilocal (boundary density) nature of gravitational energy-momentum. See [\cite{epp_momentum_2013}] for a detailed discussion and comparison between these two (local and quasilocal) views of gravitational energy-momentum.}.

A wide variety of approaches have been taken over the decades towards
formulating sensible notions of gravitational energy-momentum, with still no general
consensus among relativists today on which to qualify as ``the best'' [\cite{szabados_quasi-local_2004,jaramillo_mass_2011}]. Often the preference
for employing certain definitions over others may simply come down
to context or convenience, but in any case, there exist agreements
between the most typical definitions in various limits. A very common
feature among them is the idea of replacing a local notion of gravitational
energy-momentum, \textit{i.e.} energy-momentum as a volume density, with what is known as a \emph{quasilocal}
energy-momentum, \textit{i.e.} energy-momentum as a boundary density. The typical Hamiltonian definitions
of the (total) gravitational energy-momentum for an asymptotically-flat spacetime,
for example, are of such a form. Among the most commonly used generalizations
of these definitions to arbitrary (finite) spacetime regions was proposed
in the early 1990s by [\cite{brown_quasilocal_1993}], and follow from what is now eponymously known
as the \emph{Brown-York stress-energy-momentum tensor}. It is a quasilocal tensor, meaning it
is only defined on the boundary of an arbitrary spacetime region.
For example, using this, the total (matter plus gravitational) energy
inside a spatial volume is given up to a constant factor by
the closed two-surface (boundary) integral of the trace of the boundary extrinsic
curvature---precisely in agreement with the Hamiltonian definition of energy for the entire spacetime in the appropriate limit
(where the closed two-surface approaches a two-sphere
at asymptotically-flat spatial infinity) but, in principle, applicable
to any region in any spacetime. 

The formulation of general energy-momentum conservation laws in GR from the Brown-York tensor 
has been achieved in recent years with the use of a construction called \emph{quasilocal
frames} [\cite{epp_momentum_2013}], a concept first proposed in [\cite{epp_rigid_2009}]. Essentially, the idea is that it does not suffice to merely
specify, as in the local matter conservation laws of the form of Eq.~(\ref{eq:local_cons}), a worldtube boundary $\mathscr{B}$ (as an embedded submanifold of $\mathscr{M}$) the interior of which contains the system of interest, and
through which to measure the flux of gravitational energy-momentum. What is in
fact required is the specification of a \emph{congruence} making up this
worldtube boundary, \textit{i.e.} a two-parameter family of timelike worldlines
with some chosen four-velocity field representing the motion of a
topological two-sphere's worth of \emph{quasilocal observers}. We will motivate this construction in greater amplitude shortly, but the reason for needing it is basically to be able to meaningfully define ``time-time'' and ``time-space'' directions on $\mathscr{B}$ for our conservation laws. A congruence of this sort is
what is meant by a quasilocal frame. 

The enormous advantage in using these quasilocal conservation laws
over other approaches lies in the fact that they hold in any arbitrary
spacetime. Thus the existence of Killing vector fields---a typical
requirement in other conservation law formulations---is in no way
needed here.

This idea has been used successfully in a number of applications so far [\cite{mcgrath_quasilocal_2012,epp_existence_2012,epp_momentum_2013,mcgrath_post-newtonian_2014,mcgrath_rigid_2014,oltean_geoids_2016,oltean_geoids_2017}]. These include the resolution of a variation of Bell's spaceship paradox\footnote{~Proposed initially by [\cite{dewan_note_1959}] and later made popular by J.S. Bell's version [\cite{bell_how_1976}].} in which a box accelerates rigidly in a transverse,
uniform electric field [\cite{mcgrath_quasilocal_2012}], recovering under appropriate conditions the typical (but more limited) local matter conservation expressions of the form of Eq.~(\ref{eq:local_cons}) from the quasilocal ones [\cite{epp_momentum_2013}], application to post-Newtonian theory [\cite{mcgrath_post-newtonian_2014}] and to relativistic geodesy [\cite{oltean_geoids_2016,oltean_geoids_2017}]. 

A similar idea to quasilocal frames, called \emph{gravitational screens}, was proposed more recently in  [\cite{freidel_gravitational_2015,freidel_non-equilibrium_2015-1}]. There, the authors also make use of quasilocal ideas to develop conservation laws very similar in style and form to those obtained via quasilocal frames. A detailed comparison between these two approaches has thus far not been carried out, but it would be very interesting to do so in future work. In particular, the notion of gravitational screens has been motivated more from thermodynamic considerations, and similarly casting quasilocal frames in this language could prove quite fruitful. For example, just as these approaches have given us operational definitions of concepts like the ``energy-momentum in an arbitrary spacetime region'' (and not just for special cases such as an entire spacetime), they may help to do the same for concepts like ``entropy in an arbitrary spacetime region'' (and not just for known special cases such as a black hole).

\section{Setup: quasilocal conservation laws\label{sec:qf}}

Let $(\mathscr{M},\bm{g},\bm{\nabla})$ be any $(3+1)$-dimensional
spacetime such that, given any matter stress-energy-momentum
tensor $T_{ab}$, the Einstein equation, 
\begin{equation}
\bm{G}=\kappa\,\bm{T}\enspace\textrm{in}\enspace\mathscr{M}\,,\label{eq:Einstein_eqn}
\end{equation}
holds.  In what follows, we introduce the concept of quasilocal frames [\cite{epp_rigid_2009,epp_existence_2012,mcgrath_quasilocal_2012,epp_momentum_2013,mcgrath_post-newtonian_2014,mcgrath_rigid_2014,oltean_geoids_2016,oltean_geoids_2017}] and describe the basic steps for their construction, as well as the energy and momentum conservation
laws associated therewith.  In Subsection \ref{ssec:qf_heuristic} we offer an heuristic idea of quasilocal frames before proceeding in Subsection \ref{ssec:qf_math} to present the full mathematical construction. Then in Subsection \ref{ssec:qf_by} we motivate and discuss the quasilocal stress-energy-momentum tensor used in this work, that is, the Brown-York tensor. Finally in Subsection \ref{ssec:qf_cons_laws} we review the formulation of quasilocal conservation laws using these ingredients.

\subsection{Quasilocal frames: heuristic idea\label{ssec:qf_heuristic}}

Before we enter into the technical details, we would like to offer
a heuristic picture and motivation for defining the concept of quasilocal
frames.

We would like to show how the GSF arises from general-relativistic conservation laws. For this, we require
first the embedding into our spacetime $\mathscr{M}$ of a worldtube
boundary $\mathscr{B}\simeq\mathbb{R}\times\mathbb{S}^{2}$. The worldtube interior
contains the system the dynamics of which we are
interested in describing. In principle, such a $\mathscr{B}$ can
be completely specified by choosing an appropriate \emph{radial function}
$r(x)$ on $\mathscr{M}$ and setting it equal to a non-negative constant
(such that the $r(x)={\rm const.}>0$ Lorentzian slices of $\mathscr{M}$
have topology $\mathbb{R}\times\mathbb{S}^{2}$). This would be analogous
to defining a (Riemannian, with topology $\mathbb{R}^{3}$) Cauchy
surface by the constancy of a time function $t(x)$ on $\mathscr{M}$.

However, this does not quite suffice. As we have briefly argued in
the introduction (and will shortly elaborate upon in greater technicality),
the conservation laws appropriate to GR ought to be quasilocal in
form, that is, involving stress-energy-momentum as boundary (not volume)
densities. One may readily assume that the latter are defined by a
quasilocal stress-energy-momentum tensor living on $\mathscr{B}$,
which we denote—for the moment, generally—by $\tau_{ab}$. (Later
we give an explicit definition, namely that of the Brown-York tensor,
for $\bm{\tau}$.)

To construct conservation laws, then, one would need to project this
$\bm{\tau}$ into directions on $\mathscr{B}$, giving quantities
such as energy or momenta, and then to consider their flux through
a portion of $\mathscr{B}$ (an interval of time along the worldtube
boundary). But in this case, we have to make clear what is meant by
the energy (``time-time'') and momenta (``time-space'') components
of $\bm{\tau}$ within $\mathscr{B}$, the changes in which we are
interested in studying. For this reason, additional constructions
are required.

In particular, what we need is a \emph{congruence} of observers with
respect to which projections of $\bm{\tau}$ yield stress-energy-momentum
quantities. Since $\bm{\tau}$ is only defined on $\mathscr{B}$,
this therefore needs to be a two-parameter family of (timelike) worldlines
the union of which is $\mathscr{B}$ itself. This is analogous to
how the integral curves of a time flow vector field (as in canonical
GR) altogether constitute (``fill up'') the entire spacetime $\mathscr{M}$,
except that there we are dealing with a three- (rather than
two-) parameter family of timelike worldlines. 

We refer to any set of observers, the worldlines of which form a two-parameter
family constituting $\mathscr{B}\simeq\mathbb{R}\times\mathbb{S}^{2}$,
as \emph{quasilocal} observers. A specification of such a $2$-parameter
family, equivalent to specifying the unit four-velocity $u^{a}\in T\mathscr{B}$
of these observers (the integral curves of which ``trace out'' $\mathscr{B}$),
is what is meant by a quasilocal frame.

With this, we can now meaningfully talk about projections of $\bm{\tau}$
into directions on $\mathscr{B}$ as stress-energy-momentum quantities.
For example, $\tau_{\bm{u}\bm{u}}$ may appear immediately suggestible
as a definition for the (boundary) energy density. Indeed, later we
take precisely this definition, and we will furthermore see how momenta
(the basis of the GSF problem) can be defined as well.

\subsection{Quasilocal frames: mathematical construction\label{ssec:qf_math}}

\begin{figure}
\begin{centering}
\includegraphics[scale=0.8]{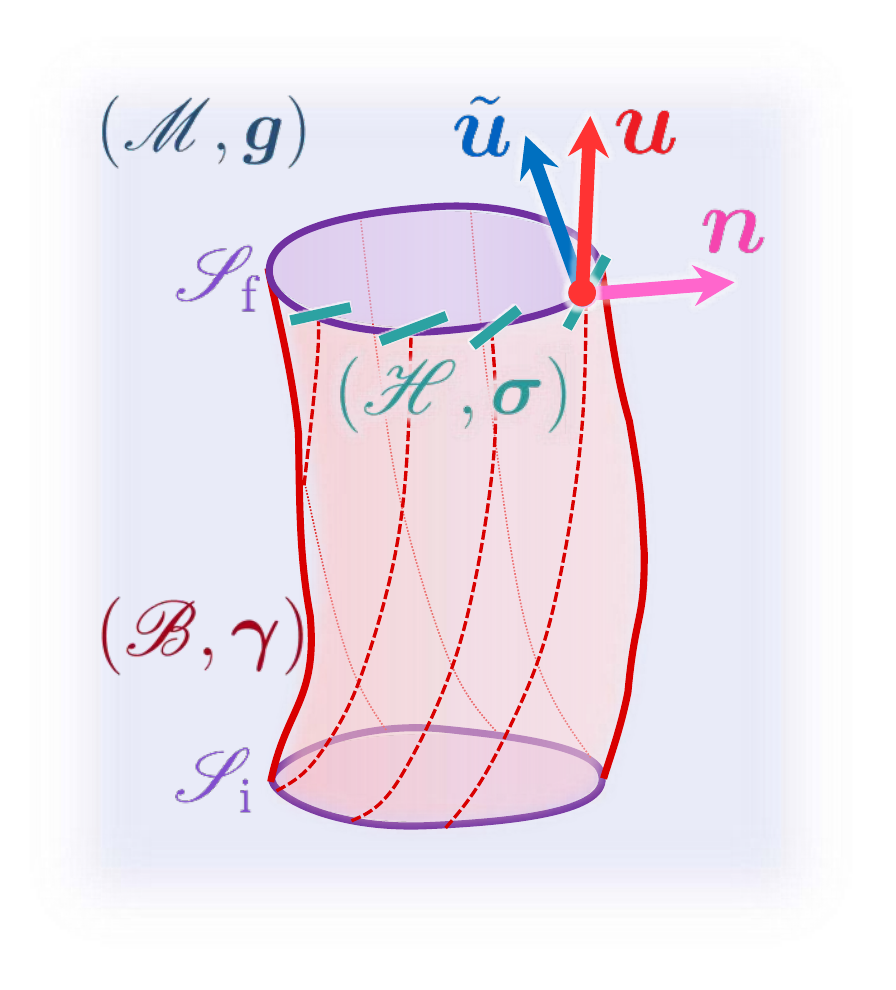}
\par\end{centering}
\caption{A portion of a quasilocal frame $(\mathscr{B};\bm{u})$ in a spacetime $\mathscr{M}$,
bounded by constant $t$ two-surfaces $\mathscr{S}_{\textrm{i}}$
and $\mathscr{S}_{\textrm{f}}$. In particular, $\mathscr{B}\simeq\mathbb{R}\times\mathbb{S}^{2}$ is the union of all integral curves (two-parameter family of timelike worldlines), depicted in the figure as dotted red lines, of the vector field $\bm{u}\in T\mathscr{B}$ which represents the unit four-velocity of quasilocal observers making up the congruence. The unit normal to $\mathscr{B}$ (in $\mathscr{M}$) is $\bm{n}$ and the normal to each constant $t$ slice $\mathscr{S}$ of $\mathscr{B}$ is $\tilde{\bm{u}}$ (not necessarily coincidental with $\bm{u}$). Finally, $\mathscr{H}$ (with induced metric $\bm{\sigma}$) is the two-dimensional subspace of $T\mathscr{B}$ consisting of the spatial vectors orthogonal to $\bm{u}$. Note that unlike $\mathscr{S}$, $\mathscr{H}$ need not be integrable (indicated in the figure by the failure of $\mathscr{H}$ to make a closed two-surface).\label{fig-qf}}
\end{figure}

Concordant with our discussion in the previous subsection, a quasilocal
frame (see Figure~\ref{fig-qf} for a graphical illustration of the construction) is defined as a two-parameter family of timelike worldlines
constituting the worldtube boundary (topologically $\mathbb{R}\times\mathbb{S}^{2}$)
of the history of a finite (closed) spatial three-volume in $\mathscr{M}$.
Let $u^{a}$ denote the timelike unit vector field tangent to these
worldlines. Such a congruence constitutes a submanifold of $\mathscr{M}$
that we call $\mathscr{B}\simeq\mathbb{R}\times\mathbb{S}^{2}$. Let
$n^{a}$ be the outward-pointing unit vector field normal to $\mathscr{B}$;
note that $\bm{n}$ is uniquely fixed once $\mathscr{B}$ is specified. There is
thus a Lorentzian metric $\bm{\gamma}$ (of signature $(-,+,+)$)
induced on $\mathscr{B}$, the components of which are given by 
\begin{equation}
\gamma_{ab}=g_{ab}-n_{a}n_{b}\,.\label{eq:gamma_ab}
\end{equation}
We denote the induced derivative operator compatible therewith by
$\bm{\mathcal{D}}$. To indicate that a topologically $\mathbb{R}\times\mathbb{S}^{2}$
submanifold $(\mathscr{B},\bm{\gamma},\bm{\mathcal{D}})$ of $\mathscr{M}$
is a quasilocal frame (that is to say, defined as a particular congruence
with four-velocity $\bm{u}$ as detailed above, and not just as an
embedded submanifold) in $\mathscr{M}$, we write $(\mathscr{B},\bm{\gamma},\bm{\mathcal{D}};\bm{u})$
or simply $(\mathscr{B};\bm{u})$.

Let $\mathscr{H}$ be the two-dimensional subspace of $T\mathscr{B}$ consisting of the ``spatial'' vectors orthogonal to $\bm{u}$. Let $\bm{\sigma}$ denote the two-dimensional (spatial) Riemannian
metric (of signature $(+,+)$) that projects tensor indices into $\mathscr{H}$,
and is induced on $\mathscr{B}$ by the choice of $\bm{u}$ (and thus
also $\bm{n}$), given by 
\begin{equation}
\sigma_{ab}=\gamma_{ab}+u_{a}u_{b}=g_{ab}-n_{a}n_{b}+u_{a}u_{b}\,.\label{eq:sigma_ab}
\end{equation}
The induced derivative operator compatible with $\bm{\sigma}$ is
denoted by $\bm{D}$. Let $\{x^{\mathfrak{i}}\}_{\mathfrak{i}=1}^{2}$
(written using Fraktur indices from the middle third of the Latin
alphabet) be spatial coordinates on $\mathscr{B}$ that label
the worldlines of the observers, and let $t$ be a time coordinate
on $\mathscr{B}$ such that surfaces of constant $t$, to which there
exists a unit normal vector that we denote by $\tilde{u}^{a}\in T\mathscr{B}$, foliate
$\mathscr{B}$ by closed spatial two-surfaces $\mathscr{S}$ (with
topology $\mathbb{S}^{2}$). Letting $N$ denote the lapse function
of $\bm{g}$, we have $\bm{u}=N^{-1}\partial/\partial t$.

Note that in general, $\mathscr{H}$ need not coincide
with the constant time slices $\mathscr{S}$. Equivalently, $\bm{u}$
need not coincide with $\tilde{\bm{u}}$. In general, there will be
a shift between them, such that 
\begin{equation}
\tilde{\bm{u}}=\tilde{\gamma}(\bm{u}+\bm{v})\,,\label{eq:u^tilde}
\end{equation}
where $v^{a}$ represents the spatial two-velocity of fiducial observers
that are at rest with respect to $\mathscr{S}$ as measured by our
congruence of quasilocal observers (the four-velocity of which is
$\bm{u}$), and $\tilde{\gamma}=1/\sqrt{1-\bm{v}\cdot\bm{v}}$ is
the Lorentz factor.

The specification of a quasilocal frame is thus equivalent to making
a particular choice of a two-parameter family of timelike worldlines comprising $\mathscr{B}$. There are,
a priori, three degrees of freedom (DoFs) available to us
for doing this. Heuristically, these can be regarded as corresponding
to the three DoFs in choosing the direction of $\bm{u}$—from which
$\bm{n}$ and all induced quantities are then computable. (Note that
$\bm{u}$ has four components, but one of the four is fixed by the
normalization requirement $\bm{u}\cdot\bm{u}=-1$, leaving three independent
direction DoFs.) Equivalently, we are in principle free to pick any
three geometrical conditions (along the congruence) to fix a quasilocal
frame. In practice, usually it is physically more natural, as well
as mathematically easier, to work with geometric quantities other
than $\bm{u}$ itself to achieve this. 

Yet, it is worth remarking that simply writing down three desired
equations (or conditions) to be satisfied by geometrical quantities
on $\mathscr{B}$ does not itself guarantee that, in general,
a submanifold $(\mathscr{B},\bm{\gamma},\bm{\mathcal{D}})$ obeying
those three particular equations will always exist—and, if
it does, that it will be the unique such submanifold—in an arbitrary
$(\mathscr{M},\bm{g},\bm{\nabla}$). Nevertheless, one choice of quasilocal
frame that is known to always exist (a claim we will qualify more
carefully in a moment) is that where the two-metric $\bm{\sigma}$
on $\mathscr{H}$ is ``rigid'' (or ``time
independent'')—these are called \emph{rigid quasilocal frames}.

Most of the past work on quasilocal frames has
in fact been done in the rigid case [\cite{epp_rigid_2009,epp_existence_2012,mcgrath_quasilocal_2012,epp_momentum_2013,mcgrath_post-newtonian_2014,mcgrath_rigid_2014}].
We know however that other quasilocal frame choices are also possible,
such as \emph{geoids}---dubbed geoid quasilocal frames [\cite{oltean_geoids_2016,oltean_geoids_2017}]:
these are the general-relativistic generalization of ``constant gravitational
potential'' surfaces in Newtonian gravity. Regardless, the quasilocal
frame choice that we will mainly consider in this chapter is the rigid
one (and we will be clear when this choice is explicitly enacted). 

Intuitively, the reason for this preference is that imposing in this
way the condition of ``spatial rigidity'' on $(\mathscr{B};\bm{u})$—a
two-dimensional (boundary) rigidity requirement, which unlike three-dimensional
rigidity, is permissible in GR—eliminates from the description of
the system any effects arising simply from the motion of the quasilocal
observers relative to each other. Thus, the physics of what is going
on inside the system (\textit{i.e.} the worldtube interior) is essentially all that affects its
dynamics. 

Technically, there is a further reason: a proof of the existence of
solutions—\textit{i.e.} the existence of a submanifold $\mathscr{B}\simeq\mathbb{R}\times\mathbb{S}^{2}$
in $\mathscr{M}$ that is also a quasilocal frame $(\mathscr{B};\bm{u})$—for
any spacetime $(\mathscr{M},\bm{g},\bm{\nabla})$ has up to now only
been fully carried out for rigid quasilocal frames\footnote{~The idea of the proof is to explicitly construct the solutions order-by-order
in an expansion in the areal radius around an arbitrary worldline
in an arbitrary spacetime [\cite{epp_existence_2012}].}. While, as we have commented, other quasilocal frame choices may
be generally possible in principle (and may be shown to be possible
to construct, case-by-case, in specific spacetimes—as we have done,
\textit{e.g.}, with geoid quasilocal frames [\cite{oltean_geoids_2016,oltean_geoids_2017}]),
they are as yet not rigorously guaranteed to exist in arbitrary spacetimes.

The quasilocal rigidity conditions can be stated in a number of ways.
Most generally, defining 
\begin{equation}
\theta_{ab}=\sigma_{ac}\sigma_{bd}\nabla^{c}u^{d}\label{eq:strain-rate}
\end{equation}
to be the \emph{strain rate tensor} of the congruence, they amount
to the requirement of vanishing expansion $\theta={\rm tr}(\bm{\theta})$
and shear $\theta_{\langle ab\rangle}=\theta_{(ab)}-\frac{1}{2}\theta\sigma_{ab}$,
\textit{i.e.} 
\begin{equation}
\theta=0=\theta_{\langle ab\rangle}\Leftrightarrow0=\theta_{(ab)}\,.\label{eq:RQF-conditions}
\end{equation}
In the adapted coordinates, these three conditions are expressible
as the vanishing of the time derivative of the two-metric on $\mathscr{H}$,
\textit{i.e.} $0=\partial_{t}\bm{\sigma}$. Both of these two equivalent mathematical
conditions, $\theta_{(ab)}=0=\partial_{t}\bm{\sigma}$, capture physically
the meaning of the quasilocal observers moving rigidly with
respect to each other (\textit{i.e.} the ``radar-ranging'' distances between
them does not change in time).

\subsection{The quasilocal stress-energy-momentum tensor\label{ssec:qf_by}}

Before we consider the formulation of conservation laws with the use
of quasilocal frames (from which our analysis of the GSF will eventually emerge),
we wish to address in a bit more detail an even more fundamental question: what are conservation
laws in GR actually supposed to be about? At the most basic level, they should
express changes (over time) in some appropriately defined notion of
energy-momentum. As we are interested in gravitational
systems (and specifically, those driven by the effect of the GSF),
this energy-momentum must include that of the gravitational field, in addition
to that of any matter fields if present.

Hence, we may assert from the outset that it does not make much sense
in GR to seek conservation laws based solely on the matter stress-energy-momentum tensor
$\bm{T}$, such as Eq.~(\ref{eq:local_cons}). It is evident that these would, by construction, account
for matter only---leaving out gravitational effects in general (which
could exist in the complete absence of matter, \textit{e.g.} gravitational
waves), and thus the GSF in particular. What is more, such conservation
laws are logically inconsistent from a general-relativistic point
of view: a non-vanishing $\bm{T}$ implies a non-trivial gravitational
field (through the Einstein equation) and thus a necessity of taking
into account that field along with the matter one(s) for a proper
accounting of energy-momentum transfer. A further technical problem is also that
the formulation of conservation laws of this sort is typically predicated
upon the existence of Killing vector fields or other types of symmetry generators in $\mathscr{M}$, which
one does not have in general---and which do not exist in spacetimes
pertinent for the GSF problem in particular. 

We are therefore led to ask: how can we meaningfully define a total---\emph{gravity
plus matter---}stress-energy-momentum tensor in GR? It turns out that the precise answer
to this question, while certainly not intractable, is unfortunately
also not unique---or at least, it lacks a clear consensus among relativists,
even today. See, \textit{e.g.}, [\cite{jaramillo_mass_2011,szabados_quasi-local_2004}]
for reviews of the variety of proposals that have been put forward
towards addressing this question. Nonetheless, for reasons already touched upon and to be elaborated presently, what is clear and generally accepted is
that such a tensor cannot be local in nature (as $\bm{T}$ is), and
for this reason is referred to as \emph{quasilocal}.

Let $\tau_{ab}$ denote this quasilocal, total (matter plus gravity) stress-energy-momentum tensor that we
eventually seek to use for our conservation laws. It has long been
understood (see, \emph{e.g.}, Section 20.4 of [\cite{misner_gravitation_1973}]) that whatever the notion
of ``gravitational energy-momentum'' (defined by $\bm{\tau}$) might
mean, it is not something localizable: in other words, there is no
way of meaningfully defining an ``energy-momentum volume density''
for gravity. This is, ultimately, due to the equivalence principle:
locally, one can always find a reference frame in which all local
``gravitational fields'' (the connection coefficients), and thus
any notion of ``energy-momentum volume density'' associated therewith,
disappear. The remedy is to make $\bm{\tau}$ quasilocal: meaning that, rather
than volume densities, it should define surface densities (of energy,
momentum \textit{etc.})---a type of construction which \emph{is} mathematically realizable
and physically sensible in general.

The specific choice we make for how to define this total (matter plus
gravity), quasilocal energy-momentum tensor $\bm{\tau}$ is the so-called Brown-York
tensor, first put forward by  [\cite{brown_quasilocal_1993}]; see also [\cite{brown_action_2002}]
for a detailed review. This proposal was based originally
upon a Hamilton-Jacobi analysis; here we will offer a simpler argument
for its definition, sketched out initially in  [\cite{epp_momentum_2013}].

Consider the standard gravitational action $S_{\textrm{G}}$ for a
spacetime volume $\mathscr{V}\subset\mathscr{M}$ such that 
$\partial \mathscr{V}=\mathscr{B}\simeq\mathbb{R}\times\mathbb{S}^{2}$ is a worldtube
boundary as in the previous subsection (possibly constituting a quasilocal
frame, but not necessarily). This action is given by the sum of two
terms, a bulk and a boundary term respectively (see Section \ref{sec:2.2-lagrangian}):
\begin{equation}
S_{\textrm{G}}\left[\bm{g}\right]=S_{\textrm{EH}}\left[\bm{g}\right]+S_{\textrm{GHY}}\left[\bm{\gamma},\bm{n}\right]\,.\label{eq:action_gravitational}
\end{equation}
In particular, the first is the Einstein-Hilbert bulk term, 
\begin{equation}
S_{\textrm{EH}}\left[\bm{g}\right]=\frac{1}{2\kappa}\intop_{\mathscr{V}}\bm{\epsilon}_{\mathscr{M}}^{\,}\,R\,,
\end{equation}
and the second is the Gibbons-Hawking-York boundary term [\cite{york_role_1972,gibbons_action_1977}],
\begin{equation}
S_{\textrm{GHY}}\left[\bm{\gamma},\bm{n}\right]=-\frac{1}{\kappa}\intop_{\partial\mathscr{V}}\bm{\epsilon}_{\mathscr{B}}^{\,}\,\Theta\,.
\end{equation}
Here, $\bm{\epsilon}_{\mathscr{M}}^{\,}={\rm d}^{4}x\sqrt{-g}$ is
the volume form on $\mathscr{M}$ with $g={\rm \det}(\bm{g})$, $\bm{\epsilon}_{\mathscr{B}}^{\,}={\rm d}^{3}x\sqrt{-\gamma}$
is the volume form on $\mathscr{B}$ with $\gamma={\rm \det}(\bm{\gamma})$,
and $\Theta={\rm tr}(\bm{\Theta})$ is the trace of the extrinsic
curvature $\Theta_{ab}=\gamma_{ac}\nabla^{c}n_{b}$ of $\mathscr{B}$
in $\mathscr{M}$. Additionally, the matter action $S_{\textrm{M}}$
for any set of matter fields $\Psi$ described by a Lagrangian $L_{\textrm{M}}$
is 
\begin{equation}
S_{\textrm{M}}\left[\Psi\right]=\intop_{\mathscr{V}}\bm{\epsilon}_{\mathscr{M}}^{\,}\,L_{\textrm{M}}\left[\Psi\right]\,.\label{eq:action_matter}
\end{equation}

The definition of the total (quasilocal) stress-energy-momentum tensor $\bm{\tau}$
for gravity plus matter can be obtained effectively in the same way
as that of the (local) stress-energy-momentum tensor $\bm{T}$ for matter
alone—from the total action in Eq.~(\ref{eq:action_gravitational}) rather
than just, respectively, the matter action in Eq.~(\ref{eq:action_matter}).
In particular, $\bm{T}$ is defined by computing the variation $\delta$
(with respect to the spacetime metric) of the matter action: 
\begin{equation}
\delta S_{\mathrm{M}}\left[\Psi\right]=-\frac{1}{2}\intop_{\mathscr{V}}\bm{\epsilon}_{\mathscr{M}}^{\,}\,T_{ab}\delta g^{ab}\,.\label{eq:dS_matter}
\end{equation}
In other words, one defines the matter stress-energy-momentum tensor
as the functional derivative, 
\begin{equation}
T_{ab}=-\frac{2}{\sqrt{-g}}\frac{\delta S_{\textrm{M}}}{\delta g^{ab}}\,.\label{eq:T_ab_definition}
\end{equation}
The definition of the Brown-York tensor follows completely analogously,
except that now gravity is also included. That is, for the total action
of gravity (minimally) coupled to matter, 
\begin{equation}
S_{\textrm{G}+\textrm{M}}\left[\bm{g},\Psi\right]=S_{\textrm{G}}\left[\bm{g}\right]+S_{\textrm{M}}\left[\Psi\right]\,,\label{eq:GM-action}
\end{equation}
we have that the metric variation is: 
\begin{align}
\delta S_{\textrm{G}+\textrm{M}}\left[\bm{g},\Psi\right]=\, & \frac{1}{2}\Bigg\{\intop_{\mathscr{V}}\bm{\epsilon}_{\mathscr{M}}\,\left(\frac{1}{\kappa}G_{ab}-T_{ab}\right)\delta g^{ab}-\intop_{\partial\mathscr{V}}\bm{\epsilon}_{\mathscr{B}}^ {}\,\left(-\frac{1}{\kappa}\Pi_{ab}\right)\delta\gamma^{ab}\Bigg\}\label{eq:dS_G+M_1}\\
=\, & -\frac{1}{2}\intop_{\partial\mathscr{V}}^{\,}\bm{\epsilon}_{\mathscr{B}}^ {}\,\tau_{ab}\delta\gamma^{ab}\,.\label{eq:dS_G+M_2}
\end{align}
In Eq.~(\ref{eq:dS_G+M_1}), $\bm{\Pi}$ is the canonical
momentum of $(\mathscr{B},\bm{\gamma},\bm{\mathcal{D}})$, given by
$\bm{\Pi}=\bm{\Theta}-\Theta\bm{\gamma}$. It follows from direct
computation using Eqs.~(\ref{eq:action_gravitational}), (\ref{eq:action_matter})
and (\ref{eq:dS_matter}); for a review of this derivation carefully
accounting for the boundary term see, \textit{e.g.}, Chapter 12 of [\cite{padmanabhan_gravitation:_2010}].
In the equality of Eq.~(\ref{eq:dS_G+M_2}), the Einstein equation
$\bm{G}=\kappa\bm{T}$ has been invoked (in other words, we impose the Einstein equation to be satisfied in the bulk),
thus leading to the vanishing of the bulk term; meanwhile in the boundary
term, a gravity plus matter stress-energy-momentum tensor $\bm{\tau}$ (the
Brown-York tensor) has been defined in direct analogy with the definition
of the matter energy-momentum tensor $\bm{T}$ in Eq.~(\ref{eq:dS_matter}).
Hence just as Eq.~(\ref{eq:dS_matter}) implies Eq.~(\ref{eq:T_ab_definition}),
Eq.~(\ref{eq:dS_G+M_2}) implies 
\begin{equation}
\bm{\tau}=-\frac{1}{\kappa}\bm{\Pi}\,.\label{eq:tau_ab_definition}
\end{equation}
Henceforth, $\bm{\tau}$ refers strictly to this (Brown-York) quasilocal stress-energy-momentum tensor of Eq.~(\ref{eq:tau_ab_definition}), and not to any other definition.

It is useful to decompose  $\bm{\tau}$ in a similar way as is
ordinarily done with $\bm{T}$, so we define: 
\begin{align}
\mathcal{E}=\, & u^{a}u^{b}\tau_{ab}\,,\\
\mathcal{P}^{a}=\, & -\sigma^{ab}u^{c}\tau_{bc}\,,\\
\mathcal{S}^{ab}=\, & -\sigma^{ac}\sigma^{bd}\tau_{cd}\,,
\end{align}
as the quasilocal energy, momentum and stress, respectively, with
units of energy per unit area, momentum per unit area and force per
unit length. Equivalently, 
\begin{equation}
\tau^{ab}=u^{a}u^{b}\mathcal{E}+2u^{(a}\mathcal{P}^{b)}-\mathcal{S}^{ab}\,.
\end{equation}

\subsection{Conservation laws\label{ssec:qf_cons_laws}}

The construction of general conservation laws from $\bm{\tau}$
was first achieved in  [\cite{mcgrath_quasilocal_2012,epp_momentum_2013}], and proceeds along the following lines.  Let $\boldsymbol{\psi}\in T\mathscr{B}$ be an arbitrary vector field
in $\mathscr{B}$. We begin by considering a projection of $\bm{\Pi}$
in the direction of $\bm{\psi}$ (in one index), \textit{i.e.} $\Pi^{ab}\psi_{b}$, and computing its divergence in $\mathscr{B}$. By using the Leibnitz rule, we simply have
\begin{equation}
\mathcal{D}_{a}\left(\Pi^{ab}\psi_{b}\right)=\left(\mathcal{D}_{a}\Pi^{ab}\right)\psi_{b}+\Pi^{ab}\left(\mathcal{D}_{a}\psi_{b}\right)\,.
\end{equation}
Next, we integrate this equation over a portion $\Delta\mathscr{B}$
of $\mathscr{B}$ bounded by initial and final constant $t$ surfaces
$\mathscr{S}_{\textrm{i}}$ and $\mathscr{S}_{\textrm{f}}$, as depicted in Figure~\ref{fig-qf}. On the
resulting LHS we apply Stokes' theorem, and on the first term on the
RHS we use the Gauss-Codazzi identity: $\mathcal{D}_{a}\Pi^{ab}=n_{a}\gamma^{b}_{\;\;c}G^{ac}$.
Thus, using the notation for tensor projections in certain directions introduced in Section \ref{sec:1-2} (see also the Notation and Conventions) for ease of readability (\textit{e.g.}, $G_{ab}n^{a}\psi^{b}=G_{\bm{n}\bm{\psi}}$ and similarly for other contractions), we obtain:
\begin{equation}
\intop_{\mathscr{S}_{\textrm{f}}-\mathscr{S}_{\textrm{i}}}\!\!\!\!\bm{\epsilon}^{\,}_{\mathscr{S}}\,\Pi_{\bm{\tilde{u}}\bm{\psi}}=-\intop_{\Delta\mathscr{B}}\!\bm{\epsilon}^{\,}_{\mathscr{B}}\,\left(G_{\bm{n}\bm{\psi}}+\Pi^{ab}\mathcal{D}_{a}\psi_{b}\right)\,,\label{eq:geometrical_identity}
\end{equation}
where $\bm{\epsilon}^{\,}_{\mathscr{S}}$ denotes the volume form on the constant time closed two-surfaces $\mathscr{S}$,
and we have used the notation: $\int_{\mathscr{S}_{\textrm{f}}-\mathscr{S}_{\textrm{i}}}(\cdot)=\int_{\mathscr{S}_{\textrm{f}}}(\cdot)-\int_{\mathscr{S}_{\textrm{i}}}(\cdot)$.
We also remind the reader that $\tilde{\bm{u}}$ represents the unit normal to each constant time closed two-surface, which in general need not coincide with the quasilocal observers' four-velocity $\bm{u}$ but is related to it by a Lorentz transformation, Eq.~(\ref{eq:u^tilde}); see also Figure~\ref{fig-qf}.

We stress that so far, Eq.~(\ref{eq:geometrical_identity}) is a purely
geometrical identity, completely general for any Lorentzian manifold
$\mathscr{M}$; in other words, thus far we have said nothing about
physics.

Now, to give this identity physical meaning, we invoke the definition
of the Brown-York tensor in Eq.~(\ref{eq:tau_ab_definition}) (giving
the boundary extrinsic geometry its meaning as stress-energy-momentum) as well as the
Einstein equation [Eq.~(\ref{eq:Einstein_eqn})], giving the spacetime
curvature its meaning as the gravitational field. With these, Eq.~(\ref{eq:geometrical_identity}) turns into:
\begin{equation}
\intop_{\mathscr{S}_{\textrm{f}}-\mathscr{S}_{\textrm{i}}}\!\!\!\!\bm{\epsilon}^{\,}_{\mathscr{S}}\,\tilde{\gamma}\left(\tau_{\bm{u}\bm{\psi}}+\tau_{\bm{v}\bm{\psi}}\right)=\intop_{\Delta\mathscr{B}}\!\bm{\epsilon}^{\,}_{\mathscr{B}}\,\left(T_{\bm{n}\bm{\psi}}-\tau^{ab}\mathcal{D}_{(a}\psi_{b)}\right)\,.\label{eq:cons_law_general}
\end{equation}
On the LHS we have inserted the relation $\tilde{\bm{u}}=\tilde{\gamma}(\bm{u}+\bm{v})$,
with $v^{a}$ representing the spatial two-velocity of fiducial observers
that are at rest with respect to $\mathscr{S}$ (the hypersurface-orthogonal
four-velocity of which is $\tilde{\bm{u}}$) as measured by our congruence
of quasilocal observers (the four-velocity of which is $\bm{u}$), and $\tilde{\gamma}=1/\sqrt{1-\bm{v}\cdot\bm{v}}$
is the Lorentz factor.

Observe that Eq.~(\ref{eq:cons_law_general}) expresses the change of
some component of the quasilocal stress-energy-momentum tensor integrated over two different $t = const.$ closed two-surfaces $\mathscr{S}$ as a flux through
the worldtube boundary $\Delta\mathscr{B}$ between them. The identification
of the different components of $\bm{\tau}$ as the various components
of the total energy-momentum of the system thus leads to the understanding of Eq.~(\ref{eq:cons_law_general}) as a general conservation law for the
system contained inside of $\Delta\mathscr{B}$. Thus, depending on
our particular choice of $\bm{\psi}\in T\mathscr{B}$, Eq.~(\ref{eq:cons_law_general})
will represent a conservation law for the total energy, momentum,
or angular momentum of this system [\cite{epp_momentum_2013}]. 

Let us now assume that $(\mathscr{B};\bm{u})$ is a rigid quasilocal frame. If we choose
$\bm{\psi}=\bm{u}$, then Eq.~(\ref{eq:cons_law_general}) becomes the
energy conservation law:
\begin{equation}
\intop_{\mathscr{S}_{\textrm{f}}-\mathscr{S}_{\textrm{i}}}\!\!\!\!\bm{\epsilon}^{\,}_{\mathscr{S}}\,\tilde{\gamma}\left(\mathcal{E}-\mathcal{P}_{\bm{v}}\right)=\intop_{\Delta\mathscr{B}}\!\bm{\epsilon}^{\,}_{\mathscr{B}}\,\left(T_{\bm{n}\bm{u}}-\bm{\alpha}\cdot\bm{\mathcal{P}}\right)\,,\label{eq:cons_law_E}
\end{equation}
where $\alpha^{a}=\sigma^{ab}a_{b}$ is the $\mathscr{H}$ projection
of the acceleration of the quasilocal observers, defined by $a^{a}=\nabla_{\bm{u}}u^{a}$. 

Now suppose, on the other hand, that we instead choose $\bm{\psi}=-\bm{\phi}$
where $\bm{\phi}\in \mathscr{H}$ is orthogonal to $\bm{u}$ (with the minus sign introduced for convenience), and
represents a stationary conformal Killing vector field with
respect to $\bm{\sigma}$. This means that $\bm{\phi}$ is chosen such that it satisfies the conformal Killing equation, $\mathcal{L}_{\bm{\phi}}\bm{\sigma}=(\bm{D}\cdot\bm{\phi})\bm{\sigma}$, with $\mathcal{L}$ the Lie derivative and $\bm{D}$ the derivative on $\mathscr{H}$ (compatible with $\bm{\sigma}$). A set of six such conformal Killing vectors always exist: three for translations
and three for rotations, respectively generating the action of boosts
and rotations of the Lorentz group on the two-sphere [\cite{epp_momentum_2013}]. The idea, then, is that the contraction of these vectors with the quasilocal momentum integrated over a constant-time topological two-sphere boundary expresses, respectively, the total linear and angular momentum (in the three ordinary spatial directions each) at that time instant. Thus, Eq.~(\ref{eq:cons_law_general})
becomes the (respectively, linear and angular) momentum conservation
law:
\begin{equation}
\intop_{\mathscr{S}_{\textrm{f}}-\mathscr{S}_{\textrm{i}}}\!\!\!\!\bm{\epsilon}^{\,}_{\mathscr{S}}\,\tilde{\gamma}\left(\mathcal{P}_{\bm{\phi}}+\mathcal{S}_{\bm{v}\bm{\phi}}\right)=-\intop_{\Delta\mathscr{B}}\!\bm{\epsilon}^{\,}_{\mathscr{B}}\,\left(T_{\bm{n}\bm{\phi}}+\mathcal{E}\alpha_{\bm{\phi}}+2\nu\epsilon_{ab}\mathcal{P}^{a}\phi^{b}+{\rm P}\bm{D}\cdot\bm{\phi}\right)\,,\label{eq:cons_law_Pa}
\end{equation}
where $\nu=\frac{1}{2}\epsilon^{ab}_{\mathscr{H}}\mathcal{D}_{a}u_{b}$ is the twist
of the congruence (with $\epsilon_{ab}^{\mathscr{H}}=\epsilon_{abcd}^{\mathscr{M}}u^{c}n^{d}$ the induced volume form on $\mathscr{H}$), and ${\rm P}=\frac{1}{2}\bm{\sigma}:\bm{\mathcal{S}}$
is the quasilocal pressure (force per unit length) between the worldlines
of $\mathscr{B}$. We remark that the latter can be shown to satisfy
the very useful general identity (which we will expediently invoke
in our later calculations):
\begin{equation}
\mathcal{E}-2{\rm P}=\frac{2}{\kappa}a_{\bm{n}}\,.\label{eq:E-P_relation}
\end{equation}

An analysis of the gravitational self-force problem should consider the conservation law in Eq.~(\ref{eq:cons_law_Pa}) for \emph{linear momentum}. Thus, we will use the fact, described in greater detail in the appendix of this chapter (Section \ref{sec:5.A}), that the conformal Killing vector $\bm{\phi}\in \mathscr{H}$ for linear momentum admits a multipole decomposition
of the following form:
\begin{align}
\phi^{\mathfrak{i}}=\, & \frac{1}{r}\,D^{\mathfrak{i}}\left(\Phi^{I}r_{I}+\Phi^{IJ}r_{I}r_{J}+\cdots\right)\label{eq:phi_multipole}\\
=\, & \frac{1}{r}\left(\Phi^{I}\mathfrak{B}_{I}^{\mathfrak{i}}+2\Phi^{IJ}\mathfrak{B}_{I}^{\mathfrak{i}}r_{J}+\cdots\right)\,,
\end{align}
with the dots indicating higher harmonics. Here, $r$ is the area radius of the quasilocal frame (such that $\mathscr{B}$
is a constant $r$ hypersurface in $\mathscr{M}$), $r^{I}$ denotes
the the standard direction cosines of a radial unit vector in $\mathbb{R}^{3}$
and $\mathfrak{B}_{I}^{\mathfrak{i}}=\partial^{\mathfrak{i}}r_{I}$
are the boost generators on the two-sphere. See this chapter's appendix (Section \ref{sec:5.A}) for a detailed
discussion regarding conformal Killing vectors and the two-sphere. In spherical coordinates
$\{\theta,\phi\}$, we have $r^{I}=(\sin\theta\cos\phi,\sin\theta\sin\phi,\cos\theta)$.
Thus Eq.~(\ref{eq:phi_multipole}) gives us a decomposition of $\bm{\phi}$
in terms of multipole moments, with the $\ell=1$ coefficients $\Phi^{I}$
simply representing vectors in $\mathbb{R}^{3}$ in the direction of which
we are considering the conservation law.

\section{General derivation of the gravitational self-force from quasilocal conservation laws\label{sec:general-analysis}}

In this section, we will show how the GSF is a general
consequence of the momentum conservation law in Eq.~(\ref{eq:cons_law_Pa})
for any system which is sufficiently localized. By that, we mean something
very simple: taking the $r\rightarrow0$ limit of a quasilocal frame around the
moving object which is treated as ``small'', \textit{i.e.} as a formal perturbation
about some background. No further assumptions are for the moment needed.
In particular, we do not even need to enter into the precise details
of how to specify the perturbation family for this problem; that will
be left to the following section, where we will carefully define and
work with the family of perturbed spacetimes typically employed for
applications of the GSF. For now, we proceed to show that the
first-order perturbation of the momentum conservation law in Eq.~(\ref{eq:cons_law_Pa})
always contains the GSF, and that it dominates the dynamics for localized systems.

Let $\{(\mathscr{B}_{(\lambda)};\bm{u}_{(\lambda)})\}_{\lambda\geq0}$
be an arbitrary one-parameter family of quasilocal frames (defined as in Section
\ref{sec:qf}) each of which is embedded, respectively, in the corresponding
element of the family of perturbed spacetimes $\{(\mathscr{M}_{(\lambda)},\bm{g}_{(\lambda)},\bm{\nabla}_{(\lambda)})\}_{\lambda\geq0}$
described in the previous subsection. Consider the general geometrical
identity (\ref{eq:geometrical_identity}) in $\mathscr{M}_{(\lambda)}$,
$\forall\lambda\geq0$:
\begin{equation}
\intop_{\mathscr{S}_{\textrm{f}}^{(\lambda)}-\mathscr{S}_{\textrm{i}}^{(\lambda)}}\!\!\!\!\!\!\!\!\bm{\epsilon}_{\mathscr{S}_{(\lambda)}}\,\Pi_{\bm{\tilde{u}}_{(\lambda)}\bm{\psi}_{(\lambda)}}^{(\lambda)}=-\!\!\intop_{\Delta\mathscr{B}_{(\lambda)}}\!\!\!\bm{\epsilon}_{\mathscr{B}_{(\lambda)}}\,\left(G_{\bm{n}_{(\lambda)}\bm{\psi}_{(\lambda)}}^{(\lambda)}+\Pi_{(\lambda)}^{ab}\mathcal{D}_{a}^{(\lambda)}\psi_{b}^{(\lambda)}\right)\,.\label{eq:general_cons_law_lambda}
\end{equation}
For $\lambda=0$ this gives us our conservation laws in the background,
and for any $\lambda>0$, those in the corresponding perturbed spacetime. It is the
latter that we are interested in, but since we do not know how to
do calculations in $\mathscr{M}_{(\lambda)}$ $\forall\lambda>0$,
we have to work with Eq. (\ref{eq:general_cons_law_lambda}) transported
to $\mathring{\mathscr{M}}$. This is easily achieved by using the
fact that for any diffeomorphism $f:\mathscr{U}\rightarrow\mathscr{V}$
between two (oriented) smooth $n$-dimensional manifolds $\mathscr{U}$ and $\mathscr{V}$
and any (compactly supported) $n$-form $\bm{\omega}$ in $\mathscr{V}$,
we have that $\int_{\mathscr{V}}\bm{\omega}=\int_{\mathscr{U}}f^{*}\bm{\omega}$.
Applying this to the LHS and RHS of Eq. (\ref{eq:general_cons_law_lambda})
respectively, we simply get
\begin{equation}
\nn\nn\nn\nn\nn\nn\nn\nn\intop_{\quad\quad\quad\quad\quad\quad\quad\varphi_{(\lambda)}^{-1}(\mathscr{S}_{\textrm{f}}^{(\lambda)})-\varphi_{(\lambda)}^{-1}(\mathscr{S}_{\textrm{i}}^{(\lambda)})}\nn\nn\nn\nn\nn\nn\nn\nn\left(\varphi_{(\lambda)}^{*}\bm{\epsilon}_{\mathscr{S}_{(\lambda)}}\right)\,\varphi_{(\lambda)}^{*}\Pi_{\bm{\tilde{u}}_{(\lambda)}\bm{\psi}_{(\lambda)}}^{(\lambda)}=\nn\nn\nn\nn\intop_{\quad\quad\quad\varphi_{(\lambda)}^{-1}(\Delta\mathscr{B}_{(\lambda)})}\nn\nn\nn\nn\left(\varphi_{(\lambda)}^{*}\bm{\epsilon}_{\mathscr{B}_{(\lambda)}}\right)\,\varphi_{(\lambda)}^{*}\left(G_{\bm{n}_{(\lambda)}\bm{\psi}_{(\lambda)}}^{(\lambda)}+\Pi_{(\lambda)}^{ab}\mathcal{D}_{a}^{(\lambda)}\psi_{b}^{(\lambda)}\right)\,.
\end{equation}
Denoting $\mathscr{S}=\varphi_{(\lambda)}^{-1}(\mathscr{S}_{(\lambda)})\subset\mathring{\mathscr{M}}$
as the inverse image of a constant time two-surface and similarly
$\mathscr{B}=\varphi_{(\lambda)}^{-1}(\mathscr{B}_{(\lambda)})\subset\mathring{\mathscr{M}}$
as the inverse image of the worldtube boundary (quasilocal frame) in the background
manifold, and using the fact that the tensor transport commutes with
contractions, the above can simply be written in the notation we have
established as
\begin{equation}
\intop_{\mathscr{S}_{\textrm{f}}-\mathscr{S}_{\textrm{i}}}\left(\varphi_{(\lambda)}^{*}\bm{\epsilon}_{\mathscr{S}_{(\lambda)}}\right)\Pi_{\bm{\tilde{u}}\bm{\psi}}=\intop_{\Delta\mathscr{B}}\left(\varphi_{(\lambda)}^{*}\bm{\epsilon}_{\mathscr{B}_{(\lambda)}}\right)\,\left(G_{\bm{n}\bm{\psi}}+\Pi^{ab}\mathcal{D}_{a}\psi_{b}\right)\,.\label{eq:general_cons_law_background}
\end{equation}

\begin{figure}
\begin{centering}
\includegraphics[scale=0.7]{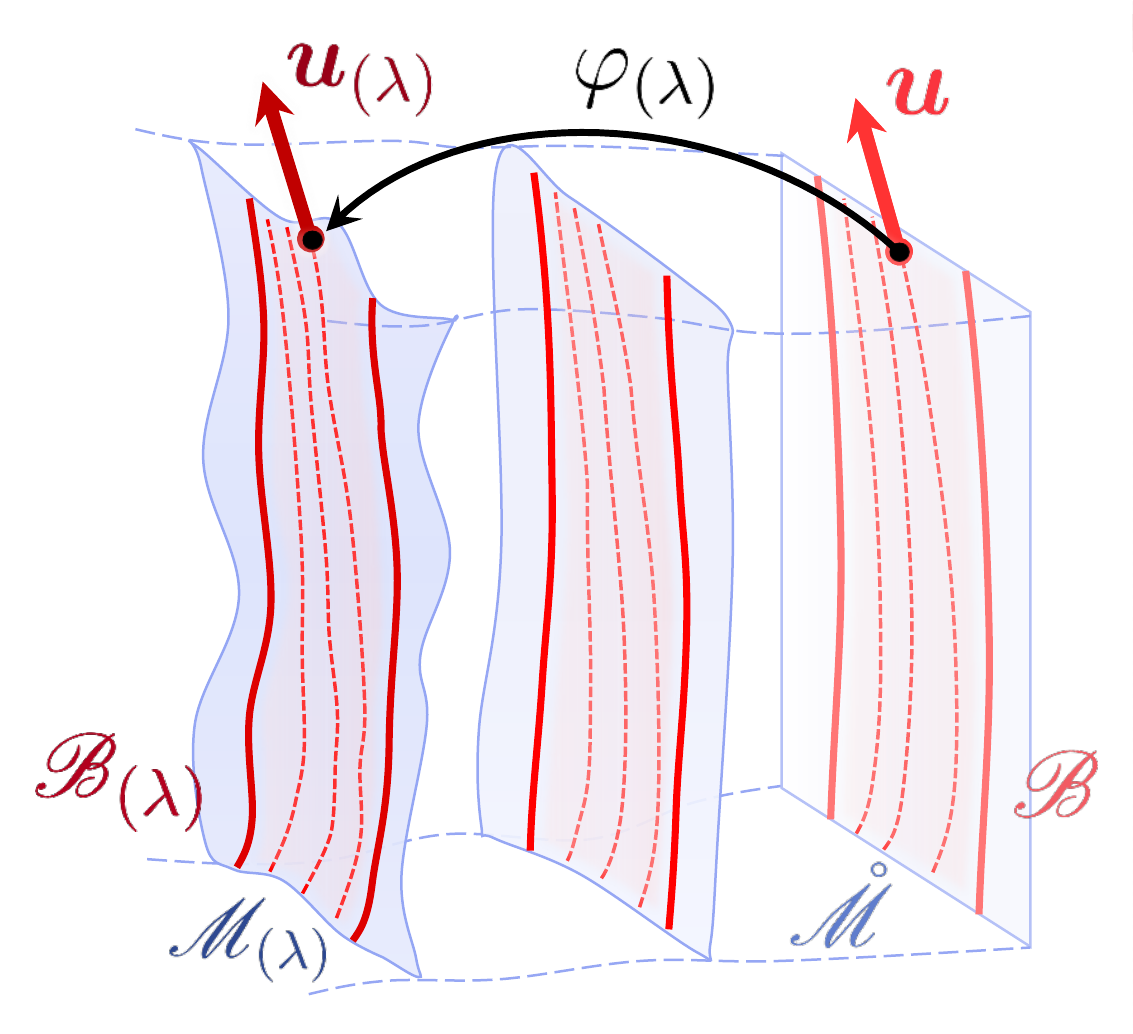}
\par\end{centering}
\caption{Representation of a one-parameter family of quasilocal frames $\{(\mathscr{B}_{(\lambda)};\bm{u}_{(\lambda)})\}_{\lambda\geq0}$
embedded correspondingly in a family of spacetimes $\{\mathscr{M}_{(\lambda)}\}_{\lambda\geq0}$.}
\end{figure}

So far we have been completely general. Now, let us restrict our attention to the momentum conservation law ($\bm{\psi}=-\bm{\phi}\in \mathscr{H}$)
given by Eq. (\ref{eq:general_cons_law_background}), and let us assume
that we do not have any matter on $\Delta\mathscr{B}$ (hence, by
the Einstein equation, $G_{\bm{n}\bm{\phi}}|_{\Delta\mathscr{B}}=\kappa T_{\bm{n}\bm{\phi}}|_{\Delta\mathscr{B}}=0$),
or even simply that any matter if present there is subdominant to
the linear perturbation, \textit{i.e.} $\bm{T}|_{\Delta\mathscr{B}}=\mathcal{O}(\lambda^{2})$.
The LHS then expresses the change in momentum of the system (inside
the worldtube interval in the perturbed spacetime) between
some initial and final time slices; for notational ease, we will simply
denote this by $\Delta\mathtt{p}^{(\bm{\phi})}$. (Note that we prefer to use typewriter font for the total quasilocal momentum, so as to avoid any confusion with matter four-momentum defined in the typical way from $T_{ab}$ and traditionally labelled by $P^{a}$, as \textit{e.g.} in Eq. \ref{eq:local_cons}.) Then, inserting
also the definition of the Brown-York tensor [Eq. (\ref{eq:tau_ab_definition})]
on the RHS and replacing $\bm{\mathcal{D}}$ with $\bm{\nabla}$ since
it does not affect the contractions, Eq. (\ref{eq:general_cons_law_background})
becomes:
\begin{equation}
\Delta\mathtt{p}^{(\bm{\phi})}=\intop_{\Delta\mathscr{B}}\left(\varphi_{(\lambda)}^{*}\bm{\epsilon}_{\mathscr{B}_{(\lambda)}}\right)\,\tau^{ab}\nabla_{a}\phi_{b}\,.\label{eq:Delta_p_general}
\end{equation}
We claim, and will now demonstrate, that the $\mathcal{O}(\lambda)$
part of this always contains the GSF.

Let us consider Eq. (\ref{eq:Delta_p_general}) term by term. First
we have the transport---in this case, the pullback---under $\varphi_{(\lambda)}$
of the volume form of $\mathscr{B}_{(\lambda)}$. Now, we know that
the pullback under a diffeomorphism of the volume form of a manifold is,
in general, not simply the volume form of the inverse
image of that manifold under the diffeomorphism. However, it is always true (see, \textit{e.g.},
Chapter 7 of  [\cite{abraham_manifolds_2001}]) that they are proportional,
with the proportionality given by a smooth function called the Jacobian
determinant and usually denoted by $J$. That is, in our case we have
$\varphi_{(\lambda)}^{*}\bm{\epsilon}_{\mathscr{B}_{(\lambda)}}=J\bm{\text{\ensuremath{\epsilon}}}_{\mathscr{B}}$,
with $J\in C^{\infty}(\mathscr{B})$. In particular, this function
is given by $J(p)=\det(T_{p}\varphi_{(\lambda)})$, $\forall p\in\mathscr{B}$,
where $T_{p}\varphi_{(\lambda)}=(\varphi_{(\lambda)})_{*}:T_{p}\mathscr{B}\rightarrow T_{\varphi_{(\lambda)}}\mathscr{B}_{(\lambda)}$
is the pushforward, and the determinant is computed with respect to
the volume forms $\bm{\text{\ensuremath{\epsilon}}}_{\mathscr{B}}(p)$
on $T_{p}\mathscr{B}$ and $\bm{\epsilon}_{\mathscr{B}_{(\lambda)}}(\varphi_{(\lambda)}(p))$
on $T_{\varphi_{(\lambda)}(p)}\mathscr{B}_{(\lambda)}$. 
Now, it is clear that we have $J=1+\mathcal{O}(\lambda)$, as $\varphi_{(0)}$
is simply the identity map. Therefore, we have
\begin{equation}
\varphi_{(\lambda)}^{*}\bm{\epsilon}_{\mathscr{B}_{(\lambda)}}=\left(1+\mathcal{O}\left(\lambda\right)\right)\bm{\text{\ensuremath{\epsilon}}}_{\mathscr{B}}\,.\label{eq:volume_form_pullback}
\end{equation}
As for the other terms in the integrand of Eq.  (\ref{eq:Delta_p_general}),
we simply have
\begin{align}
\tau^{ab}=\, & \mathring{\tau}^{ab}+\lambda\delta\tau^{ab}+\mathcal{O}(\lambda^{2})\,,\label{eq:tau_series}\\
\nabla_{a}\phi_{b}= & \mathring{\nabla}_{a}\phi_{b}+\lambda\delta\left(\nabla_{a}\phi_{b}\right)+\mathcal{O}(\lambda^{2})\,.\label{eq:Dphi_series}
\end{align}

Hence we can see that there will be three contributions to the $\mathcal{O}(\lambda)$
RHS of Eq. (\ref{eq:Delta_p_general}). Respectively, from Eqs.
(\ref{eq:volume_form_pullback})-(\ref{eq:Dphi_series}), these are
the $\mathcal{O}(\lambda)$ parts of: the volume form pullback, which
may not be easy to compute in practice; the Brown-York tensor $\bm{\tau}$,
which may be computed from its definition [Eq. (\ref{eq:tau_ab_definition})];
and the derivative of the conformal Killing vector $\bm{\phi}$, which may be readily carried
out and, as we will presently show, always contains the GSF. Thus
we denote this contribution to the $\mathcal{O}(\lambda)$ part of
$\Delta\mathtt{p}^{(\bm{\phi})}$ as $\Delta\mathtt{p}_{\textrm{self}}^{(\bm{\phi})}$,
\begin{equation}
\Delta\mathtt{p}_{\textrm{self}}^{(\bm{\phi})}=\lambda\intop_{\Delta\mathscr{B}}\bm{\text{\ensuremath{\epsilon}}}_{\mathscr{B}}\,\mathring{\tau}^{ab}\delta\left(\nabla_{a}\phi_{b}\right)\,.\label{eq:Delta_p_GSF}
\end{equation}

Now we proceed with the computation of Eq. (\ref{eq:Delta_p_GSF}). In
particular, let us consider the series expansion of Eq. (\ref{eq:Delta_p_GSF})
in the areal radius $r$ of $\mathscr{B}$. This can be defined for
any time slice by $r=(\frac{1}{4\pi}\int_{\mathscr{S}}\bm{\epsilon}^{\,}_{\mathscr{S}})^{1/2}$,
such that a constant $r$ slice of $\mathring{\mathscr{M}}$ defines
$\mathscr{B}$ (and $\bm{n}=M\mathring{\bm{\nabla}}r$ for some positive
function $M$ on $\mathscr{B}$). It has been shown [\cite{epp_existence_2012}] that the Brown-York
tensor has, in general, the following expansion in $r$:
\begin{equation}
\mathring{\tau}^{ab}=\mathring{u}^{a}\mathring{u}^{b}\mathcal{E}_{\textrm{vac}}-\mathring{\sigma}^{ab}{\rm P}_{\textrm{vac}}+\mathcal{O}(r)\,,\label{eq:tau_dominant}
\end{equation}
where 
\begin{align}
\mathcal{E}_{\textrm{vac}}=\, & -\frac{2}{\kappa r}\,,\label{eq:E_vac}\\
{\rm P}_{\textrm{vac}}=\, & -\frac{1}{\kappa r}\,,\label{eq:P_vac}
\end{align}
are called the \emph{vacuum energy} and \emph{vacuum pressure} respectively. Some
remarks regarding these are warranted before we move on. In particular,
these are terms which have sometimes been argued to play the role
of ``subtraction terms'' (to be removed from the quasilocal energy-momentum
tensor); see \textit{e.g.}  [\cite{brown_action_2002}]. From this point of view, the definition of the Brown-York tensor [Eq. (\ref{eq:tau_ab_definition})]
may be regarded as carrying a certain amount of freedom, inasmuch as any
freedom may be assumed to exist to define a ``reference'' action
$S_{0}$ to be subtracted from the total (gravitational plus matter)
action $S_{\textrm{G+M}}$ in the variational principle discussed
in Subsection \ref{ssec:qf_by}. Such a subtraction of a ``reference'' action, while
common practice in gravitational physics, has the sole function of
shifting the numerical value of the action such that, ultimately,
the numerical value of the Hamiltonian constructed from the modified
action $S_{\textrm{G+M}}-S_{0}$ may be interpreted as the ADM energy.
However, this essentially amounts to a presumption that we are free
to pick the zero of the energy---in other words, that the vacuum
energy may be freely subtracted away without affecting the physics.
Though we refrain from entering into much further detail here, it
has been shown [\cite{epp_momentum_2013}] that these vacuum terms, Eqs. (\ref{eq:E_vac})-(\ref{eq:P_vac}),
are in fact crucial for our conservation laws to yield physically
reasonable answers and to make mathematical sense---evidencing that
the vacuum energy/pressure should be taken seriously as having physically
real significance. We will now lend further credibility to this by
showing that they are precisely the energy (and pressure) associated
with the momentum flux that are typically interpreted as the GSF. Actually, we argue in this chapter that
the term implicating the vacuum energy yields the standard form
of the GSF, and the vacuum pressure term is novel in our analysis.

Now that we have an expansion [Eq. (\ref{eq:tau_dominant})] of $\mathring{\bm{\tau}}$
in $r$, let us consider the $\delta(\bm{\nabla}\bm{\phi})$ term.
We see that
\begin{equation}
\delta\left(\nabla_{a}\phi_{b}\right)=\delta\left(\mathring{\nabla}_{a}\phi_{b}-C^{d}\,_{ab}\phi_{d}\right)=-\delta C^{c}\,_{ab}\phi_{c}\,.\label{eq:delta_nabla_phi}
\end{equation}
Collecting all of our results so far---inserting Eqs. (\ref{eq:tau_dominant})-(\ref{eq:delta_nabla_phi})
into Eq. (\ref{eq:Delta_p_GSF})---we thus get:
\begin{equation}
\Delta\mathtt{p}_{\textrm{self}}^{(\bm{\phi})}=\lambda\frac{2}{\kappa}\intop_{\Delta\mathscr{B}}\bm{\text{\ensuremath{\epsilon}}}_{\mathscr{B}}\,\frac{1}{r}\left(\mathring{u}^{a}\mathring{u}^{b}-2\mathring{\sigma}^{ab}\right)\delta C^{c}\,_{ab}\phi_{c}+\mathcal{O}\left(r\right)\,.\label{eq:Delta_p_GSF_2}
\end{equation}

Let us now look at the contractions in the integrand. For the first
(energy) term, inserting the connection coefficient (\ref{eq:connection_coeff}),
we have by direct computation:
\begin{align}
\mathring{u}^{a}\mathring{u}^{b}\delta C^{c}\,_{ab}\phi_{c}=\, & \mathring{g}^{cd}\left(\mathring{\nabla}_{a}h_{bd}-\frac{1}{2}\mathring{\nabla}_{d}h_{ab}\right)\mathring{u}^{a}\mathring{u}^{b}\phi_{c}\\
=\, & -F^{c}[\bm{h};\mathring{\bm{u}}]\phi_{c}\,,
\end{align}
where the functional $\bm{F}$ is precisely the GSF four-vector functional defined
in the introduction [Eq. \ref{eq:intro_GSF_functional}], and to write the final equality we have used the orthogonality
property $\phi_{\mathring{\bm{u}}}=0$. Thus we see that this is indeed
the term that yields the GSF. For the second (pressure) term in Eq. (\ref{eq:Delta_p_GSF_2}),
we similarly obtain by direct computation:
\begin{equation}
\mathring{\sigma}^{ab}\delta C^{c}\,_{ab}\phi_{c}=2\wp^{c}[\bm{h};\mathring{\bm{\sigma}}]\phi_{c}\,,
\end{equation}
where in expressing the RHS, it is convenient to define a general
functional of two $(0,2)$-tensors similar to the GSF functional:
\begin{equation}
\wp^{c}[\bm{H};\bm{S}]=\frac{1}{2}\mathring{g}^{cd}\left(\mathring{\nabla}_{a}H_{bd}-\frac{1}{2}\mathring{\nabla}_{d}H_{ab}\right)S^{ab}\,.\label{eq:pressure_functional}
\end{equation}
We call this novel term the \emph{gravitational self-pressure force}.

Now we can collect all of the above and insert them into (\ref{eq:Delta_p_GSF_2}).
Before writing down the result, it is convenient to define a total
functional $\bm{\mathcal{F}}$ as the sum of $\bm{F}$ and $\bm{\wp}$,
\begin{equation}
\mathcal{F}^{a}[\bm{h};\mathring{\bm{u}}]=F^{a}[\bm{h};\mathring{\bm{u}}]+\wp^{a}[\bm{h};\mathring{\bm{\sigma}}]\,.\label{eq:generalized_GSF}
\end{equation}
We refer to this as the \emph{extended GSF functional}. Note that
for $\bm{\mathcal{F}}$ we write only the functional dependence on
$\bm{h}$ and $\mathring{\bm{u}}$ since the two-metric $\mathring{\bm{\sigma}}$
is determined uniquely by $\mathring{\bm{u}}$. With this, and setting the perturbation
parameter to unity, Eq. (\ref{eq:Delta_p_GSF_2})
becomes:
\begin{equation}
\boxed{\Delta\mathtt{p}_{\textrm{self}}^{(\bm{\phi})}=-\frac{1}{4\pi}\intop_{\Delta\mathscr{B}}\!\bm{\text{\ensuremath{\epsilon}}}_{\mathscr{B}}\,\frac{1}{r}\bm{\phi}\cdot\boldsymbol{\mathcal{F}}[\bm{h};\mathring{\bm{u}}]+\mathcal{O}\left(r\right)\,.}\label{eq:Delta_p_GSF_general}
\end{equation}

This is to be compared with Gralla's formula [\cite{gralla_gauge_2011}] discussed in the thesis introduction, Eq. (\ref{eq:intro_Gralla}).
While the equivalence thereto is immediately suggestive based on
the general form of our result, we have to do a bit more work to show
that indeed Eqn. (\ref{eq:Delta_p_GSF_general}), both on the LHS
and the RHS, recovers---though in general will, evidently
at least from our novel gravitational self-pressure force, also have extra terms added to---Eq.  (\ref{eq:intro_Gralla}).
We leave this task to the following section, the purpose of which is to
consider in detail the application of our conservation law formulation to a concrete
example of a perturbative family of spacetimes defined for a self-force
analysis, namely the Gralla-Wald family.

Concordantly, we emphasize that the result above  [Eq. (\ref{eq:Delta_p_GSF_general})]
holds for any family of perturbed manifolds $\{\mathscr{M}_{(\lambda)}\}_{\lambda\geq0}$
and is completely independent of the internal description of our system,
\textit{i.e.} the worldtube interior.
In other words, what we have just demonstrated---provided only that
one accepts a quasilocal notion of energy-momentum---is that the (generalized)
GSF is a completely generic perturbative effect in GR for localized
systems: it arises as a linear order contribution of any spacetime
perturbation to the momentum flux of a system in the limit where its
areal radius is small. 

This view of the self-force may cast fresh conceptual light on the
old and seemingly arcane problem of deciphering its physical origin and meaning.
In particular, recall the common view that the GSF is caused by the
backreaction of the ``mass'' of a small object upon its own motion. Yet
what we have seen here is that it is actually the vacuum ``mass'',
or vacuum energy that is responsible for the GSF. We may still regard
the effect as a ``backreaction'', in the sense that it is the boundary metric perturbations of the system---the $\bm{h}$ on $\mathscr{B}$---which
determine its momentum flux, but the point is that this flux is inexorably
present and given by Eq. (\ref{eq:Delta_p_GSF_general}) regardless of
where exactly this $\bm{h}$ is coming from. Presumably, the dominant
part of $\bm{h}$ would arise from the system itself---if we further
assume that the system itself is indeed what is being treated perturbatively
by the family $\{\mathscr{M}_{(\lambda)}\}_{\lambda\geq0}$, as is
the case with typical self-force analyses---but in principle $\bm{h}$
can comprise absolutely any perturbations, \textit{i.e.} its physical origin
doesn't even have to be from inside the system.

In this way, we may regard the GSF as a completely geometrical, purely
general-relativistic backreaction of the mass (and pressure) of the
spacetime vacuum---\emph{not} of the object inside---upon the motion of a localized
system (\textit{i.e.} its momentum flux). This point of view frees
us from having to invoke such potentially ambiguous notions as ``mass
ratios'' (in a two-body system for example), let alone ``Coulombian
$m/r$ fields'', to make basic sense of self-force effects. They
simply---and always---happen from the interaction of the vacuum
with any boundary perturbation, and are  dominant if that boundary
is not too far out. 

\section{Application to the Gralla-Wald approach to the gravitational self-force\label{sec:gralla-wald-analysis}}

In this section we will consider in detail the application of our
ideas to a particular approach to the self-force: that is to say, a particular
specification of $\{(\mathscr{M}_{(\lambda)},\bm{g}_{(\lambda)})\}$ via a few additional assumptions aimed
at encoding the notion of a ``small'' object being ``scaled down''
to zero ``size'' and ``mass'' as $\lambda\rightarrow0$. In other
words, we now identify the perturbation (which has up to this point
been treated completely abstractly) defined by $\{(\mathscr{M}_{(\lambda)},\bm{g}_{(\lambda)})\}$
as actually being that caused by the presence of the ``small'' object:
that could mean regular matter (in particular, a compact object such
as a neutron star) or a black hole. 

The assumptions (on $\{\bm{g}_{(\lambda)}\}$) that we choose to work
with here are those of the approach of [\cite{gralla_rigorous_2008}]. Certainly,
the application of our perturbed quasilocal conservation laws could
just as well be carried out in the context of any other self-force
analysis---such as, \textit{e.g.}, the self-consistent approximation of  [\cite{pound_self-consistent_2010}] (the
mathematical correspondence of which to the Gralla-Wald approach has,
in any case, been shown in  [\cite{pound_gauge_2015}]). 

Our motivation for starting with the Gralla-Wald
approach in particular is two-fold. On the one hand, it furnishes
a mathematically rigorous and physically clear picture (which we show in Fig. \ref{fig-gralla-wald-family})---arguably
more so than any other available GSF treatment---of what it means
to ``scale down'' a small object to zero ``size'' and ``mass''
(or, equivalently, of perturbing any spacetime by the presence of
an object with small ``size'' and ``mass''---we will be more
precise momentarily). On the other hand, it is within this approach
that the formula for the GSF has been obtained (in  [\cite{gralla_gauge_2011}]) as a closed two-surface
(small two-sphere) integral around the object (in \textit{lieu} of evaluating
the GSF at a spacetime point identified as the location of the object),
in the form of the Gralla ``angle averaging'' formula [Eq. (\ref{eq:intro_Gralla})]---with which our
extended GSF formula (\ref{eq:Delta_p_GSF_general}) is to be compared.

In Subsection \ref{ssec:gralla-wald_review}, we provide an overview of the assumptions and consequences
of the Gralla-Wald approach to the GSF. Afterwards, in Subsection \ref{ssec:gralla-wald_rqf_general}, we describe the general embedding of rigid quasilocal frames in the Gralla-Wald family of spacetimes, and then in Subsection \ref{ssec:gralla-wald_rqf_background} we describe their detailed construction in the background spacetime in this family. Having established this, we then proceed to derive equations of motion in two ways. In particular, we carry out the analysis with two separate choices of rigid quasilocal frames (``frames of reference''): first, inertially with the point particle approximation of the moving object in the background in Subsection \ref{ssec:gralla-wald_pp-inertial}, and second, inertially with the object itself in the perturbed spacetime in Subsection \ref{ssec:gralla-wald_sco-inertial}.

\subsection{The Gralla-Wald approach to the GSF\label{ssec:gralla-wald_review}}

The basic idea of  [\cite{gralla_rigorous_2008}] for defining a family $\{(\mathscr{M}_{(\lambda)},\bm{g}_{(\lambda)})\}_{\lambda\geq0}$
such that $\lambda>0$ represents the inclusion of perturbations generated
by a ``small'' object is the following one. One begins by imposing
certain smoothness conditions on $\{\bm{g}_{(\lambda)}\}_{\lambda\geq0}$
corresponding to the existence of certain limits of each $\bm{g}_{(\lambda)}$. In particular, two limits are sought corresponding intuitively to two limiting views of the
system: first, a view from ``far away'' from which the ``motion''
of the (extended but localized) object reduces to a worldline; second,
a view from ``close by'' the object from which the rest of the universe
(and in particular, the MBH it might be orbiting as in an EMRI) looks
``pushed away'' to infinity. A third requirement must be added to this, namely that both
of these limiting pictures nonetheless coexist in the same spacetime,
\textit{i.e.} the two limits are smoothly related (or, in other words, there is no pathological behaviour when taking these limits along different directions). While in principle this
may sound rather technical, one can actually motivate each of these
conditions with very sensible physical arguments as we shall momentarily
elaborate further upon. From them, Gralla and Wald have shown [\cite{gralla_rigorous_2008}] that
it is possible to derive a number of consequences, including geodesic
motion in the background at zeroth order and the MiSaTaQuWa equation [\cite{mino_gravitational_1997,quinn_axiomatic_1997}], Eq. (\ref{eq:intro_MiSaTaQuWa}),
for the GSF at first order in $\lambda$.

\begin{figure}
\begin{centering}
\includegraphics[scale=0.6]{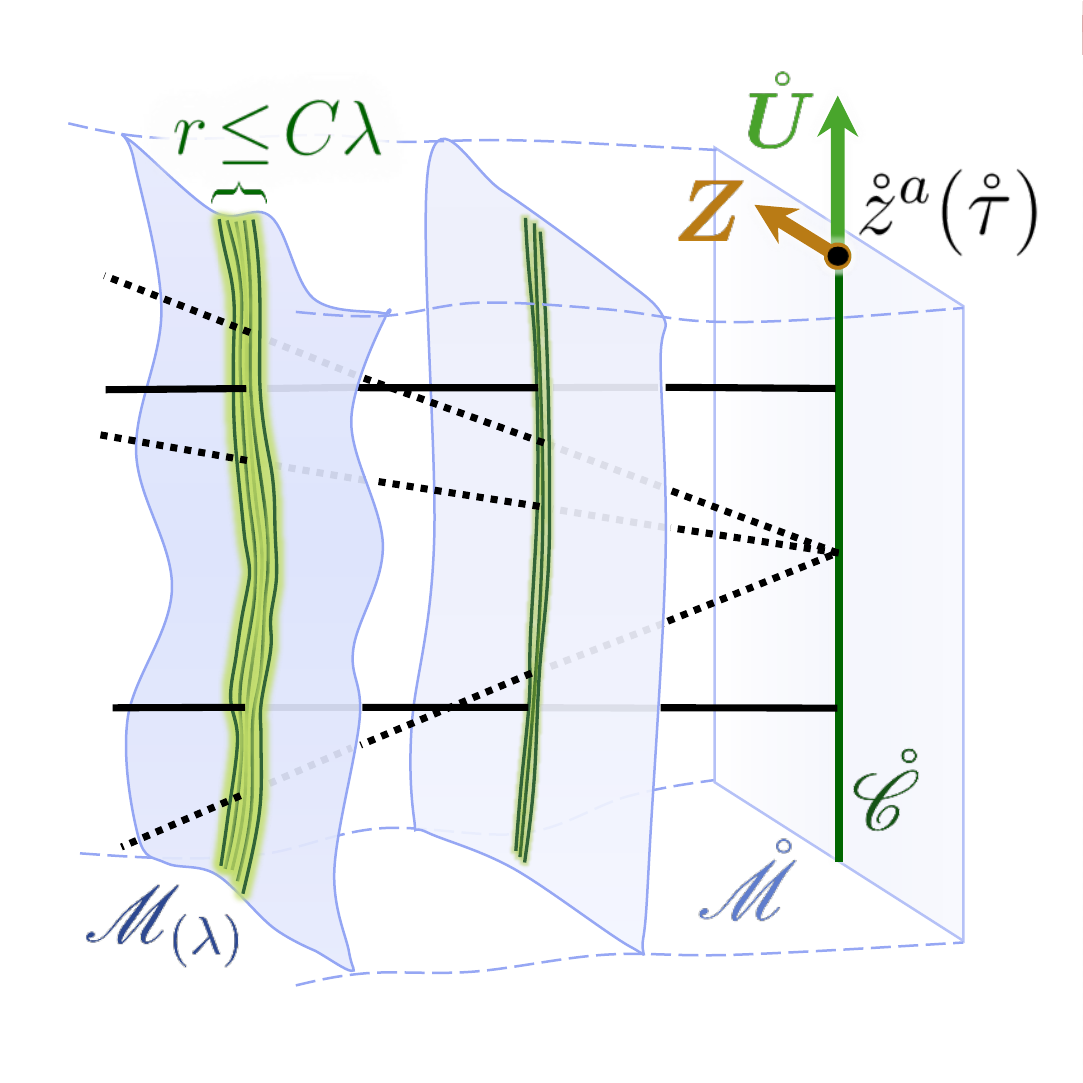}
\par\end{centering}
\caption{Representation of the Gralla-Wald family of spacetimes $\{\mathscr{M}_{(\lambda)}\}_{\lambda\geq0}$. (This is an adaptation of Fig. 1 of [\cite{gralla_rigorous_2008}].)
The lined green region that ``fills in'' $\mathscr{M}_{(\lambda)}$
for $r\leq C\lambda$ is the ``small'' object which ``scales down''
to zero ``size'' and ``mass'' in the background $\mathring{\mathscr{M}}$.
The solid black lines represent taking the ``ordinary limit'' (the
``far away'' view where the motion appears reduced to a worldline)
and the dashed black lines the ``scaled limit'' (the ``close by''
view where the rest of the universe appears ``pushed away'' to infinity).
The worldline $\mathring{\mathscr{C}}$, which can be proven to be
a geodesic, is parametrized by $\mathring{z}^{a}(\mathring{\tau})$
and has four-velocity $\mathring{\bm{U}}$. The deviation vector
$\bm{Z}$ on $\mathring{\mathscr{C}}$ is used for formulating the
first-order correction to the motion.}\label{fig-gralla-wald-family}
\end{figure}

Let us now be more precise. Let $\{(\mathscr{M}_{(\lambda)},\bm{g}_{(\lambda)})\}_{\lambda\geq0}$
be a perturbative one-parameter family of spacetimes as in the previous
section. We assume that $\{\bm{g}_{(\lambda)}\}_{\lambda\geq0}$ satisfies
the following conditions, depicted visually in Fig. \ref{fig-gralla-wald-family}:

\bigskip{}

\noindent \emph{(i) Existence of an ``ordinary limit'':} There exist
coordinates $\{x^{\alpha}\}$ in $\mathscr{M}_{(\lambda)}$ such that
$g_{\beta\gamma}^{(\lambda)}(x^{\alpha})$ is jointly smooth in $(\lambda,x^{\alpha})$
for $r>C\lambda$ where $C>0$ is a constant and $r=(x_{i}x^{i})^{1/2}$.
For all $\lambda\geq0$ and $r>C\lambda$, $\bm{g}_{(\lambda)}$ is
a vacuum solution of the Einstein equation. Furthermore, $\mathring{g}_{\beta\gamma}(x^{\alpha})$
is smooth in $x^{\alpha}$ including at $r=0$, and the curve $\mathring{\mathscr{C}}=\{r=0\}\subset\mathring{\mathscr{M}}$
is timelike.

\bigskip{}

\noindent \emph{(ii) Existence of a ``scaled limit'':} For all $t_{0}$,
define the ``scaled coordinates'' $\{\bar{x}^{\alpha}\}=\{\bar{t},\bar{x}^{i}\}$
by $\bar{t}=(t-t_{0})/\lambda$ and $\bar{x}^{i}=x^{i}/\lambda$.
Then the ``scaled metric'' $\bar{g}_{\bar{\beta}\bar{\gamma}}^{(\lambda)}(t_{0};\bar{x}^{\alpha})=\lambda^{-2}g_{\bar{\beta}\bar{\gamma}}^{(\lambda)}(t_{0};\bar{x}^{\alpha})$
is jointly smooth in $(\lambda,t_{0};\bar{x}^{\alpha})$ for $\bar{r}=r/\lambda>C$.

\bigskip{}

\noindent \emph{(iii) Uniformity condition:} Define $A=r$, $B=\lambda/r$
and $n^{i}=x^{i}/r$. Then each $g_{\beta\gamma}^{(\lambda)}(x^{\alpha})$
is jointly smooth in $(A,B,n^{i},t)$.

\bigskip{}

Mathematically, the first two conditions respectively ensure the existence
of an appropriate Taylor expansion (in $r$ and $\lambda$) of the
metric in a ``far zone'' (on length scales comparable with the mass
of the MBH in an EMRI, $r\sim M$) and a ``near zone'' (on length
scales comparable with the mass of the object, $r\sim m$) . Meanwhile,
the third is simply a consistency requirement ensuring the existence
of a ``buffer zone'' ($m\ll r\ll M$) where both expansions are
valid. (This idea is in many ways similar to the method of ``matched asymptotic expansions'' [\cite{mino_gravitational_1997}]). 

From a physical point of view, what is happening in the first (``ordinary'')
limit is that the body is shrinking down to a worldline $\mathring{\mathscr{C}}$
with its ``mass'' (understood as defining the perturbation) going
to zero at least as fast as its radius. (As we increase the perturbative
parameter $\lambda$ from zero, the radius is not allowed to grow
faster than linearly with $\lambda$; viewed conversely, this condition
ensures that the object does not collapse to a black hole if it was
not one already before reaching the point particle limit.) In the
second (``scaled'') limit, the object is shrinking down to zero
size in an asymptotically self-similar manner (its mass is proportional
to its size, and its ``shape'' is not changing). Finally, the uniformity
condition ensures that there are no ``bumps of curvature'' in the
one-parameter family. (Essentially, this guarantees that there are
no inconsistencies in evaluating the limits along different directions.)

From these assumptions alone, [\cite{gralla_rigorous_2008}] are able to derive the
following consequences: 

\bigskip{}

\noindent \emph{(a) Background motion:} The worldline $\mathring{\mathscr{C}}$
is a geodesic in $\mathring{\mathscr{M}}$; writing its parametrization
in terms of proper time $\mathring{\tau}$ as $\mathring{\mathscr{C}}=\{\mathring{z}^{a}(\mathring{\tau})\}_{\mathring{\tau}\in\mathbb{R}}$
and denoting its four-velocity by $\mathring{U}^{a}={\rm d}\mathring{z}^{a}(\mathring{\tau})/{\rm d}\mathring{\tau}$,
this means that
\begin{equation}
\mathring{\nabla}_{\mathring{\bm{U}}}\mathring{\bm{U}}=0\,.\label{eq:background_geodesic}
\end{equation}

\bigskip{}

\noindent \emph{(b) Background ``scaled'' metric:} $\mathring{\bar{\bm{g}}}$
is stationary and asymptotically flat.

\bigskip{}

\noindent \emph{(c) First-order field equation:} At $\mathcal{O}(\lambda)$,
the Einstein equation is sourced by the matter energy-momentum tensor of a point
particle $\bm{T}^{\textrm{PP}}$ supported on $\mathring{\mathscr{C}}$,
\textit{i.e.} the field equation is
\begin{equation}
\delta G_{ab}\left[\bm{h}\right]=\kappa T_{ab}^{\textrm{PP}}\label{eq:first-order_Einstein_equation}
\end{equation}
where
\begin{equation}
T_{ab}^{\textrm{PP}}=m\intop_{\mathring{\mathscr{C}}}{\rm d}\mathring{\tau}\,\mathring{U}_{a}\left(\mathring{\tau}\right)\mathring{U}_{b}\left(\mathring{\tau}\right)\delta_{4}\left(x^{c}-z^{c}\left(\mathring{\tau}\right)\right)\,.\label{eq:T_ab^PP}
\end{equation}
Here, $m$ is a constant along $\mathring{\mathscr{C}}$ and is interpreted
as representing the mass of the object---or, more precisely,
the mass of the point particle which approximates the object in the
background. (This is a subtle point that should be kept in mind, and which will be better elucidated in our analysis further on.)

\bigskip{}

\noindent \emph{(d) First-order equation of motion:} At $\mathcal{O}(\lambda)$,
the correction to the motion in the Lorenz gauge---corresponding
to the choice of a certain gauge vector $\bm{\mathsf{L}}\in T\mathscr{N}$
defined by the condition
\begin{equation}
\mathring{\nabla}^{b}(h_{ab}^{\bm{\mathsf{L}}}-\frac{1}{2}h^{\bm{\mathsf{L}}}\mathring{g}_{ab})=0\,,
\end{equation}
where $h={\rm tr}(\bm{h})$---is given by the MiSaTaQuWa equation [\cite{mino_gravitational_1997,quinn_axiomatic_1997}],
\begin{equation}
\mathring{\nabla}_{\mathring{\bm{U}}}\mathring{\nabla}_{\mathring{\bm{U}}}Z^{a}=-\mathring{E}_{b}\,^{a}Z^{b}+F^{a}[\bm{h}^{\textrm{tail}};\mathring{\bm{U}}]\,,\label{eq:MiSaTaQuWa}
\end{equation}
where $\mathring{E}_{b}\,^{a}=\mathring{R}_{cbd}\,^{a}\mathring{U}^{c}\mathring{U}^{d}$ is the electric part of the Weyl tensor
and $\bm{h}^{\textrm{tail}}$ is a ``tail'' integral of the retarded Green's
functions of $\bm{h}$. The above is an equation for a four-vector
$\bm{Z}$ called the ``deviation'' vector; the LHS is the acceleration
associated therewith and the RHS is a geodesic deviation term plus
the GSF. This deviation vector is defined on $\mathring{\mathscr{C}}$
and represents the first-order correction needed to move off $\mathring{\mathscr{C}}$
and onto the worldline representing the ``center of mass'' of the
perturbed spacetime, defined as in the Hamiltonian analysis of  [\cite{regge_role_1974}].\bigskip{}

Let us make a few comments on these results, specifically concerning
\emph{(a)} and \emph{(c)}. On the one hand, it is quite remarkable
that geodesic motion can be recovered as a consequence\footnote{~See again footnote 2 and the references mentioned therein for more on this topic.} of this analysis---\textit{i.e.}
without having to posit it as an assumption---just from smoothness
properties (existence of appropriate limits) of our family of metrics
$\{\bm{g}_{(\lambda)}\}$; and on the other, this analysis offers
sensible meaning to the usual ``delta function cartoon'' (ubiquitous
in essentially all self-force analyses) of the matter stress-energy-momentum tensor describing
the object in the background spacetime. The point is that the description
of the object is completely arbitrary inside the region that is not
covered by the smoothness conditions of the family $\{\bm{g}_{(\lambda)}\}$,
\textit{i.e.} for $r\leq C\lambda$ when $\lambda>0$. (Indeed, this region
can be ``filled in'' even with exotic matter, \textit{e.g.} failing to satisfy
the dominant energy condition, or a naked singularity, as long as
a well-posed initial value formulation exists.) Regardless of what
this description is, the smoothness conditions essentially ensure
that its ``reduction'' to $\mathring{\mathscr{M}}$ (or, more precisely,
the transport of any effect thereof with respect to the family $\{\bm{g}_{(\lambda)}\}$)
simply becomes that of a point particle sourcing the field equation
at $\mathcal{O}(\lambda)$. In this way, the background ``point particle
cartoon'' is justified as the simplest possible idealization of a
``small'' object.

What we are going to do, essentially, is to accept consequences \emph{(a)}-\emph{(c)}
(in fact, we will not even explicitly need \emph{(b)}), the proofs
of which do not rely upon any further limiting conditions such as
a restriction of the perturbative gauge, and to obtain, using our
perturbed momentum conservation law, a more general version of the
EoM, \textit{i.e.} consequence \emph{(d)}. For the latter, [\cite{gralla_rigorous_2008}]
instead rely on the typical but laborious Hadamard expansion techniques of [\cite{dewitt_radiation_1960}],
wherein the ``mass dipole moment'' of the object is set to zero.
It is possible [\cite{regge_role_1974}] to have such a notion in a well-defined Hamiltonian
sense by virtue of \emph{(b)}. While mathematically rigorous and conducive
to obtaining the correct known form of the MiSaTaQuWa equation, their
derivation and final result suffer not only from the limitation of
having to fix the perturbative gauge, but also from the (as we shall
see, potentially avoidable) technical complexity of arriving at the
final answer---including the evaluation of $\boldsymbol{h}^{\textrm{tail}}$
(or otherwise taking recourse to a regularization procedure).

The link between this approach and our conservation law derivation
of the EoM which we are about to carry out is established by the work
of [\cite{gralla_gauge_2011}], who discovered that Eq. (\ref{eq:MiSaTaQuWa}) can be equivalently
written as:
\begin{equation}
\mathring{\nabla}_{\mathring{\bm{U}}}\mathring{\nabla}_{\mathring{\bm{U}}}Z^{a}=-\mathring{E}_{b}\,^{a}Z^{b}+\frac{1}{4\pi}\lim_{r\rightarrow0}\intop_{\mathbb{S}_{r}^{2}}\bm{\epsilon}^{}_{\mathbb{S}^{2}}\,F^{a}[\mathring{\bm{U}},\bm{h}]\,.\label{eq:Gralla_ange-average}
\end{equation}
Here, the GSF term $\bm{F}[\bm{h}^{\textrm{tail}};\mathring{\bm{U}}]$
in the MiSaTaQuWa equation [Eq. (\ref{eq:MiSaTaQuWa})] is substituted by
an integral expression---an average over the angles---of $\bm{F}$.
In particular (as, strictly speaking, one cannot define integrals of vectors as such), this is evaluated by using the exponential map based on
$\mathring{\mathscr{C}}$ to associate a flat metric, in terms of
which the integration is performed over a two-sphere of radius $r$,
$\mathbb{S}_{r}^{2}$, with $\bm{\epsilon}^{}_{\mathbb{S}^{2}}$ denoting the volume
form of $\mathbb{S}^{2}$. 

Observe that, here, the functional dependence of $\bm{F}$ is on $\bm{h}$
itself (and not on $\bm{h}^{\textrm{tail}}$ or any sort of regularized
$\bm{h}$) and for this reason is referred to as the ``bare'' GSF.
Moreover, this formula is actually valid in a wider class of
gauges than just the Lorenz gauge: in particular, it holds in what
are referred to as ``parity-regular'' gauges [\cite{gralla_gauge_2011}]. We refrain from entering
here into the technical details of exactly how such gauges are defined,
except to say that the eponymous ``parity condition'' that they
need to satisfy has its ultimate origin in the Hamiltonian analysis
of [\cite{regge_role_1974}] and is imposed so as to make certain Hamiltonian
definitions---and in particular for Gralla's analysis [\cite{gralla_gauge_2011}], the Hamiltonian
``center of mass''---well defined. These, however, are \emph{not} limitations
of our quasilocal formalism, where we know how to define energy-momentum notions
more generally than any Hamiltonian approach. Thus, in our result,
there will be \emph{no restriction} on the perturbative gauge. This may constitute
a great advantage, as the ``parity-regular'' gauge class---though
an improvement from being limited to the Lorenz gauge in formulating
the EoM---still excludes entire classes of perturbative gauges convenient for formulating black hole perturbation theory (\textit{e.g.} the Regge-Wheeler gauge in Schwarzschild-Droste) and hence for carrying out practical EMRI
calculations.

We proceed to apply our quasilocal analysis to the Gralla-Wald family of spacetimes, beginning with a general setup in this family of rigid quasilocal frames.

\subsection{General setup of rigid quasilocal frames in the Gralla-Wald family\label{ssec:gralla-wald_rqf_general}}

Let $(\mathscr{B}_{(\lambda)};\bm{u}_{(\lambda)})$ be a quasilocal frame in $(\mathscr{M}_{(\lambda)},\bm{g}_{(\lambda)},\bm{\nabla}_{(\lambda)})$,
for any $\lambda>0$, constructed just as described in Section \ref{sec:qf}: with unit four-velocity $\bm{u}_{(\lambda)}$, unit normal
$\bm{n}_{(\lambda)}$, induced metric $\bm{\gamma}_{(\lambda)}$ and
so on. Using the fact that the tensor transport is linear and commutes
with tensor products, we can compute the transport (in the five-dimensional ``stacked'' manifold $\mathscr{N}=\mathscr{M}_{(\lambda)}\times\mathbb{R}^{\geq}$ used in our perturbative setup, as in Subsection \ref{sec:3.1-general-perturbations}) of any geometrical
quantity of interest to the background. For example,
\begin{align}
\gamma_{ab}=\, & \varphi_{(\lambda)}^{*}\gamma_{ab}^{(\lambda)}\\
=\, & \varphi_{(\lambda)}^{*}(g_{ab}^{(\lambda)}-n_{a}^{(\lambda)}n_{b}^{(\lambda)})\\
=\, & g_{ab}-n_{a}n_{b}\\
=\, & \mathring{\gamma}_{ab}+\lambda\delta\gamma_{ab}+\mathcal{O}\left(\lambda^{2}\right)\,,
\end{align}
where
\begin{align}
\mathring{\gamma}_{ab}=\, & \mathring{g}_{ab}-\mathring{n}_{a}\mathring{n}_{b}\,,\\
\delta\gamma_{ab}=\, & h_{ab}-2\mathring{n}_{(a}\delta n_{b)}\,.
\end{align}
Similarly, 
\begin{equation}
\sigma_{ab}=\mathring{\sigma}_{ab}+\lambda\delta\sigma_{ab}+\mathcal{O}(\lambda^{2})\,,
\end{equation}
 where
\begin{align}
\mathring{\sigma}_{ab}=\, & \mathring{\gamma}_{ab}+\mathring{u}_{a}\mathring{u}_{b}\,,\\
\delta\sigma_{ab}=\, & \delta\gamma_{ab}+2\mathring{u}_{(a}\delta u_{b)}\,.
\end{align}

Now let us assume that $(\mathscr{B}_{(\lambda)};\bm{u}_{(\lambda)})$
is a rigid quasilocal frame, meaning that the congruence defining it has a vanishing
symmetrized strain rate tensor in $\mathscr{M}_{(\lambda)}$,
\begin{equation}
\theta_{(ab)}^{(\lambda)}=0\,.\label{eq:RQF_lambda}
\end{equation}

Let $\mathscr{B}=\varphi_{(\lambda)}^{-1}(\mathscr{B}_{(\lambda)})$
be the inverse image of $\mathscr{B}_{(\lambda)}$ in the background
$\mathring{\mathscr{M}}$, with $\boldsymbol{u}=\varphi_{(\lambda)}^{*}\boldsymbol{u}_{(\lambda)}=\mathring{\bm{u}}+\lambda\delta\bm{u}+\mathcal{O}(\lambda^{2})$
giving the transport of the quasilocal observers' four-velocity, $\bm{n}=\mathring{\bm{n}}+\lambda\delta\bm{n}+\mathcal{O}(\lambda^{2})$
the unit normal and so on. In other words, $(\mathscr{B};\bm{u})$
is the background mapping of the perturbed congruence $(\mathscr{B}_{(\lambda)};\bm{u}_{(\lambda)})$,
and so  will itself constitute a congruence (in the background),
\textit{i.e.} a quasilocal frame defined by a two-parameter family of worldlines with unit
four-velocity $\bm{u}$ in $\mathring{\mathscr{M}}$. 

However, although $(\mathscr{B}_{(\lambda)};\bm{u}_{(\lambda)})$
is a rigid quasilocal frame in $\mathscr{M}_{(\lambda)}$, $(\mathscr{B};\bm{u})$ is
\emph{not} in general a rigid quasilocal frame in $\mathring{\mathscr{M}}$ (with
respect to the background metric $\mathring{\bm{g}}$). One can see
this easily as follows. Let $\bm{\vartheta}\in\mathscr{T}^{0}\,_{2}(\mathring{\mathscr{M}})$
be the strain rate tensor of $(\mathscr{B};\bm{u})$, so that it is
given by 
\begin{equation}
\vartheta_{ab}=\sigma_{ca}\sigma_{bd}\mathring{\nabla}^{c}u^{d}\,.
\end{equation}
The RHS is an series in $\lambda$, owing to the fact that $\bm{u}$
(and therefore $\bm{\sigma}$, the two-metric on the space $\mathscr{H}$ orthogonal to $\bm{u}$ in $\mathscr{B}$) are transported from
a perturbed congruence in $\mathscr{M}_{(\lambda)}$. Upon expansion
we obtain
\begin{equation}
\vartheta_{ab}=\mathring{\vartheta}_{ab}+\lambda\delta\vartheta_{ab}+\mathcal{O}\left(\lambda^{2}\right)\,,\label{eq:vartheta_espansion}
\end{equation}
where
\begin{equation}
\mathring{\vartheta}_{ab}=\mathring{\sigma}_{c(a}\mathring{\sigma}_{b)d}\mathring{\nabla}^{c}\mathring{u}^{d}
\end{equation}
is just the strain rate tensor of the background congruence---\textit{i.e.}
the congruence defined by $\mathring{\bm{u}}$---and 
\begin{equation}
\delta\vartheta_{ab}=2\mathring{\sigma}^{(c}\,_{(a}\delta\sigma^{d)}\,_{b)}\mathring{\nabla}_{c}\mathring{u}_{d}+\mathring{\sigma}^{c}\,_{(a}\mathring{\sigma}_{b)d}\mathring{\nabla}_{c}\delta u^{d}
\end{equation}
is the first-order term in $\lambda$. Note that we are abusing our
established notation slightly in writing Eq. (\ref{eq:vartheta_espansion}),
as there exists no $\bm{\vartheta}_{(\lambda)}$ in $\mathscr{M}_{(\lambda)}$
the transport (to $\mathring{\mathscr{M}}$) of which yields such
a series expansion; instead $\bm{\vartheta}$ is defined directly
on $\mathring{\mathscr{M}}$ (relative to the metric $\mathring{\bm{g}}$)
as the strain rate tensor of a conguence with four-velocity $\bm{u}$---which
itself contains the expansion in $\lambda$.

Now let us compute the transport of the rigidity condition on $(\mathscr{B}_{(\lambda)};\bm{u}_{(\lambda)})$
[Eq. (\ref{eq:RQF_lambda})] to $\mathring{\mathscr{M}}$: we have
\begin{align}
0=\, & \varphi_{(\lambda)}^{*}\theta_{(ab)}^{(\lambda)}\\
=\, & \varphi_{(\lambda)}^{*}(\sigma_{c(a}^{(\lambda)}\sigma_{b)d}^{(\lambda)}\nabla_{(\lambda)}^{c}u_{(\lambda)}^{d})\\
=\, & \sigma_{c(a}\sigma_{b)d}\nabla^{c}u^{d}\\
=\, & \mathring{\theta}_{(ab)}+\lambda\delta\theta_{(ab)}+\mathcal{O}(\lambda^{2})\,,\label{eq:theta_expansion}
\end{align}
where
\begin{align}
\mathring{\theta}_{(ab)}=\, & \mathring{\vartheta}_{(ab)}\,,\\
\delta\theta_{(ab)}=\, & \delta\vartheta_{(ab)}+\mathring{\sigma}^{c}\,_{(a}\mathring{\sigma}_{b)d}\delta C^{d}\,_{ce}\mathring{u}^{e}\,.
\end{align}
Since $0=\theta_{(ab)}^{(\lambda)}$ identically in $\mathscr{M}_{(\lambda)}$
(as we demand that $(\mathscr{B}_{(\lambda)};\bm{u}_{(\lambda)})$
is a rigid quasilocal frame), Eq. (\ref{eq:theta_expansion}) must vanish order by order
in $\lambda$. That implies, in particular, that the zeroth-order
congruence (defined by $\mathring{\bm{u}}$) is a rigid quasilocal frame, and that the
symmetrized strain rate tensor of the background-mapped perturbed
congruence (defined by $\bm{u}$) is given by
\begin{equation}
\vartheta_{(ab)}=-\lambda\mathring{\sigma}^{c}\,_{(a}\mathring{\sigma}_{b)d}\delta C^{d}\,_{ce}\mathring{u}^{e}+\mathcal{O}\left(\lambda^{2}\right)\,.
\end{equation}

This tells us that the deviation from rigidity of $(\mathscr{B};\bm{u})$
in $\mathring{\mathscr{M}}$ occurs only at $\mathcal{O}(\lambda)$
(and, in particular, is caused by the same perturbed connection coefficient
term that is responsible for the GSF). In other words, we can treat
$(\mathscr{B};\bm{u})$ as a rigid quasilocal frame at zeroth order. This
zeroth order congruence actually makes up a different worldtube boundary
$\mathring{\mathscr{B}}\neq \mathscr{B}$ in $\mathring{\mathscr{M}}$, \textit{i.e.} one defined
by a congruence with four-velocity $\mathring{\bm{u}}\neq\bm{u}$ in general. Clearly,
for a rigid quasilocal frame with a small areal radius $r$ constructed around a worldline $\mathscr{G}$
in $\mathring{\mathscr{M}}$ with four-velocity $\bm{U}_{\mathscr{G}}$,
we would simply have $\mathring{\bm{u}}=\bm{U}_{\mathscr{G}}$ (where
the RHS is understood to be transported off $\mathscr{G}$ and onto
$\mathring{\mathscr{B}}$ via the exponential map), and $\mathring{\bm{\sigma}}=r^{2}\bm{\mathfrak{S}}$,
\textit{i.e.} it is the metric of $\mathbb{S}_{r}^{2}$. This is the most trivial
possible rigid quasilocal frame: at any instant of time, a two-sphere worth of quasilocal
observers moving with the same four-velocity as is the point at its
center (parametrizing the given worldline).

At first order, the equation $0=\delta\theta_{(ab)}$ can be regarded
as the constraint on the linear perturbations ($\delta\bm{u}$) in
the motion of the quasilocal observers in terms of the metric perturbations
guaranteeing that the perturbed congruence is rigid in the perturbed
spacetime. (So presumably, going to $n$-th order in $\lambda$ would
yield equations for every term up to the $n$-th order piece of the
motion of the quasilocal observers, $\delta^{n}\bm{u}$.)

Now recall the momentum conservation
law for rigid quasilocal frames, Eq. (\ref{eq:cons_law_Pa}). This holds for $(\mathscr{B}_{(\lambda)};\bm{u}_{(\lambda)})$
in $\mathscr{M}_{(\lambda)}$. Just as we did in the previous section
with the general conservation law, we can use $\varphi_{(\lambda)}$
to turn this into an equation in $\mathring{\mathscr{M}}$:
\begin{equation}
\Delta\mathtt{p}^{(\bm{\phi})}=-\intop_{\Delta\mathscr{B}}\varphi_{(\lambda)}^{*}\bm{\epsilon}_{\mathscr{B}_{(\lambda)}}\left(\mathcal{E}\alpha_{\bm{\phi}}+2\nu\epsilon_{ab}\mathcal{P}^{a}\phi^{b}+{\rm P}\bm{D}\cdot\bm{\phi}\right)\,.
\end{equation}

Let us now further assume that we can ignore the Jacobian determinant
(discussed in the previous section) as well as the shift $\bm{v}$
of the quasilocal observers (relative to constant time surfaces).
Then, dividing the above equation by $\Delta t$, where $t$ represents
the adapted time coordinate on $\mathscr{B}$, and taking the
$\Delta t\rightarrow0$ limit, we get the time rate of change of the
momentum,
\begin{equation}
\dot{\mathtt{p}}^{(\bm{\phi})}=-\intop_{\mathscr{S}}\bm{\epsilon}^{\,}_{\mathscr{S}}N\tilde{\gamma}\left(\mathcal{E}\alpha_{\bm{\phi}}+2\nu\epsilon_{ab}\mathcal{P}^{a}\phi^{b}+{\rm P}\bm{D}\cdot\bm{\phi}\right)\,.\label{eq:dpdt_RQF}
\end{equation}
where $\dot{\mathtt{p}}^{(\bm{\phi})}={\rm d}\mathtt{p}^{(\bm{\phi})}/{\rm d}t$,
and we must keep in mind that the derivative is with respect to the
adapted time on (the inverse image on the background of) our congruence.

\subsection{Detailed construction of background rigid quasilocal frames\label{ssec:gralla-wald_rqf_background}}

Let $\mathscr{G}$ be any timelike worldline in $\mathring{\mathscr{M}}$.
Any background metric $\mathring{\bm{g}}$ on $\mathring{\mathscr{M}}$
in a neighborhood of $\mathscr{G}$ admits an expression in Fermi
normal coordinates [\cite{misner_gravitation_1973,poisson_motion_2011}], which we label by $\{X^{\alpha}\}=\{T=X^{0},X^{I}\}_{I=1}^{3}$,
as a power series in the areal radius. Denoting by $A_{K}(T)$ and
$W_{K}(T)$ the proper acceleration and proper rate of rotation of
the spatial axes (triad) along $\mathscr{G}$ (as functions of the
proper time $T$ along $\mathscr{G}$), respectively, this is given
by:
\begin{align}
\mathring{g}_{00}=\, & -\left(1+A_{K}X^{K}\right)^{2}+R^{2}W_{K}W_{L}P^{KL}-\mathring{R}_{0K0L}X^{K}X^{L}+\mathcal{O}\left(R^{3}\right)\,,\label{eq:FNC_g00}\\
\mathring{g}_{0J}=\, & \epsilon_{JKL}W^{K}X^{L}-\frac{2}{3}\mathring{R}_{0KJL}X^{K}X^{L}+\mathcal{O}\left(R^{3}\right)\,,\label{eq:FNC_g0J}\\
\mathring{g}_{IJ}=\, & \delta_{IJ}-\tfrac{1}{3}\mathring{R}_{IKJL}X^{K}X^{L}+\mathcal{O}\left(R^{3}\right)\,,\label{eq:FNC_gIJ}
\end{align}
where $R^{2}=\delta_{IJ}X^{I}X^{J}$ is the square of the radius in
these coordinates (not the square of the Ricci scalar) and $P^{KL}=\delta^{KL}-X^{K}X^{L}/R^{2}$
projects vectors perpendicular to the radial direction $X^{I}/R$. Here we have
to remember that the Riemann tensor $\mathring{R}_{IJKL}$ (along with
$\bm{A}$ and $\bm{W}$) are understood to be evaluated on $\mathscr{G}$.

For all cases that we will be interested in, we will ignore the possibility
of rotation so we set $W_{I}=0$ from now on.

Let us now assume that our background rigid quasilocal frame $(\mathring{\mathscr{B}};\mathring{\bm{u}})$
is constructed around $\mathscr{G}$: that is to say, into this coordinate
system there is embedded a two-parameter family of worldlines representing
a topological two-sphere worth of observers, \textit{i.e.} a fibrated timelike worldtube
$\mathring{\mathscr{B}}$ surrounding $\mathscr{G}$. This may be
conveniently described, as detailed in Subsection \ref{ssec:qf_math}, by defining a new set
of coordinates $\{x^{\alpha}\}=\{t,r,x^{\mathfrak{i}}\}_{\mathfrak{i}=1}^{2}$
given simply by the adapted coordinates $\{t,x^{\mathfrak{i}}\}_{\mathfrak{i}=1}^{2}$
on $\mathring{\mathscr{B}}$ supplemented with a radial coordinate
$r$. Then denoting $\{x^{\mathfrak{i}}\}=\{\theta,\phi\}$ we introduce,
as done in previous calculations with rigid quasilocal frames in Fermi normal coordinates [\cite{epp_existence_2012}], the following coordinate
transformation: 
\begin{align}
T\left(t,r,\theta,\phi\right)=\, & t+\mathcal{O}\left(r^{2},\mathcal{R}\right),\,\label{eq:FNC_T}\\
X^{I}\left(t,r,\theta,\phi\right)=\, & rr^{I}\left(\theta,\phi\right)+\mathcal{O}\left(r^{2},\mathcal{R}\right),\,\label{eq:FNC_XI}
\end{align}
where 
\begin{equation}
r^{I}(\theta,\phi)=(\sin\theta\cos\phi,\sin\theta\sin\phi,\cos\theta)
\end{equation}
are the standard direction cosines of a radial unit vector in spherical
coordinates in $\mathbb{R}^{3}$, and $\mathcal{R}$ here represents the
order of the perturbations of the quasilocal frame away from the round two-sphere
due to the background curvature effects. In particular, for rigid quasilocal frames, we know that this is in fact simply the order of the Riemann
tensor on $\mathscr{G}$, \textit{i.e.} $\mathring{R}_{IJKL}=\mathcal{O}(\mathcal{R})$.
Thus, one may ultimately desire to take $\mathcal{O}(\mathcal{R})$
effects into account for a full calculation, but for the moment---since,
in principle, this $\mathcal{R}$ is unrelated to $\lambda$ and we
can assume it to be subdominant thereto---we simply omit them. Thus
we can simply take $\mathring{\mathscr{S}}=\mathbb{S}^{2}_{r}$, and we
can assume that there is no shift, so that $\tilde{\gamma}=1$.

Applying the coordinate transformation in Eqs. (\ref{eq:FNC_T})-(\ref{eq:FNC_XI})
to the background metric given by Eqs. (\ref{eq:FNC_g00})-(\ref{eq:FNC_gIJ}) with $\bm{W}=0$, and then
using all of the definitions that we have established so far, it is
possible to obtain by direct computation all of the quantities appearing in the integrand of the conservation law [Eq. (\ref{eq:dpdt_RQF})] as series in $r$. We display the results
only up to leading order in $r$, including the possibility of setting
$\bm{A}=0$:
\begin{align}
\mathring{N}=\, & 1+rA_{I}r^{I}+\frac{1}{2}r^{2}\mathring{E}_{IJ}r^{I}r^{J}+\mathcal{O}\left(r^{3}\right)\,,\label{eq:N0}\\
\mathring{\mathcal{E}}=\, & \mathcal{E}_{\textrm{vac}}+\mathcal{O}\left(r\right)\nonumber \\
=\, & -\frac{2}{\kappa r}+\mathcal{O}\left(r\right)\,,\label{eq:E0}\\
\mathring{\alpha}_{\mathfrak{i}}=\, & rA_{I}\mathfrak{B}_{\mathfrak{i}}^{I}+r^{2}\left(\mathring{E}_{IJ}-A_{I}A_{J}\right)\mathfrak{B}_{\mathfrak{i}}^{I}r^{J}+\mathcal{O}\left(r^{3}\right)\,,\label{eq:alpha0}\\
\mathring{\nu}=\, & -r\mathring{B}_{IJ}r^{I}r^{J}+\mathcal{O}\left(r^{2}\right)\,,\label{eq:nu0}\\
\mathring{\mathcal{P}}_{\mathfrak{i}}=\, & -\frac{1}{\kappa}r^{2}\mathring{B}_{IJ}\mathfrak{R}_{\mathfrak{i}}^{I}r^{J}+\mathcal{O}\left(r^{3}\right)\,,\label{eq:P0_i}\\
\mathring{{\rm P}}=\, & {\rm P}_{\textrm{vac}}-\frac{1}{\kappa}A_{I}r^{I}+\mathcal{O}\left(r\right)\nonumber \\
=\, & -\frac{1}{\kappa r}-\frac{1}{\kappa}A_{I}r^{I}+\mathcal{O}\left(r\right)\,.\label{eq:P0}
\end{align}
Here, $\mathring{E}_{IJ}=\mathring{C}_{0I0J}|_{\mathscr{G}}$ and
$\mathring{B}_{IJ}=\frac{1}{2}\epsilon_{I}\,^{KL}\mathring{C}_{0JKL}|_{\mathscr{G}}$
are respectively the electric and magnetic parts of the Weyl tensor evaluated on
on $\mathscr{G}$. Also, $\mathfrak{B}_{\mathfrak{i}}^{I}=\partial_{\mathfrak{i}}r^{I}$
and $\mathfrak{R}_{\mathfrak{i}}^{I}=\epsilon_{\,\,\mathfrak{i}}^{\mathbb{S}^2}\,^{\mathfrak{j}}\mathfrak{B}_{\mathfrak{j}}^{I}$
are respectively the boost and rotation generators of $\mathbb{S}^{2}$. See this chapter's appendix (Section \ref{sec:5.A}) for more technical details on this.
We remind the reader that $\mathcal{E}_{\textrm{vac}}$
and ${\rm P}_{\textrm{vac}}$ are respectively
the vacuum energy and pressure, Eqs. (\ref{eq:E_vac})-(\ref{eq:P_vac}) respectively.

The way to proceed is now clear: we expand Eq. (\ref{eq:dpdt_RQF})
as a series in $\lambda$,
\begin{equation}
\dot{\mathtt{p}}^{(\bm{\phi})}=(\dot{\mathtt{p}}^{(\bm{\phi})})_{(0)}+\lambda\delta\dot{\mathtt{p}}^{(\bm{\phi})}+\mathcal{O}\left(\lambda^{2}\right)\,,
\end{equation}
using the zeroth-order parts of the various terms written above. We
need only to specify the worldline $\mathscr{G}$ in $\mathring{\mathscr{M}}$
about which we are carrying out the Fermi normal coordinate expansion (in $r$). We will consider
two cases: $\mathscr{G}=\mathring{\mathscr{C}}$ (the geodesic, such
that $\mathscr{B}$ is inertial with the point particle in $\mathring{\mathscr{M}}$)
and $\mathscr{G}=\mathscr{C}$ (an accelerated worldline such that
$\mathscr{B}_{(\lambda)}$ is inertial with the object in $\mathscr{M}_{(\lambda)}$,
\textit{i.e.} it is defined by a constant $r>C\lambda$ in $\mathscr{M}_{(\lambda)}$).
These will give us equivalent descriptions of the dynamics of the
system, from two different ``points of view'', or (quasilocal) frames of reference.

Before entering into the calculations, we can simplify things further
by remarking that the zeroth order expansions in Eqs. (\ref{eq:N0})-(\ref{eq:P0})
will always make the twist ($\nu$) term in the conservation law [Eq. (\ref{eq:dpdt_RQF})] appear at
$\mathcal{O}(r)$ or higher, both in $(\dot{\mathtt{p}}^{(\bm{\phi})})_{(0)}$
and $\delta\dot{\mathtt{p}}^{(\bm{\phi})}$, regardless of our choice
of $\mathscr{G}$. Hence we can safely ignore it, as we are interested (at least for this work)
only in the part of the conservation law which is zeroth-order in $r$. Thus we simply work
with
\begin{equation}
\dot{\mathtt{p}}^{(\bm{\phi})}=-\intop_{\mathbb{S}_{r}^{2}}\bm{\epsilon}^{}_{\mathbb{S}^{2}}\,r^{2}N\left(\mathcal{E}\alpha_{\bm{\phi}}+{\rm P}\bm{D}\cdot\bm{\phi}\right)\,.\label{eq:dpdt_RQF_simplified}
\end{equation}

Into this, we furthermore have to insert the multipole expansion of
the conformal Killing vector $\bm{\phi}$ given by Eq. (\ref{eq:phi_multipole}). We correspondingly write
\begin{equation}
\dot{\mathtt{p}}^{(\bm{\phi})}=\sum_{\ell\in\mathbb{N}}\dot{\mathtt{p}}^{(\bm{\phi}_{\ell})}\,,
\end{equation}
such that for any $\ell\in\mathbb{N}$, we have
\begin{equation}
\dot{\mathtt{p}}^{(\bm{\phi}_{\ell})}=-\Phi^{I_{1}\cdots I_{\ell}}\intop_{\mathbb{S}_{r}^{2}}\bm{\epsilon}^{}_{\mathbb{S}^{2}}\,rN\left(\mathcal{E}\alpha_{\mathfrak{i}}+{\rm P}D_{\mathfrak{i}}\right)D^{\mathfrak{i}}\left(\prod_{n=1}^{\ell}r_{I_{n}}\right)\,.
\end{equation}
Explicitly, the first two terms are
\begin{align}
\dot{\mathtt{p}}^{(\bm{\phi}_{\ell=1})}=\, & -\Phi^{I}\intop_{\mathbb{S}_{r}^{2}}\bm{\epsilon}^{}_{\mathbb{S}^{2}}\,rN\left(\mathcal{E}\alpha_{\mathfrak{i}}+{\rm P}D_{\mathfrak{i}}\right)\mathfrak{B}_{I}^{\mathfrak{i}}\,,\label{eq:p^dot_l1}\\
\dot{\mathtt{p}}^{(\bm{\phi}_{\ell=2})}=\, & -2\Phi^{IJ}\intop_{\mathbb{S}_{r}^{2}}\bm{\epsilon}^{}_{\mathbb{S}^{2}}\,rN\left(\mathcal{E}\alpha_{\mathfrak{i}}+{\rm P}D_{\mathfrak{i}}\right)\left(\mathfrak{B}_{I}^{\mathfrak{i}}r_{J}\right)\,.\label{eq:p^dot_l2}
\end{align}

\subsection{Equation of motion inertial with the background point particle\label{ssec:gralla-wald_pp-inertial}}

\begin{figure}
\noindent \begin{centering}
\includegraphics[scale=0.55]{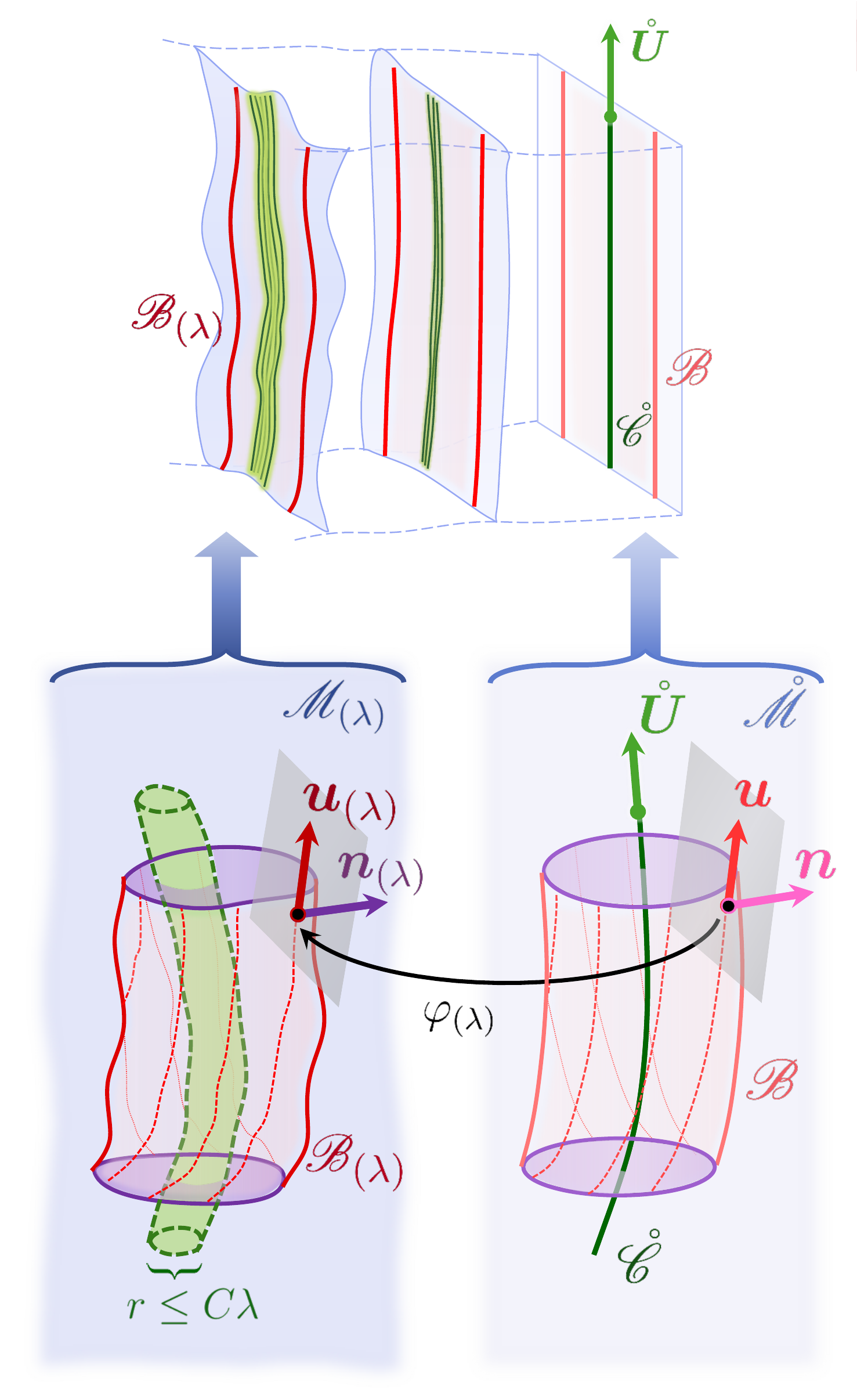}
\par\end{centering}
\caption{A family of rigid quasilocal frames $\{(\mathscr{B}_{(\lambda)};\bm{u}_{(\lambda)})\}$ embedded in the Gralla-Wald family of spacetimes $\{\mathscr{M}_{(\lambda)}\}$ such that the inverse image of any such perturbed quasilocal frame in the background is inertial with the point particle approximation of the moving object, \textit{i.e.} is centered on the geodesic $\mathring{\mathscr{C}}$.}\label{fig-pp}

\end{figure}

Let $\mathscr{G}=\mathring{\mathscr{C}}$. Then $\bm{A}=0$. We will
take this to be the case for the rest of this subsection---corresponding, as discussed, to a rigid quasilocal frame the inverse image in the background of which is inertial with the point particle approximation of the moving object in the background spacetime. This situation is displayed visually in Fig. \ref{fig-pp}.

Let us first compute the zeroth-order (in $\lambda$) part of $\dot{\mathtt{p}}^{(\bm{\phi})}$.
Inserting (\ref{eq:N0})-(\ref{eq:P0}) into the zeroth-order part
of (\ref{eq:p^dot_l1})-(\ref{eq:p^dot_l2}), and making
use of the various properties in this chapter's appendix (Section \ref{sec:5.A}), we find by direct computation:
\begin{align}
\left(\dot{\mathtt{p}}^{(\bm{\phi}_{\ell=1})}\right)_{(0)}=\, & \mathcal{O}\left(r^{2}\right)\,,\label{eq:PP-p^dot_l1_0}\\
\left(\dot{\mathtt{p}}^{(\bm{\phi}_{\ell=2})}\right)_{(0)}=\, & \mathcal{O}\left(r^{2}\right)\,.\label{eq:PP-p^dot_l2_0}
\end{align}
We provide the steps of the calculation in Appendix B of [\cite{oltean_motion_2019}].

Let us now compute the $\mathcal{O}(\lambda)$, $\ell=1$ part of $\dot{\mathtt{p}}^{(\bm{\phi})}$,
\textit{i.e.} the $\mathcal{O}(\lambda)$ part of Eq. (\ref{eq:p^dot_l1})
which as usual we denote by $\delta\dot{\mathtt{p}}^{(\bm{\phi}_{\ell=1})}$.
One can see that this will involve contributions from five $\mathcal{O}(\lambda)$
terms, respectively containing $\delta N$, $\delta\mathcal{E}$, $\delta\bm{\alpha}$,
$\delta{\rm P}$ and $\delta\bm{D}$. For convenience, we will use
the notation $(\dot{\mathtt{p}}_{(Q)}^{(\bm{\phi}_{\ell})})_{(n)}$
to indicate the term of $\delta ^{n}(\dot{\mathtt{p}}^{(\bm{\phi}_{\ell})})$
that is linear in $Q$, for any $\ell,n$. Thus we write
\begin{equation}
\delta\dot{\mathtt{p}}^{(\bm{\phi}_{\ell=1})}=\sum_{Q\in\{\delta N,\delta\mathcal{E},\delta\bm{\alpha},\delta{\rm P},\delta\bm{D}\}}\delta\dot{\mathtt{p}}_{(Q)}^{(\bm{\phi}_{\ell=1})}\,.
\end{equation}

All of the computational steps are again in Appendix B of [\cite{oltean_motion_2019}]. We find:
\begin{equation}
\delta\dot{\mathtt{p}}_{(\delta N)}^{(\bm{\phi}_{\ell=1})}=-\frac{2}{\kappa}\Phi_{I}\intop_{\mathbb{S}_{r}^{2}}\bm{\epsilon}^{}_{\mathbb{S}^{2}}\,\delta Nr^{I}+\mathcal{O}\left(r^{2}\right)\,.
\end{equation}
If $\delta N$ does not vary significantly over $\mathbb{S}_{r}^{2}$,
the $\mathcal{O}(r^{0})$ part of the above would be negligible owing
to the fact that $\intop_{\mathbb{S}_{r}^{2}}\bm{\epsilon}^{}_{\mathbb{S}^{2}}\,r^{I}=0$.

\begin{figure}
\noindent \begin{centering}
\includegraphics[scale=0.55]{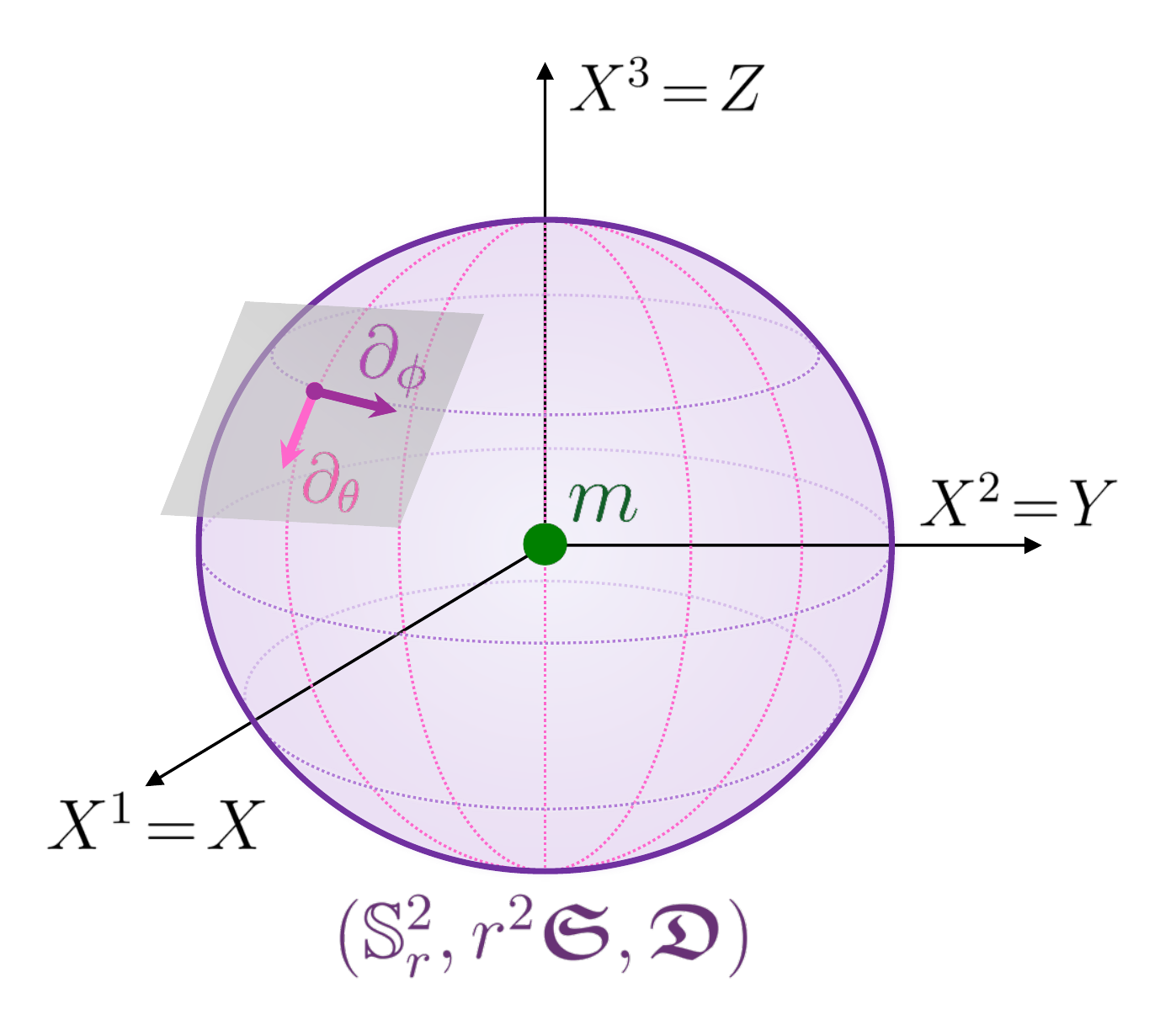}
\par\end{centering}
\caption{An instantaneous rigid quasilocal frame $(\mathbb{S}^{2}_{r},r^{2}\bm{\mathfrak{S}},\bm{\mathfrak{D}})$ (where $\bm{\mathfrak{S}}$ and $\bm{\mathfrak{D}}$ respectively are the metric and derivative compatible with the unit two-sphere) inertial with the background ``point particle''. This means that the latter is located at the center of our Fermi normal coordinate system.}\label{fig-pp-qf}
\end{figure}

Next, let us consider the $\delta\mathcal{E}$ and $\delta{\rm P}$
terms. For this, we find it useful to depict the instantaneous quasilocal frame $(\mathbb{S}^{2}_{r},r^{2}\bm{\mathfrak{S}},\bm{\mathfrak{D}})$ embedded in a constant-time three-slice of $\mathring{\mathscr{M}}$ in Fig. \ref{fig-pp-qf}.

The $\delta\mathcal{E}$ term can be easily determined by realizing that in our
current choice of quasilocal frame, the only background matter is the point particle
which is always at the center of our present coordinate system, \textit{i.e.}
it is always on $\mathring{\mathscr{C}}$ (on which we are here centering
our Fermi normal coordinates). Interpreting the constant $m$ as in the approach of [\cite{gralla_rigorous_2008}]
to be the ``mass'' of this point particle, this simply means
that
\begin{equation}
\delta\mathcal{E}=\frac{m}{4\pi r^{2}}\,,\label{eq:C0_deltaE}
\end{equation}
so that when this is integrated (as a surface energy density) over
$\mathbb{S}_{r}^{2}$, we simply recover the mass: $\int_{\mathbb{S}_{r}^{2}}r^{2}\bm{\epsilon}^{}_{\mathbb{S}^{2}}\,\delta\mathcal{E}=m$.
We remark that, by definition, it is possible to express the quasilocal
energy as $\mathcal{E}=u^{a}u^{b}\tau_{ab}=-\tfrac{1}{\kappa}k$ with
$k=\bm{\sigma}:\bm{\Theta}$ the trace of the two-dimensional boundary extrinsic curvature. Notice that the integral of this over
a closed two-surface in the $r\rightarrow\infty$ limit is in fact the same as the usual ADM definition
of the mass/energy; thus $\delta\mathcal{E}=-\tfrac{1}{\kappa}\delta k$,
and so it makes sense to interpret $m$ as the ADM mass of the object.
So now, using Eq. (\ref{eq:C0_deltaE}), we can find that the $\delta\mathcal{E}$
contribution to $\delta\dot{\mathtt{p}}^{(\bm{\phi}_{\ell=1})}$ is
also at most quadratic in $r$:
\begin{equation}
\delta\dot{\mathtt{p}}_{(\delta\mathcal{E})}^{(\bm{\phi}_{\ell=1})}=\mathcal{O}\left(r^{2}\right)\,.
\end{equation}

To compute the $\delta{\rm P}$ term, we now employ the useful identity in Eq.
(\ref{eq:E-P_relation}), which tells us that
\begin{equation}
\delta{\rm P}=\frac{1}{2}\delta\mathcal{E}-\frac{1}{\kappa}\delta a_{\bm{n}}\,.
\end{equation}
Using this, into which we insert the $\delta\mathcal{E}$ from Eq. (\ref{eq:C0_deltaE}),
we find that the $\delta{\rm P}$ contribution to $\delta\dot{\mathtt{p}}^{(\bm{\phi}_{\ell=1})}$
is at most quadratic in $r$ as well,
\begin{equation}
\delta\dot{\mathtt{p}}_{(\delta{\rm P})}^{(\bm{\phi}_{\ell=1})}=\mathcal{O}\left(r^{2}\right)\,.
\end{equation}

Note that the above results may in fact be higher order in $r$ than quadratic.
We have only explicitly checked that they vanish up to linear order
inclusive.

Finally we are left with the $\delta\bm{\alpha}$ and $\delta\bm{D}$
contributions to $\delta\dot{\mathtt{p}}^{(\bm{\phi}_{\ell=1})}$.
By direct computation, it is possible to show that their sum is in
fact precisely what we have referred to as the extended GSF in our general
analysis of the preceding section, \textit{i.e.} it is the $\ell=1$ part of Eq.
(\ref{eq:Delta_p_GSF_general}),
\begin{equation}
\delta\dot{\mathtt{p}}_{(\delta\bm{\alpha})}^{(\bm{\phi}_{\ell=1})}+\delta\dot{\mathtt{p}}_{(\delta\bm{D})}^{(\bm{\phi}_{\ell=1})}=\frac{{\rm d}}{{\rm d}t}\left(\Delta\mathtt{p}_{\textrm{self}}^{(\bm{\phi}_{\ell=1})}\right)\,.
\end{equation}
In particular, they respectively contribute the usual GSF (from $\delta\bm{\alpha}$)
and the gravitational self-pressure force (from $\delta\bm{D}$). 

Thus, we have found that the total $\mathcal{O}(\lambda)$, $\ell=1$
part of the momentum time rate of change is given at leading (zeroth)
order in $r$ by nothing more than the generalized GSF. In other words,
\begin{equation}
\boxed{\delta\dot{\mathtt{p}}^{(\bm{\phi}_{\ell=1})}=-\Phi_{I}\mathtt{F}^{I}+\mathcal{O}\left(r\right)}\,,\label{eq:C0_delta_p^dot_1}
\end{equation}
where we have defined
\begin{equation}
\mathtt{F}^{I}=-\frac{2}{\kappa}\intop_{\mathbb{S}_{r}^{2}}\bm{\epsilon}^{}_{\mathbb{S}^{2}}\,\mathfrak{S}^{\mathfrak{ij}}\mathfrak{B}_{\mathfrak{i}}^{I}\mathcal{F}_{\mathfrak{j}}[\bm{h};\mathring{\bm{u}}]+\mathcal{O}\left(r\right)\,.
\end{equation}

Without loss of generality, let us now pick $\Phi^{I}=(0,0,1)$ to
be the unit vector in the Cartesian $X^{3}=Z$ direction, and denote
its corresponding conformal Killing vector as $\bm{\phi}_{\ell=1}=\bm{\phi}_{\ell=1}^{Z}$.
(Alternately, pick the $Z$-axis to be oriented along $\Phi^{I}$.)
We know $\mathfrak{S}^{\mathfrak{ij}}\mathfrak{B}_{\mathfrak{j}}^{Z}=(-1/\sin\theta,0)$;
moreover, by the coordinate transformation $\mathcal{F}_{\mathfrak{i}}=(\partial x^{J}/\partial x^{\mathfrak{i}})\mathcal{F}_{J}$
we have $\mathcal{F}_{\theta}=\cos\theta(\cos\phi\mathcal{F}_{X}+\sin\phi\mathcal{F}_{Y})-\sin\theta\mathcal{F}_{Z}$.
Inserting these into Eq. (\ref{eq:C0_delta_p^dot_1}) we get
\begin{equation}
\delta\dot{\mathtt{p}}^{(\bm{\phi}_{\ell=1}^{Z})}=\,-\frac{2}{\kappa}\intop_{\mathbb{S}_{r}^{2}}\bm{\epsilon}^{}_{\mathbb{S}^{2}}\,\mathcal{F}_{Z}[\bm{h};\mathring{\bm{u}}]+\frac{2}{\kappa}\intop_{\mathbb{S}_{r}^{2}}{\rm d}\theta\wedge{\rm d}\phi\,\cos\theta\left(\cos\phi\mathcal{F}_{X}+\sin\phi\mathcal{F}_{Y}\right)\,.
\end{equation}
The first integral on the RHS is precisely in the form of the GSF term from the Gralla formula, Eq. (\ref{eq:Gralla_ange-average}) [\cite{gralla_gauge_2011}], except here in the integrand we have (the $Z$-component of) our extended GSF $\bm{\mathcal{F}}$ [Eq. (\ref{eq:generalized_GSF})]: the usual GSF $\bm{F}$ (the only self-force term in Gralla's formula) plus our self-pressure term, $\bm{\wp}$.
The second integral contains additional terms involving the extended GSF in the
other two (Cartesian) spatial directions. Notice however that $\intop_{\mathbb{S}_{r}^{2}}{\rm d}\theta\wedge{\rm d}\phi\,\cos\theta\cos\phi=0=\intop_{\mathbb{S}_{r}^{2}}{\rm d}\theta\wedge{\rm d}\phi\,\cos\theta\sin\phi$,
so if $\mathcal{F}_{X}$ and $\mathcal{F}_{Y}$ do not vary significantly
over $\mathbb{S}_{r}^{2}$, their contribution will be subdominant
to that of $\mathcal{F}_{Z}$.

Thus, we have shown that our EoM (\ref{eq:C0_delta_p^dot_1}) always contains Gralla's ``angle average'' of the ``bare'' (usual) GSF. However, the form of (\ref{eq:C0_delta_p^dot_1}) (expressing the perturbative change in the quasilocal momentum) still cannot be \emph{directly} compared, as such, with Gralla's EoM (\ref{eq:Gralla_ange-average}) (expressing the change in a deviation vector representing the perturbative ``correction to the motion''). In the following subsection, we clarify the correspondence by repeating the calculation using a quasilocal frame inertial with the moving extended object in the perturbed spacetime (rather than with the geodesic in the background, as here). Furthermore, we conjecture that a careful imposition of the parity condition on the perturbative gauge---of which we have made no explicit use so far---would make the contribution from our ``self-pressure'' term vanish, but a detailed proof is required and remains to be carried out.

\subsection{Equation of motion inertial with the moving object in the perturbed spacetime\label{ssec:gralla-wald_sco-inertial}}

\begin{figure}
\noindent \begin{centering}
\includegraphics[scale=0.55]{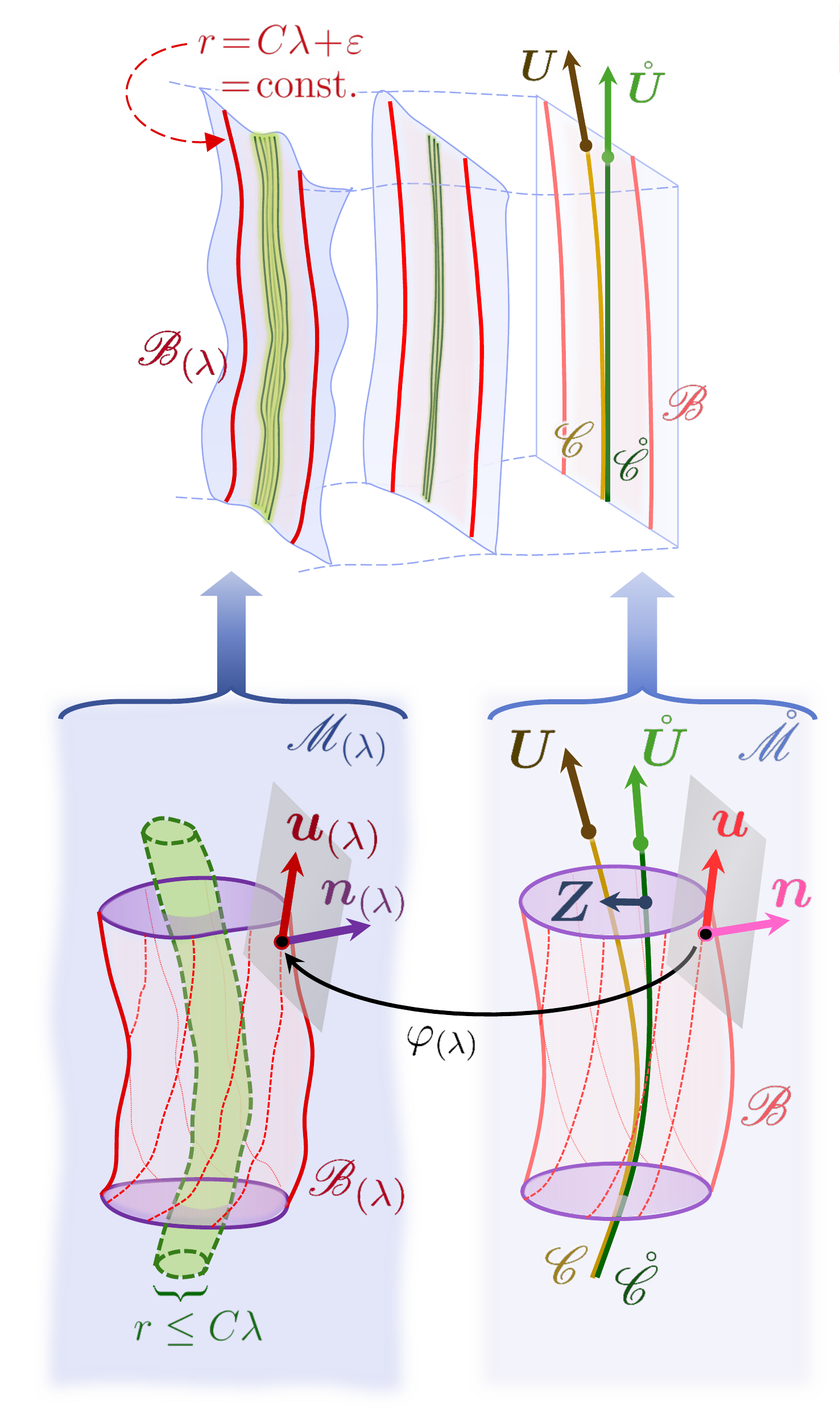}
\par\end{centering}
\caption{A family of rigid quasilocal frames $\{(\mathscr{B}_{(\lambda)};\bm{u}_{(\lambda)})\}$ embedded in the Gralla-Wald family of spacetimes $\{\mathscr{M}_{(\lambda)}\}$ inertial with the moving object in $\mathscr{M}_{(\lambda)}$. This means that $\mathscr{B}_{(\lambda)}$ is defined by the constancy of the Gralla-Wald $r$ coordinate in $\mathscr{M}_{(\lambda)}$, for any $r>C\lambda$. Thus, the inverse image $\mathscr{B}$ of $\mathscr{B}_{(\lambda)}$ in the background $\mathring{\mathscr{M}}$ is centered, in general, \emph{not} on the geodesic $\mathring{\mathscr{C}}$ followed by the point particle background approximation of the object, but on some timelike worldline $\mathscr{C}\neq\mathring{\mathscr{C}}$, with four-velocity $\bm{U}\neq \mathring{\bm{U}}$, which may be regarded as an approximation on $\mathring{\mathscr{M}}$ of the ``true motion'' of the object in $\mathscr{M}_{(\lambda)}$. Between $\mathring{\mathscr{C}}$ and $\mathscr{C}$ there is a deviation vector $\bm{Z}$, which is to be compared with the deviation vector (``correction to the motion'') in the Gralla-Wald approach.}\label{fig-sco}
\end{figure}

Now let $\mathscr{G}=\mathscr{C}\neq\mathring{\mathscr{C}}$ (so $\bm{A}\neq0$
in general) such that the quasilocal frame $(\mathscr{B};\bm{u})$ centered on $\mathscr{C}$
(in $\mathring{\mathscr{M}}$) is the inverse image of the rigid quasilocal frame $(\mathscr{B}_{(\lambda)};\bm{u}_{(\lambda)})$
defined by $r=C\lambda+\varepsilon={\rm const.}$, $\forall\varepsilon>0$,
in $\mathscr{M}_{(\lambda)}$. The meaning of the $r$ coordinate
in the latter is as given in the Gralla-Wald assumptions (Subsection \ref{ssec:gralla-wald_review}). This situation is displayed in Fig. \ref{fig-sco}. 

We now proceed to calculate, in the same way as we did for the point-particle-inertial
case, the various terms in the expansion of the momentum conservation
law, Eqs. (\ref{eq:p^dot_l1})-(\ref{eq:p^dot_l2}). At zeroth order we obtain:
\begin{align}
\left(\dot{\mathtt{p}}^{(\bm{\phi}_{\ell=1})}\right)_{(0)}=\, & \mathcal{O}\left(r^{2}\right)\,,\\
\left(\dot{\mathtt{p}}^{(\bm{\phi}_{\ell=2})}\right)_{(0)}=\, & \mathcal{O}\left(r^{2}\right)\,.
\end{align}
The steps of all these computations are again shown in Appendix B of [\cite{oltean_motion_2019}].

Let us now compute the $\mathcal{O}(\lambda)$, $\ell=1$ part of $\dot{\mathtt{p}}^{(\bm{\phi})}$.
First, we find that $\delta\dot{\mathtt{p}}_{(\delta N)}^{(\bm{\phi}_{\ell=1})}$
is the same as in the point-particle-inertial case, so if $\delta N$ does not
vary significantly over $\mathbb{S}_{r}^{2}$, the $\mathcal{O}(r^0)$
part thereof is negligible.

\begin{figure}
\noindent \begin{centering}
\includegraphics[scale=0.55]{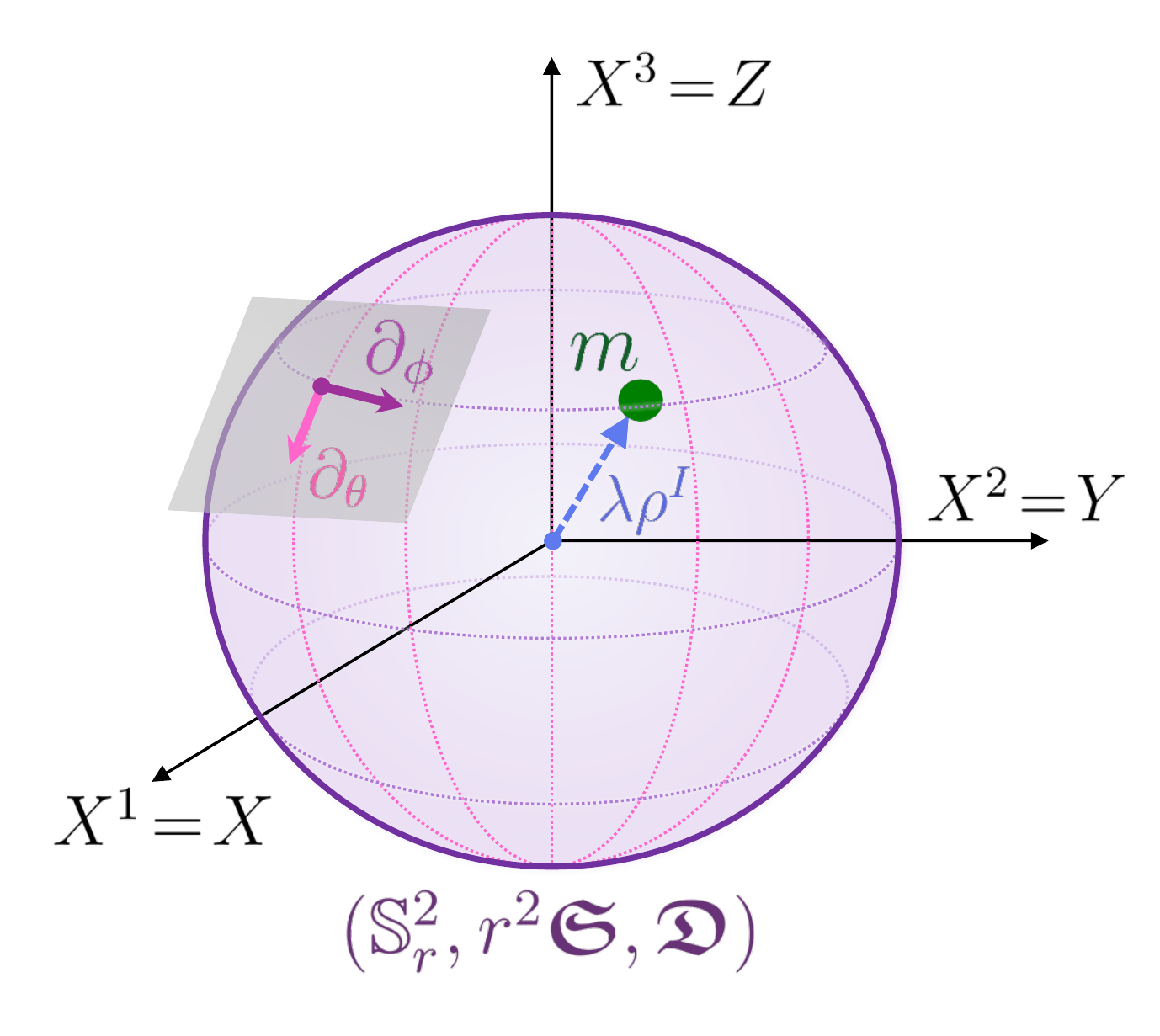}
\par\end{centering}
\caption{An instantaneous rigid quasilocal frame $(\mathbb{S}^{2}_{r},r^{2}\bm{\mathfrak{S}},\bm{\mathfrak{D}})$ (where $\bm{\mathfrak{S}}$ and $\bm{\mathfrak{D}}$ respectively are the metric and derivative compatible with the unit two-sphere) inertial with the moving object in the perturbed spacetime. This means that the point particle approximation of this object in the background spacetime is \emph{not} located at the center of our Fermi normal coordinate system. Instead, it is displaced in some direction $\rho^{I}$, which must be $\mathcal{O}(\lambda)$.}\label{fig-sco-qf}
\end{figure}

Next let us look at the $\delta\mathcal{E}$ and $\delta{\rm P}$
parts. Again, it is useful to consider in this case the visual depiction of the instantaneous quasilocal frame, shown in Fig. \ref{fig-sco-qf}. 

In this case, the particle (delta function) will \emph{not}
be at the center of our coordinate system but instead displaced in
some direction $\rho^{I}$ relative thereto. Nonetheless, we know
that this displacement must itself be $\mathcal{O}(\lambda)$ which
means that it will only contribute $\mathcal{O}(\lambda)$ corrections
to the $\delta\mathcal{E}$ having $m$ exactly at the center, \textit{i.e.} we have
\begin{equation}
\delta\mathcal{E}=\frac{m}{4\pi\left(X^{I}-\lambda\rho^{I}\right)\left(X_{I}-\lambda\rho_{I}\right)}=\frac{m}{4\pi r^{2}}+\mathcal{O}\left(\lambda\right)\,,
\end{equation}
and as before, $\delta{\rm P}=\frac{1}{2}\delta\mathcal{E}-\frac{1}{\kappa}\delta a_{\bm{n}}$.
Using these, we find:
\begin{align}
\delta\dot{\mathtt{p}}_{(\delta\mathcal{E})}^{(\bm{\phi}_{\ell=1})}=\, & -\frac{2}{3}m\Phi_{I}A^{I}+\mathcal{O}\left(r\right)\,,\\
\delta\dot{\mathtt{p}}_{(\delta{\rm P})}^{(\bm{\phi}_{\ell=1})}=\, & +\frac{1}{3}m\Phi_{I}A^{I}+\mathcal{O}\left(r\right)\,,
\end{align}
with the steps shown in Appendix B of [\cite{oltean_motion_2019}]. Thus,
\begin{equation}
\delta\dot{\mathtt{p}}_{(\delta\mathcal{E})}^{(\bm{\phi}_{\ell=1})}+\delta\dot{\mathtt{p}}_{(\delta{\rm P})}^{(\bm{\phi}_{\ell=1})}=-\frac{1}{3}m\Phi_{I}A^{I}+\mathcal{O}\left(r\right)\,.
\end{equation}
Meanwhile, we still have, exactly as in the point-particle-inertial case,
\begin{equation}
\delta\dot{\mathtt{p}}_{(\delta\bm{\alpha})}^{(\bm{\phi}_{\ell=1})}+\delta\dot{\mathtt{p}}_{(\delta\bm{D})}^{(\bm{\phi}_{\ell=1})}=-\Phi_{I}\mathtt{F}^{I}+\mathcal{O}\left(r\right)\,.
\end{equation}

Now, by construction, we know that here $\delta\dot{\mathtt{p}}^{(\bm{\phi}_{\ell=1})}=0$,
as we are inertial with the moving object (in the ``actual'' spacetime $\mathscr{M}_{(\lambda)}$).
Thus summing the above and equating them to zero, we get
\begin{equation}
0=\Phi_{I}\left(-mA^{I}-3\mathtt{F}^{I}\right)+\mathcal{O}\left(r\right)\,.
\end{equation}
Since $\Phi^{I}$ is arbitrary, we thus get the EoM 
\begin{equation}
mA^{I}=-3\mathtt{F}^{I}\,\label{eq:mAI}    
\end{equation}
in the $r\rightarrow0$ limit. 

Finally, to cast this EoM into the same form as do  [\cite{gralla_rigorous_2008}], \textit{i.e.}
in terms of a deviation vector $\bm{Z}$ on $\mathring{\mathscr{C}}$
rather than in terms of the proper acceleration $\bm{A}$ of $\mathscr{C}$, we
use the generalized deviation equation (as the name suggests, the deviation equation between arbitrary worldlines, not necessarily geodesics), Eq. (37) of [\cite{puetzfeld_generalized_2016}]. In our case, this reads
$\lambda\ddot{Z}^{I}=\lambda A^{I}-\lambda Z^{J}\mathring{E}^{I}\,_{J}+\mathcal{O}(\lambda^{2})$.
Combining this with Eq. (\ref{eq:mAI}), we finally recover the $\mathcal{O}(\lambda)$ EoM
\begin{equation}
\boxed{\lambda m\ddot{Z}^{I}=-3\lambda\mathtt{F}^{I}-\lambda\mathring{E}^{I}\,_{J}Z^{J}+\mathcal{O}\left(\lambda^{2},r\right)}\,.
\end{equation}

Note that the factor of $3$ multiplying the self-force term is in fact present in the EoM in Gralla's Appendix B, that is Eq. (B3) of [\cite{gralla_gauge_2011}]. The latter, in this case, expresses the time evolution not of the deviation vector itself, but of the change in this deviation vector due to a gauge transformation, possibly including extra terms in case that transformation is out of the ``parity-regular'' class. We conjecture that a detailed analysis of the precise correspondence between our deviation vector definition and that of Gralla-Wald (which, while encoding the same intuitive notion of a perturbative ``correction to the motion'', may not be completely identical in general), together with a relation of their gauge transformation properties, would make it possible to relate these EoM's exactly.

\section{Discussion and conclusions\label{sec:concl}}

In this chapter, we have used quasilocal conservation laws to develop
a novel formulation of self-force effects in general relativity, one
that is independent of the choice of the perturbative gauge and applicable
to any perturbative scheme designed to describe the correction to
the motion of a localized object. In particular, we have shown that
the correction to the motion of any finite spatial region, due to
any perturbation of any spacetime metric, is dominated when that region
is ``small'' (\textit{i.e.} at zero-th order in a series expansion in its
areal radius) by an \emph{extended }gravitational self-force: this
is the standard gravitational self-force term known up to now plus
a new term, not found in previous analyses and attributable to a gravitational
pressure effect with no analogue in Newtonian gravity, which we have
dubbed the gravitational \emph{self-pressure force}. Mathematically, we have found that the total change
in momentum $\Delta\mathtt{p}^{(\bm{\phi})}=\mathtt{p}_{\textrm{final}}^{(\bm{\phi})}-\mathtt{p}_{\textrm{initial}}^{(\bm{\phi})}$
between an initial and final time of any (gravitational plus matter)
system subject to any metric perturbation $\bm{h}$ is given, in a
direction determined by a conformal Killing vector $\bm{\phi}$ (see Subsection \ref{ssec:qf_cons_laws}), by the following
flux through the portion of the quasilocal frame (worldtube boundary)
$(\mathscr{B};\mathring{\bm{u}})$ delimited thereby:
\begin{equation}
\Delta\mathtt{p}^{(\bm{\phi})}_{\textrm{self}}=-\frac{c^{4}}{4\pi G}\intop_{\Delta\mathscr{B}}\bm{\epsilon}_{\mathscr{B}}^{\,}\frac{1}{r}\bm{\phi}\cdot\bm{\mathcal{F}}[\bm{h};\mathring{\bm{u}}]+\mathcal{O}\left(r\right)\,,\label{eq:concl_eom}
\end{equation}
where we have restored units, $r$ is the areal radius, and $\bm{\mathcal{F}}$
is the extended self-force functional. In particular, $\bm{\mathcal{F}}=\bm{F}+\bm{\wp}$
where $\bm{F}$ is the usual ``bare'' self-force [determined by the functional in Eq. (\ref{eq:intro_GSF_functional})] and
$\bm{\wp}$ is our novel self-pressure force [determined by the functional in Eq. (\ref{eq:pressure_functional})].

The most relevant practical application of the self-force is in the
context of modeling EMRIs. Ideally, one would like  to compute
the ``correction to the motion'' at the location of the moving object
(SCO). Yet, once a concrete perturbative procedure is established,
the latter usually ends up being described by a distribution (Dirac
delta function), rendering such a computation ill-defined unless additional
tactics (typically in the form of regularizations or Green's functions
methods) are introduced. However, if one takes a step back from the
exact point denoting the location of the ``particle'' (the distributional
support), and instead considers a flux around it, any singularities
introduced in such a model are avoided by construction.

We have, moreover,
shown that our formulation, when applied in the context of one particular and
very common approach to the self-force---namely that of [\cite{gralla_rigorous_2008}]---yields equations of motion of the same form as those known up to now; in particular, they always contain, in the appropriate limit, the ``angle
average'' self-force term of [\cite{gralla_gauge_2011}]. We conjecture that a more rigorous study of these equations of motion and their gauge transformation properties would prove their exact correspondence under appropriate conditions.

We would like here to offer a concluding discussion on our results
in this chapter in Subsection \ref{ssec:discussion}, as well as outlook towards future work
in Subsection \ref{ssec:outlook}.

\subsection{Discussion of results\label{ssec:discussion}}

From a physical point of view, our approach offers a fresh and conceptually
clear perspective on the basic mechanism responsible for the emergence
of self-force effects in general relativity. In particular, we have
demonstrated that the self-force may be regarded as nothing more than the manifestation
of a \emph{physical flux of gravitational momentum} passing through
the boundary enclosing the ``small'' moving object. This gravitational
momentum, and gravitational stress-energy-momentum in general, cannot
be defined locally in general relativity. As we have argued at length
in this chapter, such notions must instead be defined quasilocally, \textit{i.e.} as
boundary rather than as a volume densities. This is why the self-force
appears mathematically as a boundary integral
around the moving object [Eq. (\ref{eq:Delta_p_GSF_general})], dominant in the limit where the areal radius
is small.  

The interpretation of the physical meaning of the self-force as a
consequence of conservation principles leads to many interesting implications.
As we have seen, the ``mass'' of the moving object---\textit{e.g.}, the
mass $m$ of the SCO in the EMRI problem---seems to have nothing
to do \emph{fundamentally} with the general existence of a self-force
effect. Indeed, according to our analysis, the self-force is in fact
generically present as a correction to the motion---and dominant
when the moving region is ``small''---whenever one has \emph{any}
perturbation $\boldsymbol{h}$ to the spacetime metric that is non-vanishing
on the boundary of the system.

The usual way to understand the gravitational self-force up
to now has been to regard it as a backreaction of $m$ on the metric,
\textit{i.e.} on the gravitational field, and thus in turn upon its own motion
through that field. Schematically, one thus imagines that the linear
correction to the motion is ``linear in $m$'' (or more generally,
that the full correction is an infinite series in $m$), \textit{i.e.}
that it has the form $\delta\dot{\mathtt{p}}\sim m\delta\mathtt{a}$,
with a ``perturbed acceleration'' $\delta\mathtt{a}$ determined
by $\bm{h}$ (according to some perturbative prescription) causing
a correction to the momentum $\delta\dot{\mathtt{p}}$ by a (linear)
coupling to the mass $m$.

Our analysis, instead, shows that this momentum correction $\delta\dot{\mathtt{p}}$
actually arises fundamentally in the schematic form 
\begin{equation}
\delta\dot{\mathtt{p}}\sim\mathcal{E}_{\textrm{vac}}\delta\mathtt{a}+{\rm P}_{\textrm{vac}}\delta D\,,\label{eq:quasilocal_pdot_schematic}
\end{equation}
where $\mathcal{E}_{\textrm{vac}}$ and ${\rm P}_{\textrm{vac}}$
are the \emph{vacuum} energy and pressure [Eqs. (\ref{eq:E_vac})-(\ref{eq:P_vac}) respectively], and $\delta\mathtt{a}$
and $\delta D$ are perturbed acceleration and gradient terms determined
by $\bm{h}$. Thus it is the vacuum energy (or ``mass'') and vacuum
pressure, \emph{not} the ``mass'' of the moving object, which are
responsible for the backreaction that produces self-force corrections. 

Certainly, the metric perturbation $\bm{h}$ on the system boundary
determining the perturbed acceleration and gradient terms in (\ref{eq:quasilocal_pdot_schematic}) may
in turn be sourced by a ``small mass'' present in the interior of
the system. In fact, if indeed the system is ``small'', there may
well be little physical reason for expecting that (the dominant part
of) $\bm{h}$ would come for anything \emph{other than} the presence
of the ``small'' system itself. Concordantly, the aim of any concrete
self-force analysis is to prescribe exactly \emph{how} $\bm{h}$ is
sourced thereby. Nevertheless, the correction (\ref{eq:concl_eom}) is valid \emph{regardless}
of where $\bm{h}$ comes from, and regardless of the interior description
of the system, which may very well be completely empty of matter or even contain ``exotic'' matter (as long as a well-posed initial value formulation exists). The EMRI problem is just a special case, where $\bm{h}$ is sourced in the background, according to the approach considered here, by a rudimentary point particle of mass $m$.

This opens up many interesting conceptual questions, especially with
regards to the meaning of the quasilocal vacuum energy and pressure.
While traditionally these have often been regarded as unphysical,
to be ``subtracted away'' as reference terms (for the same reason
that a ``reference action'' is often subtracted from the total gravitational
action in Lagrangian formulations of GR), our analysis in this chapter
reveals instead that they are absolutely indispensable to accounting for
self-force effects. (Indeed, the initial work [\cite{epp_momentum_2013}] on the formulation of the quasilocal momentum conservation laws had similarly revealed the necessity of keeping these terms for a proper accounting of gravitational energy-momentum transfer in general.) To put it simply, the vacuum energy is what seems
to play the role of the ``mass'' in the ``mass times acceleration''
of the self-force; the pressure term, leading to what we have called
the self-pressure force, has no Newtonian analogue.

Now let us comment on our results from a more mathematical and technical
point of view. When applied to a specific self-force analysis, namely
that of [\cite{gralla_rigorous_2008}], we have been able to recover the ``angle
average'' formula of  [\cite{gralla_gauge_2011}]. The latter was put forward on the basis
of a convenient mathematical argument in a Hamiltonian setting.
As the quasilocal stress-energy-momentum definitions that we have
been working with (namely, as given by the Brown-York tensor) recover
the usual Hamiltonian definitions under appropriate conditions (stationary
asymptotically-flat spacetimes with a parity condition), it is reasonable
that our general equation of motion [Eq. (\ref{eq:Delta_p_GSF_general})]---expressing the physical flux
of gravitational momentum---should thereby recover that of Gralla [Eq. (\ref{eq:Gralla_ange-average})]---expressing
an ``angle average'' in a setting where certain surface integral
definitions of general-relativistic Hamiltonian notions (in particular, a Hamiltonian ``center of mass'') can be well-defined.
The limitation of Gralla's equation of motion (\textit{e.g.} in terms of the
perturbative gauge restriction attached to it) \textit{vis-à-vis} our general
equation of motion is therefore essentially the reflection of the
general limitation of Hamiltonian notions of gravitational stress-energy-momentum
(as defined for a total, asymptotically-flat spacetime with parity
conditions) \textit{vis-à-vis} general quasilocal notions of such concepts---of which the Hamiltonian ones arise simply as a special case.

For carrying out practical EMRI computations, there is a manifest
advantage in formulating the self-force as a closed two-surface integral
around the moving ``particle'' versus standard approaches. In the
latter, one typically attempts to formulate the problem \emph{at}
the ``particle location'', \textit{i.e.} the support of the distributional
matter stress-energy-momentum tensor modeling the moving object (SCO)
in the background spacetime. Of course, due to the distributional
source, $\bm{h}$ actually diverges on its support, and so regularization
or Green's function methods are typically employed in order to make
progress. However, in principle, no such obstacles are encountered
(nor the aforementioned technical solutions needed) if the self-force
is evaluated on a boundary around---very close to, but at a finite
distance away from---the ``particle'', where no formal singularity
is ever encountered: $\bm{h}$ remains everywhere finite over the integration,
and therefore so does the (extended) self-force functional [Eq. (\ref{eq:generalized_GSF})] with it
directly as its argument. 

\subsection{Outlook to future work\label{ssec:outlook}}

A numerical implementation of a concrete self-force computation using
the approach developed in this chapter would be arguably the most salient
next step  to take. To our knowledge, no numerical work
has been put forth even using Gralla's ``angle average'' integral
formula [\cite{gralla_gauge_2011}] (which would further require gauge transformations away from
``parity-regular'' gauges).

We stress here that our proposed equation of motion involving the
gravitational self-force is entirely formulated and in principle valid
in any choice of perturbative gauge. To our knowledge, this is the first such proposal bearing this feature. This may provide a great advantage
for numerical work, as black hole metric perturbations $\bm{h}$ are
often most easily computed (by solving the linearized Einstein equation,
usually with a delta-function source motivated as in or similarly
to the approach of [\cite{gralla_rigorous_2008}] described in Subection \ref{ssec:gralla-wald_review}) in gauges that
are \emph{not} in the ``parity-regular'' class restricting Gralla's
formula [\cite{gralla_gauge_2011}]. In other words, we claim that one may solve the linearized
Einstein equation [Eq. (\ref{eq:first-order_Einstein_equation})] for $\bm{h}^{\bm{\mathsf{X}}}$ in any desired choice
of gauge $\bm{\mathsf{X}}$, insert this $\bm{h}^{\bm{\mathsf{X}}}$
into our extended GSF functional [Eq. (\ref{eq:generalized_GSF})] to obtain $\mathcal{\bm{F}}^{\bm{\mathsf{X}}}[\bm{h}^{\bm{\mathsf{X}}};\mathring{\bm{u}}^{\bm{\mathsf{X}}}]$
(for some choice of background quasilocal frame with four-velocity
$\mathring{\bm{u}}$), and then to integrate this over a ``small
radius'' topological two-sphere surrounding the ``particle'' (so
that $\mathring{\bm{u}}$ can be approximated by the background
geodesic four-velocity of the particle, $\mathring{\bm{U}}$), to obtain the full
\emph{extended} gravitational self-force (or ``correction to the
motion'') \emph{directly in that gauge} $\bm{\mathsf{X}}$. It is
easy to speculate that this may simplify some numerical issues tremendously
\textit{vis-à-vis} current approaches, where much technical machinery is needed
to handle (and to do so in a sufficiently efficient way for future
waveform applications) the necessary gauge transformations involving
distributional source terms.

Nevertheless, further work is needed to bring the relatively abstract
analysis developed in this chapter into a form more readily suited for
practical numerics. The most apparent technical issue to be tackled
involves the fact that $\bm{h}$ is usually computed (in some kind
of harmonics) in angular coordinates centered on the MBH, while the
functional $\bm{\mathcal{F}}[\bm{h};\mathring{\bm{u}}]$ is evaluated in
angular coordinates (on a ``small'' topological two-sphere) centered
on the moving ``particle'', \textit{i.e.} the SCO. A detailed understanding
of the transformation between the two sets of angular coordinates
is thus essential to formulate this problem numerically. This issue
is discussed a bit further in Gralla's paper [\cite{gralla_gauge_2011}], but a detailed implementation
of such a computation remains to be attempted.

The abstraction and generality of our approach may, on the other hand,
also provide useful ways to address some other technical issues surrounding
the self-force problem. For example, all the calculations in this
chapter may be carried on to second order (in the formal expansion parameter
$\lambda$)---which is conceptually straightforward given our basic
perturbative setup, but of course which requires an analysis in its
own right. Nonetheless, one may readily see that any higher-order
correction to the motion manifestly remains here in the form of a
boundary flux---only now involving nonlinear terms in the integrand.
Thus any sort of singular behaviour is avoided at the level of the
equations of motion in our approach, up to any order.

As another example, if ever desired (\textit{e.g.} for astrophysical reasons),
linear or any higher-order in $r$ (the areal radius of the SCO boundary)
effects on the correction to the motion can also be computed using
our approach. Moreover, any matter fluxes (described by the usual
matter stress-energy-momentum tensor, $\bm{T}$) can also be accommodated
thanks to our general (gravity plus matter) conservation laws [Eq. (\ref{eq:cons_law_Pa})]. 

Furthermore, while we have applied our ideas in this chapter to a specific
self-force approach---that of [\cite{gralla_rigorous_2008}]---our general formulation
(Section \ref{sec:general-analysis}) can just as well be used in any other approach to the gravitational
self-force, \textit{i.e.} any other specification of a perturbative procedure
(of a family of perturbed spacetimes $\{(\mathscr{M}_{(\lambda)},\bm{g}_{(\lambda)}\}$)
for this problem. In other words, our approach permits any alternative
specification of what is meant by a (sufficiently) ``localized source''
in general relativity, as our conservation expressions always involve
fluxes on their boundaries and are not conditioned in any way by the
exact details of their interior modeling. Thus our equation of motion [Eq. (\ref{eq:concl_eom})] could be used not only for a ``self-consistent'' computation (using, \textit{e.g.}, an approach such as that of  [\cite{spallicci_fully_2014,ritter_indirect_2016}] for solving the field equations in this context)
within the Gralla-Wald approach, but also, for example, in the context of the (mathematically equivalent) self-consistent formulation of [\cite{pound_self-consistent_2010}].

Beyond the gravitational self-force, another avenue to explore from
here---of interest at the very least for conceptual consistency---is
how our approach handles the electromagnetic self-force problem. Although
undoubtedly some conceptual parallels may be drawn between the gravitational
and electromagnetic self-force problems (see \textit{e.g.} [\cite{barack_self-force_2018}]), foundationally
they are usually treated as separate problems. Indeed, shortly after
the paper of [\cite{gralla_rigorous_2008}] detailing
the self-force approach used in this work, Gralla, Harte and Wald
[\cite{gralla_rigorous_2009}] put forth a similar analysis, with an
analogous approach and level of rigour, of the electromagnetic self-force.
It would be of great interest to apply our quasilocal conservation
laws in this setting, as they can be used to account not just for gravitational
but also (and in a consistent way) matter fluxes as well. It may thus prove insightful to study how
the transfer of energy-momentum is actually accounted for (between
the gravitational and the matter sector), as in our approach we are
not restricted to fixing a non-dynamical metric in the spacetime.
In other words, the conservation laws account completely for fluxes
due to a dynamical geometry as well as matter.

\section{Appendix: conformal Killing vectors and the two-sphere}\label{sec:5.A}

In this appendix we review some basic properties of \emph{conformal Killing vectors} (CKVs), and in particular
CKVs on the two-sphere.

A vector field $\bm{X}$ on any   $n$-dimensional Riemannian manifold $(\mathscr{U},\bm{g}_{\mathscr{U}},\bm{\nabla}_{\mathscr{U}})$
is a CKV if and only if it satisfies the \emph{conformal Killing equation},
\begin{equation}
\mathcal{L}_{\bm{X}}\bm{g}_{\mathscr{U}}=\psi\bm{g}_{\mathscr{U}}\,,\label{eq:A-CKV_equation_general}
\end{equation}
where $\psi\in C^{\infty}(\mathscr{U})$. This function can be determined
uniquely by taking the trace of this equation, yielding
\begin{equation}
\psi=\frac{2}{n}\bm{\nabla}_{\mathscr{U}}\cdot\bm{X}\,.\label{eq:A-CKV_equation_factor}
\end{equation}

Let us now specialize to the $r$-radius two-sphere $(\mathbb{S}_{r}^{2},r^{2}\bm{\mathfrak{S}},\bm{\mathfrak{D}})$,
where we denote our CKV by $\bm{\phi}$. Moreover for ease of notation in this appendix, the two-sphere volume form $\bm{\epsilon}_{\mathbb{S}^{2}}^{\,}$ [Eq. (\ref{eq:S2_volume_form})] is equivalently denoted by
$\bm{\mathfrak{E}}$, \textit{i.e.} $\mathfrak{E}_{\mathfrak{ij}}=\epsilon_{\mathfrak{ij}}^{\mathbb{S}^{2}}$.

In this case, the conformal
Killing equation (\ref{eq:A-CKV_equation_general}) is
\begin{equation}
\mathfrak{D}^{(\mathfrak{i}}\phi^{\mathfrak{j})}=\frac{1}{2r^{2}}\mathfrak{S}^{\mathfrak{ij}}\mathfrak{D}_{\mathfrak{k}}\phi^{\mathfrak{k}}\Leftrightarrow\mathfrak{D}^{\langle\mathfrak{i}}\phi^{\mathfrak{j}\rangle}=0\,,\label{eq:A-CKV_equation_two-sphere}
\end{equation}
where $\langle\cdot\cdot\rangle$ on two indices indicates taking
the STF part. The solution to this equation can be usefully expressed
in the form of a spherical harmonic decomposition in terms of the
standard direction cosines of a radial unit vector in $\mathbb{R}^{3}$,
which we denote by $r^{I}$. In spherical coordinates $\{x^{\mathfrak{i}}\}=\{\theta,\phi\}$,
it is simply given by
\begin{equation}
r^{I}(\theta,\phi)=(\sin\theta\cos\phi,\sin\theta\sin\phi,\cos\theta)\,.\label{eq:A-rI}
\end{equation}
This satisfies the following useful identity:
\begin{equation}
\intop_{\mathbb{S}_{r}^{2}}\bm{\epsilon}^{}_{\mathbb{S}^{2}}\prod_{n=1}^{\ell}r^{I_{\ell}}=\begin{cases}
0\,, & \textrm{for }\ell\textrm{ odd}\,,\\
\tfrac{4\pi}{(\ell+1)!!}\delta^{\{I_{1}I_{2}}\cdots\delta^{I_{\ell-1}I_{\ell}\}}\,, & \textrm{for }\ell\textrm{ even}\,,
\end{cases}
\end{equation}
where $(\ell+1)!!=(\ell+1)(\ell-1)\cdots1$
and the curly brackets on the indices denote the smallest set of permutations
that make the result symmetric. In particular, the $\ell=2$ and $\ell=4$
cases (which suffice for the calculations presented in this chapter)
are:
\begin{align}
\intop_{\mathbb{S}_{r}^{2}}\bm{\epsilon}^{}_{\mathbb{S}^{2}}\,r^{I}r^{J}=\, & \frac{4\pi}{3}\delta^{IJ}\,,\\
\intop_{\mathbb{S}_{r}^{2}}\bm{\epsilon}^{}_{\mathbb{S}^{2}}\,r^{I}r^{J}r^{K}r^{L}=\, & \frac{4\pi}{15}\left(\delta^{IJ}\delta^{KL}+\delta^{IK}\delta^{JL}+\delta^{IL}\delta^{KJ}\right)\,.
\end{align}

We can construct from Eq. (\ref{eq:A-rI}) two sets of $\ell=1$ spherical
harmonic forms on $\mathbb{S}_{r}^{2}$, namely the \emph{boost generators},
\begin{equation}
\mathfrak{B}_{\mathfrak{i}}^{I}(\theta,\phi)=\mathfrak{D}_{\mathfrak{i}}r^{I}=\left(\begin{array}{ccc}
\cos\theta\cos\phi & \cos\theta\sin\phi & -\sin\theta\\
-\sin\theta\sin\phi & \sin\theta\cos\phi & 0
\end{array}\right)\,,
\end{equation}
and the \emph{rotation generators},
\begin{equation}
\mathfrak{R}_{\mathfrak{i}}^{I}(\theta,\phi)=\mathfrak{E}_{\mathfrak{i}}\,^{\mathfrak{j}}\mathfrak{B}_{\mathfrak{j}}^{I}=\epsilon^{I}\,_{JK}r^{J}\mathfrak{B}_{\mathfrak{i}}^{K}=\left(\!\!\begin{array}{ccc}
-\sin\phi & \cos\phi & 0\\
-\sin\theta\cos\theta\cos\phi & -\sin\theta\cos\theta\sin\phi & \sin^{2}\theta
\end{array}\!\!\right),
\end{equation}
where $\epsilon_{IJK}$ is the volume form of $\mathbb{R}^{3}$. We
can obtain from these the vector fields $\mathfrak{B}_{I}^{\mathfrak{i}}=\frac{1}{r^{2}}\delta_{IJ}\mathfrak{S}^{\mathfrak{ij}}\mathfrak{B}_{\mathfrak{j}}^{J}=\mathfrak{D}^{\mathfrak{j}}r_{I}$
and $\mathfrak{R}_{I}^{\mathfrak{i}}=\frac{1}{r^{2}}\delta_{IJ}\mathfrak{S}^{\mathfrak{ij}}\mathfrak{R}_{\mathfrak{j}}^{J}=\mathfrak{E}^{\mathfrak{i}}\,_{\mathfrak{j}}\mathfrak{B}_{I}^{\mathfrak{j}}$,
which satisfy the Lorentz algebra
\begin{align}
\left[\bm{\mathfrak{B}}_{I},\bm{\mathfrak{B}}_{J}\right]=\, & \epsilon_{IJ}\,^{K}\bm{\mathfrak{R}}_{K}\,,\\
\left[\bm{\mathfrak{R}}_{I},\bm{\mathfrak{B}}_{J}\right]=\, & -\epsilon_{IJ}\,^{K}\bm{\mathfrak{B}}_{K}\,,\\
\left[\bm{\mathfrak{R}}_{I},\bm{\mathfrak{R}}_{J}\right]=\, & -\epsilon_{IJ}\,^{K}\bm{\mathfrak{R}}_{K}\,.
\end{align}

From the above, it is possible to derive a number of useful properties:
\begin{align}
r_{I}\mathfrak{B}_{\mathfrak{i}}^{I}=0\,,\quad & \mathfrak{D}_{\mathfrak{i}}\mathfrak{B}_{\mathfrak{j}}^{I}=-\mathfrak{S}_{\mathfrak{ij}}r^{I}\Rightarrow\mathfrak{S}^{\mathfrak{ij}}\mathfrak{D}_{\mathfrak{i}}\mathfrak{B}_{\mathfrak{j}}^{I}=-2r^{I}\,,\\
r_{I}\mathfrak{R}_{\mathfrak{i}}^{I}=0\,,\quad & \mathfrak{D}_{\mathfrak{i}}\mathfrak{R}_{\mathfrak{j}}^{I}=\mathfrak{E}_{\mathfrak{ij}}r^{I}\Rightarrow\mathfrak{S}^{\mathfrak{ij}}\mathfrak{D}_{\mathfrak{i}}\mathfrak{R}_{\mathfrak{j}}^{I}=0\,.
\end{align}
Using these, one can show that the sets of $\ell=1$ vector fields
$\bm{\mathfrak{B}}_{I}$ and $\bm{\mathfrak{R}}_{I}$ all satisfy
the conformal Killing equation, \textit{i.e.}
\begin{equation}
\mathfrak{D}^{\langle\mathfrak{i}}\mathfrak{B}_{I}^{\mathfrak{j}\rangle}=0=\mathfrak{D}^{\langle\mathfrak{i}}\mathfrak{R}_{I}^{\mathfrak{j}\rangle}\,.
\end{equation}
Finally, we give a list of useful relations for various contractions
involving these vector fields:
\begin{align}
\mathfrak{S}^{\mathfrak{ij}}\mathfrak{B}_{\mathfrak{i}}^{I}\mathfrak{B}_{\mathfrak{j}}^{J}=\mathfrak{S}^{\mathfrak{ij}}\mathfrak{R}_{\mathfrak{i}}^{I}\mathfrak{R}_{\mathfrak{j}}^{J}=-\mathfrak{E}^{\mathfrak{ij}}\mathfrak{B}_{\mathfrak{i}}^{I}\mathfrak{R}_{\mathfrak{j}}^{J}=\, & P^{IJ}\,,\\
\mathfrak{S}^{\mathfrak{ij}}\mathfrak{B}_{\mathfrak{i}}^{I}\mathfrak{R}_{\mathfrak{j}}^{J}=\mathfrak{E}_{\mathfrak{ij}}\mathfrak{B}_{I}^{\mathfrak{i}}\mathfrak{B}_{J}^{\mathfrak{j}}=\mathfrak{E}_{\mathfrak{ij}}\mathfrak{R}_{I}^{\mathfrak{i}}\mathfrak{R}_{J}^{\mathfrak{j}}=\, & \epsilon^{IJK}r_{K}\,,\\
\delta_{IJ}\mathfrak{B}_{\mathfrak{i}}^{I}\mathfrak{B}_{\mathfrak{j}}^{J}=\delta_{IJ}\mathfrak{R}_{\mathfrak{i}}^{I}\mathfrak{R}_{\mathfrak{j}}^{J}=\, & \mathfrak{S}_{\mathfrak{ij}}\,,\\
\delta_{IJ}\mathfrak{B}_{\mathfrak{i}}^{I}\mathfrak{R}_{\mathfrak{j}}^{J}=\, & -\mathfrak{E}_{\mathfrak{ij}}\,,\\
\epsilon_{IJK}\mathfrak{B}_{\mathfrak{i}}^{I}\mathfrak{B}_{\mathfrak{j}}^{J}=\epsilon_{IJK}\mathfrak{R}_{\mathfrak{i}}^{I}\mathfrak{R}_{\mathfrak{j}}^{J}=\, & \mathfrak{E}_{\mathfrak{ij}}r_{K}\,,\\
\epsilon_{IJK}\mathfrak{B}_{\mathfrak{i}}^{I}\mathfrak{R}_{\mathfrak{j}}^{J}=\, & \mathfrak{S}_{\mathfrak{ij}}r_{K}\,,
\end{align}
where $P^{IJ}=\delta^{IJ}-r^{I}r^{J}$ projects vectors perpendicular
to the radial direction.

Now we have everything in hand to formulate the general solution to
the conformal Killing equation (\ref{eq:A-CKV_equation_two-sphere})
on $\mathbb{S}_{r}^{2}$; it can be expanded as:
\begin{align}
\phi^{\mathfrak{j}}=\, & \frac{1}{r}\bigg[\mathfrak{D}^{\mathfrak{j}}\left(\sum_{\ell\in\mathbb{N}}\Phi^{I_{1}\cdots I_{\ell}}\prod_{n=1}^{\ell}r_{I_{n}}\right)+\mathfrak{E}^{\mathfrak{j}}\,_{\mathfrak{k}}\mathfrak{D}^{\mathfrak{k}}\left(\sum_{\ell\in\mathbb{N}}\Psi^{I_{1}\cdots I_{\ell}}\prod_{n=1}^{\ell}r_{I_{n}}\right)\bigg]\,,\label{eq:A-phi_general_solution}\\
=\, & \frac{1}{r}\bigg[\left(\Phi^{I}\mathfrak{B}_{I}^{\mathfrak{j}}+\Psi^{I}\mathfrak{R}_{I}^{\mathfrak{j}}\right)+\sum_{\ell\geq2}\ell\Big(\Phi^{I_{1}\cdots I_{\ell}}\mathfrak{B}_{I_{1}}^{\mathfrak{j}}+\Psi^{I_{1}\cdots I_{\ell}}\mathfrak{R}_{I_{1}}^{\mathfrak{j}}\Big)\prod_{n=2}^{\ell}r_{I_{n}}\bigg]\,,
\end{align}
where to write the second equality we have used the fact that $\Phi^{I_{1}\cdots I_{\ell}}$
and $\Psi^{I_{1}\cdots I_{\ell}}$ are symmetric in their indices.

We are interested in working with the $\ell=1$ and $\ell=2$ parts of
$\bm{\phi}$ corresponding to linear momentum only ($\bm{\Psi}=0$):
\begin{align}
\phi_{\ell=1}^{\mathfrak{i}}=\, & \frac{1}{r}\Phi^{I}\mathfrak{B}_{I}^{\mathfrak{i}}\,,\\
\phi_{\ell=2}^{\mathfrak{i}}=\, & \frac{2}{r}\Phi^{IJ}\mathfrak{B}_{I}^{\mathfrak{i}}r_{J}\,.
\end{align}


\chapter[A Frequency-Domain Implementation of the PwP Approach to EMRIs]{A Frequency-Domain Implementation of the Particle-without-Particle Approach to EMRIs\label{6-fd}}
\newrefsegment

\subsection*{Chapter summary}

This chapter is based on the conference proceeding [\cite{oltean_frequency-domain_2017}] and ongoing work.

We present here a frequency-domain implementation of the Particle-without-Particle (PwP) technique previously developed for the computation of the scalar self-force, a helpful testbed for the gravitational case.

We offer a short introduction in Section \ref{sec:6-Intro}, commenting briefly on the choice between time and frequency domain methods in carrying out numerics for the EMRI problem.

In Section \ref{sec:6-Scalar}, we formulate the problem of the scalar self-force in a non-spinning black hole spacetime in full mathematical detail. In particular, the moving particle here possesses a scalar charge due to a scalar field which does not back-react on the geometry (\textit{i.e.} the background remains fixed). We comment on the widely-used mode-sum regularization procedure for devising a numerical implementation.

Then in Section \ref{sec:6-PwP}, we discuss the Particle-without-Particle (PwP) method, a pseudospectral collocation method previously used for the computation of the scalar self-force in the time domain. The idea is to decompose quantities into linear combinations of Heaviside functions (supported, in this case, at inner and outer radii relative to the particle orbit), turning the distributionally-sourced field equations into systems of homogeneous equations (away from the particle) supplemented by “jump” (boundary) conditions connecting them (at the particle location).

In Section \ref{sec:6-FD}, we present the frequency-domain formulation of the scalar self-force problem, including the appropriate boundary and jump conditions.

Finally, in Section \ref{sec:6-Numerical}, we discuss in basic outline of our numerical implementation using a hyperbolic compactification and multidomain splitting of the computational grids, omitting much of the technical detail. We also present some results on circular orbits, with work on eccentric orbits in progress.

\subsection*{Una implementació en el domini de freqüències del mètode Partícula-sense-Partícula als \textit{EMRIs} \normalfont{(chapter summary translation in Catalan)}}

Aquest capítol es basa en la acta [\cite{oltean_frequency-domain_2017}] i treball en curs.

Aquí presentem una implementació en el domini de freqüència de la tècnica Partícula-sense-Partícula (Particle-without-Particle, PwP) desenvolupada anteriorment per a la computació de la força pròpia escalar, una prova útil per al cas gravitatori.

Oferim una breu introducció a la secció \ref{sec:6-Intro}, comentant breument sobre l’elecció entre mètodes de domini de temps i domini de freqüència en la solució numèrica del problema EMRI.

A la secció \ref{sec:6-Scalar}, formulem el problema de la força pròpia escalar en un espai-temps de un forat negre que no gira, amb tot el detall matemàtic. En particular, la partícula en moviment aquí té una càrrega escalar a causa d'un camp escalar que no retroacciona sobre la geometria (és a dir, el fons queda fixat). Comentem el procediment de regularització de sumes de modes, molt utilitzat per idear implementacions numèriques.

A continuació, a la secció \ref{sec:6-PwP}, es discuteix el mètode Partícula-sense- Partícula (PwP), un mètode de col·locació pseudospectral usat anteriorment per al càlcul de la força pròpia escalar en el domini temporal. La idea és descompondre quantitats en combinacions lineals de funcions Heaviside (suportades, en aquest cas, en els radis interns i externs respecte a l’òrbita de la partícula), convertint les equacions de camp amb fonts distributives en sistemes d’equacions homogènies (allunyades de la partícula) complementades mitjançant les condicions de ``salt'' (límit) que els connecten (a la ubicació de la partícula).  

A la secció \ref{sec:6-FD}, es presenta la formulació de dominis de freqüència del problema de la força pròpia escalar, incloent-ne els límits i les condicions de salt adequades. 

Finalment, a la secció \ref{sec:6-Numerical}, es discuteix en l'esquema bàsic de la nostra implementació numèrica mitjançant una compactació hiperbòlica i una divisió multidomànica de les reixes computacionals, ometent gran part del detall tècnic. També presentem alguns resultats sobre òrbites circulars, amb treballs sobre òrbites excèntriques en marxa.

\subsection*{Une implémentation dans le domaine fréquentiel de l'approche Particule-sans-Particule aux \textit{EMRIs}  \normalfont{(chapter summary translation in French)}}
 
Ce chapitre est basé sur l’acte de congrès [\cite{oltean_frequency-domain_2017}] et travaux en cours.

Nous présentons ici une implémentation dans le domaine fréquentiel de la technique Particule-sans-Particule (Particle-without-Particle, PwP) développée précédemment pour le calcul de la force propre scalaire - un test utile pour le cas gravitationnel.

Nous proposons une brève introduction à la section \ref{sec:6-Intro}, en commentant brièvement sur le choix entre les méthodes de domaine temporel et fréquentiel pour la réalisation des calculs pour le problème des EMRIs.

Dans la section \ref{sec:6-Scalar}, nous formulons le problème de la force propre scalaire dans un espace-temps de trou noir que ne tourne pas, avec tous les détails mathématiques. En particulier, la particule en mouvement possède ici une charge scalaire due à un champ scalaire que ne rétroactionne pas sur la géométrie (c'est-à-dire que le fonde reste fixe). Nous commentons la procédure de régularisation des sommes des modes largement utilisée pour concevoir une implémentation numérique.

Ensuite, dans la section \ref{sec:6-PwP}, nous discutons de la méthode PwP, une méthode de collocation pseudospectrale précédemment utilisée pour le calcul de la force propre scalaire dans le domaine temporel. L’idée est de décomposer les quantités en combinaisons linéaires de fonctions de Heaviside (supportées, dans ce cas, aux rayons intérieurs et extérieurs par rapport à l’orbite de la particule), en transformant les équations du champ avec sources distributionnelles en systèmes d’équations homogènes (loin de la particule) complétées par des conditions de « saut » (aux limites) que les connecte (à l’emplacement de la particule).

Dans la section \ref{sec:6-FD}, nous présentons la formulation dans le domaine fréquentiel du problème de la force propre scalaire, y compris les conditions de limite et de saut appropriées.

Enfin, dans la section \ref{sec:6-Numerical}, nous discutons dans les grandes lignes de notre implémentation numérique en utilisant une compactification hyperbolique et une division en plusieurs domaines des grilles de calcul, en omettant une grande partie des détails techniques. Nous présentons également quelques résultats sur des orbites circulaires, avec des travaux sur des orbites excentriques en cours.

\section{\label{sec:6-Intro}Introduction}

The computation of the self-force and waveforms, and any other physical
relevant information related to the inspiral due to radiation reaction
constitute the main challenge of the EMRI problem. One possible strategy
is to resort to analytic techniques by adding extra approximations
to the problem, similar to those from post-Newtonian methods. However,
the results may not be applicable to situations of physical relevance
involving highly spinning MBHs and very eccentric orbits. To make
computations without making further simplifications of the problem,
numerical techniques appear to be a necessary tool. 

Broadly speaking, one faces a choice in how to proceed between frequency-domain
and time-domain calculations. The frequency domain approach has been
used for a long time; it provides accurate results for the computation
of quasinormal modes and frequencies~[\cite{vishveshwara_stability_1970,Chandrasekhar:1975qn}].
However, this approach encounters greater difficulties when one is
interested in computing the waves originated from highly eccentric
orbits since one has to sum over a large number of modes to obtain
a good accuracy. In this sense, calculations in the time-domain can
be better adapted for obtaining accurate waveforms for the physical
situations of relevance. Nevertheless, overall, time-domain methods
can be much slower than working in the frequency domain.

We consider in this chapter a frequency-domain implementation of a
simplified EMRI model, corresponding to a charged scalar particle
orbiting a non-rotating MBH. There is in this case no (gravitational)
backreaction upon the background geometry (which therefore remains
fixed). This offers a very useful setting to test different numerical
implementations, with a view towards using those which prove most
successful in the full gravitational self-force problem. 

The method that we use here, and which has been developed in the past
in the time domain, is called the Particle-without-Particle (PwP)
method [\cite{canizares_extreme-mass-ratio_2011,canizares_efficient_2009,canizares_pseudospectral_2010,canizares_time-domain_2011,canizares_tuning_2011,jaramillo_are_2011,canizares_overcoming_2014}].
The basic idea is to split the computational domain into two (or more)
disjoint regions whereby any non-singular quantity $Q$ is decomposed
as $Q=Q_{-}\Theta_{p}^{-}+Q_{+}\Theta_{p}^{+}$, where $\Theta_{p}^{\pm}=\Theta(\pm(r-r_{p}))$
is the Heaviside step function with the step at the particle's radial
location $r_{p}$. Quantities that are not continuous will have jumps
across the SCO trajectory, which we denote by $[Q]_{p}=\lim_{r\rightarrow r_{p}(t)}(Q_{+}-Q_{-})$.
In this setup then, any differential equation with a singular (distributional)
source is effectively replaced with homogeneous equations to the left
and right of the SCO, subject to certain jump conditions across it.
See also Appendix B.

\section{\label{sec:6-Scalar}The scalar self-force}

In our simplified EMRI model, the SCO is represented as a scalar particle,
\emph{i.e.} a body the charge distribution of which has support only
on its (timelike) worldline $\mathscr{C}$, parametrized by $z^{a}(\tau)$,
with a charge $q$ associated to a scalar field $\Phi$; meanwhile,
the MBH is described by a fixed Schwarzschind-Droste spacetime $(\mathring{\mathscr{M}},\mathring{\bm{g}},\mathring{\bm{\nabla}}$),
\emph{i.e.} a background not affected by the charged particle, with
the following metric metric (for more details, see Section 3.3): 
\begin{align}
\mathring{g}_{ab}{\rm d}x^{a}{\rm d}x^{b} & =-f{\rm d}t^{2}+f^{-1}{\rm d}r^{2}+r^{2}{\rm d}\Omega^{2}\\
 & =f\left(-{\rm d}t^{2}+{\rm d}r_{*}^{2}\right)+r^{2}{\rm d}\Omega^{2}\,,
\end{align}
where $f(r)=1-2M/r$, ${\rm d}\Omega^{2}={\rm d}\theta^{2}+\sin^{2}\theta{\rm d}\varphi^{2}$
is the two-sphere line element and $r_{*}=r+2M\ln(r/2M-1)$ is the
so-called radial \emph{tortoise} coordinate.

The dynamics are in this case determined by the following action:
\begin{equation}
\mathcal{S}\left[\Phi,z\right]=\int\boldsymbol{\epsilon}_{\mathscr{\mathring{M}}}^{\,}\bigg\{\boldsymbol{\nabla}\Phi\cdot\boldsymbol{\nabla}\Phi+\int_{\mathscr{C}}{\rm d}\tau\,\delta_{4}(x-z(\tau))\bigg[\frac{m}{2}\,\boldsymbol{u}\cdot\bm{u}+q\Phi\bigg]\bigg\}\,,\label{action}
\end{equation}
where $\bm{u}=\dot{\bm{z}}$ is the particle four-velocity and $m$
is its (time-dependent) ``mass''. The first term is the kinetic
term for the scalar field, the one proportional to $m$ is the standard
geodesic action for the particle, and the one proportional to $q$
is a coupling between the field and the particle motion---leading
to nontrivial sources in both the field equation and equation of motion.

The scalar field satisfies the following wave-like equation, obtained
by extremizing the action (\ref{action}) with respect to $\Phi$
(see, e.g.~[\cite{poisson_motion_2011}]): 
\begin{equation}
\mathring{\square}\Phi(x)=-4\pi q\int_{\mathscr{C}}{\rm d}\tau\,\delta_{4}(x-z(\tau))\,,\label{geo}
\end{equation}
where $\mathring{\square}=\mathring{\nabla}^{a}\mathring{\nabla_{a}}$
is the wave operator.

The field equation~(\ref{geo}) has to be complemented with the equation
of motion for the scalar charged particle, obtained by extremizing
the action (\ref{action}) with respect to $\boldsymbol{z}$: 
\begin{equation}
\nabla_{\bm{u}}(mu^{a})=\mathfrak{F}^{a}=q\mathring{g}^{ab}\left.\left(\mathring{\nabla}_{b}\Phi\right)\right|_{\mathscr{C}}\,,\label{particlemotion}
\end{equation}

The coupled set of equations formed by the PDE for the scalar field~(\ref{geo})
and the ODE for the particle trajectory~(\ref{particlemotion}) constitute
our testbed model for an EMRI. The SCO generates (sources) a scalar
field according to (\ref{geo}), which in turn affects to the SCO
motion according to~(\ref{particlemotion}), that is, through the
action of a local self-force $\mathfrak{F}^{a}$. This mechanism is
the (scalar) analogue of the gravitational backreaction mechanism
that produces the inspiral via the gravitational self-force.

Now, the retarded solution of~(\ref{geo}) is singular at the particle
location, while the force in Eq.~(\ref{particlemotion}) involves
the gradient of the field evaluated at the particle location. Therefore,
as they stand, Eqs.~(\ref{geo}) and~(\ref{particlemotion}) are
formal equations that require an appropriate regularization to become
fully meaningful. Following~[\cite{detweiler_self-force_2003}], the retarded
field can be split into two parts: a singular piece, $\Phi^{\textrm{S}}$,
which contains the singular structure of the field and satisfies the
same wave equation as the retarded field, \textit{i.e.} Eq.~(\ref{geo}),
and a regular part, $\Phi^{\textrm{R}}$, that satisfies the homogeneous
equation associated with Eq.~(\ref{geo}). As it turns out, $\Phi^{\textrm{R}}$
is regular and differentiable at the particle position and is solely
responsible for the scalar self-force~[\cite{detweiler_self-force_2003}]. We
can therefore write 
\begin{equation}
\mathfrak{F}_{\textrm{R}}^{a}=q\mathring{g}^{ab}\left.\left(\mathring{\nabla}_{b}\Phi^{\textrm{R}}\right)\right|_{\mathscr{C}}\,,\label{regparticlemotion}
\end{equation}
which gives a definite sense to the equations of motion of the system.

We can solve the field equation~(\ref{geo}) by expanding the scalar
field in (scalar) spherical harmonics: 
\begin{eqnarray}
\Phi=\sum_{\ell=0}^{\infty}\sum_{m=-\ell}^{\ell}\Phi^{\ell m}(t,r)Y^{\ell m}(\theta,\varphi)\;.\label{phiex}
\end{eqnarray}
The equations for each harmonic mode, $\Phi^{\ell m}(t,r)$, are decoupled
from the rest and take the form of the following ($1+1$)-dimensional
wave equation for $\psi^{\ell m}=r\Phi^{\ell m}$: 
\begin{eqnarray}
\left(-\partial_{t}^{2}+\partial_{r_{*}}^{2}-V_{\ell}(r)\right)\psi^{\ell m}=S^{\ell m}\delta(r-r_{p}(t))\,,\label{master}
\end{eqnarray}
where 
\begin{equation}
V_{\ell}(r)=f(r)\left[\frac{\ell(\ell+1)}{r^{2}}+\frac{2M}{r^{3}}\right]\,,
\end{equation}
is the Regge-Wheeler potential for scalar (spin-zero) fields on the
Schwarzschild-Droste geometry, and 
\begin{eqnarray}
S^{\ell m}=-\frac{4\pi qf_{p}}{r_{p}u^{t}}\,\bar{Y}^{\ell m}\left(\frac{\pi}{2},\varphi_{p}(t)\right)\,,\label{source}
\end{eqnarray}
is the coefficient of the singular source term due to the presence
of the particle, $f_{p}=f(r_{p})$, and the bar denotes complex conjugation.
Here we have assumed, without loss of generality, that the particle's
orbit takes place in the equatorial plane $\theta=\pi/2$. Moreover,
$r_{p}$ and $\varphi_{p}$ denote the radial and azimuthal coordinates
of the particle, and are functions of the coordinate time $t$.

The expansion in spherical harmonics is also very useful to construct
the regular field, $\Phi^{\textrm{R}}$. Indeed, it turns out that
each harmonic mode of the retarded field, $\Phi^{\ell m}(t,r)$, is
finite and continuous at the location of the particle; it is the sum
over $\ell$ what diverges there.

Here, the \emph{mode-sum} regularization scheme~[\cite{Barack:1999wf,Barack:2001gx,Barack:2002mha}]
comes into play: it provides analytic expressions for the singular
part of each $\ell$-mode of the retarded field. These expressions
for the singular field are valid only near the particle location.
The regularized self-force is thus obtained by computing numerically
each harmonic mode of the self-force and subtracting the singular
part provided by the mode-sum scheme.

The regular part of the gradient of the field, which in coordinates
$x^{\alpha}$ we denote simply as $\nabla_{\alpha}\Phi^{\textrm{R}}\equiv\Phi_{\alpha}^{\textrm{R}}$,
is given by 
\begin{equation}
\Phi_{a}^{\textrm{R}}(\bm{z}(\tau))=\lim_{x^{\mu}\to z^{\mu}(\tau)}\sum_{\ell=0}^{\infty}\left(\Phi_{\alpha}^{\ell}(x^{\mu})-\Phi_{\alpha}^{\textrm{S},\ell}(x^{\mu})\right)\,.\label{regular2}
\end{equation}
where 
\begin{equation}
\Phi_{\alpha}^{\ell}(x^{\mu})=\sum_{m=-\ell}^{\ell}\nabla_{\alpha}(\Phi^{\ell m}(t,r)Y^{\ell m}(\theta,\varphi))\,,\label{fullphiell}
\end{equation}
and the structure of the singular field can can be written as: 
\begin{eqnarray}
\Phi_{\alpha}^{\textrm{S},\ell} & = & q\left[\left(\ell+\frac{1}{2}\right)A_{\alpha}+B_{\alpha}+\frac{C_{\alpha}}{\ell+\frac{1}{2}}+\frac{D_{\alpha}}{(\ell-\frac{1}{2})(\ell+\frac{3}{2})}+\ldots\right]\,.\label{singular}
\end{eqnarray}
The expressions for the regularization parameters $A_{\alpha}$, $B_{\alpha}$,
$C_{\alpha}$, and $D_{\alpha}$, can be found in the literature for
generic orbits~[\cite{Barack:2002mha,Kim:2004yi,hoon:2005dh,Haas:2006ne}].
They do not depend on $\ell$, but on the trajectory of the particle.
The three first coefficients of~(\ref{singular}) are responsible
for the divergences, whereas the remaining terms converge to zero
once they are summed over $\ell$. The expressions for the regularization
parameters that we use in this chapter are listed in the Appendix of~[\cite{canizares_pseudospectral_2010}].


\section{\label{sec:6-PwP}The Particle-without-Particle method}

\begin{figure}
\centering{}\includegraphics[scale=0.26]{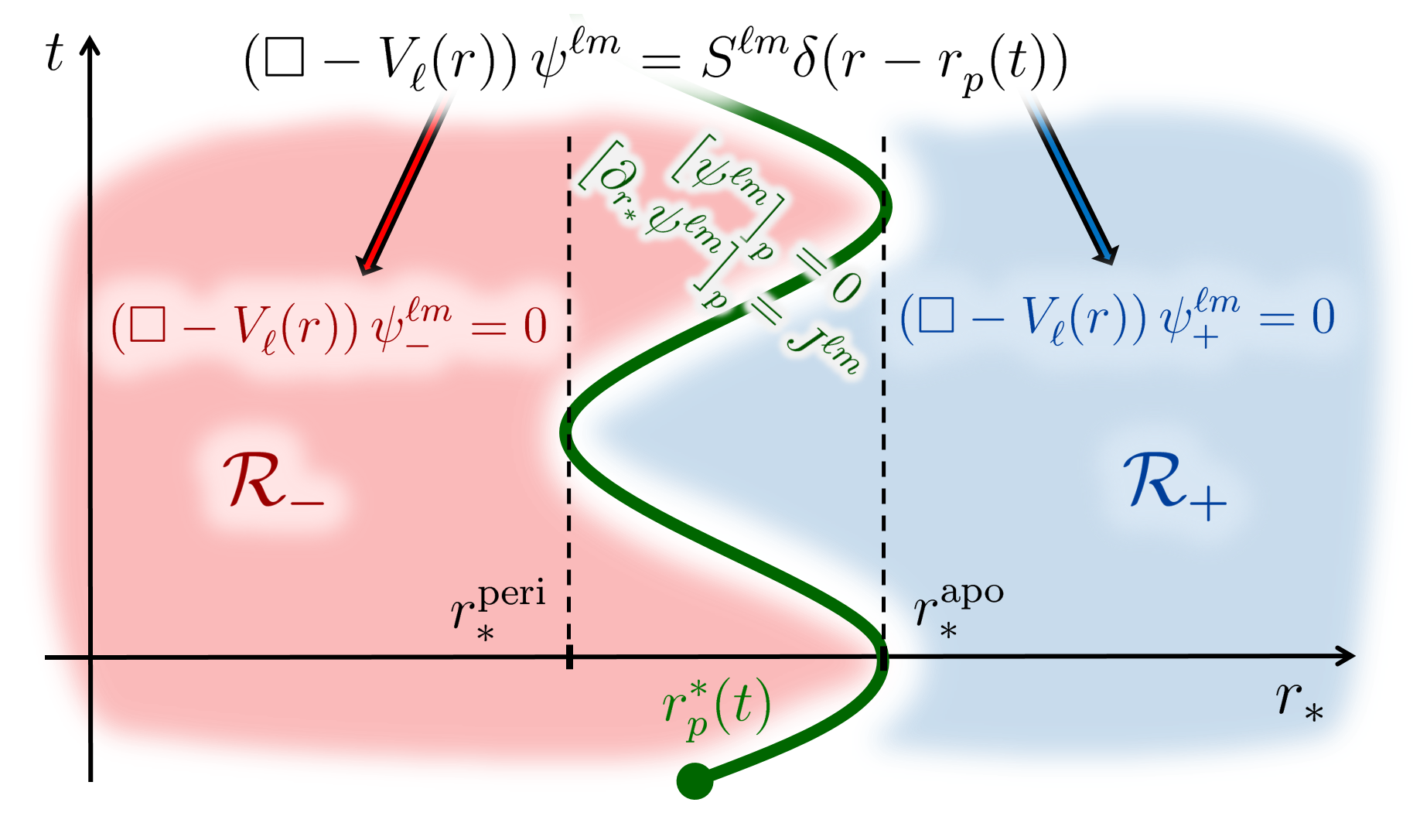} \caption{Schematic representation of the PwP formulation. The field equations
with singular source terms become homogeneous equations at each side
of the particle worldline together with a set of jump conditions to
communicate their solutions.}
\label{FIG6.1} 
\end{figure}

The full retarded solution has to be found numerically and hence,
it is very convenient to formulate the equations so that we obtain
smooth solutions. However, the presence of singularities in Eqs.~(\ref{master})
makes the task difficult in principle. To overcome these problems,
the Particle-without-Particle (PwP) method~[\cite{Canizares:2008dp,canizares_efficient_2009,canizares_pseudospectral_2010,canizares_tuning_2011}]
splits the computational domain (in the $(t,r)$ space) into two disjoin
regions (see Fig.~\ref{FIG6.1}): Region $\mathcal{R}_{-}$ to the left
of the SCO trajectory ($r<r_{p}(t)$) and region $\mathcal{R}_{+}$ to
the right ($r>r_{p}(t)$). Then, any non-singular quantity $Q(t,r)$
admits a decomposition 
\begin{equation}
Q=Q_{-}\Theta_{p}^{-}+Q_{+}\Theta_{p}^{+}\,,\label{PwPsplit}
\end{equation}
where $\Theta_{p}^{-}\equiv\Theta(r_{p}-r)$ and $\Theta_{p}^{+}\equiv\Theta(r-r_{p})$,
and $\Theta$ is the Heaviside step function. Quantities that are
not continuous will have jumps across the SCO trajectory. The jump
in a quantity $\mathcal{Q}$ is a time-only dependent quantity defined
as: $[Q](t)=\lim_{r\rightarrow r_{p}(t)}(Q_{+}(t,r)-Q_{-}(t,r))\equiv\left[Q\right]_{p}\,$.

Applying the PwP formulation to the scalar equation~(\ref{master}),
\emph{i.e.} introducing 
\begin{equation}
\psi^{\ell m}=\psi_{-}^{\ell m}\,\Theta_{p}^{-}+\psi_{+}^{\ell m}\,\Theta_{p}^{+}\,,\label{PwPsplitrPhi}
\end{equation}
it transforms into homogeneous equations (with no matter source terms)
at each region $\mathcal{R}_{\pm}$: 
\begin{equation}
\left(-\partial_{t}^{2}+\partial_{r_{*}}^{2}-V_{\ell}(r)\right)\psi_{\pm}^{\ell m}=0\,,\label{mastervacuum}
\end{equation}
plus a set of jump conditions on $\psi_{\pm}^{\ell m}$ and $\partial_{r_{*}}\psi_{\pm}^{\ell m}\,,$
which read: 
\begin{align}
\left[\psi^{\ell m}\right]_{p} & =0\,,\label{jumppsi}\\
\left[\partial_{r_{*}}\psi_{\pm}^{\ell m}\right]_{p} & ={\displaystyle \frac{S^{\ell m}}{(1-(\dot{r}_{*}^{2})_{p})f_{p}}}\equiv J^{\ell m}\,.\label{jumppsirsu}
\end{align}
In summary, at each region we have equations without the singular
terms induced by the SCO. Then, since these equations are strongly
hyperbolic, we obtain smooth solutions. Finally, the SCO appears in
the communication between the two regions by enforcing the jump conditions.
The spherical symmetry of the MBH background leads to jumps only in
time and radial derivatives. For instance, for first order derivatives
we find: $[\partial_{t}\mathcal{Q}^{\ell m}]_{p}={\rm d}[\mathcal{Q}^{\ell m}]_{p}/{\rm d}t-\dot{r}_{p}[\partial_{r}\mathcal{Q}^{\ell m}]_{p}$;
the same happens for derivatives of higher order. In particular, we
get 
\begin{equation}
\left[\partial_{t}\psi^{\ell m}\right]_{p}=-\frac{(\dot{r}_{*})_{p}S^{\ell m}}{(1-(\dot{r}_{*}^{2})_{p})f_{p}}\,.
\end{equation}


\section{\label{sec:6-FD}Frequency domain analysis}


We now turn the analysis to the frequency domain. We Fourier decompose
our solution: 
\begin{equation}
\tilde{\psi}_{\pm}^{\ell m}(\omega,r)=\frac{1}{\sqrt{2\pi}}\int_{\mathbb{R}}{\rm d}t\,e^{{\rm i}\omega t}\,\psi_{\pm}^{\ell m}(t,r)\,,
\end{equation}
\begin{equation}
\psi_{\pm}^{\ell m}(t,r)=\frac{1}{\sqrt{2\pi}}\int_{\mathbb{R}}{\rm d}\omega\,e^{-{\rm i}\omega t}\,\tilde{\psi}_{\pm}^{\ell m}(\omega,r)\,.
\end{equation}
In this work we are interested only on bounded trajectories around
the MBH, \textit{i.e.} trajectories such that the SCO radial coordinate ranges
over a finite interval neither crossing the event horizon nor escaping
to spatial infinity~[\cite{Wilkins:1972rs}]. In that case, the motion
in the radial coordinate is periodic and as a consequence, the jump
of Eq.~(\ref{jumppsirsu}), can be expanded as a discrete Fourier
series: 
\begin{equation}
\left[\partial_{r_{*}}\psi^{\ell m}\right]_{p}(t)=\sum_{n=-\infty}^{n=+\infty}c_{\ell mn}e^{-{\rm i}\omega_{n\ell m}t}\,,\label{fourierjump}
\end{equation}
with 
\begin{equation}
\omega_{\ell mn}=n\omega_{r}+m\omega_{\varphi}\equiv\omega_{nm}\,,
\end{equation}
where $\omega_{r}$ and $\omega_{\varphi}$ are the frequencies associated
with the radial motion (going from periapsis to apoapsis and back)
and the azimuthal motion (going $2\pi$ around the polar axis), respectively.
We only need to sum over $n$ because the only dependence of the jump
on the $\omega_{\varphi}$ frequency comes from the spherical harmonics,
in the term $\exp\{-{\rm i}m\varphi_{p}(t)\}$ {[}see Eqs.~(\ref{jumppsi})-(\ref{jumppsirsu})
and~(\ref{source}){]}. Since we have bounded orbits, we can expand
the fields $\psi_{\pm}^{\ell m}$ in discrete Fourier series as
\begin{equation}
\psi_{\pm}^{\ell m}(t,r)=e^{-{\rm i}m\omega_{\varphi}t}\sum_{n=-\infty}^{+\infty}e^{-{\rm i}n\omega_{r}t}\,R_{\ell mn}^{\pm}(r)\,.
\end{equation}
This leads to ODEs for the components of the series, that is, for
the functions $R_{\ell mn}(r)$\footnote{~In the continuum, when the spectrum of $\omega$ is not discrete as
in this case where we have bounded orbits around a Schwarzschild black
hole, the radial functions are denoted by $R_{\ell m\omega}^ {}$,
so the corresponding notation should be $R_{\ell m\omega_{mn}^ {}}^ {}$
as in~[\cite{Barack:2008ms}]. However, for the sake of simplicity
we will use $R_{\ell mn}^ {}$.}. These are {[}from Eq.~(\ref{mastervacuum}){]}: 
\begin{equation}
\left(\frac{{\rm d}^{2}}{{\rm d}r_{*}^{2}}-V_{\ell}(r)+\omega_{nm}^{2}\right)R_{\ell mn}^{\pm}=0\,.\label{masterpwp}
\end{equation}

To complete the problem, we need boundary conditions. In this case
the boundary conditions have to be prescribed both at the horizon
($r_{*}\rightarrow-\infty$) and at spatial infinity ($r_{*}\rightarrow+\infty$).
The condition at spatial infinity is that the field has to be purely
outgoing: 
\begin{equation}
\left.\left(\partial_{t}+\partial_{r_{*}}\right)\psi_{+}^{\ell m}\right|_{r_{*}\rightarrow+\infty}=0\,,\label{bcright}
\end{equation}
and at the horizon it has also be outgoing (ingoing from the point
of view of spatial infinity): 
\begin{equation}
\left.\left(\partial_{t}-\partial_{r_{*}}\right)\psi_{-}^{\ell m}\right|_{r_{*}\rightarrow-\infty}=0\,.\label{bcleft}
\end{equation}
These boundary conditions, in terms of the radial functions $R_{\ell mn}$
become: 
\begin{align}
\left.\left(-{\rm i}\omega_{mn}+\frac{{\rm d}}{{\rm d}r_{*}}\right)R_{\ell mn}^{+}\right|_{r_{*}\rightarrow+\infty} & =0\,,\label{bcatspatialinfinity}\\
\left.\left({\rm i}\omega_{mn}+\frac{{\rm d}}{{\rm d}r_{*}}\right)R_{\ell mn}^{-}\right|_{r_{*}\rightarrow-\infty} & =0\,.\label{bcathorizon}
\end{align}
These equations have to be solved simultaneously with the junction
conditions for the radial functions $R_{\ell nm}^{\pm}$, which are
given {[}from Eqs.~(\ref{jumppsi})-(\ref{jumppsirsu}) and~(\ref{fourierjump}){]}
by 
\begin{align}
\sum_{n=-\infty}^{+\infty}e^{-{\rm i}n\omega_{r}t}\,\left[R_{\ell mn}\right]_{p} & =0\,,\label{jumpsR}\\
e^{-{\rm i}m\omega_{\varphi}t}\sum_{n=-\infty}^{+\infty}e^{-{\rm i}n\omega_{r}t}\,\left[\frac{{\rm d}R_{\ell mn}}{{\rm d}r_{*}}\right] & =J^{\ell m}\,.\label{jumpsRrsu}
\end{align}
Note that for circular orbits, these reduce to $[R_{\ell mn}]_{p}=0$
and $[{\rm d}R_{\ell mn}/{\rm d}r_{*}]_{p}=c_{\ell mn}$ respectively,
where $\ensuremath{c_{\ell mn}}$ {[}determined from Eq.~(\ref{jumppsirsu}){]}
are just the Fourier components of the jump in the gradient of the
scalar field at the particle location.


\section{\label{sec:6-Numerical}Numerical implementation and results}

Now, let us discuss the strategy to solve the set of equations~(\ref{masterpwp}),
(\ref{bcatspatialinfinity}), (\ref{bcathorizon}), and~(\ref{jumpsR}).
We offer here only an outline of our procedure and omit entering fully
into the technicalities.

The first ingredient we consider is the type of grid and how many
domains would be adequate to use. The tortoise coordinate has an unbounded
range, $r_{*}\in(-\infty,+\infty)$, and hence we either truncate
the physical domain or we use some different coordinate system in
which we cover the horizon and spatial infinity with a coordinate
with a finite range. The first option is the most widely used solution
in many problems. The drawback of this choice is that if we still
use the same boundary conditions, Eqs.~(\ref{bcright}) and~(\ref{bcleft}),
we are making an error since these boundary conditions are not exact
at a finite value of $r_{*}$. Of course, the error made will depend
on how far from the particle location we truncate the domain. In a
time-domain setup we can always choose the truncation locations in
such a way that the boundaries remain out of causal contact with the
particle, avoiding the contamination of the solution around the particle
from the boundaries. But in the frequency domain we are solving elliptic
equations, which care about the boundaries. A possible solution is
to obtain precise boundary conditions from an expansion of our equation
near the horizon and spatial infinity (see~[\cite{Barack:2008ms}]). 

However, here we are going to use the second option, that is, to use
a compactified coordinate so that both the horizon and spatial infinity
are located at finite values of the new coordinate. We are not going
to use just a compactification of the radial coordinate, as this would
solve the problem of the boundary conditions but would create another
problem, namely that many cycles of the radiation would accumulate
near the boundaries and it would not be possible to resolve them appropriately.
A solution to this is to use a hyperboloidal compactification [\cite{Zenginoglu:2010cq}],
where we also change the time coordinate (the slicing of the spacetime
in $t=\mbox{const.}$ hypersurfaces) so that we avoid the problem
just mentioned, since in the new slicing we will only have a few number
of cycles in such a way that they can be resolvable numerically with
a reasonable amount of computational resources. In particular, we
follow essentially the method of [\cite{Zenginoglu:2010cq}], with a
multidomain splitting of our computational grid so that the hyperboloidal
compactification is applied only to suitable boundary regions (extending
from a certain point to the horizon and spatial infinity respectively).

In the frequency-domain, as we have seen, the $\ell$-harmonics of
the SF are found by decomposing the retarded field in a Fourier series
with indices $(\ell,m,\omega)$ (where the frequency $\omega$ in
the case of bounded orbits can be labeled by $m$ and an integer $n$,
$\omega\equiv\omega_{mn}$). Then, summing over $\omega$ and $m$
(\textit{i.e.}, over $m$ and $n$ in the case of bounded orbits) we obtain
the different $\ell$-harmonics of the SF. The main advantage, as
discussed, is that the computation involves only ODEs. The drawback,
however, is that it was found that for the case of eccentric orbits,
the sum over $\omega$ ($n$) has bad convergence properties as a
consequence of the discontinuities at the particle location. In practice,
the problem is analogous to the well-known Gibbs phenomenon that arises
in standard Fourier analysis. In~[\cite{Barack:2008ms}], a solution
to this problem was proposed; the key point of the method was to use
the homogeneous solutions to construct the modes of the SF instead
of the inhomogeneous ones, and hence was named the method of {\em
extended homogeneous solutions}. The method leads to spectral convergence
to the value of the SF modes.

Our implementation of a frequency-domain solver for the SF modes follows
very close lines to the method of extended homogeneous solutions of~[\cite{Barack:2008ms}].
Indeed, the PwP already works with homogeneous solutions since it
eliminates the explicit presence of the particle in the equations
by moving it to the boundary conditions across the interface between
two domains (the method is designed in such a way that the particle
is always located at the interface, even for eccentric orbits), {\em
i.e.} the jump conditions. However, the method we will develop here
has some differences with the method of extended homogeneous solutions.
First of all, we use a multidomain splitting with a hyperboloidal
compactification. But we are going to introduce the following modification
from previous implementations of the PwP: Instead of using complementary
domains as has always been done until now (including in the time domain),
we are going to solve the homogeneous problems on the domains $\mathcal{D}_{-}=\{r_{*}|-\infty<r_{*}\leq r_{*}^{\textrm{apo}}\}$
and $\mathcal{D}_{+}=\{r_{*}|r_{*}^{\textrm{peri}}\leq r_{*}<+\infty\}$
that have an overlap. In contrast with the complementary regions $\mathcal{R}_{\pm}$
(shown in Figure~\ref{FIG6.1}), the regions $\mathcal{D}_{\pm}$ have
a nonempty intersection in general. Thus $\mathcal{D}_{-}\cap\mathcal{D}_{+}=[r_{*}^{\textrm{peri}},r_{*}^{\textrm{apo}}]$.
Only in the circular case the two setups, the one based on disjoint
regions as in Figure~\ref{FIG6.1} and the one based on the regions
$\mathcal{D}_{\pm}$, coincide in the sense that there is no intersection
(or just a point, the particle location).

Then, we proceed by solving for the $R_{\ell mn}^{\pm}$ with arbitrary
Dirichlet boundary conditions at the pericenter and apocenter respectively,
\begin{equation}
R_{\ell mn}^{\pm}(r_{*}^{\textrm{peri/apo}})=\lambda^{\pm}\,,\label{eq:fakebc}
\end{equation}
for some free (non-zero) choice of $\lambda^{\pm}$. Let us call the
solutions thus obtained $\hat{R}_{\ell mn}^{\pm}$ on $\mathcal{D}_{\pm}$.

Now the question is how, from these solutions (with the particular
boundary conditions that we have used), we can find the solutions
that we are actually interested in (taking into account the presence
of the particle). Here we are going to take advantage of the linearity
of our problem. Given the solutions $\hat{R}_{\ell mn}^{\pm}$ of
the problem described above, \textit{i.e.} for a single Fourier mode, with
boundary conditions~(\ref{eq:fakebc}), the solution for our actual
problem ({\em i.e.} including the particle) will be: 
\begin{equation}
R_{\ell mn}(r)=\left\{ \begin{array}{ll}
C_{\ell mn}^{-}\hat{R}_{\ell mn}^{-}(r) & \mbox{if}\quad r<r_{p}(t)\,,\\
C_{\ell mn}^{+}\hat{R}_{\ell mn}^{+}(r) & \mbox{if~}\quad r>r_{p}(t)\,.
\end{array}\right.\label{actual solution}
\end{equation}
where the coefficients $C_{\ell mn}^{\pm}$ are constants to be determined.
What allows us to do this is the linearity of the equations, since
by multiplying the solutions $\hat{R}_{\ell mn}^{\pm}$ of the two
problems defined on $\mathcal{D}_{\pm}$ by a constant, we obtain again
a solution of the same equations, just with different boundary conditions
than~(\ref{eq:fakebc}). The coefficients $C_{\ell mn}^{\pm}$ are
then determined uniquely by enforcing the jump conditions~(\ref{jumpsR})
and~(\ref{jumpsRrsu}) across the particle location.

While our work on eccentric orbits is still in progress, we present
some results on circular orbits, where the problem simplifies a bit
as discussed (the $\mathcal{D}_{\pm}$ and $\mathcal{R}_{\pm}$ regions
coincide). The results are obtained from codes developed either in
Matlab or in Python, and are shown in Table \ref{SF-circular}.

\begin{table*}
\begin{minipage}{\textwidth}
\caption{Numerical values of the components of the gradient of the regularized field ($f_{r}$) for circular orbits. Here, $N$ is the number of collocation grid points used and $\ell^{}_{\rm max}$ the highest $\ell$-harmonic used in the summation. For reference, the values for a circular orbit at the last stable circular orbit ($r=6\,M$) obtained, using frequency-domain methods, in~\cite{DiazRivera:2004ik} was: $1.677 283 4 \times 10^{-4}$.
\label{SF-circular}} 
\centering    
\begin{tabular}{ccc}  
    $N$  & $\ell^{}_{\rm max}$ & $f_r$ \\[1mm]
\hline
\hline\\[-2mm]
 $50$  & $20$ &  $1.674125346413219 \times 10^{-4}$ \\[1mm]
 $50$  & $30$ &  $1.680135078016693 \times 10^{-4}$ \\[1mm]
 $80$  & $20$ &  $1.673179411442940  \times 10^{-4}$ \\[1mm]
 $80$  & $30$ &  $1.675719825341073  \times 10^{-4}$ \\[1mm]
 $80$  & $40$ &  $1.676358608421948 \times 10^{-4}$ \\[1mm]
 $80$  & $50$ &  $1.676586279346283 \times 10^{-4}$ \\[1mm]
 $80$  & $60$ &  $1.675397087197872 \times 10^{-4}$ \\[1mm]
 $100$ & $20$ &  $1.673697449614004 \times 10^{-4}$ \\[1mm]
 $100$ & $30$ &  $1.676237863251235 \times 10^{-4}$ \\[1mm]
 $100$ & $40$ &  $1.676876512268239 \times 10^{-4}$ \\[1mm]
 $100$ & $50$ &  $1.677108498387443 \times 10^{-4}$ \\[1mm]
$150$ & $80$ &  $1.677232315338774 \times 10^{-4}$ \\[1mm]
$200$ & $20$ &  $1.673570555862998 \times 10^{-4}$ \\[1mm]
$200$ & $40$ &  $1.676749618219680 \times 10^{-4}$ \\[1mm]
$200$ & $50$ &  $1.676981603683444 \times 10^{-4}$ \\[1mm]
$200$ & $60$ &  $1.677085514921244 \times 10^{-4}$ \\[1mm]
\end{tabular} 
\end{minipage}
\end{table*}


\chapter{Conclusions\label{7-concl}}
\newrefsegment

\subsection*{Conclusion summary}

The theory of general relativity has now withstood its first century of existence, one at the end of which it has decisively opened the door to a new era in astronomy. The electromagnetic waves that once told us nearly everything we knew of the Universe beyond our Earth are now only one---albeit a still largely dominant---voice in the story. Gravitational waves have begun to tell their own story, the first pages of which are being written as we speak.

In this thesis, we have investigated two-body gravitational systems in the strong-field—that is, fully general relativistic—and extreme-mass-ratio regime, known as extreme-mass-ratio inspirals. These are expected to be among the main and most interesting sources of the future space-based gravitational-wave detector LISA. Prospective observations of such systems will furnish us with a wealth of opportunities to probe strong gravity, as the complicated orbits of the inspiraling object (stellar-mass black hole or neutron star) will effectively “map out” the gravitational field around the more massive one (the massive black hole at a galactic center). The problem of modeling such systems to sufficient accuracy—that is, for producing the theoretical waveform templates needed by LISA in its envisioned search for them—has witnessed significant progress over the last few decades, yet remains today an open one.

The understanding of this problem is intimately connected with concepts such as gravitational energy-momentum and mathematical techniques such as spacetime decompositions---for example, via canonical or quasilocal approaches---as well as perturbation theory. In the first half of this thesis, we have developed in detail the basic methods needed for dealing with these. In the second half, we have presented our novel contributions in these areas, notably on the issues of entropy, motion and the self-force in general relativity. 

In what follows, we summarize briefly the results obtained in this work. This then leads us into offering some closing reflections on the broad conceptual issues that have historically been at the basis of the interpretation of general relativity. In view of the intrinsic dichotomy of the theory, as Einstein himself saw it, between “measuring rods and clocks [and] all other things”, it is perhaps unsurprising that more subtle notions such as entropy, energy-momentum and the self-force continue to elude a clear consensus among relativists to this day. Our contributions in this thesis have sought to offer some fresh perspectives on these basic issues.

\subsection*{Conclusions \normalfont{(conclusion summary translation in Catalan)}}

 La teoria de la relativitat general ha viscut ara el seu primer segle d’existència, un al final del qual ha obert decisivament la porta a una nova era en l’astronomia. Les ones electromagnètiques que abans ens van dir gairebé tot el que sabíem de l’Univers fora de la nostra Terra són ara només una veu de la història. Les ones gravitacionals han començat a transmetre la seva pròpia història, les primeres pàgines de la qual estan sent escrites en aquests mateixos moments.

En aquesta tesi, hem investigat sistemes gravitacionals de dos cossos en el règim de camps forts - és a dir, en la teoria completa de la relativitat general - i raons de masses extremes (conegudes com a caigudes en espiral amb raó de masses extrema, \textit{EMRIs}), que s’esperaven figurar entre les principals fonts del futur detector d’ones gravitacionals LISA, situada en l’espai. Les observacions possibles d'aquests sistemes ens proporcionaran una gran varietat d'oportunitats per provar la gravetat en el règim fort, ja que les òrbites complicades de l'objecte caient en espiral (un forat negre de massa estel·lar o una estrella de neutrons) realitzaran un “mapa” del camp gravitatori al voltant del masiu (un forat negre massiu d’un centre galàctic). El problema de modelar aquests sistemes amb una precisió suficient (és a dir, per produir les plantilles teòriques de formas d'onas necessàries per LISA en la seva cerca prevista) ha vist progressos significatius durant les últimes dècades, encara que avui en dia queda obert.

La comprensió d’aquest problema està íntimament relacionada amb conceptes com ara l’energia i la quantitat de moviment gravitatòria, i tècniques matemàtiques com les descomposicions de l’espai-temps - per exemple, mitjançant enfocaments canònics o quasilocals - i també amb la teoria de pertorbacions. En la primera meitat d'aquesta tesi, hem desenvolupat en detall els mètodes bàsics necessaris per tractar-los. En la segona meitat, hem presentat les nostres contribucions en aquestes àrees, en particular sobre els temes de l’entropia, el moviment i la força pròpia en la relativitat general.

A continuació, resumim breument els resultats obtinguts en aquest treball. Això ens porta a oferir algunes reflexions tancades sobre els grans temes conceptuals que històricament han estat a la base de la interpretació de la relativitat general. A la vista de la dicotomia intrínseca de la teoria, tal com ho va veure el mateix Einstein, entre ``varetes i rellotges de mesurament [i] totes les altres coses'', potser no és sorprenent que nocions més subtils com l'entropia, l’energia i la quantitat de moviment gravitatòria i la força pròpia actualment continuen eludint un consens clar entre els relativistes. Les nostres contribucions en aquesta tesi han buscat oferir algunes perspectives noves sobre aquests temes bàsics.

\subsection*{Conclusions \normalfont{(conclusion summary translation in French)}}

La théorie de la relativité générale a maintenant traversé son premier siècle d'existence, à l'issue duquelle elle a ouvert de manière décisive la porte d'une nouvelle ère dans l’astronomie. Les ondes électromagnétiques qui nous disaient à peu près tout ce que nous savions de l'univers en dehors de notre Terre ne sont plus qu'une voix dans l'histoire, même si les différents fréquences de la lumière restent le messager principal aujourd’hui. Les ondes gravitationnelles ont commencé à transmettre leur propre histoire, dont les premières pages sont tout de suite en cours d’écriture.

Dans cette thèse, nous avons étudié les systèmes gravitationnels à deux corps dans le régime des champs forts - c’est-à-dire, dans la théorie complète de la relativité générale - et les quotients extrêmes des masses (appelés inspirals avec quotients extrêmes des masses, \textit{EMRIs}), qui devraient être parmi les principales sources du futur détecteur spatial d'ondes gravitationnelles LISA. Les observations prospectives de tels systèmes nous fourniront une grande variété de posibilités pour tester la gravité forte, car les orbites compliquées de l'objet spirallant (un trou noir à masse stellaire ou une étoile à neutrons) « cartographieront » effectivement le champ gravitationnel autour du plus massif (un trou noir massif au centre galactique). Le problème de la modélisation de tels systèmes avec une précision suffisante - c'est-à-dire pour la production des modèles de formes des ondes théoriques requis par LISA dans sa recherche envisagée - a connu des progrès significatifs au cours des dernières décennies, mais reste aujourd'hui ouvert.  

La compréhension de ce problème est intimement liée aux concepts tels que l’énergie et la quantité de mouvement gravitationnelles et les techniques mathématiques telles que les décompositions de l’espace-temps - par exemple, en usant des approches canoniques ou quasi-locales - ainsi que la théorie des perturbations. Dans la première partie de cette thèse, nous avons développé en détail les méthodes de base nécessaires pour y faire face. Dans la deuxième partie, nous avons présenté nos nouvelles contributions dans ces domaines, en particulier sur les problèmes de l'entropie, du mouvement et de la force propre dans la relativité générale.

Dans ce qui suit, nous résumons brièvement les résultats obtenus dans ce travail. Cela nous amène ensuite à proposer des réflexions finales sur les grandes questions conceptuelles qui ont toujours été à la base de l'interprétation de la relativité générale. Compte tenu de la dichotomie intrinsèque de la théorie, telle que l’a vue Einstein lui-même, entre ``bâtonnets de mesure et horloges [et] tout le reste'', il n’est peut-être pas surprenant que des notions plus subtiles telles que l’entropie, l’énergie et la quantité de mouvement gravitationnelles et la force propre continuent à éluder un consensus clair parmi les relativistes à ce jour. Nos contributions dans cette thèse ont cherché à offrir de nouvelles perspectives sur ces questions fondamentales.

\begin{center}
{*}
\par\end{center}

$\,$

$\,$

We now offer a brief concluding summary of the novel contributions
of this thesis.

In Chapter \ref{4-entropy}, we have studied entropy theorems in classical mechanics
and general relativity, with a focus on the gravitational two-body
problem. In particular, we have proved that canonical theories of
classical particles for certain classes of Hamiltonians, as well as
of some typical matter (in particular, scalar and electromagnetic)
fields in curved spacetime, do not admit any monotonically increasing
function of phase space (along trajectories of the Hamiltonian flow).
Thus, such theories preclude the existence of entropy in what we have
referred to as a ``mechanical'' sense, \textit{i.e.} as a phase space functional.
We have then looked at why these proofs do not carry over to general
relativity, which we do know to manifest the existence of entropy
in such a sense. We have furthermore discussed another method of proof
based on a topological argument, in particular, phase space compactness,
and have investigated the meaning of these results for the gravitational
two-body problem, in particular, by proving the non-compactness of
the phase space of perturbed Schwarzschild-Droste spacetimes. In the
absence today of a general formula for gravitational entropy, an understanding
of why general relativity differs from, for example, classical mechanics
or Maxwellian electromagnetism in this sense can give helpful indications
for future progress.

In Chapter \ref{5-motion}, we have presented a novel derivation, based on conservation
laws, of the basic equations of motion for the EMRI problem. They
are formulated with the use of a quasilocal (rather than matter) stress-energy-momentum
tensor---in particular, the Brown-York tensor---so as to capture
gravitational effects in the momentum flux of the object, including
the gravitational self-force. Our formulation and resulting equations
of motion are independent of the choice of the perturbative gauge.
We have shown that, in addition to the usual gravitational self-force
term, they also lead to an additional  ``self-pressure''
force not found in previous analyses, and the effects of which warrant further investigation. Our approach
thus offers a fresh geometrical picture from which to understand the
self-force fundamentally, and potentially useful new avenues for computing
it practically.

In Chapter \ref{6-fd}, we have presented some numerical work based on the Particle-without-Particle
(PwP) approach, a pseudospectral collocation method previously developed
for the computation of the scalar self-force---a helpful testbed
for the gravitational case. The basic idea of this method is to discretize
the computational domain into two (or more) disjoint grids such that
the ``particle''---the distributional source in the field equations
of the self-force problem---is always at the interface between them;
thus, one only needs to solve homogeneous equations in each domain,
with the source effectively replaced by jump (boundary) conditions
thereon. Here we have presented some results on the numerical computation
of the scalar self-force, using this method, for circular orbits in
the frequency domain. Moreover, in Appendix \ref{b-pwp}, we present a generalization
of this method to general partial differential equations with distributional
sources, including also applications to other areas of applied mathematics.
We generically obtain improved convergence rates relative to other
implementations in these areas, typically relying on delta function
approximations on the computational grid.

\begin{center}
{*}
\par\end{center}

As we have seen, the EMRI problem is intimately connected with conceptual
as well as technical questions regarding entropy, energy-momentum
and motion in general relativity. It is remarkable that, despite the
multiplicity of fruitful insights which have so far been achieved
towards their understanding, relativists today continue to lack a
clear, general consensus on the conceptual interpretation and, strictly
speaking, even the formal mathematical expression of such notions.

These considerations naturally invite us to reflect back upon our
discussion in the introduction, specifically regarding the interpretation
of general relativity and more generally the evolution of our ideas
about gravitation in physics. 

It may be argued that, with regard to the basic content of his theory,
Einstein's key physical insight was to realize what sort of object
it should be that the gravitational field equations describe (in particular,
the metric tensor of spacetime, or something like it), much more so,
in a certain sense, than eventually obtaining the exact final form
of these equations---an effort which relied essentially on mathematical
reasoning and consistency with the Newtonian theory once the spacetime
geometry was understood to be the basic object of study.

There is a simple \emph{gedankenexperiment} that Einstein frequently
used to illustrate how the local effects of special relativity plus
the requirement that physical laws be formulated in any coordinate
frame of reference together logically imply that our spacetime must
be, in general, globally curved. It is worthwhile to recount it here,
from his 1921 lecture series [\cite{einstein_four_1922}] (taken
from [\cite{einstein_four_2002}]):
\begin{quote}
\begin{spacing}{1.05}
{\small{}{[}Let $K$ be an inertial coordinate system, with spatial
Cartesian coordinates $x,y,z$.{]} Imagine a circle drawn about the
origin in the {[}Cartesian{]} $x'y'$ plane of {[}another coordinate
system{]} $K'$ {[}the $z'$ axis of which coincides with the $z$
axis of $K${]}, and a diameter of this circle. Imagine, further,
that we have given a large number of rigid rods, all equal to each
other. We suppose these laid in series along the periphery and the
diameter of the circle, at rest relatively to $K'$. If $U$ is the
number of these rods along the periphery, $D$ the number along the
diameter, then, if $K'$ does not rotate relatively to $K$ we shall
have
\[
\frac{U}{D}=\pi\,.
\]
But if $K'$ rotates we get a different result. Suppose that at a
definite time $t$, of $K$ we determine the ends of all the rods.
With respect to $K$ all the rods upon the periphery experience the
Lorentz contraction, but the rods upon the diameter do not experience
this contraction (along their lengths!).{*} It therefore follows that
\[
\frac{U}{D}>\pi\,.
\]
}{\small \par}

{\small{}{[}...{]} Space and time, therefore, cannot be defined with
respect to $K'$ as they were in the special theory of relativity
with respect to inertial systems. But, according to the principle
of equivalence, $K'$ may also be considered as a system at rest,
with respect to which there is a gravitational field (field of centrifugal
force, and force of Coriolis). We therefore arrive at the result:
the gravitational field influences and even determines the metrical
laws of the space-time continuum. If the laws of configuration of
ideal rigid bodies are to be expressed geometrically, then in the
presence of a gravitational field the geometry is not Euclidean. }{\small \par}

\noindent{\small{}\_\_\_\_\_\_\_\_\_\_\_\_}{\small \par}
\tiny{}$\,$
\end{spacing}
\begin{spacing}{0.92}

\noindent{\small{}{*} These considerations assume that the behavior of rods
and clocks depends only upon velocities, and not upon accelerations,
or, at least, that the influence of acceleration does not counteract
that of velocity.}{\small \par}
\end{spacing}
\end{quote}
Notwithstanding the elegant simplicity of the above argument in capturing
the essence of the motivation for general relativity (that is, for
devising a theory of gravitation in terms of spacetime geometry),
Einstein was certainly aware that there is greater subtlety here than
first meets the eye. Referring in particular to special relativity,
he writes in his \emph{Autobiographical Notes} [\cite{einstein_autobiographical_1949}]:
\begin{quote}
\begin{spacing}{1.05}
{\small{}One is struck {[}by the fact{]} that the theory (except for
the four-dimensional space) introduces two kinds of physical things,
i.e., (1) measuring rods and clocks, (2) all other things, e.g., the
electro-magnetic field, the material point, etc. This, in a certain
sense, is inconsistent; strictly speaking measuring rods and clocks
would have to be represented as solutions of the basic equations (objects
consisting of moving atomic configurations), not, as it were, as theoretically
self-sufficient entities. However, the procedure justifies itself
because it was clear from the very beginning that the postulates of
the theory are not strong enough to deduce from them sufficiently
complete equations for physical events sufficiently free from arbitrariness,
in order to base upon such a foundation a theory of measuring rods
and clocks. }{\small \par}
\end{spacing}
\end{quote}
It would be fair to say that this basic dichotomy in the physical
foundations of the theory---between ``measuring rods and clocks''
and ``all other things''---has never been fully resolved, at least
not at the level that Einstein would have regarded as ``sufficiently
complete'' within this discussion. Nevertheless, this sort of ``inherent
contradiction'' of general relativity, if one is inclined to regard
it as such, is one which he certainly saw, at the very least, as a
reasonable exchange for the Newtonian ones it has come to replace.
With what Arthur Koestler might have called ``sleepwalker's assurance'',
Einstein writes a few years after the discovery of general relativity
[\cite{einstein_geometrie_1921}] (English translation taken from
[\cite{goenner_expanding_1999}]):
\begin{quote}
\begin{spacing}{1.05}
{\small{}The concept of the measuring-rod and the concept of the clock
coordinated with it in the theory of relativity do not find an exactly
corresponding object in the real world {[}there are no perfectly rigid
rods{]}. It is also clear that the solid body and clock do not play
the role of irreducible elements in the conceptual edifice of physics,
but that of composite structures, which may not play any independent
role in theoretical physics. But it is my conviction that in the present
stage of development of theoretical physics these concepts must still
be employed as independent concepts; for we are still far from possessing
such certain knowledge of the theoretical foundations as to be able
to give theoretical constructions of such structures.}{\small \par}
\end{spacing}
\end{quote}
Early in our introduction to this thesis, we briefly traced Johannes
Kepler's struggle with the idea of the ``force'' governing planetary
motion in his emerging vision of a clockwork universe.
At that time, he could do no better than to visualize it as a sort
of vortex, ``a raging current which tears all the planets, and perhaps
all the celestial ether, from West to East'' [\cite{kepler_astronomia_1609}] (from [\cite{koestler_sleepwalkers_1959}]);
it took the arrival of Newton for this concept to encounter its first
clear formulation. Today, our technically-advanced struggle with the
``self-force'' and related concepts should not obscure the fact
that we are still, to a large extent, following in the inertia of
sleepwalking---not least evinced by the manifestly neo-Newtonian
nomenclature to which we still stubbornly cling. We may take with
a good dose of welcome encouragement Arthur Koestler's remark [\cite{koestler_sleepwalkers_1959}], no less resonant today than half a century ago,
that ``{[}t{]}he contemporary physicist grappling with the paradoxa
of relativity and quantum mechanics will find {[}in Kepler's own struggle{]}
an echo of his perplexities''.


\begin{appendix}
\fancyhead[RO,LE]{\rule[-1ex]{0pt}{1ex} \fontsize{11}{12}\selectfont \thepage}
\fancyhead[RE]{\fontsize{11}{12}\sl\selectfont\nouppercase Appendix A. Topics in Differential Geometry}
\fancyhead[LO]{\fontsize{11}{12}\sl\selectfont\nouppercase \rightmark} 
\chapter{Topics in Differential Geometry:\\
Maps on Manifolds, Lie Derivatives, Forms and Integration\label{a-geometry}}
\newrefsegment

\subsection*{Appendix summary}

In this appendix, $(\mathscr{M},\bm{g},\bm{\nabla})$ is any $n$-dimensional (oriented, smooth, topological [\cite{lee_introduction_2002}]) manifold of any signature, with metric $\bm{g}$ and compatible derivative $\bm{\nabla}$. 

We define and develop here four broad geometrical notions used amply throughout this thesis: maps on manifolds in Section \ref{sec:A.1-maps}, Lie derivatives in Section \ref{sec:A.2-lie}, differential forms in Section \ref{sec:A.3-forms}, and finally integration on manifolds in Section \ref{sec:A.4-integration}. At the end of each of these sections we offer a brief example from physics. The exposition is mainly based on Appendices B and C of [\cite{wald_general_1984}] and [\cite{lee_introduction_2002}].



\subsection*{Temes en geometria diferencial \normalfont{(appendix summary translation in Catalan)}}

En aquest apèndix, $(\mathscr{M},\bm{g},\bm{\nabla})$ és qualsevol varietat $n$-dimensional (orientada, suau, topològica [\cite{lee_introduction_2002}]) de qualsevol signatura, amb $\bm{g}$ el tensor mètric i derivada compatible $\bm{\nabla}$.

Definim i desenvolupem aquí quatre nocions geomètriques àmplies que s’utilitzen de forma extensiva al llarg de aquesta tesi: fonctions sobre varietats a la secció \ref{sec:A.1-maps}, derivats de Lie a la secció \ref{sec:A.2-lie}, formes diferencials de la secció \ref{sec:A.3-forms} i, finalment, integració sobre varietats de la secció \ref{sec:A.4-integration}. Al final de cadascuna d'aquestes seccions oferim un breu exemple de física. L’exposició es basa principalment en els apèndixs B i C de [\cite{wald_general_1984}] i [\cite{lee_introduction_2002}].

\subsection*{Sujets dans la géométrie différentielle \normalfont{(appendix summary translation in French)}}

Dans cette annexe, $(\mathscr{M},\bm{g},\bm{\nabla})$ nous décrivons n’importe quelle variété $n$-dimensionnelle (orientée, lisse, topologique [\cite{lee_introduction_2002}]) de n’importe quelle signature, avec $\bm{g}$ le tenseur métrique et dérivé compatible $\bm{\nabla}$.

Nous définissons et développons ici quatre grandes notions géométriques largement utilisées tout au long de cette thèse : applications sur le variétés dans la section \ref{sec:A.1-maps}, dérivées de Lie dans la section \ref{sec:A.2-lie}, formes différentielles dans la section \ref{sec:A.3-forms} et enfin intégration sur variétés dans la section \ref{sec:A.4-integration}. À la fin de chacune de ces sections, nous proposons un bref exemple tiré de la physique. L'exposition est principalement basée sur les annexes B et C de [\cite{wald_general_1984}] et [\cite{lee_introduction_2002}].

\section{Maps on manifolds\label{sec:A.1-maps}}

Let $(\mathscr{M},\bm{g},\bm{\nabla})$ be an $n$-dimensional manifold
and $(\tilde{\mathscr{M}},\tilde{\bm{g}},\tilde{\bm{\nabla}})$ an
$\tilde{n}$-dimensional manifold. They could be of the same dimension,
and could even be the same manifold, but not necessarily.

An important question, one that often arises in physics and especially in GR, is how to
establish an identification of points between manifolds (or between points on the same manifold), \textit{i.e.} how
to relate a point $p\in\mathscr{M}$ with a point $\tilde{p}\in\tilde{\mathscr{M}}$,
and more generally, an arbitrary tensor $\bm{A}\in\mathscr{T}^{k}\,_{l}(\mathscr{M})$
at $p\in\mathscr{M}$ with another tensor $\tilde{\bm{A}}\in\mathscr{T}^{k}\,_{l}(\tilde{\mathscr{M}})$
at $\tilde{p}\in\tilde{\mathscr{M}}$. In this section, overset tildes
will generally be used to indicate objects living on $\tilde{\mathscr{M}}$. 

\begin{figure}
\begin{centering}
\includegraphics[scale=0.8]{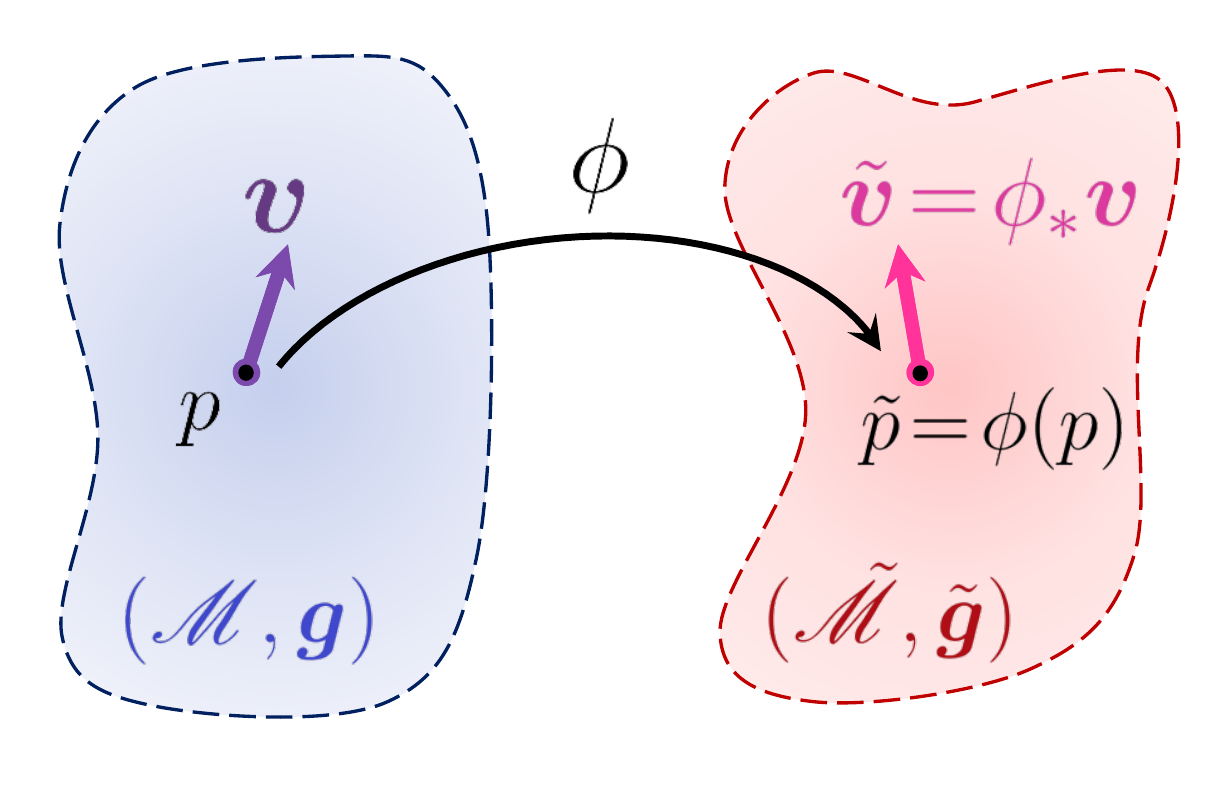} 
\par\end{centering}
\caption{An illustration of two manifolds $(\mathscr{M},\bm{g})$ and $(\tilde{\mathscr{M}},\tilde{\bm{g}})$
with a map $\phi:\mathscr{M}\rightarrow\tilde{\mathscr{M}}$ between
them. This identifies any point $p\in\mathscr{M}$ with $\tilde{p}=\phi(p)\in\tilde{\mathscr{M}}$,
and can be used, for example, to push-forward the vector $\bm{v}\in T_{p}\mathscr{M}$
to $\phi_{*}\bm{v}\in T_{\tilde{p}}\tilde{\mathscr{M}}$. If $\phi$
is a diffeomorphism, a general transport of tensors from one manifold
to the other can be defined. Note that in this notation, the metric
$\tilde{\bm{g}}$ of $\tilde{\mathscr{M}}$ is not necessarily the
same as the metric transported from $\mathscr{M}$, \textit{i.e.}  $\phi_{*}\bm{g}$.
If indeed $\phi_{*}\bm{g}=\tilde{\bm{g}}$, then $\phi$ is called
an \emph{isometry}---a symmetry of the metric.}
\label{a-fig-map} 
\end{figure}

First, in order to identify the points themselves, we suppose that
there exists a smooth map between these manifolds,
\begin{align}
\phi:\mathscr{M}\, & \rightarrow\tilde{\mathscr{M}}\label{eq:phi-map}\\
p\, & \mapsto\phi\left(p\right)=\tilde{p}\,,\label{eq:phi-mapsto}
\end{align}
such that any point $p\in\mathscr{M}$ is identified with its image
under this map, $\tilde{p}=\phi(p)$. See Fig. \ref{a-fig-map}.

Consider now any function $\tilde{f}:\tilde{\mathscr{M}}\rightarrow\mathbb{R}$.
The map $\phi:\mathscr{M}\rightarrow\tilde{\mathscr{M}}$ can be used
to define a new function $f$ on $\mathscr{M}$, referred to as the
\emph{pull-back} of $\tilde{f}$, via simple composition. We denote
this as $f=\phi^{*}\tilde{f}:\mathscr{M}\rightarrow\mathbb{R}$, and
it is simply given by:
\begin{equation}
\phi^{*}\tilde{f}=\tilde{f}\circ\phi\,.\label{eq:pullback-function}
\end{equation}

This idea can be extended from functions to tensors of higher rank.
First, suppose $v^{a}\in T\mathscr{M}$ is a vector field in the tangent
bundle of $\mathscr{M}$. The map $\phi$ can be used to ``carry
along'' this vector field to another vector field $\tilde{v}^{a}=\phi_{*}v^{a}\in T\tilde{\mathscr{M}}$
in the tangent bundle of $\tilde{\mathscr{M}}$, from point to point,
via a map called the \emph{push-forward}, $\phi_{*}:T_{p}\mathscr{M}\rightarrow T_{\tilde{p}}\tilde{\mathscr{M}}$.
Its action is defined by
\begin{equation}
\left(\phi_{*}\bm{v}\right)(\tilde{f})=\bm{v}(\phi^{*}\tilde{f})\,,\label{eq:pushforward}
\end{equation}
for any function $\tilde{f}:\tilde{\mathscr{M}}\rightarrow\mathbb{R}$,
with its pull-back $\phi^{*}\tilde{f}=\tilde{f}\circ\phi$ as given
by (\ref{eq:pullback-function}). See again Fig. \ref{a-fig-map}. One can push-forward vector fields
from $T\tilde{\mathscr{M}}$ to $T\mathscr{M}$ in the same way using
instead the inverse map $\phi^{-1}:\tilde{\mathscr{M}}\rightarrow\mathscr{M}$
if it exists.

One can continue in this way to define the pull-back of co-vectors
($(0,1)$-tensors) from the co-tangent bundle of $\tilde{\mathscr{M}}$
to that of $\mathscr{M}$. Let $\tilde{w}_{a}\in T^{*}\tilde{\mathscr{M}}$.
Then the pull-back $\phi^{*}:T_{\tilde{p}}^{*}\tilde{\mathscr{M}}\rightarrow T_{p}^{*}\mathscr{M}$
is defined by
\begin{equation}
\left(\phi^{*}\tilde{\bm{w}}\right)_{a}v^{a}=\tilde{w}_{a}\left(\phi_{*}\bm{v}\right)^{a}\,.\label{eq:pullback}
\end{equation}

Suppose now that $\phi:\mathscr{M}\rightarrow\tilde{\mathscr{M}}$
is a \emph{diffeomorphism}, meaning that it is bijective (one-to-one
and onto) and has a smooth inverse $\phi^{-1}:\tilde{\mathscr{M}}\rightarrow\mathscr{M}$.
Then a generalization of the push-forward and pull-back maps can be
defined to relate an arbitrary tensor $\bm{A}\in\mathscr{T}^{k}\,_{l}(\mathscr{M})$
at $p\in\mathscr{M}$ with another tensor $\tilde{\bm{A}}\in\mathscr{T}^{k}\,_{l}(\tilde{\mathscr{M}})$
at $\tilde{p}\in\tilde{\mathscr{M}}$. In particular, we write $\tilde{\bm{A}}=\phi_{*}\bm{A}$
and refer to this generally as the \emph{transport} of $\bm{A}$ (under
$\phi$) from $\mathscr{M}$ to $\tilde{\mathscr{M}}$. (Note that
in this notation, the metric $\tilde{\bm{g}}$ of $\tilde{\mathscr{M}}$
is not necessarily $\phi_{*}\bm{g}$. If it is, then $\phi$ is called
an \emph{isometry}---a symmetry of the metric.) For any set of $k$
co-vectors $\{\tilde{w}_{a}^{(j)}\}_{j=1}^{k}$ in $T^{*}\tilde{\mathscr{M}}$
and any set of $l$ vectors $\{\tilde{v}_{(j)}^{a}\}_{j=1}^{l}$ in
$T\tilde{\mathscr{M}}$, the transport map 
\begin{equation}
\phi_{*}:\left(\mathscr{T}^{k}\,_{l}\left(\mathscr{M}\right)\right)_{p}\rightarrow\left(\mathscr{T}^{k}\,_{l}(\tilde{\mathscr{M}})\right)_{\tilde{p}}\label{eq:transport-forward}
\end{equation}
is defined by
\begin{equation}
\left(\phi_{*}\bm{A}\right)^{a_{1}\cdots a_{k}}\,_{b_{1}\cdots b_{l}}\tilde{w}_{a_{1}}^{(1)}\cdots\tilde{w}_{a_{k}}^{(k)}\tilde{v}_{(1)}^{b_{1}}\cdots\tilde{v}_{(l)}^{b_{l}}=A^{a_{1}\cdots a_{k}}\,_{b_{1}\cdots b_{l}}w_{a_{1}}^{(1)}\cdots v_{(l)}^{b_{l}}\,,\label{eq:transport-defn}
\end{equation}
where $\bm{w}^{(j)}=\phi^{*}\tilde{\bm{w}}^{(j)}$ is the pull-back
(under $\phi$) of each co-vector and $\bm{v}_{(j)}=(\phi^{-1})_{*}\tilde{\bm{v}}_{(j)}$
the push-forward (under $\phi^{-1}$) of each vector, from $\tilde{\mathscr{M}}$
to $\mathscr{M}$. 

The transport of any tensor from $\tilde{\mathscr{M}}$ to $\mathscr{M}$,
\textit{i.e.} the transport under $\phi^{-1}$, is denoted by super-scripting the star,
\begin{equation}
\left(\phi^{-1}\right)_{*}=\phi^{*}:\left(\mathscr{T}^{k}\,_{l}(\tilde{\mathscr{M}})\right)_{\tilde{p}}\rightarrow\left(\mathscr{T}^{k}\,_{l}\left(\mathscr{M}\right)\right)_{p}\,.\label{eq:transport-back}
\end{equation}

We now enumerate some useful properties of the tensor transport (\ref{eq:transport-forward}).
See Theorem 10.6 of [\cite{felsager_geometry_2012}]. Let $\bm{A},\bm{B}\in\mathscr{T}^{k}\,_{l}(\mathscr{M})$
and $\bm{B}'\in\mathscr{T}^{k'}\,_{l'}(\mathscr{M})$ be any tensors
in $\mathscr{M}$ and $c\in\mathbb{R}$ a constant. Then we have the
following:
\begin{enumerate}
\item $\phi_{*}$ \emph{is linear}, \textit{i.e.} 
\begin{equation}
\phi_{*}\left(\bm{A}+\bm{B}\right)=\phi_{*}\bm{A}+\phi_{*}\bm{B}\,,\quad\phi_{*}\left(c\bm{A}\right)=c\phi_{*}\bm{A}\,.
\end{equation}
\item $\phi_{*}$ \emph{commutes with the tensor product}, \textit{i.e.}
\begin{equation}
\phi_{*}\left(\bm{A}\otimes\bm{B}'\right)=\left(\phi_{*}\bm{A}\right)\otimes\left(\phi_{*}\bm{B}'\right)\,.
\end{equation}
\item $\phi_{*}$ \emph{commutes with contractions}, \textit{i.e.}
\begin{equation}
\phi_{*}\left(A^{a_{1}\cdots c\cdots a_{k-1}}\,_{b_{1}\cdots c\cdots b_{l-1}}\right)=\left(\phi_{*}\bm{A}\right)^{a_{1}\cdots c\cdots a_{k-1}}\,_{b_{1}\cdots c\cdots b_{l-1}}\,.
\end{equation}
\end{enumerate}

\subsection*{Example: gauge freedom in GR}

If the two manifolds $(\mathscr{M},\bm{g})$ and $(\tilde{\mathscr{M}},\tilde{\bm{g}})$
are (four-dimensional, Lorentzian) spacetimes, the existence of a
diffeomorphism $\phi:\mathscr{M}\rightarrow\tilde{\mathscr{M}}$ is
interpreted as signifying that the spacetimes describe \emph{the same
physical situation}. In other words, any solution of the field equations
of a theory for some collection of fields $\psi$ is considered to
be physically indistinguishable from the solution $\phi_{*}\psi$.
(Thus, we may speak of an equivalence class of solutions with the
equivalence relation $\psi\sim\phi_{*}\psi$.) Conversely, if there
exists no diffeomorphism $\phi:\mathscr{M}\rightarrow\tilde{\mathscr{M}}$,
then the two spacetimes (and the correspondent solutions to the field
equations thereon) are seen to represent physically different situations.

The existence in GR of the freedom to transform the spacetime metric
$\bm{g}$ by a diffeomorphism, (such that $\bm{g}\sim\tilde{\bm{g}}$),
is often referred to as the ``active view'' of gauge freedom. Equivalently,
one may take the ``passive view'', where gauge freedom can be seen to
manifest itself as coordinate transformations. Concretely, suppose $\{x^{\alpha}\}$
is a coordinate system covering a neighborhood $\mathscr{U}$ of a
point $p\in\mathscr{M}$, and $\{y^{\alpha}\}$ one covering a neighborhood
$\mathscr{V}$ of $\tilde{p}=\phi(p)\in\tilde{\mathscr{M}}$. One
can then define a new coordinate system $\{x'^{\alpha}\}$ in a neighborhood
$\phi^{-1}(\mathscr{V})$ of $p\in\mathscr{M}$ by setting $x'^{\alpha}(q)=y^{\alpha}(\phi(q))$,
for all $q\in\phi^{-1}(\mathscr{V})$. From this point of view, one
may thus regard the effect of $\phi$ as leaving $p$ and all tensors
at $p$ unchanged, but instead inducing a local coordinate transformation
$x^{\alpha}\mapsto x'^{\alpha}$. In other words, the components of
any $\phi_{*}\bm{A}$ at $\tilde{p}=\phi(p)$ in the coordinates $\{y^{\alpha}\}$
(in the ``active'' viewpoint) are the same as those of $\bm{A}$
at $p$ in the coordinates $\{x'^{\alpha}\}$ (in the ``passive''
viewpoint).

\section{Lie derivatives\label{sec:A.2-lie}}

Let $(\mathscr{M},\bm{g},\bm{\nabla})$ be any manifold, and let $v^{a}\in T\mathscr{M}$
be any vector field. An important question to address is: how do tensors
change ``in the direction'' of $\bm{v}$? Or, more precisely, how
do they change (from point to point) along the curves in $\mathscr{M}$
to which $\bm{v}$ is tangent? Firstly, it is necessary to formalize
the meaning of the latter concept: the set of curves in $\mathscr{M}$
to which $\bm{v}$ is tangent are referred to as the \emph{integral
curves} of $\bm{v}$. These are defined by a \emph{one-parameter group}
of diffeomorphisms in $\mathscr{M}$, referred to as the \emph{flow}
of $\bm{v}$,
\begin{equation}
\phi_{t}^{\left(\bm{v}\right)}:\mathscr{M}\times\mathbb{R}\rightarrow\mathscr{M}\,,
\end{equation}
which are solutions to the ODE
\begin{equation}
\frac{{\rm d}\phi_{t}^{\left(\bm{v}\right)}}{{\rm d}t}=\bm{v}\circ\phi_{t}^{\left(\bm{v}\right)}\,.
\end{equation}
In this case, $\bm{v}$ is referred to as the \emph{generator} of
the flow. 

Thanks to our discussion in the previous section, we have a precise
way of ``comparing tensors'' at different points on a manifold.
In particular, we can compare the values of tensors at different points
along the integral curves of $\bm{v}$ simply by transporting them
under the flow $\phi_{t}^{\left(\bm{v}\right)}$.

\begin{figure}
\begin{centering}
\includegraphics[scale=0.75]{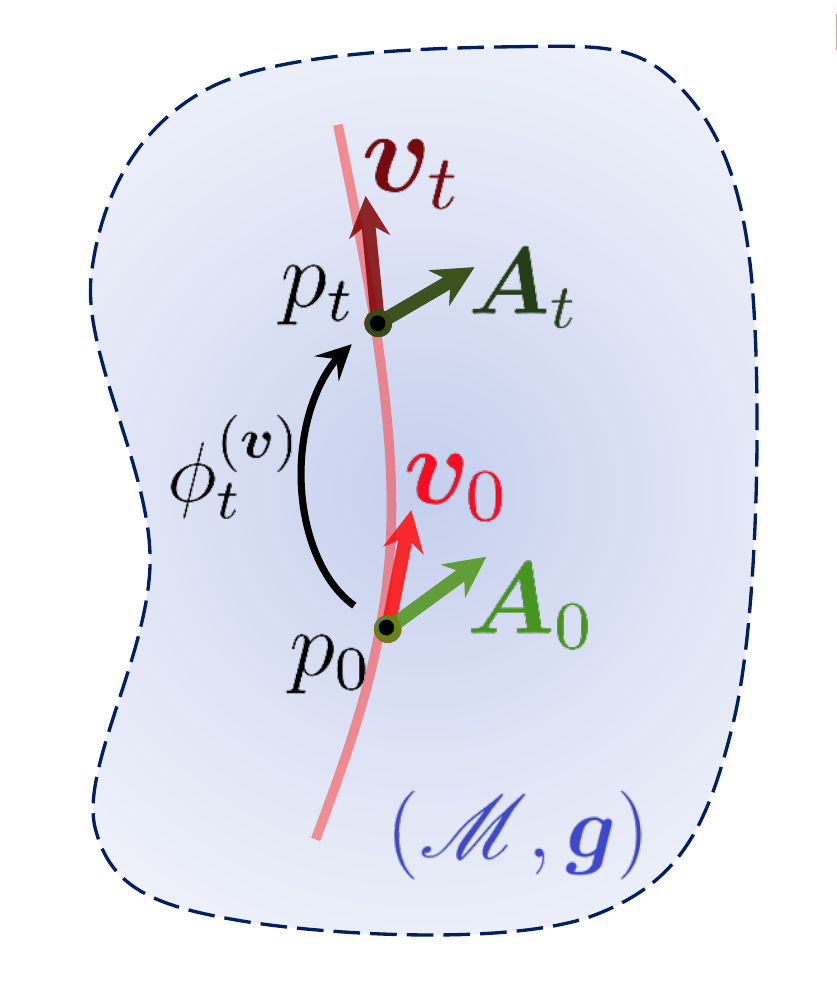} 
\par\end{centering}
\caption{An illustration of the meaning of the Lie derivative along a
vector field $\bm{v}$ in $(\mathscr{M},\bm{g})$ of a tensor $\bm{A}$,
depicted here for ease of visualization in the case where $\bm{A}$
is a vector. In particular, one compares $\bm{A}_{0}=(\bm{A})_{p_{0}}$
at the point $p_{0}$ corresponding to $t=0$ in the flow $\phi_{t}^{\left(\bm{v}\right)}$
with its value $\bm{A}_{t}=(\bm{A})_{p_{t}}$ at the point $p_{t}=\phi_{t}^{\left(\bm{v}\right)}(p_{0})$
for some $t>0$ by transporting the latter back to $p_{0}$, i.e.
one compares $\bm{A}_{0}$ with $(\phi_{-t}^{\left(\bm{v}\right)})_{*}\bm{A}$
at $p_{0}$. Their difference divided by $t$, in the small $t$ limit,
is the Lie derivative.}
\label{a-fig-lie} 
\end{figure}

To be more precise, let $\bm{A}$ be any $(k,l)$-tensor. We may ask,
for example, how its value at a point $p_{0}\in\mathscr{M}$ on the
manifold corresponding to $t=0$ in the flow parametrization changes
relative that at a point $\phi_{t}^{\left(\bm{v}\right)}(p_{0})=p_{t}\in\mathscr{M}$
at some parameter value $t>0$. In this case, one needs to compare
$\bm{A}$ at $p_{0}$ with the transport of $\bm{A}$ from $p_{t}$
to $p_{0}$, \textit{i.e.} with $((\phi_{t}^{\left(\bm{v}\right)})^{-1})_{*}\bm{A}=(\phi_{-t}^{\left(\bm{v}\right)})_{*}\bm{A}$. See Fig. \ref{a-fig-lie}.
The limit in which $t$ is small gives rise precisely to the notion
of the \emph{Lie derivative},
\begin{equation}
\mathcal{L}_{\bm{v}}\bm{A}=\lim_{t\rightarrow0}\frac{(\phi_{-t}^{\left(\bm{v}\right)})_{*}\bm{A}-\bm{A}}{t}\,.\label{eq:Lie-deriv}
\end{equation}

To make this more concrete, notice firstly that when applied to a
function $\bm{A}=f:\mathscr{M}\rightarrow\mathbb{R}$, (\ref{eq:Lie-deriv})
immediately recovers the usual notion of the ``directional derivative'',
\begin{equation}
\mathcal{L}_{\bm{v}}f=\bm{v}\left(f\right)\,.
\end{equation}
Moreover, for any $\bm{A}$, it is to instructive to consider the
result of the formula (\ref{eq:Lie-deriv}) in a choice of coordinates
$\{x^{\alpha}\}$ adapted to $\bm{v}$. This means that the action
of $\phi_{t}$ on a point corresponds to a transformation in one coordinate
$x^{1}\rightarrow x^{1}+t$ with the rest $x^{2},\ldots,x^{n}$ being
held fixed. In such a coordinate system, it is possible to show (see
Appendix C of [\cite{wald_general_1984}]) that (\ref{eq:Lie-deriv})
yields:
\begin{equation}
\mathcal{L}_{\bm{v}}A^{\alpha_{1}\cdots\alpha_{k}}\,_{\beta_{1}\cdots\beta_{l}}=\frac{\partial A^{\alpha_{1}\cdots\alpha_{k}}\,_{\beta_{1}\cdots\beta_{l}}}{\partial x^{1}}\,,\label{eq:Lie-adapted}
\end{equation}
again recovering the notion of a general ``directional derivative'',
expressed here in the coordinates adapted to the direction of change.
From (\ref{eq:Lie-adapted}), it is then possible (see Appendix C
of [\cite{wald_general_1984}]) to obtain a completely general, \emph{coordinate-independent}
abstract index formula for (\ref{eq:Lie-deriv}),
\begin{align}
\mathcal{L}_{\bm{v}}A^{a_{1}\cdots a_{k}}\,_{b_{1}\cdots b_{l}}=\, & \nabla_{\bm{v}}A^{a_{1}\cdots a_{k}}\,_{b_{1}\cdots b_{l}}-\sum_{i=1}^{k}A^{a_{1}\cdots c\cdots a_{k}}\,_{b_{1}\cdots b_{l}}\nabla_{c}v^{a_{i}}\nonumber \\
 & +\sum_{j=1}^{l}A^{a_{1}\cdots a_{k}}\,_{b_{1}\cdots c\cdots b_{l}}\nabla_{b_{j}}v^{c}\,.\label{eq:Lie-abs-index}
\end{align}

\subsection*{Example: perturbative gauge freedom and Killing vectors in GR}

As developed in Chapter \ref{3-perturbations}, a perturbative gauge transformation generated
by a vector field $\bm{\xi}\in T\mathscr{M}$ changes the linear perturbation
$\bm{h}$ according to $\bm{h}\mapsto\bm{h}+\mathcal{L}_{\bm{\xi}}\mathring{\bm{g}}$,
where $\mathring{\bm{g}}$ is the background metric. Using (\ref{eq:Lie-abs-index}),
we see that
\begin{align}
\mathcal{L}_{\bm{\xi}}\mathring{g}_{ab}=\, & \xi^{c}\mathring{\nabla}_{c}\mathring{g}_{ab}+\mathring{g}_{cb}\mathring{\nabla}_{a}\xi^{c}+\mathring{g}_{ac}\mathring{\nabla}_{b}\xi^{c}\\
=\, & 2\mathring{\nabla}_{(a}\xi_{b)}\,.
\end{align}

The Lie derivative is also used to define \emph{Killing vector fields}
in any manifold $(\mathscr{M},\bm{g},\bm{\nabla})$ as those vector
fields $\bm{\Xi}\in T\mathscr{M}$ satisfying the \emph{Killing equation},
$\mathcal{L}_{\bm{\Xi}}\bm{g}=0$. This is equivalent to the statement
that the flow generated by such vector fields, $\phi_{t}^{\left(\bm{\Xi}\right)}$,
are \emph{isometries} of the metric, \textit{i.e.}  $(\phi_{t}^{\left(\bm{\Xi}\right)})_{*}\bm{g}=\bm{g}$.

\section{Differential forms\label{sec:A.3-forms}}

Let $(\mathscr{M},\bm{g},\bm{\nabla})$ be any $n$-dimensional manifold
(of any signature). If a $(0,k)$-tensor $\bm{\alpha}\in\mathscr{T}^{0}\,_{k}(\mathscr{M})$
is totally antisymmetric, \textit{i.e.} if
\begin{equation}
\alpha_{a_{1}\cdots a_{k}}=\alpha_{[a_{1}\cdots a_{k}]}\,,\label{eq:p-form-defn}
\end{equation}
then $\bm{\alpha}$ is referred to as a (\emph{differential}) \emph{$k$-form}.
The notion of a form is a crucial one in geometry, and was first developed
in the work of [\cite{cartan_sur_1899}]. As we shall see in the following
section, forms serve as the basis for defining integration over (regions
of) manifolds. For the remainder of the current section, we summarize
some useful definitions and properties.

The set of $k$-forms on $\mathscr{M}$ is typically denoted $\Lambda^{k}(\mathscr{M})$.
As the simplest examples, $\Lambda^{0}(\mathscr{M})=\mathscr{F}(\mathscr{M})$
is the set of smooth functions on $\mathscr{M}$, and $\Lambda^{1}(\mathscr{M})=T^{*}\mathscr{M}$
is just the cotangent bundle. Any $k$-form for $k>n$ vanishes identically
due to the antisymmetry. 

A useful operation between forms is the \emph{wedge product}, $\wedge$.
It can be applied between any $k$-form $\bm{\alpha}$ and any $l$-form
$\bm{\beta}$ to produce a $(k+l)$-form $\bm{\alpha}\wedge\bm{\beta}$,
given by their completely antisymmetrized outer product. That is,
\begin{align}
\wedge:\Lambda^{k}\left(\mathscr{M}\right)\times\Lambda^{l}\left(\mathscr{M}\right)\, & \rightarrow\Lambda^{k+l}\left(\mathscr{M}\right)\\
\left(\alpha_{a_{1}\cdots a_{k}},\beta_{b_{1}\cdots b_{l}}\right)\, & \mapsto\left(\bm{\alpha}\wedge\bm{\beta}\right)_{a_{1}\cdots a_{k}b_{1}\cdots b_{l}}=\frac{\left(k+l\right)!}{k!l!}\alpha_{[a_{1}\cdots a_{k}}\beta_{b_{1}\cdots b_{l}]}\,.
\end{align}
This definition implies $\bm{\alpha}\wedge\bm{\beta}=(-1)^{kl}\bm{\beta}\wedge\bm{\alpha}$.
(So in particular, $\bm{\alpha}\wedge\bm{\alpha}=0$ for $k$ odd.)

Another useful operation is the \emph{exterior derivative} ${\rm d}$,
which takes $k$-forms to $(k+1)$-forms, defined as their completely
antisymmetrized derivative. That is,
\begin{align}
{\rm d}:\Lambda^{k}\left(\mathscr{M}\right)\, & \rightarrow\Lambda^{k+1}\left(\mathscr{M}\right)\\
\alpha_{a_{1}\cdots a_{k}}\, & \mapsto\left({\rm d}\bm{\alpha}\right)_{a_{0}a_{1}\cdots a_{k}}=(k+1)\partial_{[a_{0}}\alpha_{a_{1}\cdots a_{k}]}\,.
\end{align}
In fact, due to the antisymmetry of forms and the symmetry of the
connection coefficient (between any two derivative operators on $\mathscr{M}$),
one can show that the above definition is independent of the choice
of the derivative operator. (So we simply write it in terms of the
partial derivative $\bm{\partial}$.) 

Thus, given a basis $\{{\rm d}x^{\alpha}\}$ of the cotangent space
$T_{p}^{*}\mathscr{M}$ at any point $p\in\mathscr{M}$, any $k$-form
$\bm{\alpha}$ and its exterior derivative ${\rm d}\bm{\alpha}$ can
locally always be written, respectively, as
\begin{align}
\boldsymbol{\alpha}=\, & \alpha_{\alpha_{1}\cdots\alpha_{k}}{\rm d}x^{\alpha_{1}}\wedge\cdots\wedge{\rm d}x^{\alpha_{k}}\,,\label{eq:alpha-local}\\
{\rm d}\bm{\alpha}=\, & \frac{\partial\alpha_{\alpha_{1}\cdots\alpha_{k}}}{\partial x^{\alpha_{0}}}{\rm d}x^{\alpha_{0}}\wedge{\rm d}x^{\alpha_{1}}\wedge\cdots\wedge{\rm d}x^{\alpha_{k}}\,.
\end{align}

A $k$-form $\bm{\alpha}$ is called \emph{closed} if ${\rm d}\bm{\alpha}=0$,
and \emph{exact} if there exists a $(k-1)$-form $\bm{\beta}$ such
that $\boldsymbol{\alpha}={\rm d}\boldsymbol{\beta}$. All exact forms
are closed, as the definition of the exterior derivative implies ${\rm d}^{2}={\rm d}\circ{\rm d}=0$,
a result known as the \emph{Poincaré lemma}. The converse is true,
however, only \emph{locally}: all closed forms are locally exact,
but globally they may not be in general\footnote{~The study of the implications of this leads to a very useful set of
geometrical invariants called \emph{de Rham cohomology groups}. See
Chapter 11 of [\cite{lee_introduction_2002}].}.

In the study of integration in the next section, we will work with
forms of the same rank as the dimension of the manifold, $n$. These
are sometimes referred to as \emph{top forms}. (As we shall see, integration
in any $n$-dimensional space is defined for $n$-forms.) Let $\bm{\alpha}\in\Lambda^{n}(\mathscr{M})$
be any $n$-form. Its antisymmetry in all $n$ indices implies that,
locally, its expansion (\ref{eq:alpha-local}) is simply
\begin{equation}
\bm{\alpha}=\alpha\left(x\right){\rm d}x^{1}\wedge\cdots\wedge{\rm d}x^{n}\,,\label{eq:n-form}
\end{equation}
where $x=(x^{1},\ldots,x^{n})$ and $\alpha(x)$ is a function on
$\mathscr{M}$. 

Recall that zero-forms are also functions. Indeed, it is generally
true, for any $0\leq k\leq n$, that $\Lambda^{k}(\mathscr{M})$ is
isomorphic to $\Lambda^{n-k}(\mathscr{M})$. Moreover, this isomorphism
is provided by another famous and very useful operator, the \emph{Hodge
star} $\star:\Lambda^{k}(\mathscr{M})\rightarrow\Lambda^{n-k}(\mathscr{M})$
defined uniquely, for any two $k$-forms $\bm{\alpha}$ and $\bm{\beta}$,
by the relation $\bm{\alpha}\wedge\star\bm{\beta}=\alpha_{\alpha_{1}\cdots\alpha_{k}}\beta^{\beta_{1}\cdots\beta_{k}}\sqrt{{\rm det}(\bm{g})}{\rm d}x^{1}\wedge\cdots\wedge{\rm d}x^{n}$.
For more on this, see Chapter 7 of [\cite{nakahara_geometry_2003}]. 

\subsection*{Example: electromagnetism}

Consider the electromagnetic four-vector potential $A^{a}$ (traditionally
written as $(\phi,A^{i}$), with $\phi$ the scalar potential and
$A^{i}$ the vector potential). One can (using the spacetime metric
to lower the index) think of this instead as a one-form $\bm{A}=A_{a}{\rm d}x^{a}$.
Then, the Faraday tensor $F_{ab}=2\partial_{[a}A_{b]}$ is a two-form
(as $F_{ab}=F_{[ab]})$, given simply by the exterior derivative of
$\bm{A}$, \textit{i.e.} $\bm{F}={\rm d}\boldsymbol{A}$. Automatically, the
Poincaré lemma implies 
\begin{equation}
{\rm d}\bm{F}={\rm d}^{2}\boldsymbol{A}=0\,.
\end{equation}
In traditional notation, with $\vec{E}$ denoting the electric field
and $\vec{B}$ the magnetic field (from $\bm{A}$), this is equivalent
to two Maxwell equations: the Gauss law for magnetism $\vec{\nabla}\cdot\vec{B}=0$
and the Faraday induction law $\vec{\nabla}\times\vec{E}=-\partial_{t}\vec{B}$.
These are in fact the constraints of the theory, manifested here as
conditions of geometrical consistency. (Recall from Chapter \ref{2-canonical} that
the constraints of GR are also conditions of geometrical consistency.) 

The other two Maxwell equations are the dynamical evolution equations:
the Gauss law for electrostatics $\vec{\nabla}\cdot\vec{E}=\rho$
and the Ampère-Maxwell law $\vec{\nabla}\times\vec{B}=\vec{j}+\partial_{t}\vec{E}$,
where $\rho$ and $\vec{j}$ are the charge and current densities.
These are obtained from the Maxwell Lagrangian,
given by $\mathcal{L}=-\frac{1}{4}\bm{F}:\bm{F}+\bm{A}\cdot\bm{J}$,
where here we denote $\bm{J}=J_{a}{\rm d}x^{a}$ with $J_{a}=(\rho,j_{i})$
the four-current written as a one-form. (See Chapter \ref{4-entropy}.) One obtains
from the stationary action principle the dynamical Maxwell equations,
\begin{equation}
\star{\rm d}\star\bm{F}=\bm{J}\,.
\end{equation}

\section{Integration on manifolds\label{sec:A.4-integration}}

In this section we review integration over (regions of) $\mathscr{M}$.

First, suppose $\{(\mathscr{U}_{i},\varphi_{i})\}$ is an atlas for
$\mathscr{M}$, where $\mathscr{U}_{i}\subset\mathscr{M}$ are the
open subsets covering $\mathscr{M}$ (such that $\bigcup_{i}\mathscr{U}_{i}=\mathscr{M}$),
and the homeomorphisms $\varphi_{i}:\mathscr{U}_{i}\rightarrow\mathbb{R}^{n}$
are the associated coordinate maps (or charts). 

We begin by defining the integral of an $n$-form $\bm{\alpha}$ over
any $\mathscr{U}_{i}$. In particular, this is defined to be the same
as the integral of the transported (pushed-forward) form over the
image of $\mathscr{U}_{i}$ in Euclidean space under $\varphi_{i}$,
i.e.
\begin{equation}
\int_{\mathscr{U}_{i}}\bm{\alpha}=\int_{\varphi_{i}\left(\mathscr{U}_{i}\right)}\left(\varphi_{i}\right)_{*}\bm{\alpha}=\int_{\varphi_{i}\left(\mathscr{U}_{i}\right)}\alpha\left(x\right){\rm d}x^{1}\wedge\cdots\wedge{\rm d}x^{n}\,,\label{eq:integration_Ui}
\end{equation}
where to write the last equality we have used the expansion (\ref{eq:n-form}).
This can now be made sense of as a usual integral in $\mathbb{R}^{n}$
by identifying the wedge product of basis forms ${\rm d}x^{1}\wedge\cdots\wedge{\rm d}x^{n}$
with the usual Riemann (or Lebesgue) measure ${\rm d}\mu_{\mathbb{R}^{n}}^{\,}={\rm d}x^{1}\cdots{\rm d}x^{n}$
on $\mathbb{R}^{n}$, defined in the usual way. This finally gives
us:
\begin{equation}
\int_{\mathscr{U}_{i}}\bm{\alpha}=\int_{\varphi_{i}\left(\mathscr{U}_{i}\right)}{\rm d}\mu_{\mathbb{R}^{n}}^{\,}\alpha\left(x\right)\,.\label{eq:integration_Ui_2}
\end{equation}

The extension of this definition to integration over the entire manifold
$\mathscr{M}$ is not quite so straightforward. In particular, it
requires a result which is rather technical, and the proof
of which we omit here (see, \textit{e.g.}, Chapter 2 of [\cite{lee_introduction_2002}]): namely the existence in any manifold of \emph{partitions
of unity}. These are a set of functions $\{\psi_{i}:\mathscr{M}\rightarrow[0,1]\subset\mathbb{R}\}$,
said to be \emph{subordinate} to $\{\mathscr{U}_{i}\}$, satisfying
the following conditions: \emph{(i)} they are supported entirely within
each $\mathscr{U}_{i}$ (\textit{i.e.} ${\rm supp}(\psi_{i})\subset\mathscr{U}_{i}$);
\emph{(ii)} at any point $p\in\mathscr{M}$, they are nonzero for
a finite number of $i$ and add up to unity, i.e.
\begin{equation}
\sum_{i}\psi_{i}(p)=1\,.\label{eq:partitions-of-unity}
\end{equation}

With this in hand, integration over the entire manifold $\mathscr{M}$
of a $k$-form $\bm{\alpha}$ can be defined. In particular, one ``inserts
the identity'' (\ref{eq:partitions-of-unity}) into the integrand
of $\int_{\mathscr{M}}\bm{\alpha}$ to transform it into a sum of
integrals over each $\mathscr{U}_{i}$ (given that each $\psi_{i}$
is only supported therein), which are themselves in turn given by
(\ref{eq:integration_Ui_2}). That is, we define:
\begin{equation}
\int_{\mathscr{M}}\bm{\alpha}=\sum_{i}\int_{\mathscr{U}_{i}}\psi_{i}\bm{\alpha}\,.
\end{equation}

An important and often used result is \emph{Stokes' theorem}\footnote{~This has evolved through different versions throughout history. See
[\cite{katz_history_1979}] for a detailed account. The first record
of its appearance is in an 1850 letter by Lord Kelvin to Stokes, who
then used it for several years as a problem in the Smith's Prize exam
at Cambridge. (It is unknown if any of the students managed to prove
it.) The first published proof is in [\cite{hankel_zur_1861}].}. The modern geometrical version of this theorem was first formulated
by [\cite{cartan_les_1945}]. It states that for any $(n-1)$-form
$\bm{\alpha}$,
\[
\int_{\mathscr{M}}{\rm d}\bm{\alpha}=\int_{\partial\mathscr{M}}\bm{\alpha}\,,
\]
where $\partial\mathscr{M}$ is the boundary\footnote{~Formally, this can be defined if one of the coordinates in the charts
$\varphi_{i}$ of $\mathscr{M}$, say $x^{n}$, is always non-negative.
Then one takes $\partial\mathscr{M}=\{p\in\mathscr{M}|\varphi_{i}(p)=(x^{1},\ldots,x^{n-1},0)\}$.} of $\mathscr{M}$.

Now that we know how to make sense of integration of forms, we would
also like to give meaning to the integration of a function $f:\mathscr{M}\rightarrow\mathbb{R}$
over (regions of) $\mathscr{M}$. In order to do this, one first requires
the definition of a \emph{volume form}. This is a nowhere-vanishing
$n$-form, denoted as $\bm{\epsilon}_{\mathscr{M}}^{\,}\in\Lambda^{n}(\mathscr{M})$,
such that the total volume or \emph{total measure} $\mu(\mathscr{M})$
of the manifold is given by its integral thereover, 
\begin{equation}
\mu\left(\mathscr{M}\right)=\int_{\mathscr{M}}\bm{\epsilon}_{\mathscr{M}}^{\,}\,.\label{eq:measure}
\end{equation}

The volume form is usually defined by the condition 
\begin{equation}
\frac{1}{n!}\epsilon_{a_{1}\cdots a_{n}}^{\mathscr{M}}\epsilon_{\mathscr{M}}^{a_{1}\cdots a_{n}}=\left(-1\right)^{s}\,,\label{eq:vol-form-def}
\end{equation}
where $s$ denotes the number of minus signs in the signature of $\bm{g}$
(so $s=0$ if it is Riemannian and $s=1$ if it is Lorentzian). It
can be shown (see, \textit{e.g.}, Appendix B of [\cite{wald_general_1984}])
that (\ref{eq:vol-form-def}) implies that $\bm{\epsilon}_{\mathscr{M}}^{\,}$
has a local expansion (\ref{eq:alpha-local}) given by:
\begin{equation}
\bm{\epsilon}_{\mathscr{M}}^{\,}=\sqrt{|g|}{\rm d}x^{1}\wedge\cdots\wedge{\rm d}x^{n}\,,\label{eq:vol-form-def-coords}
\end{equation}
where $g=\textrm{det}(\bm{g})$.

With this, we can now define the integral of a function $f:\mathscr{M}\rightarrow\mathbb{R}$
over $\mathscr{M}$ as
\begin{equation}
\int_{\mathscr{M}}\bm{\epsilon}_{\mathscr{M}}^{\,}f\,.
\end{equation}

One final useful result that we state here concerns the transport
under a diffeomorphism $\phi:\mathscr{M}\rightarrow\tilde{\mathscr{M}}$
of the volume form of $\mathscr{M}$ to another manifold $\tilde{\mathscr{M}}$.
Indeed, this does not simply yield the volume from of $\tilde{\mathscr{M}}$
itself; however, they are proportional, with the proportionality factor
given by a smooth function on $\tilde{\mathscr{M}}$ called the \emph{Jacobian
determinant}, $J\in\mathscr{F}(\tilde{\mathscr{M}})$. To be more
precise, let $\bm{\epsilon}_{\tilde{\mathscr{M}}}^{\,}$ denote the
volume form of $\tilde{\mathscr{M}}$. Then we have (see Chapter 7
of [\cite{abraham_manifolds_2001}]):
\begin{equation}
\phi_{*}\bm{\epsilon}_{\mathscr{M}}^{\,}=J\bm{\epsilon}_{\tilde{\mathscr{M}}}^{\,}\,,\label{eq:vol-form-trans}
\end{equation}
where $J=\det(\phi^{*})$, with $\phi*:T\tilde{\mathscr{M}}\rightarrow T\mathscr{M}$
here indicating the push-forward.

~

\subsection*{Example: diffeomorphism-invariant action functionals}

A useful result (see Proposition 10.20 of [\cite{lee_introduction_2002}])
is that for any diffeomorphism $\phi:\mathscr{M}\rightarrow\tilde{\mathscr{M}}$
and any compactly supported $n$-form $\bm{\alpha}$ on $\mathscr{M}$,
we have
\begin{equation}
\int_{\mathscr{M}}\bm{\alpha}=\int_{\tilde{\mathscr{M}}}\phi_{*}\bm{\alpha}\,.\label{eq:diff-inv}
\end{equation}

Suppose $S[\psi]=\int_{\mathscr{V}}\bm{\epsilon}_{\mathscr{M}}^{\,}f[\psi]$
is an action functional of a field theory, as defined in Chapter \ref{2-canonical}.
The above result can then be regarded as a statement of diffeomorphism
invariance: we have
\begin{align}
S\left[\psi\right]=\, & \int_{\mathscr{V}}\bm{\epsilon}_{\mathscr{M}}^{\,}f\,,\\
=\, & \int_{\phi(\mathscr{V})}\phi_{*}\left(\bm{\epsilon}_{\mathscr{M}}^{\,}f\right)\\
=\, & \int_{\phi(\mathscr{V})}\left(\phi_{*}\bm{\epsilon}_{\mathscr{M}}^{\,}\right)\left(f\circ\phi^{-1}\right)\\
=\, & \int_{\phi(\mathscr{V})}\bm{\epsilon}_{\tilde{\mathscr{M}}}^{\,}J\cdot f\circ\phi^{-1}\,.
\end{align}
In the second line we have used (\ref{eq:diff-inv}), in the third
line the fact that the transport commutes with tensor products as well as (\ref{eq:pullback-function}), and finally
in the last line (\ref{eq:vol-form-trans}). The latter may readily
be recognized simply as a general version of the ``change of coordinates''
formula from standard multi-variable calculus in Euclidean space.


\fancyhead[RO,LE]{\rule[-1ex]{0pt}{1ex} \fontsize{11}{12}\selectfont \thepage}
\fancyhead[RE]{\fontsize{11}{12}\sl\selectfont\nouppercase Appendix B. Particle-without-Particle: A Practical Pseudospectral Collocation Method}
\fancyhead[LO]{\fontsize{11}{12}\sl\selectfont\nouppercase \rightmark}

\chapter{Particle-without-Particle:\\
A Practical Pseudospectral Collocation Method for Linear Partial Differential Equations with Distributional Sources\label{b-pwp}}
\newrefsegment

\subsection*{Appendix summary}

This appendix is based on the publication [\cite{oltean_particle-without-particle:_2019}].
 
Partial differential equations with distributional sources---involving (derivatives of) delta distributions---have become increasingly ubiquitous in numerous areas of physics and applied mathematics. It is often of considerable interest to obtain numerical solutions for such equations, but any singular (``particle''-like) source modeling invariably introduces nontrivial computational obstacles. A common method in the literature used to circumvent these is through some form of delta function approximation procedure on the computational grid; however, this often carries significant limitations on the efficiency of the numerical convergence rates, or sometimes even the resolvability of the problem at all.
 
In this appendix, we present an alternative technique for tackling such equations which avoids the singular behavior entirely: the Particle-without-Particle method. Previously introduced in the context of the self-force problem in gravitational physics, the idea is to discretize the computational domain into two (or more) disjoint pseudospectral (Chebyshev-Lobatto) grids such that the ``particle'' is always at the interface between them; thus, one only needs to solve homogeneous equations in each domain, with the source effectively replaced by jump (boundary) conditions thereon. We prove here that this method yields solutions to any linear PDE the source of which is any linear combination of delta distributions and derivatives thereof supported on a one-dimensional subspace of the problem domain. We then implement it to numerically solve a variety of relevant PDEs with applications in neuroscience, finance and acoustics. We generically obtain improved convergence rates relative to typical past implementations relying on delta function approximations.
 
Following an introduction in Section \ref{intro} and some mathematical preliminaries in Section \ref{setup}, we prove in Section \ref{pwp} how the Particle-without-Particle method can be formulated and applied to problems with the most general possible “point” source, that is, one containing an arbitrary number of (linearly combined) one-dimensional delta functions and derivatives supported at an arbitrary number of points. Thus, one can use it on any type of (linear) PDE involving such sources.
 
Then in Sections \ref{first-order-hyperbolic-pdes}-\ref{elliptic-pdes}, we illustrate the application of this method, respectively, to first-order hyperbolic problems (with applications in neuroscience), parabolic problems (with applications in finance), second-order hyperbolic problems (with applications in acoustics), and finally elliptic problems.
 
Section \ref{conclusions} offers some concluding remarks.

\subsection*{Partícula-sense-Partícula \normalfont{(appendix summary translation in Catalan)}}

Aquest apèndix es basa en la publicació [\cite{oltean_particle-without-particle:_2019}].
 
Les equacions diferencials parcials amb fonts distributives - en particular, que impliquen (derivats de) distribucions delta - s'han tornat cada vegada més omnipresents en nombroses àrees de la física i les matemàtiques aplicades. Sovint és d’interès considerable obtenir solucions numèriques per a aquestes equacions, però qualsevol model de font singular (de tipus ``partícula'') introdueix invariablement obstacles computacionals no privats. Un mètode comú per evitar-les és mitjançant una forma d’aproximació de la funció delta a la graella computacional; no obstant això, sovint comporta limitacions importants en l’eficiència de les taxes de convergència numèrica, o, fins i tot, en la possibilitat de resoldre el problema en si mateix.
 
En aquest apèndix, presentem una tècnica alternativa per abordar aquestes equacions que evita completament el comportament singular: el mètode Partícula-sense-Partícula (Particle-without-Particle). Anteriorment introduïda en el context del problema de la força pròpia en la física gravitatòria, la idea és discretitzar el domini computacional en dues (o més) reixes disjuntes pseudospectrals (Chebyshev-Lobatto) de manera que la ``partícula'' sempre estigui a la interfície entre elles. Per tant, només cal resoldre equacions homogènies en cada domini, efectivament substituint la font per condicions de salt (de frontera). Aquí demostrem que aquest mètode produeix solucions a qualsevol equació diferencial parcial lineal la font de la qual és qualsevol combinació lineal de distribucions delta i derivats de les mateixes suportades en un subespai unidimensional del domini de la problema. A continuació, l’implementem per resoldre numèricament diverses equacions diferencials parcials rellevants amb aplicacions en neurociència, finances i acústica. Obtenim genèricament taxes de convergència millorades respecte a les implementacions anteriors típiques basades en aproximacions de la funció delta.
 
Després d'una introducció a la secció \ref{intro} i d’alguns preliminaris matemàtics a la secció \ref{setup}, demostrem a la secció \ref{pwp} com es pot formular i aplicar el mètode Partícula-sense-Partícula a problemes amb la font de punt més general possible, és a dir, que conté un nombre arbitrari de funcions delta unidimensionals (linealment combinades) i derivades suportades en un nombre arbitrari de punts. Per tant, es pot utilitzar en qualsevol tipus d’equació diferencial parcial (lineal) que impliqui aquestes fonts.
 
A continuació, a les Seccions \ref{first-order-hyperbolic-pdes}-\ref{elliptic-pdes}, il·lustrem l’aplicació d’aquest mètode, respectivament, a problemes hiperbòlics de primer ordre (amb aplicacions en neurociència), problemes parabòlics (amb aplicacions en finances), problemes hiperbòlics de segon ordre (amb aplicacions en acústica) i, finalment, problemes el·líptics.

La secció \ref{conclusions} ofereix algunes observacions finals.

\subsection*{Particule-sans-Particule \normalfont{(appendix summary translation in French)}}

Cette annexe est basée sur la publication [\cite{oltean_particle-without-particle:_2019}].
 
Les équations aux dérivées partielles (EDP) avec sources distributionnelles - en particulier, impliquant (dérivées de) distributions delta - sont devenues de plus en plus omniprésentes dans de nombreux domaines de la physique et des mathématiques appliquées. Il est souvent d’un intérêt considérable d’obtenir des solutions numériques pour de telles équations, mais toute modélisation de source singulière (semblable à une « particule ») introduit invariablement des obstacles de calcul non triviaux. Une méthode possible pour les contourner consiste à utiliser une procédure d'approximation de la fonction delta sur la grille de calcul ; cependant, cela limite souvent considérablement l'efficacité des taux de convergence numérique, voire parfois même la posibilité de resoudre le problème.

Dans cette annexe, nous présentons une technique alternative pour traiter de telles équations, qui évite totalement le comportement singulier : la méthode Particule-sans-Particule (Particle-without-Particle, PwP). Auparavant introduite dans le contexte du problème de la force propre dans la physique gravitationnelle, l’idée est de discrétiser le domaine de calcul en deux (ou plus) grilles disjointes pseudospectraux (Chebyshev-Lobatto) de telle sorte que la « particule » soit toujours à l’interface entre eux ; il suffit donc de résoudre des équations homogènes dans chaque domaine, la source étant effectivement remplacée par des conditions de saut (aux limites). Nous montrons ici que cette méthode fournit des solutions à toute EDP linéaire dont la source est quelque combinaison linéaire de distributions delta et de leurs dérivées supportées sur un sous-espace unidimensionnel du domaine du problème. Nous l’implémentons ensuite pour résoudre numériquement divers types des EDP pertinentes dans les domaines des neurosciences, de la finance et de l’acoustique. Nous obtenons de manière générique des taux de convergence meilleurs par rapport aux implémentations passées typiques reposant sur des approximations de fonctions delta.

Après une introduction dans la section \ref{intro} et quelques préliminaires mathématiques dans la section \ref{setup}, nous montrons à la section \ref{pwp} comment la méthode Particule-sans-Particule peut être formulée et appliquée aux problèmes avec la source « ponctuelle » la plus générale possible, c’est-à-dire contenant un nombre arbitraire de fonctions delta unidimensionnelles (combinées linéairement) et des dérivées avec support à un nombre arbitraire de points. Ainsi, on peut l’utiliser sur n’importe quel type d’EDP (linéaire) impliquant de telles sources. 

Ensuite, dans les sections \ref{first-order-hyperbolic-pdes} à \ref{elliptic-pdes}, nous illustrons l’application de cette méthode, respectivement, aux problèmes hyperboliques du premier ordre (avec applications dans la neuroscience), aux problèmes paraboliques (avec des applications dans la finance), aux problèmes hyperboliques du second ordre (avec applications dans l’acoustique) et enfin des problèmes elliptiques.

La section \ref{conclusions} propose quelques remarques de conclusion.

\section{Introduction}
\label{intro}

Mathematical models often have to resort---be it out of expediency
or mere ignorance---to deliberately idealized descriptions of their
contents. A common idealization across different fields of applied
mathematics is the use of the Dirac delta distribution, often simply
referred to as the \textsl{delta ``function''}, for the purpose of describing
highly localized phenomena: that is to say, phenomena the length
scale of which is significantly smaller, in some suitable sense, than that
of the problem into which they figure, and the (possibly complicated)
internal structure of which can thus be safely (or safely enough) ignored in
favour of a simple ``point-like'' cartoon. Canonical examples of this
from physics are notions such as ``point masses'' in gravitation or
``point charges'' in electromagnetism. 

Yet, despite their potentially powerful conceptual simplifications, introducing distributions into any mathematical
model is something that must be handled with
great technical care. In particular, let us suppose that our problem of interest
has the very general form
\begin{equation}
\mathcal{L}u=S\quad\mbox{in}\enskip\mathscr{U}\subseteq\mathbb{R}^{n}\,,\label{eq:intro_general_pde}
\end{equation}
where $\mathcal{L}$ is an $n$-dimensional (partial, if $n>1$) differential
operator (of arbitrary order $m$), $u$ is a quantity to be solved
for (a function, a tensor \textit{etc.}) and we assume that $S$---the ``source''---is
distributional in nature, \textit{i.e.} we have $S:\mathcal{D}(\mathscr{U})\rightarrow\mathbb{R}$, where we use the common notation $\mathcal{D}(\mathscr{U})$ to
refer to the set of smooth compactly-supported functions, \textit{i.e.} ``test
functions'', on $\mathscr{U}$. It follows, therefore, that $u$---if
it exists---must also be distributional in nature. So strictly speaking,
from the point of view of the classic theory of distributions [\cite{schwartz_theorie_1957}], the problem (\ref{eq:intro_general_pde})
is only well-defined---and hence may admit distributional solutions
$u$---provided that $\mathcal{L}$ is linear\footnote{~Here the terms ``linear''/``nonlinear'' have their standard meaning from the theory of partial differential equations.}. 

The problem with a nonlinear
$\mathcal{L}$ is essentially that, classically, products of distributions
do not make sense [\cite{schwartz_sur_1954}]. While there has certainly been work by mathematicians
aiming to generalize the theory of distributions so as to accommodate
this possibility [\cite{li_review_2007,colombeau_nonlinear_2013,bottazzi_grid_2019}], in the standard setting we are only really allowed
to talk of \textsl{linear} problems of the form (\ref{eq:intro_general_pde}).
Opportunely, very many of the typical problems in physics and applied
mathematics involving distributions take precisely this form. 

The inspiration for considering (\ref{eq:intro_general_pde}) in general
in this appendix actually comes from a setting where one does, in fact,
encounter non-linearities a priori: namely, gravitational physics. (For a general discussion regarding the treatment of distributions therein, see [\cite{geroch_strings_1987}].)
In particular, equations such as (\ref{eq:intro_general_pde}) arise
when attempting to describe the backreaction of a body with a ``small''
mass upon the spacetime through which it moves---known as its \textsl{self-force} [\cite{mino_gravitational_1997,quinn_axiomatic_1997,detweiler_self-force_2003,gralla_rigorous_2008,gralla_note_2011,poisson_motion_2011,blanchet_mass_2011,spallicci_self-force_2014,pound_motion_2015,wardell_self-force:_2015}].
(A similar version of this problem exists in electromagnetism, where
a ``small'' charge backreacts upon the electromagnetic field that
determines its motion [\cite{dirac_classical_1938,dewitt_radiation_1960,barut_electrodynamics_1980,poisson_motion_2011}].) In the full Einstein equations of general
relativity, which can be regarded as having the schematic form (\ref{eq:intro_general_pde})
with $u$ describing the gravitational field (that is, the spacetime
geometry, in the form of the metric) and $S$ denoting the matter
source (the stress-energy-momentum tensor), $\mathcal{L}$ is a nonlinear
operator. Nevertheless, for a distributional $S$ (representing the
``small'' mass as a ``point particle'' source) one \textsl{can} legitimately
seek solutions to a \textsl{linearized} version of (\ref{eq:intro_general_pde})
in the context of perturbation theory, \textit{i.e.} at first order in an expansion
of $\mathcal{L}$ in the mass. The detailed problem, in this case,
turns out to be highly complex, and in practice, $u$ must be computed
numerically. The motivation for this, we may add, is not just out of purely theoretical or foundational concern---the calculation of the self-force is also of significant applicational value for gravitational wave astronomy. To wit, it will in fact be indispensable for generating accurate enough waveform templates for future space-based gravitational wave detectors such as LISA [\cite{amaro-seoane_et_al._gravitational_2013,amaro-seoane_et_al._laser_2017}] \textit{vis-à-vis} extreme-mass-ratio binary systems, which are expected to be among the most fruitful sources thereof. For these reasons, having at our disposal a practical and
efficient numerical method for handling equations of the form (\ref{eq:intro_general_pde})
is of consequential interest.

What is more, these sorts of partial differential equations (PDEs)
arise frequently in other fields as well; indeed, (\ref{eq:intro_general_pde})
can adequately characterize quite a wide variety of (linear) mathematical
phenomena assumed to be driven by ``localized sources''. A few examples,
which we will consider one by one in different sections of this appendix,
are the following:

\begin{enumerate}[label=(\roman*)]

\item \textsl{First-order hyperbolic PDEs}: in neuroscience, advection-type
PDEs with a delta function source can be used in the modeling of neural
populations [\cite{haskell_population_2001,casti_population_2002,caceres_analysis_2011,caceres_blow-up_2016}];

\item \textsl{Parabolic PDEs}: in finance, heat-type PDEs with delta
function sources are sometimes used to model price formation [\cite{lasry_mean_2007,markowich_parabolic_2009,caffarelli_price_2011,burger_boltzmann-type_2013,achdou_partial_2014,pietschmann_partial_2012}];

\item \textsl{Second-order hyperbolic PDEs}: in acoustics, wave-type
PDEs with delta function (or delta derivative) sources are used to
model monopoles (or, respectively, multipoles) [\cite{petersson_stable_2010,kaltenbacher_computational_2017}]; more complicated equations of this form also appear, for example, in seismology models [\cite{romanowicz_seismology_2007,aki_quantitative_2009,shearer_introduction_2009,madariaga_seismic_2007,petersson_stable_2010}], which we will briefly comment upon.

\item \textsl{Elliptic PDEs}: Finally, we will look at a simple Poisson equation with a singular source [\cite{tornberg_numerical_2004}]; such equations can describe, for example, the potential produced by a very localized charge in electrostatics.

\end{enumerate}

\subsection{Scope of this work}

The purpose of this work is to explicate and generalize a practical method for numerically solving equations
like (\ref{eq:intro_general_pde}), as well as to illustrate its broad
applicability to the various problems listed in (i)-(iv) above. Previously
implemented with success only in the specific context of the self-force
problem [\cite{canizares_extreme-mass-ratio_2011,canizares_efficient_2009,canizares_pseudospectral_2010,canizares_time-domain_2011,canizares_tuning_2011,jaramillo_are_2011,canizares_overcoming_2014,oltean_frequency-domain_2017}], we dub it the ``Particle-without-Particle'' (PwP) method.
(Other methods for the computation of the self-force have also been developed based on matching the properties of the solutions on the sides of the delta distributions---see, \textit{e.g.}, the indirect (source-free) integration method of  [\cite{aoudia_source-free_2011,ritter_fourth-order_2011,spallicci_towards_2012,spallicci_fully_2014,ritter_indirect_2015,ritter_indirect_2016}].)
The basic idea of the PwP approach is the following: One begins by writing $u$ as a sum of distributions
each of which has support outside (plus, if necessary, at the location
of) the points where $S$ is supported; one then solves the equations
for each of these pieces of $u$ and finally matches them in such
a way that their sum satisfies the original problem (\ref{eq:intro_general_pde}).
In fact, as we shall soon elaborate upon, this approach will not work
in general for all possible problems of the form (\ref{eq:intro_general_pde}).
However, we will prove that it will \textsl{always} work if, rather
than the source being a distribution defined on all of $\mathscr{U}$,
we have instead $S:\mathcal{D}(\mathscr{I})\rightarrow\mathbb{R}$
with $\mathscr{I}\subseteq\mathbb{R}$ representing a \textsl{one-dimensional subspace}
of $\mathscr{U}$.

To make things more concrete, let us briefly describe this procedure
using the simplest possible example: let $f:\mathscr{U}\rightarrow\mathbb{R}$
be an arbitrary given function and suppose $S=f\delta$ where $\delta:\mathcal{D}(\mathscr{I})\rightarrow\mathbb{R}$
is the delta function supported at some point $x_{p}\in\mathscr{I}$.
Then, to solve (\ref{eq:intro_general_pde}), one would assume the
decomposition (or ``ansatz'') $u=u^{-}\Theta^{-}+u^{+}\Theta^{+}$ with
$\Theta^{\pm}:\mathcal{D}(\mathscr{I})\rightarrow\mathbb{R}$ denoting
appropriately defined Heaviside distributions (supported to the right/left
of $x_{p}$, respectively), and $u^{\pm}:\mathscr{U}\rightarrow\mathbb{R}$ being simple functions
(not distributions) to be solved for. Inserting such a decomposition for $u$ into
(\ref{eq:intro_general_pde}), one obtains \textsl{homogeneous} equations
$\mathcal{L}u^{\pm}=0$ on the appropriate domains, supplemented by
the necessary boundary conditions (BCs) for these equations at $x_{p}\in\mathscr{I}$,
explicitly determined by $f$. Generically, the latter arise in the
form of relations between the limits of $u^{-}$ and $u^{+}$ (and/or
the derivatives thereof) at $x_{p}$, and for this reason are called
``jump conditions'' (JCs). Effectively, the latter completely replace
the ``point'' source $S$ in the original problem, now simply reduced
to solving sourceless equations---hence the nomenclature of the method.

While in principle one can certainly contemplate the adaptation of these ideas into a variety of established approaches for the numerical solution of PDEs, we will focus specifically on their implementation through pseudospectral collocation (PSC) methods on Chebyshev-Lobatto (CL) grids. The principal advantages thereof lie in their typically very efficient (exponential) rates of numerical convergence as well as the ease of incorporating and modifying BCs (JCs) throughout the evolution. Indeed, PSC methods have enjoyed very good success in past work [\cite{canizares_extreme-mass-ratio_2011,canizares_efficient_2009,canizares_pseudospectral_2010,canizares_time-domain_2011,canizares_tuning_2011,jaramillo_are_2011,canizares_overcoming_2014,oltean_frequency-domain_2017}] on the PwP approach for self-force calculations (and in gravitational physics more generally [\cite{grandclement_spectral_2009}], including arbitrary precision implementations [\cite{santos-olivan_pseudo-spectral_2018}]), and so we shall not deviate very much from this recipe in the models considered in this appendix. Essentially the main difference will be that here, instead of the method of lines which featured in most of the past PwP self-force work, we will for the most part carry out the time evolution using the simplest first-order forward finite difference scheme; we do this, on the one hand, so that we may illustrate the principle of the method explicitly in a very elementary way without too many technical complications, and on the other, to show how well it can work even with such basic tactics. Depending on the level of accuracy and computational efficiency required for any realistic application, these procedures can naturally be complexified (to higher order, more domains, more complicated domain compactifications \textit{etc.}) for properly dealing with the sophistication of the problem at hand.

To summarize, past work using the PwP method only solved a specific
form of Eq. (\ref{eq:intro_general_pde}) pertinent to the self-force
problem: that is, with a particular choice of $\mathcal{L}$ and $S$
(upon which we will comment more later). It did \textit{not} consider
the question of the extent to which the idea of the method could be
useful in general for solving distributionally-sourced PDEs. These
appear, as enumerated above, in many other fields of study---and
we submit that a method such as this would be of valuable benefit
to researchers working therein. The novelty of the present work will
thus be to formulate a \textit{completely general} PwP method for
\textit{any} distributionally-sourced (linear) problem of the form
(\ref{eq:intro_general_pde}) with the single limiting condition that
${\rm supp}(S)\subset\mathscr{I}\subseteq\mathscr{U}$ where ${\rm dim}(\mathscr{I})=1$.
We will prove rigorously why and how the method works for such problems,
and then we will implement it to obtain numerical solutions to the
variety of different applications mentioned earlier in order to illustrate
its broad practicability. We will see that, in general, this method
either matches or improves upon the results of other methods existent in the literature
for tackling distributionally-sourced PDEs---and we turn to a more
detailed discussion of this topic in the next subsection.

\subsection{Comparison with other methods in the literature}

Across all areas of application, the most commonly encountered---and,
perhaps, most naively suggestible---strategy for numerically solving
equations of the form (\ref{eq:intro_general_pde}) is to rely upon
some sort of delta function approximation procedure on the computational
grid [\cite{tornberg_numerical_2004,jung_collocation_2009,jung_note_2009,petersson_discretizing_2016}].
For instance, the simplest imaginable choice in this vein is just
a narrow hat function (centered at the point where the delta function
is supported, and having total measure $1$) which, for better accuracy,
one can upgrade to higher-order polynomials, or even trigonometric
functions. Another readily evocable possibility is to use a narrow
Gaussian---and indeed, this is one option that has in fact been tried
in self-force computations as well (see [\cite{lopez-aleman_perturbative_2003}],
for example). However, this unavoidably introduces
into the problem an additional, artificial length scale: that is,
the width of the Gaussian, which a priori need not have anything to
do with the actual (\textquotedblleft physical\textquotedblright )
length scale of the source. Moreover, there is the evident drawback
that no matter how small this artificial length scale is chosen, the
solutions will never be well-resolved close to the distributional
source location: there will always be some sort of Gibbs-type phenomenon\footnote{~The Gibbs phenomenon, originally discovered by Henry Wilbraham [\cite{wilbraham_certain_1848}] and rediscovered by J. Willard Gibbs [\cite{gibbs_letter_1899}], refers generally to an overshoot in the approximation of a piecewise continuously differentiable function near a jump discontinuity.}
there.

Methods for solving (\ref{eq:intro_general_pde}) which are closer
in spirit to our PwP method have been explored in  [\cite{field_discontinuous_2009}]
and [\cite{shin_spectral_2011}]. In particular, both of these works
have used the idea of placing the distributional source at the interface
of computational grids---however, they tackle the numerical implementation
differently than we do.

In the case of  [\cite{field_discontinuous_2009}]---which, incidentally,
is also concerned with the self-force problem---the difference is
that the authors use a discontinuous Galerkin method (rather than
spectral methods, as in our PwP approach), and the effect of the distributional
source is accounted for via a modification of the numerical flux at
the ``particle'' location. This relies essentially upon a weak formulation
of the problem, wherein a choice has to be made about how to assign
measures to the distributional terms over the relevant computational
domains. In contrast, we directly solve only for smooth solutions
supported away from the ``particle'' location, and account for the
distributional source simply by imposing adequate boundary---\textit{i.e.} jump---
conditions there.

[\cite{shin_spectral_2011}] is closer to our approach in this
sense, as the authors there also use spectral methods and also account
for the distributional source via jump conditions. However, the difference
with our method is that [\cite{shin_spectral_2011}] treat these
jump conditions as additional constraints (rather than built-in boundary
conditions) for the smooth solutions away from the distributional
source, thus over-determining the problem. That being the case, the
authors are led to the need to define a functional (expressing how
well the differential equations plus the jump conditions are satisfied)
to be minimized---constituting what they refer to as a ``least squares
spectral collocation method''. There is however no unique way to
choose this functional. Moreover, the complication of introducing
it is not at all necessary: our approach, in contrast, simply replaces
the discretization of (the homogeneous version of) the differential
equations at the ``particle'' location with the corresponding jump
conditions (\textit{i.e.} it imposes the jump conditions \textit{as} boundary
conditions, \textit{by construction}---something which PSC methods are precisely designed to be able to handle), leading to completely determined
systems in all cases which are solved directly, without further complications.

Finally, neither  [\cite{field_discontinuous_2009}] nor [\cite{shin_spectral_2011}]
analyzed to any significant extent the conditions under which their
methods might be applicable to more general distributionally-sourced
PDEs. As mentioned, in the present appendix we will devote a careful
proof entirely to this issue.

This appendix is structured as follows. Following some mathematical preliminaries
in Section \ref{setup}, we prove in Section \ref{pwp} how the PwP method can be formulated and applied to problems with the most general possible ``point'' source $S:\mathcal{D}(\mathscr{I})\rightarrow\mathbb{R}$,
that is, one containing an arbitrary number of (linearly combined) delta
derivatives and supported at an arbitrary number of points in $\mathscr{I}$.
Thus, one can use it on any type of (linear) PDE involving such sources,
which we illustrate with the applications listed in (i)-(iv) above
in Sections \ref{first-order-hyperbolic-pdes}-\ref{elliptic-pdes}
respectively. Finally, we give concluding remarks in Section \ref{conclusions}.

\section{Setup}
\label{setup}

We wish to begin by establishing some basic notation and then reviewing
some pertinent properties of distributions that we will need to make
use of later on. While we will certainly strive to maintain a fair level of mathematical
rigour here and throughout this appendix (at least, insofar as a certain
amount of formal precaution is inevitably necessary when dealing
with distributions), our principal aim remains that of presenting
practical methodologies; hence the word ``distribution'' may at times be liberally interchanged for
``function'' (\textit{e.g.} we may say ``delta function'' instead of ``delta distribution'') and some notation possibly
slightly abused, when the context is clear enough to not pose dangers
for confusion.

\subsection{Distributionally-sourced linear PDEs}

Consider the problem (\ref{eq:intro_general_pde}) with $S:\mathcal{D}(\mathscr{I})\rightarrow\mathbb{R}$,
where $\mathscr{I}\subseteq\mathbb{R}$ is a one-dimensional subspace
of $\mathscr{U}\subseteq\mathbb{R}^{n}$, as discussed in the introduction.
Then we can view $\mathscr{U}$ as a product space, $\mathscr{U}=\mathscr{I}\times\mathscr{V}$
with $\mathscr{V}=\mathscr{U}/\mathscr{I}\subseteq\mathbb{R}^{n-1}$,
and write coordinates on $\mathscr{U}$ as $\mathbf{x}=(x,\mathbf{y})$
with $x\in\mathscr{I}$ and $\mathbf{y}=(y_{1},\ldots,y_{n-1})\in\mathscr{V}$,
such that
\begin{align}
f:\mathscr{U}=\mathscr{I}\times\mathscr{V}\subseteq\mathbb{R}\times\mathbb{R}^{n-1}=\mathbb{R}^{n} & \rightarrow\mathbb{R}\nonumber\\
\mathbf{x}\,\,=\,\,\left(x,\mathbf{y}\right)\,=\,\left(x,y_{1},\ldots,y_{n-1}\right) & \mapsto f\left(\mathbf{x}\right)\label{eq:function_general} 
\end{align}
denotes any arbitrary function on $\mathscr{U}$.

It is certainly possible, in the setup we are about to describe, to
have $\mathscr{V}=\emptyset$, \textit{i.e.} problems involving just \emph{ODEs}
(on $\mathscr{U}=\mathscr{I}$) of the form (\ref{eq:intro_general_pde})---and,
in fact, our first elementary example illustrating the PwP method
in the following section will be of such a kind. For the more involved
numerical examples we will study in later sections, we will most often
be dealing with functions of two variables, $x\in\mathscr{I}$ for
``space'' (or some other pertinent parameter) and $t\in\mathscr{V}\subseteq\mathbb{R}$
for time.

For any function (\ref{eq:function_general}) involved in these problems,
we will sometimes use the notation $f'=\partial_{x}f$ for the ``spatial''
derivative; also, we may employ $\dot{f}=\partial_{t}f$ for the partial
derivative with respect to time $t$ when $\{t\}$ is (a subspace
of) $\mathscr{V}$.

Now, as in the introduction, let $\mathcal{L}$ be any general $m$-th
order \emph{linear} differential operator. The sorts of PDEs (\ref{eq:intro_general_pde})
that we will be concerned with have the basic form 
\begin{equation}
\mathcal{L}u=S=f\delta_{\left(p\right)}+g\delta_{\left(p\right)}^{'}+\cdots\,,\label{eq:setup-pde_general_1}
\end{equation}
where $f(\mathbf{x})$, $g(\mathbf{x})$ \textit{etc.} are ``source'' functions
prescribed by the problem at hand, and we employ the convenient notation
\begin{equation}
\delta_{\left(p\right)}\left(x\right)=\delta\left(x-x_{p}\left(\mathbf{y}\right)\right)
\end{equation}
to indicate the Dirac delta distribution on $\mathscr{I}$ centered
at the ``particle location'' $x_{p}(\mathbf{y})$—the functional
form of which can be either specified a priori, or determined via
some given prescription as the solution $u$ itself is evolved. When
there is no risk of confusion, we may sometimes omit the $\mathbf{y}$
dependence in our notation and simply write $x_{p}$.

In fact, our PwP method can even deal with multiple, say $M$, ``particles''. PwP computations of the self-force have actually only required $M=1$ (there being only one ``particle'' involved in the problem), so the general $M\geq1$ case has not been considered up to now.
Concordantly, to express our problem of interest (\ref{eq:setup-pde_general_1})
in the most general possible form, let us employ the typical PDE notation
for ``multi-indices'' [\cite{evans_partial_1998}], $\alpha=(\alpha_{0},\alpha_{1},\ldots,\alpha_{n-1})$
with each $\alpha_{I}\in\mathbb{Z}^{\geq}$ being a non-negative integer (indexed from $I=0$ to $I=n-1$ so as to make sense \textit{vis-à-vis} our coordinate notation on $\mathscr{U}$, instead of the more usual practice to label them from $1$ to $n$),
and $|\alpha|=\sum_{I=0}^{n-1}\alpha_{I}$. Furthermore, we define $\alpha!=\prod_{I=0}^{n-1}\alpha_{I}!$. Thus, the most general
$m$-th order linear partial differential operator can be written
as $\mathcal{L}=\sum_{|\alpha|\leq m}\xi^{\alpha}(\mathbf{x})D^{\alpha}$
where $\xi^{\alpha}:\mathscr{U}\rightarrow\mathbb{R}$ are arbitrary functions and $D^{\alpha}=\partial^{|\alpha|}/\partial x^{\alpha_{0}}\partial y_{1}^{\alpha_{1}}\cdots\partial y_{n-1}^{\alpha_{n-1}}$.
Hence, we are dealing with any problem which can be placed into the
form 
\begin{equation}
\sum_{|\alpha|\leq m}\xi^{\alpha}\left(\mathbf{x}\right)D^{\alpha}u\left(\mathbf{x}\right)=\sum_{i=1}^{M}\sum_{j=0}^{K}f^{ij}\left(\mathbf{x}\right)\delta^{(j)}\left(x-x_{p_{i}}\left(\mathbf{y}\right)\right)\,,\label{eq:setup-pde_general_2}
\end{equation}
with $f^{ij}:\mathscr{U}\rightarrow\mathbb{R}$ denoting the ``source''
functions (for the $j$-th delta derivative of the $i$-th particle)
and $K\in\mathbb{Z}^{\geq}$ the highest order of the delta function
derivatives in $S$, appropriately supplemented by initial/boundary
conditions (ICs/BCs).

Let us give a few basic examples to render this setup more palpable.
One very simple example---that which will serve as our first illustration
of the PwP method in the next section---is the simple harmonic oscillator
with a constant delta function forcing (source) term---that is, the
ODE (with $\mathscr{V}=\emptyset$): 
\begin{equation}
u''+u=a\delta_{\left(p\right)}\,,
\end{equation}
where $\delta_{\left(p\right)}(x)=\delta(x-x_{p})$ for some fixed
$x_{p}\in\mathscr{I}$, and $a\in\mathbb{R}$. Another example is
the wave equation with a moving singular source, 
\begin{equation}
\left(\partial_{t}^{2}-\partial_{x}^{2}\right)u\left(x,t\right)=f\left(x,t\right)\delta\left(x-x_{p}\left(t\right)\right)\,,
\end{equation}
with $x_{p}(t)$ specified as a function of time.

\subsection{Properties of distributions}

We now wish to remind the reader of a few basic properties of distributions
before proceeding to describe the PwP procedure; for a good detailed exposition, see \textit{e.g.}  [\cite{stakgold_greens_2011}].

Let $f:\mathscr{U}\rightarrow\mathbb{R}$ be, as before, any function
involved in the problem (\ref{eq:setup-pde_general_2}). We denote
by
\begin{align}
f_{p}:\mathscr{V}\rightarrow\, & \mathbb{R}\nonumber \\
\mathbf{y}\mapsto\, & f_{p}\left({\bf y}\right)=f\left(x_{p}\left({\bf y}\right),{\bf y}\right)
\end{align}
the function evaluated at the ``particle'' position.

Furthermore, let $\phi\in\mathcal{D}(\mathscr{I})$ be any test function
on $\mathscr{I}$. Then we define the action of the distribution associated
with $f$ as: 
\begin{equation}
\left\langle f,\phi\right\rangle =\int_{\mathscr{I}}{\rm d}x\,f\left(x,\mathbf{y}\right)\phi\left(x\right)\,.
\end{equation}
We say that two functions $f$ and $g$ are \emph{equivalent} in the
sense of distributions if 
\begin{equation}
\langle f,\phi\rangle=\langle g,\phi\rangle\Leftrightarrow f\equiv g\,.
\end{equation}

An identity which will be important for us in discussing the PwP method
is the following [\cite{cortizo_diracs_1995,li_review_2007}]: 
\begin{equation}
f\left(x,\mathbf{y}\right)\delta_{\left(p\right)}^{\left(n\right)}\left(x\right)\equiv\left(-1\right)^{n}\sum_{j=0}^{n}\left(-1\right)^{j}\binom{n}{j}f_{p}^{\left(n-j\right)}\left(\mathbf{y}\right)\delta_{\left(p\right)}^{\left(j\right)}\left(x\right)\,,\label{eq:setup-distributions_identity}
\end{equation}
where $\delta_{\left(p\right)}^{\left(n\right)}=\partial_{x}^{n}\delta_{\left(p\right)}$.
For concreteness, let us write down the first three cases explicitly
here: 
\begin{align}
f\left(x,\mathbf{y}\right)\delta_{\left(p\right)}\left(x\right)\equiv\, & f_{p}\left(\mathbf{y}\right)\delta_{\left(p\right)}\left(x\right)\,,\\
f\left(x,\mathbf{y}\right)\delta_{\left(p\right)}'\left(x\right)\equiv\, & -f_{p}'\left(\mathbf{y}\right)\delta_{\left(p\right)}\left(x\right)+f_{p}\left(\mathbf{y}\right)\delta_{\left(p\right)}'\left(x\right)\,,\\
f\left(x,\mathbf{y}\right)\delta_{\left(p\right)}''\left(x\right)\equiv\, & f_{p}''\left(\mathbf{y}\right)\delta_{\left(p\right)}\left(x\right)-2f_{p}'\left(\mathbf{y}\right)\delta_{\left(p\right)}'\left(x\right)+f_{p}\left(\mathbf{y}\right)\delta_{\left(p\right)}''\left(x\right)\,.
\end{align}
For the interested reader, we offer in Appendix A of [\cite{oltean_particle-without-particle:_2019}]
a proof by induction of the formula (\ref{eq:setup-distributions_identity}),
which is instructive for appreciating the subtleties generally involved
in manipulating distributions.

Let 
\begin{equation}
\Theta_{\left(p\right)}^{\pm}\left(x\right)=\Theta\left(\pm\left(x-x_{p}\left(\mathbf{y}\right)\right)\right)
\end{equation}
be the Heaviside function which is supported to the right/left (respectively)
of $x_{p}$. Then, we have: 
\begin{align}
\partial_{x}\Theta_{\left(p\right)}^{\pm}=\, & \pm\delta_{\left(p\right)}\,,\label{eq:setip-partial_x_Theta}\\
\partial_{y_{j}}\Theta_{\left(p\right)}^{\pm}=\, & \mp\left(\partial_{y_{j}}x_{p}\right)\delta_{\left(p\right)}\,,\label{eq:setup-partial_y_Theta}
\end{align}
and so on for higher order partials.

For notational expediency, we may sometimes omit the $(p)$ subscript
on the Heaviside functions (and derivatives thereof) when the context
is sufficiently clear.

\section{The Particle-without-Particle method}
\label{pwp}

As discussed heuristically in the introduction, the basic idea of
our method for solving (\ref{eq:setup-pde_general_2}) is to effectively
eliminate the ``point''-like source or ``particle'' from the problem
by decomposing the solution $u$ into a series of distributions: specifically,
Heaviside functions $\Theta^{i}:\mathcal{D}(\mathscr{I})\rightarrow\mathbb{R}$
supported in each of the $M+1$ disjoint regions of $\mathscr{I}\backslash{\rm supp}(S)$
(\textit{i.e.} ${\rm supp}(\Theta^{i})\cap{\rm supp}(S)=\emptyset,\forall i$
and ${\rm supp}(\Theta^{i})\cap{\rm supp}(\Theta^{j})=\emptyset,\forall i\neq j$)
and, if necessary, delta functions (plus delta derivatives) at ${\rm supp}(S)$:
\begin{equation}
u=\sum_{i=0}^{M}u^{i}\Theta^{i}+\sum_{i=1}^{M}\sum_{j=0}^{K-m}h^{ij}\delta_{\left(p_{i}\right)}^{\left(j\right)}\,,\label{eq:pwp-decomposition}
\end{equation}
where $u^{i}:\mathscr{U}\rightarrow\mathbb{R}$ and we need to include
the second sum with $h^{ij}:\mathscr{V}\rightarrow\mathbb{R}$ only
if $K\geq m$.

We will prove in this section that one can always obtain solutions
of the form (\ref{eq:pwp-decomposition}) to the problem (\ref{eq:setup-pde_general_2}).
In particular, inserting (\ref{eq:pwp-decomposition}) into (\ref{eq:setup-pde_general_2})
will always yield homogeneous equations
\begin{equation}
\mathcal{L}u^{i}=0\quad\mbox{in}\,\,\left(\mathscr{I}\backslash{\rm supp}(S)\right)\times\mathscr{V}\,,
\end{equation}
along with JCs on (the derivatives of) $u$---and possibly (derivatives
of) $h^{ij}$ if applicable. In general, we define the ``jump'' $[\cdot]_{p}:\mathscr{V}\rightarrow\mathbb{R}$
in the value of any function $f:\mathscr{U}\rightarrow\mathbb{R}$
at $x_{p}({\bf y})$ as
\begin{equation}
\left[f\right]_{p}\left({\bf y}\right)=\lim_{x\rightarrow x_{p}({\bf y})^{+}}f\left(x,{\bf y}\right)-\lim_{x\rightarrow x_{p}({\bf y})^{-}}f\left(x,{\bf y}\right)\,.\label{eq:pwp-jump}
\end{equation}
Henceforth, for convenience, we will generally omit the ${\bf y}$-dependence
and simply write $[f]_{p}$.

First we will work through a simple example in order to offer a more
concrete sense of the method, and afterwards we will show in general
how (\ref{eq:pwp-decomposition}) solves (\ref{eq:setup-pde_general_2}).

\subsection{Simple example}

We illustrate here the application of our PwP method to a very simple
ODE (and single-particle) example. We will consider the problem 
\begin{equation}
\mathcal{L}u=u''+u=a\delta+b\delta^{'}\,,\enskip x\in\mathscr{I}=\left[-L,L\right]\,,\enskip u\left(\pm L\right)=0\,,\label{eq:pwp-simple_example}
\end{equation}
where $\delta$ is simply the delta function centered at $x_{p}=0$.

We begin by decomposing $u$ as 
\begin{equation}
u=u^{-}\Theta^{-}+u^{+}\Theta^{+}\,,
\end{equation}
where $\Theta^{\pm}(x)=\Theta(\pm x)$, and we insert this into (\ref{eq:pwp-simple_example}).
Using (\ref{eq:setip-partial_x_Theta}), the LHS becomes simply 
\begin{align}
\mathcal{L}u\left(x\right)=\, & u''\left(x\right)+u\left(x\right)\\
=\, & \left\{ \mathcal{L}u^{-}\left(x\right)\right\} \Theta^{-}\left(x\right)+\left\{ \mathcal{L}u^{+}\left(x\right)\right\} \Theta^{+}\left(x\right)\nonumber \\
 & +\left\{ -2\left(u^{-}\left(x\right)\right)'+2\left(u^{+}\left(x\right)\right)'\right\} \delta\left(x\right)\nonumber \\
 & +\left\{ -u^{-}\left(x\right)+u^{+}\left(x\right)\right\} \delta'\left(x\right)\,.
\end{align}
Now before we can equate this to the distributional terms in the source (RHS), we must apply the identity (\ref{eq:setup-distributions_identity}).
In particular, we use $f(x)\delta(x)\equiv f_{p}\delta(x)$ and
$f(x)\delta'(x)\equiv-f_{p}'\delta(x)+f_{p}\delta'(x)$. Thus, the
above becomes 
\begin{align}
\mathcal{L}u\left(x\right)\equiv\, & \left\{ \mathcal{L}u^{-}\left(x\right)\right\} \Theta^{-}\left(x\right)+\left\{ \mathcal{L}u^{+}\left(x\right)\right\} \Theta^{+}\left(x\right)\nonumber \\
 & +\left\{ -2\left(u^{-}\right)'_{p}+2\left(u^{+}\right)'_{p}\right\} \delta\left(x\right)\nonumber \\
 & +\left\{ \left(u^{-}\right)'_{p}-\left(u^{+}\right)'_{p}\right\} \delta\left(x\right)+\left\{ -u_{p}^{-}+u_{p}^{+}\right\} \delta'\left(x\right)\\
=\, & \left\{ \mathcal{L}u^{-}\left(x\right)\right\} \Theta^{-}\left(x\right)+\left\{ \mathcal{L}u^{+}\left(x\right)\right\} \Theta^{+}\left(x\right)+\left[u'\right]_{p}\delta\left(x\right)+\left[u\right]_{p}\delta'\left(x\right)\,.
\end{align}
Plugging this into the DE (\ref{eq:pwp-simple_example}), we have
\begin{equation}
\left\{ \mathcal{L}u^{-}\right\} \Theta^{-}+\left\{ \mathcal{L}u^{+}\right\} \Theta^{+}+\left[u'\right]_{p}\delta+\left[u\right]_{p}\delta'\equiv a\delta+b\delta^{'}\,.
\end{equation}
Therefore the original problem is equivalent to the system of equations:
\begin{equation}
\begin{cases}
\mathcal{L}u^{-}=0, & x\in\mathscr{D}^{-}=\left[-L,0\right]\,,\enskip u^{-}\left(-L\right)=0\,,\\
\mathcal{L}u^{+}=0, & x\in\mathscr{D}^{+}=\left[0,L\right]\,,\enskip u^{+}\left(L\right)=0\,,\\
\left[u\right]_{p}=b, & \left[u'\right]_{p}=a\,.
\end{cases}\label{eq:pwp-simple_example_pwp}
\end{equation}

Let us solve (\ref{eq:pwp-simple_example_pwp}), for simplicity, taking
$L=\pi/4$. The left homogeneous equation in (\ref{eq:pwp-simple_example_pwp})
has the general solution $u^{-}=A^{-}\cos(x)+B^{-}\sin(x)$, and the
BC tells us that $0=u^{-}(-\pi/4)=\frac{1}{\sqrt{2}}(A^{-}-B^{-})$,
\textit{i.e.} 
\begin{equation}
A^{-}-B^{-}=0\,.\label{eq:pwp-simple_example_constants1}
\end{equation}
The right homogeneous equation in (\ref{eq:pwp-simple_example_pwp})
similarly has general solution $u^{+}=A^{+}\cos(x)+B^{+}\sin(x)$,
with the BC stating $0=u^{+}(\pi/4)=\frac{1}{\sqrt{2}}(A^{+}+B^{+})$,
\textit{i.e.} 
\begin{equation}
A^{+}+B^{+}=0\,.\label{eq:pwp-simple_example_constants2}
\end{equation}
So far we have two equations (\ref{eq:pwp-simple_example_constants1})-(\ref{eq:pwp-simple_example_constants2})
for four unknowns (the integration constants in the general solutions).
It is the JCs in (\ref{eq:pwp-simple_example_pwp}) that provide us
with the remaining necessary equations to fix the solution. We have
$u^{-}(0)=A^{-}$, $(u^{-})'(0)=B^{-}$, $u^{+}(0)=A^{+}$ and $(u^{+})'(0)=B^{+}$
(understood in the appropriate limit approaching $x_{p}=0$). Hence
the JCs tell us: 
\begin{align}
b=\, & \left[u\right]_{p}=u^{+}(0)-u^{-}(0)=A^{+}-A^{-}\,,\label{eq:pwp-simple_example_constants3}\\
a=\, & \left[u'\right]_{p}=(u^{+})'(0)-(u^{-})'(0)=B^{+}-B^{-}\,.\label{eq:pwp-simple_example_constants4}
\end{align}
(We can think of the JCs as a mixing of the degrees of freedom in
the homogeneous solutions in such a way that they ``link together''
to produce the solution generated by the original distributional source.)
Solving (\ref{eq:pwp-simple_example_constants1})-(\ref{eq:pwp-simple_example_constants4}),
we get $A^{-}=-\frac{a+b}{2}=B^{-}$, $A^{+}=-\frac{a-b}{2}=-B^{+}$.
We now have the full solution to our original problem (\ref{eq:pwp-simple_example}):
\begin{equation}
u\left(x\right)=-\frac{a+b}{2}\left(\cos\left(x\right)+\sin\left(x\right)\right)\Theta\left(-x\right)-\frac{a-b}{2}\left(\cos\left(x\right)-\sin\left(x\right)\right)\Theta\left(x\right)\,.
\end{equation}

\subsection{General proof}

Suppose we have $M$ ``particles'' located at ${\rm supp}(S)=\{x_{p_{i}}\}_{i=1}^{M}\subset\mathscr{I}$,
as in the problem (\ref{eq:setup-pde_general_2}), with $x_{p_{1}}<x_{p_{2}}<\cdots<x_{p_{M}}$.
(NB: For $M\geq2$, if there exists any subset of $\mathscr{V}$ where
it should happen that $x_{p_{i}}({\bf y})>x_{p_{i+1}}({\bf y})$ as
a consequence of the ${\bf y}$-evolution, we can, without loss of
generality, simply swap indices within that subset so as to always
have $x_{p_{i}}<x_{p_{i+1}},\forall i$.) Furthermore let us assume
for the moment that the maximum order of delta function derivatives
in the source is one less than the order of the PDE (or smaller),
\textit{i.e.} $K=m-1$. In this case, we do not need to consider the second
term on the RHS of (\ref{eq:pwp-decomposition}), \textit{i.e.} $u$ is just
split up into pieces which are supported only in between all the particle
locations: $u^{0}(\mathbf{x})$ to the left of $x_{p_{1}}$, $u^{1}(\mathbf{x})$
between $x_{p_{1}}$ and $x_{p_{2}}$, ..., $u^{i}(\mathbf{x})$ between
$x_{p_{i}}$ and $x_{p_{i+1}}$, ..., and finally $u^{M}$ to the
right of $x_{p_{M}}$. Thus, we take 
\begin{equation}
u=\sum_{i=0}^{M}u^{i}\Theta^{i}\,,\label{eq:pwp-decomposition_general}
\end{equation}
where we define
\begin{equation}
\Theta^{i}=\begin{cases}
\Theta_{\left(p_{1}\right)}^{-}\,, & i=0\,,\\
\Theta_{\left(p_{i}\right)}^{+}-\Theta_{\left(p_{i+1}\right)}^{+}\,, & 1\leq i\leq M-1\,,\\
\Theta_{\left(p_{M}\right)}^{+}\,, & i=M\,,
\end{cases}
\end{equation}
denoting, as before, $\Theta_{(p_{i})}^{\pm}(x)=\Theta(\pm(x-x_{p_{i}}(\mathbf{y})))$.
Another way of stating this is that we assume for $u$ a piecewise
decomposition 
\begin{equation}
u=\begin{cases}
u^{0}\,, & x\in\mathscr{D}^{0}\,,\\
 & \vdots\\
u^{M}\,, & x\in\mathscr{D}^{M}\,,
\end{cases}
\end{equation}
where the $\mathscr{D}^{i}$'s are disjoint subsets of $\mathscr{I}$
between each ``particle location'', \textit{i.e.}  
\begin{equation}
\mathscr{I}={\rm supp}\left(S\right)\cup\left(\bigcup_{i=0}^{M}\mathscr{D}^{i}\right)\,,
\end{equation}
where
\begin{equation}
\mathscr{D}^{i}=\begin{cases}
\left\{ x\in\mathscr{I}|x<x_{p_{1}}\right\} \,, & i=1\,,\\
\left\{ x\in\mathscr{I}|x_{p_{i}}<x<x_{p_{i+1}}\right\} \,, & 1\leq i\leq M-1\,,\\
\left\{ x\in\mathscr{I}|x_{p_{M}}<x\right\} \,, & i=M\,.
\end{cases}
\end{equation}

The general strategy, then, is to insert (\ref{eq:pwp-decomposition_general})
into (\ref{eq:setup-pde_general_2}), and to obtain a set of equations
by matching (regular function) terms multiplying the same derivative
order of the Heaviside distributions. Explicitly, using the Leibniz rule, we get
\begin{equation}
\mathcal{L}u=\sum_{i=0}^{M}\sum_{|\alpha|\leq m}\sum_{|\beta|\leq|\alpha|}\binom{\alpha}{\beta}\xi^{\alpha}\left(D^{\alpha-\beta}u^{i}\right)\left(D^{\beta}\Theta^{i}\right)=\sum_{i=1}^{M}\sum_{j=0}^{m-1}f^{ij}\delta_{\left(p_{i}\right)}^{(j)}\,.\label{eq:pwp-pde_decomposed}
\end{equation}

At zeroth order in derivatives of the Heaviside functions, \textit{i.e.} the sum of
all $|\beta|=0$ terms in the LHS above, we will always simply obtain—in
the absence of any Heaviside functions on the RHS—a set of $M+1$ homogeneous
equations, which constitute simply the original equation on each disjoint
subset of $\mathscr{I}$ but \emph{with no source}: 
\begin{equation}
\sum_{i=0}^{M}\Bigg(\sum_{|\alpha|\leq m}\xi^{\alpha}D^{\alpha}u^{i}\Bigg)\Theta^{i}=0\Leftrightarrow\mathcal{L}u^{i}=0\enskip\mbox{in}\enskip\mathscr{D}^{i}\times\mathscr{V},\,\forall i\,.\label{eq:pwp-homogeneous_equations}
\end{equation}

At first order and higher in the Heaviside derivatives (thus, zeroth
order and higher in delta function derivatives), \textit{i.e.} the sum of all
$|\beta|\neq0$ terms in the LHS of (\ref{eq:pwp-pde_decomposed}),
we have terms of the form 
\begin{align}
D^{\beta}\Theta^{i}=\, & \partial_{x}^{\beta_{0}}\partial_{y_{1}}^{\beta_{1}}\cdots\partial_{y_{n-1}}^{\beta_{n-1}}\begin{cases}
\Theta_{\left(p_{1}\right)}^{-}\,, & i=0\,,\\
\Theta_{\left(p_{i}\right)}^{+}-\Theta_{\left(p_{i+1}\right)}^{+}\,, & 1\leq i\leq M-1\,,\\
\Theta_{\left(p_{M}\right)}^{+}\,, & i=M\,,
\end{cases}\\
=\, & \sum_{j=0}^{|\beta|-1}\begin{cases}
F^{0j}\delta_{\left(p_{1}\right)}^{\left(j\right)}\,, & i=0\,,\\
F^{ij}\delta_{\left(p_{i}\right)}^{\left(j\right)}+G^{ij}\delta_{\left(p_{i+1}\right)}^{\left(j\right)}\,, & 1\leq i\leq M-1\,,\\
F^{Mj}\delta_{\left(p_{M}\right)}^{\left(j\right)}\,, & i=M\,,
\end{cases}\label{eq:pwp-D_beta_Theta2}
\end{align}
for some $\mathbf{y}$-dependent functions $F^{ij}:\mathscr{V}\rightarrow\mathbb{R}$
and $G^{ij}:\mathscr{V}\rightarrow\mathbb{R}$ which arise from the
implicit differentiation (\textit{e.g.}, Eqns. (\ref{eq:setip-partial_x_Theta})-(\ref{eq:setup-partial_y_Theta})),
and the precise form of which does not concern us for the present
purposes. Plugging (\ref{eq:pwp-D_beta_Theta2}) into (\ref{eq:pwp-pde_decomposed})
and manipulating the sums, we get 
\begin{equation}
\sum_{i=1}^{M}\sum_{|\alpha|\leq m}\sum_{0<|\beta|\leq|\alpha|}\sum_{j=0}^{|\beta|-1}\Phi^{\alpha,\beta,ij}\delta_{\left(p_{i}\right)}^{\left(j\right)}=\sum_{i=1}^{M}\sum_{j=0}^{m-1}f^{ij}\delta_{\left(p_{i}\right)}^{\left(j\right)}\,.\label{eq:pwp-pde_beta_nonzero}
\end{equation}
where for convenience we have defined
\begin{equation}
\Phi^{\alpha,\beta,ij}\left(\mathbf{x}\right)=\binom{\alpha}{\beta}\xi^{\alpha}\left(\mathbf{x}\right)\left(F^{ij}\left(\mathbf{y}\right)D^{\alpha-\beta}u^{i}\left(\mathbf{x}\right)+H^{ij}\left(\mathbf{y}\right)D^{\alpha-\beta}u^{i-1}\left(\mathbf{x}\right)\right)\,,
\end{equation}
for some $\mathbf{y}$-dependent functions $H^{ij}:\mathscr{V}\rightarrow\mathbb{R}$
(related to $F^{ij}$ and $G^{ij}$, and the precise form of which
is also unimportant). At this point, one must be careful: \emph{before}
drawing conclusions regarding the equality of terms (the coefficients
of the delta function derivatives) in (\ref{eq:pwp-pde_beta_nonzero}),
one should apply the identity (\ref{eq:setup-distributions_identity}).
Doing this, one obtains: 
\begin{equation}
\sum_{\alpha,\beta,i,j}\sum_{k=0}^{j}\left(-1\right)^{j+k}\binom{j}{k}\left(\partial_{x}^{j-k}\Phi^{\alpha,\beta,ij}\right)_{p_{i}}\delta_{\left(p_{i}\right)}^{\left(k\right)}=\sum_{i,j}\sum_{k=0}^{j}\left(-1\right)^{j+k}\binom{j}{k}\left(\partial_{x}^{j-k}f^{ij}\right)_{p_{i}}\delta_{\left(p_{i}\right)}^{\left(k\right)}\,,\label{eq:pwp-pde_with_identity}
\end{equation}
with the omitted summation limits as before. Thus, we see that on
the LHS, we have terms involving 
\begin{align}
\partial_{x}^{j-k}\Phi^{\alpha,\beta,ij}=\, & \partial_{x}^{j-k}\left\{ \binom{\alpha}{\beta}\xi^{\alpha}\left(F^{ij}D^{\alpha-\beta}u^{i}+H^{ij}D^{\alpha-\beta}u^{i-1}\right)\right\} \\
=\, & \binom{\alpha}{\beta}\left\{ F^{ij}\partial_{x}^{j-k}\left(\xi^{\alpha}D^{\alpha-\beta}u^{i}\right)+H^{ij}\partial_{x}^{j-k}\left(\xi^{\alpha}D^{\alpha-\beta}u^{i-1}\right)\right\} \\
=\, & \binom{\alpha}{\beta} \sum_{l=0}^{j-k}\binom{j-k}{l}\left(\partial_{x}^{j-k-l}\xi^{\alpha}\right)\left[F^{ij}\left(\partial_{x}^{l}D^{\alpha-\beta}u^{i}\right)+H^{ij}\left(\partial_{x}^{l}D^{\alpha-\beta}u^{i-1}\right)\right] \,.\label{eq:pwp-partial_x_Phi}
\end{align}
Thus, defining the $\mathbf{y}$-dependent functions 
\begin{align}
\Psi^{\alpha,\beta,ijkl}=\, & \left(-1\right)^{j+k}\binom{j}{k}\binom{j-k}{l}\binom{\alpha}{\beta}\left(\partial_{x}^{j-k-l}\xi^{\alpha}\right)_{p_{i}}\,,\\
\psi^{ijk}=\, & \left(-1\right)^{j+k}\binom{j}{k}\left(\partial_{x}^{j-k}f^{ij}\right)_{p_{i}}\,,
\end{align}
we can use (\ref{eq:pwp-partial_x_Phi}) to write (\ref{eq:pwp-pde_with_identity})
in the form: 
\begin{equation}
\sum_{\alpha,\beta,i,j,k,l}\Psi^{\alpha,\beta,ijkl}\left[F^{ij}\left(\partial_{x}^{l}D^{\alpha-\beta}u^{i}\right)_{p_{i}}+H^{ij}\left(\partial_{x}^{l}D^{\alpha-\beta}u^{i-1}\right)_{p_{i}}\right]\delta_{\left(p_{i}\right)}^{\left(k\right)}=\sum_{i,j,k}\psi^{ijk}\delta_{\left(p_{i}\right)}^{\left(k\right)}\,,\label{eq:pwp-pde_deltas_final}
\end{equation}
where the terms involving $u^{i}$ partials ``at the particle''
should be understood as the limit evaluated from the appropriate direction,
\textit{i.e.} 
\begin{align}
\left(D^{\gamma}u^{i}\left(\mathbf{x}\right)\right)_{p_{i}}=\, & \lim_{x\rightarrow x_{p_{i}}^{+}}D^{\gamma}u^{i}\left(x,\mathbf{y}\right)\,,\\
\left(D^{\gamma}u^{i-1}\left(\mathbf{x}\right)\right)_{p_{i}}=\, & \lim_{x\rightarrow x_{p_{i}}^{-}}D^{\gamma}u^{i}\left(x,\mathbf{y}\right)\,.
\end{align}
Having obtained (\ref{eq:pwp-pde_deltas_final}), we can finally match
the coefficients of each $\delta_{\left(p_{i}\right)}^{\left(k\right)}$
to obtain the JCs with which the homogeneous equations (\ref{eq:pwp-homogeneous_equations})
must be supplemented.

Let us now extend this method to problems where the maximum order
of delta function derivatives in the source equals or exceeds the
order of the PDE, \textit{i.e.} $K\geq m$, a case not previously required---and hence not yet considered---in any of the past PwP work on the self-force. To do this, we just add to our
ansatz the second term on the RHS of (\ref{eq:pwp-decomposition}),
which for convenience we denote $u^{\delta}$; that is: 
\begin{equation}
u=\sum_{i=0}^{M}u^{i}\Theta^{i}+u^{\delta}\,,\quad u^{\delta}=\sum_{i=1}^{M}\sum_{j=0}^{K-m}h^{ij}\delta_{\left(p_{i}\right)}^{\left(j\right)}\,,\label{eq:pwp-decomposition_general_with_delta}
\end{equation}
with $h^{ij}(\mathbf{y})$ to be solved for. Inserting (\ref{eq:pwp-decomposition_general_with_delta})
into (\ref{eq:setup-pde_general_2}) we get, on the LHS of the PDE,
the homogeneous problems (at zeroth order) as before, then the LHS
of (\ref{eq:pwp-pde_deltas_final}) due again to the sum of Heaviside functions
term in (\ref{eq:pwp-decomposition_general_with_delta}), plus the
following due to the sum of delta function derivatives: 
\begin{align}
\mathcal{L}u^{\delta}=\, & \sum_{|\alpha|\leq m}\xi^{\alpha}D^{\alpha}\sum_{i=1}^{M}\sum_{j=0}^{K-m}h^{ij}\delta_{\left(p_{i}\right)}^{\left(j\right)}\\
=\, & \sum_{\alpha,i,j}\sum_{|\beta|\leq|\alpha|}\binom{\alpha}{\beta}\xi^{\alpha}\left(D^{\alpha-\beta}h^{ij}\right)\left(D^{\beta}\delta_{\left(p_{i}\right)}^{\left(j\right)}\right)\,,
\end{align}
using the Leibniz rule. Next, we employ the Faà di Bruno formula [\cite{constantine_multivariate_1996}] to
carry out the implicit differentiation of the delta function derivatives;
writing $(n-1)$ dimensional multi-indices on $\mathscr{V}$ (pertaining
only to the $\mathbf{y}$ variables) with tildes, \textit{e.g.} $\tilde{\beta}=(\beta_{1},\ldots,\beta_{n-1})$,
we have the following: 
\begin{equation}
D^{\beta}\delta_{\left(p_{i}\right)}^{\left(j\right)}=D^{\tilde{\beta}}\delta_{\left(p_{i}\right)}^{(j+\beta_{0})}=\tilde{\beta}!\sum_{l=1}^{|\tilde{\beta}|}\delta_{\left(p_{i}\right)}^{(j+\beta_{0}+l)}\sum_{s=1}^{|\tilde{\beta}|}\sum_{\mathscr{P}_{s}(\tilde{\beta},l)}\prod_{k=1}^{s}\frac{\left(-D^{\tilde{\lambda}_{k}}x_{p_{i}}\right)^{q_{k}}}{q_{k}!\left(\tilde{\lambda}_{k}!\right)^{q_{k}}}\,,
\end{equation}
where $\mathscr{P}_{s}(\tilde{\beta},l)=\{(q_{1},\ldots,q_{s};\tilde{\lambda}_{1},\ldots,\tilde{\lambda}_{s}):q_{k}>0,0\prec\tilde{\lambda}_{1}\prec\cdots\prec\tilde{\lambda}_{s},\sum_{k=1}^{s}q_{k}=l$
and $\sum_{k=1}^{s}q_{k}\tilde{\lambda}_{k}=\tilde{\beta}\}$. Therefore,
with all the summation limits the same as above, we get 
\begin{equation}
\mathcal{L}u^{\delta}=\sum_{\alpha,\beta,i,j,l,s}\sum_{\mathscr{P}_{s}(\tilde{\beta},l)}\binom{\alpha}{\beta}\tilde{\beta}!\xi^{\alpha}\left(D^{\alpha-\beta}h^{ij}\right)\delta_{\left(p_{i}\right)}^{(j+\beta_{0}+l)}\prod_{k=1}^{s}\frac{\left(-D^{\tilde{\lambda}_{k}}x_{p_{i}}\right)^{q_{k}}}{q_{k}!\left(\tilde{\lambda}_{k}!\right)^{q_{k}}}\,.
\end{equation}
Finally, we use the distributional identity (\ref{eq:setup-distributions_identity})
to obtain 
\begin{align}
\mathcal{L}u^{\delta}\equiv\, & \sum_{\alpha,\beta,i,j,l,s}\sum_{\mathscr{P}_{s}(\tilde{\beta},l)}\binom{\alpha}{\beta}\tilde{\beta}!\left(\left(D^{\alpha-\beta}h^{ij}\right)\prod_{k=1}^{s}\frac{\left(-D^{\tilde{\lambda}_{k}}x_{p_{i}}\right)^{q_{k}}}{q_{k}!\left(\tilde{\lambda}_{k}!\right)^{q_{k}}}\right)_{p_{i}}\nonumber \\
 & \times\left(-1\right)^{j+\beta_{0}+l}\sum_{r=0}^{j+\beta_{0}+l}\left(-1\right)^{r}\binom{j+\beta_{0}+l}{r}\left(\partial_{x}^{j+\beta_{0}+l-r}\xi^{\alpha}\right)_{p_{i}}\delta_{\left(p_{i}\right)}^{\left(r\right)}\,,
\end{align}
with which the higher order delta function derivatives on the RHS
of (\ref{eq:setup-pde_general_2}) can be matched.

\subsection{Limitations of the method}

Let us now discuss more amply the potential issues one is liable to
encounter in any attempt to extend the PwP method further beyond the
setup we have described so far.

Firstly, we stress once more that the method is applicable only to
\emph{linear} PDEs. As pointed out in the introduction, this is simply
an inherent limitation of the classic theory of distributions. In
particular, there it has long been proved [\cite{schwartz_sur_1954}] (see also the discussion in  [\cite{bottazzi_grid_2019}]) that there does not exist
a differential algebra $(A,+,\otimes,\delta)$ wherein the real distributions
can be embedded, and: (i) $\otimes$ extends the product over $C^{0}(\mathbb{R})$;
(ii) $\delta:A\rightarrow A$ extends the distributional derivative;
(iii) $\forall u,v\in A$, the product rule $\delta(u\otimes v)=(\delta u)\otimes v+u\otimes(\delta v)$
holds. Attempts have been made to overcome this and create a sensible
nonlinear theory of distributions by defining and working with more
general objects dubbed ``generalized functions'' [\cite{colombeau_nonlinear_2013}]. Nonetheless, these
have their own drawbacks (\textit{e.g.} they sacrifice coherence between the
product over $C^{0}(\mathbb{R})$ and that of the differential
algebra), and different formulations are actively being investigated
by mathematicians [\cite{benci_ultrafunctions_2013,bottazzi_grid_2019}]. A PwP method for nonlinear problems in the context
of these formulations could be an interesting line of inquiry for
future work.

Secondly, as we have seen, the PwP method as developed here is guaranteed
to work only for those (linear) PDEs the source $S$ of which is a distribution
not on the entire problem domain $\mathscr{U}$, but only on a one-dimensional
subspace $\mathscr{I}$ of that domain. One may sensibly wonder whether
this situation can be improved, \textit{i.e.} whether a similar procedure could
succeed in tackling equations with sources involving (derivatives
of) delta functions in \emph{multiple} variables—yet, one may also
immediately realize that such an attempted extension quickly leads
to significant complications and potentially impassable problems.
Let us suppose that the source contains (derivatives of) delta functions
in $\bar{n}>1$ variables. We still define $\mathscr{I}$ such that
${\rm supp}(S)\subset\mathscr{I}$, so now we have $\mathscr{I}\subseteq\mathbb{R}^{\bar{n}}$,
and let us adapt the rest of our notation accordingly so that an arbitrary
function on $\mathscr{U}$ is
\begin{align}
f:\mathscr{U}\,=\,\mathscr{I}\times\mathscr{V}\,\,\subseteq\,\,\mathbb{R}^{\bar{n}}\times\mathbb{R}^{n-\bar{n}}\,\,=\,\,\mathbb{R}^{n} & \rightarrow\mathbb{R}\nonumber \\
\mathbf{x}=\left(\bar{{\bf x}},\mathbf{y}\right)=\left(\bar{x}_{1},...,\bar{x}_{\bar{n}},y_{1},...,y_{n-\bar{n}}\right) & \mapsto f\left(\mathbf{x}\right)\,.
\end{align}
We also adapt the multi-index notation to $\alpha=(\bar{\alpha}_{1},\bar{\alpha}_{2},\ldots,\bar{\alpha}_{\bar{n}},\alpha_{1},\alpha_{2},\ldots,\alpha_{n-\bar{n}})$.
We can still write the most general linear partial differential operator,
just as we did earlier, as $\mathcal{L}=\sum_{|\alpha|\leq m}\xi^{\alpha}(\mathbf{x})D^{\alpha}$
where now $D^{\alpha}=\partial^{|\alpha|}/\partial\bar{x}_{1}^{\bar{\alpha}_{1}}\cdots\partial\bar{x}_{\bar{n}}^{\bar{\alpha}_{\bar{n}}}\partial y_{1}^{\alpha_{1}}\cdots\partial y_{n-\bar{n}}^{\alpha_{n-\bar{n}}}$.
Moreover, in general, we use the barred boldface notation $\bar{{\bf v}}$
for any vector in $\mathscr{I}$, $\bar{{\bf v}}=(\bar{v}_{1},\ldots,\bar{v}_{\bar{n}})\in\mathscr{I}\subseteq\mathbb{R}^{\bar{n}}$.

One may first ask whether a PwP-type method could be used to handle
``point'' sources in $\mathscr{I}\subseteq\mathbb{R}^{\bar{n}}$.
In other words, can we find a decomposition of $u$ which could be
useful for a problem of the form 
\begin{equation}
\mathcal{L}u\left(\mathbf{x}\right)=f\left(\mathbf{x}\right)\delta\left(\bar{{\bf x}}-\bar{{\bf x}}_{p}\left(\mathbf{y}\right)\right)+\bar{{\bf g}}\left(\mathbf{x}\right)\cdot\bar{\boldsymbol{\nabla}}\delta\left(\bar{{\bf x}}-\bar{{\bf x}}_{p}\left(\mathbf{y}\right)\right)+\cdots\,,\label{eq:pwp-multivariable_source_Lu}
\end{equation}
(assuming for simplicity a \emph{single} point source at $\bar{{\bf x}}_{p}\in\mathscr{I}$)
with $\bar{\boldsymbol{\nabla}}=\partial/\partial\bar{{\bf x}}$ and
given functions $f:\mathscr{U\rightarrow\mathbb{R}}$, $\bar{{\bf g}}:\mathscr{U}^{\bar{n}}\rightarrow\mathbb{R}$
\textit{etc.}? Intuitively, in order to match the delta function (derivatives)
on the RHS, we might expect $u$ to contain the $\bar{n}$-dimensional
Heaviside function $\Theta:\mathcal{D}(\mathscr{I})\rightarrow\mathbb{R}$.
Thus, in the same vein as (\ref{eq:pwp-decomposition_general}), a
possible attempt (for $K<m$) might be to try a splitting such as
\begin{equation}
u\left(\mathbf{x}\right)=\sum_{\bar{\boldsymbol{\sigma}}=\mathrm{\Pi}^{\bar{n}}(\pm)}u^{\bar{\boldsymbol{\sigma}}}\left(\mathbf{x}\right)\Theta\left(\bar{\boldsymbol{\sigma}}\odot\left(\bar{{\bf x}}-\bar{{\bf x}}_{p}\left(\mathbf{y}\right)\right)\right)\,,\label{eq:pwp-multivariable_source_u}
\end{equation}
where $\mathrm{\Pi}$ is here the Cartesian product and $\odot$ the entrywise
product; \emph{but} whether or not this will work depends completely
upon the detailed form of $\mathcal{L}$. For example, the procedure
\textsl{might} work in the case where $\mathcal{L}$ contains a nonvanishing
$D^{(1,1,\ldots,1,\alpha_{1},...,\alpha_{n-\bar{n}})}$ term, so as
to produce a $\delta(\bar{{\bf x}}-\bar{{\bf x}}_{p})$ term upon
its action on $u$ (in the form (\ref{eq:pwp-multivariable_source_u})),
needed to match the $f(\mathbf{x})$ term on the RHS of (\ref{eq:pwp-multivariable_source_Lu}).
However, this still does not guarantee that \textsl{all} the distributional
terms can in the end be appropriately matched, and so in general,
one should \emph{not} expect that such an approach in these sorts
of problems will yield a workable strategy.

To render the above discussion a little less abstract, let us illustrate
what we mean by way of a very simple example. Consider a two-dimensional
Poisson equation on $\mathscr{U}=\{(x,y)\}\subseteq\mathbb{R}^{2}$:
$(\partial_{x}^{2}+\partial_{y}^{2})u=\delta_{2}(x,y)$, where the
RHS is the two-dimensional delta function supported at the origin.
An attempt to solve this via our method would begin by decomposing
the solution into a form $u=\sum_{j}u^{j}\Theta^{j}$, for some suitably-defined
Heaviside functions $\Theta^{j}$--- supported, for example,
on positive/negative half-planes in each of the two coordinates, or
perhaps on each quadrant of $\mathbb{R}^{2}$. However, the RHS of
this problem is, by definition, $\delta_{2}(x,y)=\delta(x)\delta(y)=(\partial_{x}\Theta^{+}(x))(\partial_{y}\Theta^{+}(y))$,
and there is no way to get such a term from the operator $\mathcal{L}=\partial_{x}^{2}+\partial_{y}^{2}$
acting on any linear combination of Heaviside functions. The unconvinced
reader is invited to try a few attempts for themselves, and the difficulties
with this will quickly become apparent.

That said, \emph{one} case in which a PwP-type procedure could work
is when the source contains (one-dimensional) ``string''-like singularities
(instead of $\bar{n}$-dimensional ``point''-like ones) in each of
the $\bar{{\bf x}}$ variables---in other words, when our problem
is of the form 
\begin{equation}
\mathcal{L}u\left(\mathbf{x}\right)=\sum_{a=1}^{\bar{n}}f^{a}\left(\mathbf{x}\right)\delta\left(\bar{x}_{a}-\bar{x}_{a,p}\left(\mathbf{y}\right)\right)+\sum_{a=1}^{\bar{n}}g^{a}\left(\mathbf{x}\right)\delta'\left(\bar{x}_{a}-\bar{x}_{a,p}\left(\mathbf{y}\right)\right)+\cdots\,,
\end{equation}
with $f^{a}:\mathscr{U}\rightarrow\mathbb{R}$, $g^{a}:\mathscr{U}\rightarrow\mathbb{R}$
\textit{etc.} Then, a decomposition of $u$ which can be tried in such situations
(for $K<m$) is 
\begin{equation}
u\left(\mathbf{x}\right)=\sum_{a=1}^{\bar{n}}\sum_{\sigma_{a}=\pm}u^{a,\sigma_{a}}\left(\mathbf{x}\right)\Theta\left(\sigma_{a}\left(\bar{x}_{a}-\bar{x}_{a,p}\left(\mathbf{y}\right)\right)\right)\,.
\end{equation}

\section{First order hyperbolic PDEs}
\label{first-order-hyperbolic-pdes}

We now move on to applications of the PwP method, beginning with first
order hyperbolic equations. First we look at the standard advection
equation, and then a simple neural population model from neuroscience.
Finally, we consider another popular advection-type problem with a
distributional source---namely, the shallow water equations with
discontinuous bottom topography---and briefly explain why the PwP
method cannot be used in that case.

\subsection{Advection equation}

As a first very elementary illustration of our method, let us consider
the $(1+1)$-dimensional advection equation for $u(x,t)$ with a time
function singular point source at some $x=x_{*}$: 
\begin{equation}
\begin{cases}
\partial_{t}u+\partial_{x}u=g\left(t\right)\delta\left(x-x_{*}\right)\,, & x\in\mathscr{I}=\left[0,L\right],\enskip t>0\,,\\
u\left(x,0\right)=0\,, & u\left(0,t\right)=u\left(L,t\right)\,,
\end{cases}\label{eq:advection_pde}
\end{equation}
where we assume that the source time function $g(t)$ is smooth and
vanishes at $t=0$. On an unbounded spatial domain (\textit{i.e.} $x\in\mathbb{R}$),
the exact solution of this problem is 
\begin{equation}
u_{\text{ex}}\left(x,t\right)=\left[\Theta\left(x-x_{*}\right)-\Theta\left(x-x_{*}-t\right)\right]g\left(t-\left(x-x_{*}\right)\right)\,,\label{eq:advection_exact_soln}
\end{equation}
\textit{i.e.} the forward-translated source function in the right half
of the future light cone emanating from $x_{*}$. If we suppose that the source location
satisfies $x_{*}\in(0,L/2]$, then (\ref{eq:advection_exact_soln})
is also a solution of our problem (\ref{eq:advection_pde}) for $t\in[0,L-x_{*}]$.

This precise problem is treated in [\cite{petersson_discretizing_2016}] using a (polynomial) delta function approximation procedure,
with the following: $g(t)={\rm e}^{-(t-t_{0})^{2}/2}$, $t_{0}=8$,
$L=40$ and $x_{*}=10+\pi$. We numerically implement the exact same setup, but using our
PwP method: that is, we decompose $u=u^{-}\Theta^{-}+u^{+}\Theta^{+}$ where
$\Theta^{\pm}=\Theta(\pm(x-x_{*}))$. Inserting this into (\ref{eq:advection_pde}),
we get homogeneous PDEs $\partial_{t}u^{\pm}+\partial_{x}u^{\pm}=0$
to the left and right of the singularity, \textit{i.e.} on $x\in\mathscr{D}^{-}=[0,x_{*}]$
and $x\in\mathscr{D}^{+}=[x_{*},L]$ respectively, along with a jump
in the solution $[u]_{*}=g(t)$ at the point of the source singularity.

The details of all our numerical schemes in this work are described in an appendix, Section \ref{a-psc}. In particular, for the present problem, see Subsection \ref{a-numerical-first-order-hyperbolic}. We also offer in Subsection \ref{a-psc-gen} a brief description of the PSC methods and notation used therein.

\begin{figure}
\begin{center}
\includegraphics[scale=0.6]{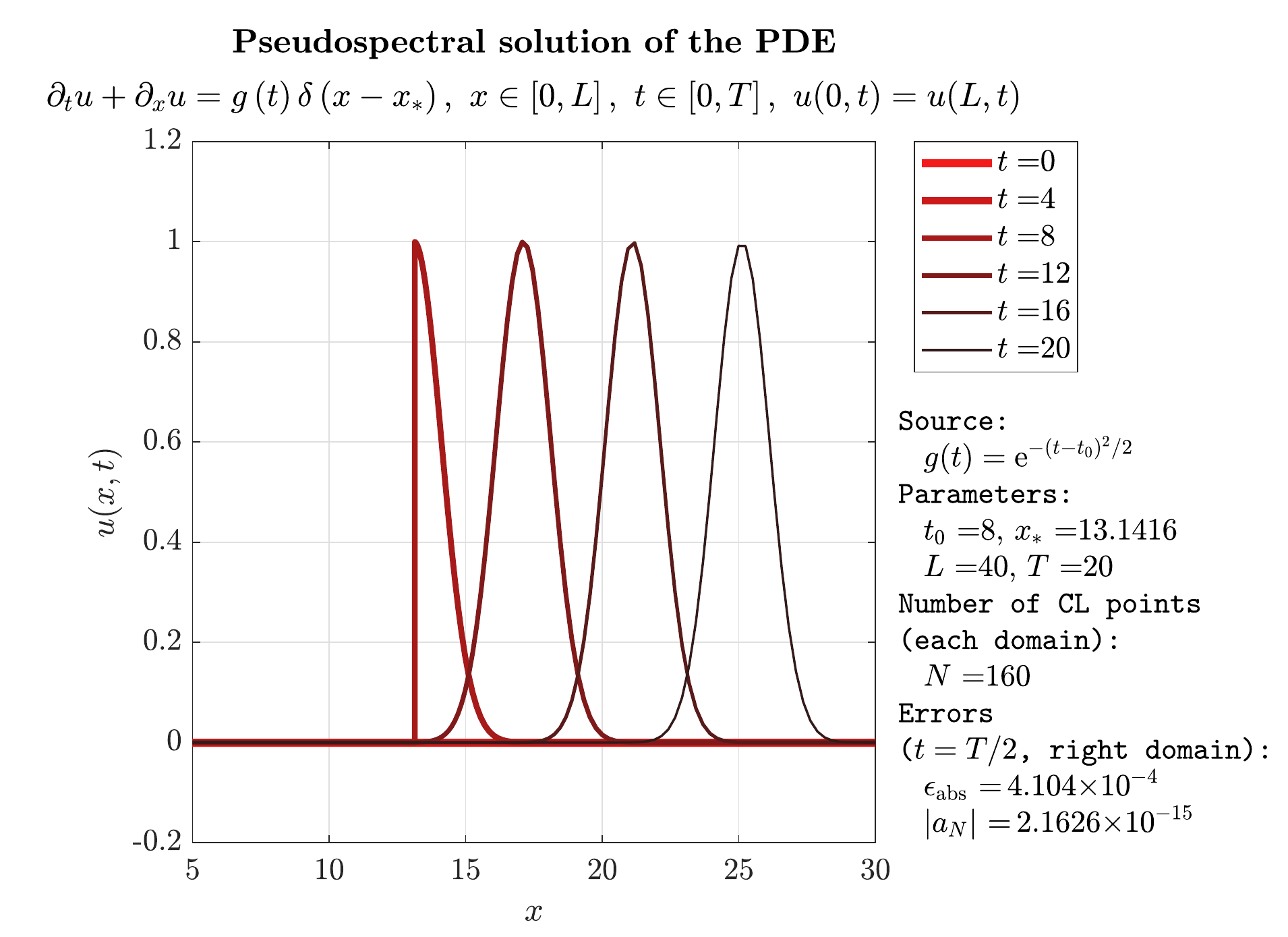}
\includegraphics[scale=0.6]{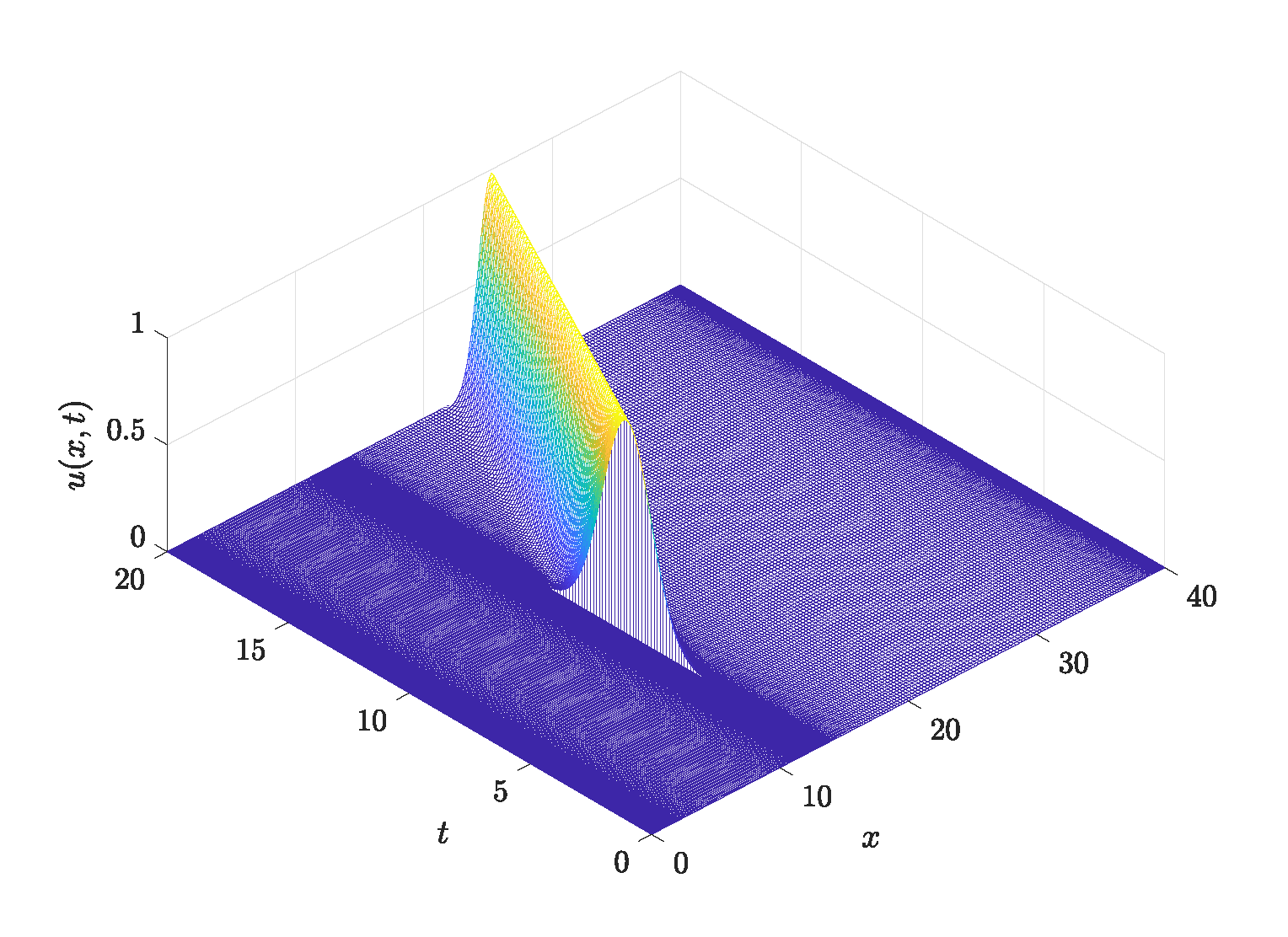}
\caption{Solution of the problem (\ref{eq:advection_pde}) with zero initial data.}\label{fig:pde_pwp_advection_soln}
\end{center}
\end{figure}

\begin{figure}
\begin{center}
\includegraphics[scale=0.6]{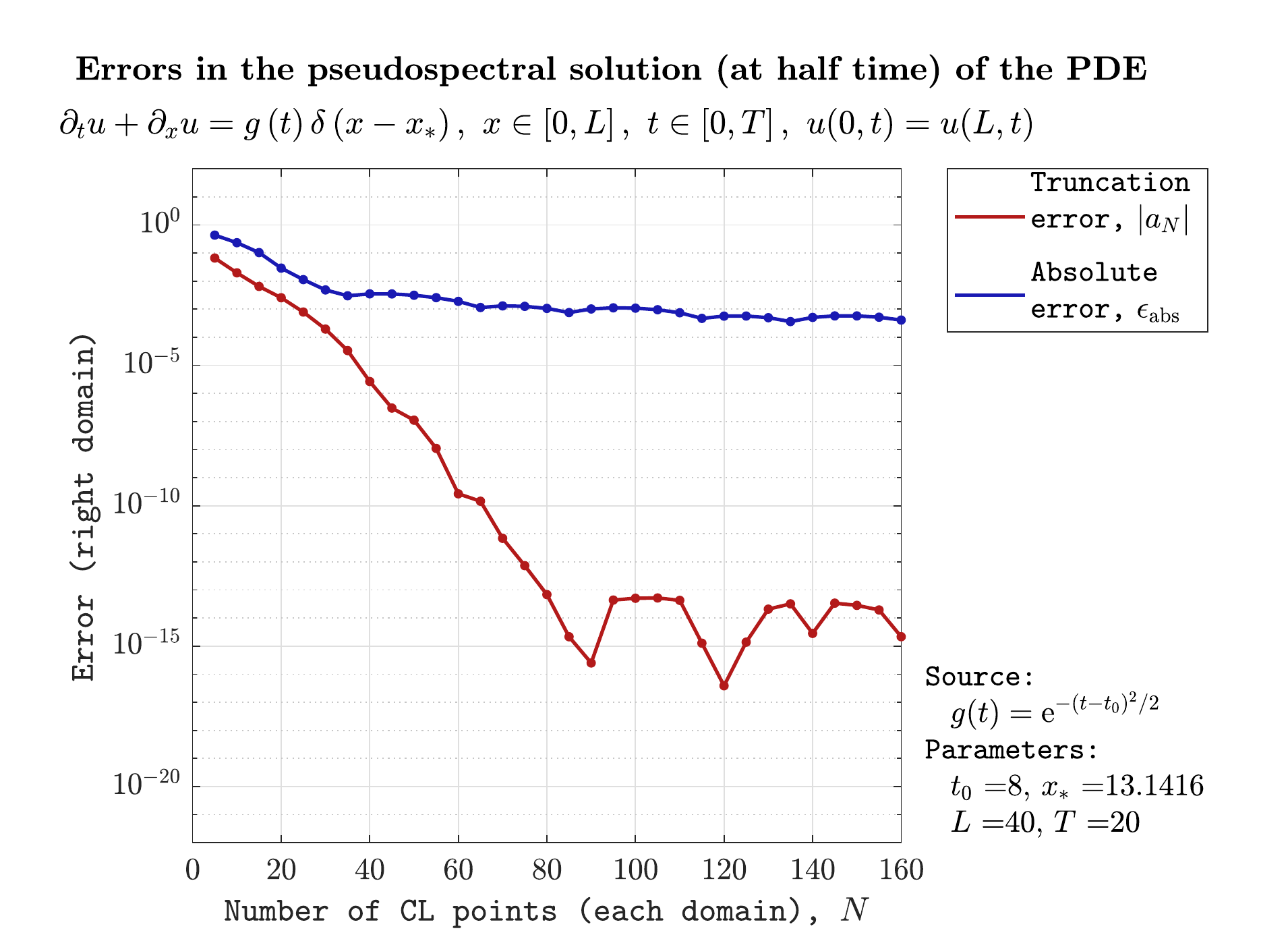}
\caption{Convergence of the numerical scheme for the problem (\ref{eq:advection_pde}).}\label{fig:pde_pwp_advection_error}
\end{center}
\end{figure}

The solution for zero initial data is displayed in Figure \ref{fig:pde_pwp_advection_soln}, and the numerical convergence in Figure \ref{fig:pde_pwp_advection_error}. For the latter, we plot---for the numerical solution $\bm{u}$
at $t=T/2$---both the absolute error (in the
$l^{2}$ norm on the CL grids, as in [\cite{petersson_discretizing_2016}]), $\epsilon_{{\text{abs}}}=||\bm{u}-\bm{u}_{{\text{ex}}}||_{2}$,
as well as the truncation error in the right CL domain $\mathscr{D}^{+}$
given simply the absolute value of the last spectral coefficient $a_{N}$
of $\bm{u}^{+}$. We see that the truncation error exhibits typical
(exponential) spectral convergence; the absolute error converges at
the same rate until $N\approx40$, after which it converges more slowly
because it becomes dominated by the $\mathcal{O}(\mathrm{\Delta} t)=\mathcal{O}(N^{-2})$
error in the finite difference time evolution scheme. Nevertheless, for the same number of grid points,
our procedure still yields a lower order of magnitude of the $l^{2}$ error as was obtained in  [\cite{petersson_discretizing_2016}] with a \emph{sixth} order finite difference scheme (relying on a a source discretization with 6 moment conditions and 6 smoothness conditions); we present a simple comparison of these in the following table:

\begin{center}
\begin{tabular}{c|c|c}
$\epsilon_{\text{abs}}$ & $N=80$ & $N=160$\tabularnewline
\hline 
[\cite{petersson_discretizing_2016}] & $\mathcal{O}(10^{-2})$ & $\mathcal{O}(10^{-3})$\tabularnewline
\hline 
PwP method & $\mathcal{O}(10^{-3})$ & $\mathcal{O}(10^{-4})$\tabularnewline
\end{tabular}
\par\end{center}

\subsection{Advection-type equations in neuroscience}
Advection-type equations with distributional sources arise in practice,
for example, in the modeling of neural populations. In particular,
among the simplest of these are the so-called ``integrate-and-fire''
models. For some of the earlier work on such models from a neuroscience
perspective, see for example [\cite{haskell_population_2001,casti_population_2002}]
and references therein; for more recent work focusing on mathematical
aspects, see [\cite{caceres_analysis_2011,caceres_blow-up_2016}]. Their
aim is to describe the probability density $\rho\left(\bm{v},t\right)$
of neurons as a function of certain state variables $\bm{v}$ and
time $t$. Often the detailed construction of these models can be
quite involved and dependent on a large number of parameters, so to
simply illustrate the principle of our method we here consider the
simple case where the single state variable is the voltage $V$. Then,
generally speaking, the dynamics of $\rho(V,t)$ takes the form of
a Fokker-Planck-type equation on $V\in(-\infty,L]$ with a singular
source at some fixed $V=V_{*}<L$,
\begin{equation}
\partial_{t}\rho+\partial_{V}\left(f\left(V,N\left(t\right)\right)\rho\right)-\frac{\sigma^{2}}{2}\partial_{V}^{2}\rho=N\left(t\right)\delta\left(V-V_{*}\right)\,.\label{eq:fokker-planck}
\end{equation}
The source time function $N(t)$ must be such that conservation
of probability, \textit{i.e.} $\partial_{t}\int{\rm d}V\rho=0$, is guaranteed
under homogeneous Dirichlet BCs.

As a simplification of this problem, let us suppose, as is sometimes done, that the diffusive
part (the second derivative term on the LHS) of (\ref{eq:fokker-planck})
is negligible. Moreover, in simple cases, the velocity function
$f$ in the advection term has the form $f=-V+{\rm constant}$, and
we just work with the constant set equal to $1$. We restrict ourselves
to a bounded domain for $V$ which for illustrative purposes we just
choose to be $\mathscr{I}=\left[0,L\right]$. Demanding homogeneous
Dirichlet BCs at the left boundary in conjunction with conservation
of probability fixes the source time function to be $N(t)=(1-L)\rho(L,t)$.
Thus, we are going to tackle the following problem: 
\begin{equation}
\begin{cases}
\partial_{t}\rho+\partial_{V}\left(\left(1-V\right)\rho\right)=\left(1-L\right)\rho\left(L,t\right)\delta\left(V-V_{*}\right)\,, & V\in\mathscr{I}=\left[0,L\right],t>0\,,\\
\rho\left(V,0\right)=\rho_{0}\left(V\right),\enskip\int_{\mathscr{I}}{\rm d}V\rho_{0}\left(V\right)=1\,, & \rho\left(0,t\right)=0\,.
\end{cases}\label{eq:neural_population}
\end{equation}

We now implement the PwP decomposition: $\rho=\rho^{-}\Theta^{-}+\rho^{+}\Theta^{+}$
with $\Theta^{\pm}=\Theta(\pm(V-V_{*}))$. Inserting this into the
PDE (\ref{eq:neural_population}), we get the homogeneous problems
$\partial_{t}\rho^{\pm}+\partial_{V}\left(\left(1-V\right)\rho^{\pm}\right)=0$
on $\mathscr{D}^{\pm}$, with $\mathscr{D}^{-}=[0,V_{*}]$ and $\mathscr{D}^{+}=[V_{*},L]$,
along with the JC $[\rho]_{*}=\frac{1-L}{1-V_{*}}\rho(L,t)$.

\begin{figure}
\begin{center}
\includegraphics[scale=0.6]{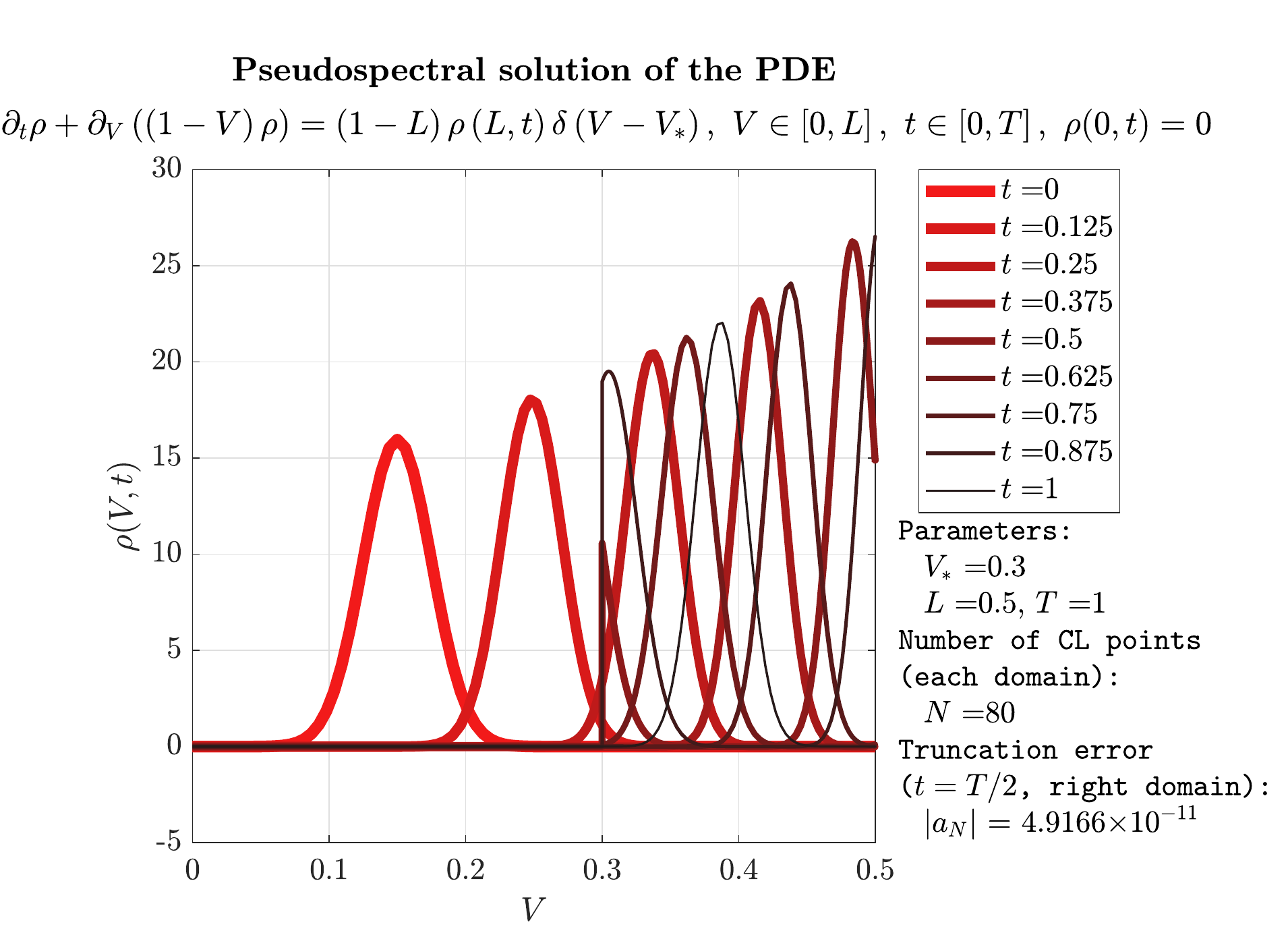}
\includegraphics[scale=0.6]{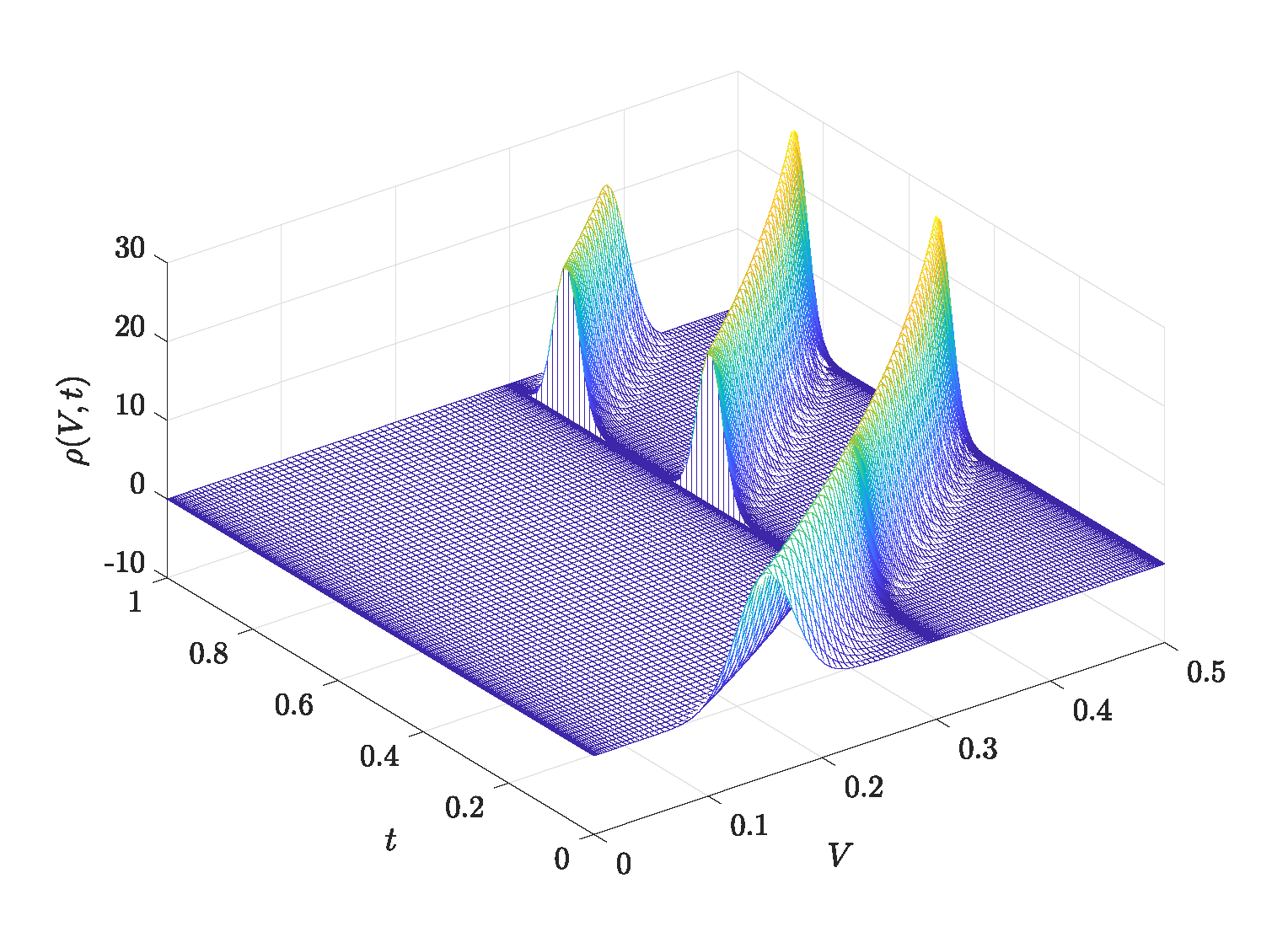}
\caption{Solution of (\ref{eq:neural_population}) with (normalized) Gaussian initial data centered at $V=0.3$.}\label{fig:pde_pwp_advection_neurosci_soln}
\end{center}
\end{figure}

\begin{figure}
\begin{center}
\includegraphics[scale=0.6]{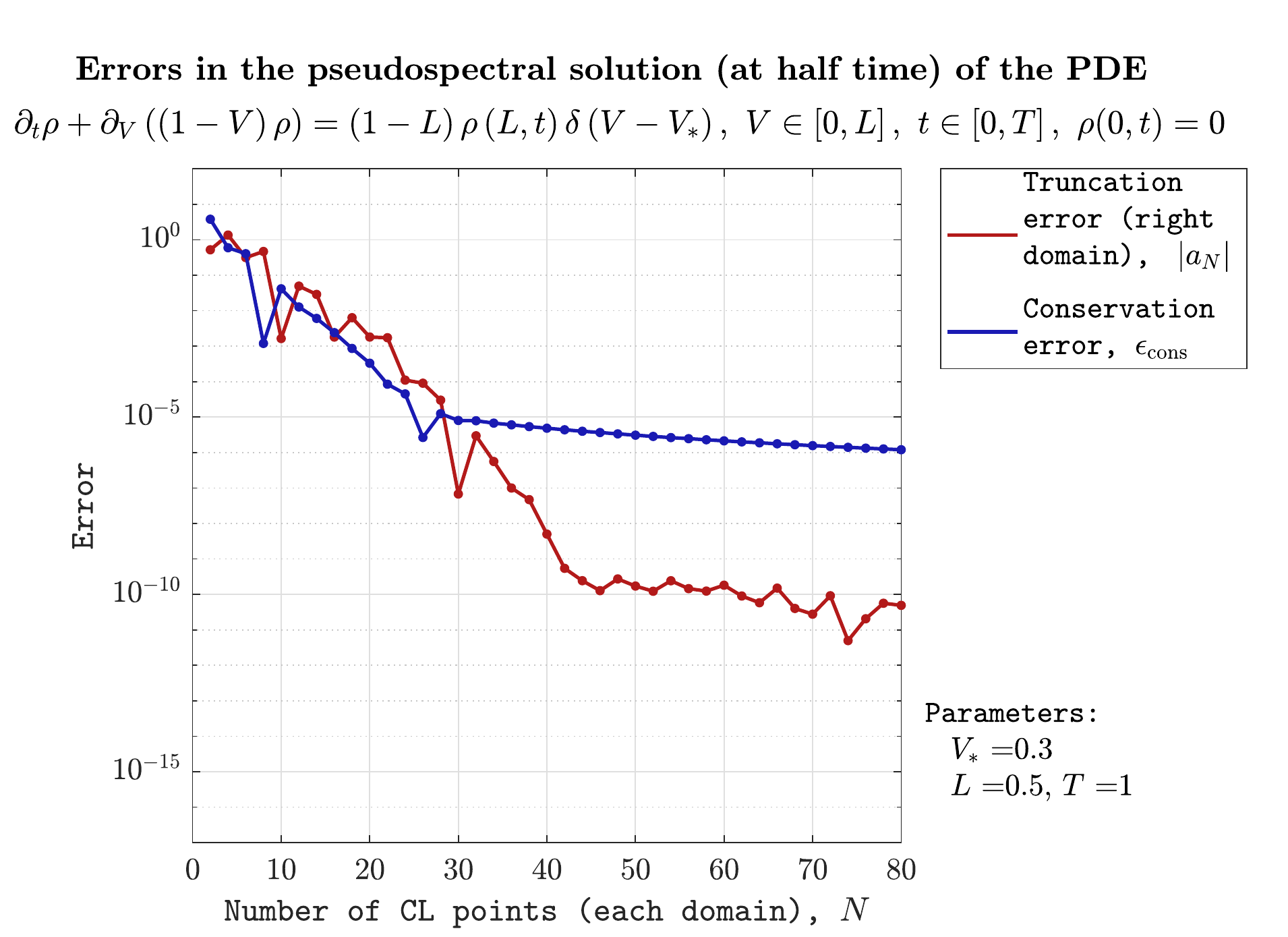}
\caption{Convergence of the numerical scheme for the problem (\ref{eq:neural_population}).}\label{fig:pde_pwp_advection_neurosci_error}
\end{center}
\end{figure}

An example solution for Gaussian initial data centered at $V=0.3$
is displayed in Figure \ref{fig:pde_pwp_advection_neurosci_soln},
and the numerical convergence in Figure \ref{fig:pde_pwp_advection_neurosci_error}.
In the latter, we plot---again for the numerical solution
$\bm{\rho}$ at the final time---the truncation error
as well as (in the absence of an exact solution) what we refer to as the conservation
error, $\epsilon_{{\text{cons}}}=|1-\int_{\mathscr{I}}{\rm d}V\rho(V,t)|$,
which simply measures how far we are from exact conservation of probability.
Both of these exhibit exponential convergence. The integral
in $\epsilon_{{\text{cons}}}$ is computed as a sum over both domains,
$\int_{\mathscr{I}}{\rm d}V\rho=\int_{\mathscr{D}^{-}}{\rm d}V\rho+\int_{\mathscr{D}^{+}}{\rm d}V\rho$,
and numerically performed on each using a standard pseudospectral
quadrature method (as in, \textit{e.g.}, Chapter 12 of  [\cite{trefethen_spectral_2001}]).

This procedure can readily be complexified with the inclusion of a
diffusion term, and indeed we will shortly turn to purely diffusion (heat-type equation)
problems in the following section.

\subsection{Advection-type equations in other applications}

Another advection-type application in which one may be tempted to
try applying some form the PwP method is the shallow water equations.
Setting the gravitational acceleration to $1$, these read:
\begin{equation}
\partial_{t}\left[\begin{array}{c}
h\\
hu
\end{array}\right]+\partial_{x}\left[\begin{array}{c}
hu\\
hu^{2}+\tfrac{1}{2}h^{2}
\end{array}\right]=\left[\begin{array}{c}
0\\
-h\partial_{x}B
\end{array}\right]\,,\label{eq:shallow_water}
\end{equation}
where $B(x)$ is the elevation of the bottom topography, $h(x,t)$
is the fluid depth above the bottom and $u(x,t)$ is the velocity.
If the topography is discontinuous, \textit{i.e.} if $B\notin C^{0}(\mathbb{R})$,
then the RHS of (\ref{eq:shallow_water}) will be distributional;
this can happen, \textit{e.g.}, if the bottom is a step (if $B=\Theta$, then
the RHS is $\icol{0\\-h\delta}$), a wall \textit{etc.} However,
the problem with applying the PwP method here is that (\ref{eq:shallow_water})
is nonlinear, and so one encounters precisely the sorts of issues
detailed at the end of the preceding section. Indeed, explicit numerical
solutions that have been obtained for (\ref{eq:shallow_water}) in
the literature [\cite{zhou_numerical_2002,bernstein_central-upwind_2016}] qualitatively indicate that a PwP-type decomposition as described
here would be inadequate (and, anyway, nonsensical mathematically)
for such problems.

\section{Parabolic PDEs}
\label{parabolic-pdes}

We begin by analyzing the standard heat equation and then move on
to an application in finance which includes two (time-dependent) singular
source terms. 

\subsection{Heat equation}

Let us consider now the $(1+1)$-dimensional heat equation for $u(x,t)$
with a constant point source at a time-dependent location $x=x_{p}(t)$,
with Dirichlet boundary conditions:
\begin{equation}
\begin{cases}
\partial_{t}u-\partial_{x}^{2}u=\lambda\delta\left(x-x_{p}\left(t\right)\right)\,, & x\in\mathscr{I}=\left[a,b\right],\enskip t>0\,,\\
u\left(x,0\right)=0\,, & u\left(a,t\right)=\alpha,\,u\left(b,t\right)=\beta\,.
\end{cases}\label{eq:heat_pde}
\end{equation}
In this case, we do not have the exact solution.

This problem is treated in [\cite{tornberg_numerical_2004}] using a delta function approximation procedure,
with the following setup: $\mathscr{I}=[0,1]$, $\alpha=0=\beta$ and $\lambda=10$; constant-valued and sinusoidal point source locations $x_{p}(t)$ are considered. We implement here the same, using our PwP method: we decompose $u=u^{-}\Theta^{-}+u^{+}\Theta^{+}$
where $\Theta^{\pm}=\Theta(\pm(x-x_{p}(t)))$. Inserting this into
(\ref{eq:heat_pde}), we get homogeneous PDEs $\partial_{t}u^{\pm}-\partial_{x}^{2}u^{\pm}=0$
to the left and right of the singularity, $x\in\mathscr{D}^{-}=[0,x_{p}(t)]$
and $x\in\mathscr{D}^{+}=[x_{p}(t),1]$ respectively; additionally,
we have the following JCs: $[u]_{p}=0$ and $[\partial_{x}u]_{p}=-\lambda$.

\begin{figure}
\begin{center}
\includegraphics[scale=0.6]{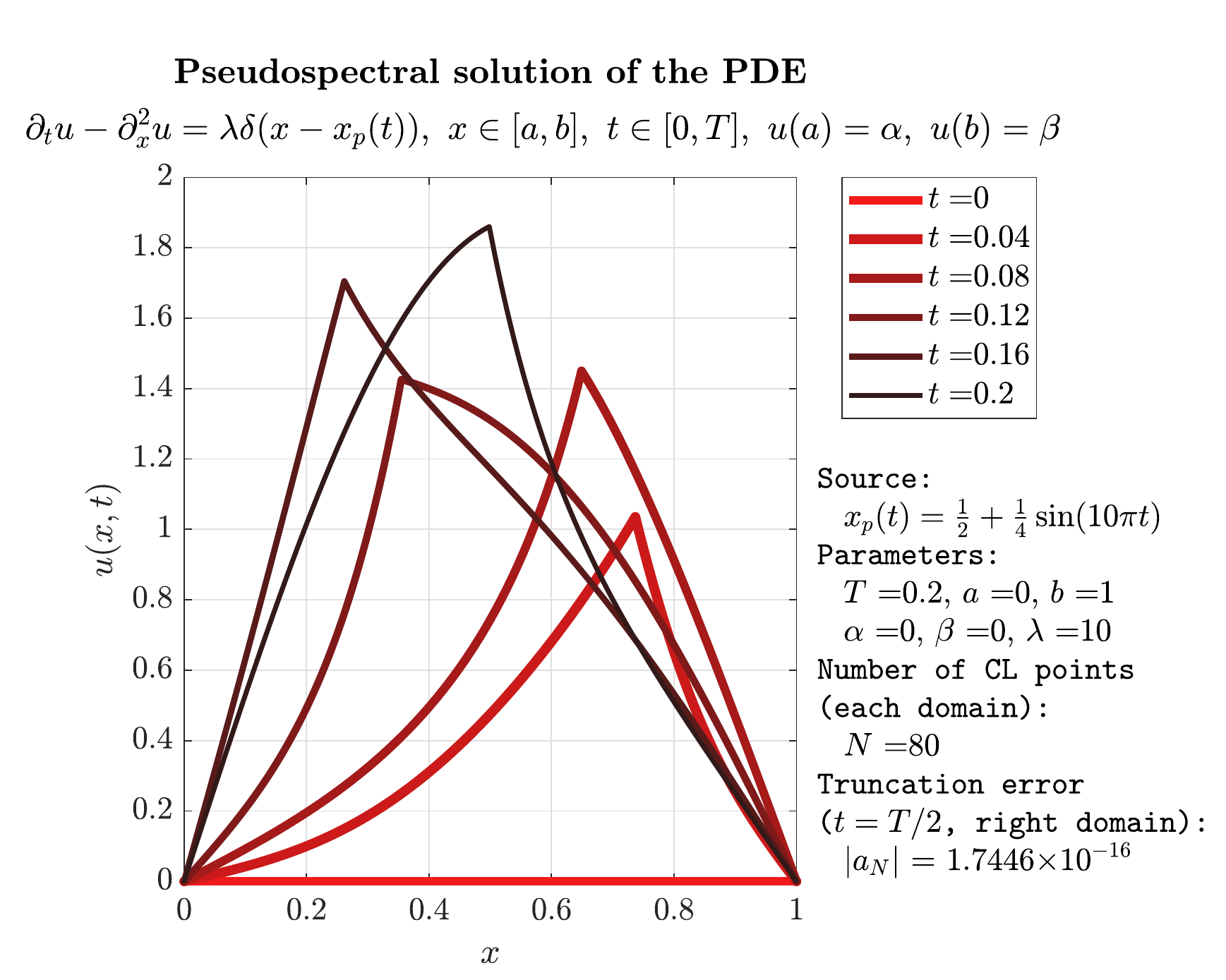}
\includegraphics[scale=0.6]{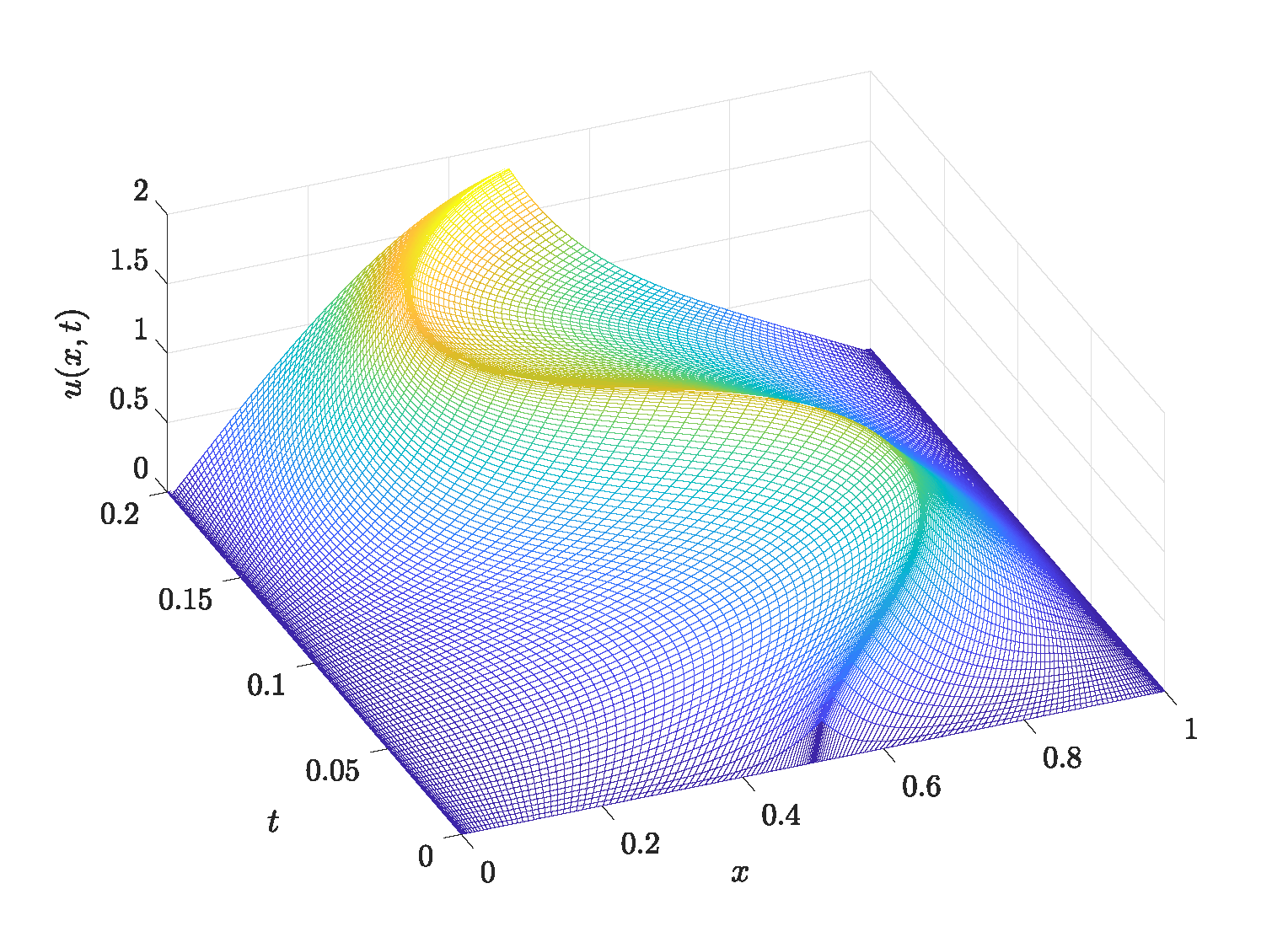}
\caption{Solution of the problem (\ref{eq:heat_pde}) with zero initial data.}\label{fig:pde_pwp_heat_soln}
\end{center}
\end{figure}

\begin{figure}
\begin{center}
\includegraphics[scale=0.6]{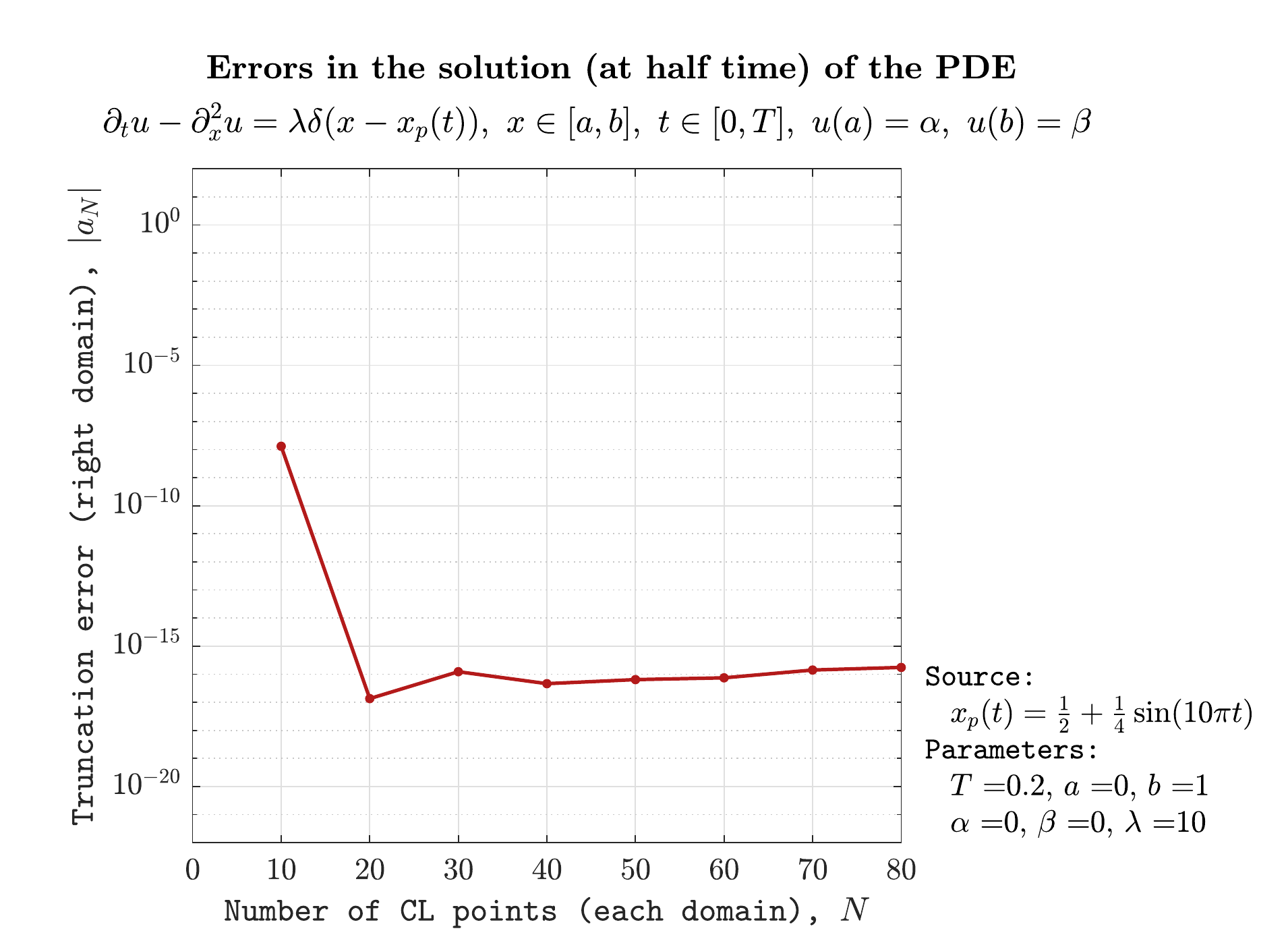}
\caption{Convergence of the numerical scheme for the problem (\ref{eq:heat_pde}).}\label{fig:pde_pwp_heat_error}
\end{center}
\end{figure}

The details of the numerical scheme are given in Subsection \ref{a-numerical-parabolic}, and results for zero initial data in Figures \ref{fig:pde_pwp_heat_soln} and \ref{fig:pde_pwp_heat_error}.

\subsection{Heat-type equations in finance}
We consider a model of price formation initially proposed in  [\cite{lasry_mean_2007}];
see also [\cite{markowich_parabolic_2009,caffarelli_price_2011,burger_boltzmann-type_2013,achdou_partial_2014,pietschmann_partial_2012}]. This model describes
the density of buyers $f_{{\text B}}(x,t)$ and the density of vendors
$f_{{\text V}}(x,t)$ in a system, as functions of of the bid or, respectively,
ask price $x\in\mathbb{R}$ for a certain good being traded between
them, and time $t\in[0,\infty)$.

The idea is that when a buyer and vendor agree on a price, the transaction
takes place; the buyer then becomes a vendor, and vice-versa. However,
it is also assumed that there exists a fixed transaction fee $a\in\mathbb{R}$.
Consequently, the actual buying price is $x+a$, and so the (former)
buyer will try to sell the good at the next trading event not for
the price $x$, but for $x+a$. Similarly, the profit for the vendor
is actually $x-a$, and so he/she would not be willing to pay more
than $x-a$ for the good at the next trading event. In time, this
system should achieve an equilibrium.

Mathematically, the dynamics of the buyer/vendor densities is assumed
to be governed by the heat equation with a certain source term. The
source term in each case is simply the (time-dependent) transaction
rate $\lambda(t)$, corresponding to the flux of buyers and vendors,
at the particular price where the trading event occurs, shifted accordingly
by the transaction cost. Thus the system is described by
\begin{equation}
\begin{cases}
\left(\partial_{t}-\partial_{x}^{2}\right)f_{{\text B}}=\lambda\left(t\right)\delta\left(x-\left(x_{p}\left(t\right)-a\right)\right)\,, & {\rm for}\,\,x<x_{p}\left(t\right)\,,\\
f_{{\text B}}=0\,, & {\rm for}\,\,x>x_{p}\left(t\right)\,,
\end{cases}
\end{equation}
and
\begin{equation}
\begin{cases}
\left(\partial_{t}-\partial_{x}^{2}\right)f_{{\text V}}=\lambda\left(t\right)\delta\left(x-\left(x_{p}\left(t\right)+a\right)\right)\,, & {\rm for}\,\,x>x_{p}\left(t\right)\,,\\
f_{{\text V}}=0\,, & {\rm for}\,\,x<x_{p}\left(t\right)\,,
\end{cases}
\end{equation}
where the free boundary $x_{p}(t)$ represents the agreed price of
trading at time $t$, and the transaction rate is $\lambda(t)=-\partial_{x}f_{{\text B}}(x_{p}(t),t)=\partial_{x}f_{{\text V}}(x_{p}(t),t)$.
(NB: The functional form of $\lambda(t)$ is uniquely fixed simply by the
requirement that the two densities are conserved, \textit{i.e.} $\partial_{t}\int{\rm d}xf_{{\text B}}=0=\partial_{t}\int{\rm d}xf_{{\text V}}$,
under the assumption that we have homogeneous Neumann BCs at the left
and right boundaries respectively.) Now, we can actually combine this
system into a single problem for the difference between buyer and
vendor densities, 
\begin{equation}
f=f_{{\text B}}\Theta\left(-\left(x-x_{p}\left(t\right)\right)\right)-f_{{\text V}}\Theta\left(x-x_{p}\left(t\right)\right)\,.
\end{equation}
The ``spatial'' (\textit{i.e.} price) domain can be taken to be bounded,
and homogeneous Neumann BCs are assumed at the boundaries. Thus the problem we are interested in is: 
\begin{equation}
\begin{cases}
\partial_{t}f-\partial_{x}^{2}f=\lambda\left(t\right)\left(\delta\left(x-x_{p_{-}}\left(t\right)\right)-\delta\left(x-x_{p_{+}}\left(t\right)\right)\right)\,, & x\in\mathscr{I}=\left[0,1\right],\enskip t>0\,,\\
f\left(x,0\right)=f_{\text{I}}\left(x\right),\enskip f\gtrless0\enskip{\rm for}\enskip x\lessgtr x_{p}\left(t\right)\,, & \partial_{x}f\left(0,t\right)=\partial_{x}f\left(1,t\right)=0\,.
\end{cases}\label{eq:pde_finance_model}
\end{equation}
where $\lambda(t)=-\partial_{x}f(x_{p}(t),t),$ and we have defined
$x_{p_{\pm}}(t)=x_{p}(t)\pm a$. Moreover, one can show that from
this setup, it follows that the free boundary evolves via
\begin{equation}
\dot{x}_{p}\left(t\right)=\frac{\partial_{x}^{2}f\left(x_{p}\left(t\right),t\right)}{\lambda\left(t\right)}\,.\label{eq:pde_finance_price_evolution}
\end{equation}

In this case, we have not one but two singular source locations on
the RHS of the PDE. Hence, in order to implement the PwP method, we
must here divide the spatial domain $\mathscr{I}$ into three disjoint
regions, with the two singularity locations at their interfaces: $\mathscr{I}=\mathscr{D}^{-}\cup\mathscr{D}^{0}\cup\mathscr{D}^{+}$
with $\mathscr{D}^{-}=[0,x_{p_{-}}(t)]$, $\mathscr{D}^{0}=[x_{p_{-}}(t),x_{p_{+}}(t)]$
and $\mathscr{D}^{+}=[x_{p_{+}}(t),1]$. Then, we decompose $f=f^{-}\Theta^{-}+f^{0}\Theta^{0}+f^{+}\Theta^{+}$
with $\Theta^{-}=\Theta(-(x-x_{p_{-}}(t)))$, $\Theta^{0}=\Theta(x-x_{p_{-}}(t))-\Theta(x-x_{p_{+}}(t))$
and $\Theta^{+}=\Theta(x-x_{p_{+}}(t))$. Inserting this into the
PDE (\ref{eq:pde_finance_model}), we get homogeneous problems $(\partial_{t}-\partial_{x}^{2})f^{\sigma}=0$
on $\mathscr{D}^{\sigma}$ for $\sigma\in\{0,\pm\}$, along with the
JCs $[f]_{p_{\pm}}=0$ and $[\partial_{x}f]_{p_{\pm}}=\pm\lambda(t)$.

Before proceeding to the numerical implementation, we note that it
is possible to derive an exact stationary (\textit{i.e.} $t\rightarrow\infty$)
solution of the problem (\ref{eq:pde_finance_model}). In particular,
denoting the (time-conserved) number of buyers and vendors, respectively,
by $N_{{\text B}}=\int_{0}^{x_{p}}{\rm d}x\:f$ and $N_{{\text V}}=-\int_{x_{p}}^{1}{\rm d}x\:f$,
one can show that in the stationary ($t\rightarrow\infty$) limit,
\begin{align}
 & \begin{cases}
N_{{\text B}}=\, & -\lambda^{\text{stat}}a(x_{p}^{\text{stat}}-a/2)\,,\\
N_{{\text V}}=\, & -\lambda^{\text{stat}}a(1-x_{p}^{\text{stat}}-a/2)\,,
\end{cases}\\
\Leftrightarrow\, & \begin{cases}
\lambda^{\text{stat}}=\, & \left[-\left(N_{{\text B}}+N_{{\text V}}\right)\right]/\left[a\left(1-a\right)\right]\,,\\
x_{p}^{\text{stat}}=\, & \left[2N_{{\text B}}+a\left(N_{{\text V}}-N_{{\text B}}\right)\right]/\left[2\left(N_{{\text B}}+N_{{\text V}}\right)\right]\,,
\end{cases}\label{eq:pde_finance_stationary_lambda_price}
\end{align}
which we can use to determine the exact stationary solution
\begin{equation}
\lim_{t\rightarrow\infty}f\left(x,t\right)=f^{\text{stat}}\left(x\right)=\begin{cases}
-\lambda^{\text{stat}}a\,, & {\rm for}\,\,0\leq x<x_{p_{-}}^{\text{stat}}\,,\\
\lambda^{\text{stat}}\left(x-x_{p}^{\text{stat}}\right)\,, & {\rm for}\,\,x_{p_{-}}^{\text{stat}}\leq x\leq x_{p_{+}}^{\text{stat}}\,,\\
\lambda^{\text{stat}}a\,, & {\rm for}\,\,x_{p_{+}}^{\text{stat}}<x\leq1\,.\label{eq:pde_finance_stationary_solution}
\end{cases}
\end{equation}

The problem (\ref{eq:pde_finance_model}) is solved numerically in  [\cite{markowich_parabolic_2009}]
(see also section 2.5.2 of [\cite{pietschmann_partial_2012}]) using (Gaussian) delta function approximations for the source on an equispaced computational grid. We implement here using our PwP method the exact same setup: in particular, we take
a transaction fee of $a=0.1$ and initial data $f_{\text{I}}(x)=\tfrac{875}{6}x^{3}-\tfrac{700}{3}x^{2}+\tfrac{175}{2}x$.
(NB: Despite the fact that this does not actually satisfy homogeneous
Neumann BCs, the numerical evolution will force it to.) Analytically,
we have $x_{p}(0)=\tfrac{3}{5}$ and $\lambda(0)=35$. Also, using (\ref{eq:pde_finance_stationary_lambda_price}), we have $\lambda^{\text{stat}}=-\frac{8855}{162}$
and $x_{p}^{\text{stat}}=\frac{731}{1012}\approx0.7223$. As we evolve
forward in time, we use Chebyshev polynomial interpolation to determine
the transaction rate $\lambda(t)$ (\textit{i.e.} the negative of the spatial
derivative of the solution at $x_{p}(t)$) as well as the evolution
of $x_{p}(t)$ via (\ref{eq:pde_finance_price_evolution}).

\begin{figure}
\begin{center}
\includegraphics[scale=0.6]{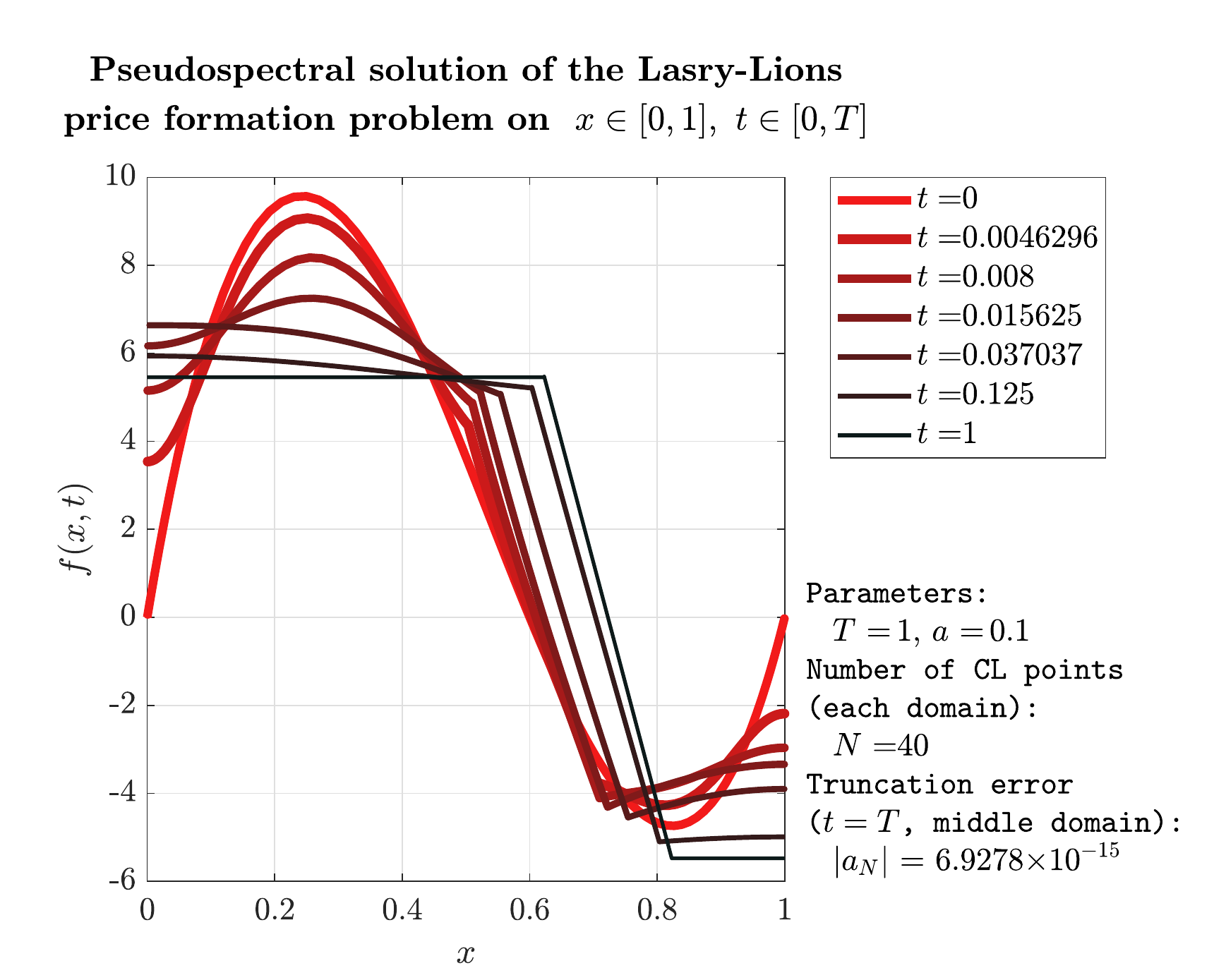}
\includegraphics[scale=0.6]{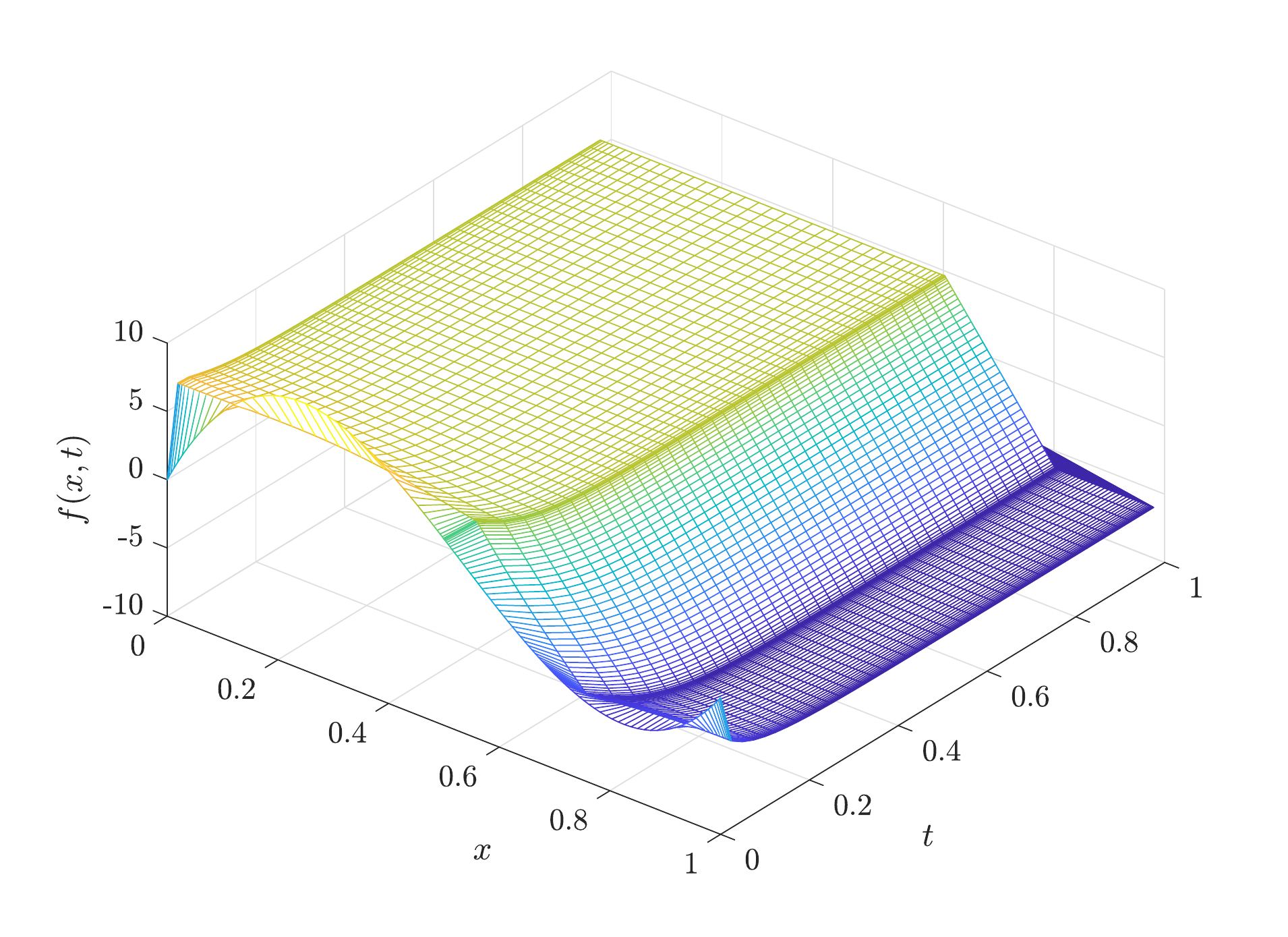}
\caption{Solution of the problem (\ref{eq:pde_finance_model}).}\label{fig:pde_pwp_price_formation_soln}
\end{center}
\end{figure}

\begin{figure}
\begin{center}
\includegraphics[scale=0.6]{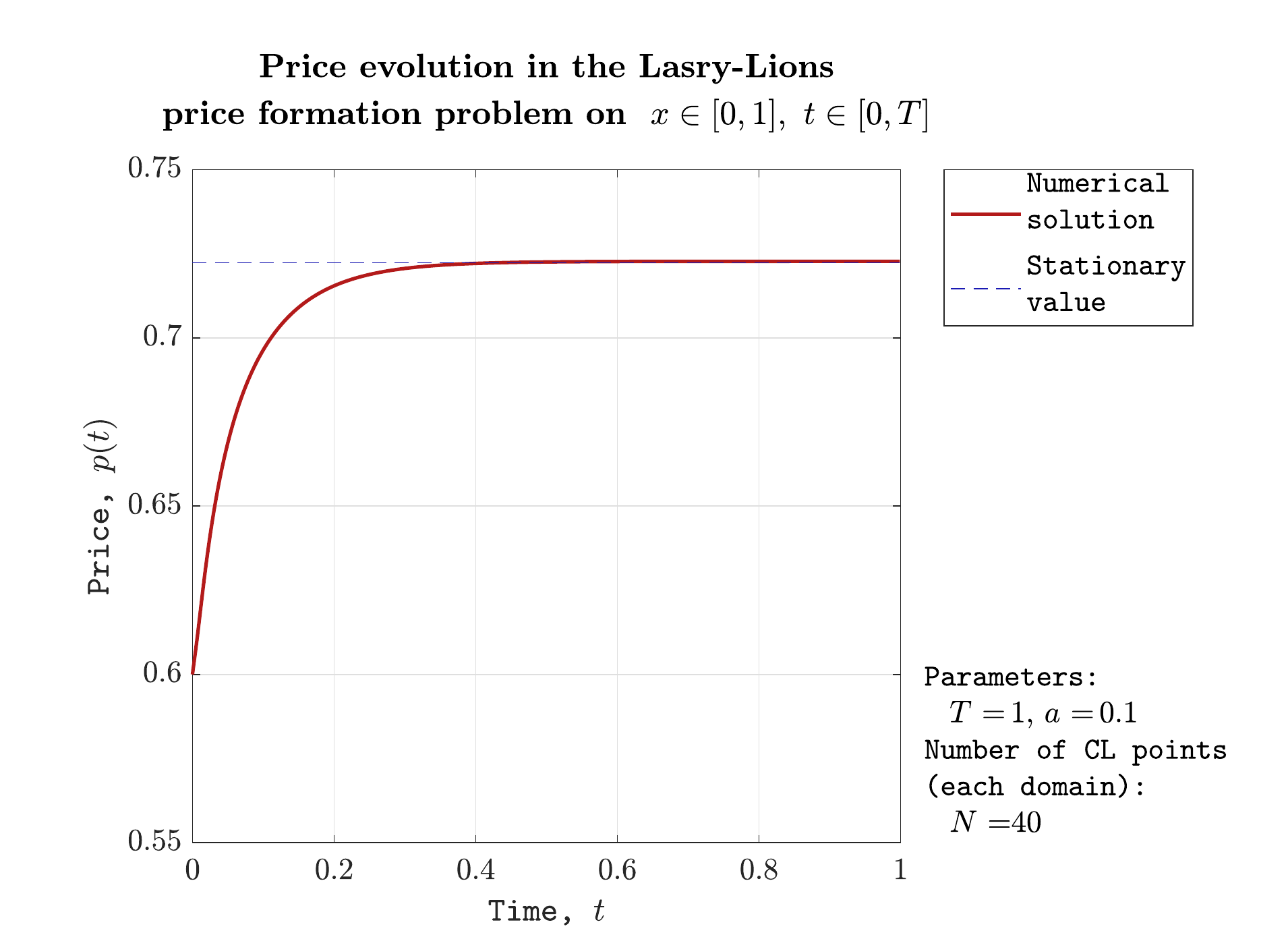}
\includegraphics[scale=0.6]{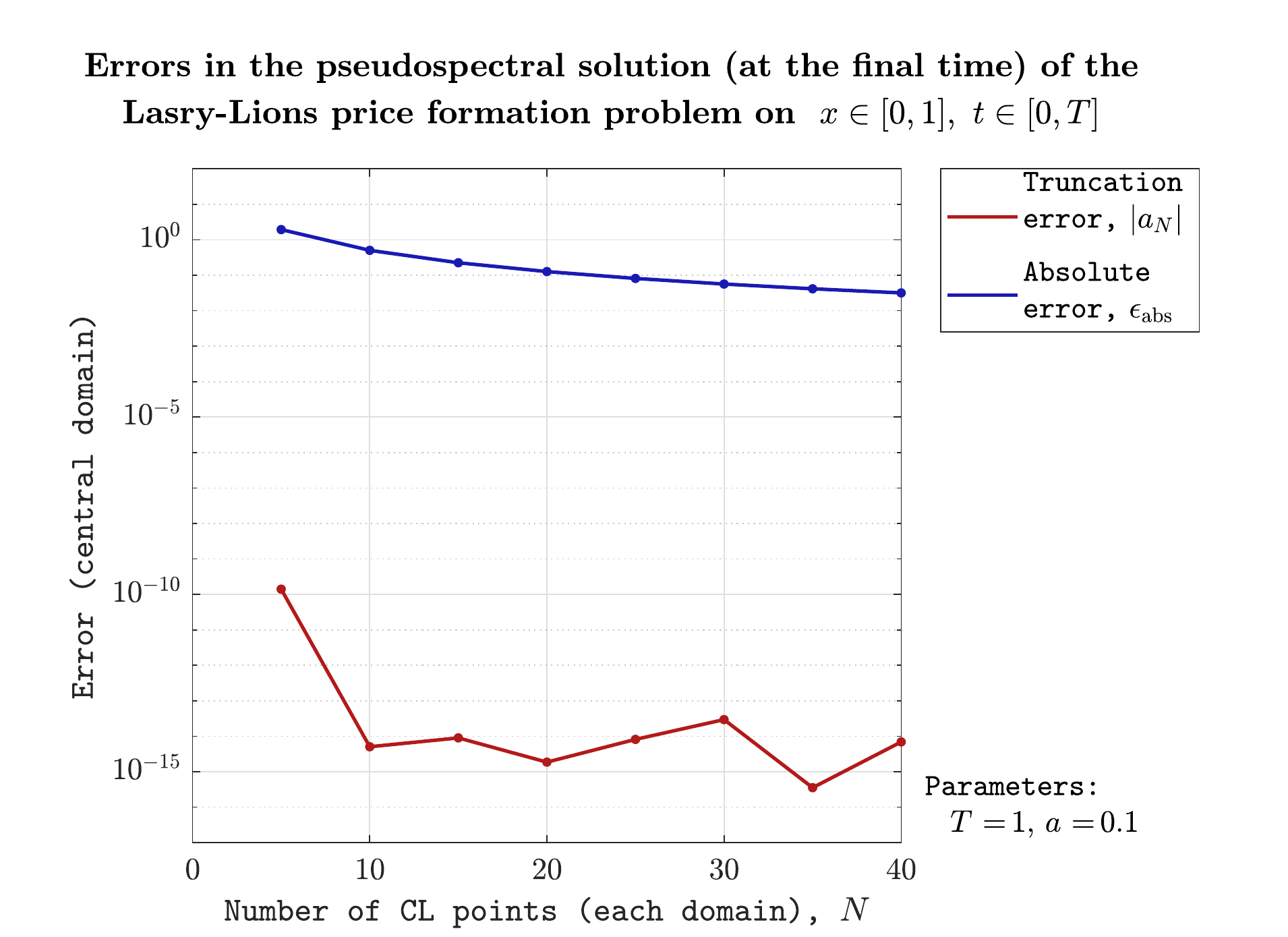}
\caption{Price evolution and convergence of the numerical scheme for the problem (\ref{eq:pde_finance_model}).}\label{fig:pde_pwp_price_formation_price_and_error}
\end{center}
\end{figure}

The numerical scheme is given in Subsection \ref{a-numerical-parabolic},
and results in Figures \ref{fig:pde_pwp_price_formation_soln} and \ref{fig:pde_pwp_price_formation_price_and_error}. In particular, in Figure \ref{fig:pde_pwp_price_formation_soln} we show the numerical solution for $f$, and in Figure \ref{fig:pde_pwp_price_formation_price_and_error}, the price as a function of time as well as the numerical convergence rates. For the latter, we plot not only the truncation error but also the absolute error with the stationary solution (\ref{eq:pde_finance_stationary_solution}), in this case, using the infinity norm: $\epsilon_{\text{abs}}=||\bm{f}-\bm{f}^{\text{stat}}||_{\infty}$. Of course, since we can only evolve the solution up to a finite time (which we choose to be $t=T=1$), we should not expect this to converge to zero; however, its decline with increasing $N$ nevertheless serves to illustrate a good validation of our results.

We remark that our numerical implementation here not only requires an order of magnitude fewer grid points than that of  [\cite{markowich_parabolic_2009}], but in fact yields convergence to the \emph{correct} stationary solution while that of  [\cite{markowich_parabolic_2009}] \emph{does not}. Indeed, in the latter, not only are more points required (essentially due to the necessity of resolving well enough the Gaussian-approximated delta functions) but the scheme actually fails, even so, to approach (\ref{eq:pde_finance_stationary_solution}) as well as ours by the same finite time, $t=T=1$. (To wit, [\cite{markowich_parabolic_2009}] obtain $x_{p}\rightarrow0.71$ in the large $t$ limit, instead of the correct value, $0.7223$, which we achieve with our PwP method as shown in Figure \ref{fig:pde_pwp_price_formation_price_and_error}.)

\section{Second order hyperbolic PDEs}
\label{second-order-hyperbolic-pdes}

We move on to consider in this section second order hyperbolic problems.
In particular, we first solve the standard $(1+1)$-dimensional elastic
wave equation, taking a delta derivative source. Afterwards, we discuss
possible physical applications of this and obstacles thereto---including
problems in gravitational physics and seismology. 

\subsection{Wave equation}

Let us consider the the elastic wave equation, in the form of the
following simplified $(1+1)$-dimensional problem for $u(x,t)$ with
a delta function derivative source at a fixed point $x_{*}\in\mathscr{I}=[0,L]$,
and homogeneous Dirichlet boundary conditions:
\begin{equation}
\begin{cases}
\partial_{t}^{2}u-\partial_{x}^{2}u=g\left(t\right)\delta'\left(x-x_{*}\right)\,, & x\in\mathscr{I}=\left[0,L\right]\,,\enskip t>0\,,\\
u\left(x,0\right)=0\,,\enskip\partial_{t}u\left(x,0\right)=0\,, & u\left(0,t\right)=0=u\left(L,t\right)\,.
\end{cases}\label{eq:elastic_wave_pde}
\end{equation}

It is actually possible to derive an exact solution for this problem
on an unbounded domain $\mathscr{I}=\mathbb{R}$. For the interested
reader, the procedure is explained in Appendix D of [\cite{oltean_particle-without-particle:_2019}].
For concreteness we take a simple sinusoidal source time function
$g(t)=\kappa\sin(\omega t)$, in which case the exact solution reads:
\begin{align}
u_{{\text{ex}}}\left(x,t\right)=\, & \kappa\bigg[\frac{1}{4}\sum_{\sigma=\pm}\sigma{\rm sgn}\left(x-x_{*}+\sigma t\right)\sin\left(\omega\left(x-x_{*}+\sigma t\right)\right)\nonumber \\
 & -\frac{1}{2}{\rm sgn}\left(x-x_{*}\right)\cos\left(\omega\left(x-x_{*}\right)\right)\sin\left(\omega t\right)\bigg]\,,\label{eq:elastic_wave_solution_exact}
\end{align}
where ${\rm sgn}(\cdot)$ is the sign function, with the property
${\rm d}({\rm sgn}(x))/{\rm d}x=2{\rm d}\Theta(x)/{\rm d}x=2\delta(x)$.

To solve (\ref{eq:elastic_wave_pde}) numerically, we implement the
now familiar PwP decomposition: $u=u^{-}\Theta^{-}+u^{+}\Theta^{+}$
where $\Theta^{\pm}=\Theta(\pm(x-x_{*}))$. Inserting this into (\ref{eq:elastic_wave_pde}),
we get homogeneous PDEs $\partial_{t}^{2}u^{\pm}-\partial_{x}^{2}u^{\pm}=0$
to the left and right of the singularity, $x\in\mathscr{D}^{-}=[0,x_{*}]$
and $x\in\mathscr{D}^{+}=[x_{*},L]$ respectively, along with the
JCs $[u]_{p}=-g(t)$ and $[\partial_{x}u]_{p}=0$. We now proceed
by recasting (\ref{eq:elastic_wave_pde}) as a first-order hyperbolic
system for $\vec{U}=[u\enskip v\enskip w]^{{\rm T}}$ with $v=\partial_{x}u$
and $w=\partial_{t}u$, as
\begin{equation}
\partial_{t}\vec{U}=\left[\begin{array}{ccc}
0 & 0 & 0\\
0 & 0 & 1\\
0 & 1 & 0
\end{array}\right]\partial_{x}\vec{U}+\left[\begin{array}{ccc}
0 & 0 & 1\\
0 & 0 & 0\\
0 & 0 & 0
\end{array}\right]\vec{U}\quad{\rm on}\enskip\mathscr{D}^{\pm}\,,\quad\left[\vec{U}\right]_{p}=\left[\begin{array}{c}
-g\\
0\\
-\dot{g}
\end{array}\right]\,.\label{eq:elastic_wave_pde_first-order}
\end{equation}

\begin{figure}
\begin{center}
\includegraphics[scale=0.6]{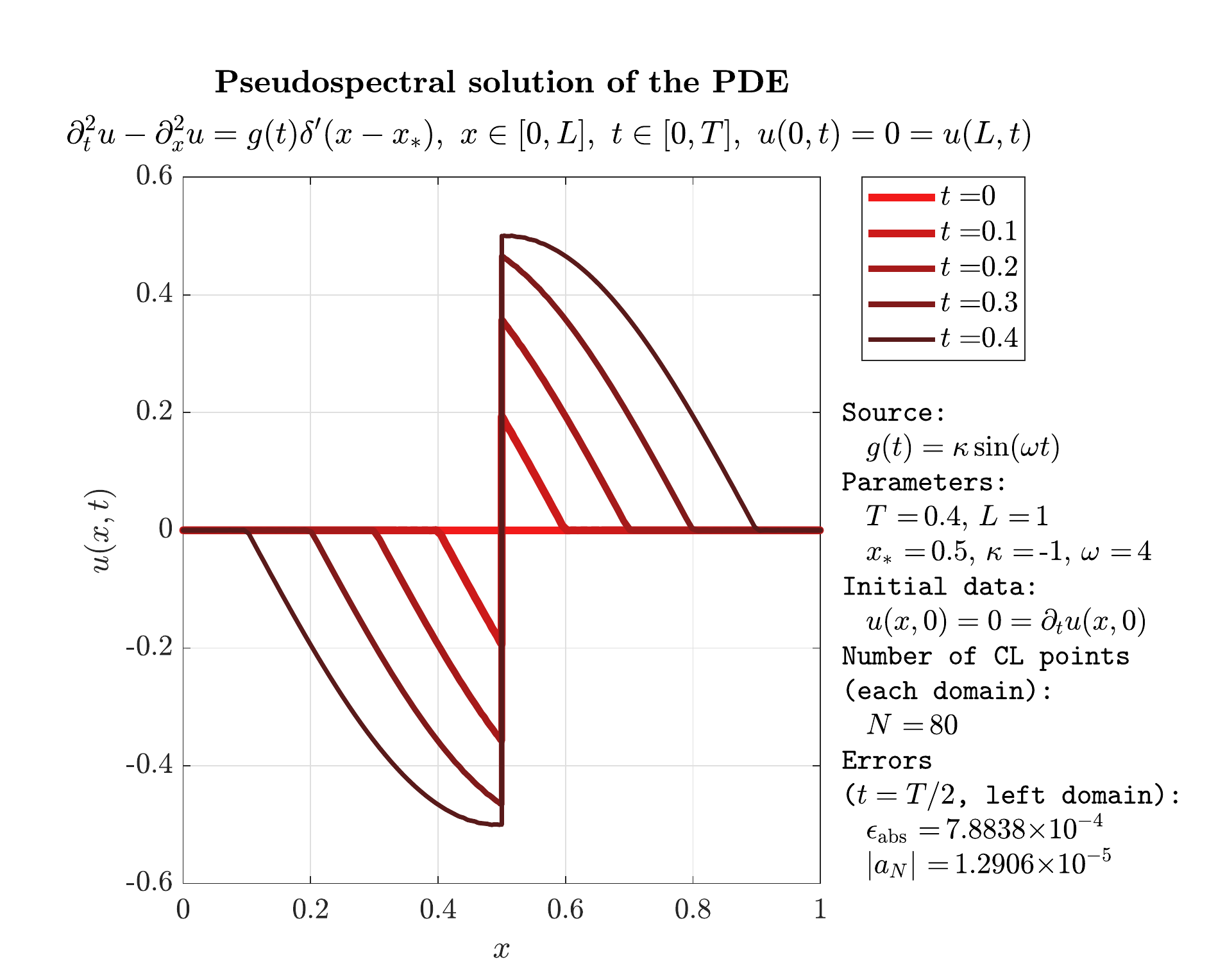}
\includegraphics[scale=0.6]{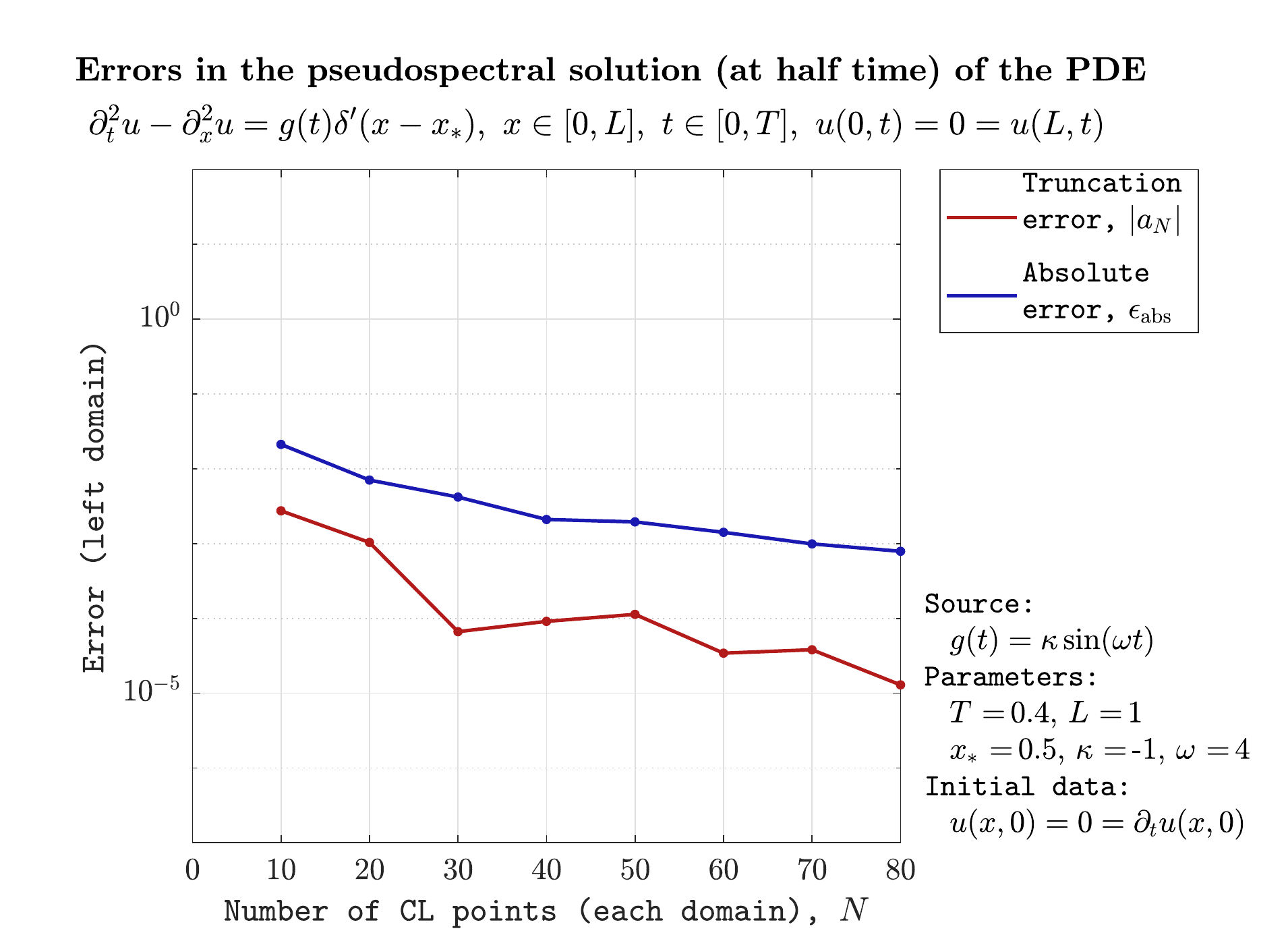}
\caption{Solution and convergence of the numerical scheme for the problem (\ref{eq:elastic_wave_pde}).}\label{fig:pde_pwp_wave_soln_and_error}
\end{center}
\end{figure}

\begin{figure}
\begin{center}
\includegraphics[scale=0.6]{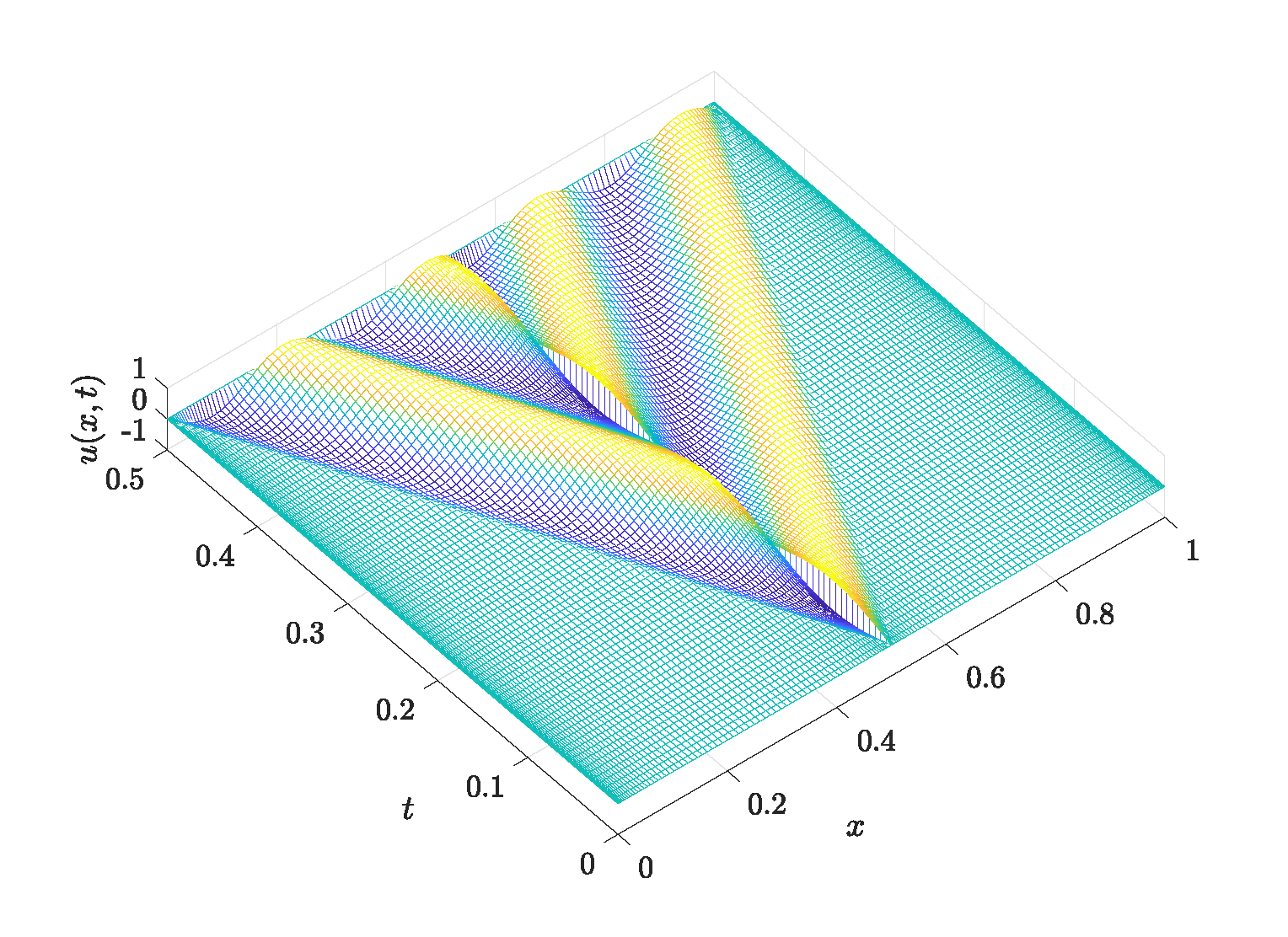}
\caption{Solution (with $N=80$) of the same problem as in Figure \ref{fig:pde_pwp_wave_soln_and_error} but using $\omega=24$.}\label{fig:pde_pwp_wave_soln3d}
\end{center}
\end{figure}

The numerical scheme is given in Subsection \ref{a-numerical-second-order-hyperbolic},
and results in Figures \ref{fig:pde_pwp_wave_soln_and_error} and \ref{fig:pde_pwp_wave_soln3d}. The absolute error is again computed in the infinity norm on the CL grids: $\epsilon_{\text{abs}}=||\bm{u}-\bm{u}_{\text{ex}}||_{\infty}$.

The same problem (\ref{eq:elastic_wave_pde}) is considered numerically in [\cite{petersson_stable_2010}], but using a different (polynomial) source function $g(t)$, and a discretization procedure for the delta function (derivatives) on the computational grid (carried out in such a way that the distributional action thereof yields the expected result on polynomials up to a given degree). With our PwP method here, we obtain the same order of magnitude of the (absolute) error in the numerical solution as that in  [\cite{petersson_stable_2010}] for the same (order of magnitude of) number of grid points; however the drawback of the ``discretized delta'' method of [\cite{petersson_stable_2010}], in contrast to the PwP method, is that the solution in the former is visibly quite poorly resolved close to the singularity.

We add that we have also carried out the solution to the problem shown in Figure \ref{fig:pde_pwp_wave_soln_and_error} using higher-order (from second up to eighth order) finite-difference time evolution schemes. These yield no visible improvement (at any order tried) in either the absolute or the truncation error relative to the first-order time evolution results. Thus the spacial pseudospectral grid appears to control the total level of the error, with a higher-order scheme for the time evolution producing, at least in this case, no greater benefits.

\subsection{Wave-type equations in physical applications}

As we have mentioned, our numerical
studies in this appendix are largely motivated by their applicability
to the computation of the self-force in gravitational physics. There
one encounters different levels of complexity of this problem, the
simplest being that of the self-force due to a scalar field---as
a conceptual testbed for the more complicated and realistic problem
of the full self-force of the gravitational field---in a fixed (non-dynamical) black
hole spacetime. This can refer to a non-spinning (Schwarzschild-Droste) black
hole, where the problem is of the form (\ref{eq:intro_general_pde})
where $\mathcal{L}=\partial_{t}^{2}-\partial_{x}^{2}+V$ is just a
simple $(1+1)$-dimensional wave operator with some known potential
$V$ and a source $S=f\delta_{(p)}$, with $\dim(\mathscr{I})=1$.
We recognize this now as quite typical for the application of our PwP
method, and indeed this has been done with success in the past [\cite{canizares_extreme-mass-ratio_2011,canizares_efficient_2009,canizares_pseudospectral_2010,canizares_time-domain_2011,canizares_tuning_2011,jaramillo_are_2011,canizares_overcoming_2014,oltean_frequency-domain_2017}]. As we briefly remarked in the Introduction, the main difference between most of these works and our numerical schemes throughout this appendix is that for the time evolution, rather than relying on finite-difference methods, the former made use of the method of lines. This can be quite well-suited especially for these types of $(1+1)$-dimensional hyperbolic problems, which can be formulated in terms of characteristic fields propagating along the two lightcone directions ($t\pm x = {\textrm{const.}}$). The imposition of the JCs is then achieved quite simply in this setting by just evolving, in the left domain, the characteristic field propagating towards the right and relating it (via the JC) to the value of the characteristic field propagating towards the left in the right domain. For the interested reader, this kind of procedure is described in detail in Chapter 3 of [\cite{canizares_extreme-mass-ratio_2011}].

We could also consider the scalar self-force problem in a spinning (Kerr)
black hole spacetime, however the issue there---owing to the existence of fewer symmetries in the problem than in the non-spinning case---is that $\dim(\mathscr{I})=2$
in the time domain (with a more complicated second-order hyperbolic
operator $\mathcal{L}$); however, this could be remedied for a possible PwP implementation by passing to the
frequency domain, which transforms (\ref{eq:intro_general_pde}) to
an ODE (with $\dim(\mathscr{I})=1$, $\mathscr{V}=\emptyset$, and
again, a simple source $S=f\delta_{(p)}$).

The application of the PwP method to the full gravitational self-force
is a subject of ongoing work, however (modulo certain technical problems
relating to the gauge choice, which we will not elaborate upon here)
in the Schwarzschild-Droste case it essentially reduces to solving the same
type of problem (\ref{eq:intro_general_pde}) with $\dim(\mathscr{I})=1$
and $S=f\delta_{(p)}+g\delta'_{(p)}$. The equivalent problem in the
Kerr case once again suffers from the issue that $\dim(\mathscr{I})=2$
in the time domain, so the PwP method cannot be applied there except
after a transformation to the frequency domain (which produces $\dim(\mathscr{I})=1$
and $S=f\delta_{(p)}+g\delta'_{(p)}+h\delta''_{(p)}$ in this case).

Outside of gravitational physics, another setting where the PwP technique
could also possibly prove useful is in seismology. There, however,
the modeling of seismic waves [\cite{romanowicz_seismology_2007,aki_quantitative_2009,shearer_introduction_2009,madariaga_seismic_2007,petersson_stable_2010}] typically involves equations of the
form (\ref{eq:pwp-multivariable_source_Lu}) with $3$-dimensional
delta functions (\textit{i.e.} $\dim(\mathscr{I})=\bar{n}=3$, usually referring
to the $3$ dimensions of ordinary space) which, as we have amply
discussed in relation thereto, are not directly amenable to a PwP-type
approach as such. However, the methods outlined in this appendix might
be of some use if symmetries or other simplifying assumptions can,
in a situation of interest, reduce the dimension of the distributional
source to $1$ (as an alternative to delta function approximation
procedures, which are common practice in this area as well).

\section{Elliptic PDEs}
    \label{elliptic-pdes}

Finally, we consider in this section the elliptical problem appearing
in section 4.3 of  [\cite{tornberg_numerical_2004}]: namely, the Poisson equation on a square
of side length $2$ centered on the origin in $\mathbb{R}^{2}$, with
a simple (negative) one-dimensional delta function source supported
on the circle of radius $r_{*}=\tfrac{1}{2}$,
\begin{equation}
\begin{cases}
\triangle_{\mathbb{R}^{2}}u=-\delta\left(r-r_{*}\right)\,, & \text{on }\mathscr{U}=\left[-1,1\right]\times\left[-1,1\right]\subset\mathbb{R}^{2}\,,\\
u=1-\tfrac{1}{2}\log\left(2r\right)\,, & \text{on }\partial\mathscr{U}\,.
\end{cases}\label{eq:poisson_pde}
\end{equation}
In this case, the polar symmetry of the PDE entails that the solution
will only depend on the radial coordinate $r$ (which in this case
notationally substitutes the $x$ coordinate in antecedent sections).
Indeed, (\ref{eq:poisson_pde}) has an exact solution which is simply
given by
\begin{equation}
u_{\text{ex}}=1-\tfrac{1}{2}\log\left(2r\right)\Theta\left(r-r_{*}\right)\,.\label{eq:poisson_solution_exact}
\end{equation}

We can use the fact that in polar coordinates, $\triangle_{\mathbb{R}^{2}}=\partial_{r}^{2}+\tfrac{1}{r}\partial_{r}+\tfrac{1}{r^{2}}\partial_{\theta}^{2}$,
and so numerically all we need to do is solve $(\partial_{r}^{2}+\tfrac{1}{r}\partial_{r})u(r)=-\delta(r-r_{*})$
for a given $\theta\in[0,2\pi]$, where the value of $\theta$ will
determine $\{r\}=\mathscr{I}=[0,L]$ and hence the BC at $u(L)$ (that
is, on $\partial\mathscr{U}$), and repeat over some set of discrete
$\theta$ values in case the entire numerical solution on the $(r,\theta)$-plane
is desired.

Thus, we simply implement the PwP method here by writing $u=u^{-}\Theta^{-}+u^{+}\Theta^{+}$
for $\Theta^{\pm}=\Theta(r-r_{*})$, whereby we obtain the homogeneous
equations $(\partial_{r}^{2}+\tfrac{1}{r}\partial_{r})u^{\pm}=0$
along with the JCs $[u]_{*}=0$ and $[\partial_{r}u]_{*}=-1$.

\begin{figure}
\begin{center}
\includegraphics[scale=0.6]{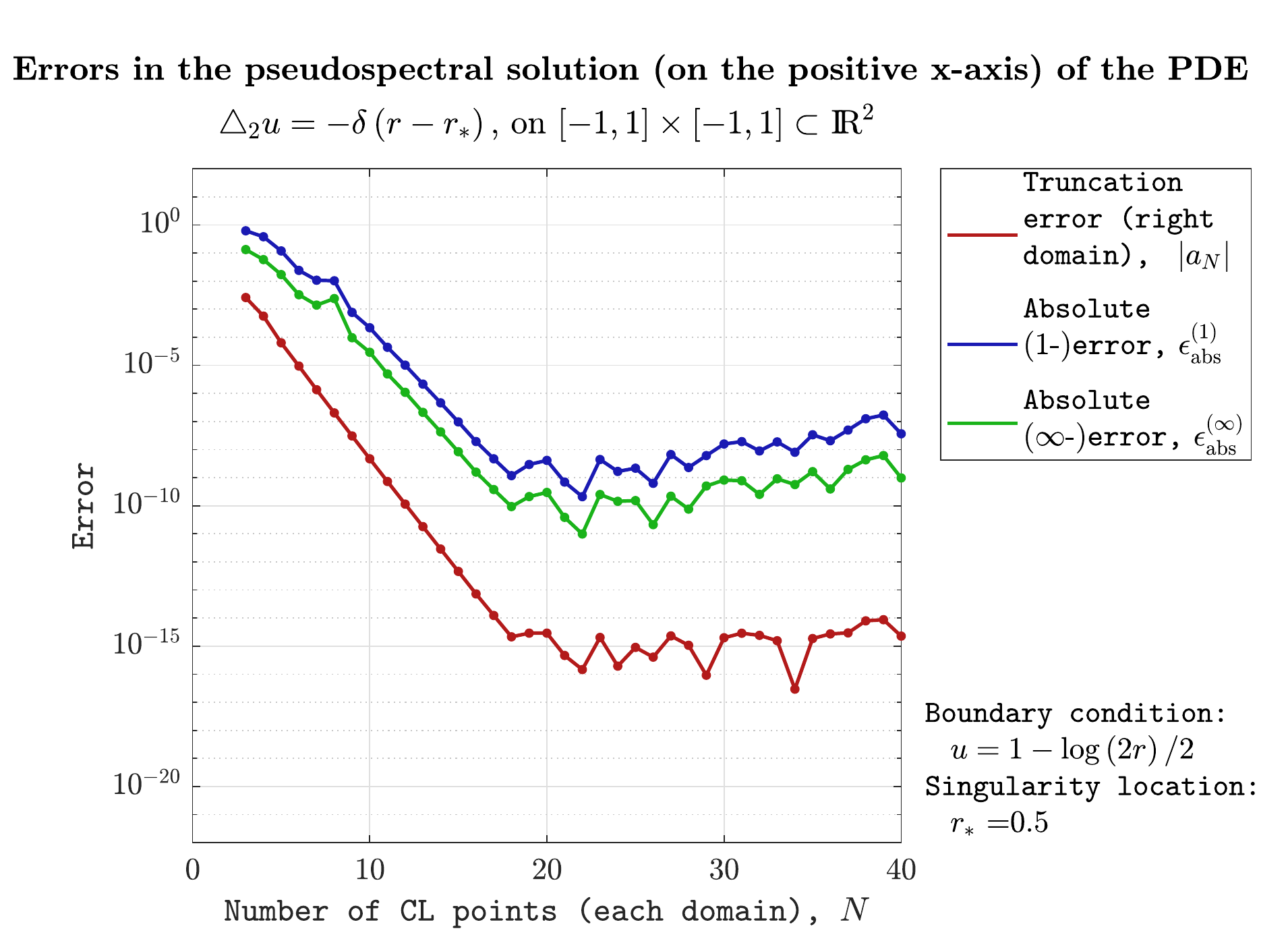}
\caption{Convergence of the pseudospectral numerical scheme for the problem (\ref{eq:poisson_pde}).}\label{fig:pde_pwp_poisson_error}
\end{center}
\end{figure}

The detailed numerical scheme is given in Subsection \ref{a-numerical-elliptic}, and results in
Figure \ref{fig:pde_pwp_poisson_error}. In this case, we simply plot the errors along the positive
$x$-axis in $\mathbb{R}^{2}$ on the CL grids: in addition to the
right-domain truncation error, we also show (as is done in [\cite{tornberg_numerical_2004}])
the absolute error in both the $l^{1}$-norm, $\epsilon_{\text{abs}}^{(1)}=||\bm{u}-\bm{u}_{\text{ex}}||_{1}$,
as well as in the infinity norm, $\epsilon_{\text{abs}}^{(\infty)}=||\bm{u}-\bm{u}_{\text{ex}}||_{\infty}$.
Up to $N\approx20$, we observe the typical (exponential) spectral
convergence of all three errors, with a significant (by a few orders
of magnitude) improvement over the results of  [\cite{tornberg_numerical_2004}] (using delta function
approximations) for the latter two.

\section{Conclusions}
    \label{conclusions}

We have expounded in this appendix a practical approach---the ``Particle-without-Particle'' (PwP) method---for numerically solving differential equations with distributional sources; to summarize, one does this by breaking up the solution into (regular function) pieces supported between---plus, if necessary, at---singularity (``particle'') locations, solving sourceless (homogeneous) problems for these pieces, and then matching them via the appropriate ``jump'' (boundary) conditions effectively substituting the original singular source. Building upon its successful prior application in the specific context of the self-force problem in general relativity, we have here generalized this method and have shown it to be viable for any linear partial differential equation of arbitrary order, with the provision that the distributional source is supported only on a one-dimensional subspace of the total problem domain. Accordingly, we have demonstrated its usefulness by solving first and second order hyperbolic problems, with applications in neuroscience and acoustics, respectively; parabolic problems, with applications in finance; and finally a simple elliptic problem. In particular, the numerical schemes we have employed for carrying these out have been based on pseudospectral collocation methods on Chebyshev-Lobatto grids. Generally speaking, our results have yielded varying degrees of improvement in the numerical convergence rates relative to other methods in the literature that have been attempted for solving these problems (typically relying on delta function approximation procedures on the computational grid).

We stress once more that the main limitations of the our PwP method as developed here are that it is only applicable to \emph{linear} problems with \emph{one}-dimensionally supported distributional sources. Thus, interesting lines of inquiry for future work might be to explore---however/if at all possible---extensions or adaptations of these ideas (a) to nonlinear PDEs, which would require working with nonlinear theories of distributions (having potential applicability to problems such as, \textit{e.g.}, the shallow-water equations with discontinuous bottom topography); (b) to more complicated sources than the sorts considered in this appendix, perhaps even containing higher-dimensional distributions but possibly also requiring additional assumptions, such as symmetries (which might be useful for problems such as, \textit{e.g.}, seismology models with three-dimensional delta function sources).


\section{Appendix: pseudospectral numerical schemes}
\label{a-psc}

\subsection{Pseudospecteal collocation methods}
\label{a-psc-gen}

We use this subsection to describe very cursorily the PSC methods used for the numerical schemes in this work and
to introduce some notation in relation thereto. For good detailed expositions see, for example, [\cite{boyd_chebyshev_2001,trefethen_spectral_2001,peyret_chebyshev_2002}].

We
work on Chebyshev-Lobatto (CL) computational grids. On any domain
$[a,b]=\mathscr{D}\subseteq\mathscr{I}$, these comprise the (\emph{non-uniformly}
spaced) set of $N$ points $\{X_{i}\}_{i=0}^{N}\subset\mathscr{D}$
obtained by projecting onto $\mathscr{D}$ those points located at
equal angles on a hypothetical semicircle having $\mathscr{D}$ as
its diameter. That is to say, the CL grid on the ``standard'' spectral
domain $\mathscr{D}^{\text{s}}=[-1,1]$ is given by
\begin{equation}
X_{i}^{\text{s}}=-\cos\left(\frac{\pi i}{N}\right)\,,\quad\forall0\leq i\leq N\,,
\end{equation}
which can straightforwardly be transformed (by shifting and stretching)
to the desired grid on $\mathscr{D}$. For any function $f:\mathscr{D}\rightarrow\mathbb{R}$
we denote via a subscript its value at the $i$-th CL point, $f(X_{i})=f_{i}$,
and in slanted boldface the vector containing all such values,
\begin{equation}
\bm{f}=\left[\begin{array}{c}
f_{0}\\
f_{1}\\
\vdots\\
f_{N}
\end{array}\right]\,.\label{eq:f_cl_grid}
\end{equation}
There exists an $(N+1)\times(N+1)$ matrix $\mathbb{D}$, the so-called
CL differentiation matrix, such that the derivative values of $f$
can be approximated simply by applying it to (\ref{eq:f_cl_grid}),
\textit{i.e.} $\bm{f}'=\mathbb{D}\bm{f}$. For convenience, we also employ
the notation $\mathbb{M}(r_{\text{i}}:r_{\text{f}},c_{\text{i}}:c_{\text{f}})$
to refer to the part of any matrix $\mathbb{M}$ from the $r_{\text{i}}$-th
to the $r_{\text{f}}$-th row and from the $c_{\text{i}}$-th to the
$c_{\text{f}}$-th column. (A simple ``$:$'' indicates taking all
rows/columns.)

\subsection{First-order hyperbolic PDEs}
\label{a-numerical-first-order-hyperbolic}

We apply a first order in time finite difference scheme to the homogeneous
PDEs; thus, prior to imposing BCs/JCs, the equations become $\frac{1}{\mathrm{\Delta} t}(\bm{u}_{k+1}^{\pm}-\bm{u}_{k}^{\pm})=-\mathbb{D}^{\pm}\bm{u}_{k}^{\pm}$,
where the vectors $\bm{u}_{k}^{\pm}$ contain the values of the solutions
on the CL grids at the $k$-th time step, $\mathbb{D}^{\pm}$ is the
CL differentiation matrix on the respective domains, and $\mathrm{\Delta} t$
is our time step. We can rewrite the discretized PDE as $\bm{u}_{k+1}^{\pm}=\bm{u}_{k}^{\pm}-\mathrm{\Delta} t\mathbb{D}^{\pm}\bm{u}_{k}^{\pm}=\bm{s}_{k}^{\pm}$.
To impose the BC and JC, we modify the equations as follows: 
\begin{equation}
\left[\begin{array}{c}
\bm{u}_{k+1}^{-}\\
\hline \bm{u}_{k+1}^{+}
\end{array}\right]=\left[\begin{array}{c}
u_{N,k}^{+}\\
\bm{s}_{k}^{-}(2:N+1)\\
\hline u_{N,k}^{-}+g_{k}\\
\bm{s}_{k}^{+}(2:N+1)
\end{array}\right].\label{eq:advection_psc}
\end{equation}

Similarly, for our neuroscience application, we discretize the PDE using a first order finite difference scheme:
$\frac{1}{\mathrm{\Delta} t}(\bm{\rho}_{k+1}^{\pm}-\bm{\rho}_{k}^{\pm})=-\mathbb{D}^{\pm}\bm{R}_{k}^{\pm}$
where $R_{i,k}^{\pm}=(1-V_{i}^{\pm})\rho_{i,k}^{\pm}$. Hence, prior
to imposing the BC/JC, we have $\bm{\rho}_{k+1}^{\pm}=\bm{\rho}_{k}^{\pm}-\mathrm{\Delta} t\mathbb{D}^{\pm}\bm{R}_{k}^{\pm}=\bm{s}_{k}^{\pm}$.
To impose the BC/JC, we just modify the equations accordingly:
\begin{equation}
\left[\begin{array}{c}
\bm{\rho}_{k+1}^{-}\\
\hline \bm{\rho}_{k+1}^{+}
\end{array}\right]=\left[\begin{array}{c}
0\\
\bm{s}_{k}^{-}(2:N+1)\\
\hline \rho_{N,k}^{-}+\frac{1-L}{1-V_{*}}\rho_{N,k}^{+}\\
\bm{s}_{k}^{+}(2:N+1)
\end{array}\right].\label{eq:neural_population_psc}
\end{equation}

\subsection{Parabolic PDEs}
\label{a-numerical-parabolic}

In these problems, we have moving boundaries for the CL grids (since
the location of the singular source is time-dependent). The mapping
for transforming the standard (fixed) spectral domain $[-1,1]$ into
an arbitrary (time-dependent) one, say $\mathscr{D}=[a(t),b(t)]$,
is given by
\begin{align}
\mathscr{V}\times\left[0,1\right]\rightarrow\, & \mathscr{V}\times\mathscr{D}\\
\left(T,X\right)\mapsto\, & \left(t\left(T\right),x\left(T,X\right)\right)\,,
\end{align}
where
\begin{align}
t\left(T\right)=\, & T\,,\\
x\left(T,X\right)=\, & \frac{b-a}{2}X+\frac{a+b}{2}\,.
\end{align}
For transforming back, we have
\begin{align}
\mathscr{V}\times\mathscr{D}\rightarrow\, & \mathscr{V}\times\left[0,1\right]\\
\left(t,x\right)\mapsto\, & \left(T\left(t\right),X\left(t,x\right)\right)\,,
\end{align}
where
\begin{align}
T\left(t\right)=\, & t\,,\label{eq:spectral_T(t)}\\
X\left(t,x\right)=\, & \frac{2x-a-b}{b-a}\,.\label{eq:spectral_X(t,x)}
\end{align}
Thus, for any function $f(t,x)$ in these problems, we must take care
to express the time partial using the chain rule as
\begin{align}
\frac{\partial f}{\partial t}=\, & \frac{\partial f}{\partial T}\frac{\partial T}{\partial t}+\frac{\partial f}{\partial X}\frac{\partial X}{\partial t}\\
=\, & \frac{\partial f}{\partial T}-\frac{2}{\left(b-a\right)^{2}}\left[\left(b-x\right)\dot{a}+\left(x-a\right)\dot{b}\right]\frac{\partial f}{\partial X}\,,\label{eq:spectral_partial_t_f}
\end{align}
where in the second line we have used (\ref{eq:spectral_T(t)})-(\ref{eq:spectral_X(t,x)}). 

Now, let us use this to formulate the numerical schemes for our problems---first,
for the heat equation. Let $\mathbb{D}_{k}^{\pm}$ denote the CL differentiation
matrices on each of the two domains at the $k$-th time step. Then,
using (\ref{eq:spectral_partial_t_f}), we have here the following
finite difference formula for the homogeneous PDEs prior to imposing
BCs/JCs: $\frac{1}{\mathrm{\Delta} t}(\bm{u}_{k+1}^{\pm}-\bm{u}_{k}^{\pm})=(\mathbb{D}_{k}^{\pm})^{2}\bm{u}_{k}^{\pm}-\mathbb{C}_{k}^{\pm}\mathbb{D}\bm{u}_{k}^{\pm}$,
where $\mathbb{D}$ is the CL differentiation matrix on $[-1,1]$
and $\mathbb{C}_{k}^{-}={\rm diag}([2/(x_{p}(t_{k}))^{2}][(-x_{i}^{-})\dot{x}_{p}(t_{k})])$,
$\mathbb{C}_{k}^{+}={\rm diag}([2/(1-x_{p}(t_{k}))^{2}][(x_{i}^{+}-1)\dot{x}_{p}(t_{k})])$.
Thus $\bm{u}_{k+1}^{\pm}=\bm{u}_{k}^{\pm}+\mathrm{\Delta} t[(\mathbb{D}_{k}^{\pm})^{2}-\mathbb{C}_{k}^{\pm}\mathbb{D}]\bm{u}_{k}^{\pm}=\bm{s}_{k}^{\pm}$.
We can implement the BCs and JCs, by modifying the first and last
equations on each domain: 
\begin{equation}
\left[\begin{array}{cccc|cccc}
1 & 0 & \cdots & 0 & 0\\
0 & 1 & \cdots & 0 &  & 0\\
\vdots & \vdots & \ddots & \vdots &  &  & \ddots\\
0 & 0 & \cdots & 1 &  &  &  & 0\\
\hline 0 &  &  &  &  &  & \mathbb{D}_{k}^{+}(1,:)\\
 & 0 &  &  & 0 & 1 & \cdots & 0\\
 &  & \ddots &  & \vdots & \vdots & \ddots & \vdots\\
 &  &  & 0 & 0 & 0 & \cdots & 1
\end{array}\right]\left[\begin{array}{c}
\\
\bm{u}_{k+1}^{-}\\
\\
\hline \\
\bm{u}_{k+1}^{+}\\
\\
\end{array}\right]=\left[\begin{array}{c}
0\\
\bm{s}_{k}^{-}(2:N)\\
u_{0,k}^{+}\\
\hline \mathbb{D}_{k}^{-}(N,:)\bm{u}_{k}^{-}-\lambda\\
\bm{s}_{k}^{+}(2:N)\\
0
\end{array}\right].\label{eq:heat_psc}
\end{equation}
Note that we are actually introducing an error by using (for convenience
and ease of adaptability) $\mathbb{D}_{k}^{+}$ instead of $\mathbb{D}_{k+1}^{+}$
on the LHS (in the equation for $u_{0,k+1}^{+}$). However, one can
easily convince oneself that $\mathbb{D}_{k+1}^{+}-\mathbb{D}_{k}^{+}=\mathcal{O}(\mathrm{\Delta} t)$,
which is already the order of the error of the finite difference scheme,
so we are not actually introducing any new error in this way. Furthermore,
because we use up the last equation for $\bm{u}_{k}^{-}$ to impose
the JC on $u$ (\textit{i.e.} we do not have an equation for $u_{N,k}^{-}$),
we must use the derivative at the previous point (\textit{i.e.}, at $u_{N-1,k}^{-}$)
in order to impose the derivative JC. Hence on the RHS, we use $\mathbb{D}_{k}^{-}(N,:)$
instead of $\mathbb{D}_{k}^{-}(N+1,:)$.

The scheme for the finance model is analogous. We use again the first-order
finite-difference method for the homogeneous equations, $\frac{1}{\mathrm{\Delta} t}(\bm{f}_{k+1}^{\sigma}-\bm{f}_{k}^{\sigma})=(\mathbb{D}_{k}^{\sigma})^{2}\bm{f}_{k}^{\sigma}-\mathbb{C}_{k}^{\sigma}\mathbb{D}\bm{f}_{k}^{\sigma}$
with the matrices $\mathbb{C}_{k}^{\sigma}$ defined similarly to
those in the heat equation problem (again using (\ref{eq:spectral_partial_t_f}));
thus $\bm{f}_{k+1}^{\sigma}=\bm{f}_{k}^{\sigma}+\mathrm{\Delta} t[(\mathbb{D}_{k}^{\sigma})^{2}-\mathbb{C}_{k}^{\sigma}\mathbb{D}]\bm{f}_{k}^{\sigma}=\bm{s}_{k}^{\sigma}$.
To impose the BCs/JCs, we modify the equations appropriately: 
\begin{align}
\left[\begin{array}{ccccc}
 &  & \mathbb{D}_{k}^{-}(1,:)\\
0 & 1 & \cdots & 0 & 0\\
\vdots & \vdots & \ddots & \vdots & \vdots\\
0 & 0 & \cdots & 1 & 0\\
0 & 0 & \cdots & 0 & 1
\end{array}\right]\left[\begin{array}{c}
\\
\\
\bm{f}_{k+1}^{-}\\
\\
\\
\end{array}\right] & =\left[\begin{array}{c}
0\\
s_{1,k}^{-}\\
\vdots\\
s_{N,k}^{-}\\
f_{0,k}^{0}
\end{array}\right],\label{eq:finance_psc_minus}\\
\left[\begin{array}{ccccc}
 &  & \mathbb{D}_{k}^{0}(1,:)\\
0 & 1 & \cdots & 0 & 0\\
\vdots & \vdots & \ddots & \vdots & \vdots\\
0 & 0 & \cdots & 1 & 0\\
0 & 0 & \cdots & 0 & 1
\end{array}\right]\left[\begin{array}{c}
\\
\\
\bm{f}_{k+1}^{0}\\
\\
\\
\end{array}\right] & =\left[\begin{array}{c}
\mathbb{D}_{k}^{-}(N,:)\bm{f}_{k}^{-}-\lambda_{k}\\
s_{1,k}^{0}\\
\vdots\\
s_{N,k}^{0}\\
f_{0,k}^{+}
\end{array}\right],\label{eq:finance_psc_zero}\\
\left[\begin{array}{ccccc}
 &  & \mathbb{D}_{k}^{+}(1,:)\\
0 & 1 & \cdots & 0 & 0\\
\vdots & \vdots & \ddots & \vdots & \vdots\\
0 & 0 & \cdots & 1 & 0\\
 &  & \mathbb{D}_{k}^{+}(N+1,:)
\end{array}\right]\left[\begin{array}{c}
\\
\\
\bm{f}_{k+1}^{+}\\
\\
\\
\end{array}\right] & =\left[\begin{array}{c}
\mathbb{D}_{k}^{0}(N,:)\bm{f}_{k}^{0}+\lambda_{k}\\
s_{1,k}^{+}\\
\vdots\\
s_{N,k}^{+}\\
0
\end{array}\right].\label{eq:finance_psc_plus}
\end{align}

\subsection{Second-order hyperbolic PDEs}
\label{a-numerical-second-order-hyperbolic}

We again apply a first order in time finite difference scheme to the
homogeneous PDEs; prior to imposing BCs/JCs, the equations become
\begin{equation}
\frac{1}{\mathrm{\Delta} t}\left(\left[\begin{array}{c}
\bm{u}_{k+1}^{\pm}\\
\bm{v}_{k+1}^{\pm}\\
\bm{w}_{k+1}^{\pm}
\end{array}\right]-\left[\begin{array}{c}
\bm{u}_{k}^{\pm}\\
\bm{v}_{k}^{\pm}\\
\bm{w}_{k}^{\pm}
\end{array}\right]\right)=\mathbb{C}^{\pm}\left[\begin{array}{c}
\bm{u}_{k}^{\pm}\\
\bm{v}_{k}^{\pm}\\
\bm{w}_{k}^{\pm}
\end{array}\right]\,,
\end{equation}
where
\begin{equation}
\mathbb{C}^{\pm}=\left[\begin{array}{ccc}
0 & 0 & 0\\
0 & 0 & \mathbb{I}\\
0 & \mathbb{I} & 0
\end{array}\right]\left[\begin{array}{ccc}
\mathbb{D}^{\pm} & 0 & 0\\
0 & \mathbb{D}^{\pm} & 0\\
0 & 0 & \mathbb{D}^{\pm}
\end{array}\right]+\left[\begin{array}{ccc}
0 & 0 & \mathbb{I}\\
0 & 0 & 0\\
0 & 0 & 0
\end{array}\right]=\left[\begin{array}{ccc}
0 & 0 & \mathbb{I}\\
0 & 0 & \mathbb{D}^{\pm}\\
0 & \mathbb{D}^{\pm} & 0
\end{array}\right]\,.
\end{equation}
We can rewrite the discretized PDE as 
\begin{equation}
\left[\begin{array}{c}
\bm{u}_{k+1}^{\pm}\\
\bm{v}_{k+1}^{\pm}\\
\bm{w}_{k+1}^{\pm}
\end{array}\right]=\left(\mathrm{\Delta} t\mathbb{C}^{\pm}+\mathbb{I}\right)\left[\begin{array}{c}
\bm{u}_{k}^{\pm}\\
\bm{v}_{k}^{\pm}\\
\bm{w}_{k}^{\pm}
\end{array}\right]=\left[\begin{array}{c}
\bm{s}_{k}^{\pm}\\
\bm{y}_{k}^{\pm}\\
\bm{z}_{k}^{\pm}
\end{array}\right]\,.
\end{equation}
To impose the BCs and JCs, we modify the equations as follows: 
\begin{align}
\left[\begin{array}{cccc}
1 & \cdots & 0 & 0\\
\vdots & \ddots & \vdots & 0\\
0 & \cdots & 1 & 0\\
 & \mathbb{D}^{-}(N+1,:)
\end{array}\right]\left[\begin{array}{c}
\\
\bm{u}_{k+1}^{-}\\
\\
\end{array}\right]=\, & \left[\begin{array}{c}
0\\
\bm{s}_{k}^{-}(2:N)\\
\mathbb{D}^{+}(1,:)\bm{u}_{k}^{+}
\end{array}\right]\,,\\
\left[\begin{array}{c}
\\
\bm{u}_{k+1}^{+}\\
\\
\end{array}\right]=\, & \left[\begin{array}{c}
u_{N-1,k+1}^{-}-g_{k}\\
\bm{s}_{k}^{+}(2:N)\\
0
\end{array}\right]\,,\\
\left[\begin{array}{c}
\bm{v}_{k+1}^{-}\\
\hline \bm{v}_{k+1}^{+}
\end{array}\right]=\, & \left[\begin{array}{c}
\mathbb{D}^{-}\bm{u}_{k+1}^{-}\\
\hline \mathbb{D}^{+}\bm{u}_{k+1}^{+}
\end{array}\right]\,,\\
\left[\begin{array}{c}
\bm{w}_{k+1}^{-}\\
\hline \bm{w}_{k+1}^{+}
\end{array}\right]=\, & \left[\begin{array}{c}
0\\
\bm{z}_{k}^{-}(2:N+1)\\
\hline w_{N,k+1}^{-}-\dot{g}_{k+1}\\
\bm{z}_{k}^{+}(2:N)\\
0
\end{array}\right]\,,
\end{align}

\subsection{Elliptic PDEs}
\label{a-numerical-elliptic}

In this case we have no time evolution, and we simply need to solve
$((\mathbb{D}^{\pm})^{2}+{\rm diag}(1/X_{i}^{\pm})\mathbb{D}^{\pm})\bm{u}^{\pm}=\mathbb{M}^{\pm}\bm{u}^{\pm}={\bf 0}$,
modified appropriately to account for the BCs and JCs. In particular,
we first solve for $\bm{u}^{+}$ using the BCs, and then for $\bm{u}^{-}$
using the solution for $\bm{u}^{+}$ to implement the JCs:
\begin{align}
\left[\begin{array}{c}
\mathbb{M}^{+}(1:N-1,:)\\
0\enskip0\cdots0\enskip1\\
\mathbb{D}^{+}(N+1,:)
\end{array}\right]\bm{u}^{+}=\, & \left[\begin{array}{c}
{\bf 0}(1:N-1)\\
1-\tfrac{1}{2}\log(2L)\\
-\tfrac{1}{2L}
\end{array}\right]\,,\\
\left[\begin{array}{c}
\mathbb{M}^{-}(1:N-1,:)\\
0\enskip0\cdots0\enskip1\\
\mathbb{D}^{-}(N+1,:)
\end{array}\right]\bm{u}^{-}=\, & \left[\begin{array}{c}
{\bf 0}(1:N-1)\\
u_{0}^{+}\\
\mathbb{D}^{+}(1,:)\bm{u}^{+}+1
\end{array}\right]\,.
\end{align}


\end{appendix}

\fancyhead[RO,LE]{\rule[-1ex]{0pt}{1ex} \fontsize{11}{12}\selectfont \thepage}
\fancyhead[RE]{\fontsize{11}{12}\sl\selectfont\nouppercase Bibliography}
\fancyhead[LO]{\fontsize{11}{12}\sl\selectfont\nouppercase \rightmark}

\printbibliography

\end{document}